\documentclass[a4paper,fleqn,usenatbib]{mnras}

\usepackage{newtxtext,newtxmath}

\usepackage[T1]{fontenc}
\usepackage{ae,aecompl}

\usepackage{graphicx}	
\usepackage{amsmath}	
\usepackage[ugly]{nicefrac}
\usepackage{pdfpages}
\usepackage{longtable}
\usepackage[maxfloats=256]{morefloats}
\usepackage{subfig}
\usepackage{caption}
\usepackage{pdflscape}
\newcommand{\angstrom}{\textup{\AA}}

\title[MASCOT]{MASCOT - An ESO-ARO legacy survey of molecular gas in nearby SDSS-MaNGA galaxies: I. first data release, and global and resolved relations between H$_{2}$ and stellar content}

\author[D. Wylezalek]{D. Wylezalek$^{1}$\thanks{E-mail:dominika.wylezalek@uni-heidelberg.de}, 
C. Cicone$^{2}$,
F. Belfiore$^{3}$,
C. Bertemes$^{1}$,
S. Cazzoli$^{4}$,
J. Wagg$^{5}$,  \newauthor
W. Wang$^{1}$,
M. Aravena$^{6}$, 
R. Maiolino$^{7, 8}$,
S. Martin$^{9, 10}$, 
M.S. Bothwell$^{7, 8}$,
J.R. Brownstein$^{11}$, \newauthor
K. Bundy$^{12, 13}$,
C. De Breuck$^{14}$
\\ 
$^{1}$Zentrum f\"{u}r Astronomie der Universit\"{a}t Heidelberg, Astronomisches Rechen-Institut, M\"{o}nchhofstr, 12-14 69120 Heidelberg, Germany\\
$^{2}$ Institute of Theoretical Astrophysics, University of Oslo, PO Box 1029, Blindern 0315, Oslo, Norway \\
$^{3}$ INAF - Osservatorio Astrofisico di Arcetri, Largo Enrico Fermi 5, I-50125 Firenze, Italy \\
$^{4}$ IAA - Instituto de Astrofísica de Andalucía (CSIC), Apdo. 3004, E-18008 Granada, Spain\\
$^{5}$ SKA Observatory, Lower Withington Macclesfield, Cheshire SK11 9FT, UK\\
$^{6}$ Núcleo de Astronomía, Facultad de Ingeniería y Ciencias, Universidad Diego Portales, Av. Ejército 441, Santiago, Chile\\
$^{7}$ Cavendish Laboratory, University of Cambridge, Cambridge CB3 0HE, UK\\
$^{8}$ Kavli Institute for Cosmology, University of Cambridge, Cambridge CB3 0HE, UK\\
$^{9}$ European Southern Observatory, Alonso de Córdova, 3107, Vitacura, Santiago, 763-0355, Chile \\
$^{10}$ Joint ALMA Observatory, Alonso de Córdova, 3107, Vitacura, Santiago, 763-0355, Chile\\
$^{11}$ University of Utah, Department of Physics and Astronomy, 115 S. 1400 E., Salt Lake City, UT 84112, USA\\
$^{12}$ Department of Astronomy and Astrophysics, University of California, 1156 High Street, Santa Cruz, CA 95064, USA \\
$^{13}$ UCO/Lick Observatory, Department of Astronomy and Astrophysics, University of California, 1156 High Street, Santa Cruz, CA 95064, US \\
$^{14}$ European Southern Observatory, Karl-Schwarzschild-Str. 2, D-85748 Garching, Germany
}
\date{Accepted XXX. Received YYY; in original form ZZZ}

\pubyear{2017}

\begin{document}
\label{firstpage}
\pagerange{\pageref{firstpage}--\pageref{lastpage}}
\maketitle

\begin{abstract}
We present the first data release of the MaNGA-ARO Survey of CO Targets (MASCOT), an ESO Public Spectroscopic Survey conducted at the Arizona Radio Observatory (ARO). We measure the CO(1-0) line emission in a sample of 187 nearby galaxies selected from the Mapping Nearby Galaxies at Apache Point Observatory (MaNGA) survey that has obtained integral field unit (IFU) spectroscopy for a sample of $\sim$ 10,000 galaxies at low redshift. The main goal of MASCOT is to probe the molecular gas content of star-forming galaxies with stellar masses $> 10^{9.5}$~M$_{\odot}$ and with associated MaNGA IFU observations and well-constrained quantities like stellar masses, star formation rates and metallicities. In this paper we present the first results of the MASCOT survey, providing integrated CO(1-0) measurements that cover several effective radii of the galaxy and present CO luminosities, CO kinematics, and estimated H$_2$ gas masses. We observe that the decline of galaxy star formation rate with respect to the star formation main sequence (SFMS) increases with the decrease of molecular gas and with a reduced star formation efficiency, in agreement with results of other integrated studies. Relating the molecular gas mass fractions with the slope of the stellar age gradients inferred from the MaNGA observations, we find that galaxies with lower molecular gas mass fractions tend to show older stellar populations close to the galactic center, while the opposite is true for galaxies with higher molecular gas mass fractions, providing tentative evidence for inside-out quenching. 

\end{abstract}




\section{Introduction}

Understanding the behaviour of the cold phase of the interstellar medium (ISM) is central to understanding the galaxy evolution process as a whole. Galaxies exist in a state of flux, being subject to a range of physical processes (including accretion, gas outflow, and star formation) which are the drivers of their evolution \citep{Tumlinson_2017, Peroux_2020}. For example, star formation is regulated by the amount of available gas and internal feedback processes \citep{kenn98b, harr17, Rupke_2018}. The tight relation between molecular gas and star formation has been studied for decades \citep{Sanders_1985, Solomon_1997, Combes_2007, Young_2011, Saintonge_2011, Saintonge_2017, Lin_2019, Colombo_2020}, and has come to represent a central pillar of our understanding of galaxy evolution as well as a critical ingredient in models \citep{Bouche_2010, Lilly_2013, Saintonge_2016}. However, disentangling the detailed physics underlying this correlation remains an important area of research. As such, attaining a better understanding of the gas content of galaxies, in particular the molecular gas content, which is the fuel for star formation, is key to understanding how galaxies evolve. 

Carbon Monoxide, $^{12}$C$^{16}$O, is the most widely used tracer of molecular gas, both in the local and in the high-$z$ Universe. The luminosities of the low-$J$ CO transitions, in particular the lowest energy CO($1-0$) emission line (rest frequency $\nu = 115.271$ GHz), can be used to measure the mass of molecular gas, modulo a conversion factor $\rm{\alpha_{CO}}$ \citep[for a review, see][]{Bolatto_2013}. Several large surveys have therefore targeted the low-$J$ transitions of CO to measure the probe gas content in the past decade. The COLD GASS survey \citep[CO Legacy Database for GASS;][]{Saintonge_2011} used the IRAM 30m telescope to measure the CO(1-0) line in a sample of $\sim $350 nearby (D=100-200 Mpc), massive galaxies with $\log (M_{*}/M_{\odot}) >10.0$ selected from the Sloan Digital Sky Survey (SDSS). The survey was later expanded to also include lower-mass galaxies \citep[xCOLDGASS;][]{Saintonge_2017}, bringing the combined surveys to a sample size of 532 galaxies probing the entire star formation rate (SFR) - stellar mass plane above $\log (M_{*}/M_{\odot}) > 9.0$. Furthermore, the ALLSMOG survey \citep[APEX low-redshift legacy survey for molecular gas;][]{Bothwell_2014, Cicone_2017} complements the xCOLDGASS survey by providing CO(2-1) emission line observations of 88 nearby, low-mass ($8.5< \log (M_{*}/M_{\odot}) < 10$) galaxies taken with the APEX-1 receiver on the APEX telescope. These surveys revealed that it is both the molecular gas mass fraction, i.e. the molecular gas to stellar mass ratio $f_{\rm{H_2}}= M_{H_2}/M_{*}$, as well as the star formation efficiency (SFE = SFR / M$_{H_2}$) that vary strongly as a function of specific star formation rate and determine a galaxy's offset from the star forming main sequence. The CO luminosity appears to correlate not only strongly with stellar mass and SFR but also varies with metallicity and HI mass. 

These surveys were able to explore the connection between the global gas content of galaxies where galaxy parameters had been obtained via single-fibre SDSS spectroscopy (and SDSS photometry). That means all information about stellar mass, star formation rate, central AGN activity, concentration parameter, metallicity was therefore restricted to a central 3$\arcsec$ aperture probing very different spatial scales depending on the redshift of the source. Integral field unit (IFU) surveys now offer new possibilities in investigating the spatial dimension of galaxy evolution and allow these quantities to be mapped in two dimensions \citep{Cappellari_2011, Sanchez_2012, Croom_2012, Bundy_2015}. 

The SDSS-IV \citep{Blanton_2017} survey Mapping Nearby Galaxies at APO \citep[MaNGA;][]{Bundy_2015, Drory_2015, Law_2015, Yan_2016a, Yan_2016b, Wake_2017} is an optical fibre-bundle IFU survey that has mapped the detailed composition and kinematic structure of 10,010 unique nearby galaxies at $0.01 < z < 0.15$. The current public data release, DR15 \citep{Aguado_2019}, includes data for 4824 MaNGA data cubes. The full MaNGA data set will be released as part of DR17, expected in December 2021. MaNGA delivers resolved optical spectroscopic data. Many critical diagnostics, which provide insight into the formation processes of galaxies, such as metallicity gradients, age gradients, and resolved AGN diagnostics, and are only available to such resolved observations. However, the cold, molecular gas phase is not probed by MaNGA observations. 

We started to fill in this gap with the MaNGA Survey of CO Targets (MASCOT, Figure 1). MASCOT was initially granted 200 hours on the Arizona Radio Observatory (ARO) in 2018 to obtain molecular gas mass measurements of MaNGA galaxies via the $^{12}$CO(1-0) transition. After that first successful set of observations, MASCOT was granted another 1200 hours with the ARO. A large survey for molecular gas in galaxies with existing IFU data represents a uniquely powerful tool for addressing a host of science questions and opens new avenues for molecular gas studies by drawing targets from the current generation of SDSS surveys. 

The MASCOT survey complements other recent spatially-resolved observational programs in the community. The EDGE (Extragalactic Database for Galaxy Evolution) survey targeted 177 infrared-bright CALIFA (the Calar Alto Integral Field Area survey) galaxies using the Combined Array for Millimeter-wave Astronomy (CARMA). The survey targeted the $^{12}$CO(1-0) line and its $^{13}$CO isotopologue providing spatially resolved CO maps with a resolution corresponding to size scales of 0.5-2 kpc matched to the CALIFA observations \citep{Bolatto_2017}. The maps have a half-power field-of-view (FOV) with radius $\sim 50 \arcsec$, covering the galaxies out to $\sim 2-3$ effective radii (R$_{\rm{eff}}$). This dataset enables studies of the relationships between molecular gas and its kinematics, stellar mass, star formation rate, metallicity, and dust extinction on kpc scales \citep{Utomo_2017, Colombo_2018, Leung_2018, Levy_2018, Dey_2019, Levy_2019, Barrera_2020, Barrera_2021}. These observations are currently being complemented by $^{12}$CO(2-1) measurements of CALIFA galaxies using the APEX 12m sub-millimetre telescope \citep{Guesten_2006}. Compared to MASCOT, the CALIFA-APEX observations target galaxies at lower redshift ($0.005 < z < 0.03$) and provide less coverage. The first set of combined CARMA and APEX observations of CALIFA galaxies comprises 472 sources \citep{Colombo_2020}. The APEX CO(2-1) beam covers the galaxies roughly out to 1 R$_{\rm{eff}}$. By dividing the sample into galaxies that are centrally quenched and star forming galaxies, first results suggest that once star formation has been significantly reduced due to the consumption of molecular gas, changes in the star formation efficiency are what drives a galaxy deeper into the red sequence. 

On the other hand, the ALMaQUEST (ALMA-MaNGA QUEnching and STar formation) survey is a program with spatially-resolved 12CO(1-0) measurements obtained with the Atacama Large Millimeter Array (ALMA) for 46 galaxies from MaNGA DR15 \citep{Lin_2020}. The aim of the ALMaQUEST survey is to investigate the dependence of star formation activity on the cold molecular gas content at kpc scales in nearby galaxies and targets not only main-sequence galaxies, but also starburst and green valley galaxies \citep{Lin_2019,  Ellison_2020a, Ellison_2020b, Lin_2020, Ellison_2021a, Ellison_2021b}. The survey spatially resolves the CO(1-0) line on scales matching the MaNGA resolution and its field-of-view is $\sim$ 50'', very similar to the MASCOT observations. The ALMaQUEST survey is very complementary to MASCOT and we include their spatially integrated CO measurements in this paper (for details, see Section 3.8). Together, the EDGE-CALIFA and ALMaQUEST surveys provide complementary strengths for studying gas and star formation with supporting optical IFU data on kpc-scales in the nearby Universe.

In this paper we present the first data release of the MASCOT survey of molecular gas mass measurements for 187 galaxies selected from the MaNGA survey. The paper is structured as follows: In Section 2 we summarise the available MaNGA data products and additional ancillary data for the galaxies. In Section 3 we present details on the MASCOT survey, including the sample selection, details on the observations with the ARO, data reduction and analysis. Section 4 investigates some first global and resolved relations between the molecular gas mass measurements and the MaNGA-derived galaxy properties and in Section 5 we conclude. For the derived quantities we assume a flat $\Lambda$CDM cosmology with H$_0$ = 70 km s$^{-1}$ Mpc$^{-1}$, $\Omega_{\rm{m}}$ =0.3,  $\Omega_{\Lambda}$ =0.7.  We furthermore use the redshifts published in the NASA Sloan Atlas catalog \citep[NSA catalog\footnote{http://nsatlas.org},][]{Blanton_2011}. \\

\section{SDSS MaNGA}

\subsection{Survey Description}

Mapping Nearby Galaxies at Apache Point Observatory (MaNGA) is a two-dimensional spectroscopic survey and is part of the Sloan Digital Sky Survey-IV (SDSS-IV). MaNGA \citep{Yan_2016b} uses the BOSS spectrographs \citep{Gunn_2006, smee13} with $R \sim 2000$ to take integral field unit (IFU) observations of each galaxy in the 3,600 -- 10,000\angstrom\ range. Fibers are arranged into hexagonal bundles. The bundles have sizes that range from 19 -- 127 fibres, depending on the apparent size of the target galaxy (which corresponds to diameters ranging between 12\arcsec\ to 32\arcsec). This leads to an average footprint of $400-500$~arcsec$^{2}$ per IFU. The fibres have a size of 2\arcsec\ aperture (2.5\arcsec\ separation between fibre centres), which corresponds to $\sim 2$ kpc at $z\sim 0.05$, although with dithering the effective sampling improves to $1.4$\arcsec \citep[see also][]{wyle18}. The current data release DR15 \citep{dr15_2019} contains 4,688 MaNGA galaxies (including ancillary targets and $\sim$ 65 repeat observations). MaNGA has observed 10,010 galaxies at $z\la 0.15$ and with stellar masses $>10^9 M_{\odot}$ and the entire data set will become public as part of DR17 (currently scheduled for December 2021). 

\subsection{MaNGA Data Products}

The data is first fed through the MaNGA Data Reduction Pipeline (DRP) which produces sky-subtracted spectrophotometrically calibrated spectra and rectified three-dimensional data cubes \citep{wyle18}. These combine individual dithered observations \citep[for details on MaNGA data reduction see][]{Law_2016} with a spatial pixel scale of 0.5~\arcsec\ pixel$^{-1}$. The median spatial resolution of the MaNGA data is 2.54~\arcsec\ FWHM while the median spectral resolution is $\sim 72$ km/s \citep{Law_2016, Law_2021}. 

The MaNGA Data Analysis Pipeline \citep[DAP, ][]{Belfiore_2019, Westfall_2019} is a project-led software package and is used to analyse the data products produced by the MaNGA DRP. The analysis results of the DAP provide the collaboration and public with survey-level quantities, such as spectral indices, kinematics and emission-line properties for 21 different emission lines. To make these calculations, the DAP first fits the stellar continuum using the Penalized Pixel-Fitting method \citep[pPXF, ][]{Cappellari_2004, Cappellari_2017} and then performs a second fitting stage for the emission lines which optimises simultaneously the continuum and the emission lines, which are added as templates in a pPXF call. For example, for DR15, the DAP provides spatially stacked spectra, stellar kinematics (V and $\sigma$) derived on a set of Voronoi-binned spectra, spaxel-based nebular emission-line properties including fluxes, equivalent widths, and kinematics (V and $\sigma$) and spectral indices from absorption-lines (e.g., H$\delta$) and bandhead (e.g., D4000) measurements. 

In this paper, we also make use of the Value Added Catalog `MaNGA Pipe3D'  \citep{Sanchez18} which is based on measurements performed with the Pipe3D pipeline \citep{Sanchez_2016}. The pipeline is designed to fit the continuum with stellar population models and to measure the nebular emission lines of IFS data. The stellar population library uses the \citep{Salpeter_1955} initial mass function. The stellar mass is then obtained based on the best-fit stellar population model.
The stellar-population model spectra are then subtracted from the original cube allowing the analysis of the ionised gas emission lines \citep{Sanchez_2016}. For example, the star formation rate was derived by using the H$\alpha$ luminosities for all the spaxels with detected ionised gas following the conversion given by \citet{kenn98} as well as from the single stellar population (SSP) fits computing the amount of stellar mass formed in the last 32Myr \footnote{Link to the datamodel of the Pipe3D catalog: \href{https://data.sdss.org/datamodel/files/MANGA_PIPE3D/MANGADRP_VER/PIPE3D_VER/manga.Pipe3D.html}{https://data.sdss.org/datamodel/files/MANGA\_PIPE3D/MANGADRP\_VER/ \\ PIPE3D\_VER/manga.Pipe3D.html}}. The Value Added Products include both a set of datacubes, containing spatially resolved stellar population properties, star formation histories, emission line fluxes and stellar absorption line indices derived using the Pipe3D pipeline, as well as a catalog with one entry per galaxy, comprising the integrated properties of those galaxies (e.g., stellar mass, star-formation rate), and the characteristic values (e.g., oxygen abundance at the effective radius, all including the associated uncertainties \footnote{Link to the Pipe3D Value Added Catalogs: \href{https://www.sdss.org/dr14/manga/manga-data/manga-pipe3d-value-added-catalog/}{https://www.sdss.org/dr14/manga/manga-data/manga-pipe3d-value-added-catalog/}}. In this paper we make use of the integrated stellar masses and SSP-based SFRs.

We also point out to the reader that a number of additional `Value Added Catalogs' publicly released by the SDSS collaboration. These catalogs have been created by MaNGA collaborators and are distributed through the SDSS Science Archive Server. In particular, we point out the `HI-MaNGA Data Release 1' catalog which presents the first catalog of HI (21 cm neutral hydrogen) follow-up for MaNGA galaxies \citep{Masters_2019}. 

In this work, we furthermore make use of Marvin, a tool specifically designed to visualise and analyse MaNGA data. It is developed and maintained by the MaNGA team. Among other tools, Marvin allows the user to access reduced MaNGA datacubes locally, remotely, or via a web interface, access and visualise data analysis products and perform powerful queries on data and metadata \footnote{see also the Marvin Documentation: \href{https://sdss-marvin.readthedocs.io/en/latest/}{https://sdss-marvin.readthedocs.io/en/latest/}}. Details about Marvin are described in \citet{Cherinka_2019} and on the \href{https://dr15.sdss.org/marvin}{Marvin website}.

\section{MASCOT Sample and Observations}

\subsection{Survey Description}

The MASCOT sample is being assembled over the course of two programs at the Arizona Radio Observatory (ARO) as part of the MASCOT 1.0 and MASCOT 2.0 programs (PI: Wylezalek). As part of the agreement for transferring ownership of the ALMA prototype antenna to the University of Arizona to install it on Kitt Peak as a telescope of the ARO, ESO distributed to its user community a total of 3600h of observing time on the ARO telescopes from 2015 to 2020 \footnote{http://www.eso.org/sci/activities/call\_for\_public\_surveys.html}. It was decided that this time should be dedicated to large Public Surveys with an important legacy value. MASCOT was chosen to be one of these surveys. 

The first part of the survey, MASCOT 1.0, was awarded 200 hours of observing time in 2018 and the survey targeted 34 sources selected from the MaNGA survey. For the sample selection, we used the then available Data Release 14 \citep[DR14, ][]{dr14_2017} which contained 2778 galaxies at $0.01 < z < 0.15$ with a mean $z \sim 0.05$.

The second part of the survey, MASCOT 2.0, was awarded 1200 hours spread over four semesters in 2019 and 2020 (P103-P106). Due to the COVID-related shutdown of the telescope in 2020, the timeline of the execution of the observations has been significantly delayed. Sources have been selected from the current MaNGA data release DR15.

\begin{figure*}
\includegraphics[ width=0.9\textwidth, trim = 4cm 2cm 5cm 2.5cm, clip= true]{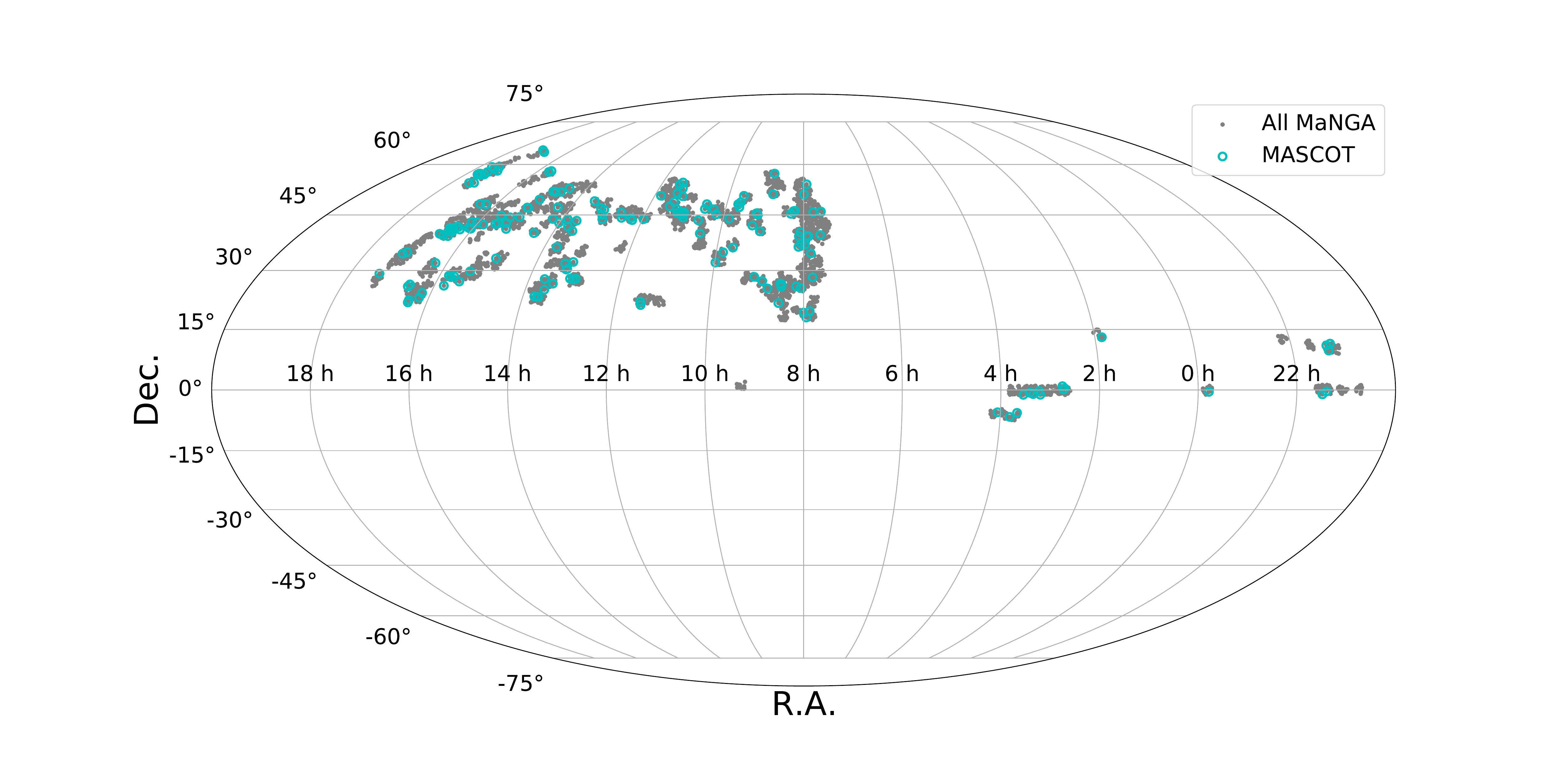}
\caption{Sky distribution of observed MASCOT sources. We also show the distribution of the DR15 MaNGA sample as small grey points.}
\label{sky_dist}
\end{figure*}

At the time of writing of this paper, observations for 153 sources have been completed as part of MASCOT 2.0, resulting in a total number of observed MASCOT targets of 187. The distribution of the observed objects on the sky and in redshift space are shown in Figure \ref{sky_dist} and Figure \ref{z_dist}, respectively.

\subsection{Sample Selection}

The aim of this survey is to provide molecular gas measurements for galaxies with existing IFU data. All targets are therefore drawn from the sample of targets with existing observations from the MaNGA survey. The current MaNGA data release at the time of the start of the MASCOT 1.0 program in early 2018 consisted of $\sim$ 3000 members, but the survey continued to observe $\sim$ 1600 galaxies/year and public data releases happened on a yearly basis. As a result, the target list for MASCOT 2.0 has evolved and will continue to evolve over the remaining part of the survey, as MaNGA releases its complete and final data set (see Section 2.1).

We select our targets from MaNGA sources with available SFR measurements and we estimate the expected CO(1-0) luminosity from the total SFR by assuming the $SFR-L_{CO(1-0)}^{\prime}$ scaling relation obtained by \citet[][shown in Fig. 8 in their paper]{Cicone_2017} for the ALLSMOG and COLDGASS samples of local star-forming galaxies \citep{Saintonge_2011, Cicone_2017}. We then derived the expected CO(1-0) integrated flux and the expected CO(1-0) peak flux density assuming a total line width of 200~km/s. These estimates were carried out to prioritise the first set of sources to be observed within MASCOT. Since we carried out observations spread over the entire year, our sample selection is independent of any R.A. constraints.

To test the (at that time unknown) telescope efficiency, in 2018, we chose to start the survey with the brightest sources with an expected CO(1-0) peak flux density $>75~mJy$, corresponding to a lower limit in molecular gas mass of $\log(M_{H_{2}}/M_{\odot}) \sim 8.9$ for a galaxy with a redshift of $z=0.02$ (which corresponds to the mean redshift of our sample, see Figure \ref{z_dist}). This selection ensured that we could either detect the CO(1-0) emission line in these targets at a $S/N>3$ or infer valuable upper limits on their molecular gas content. We integrated until the CO line is detected, or until we reached a 1-$\sigma_{\rm{rms}}$ sensitivity of 0.5~mK in $\delta v= 50~km~s^{-1}$-wide channels, corresponding to $\sim 13$ mJy. That selection resulted in a sample of star forming galaxies located with a mean of $\sim$~0.15 dex above the star formation main sequence. 

\begin{figure}
\includegraphics[ width=0.45\textwidth, trim = 0.1cm 0.4cm 0.9cm 0.5cm, clip= true]{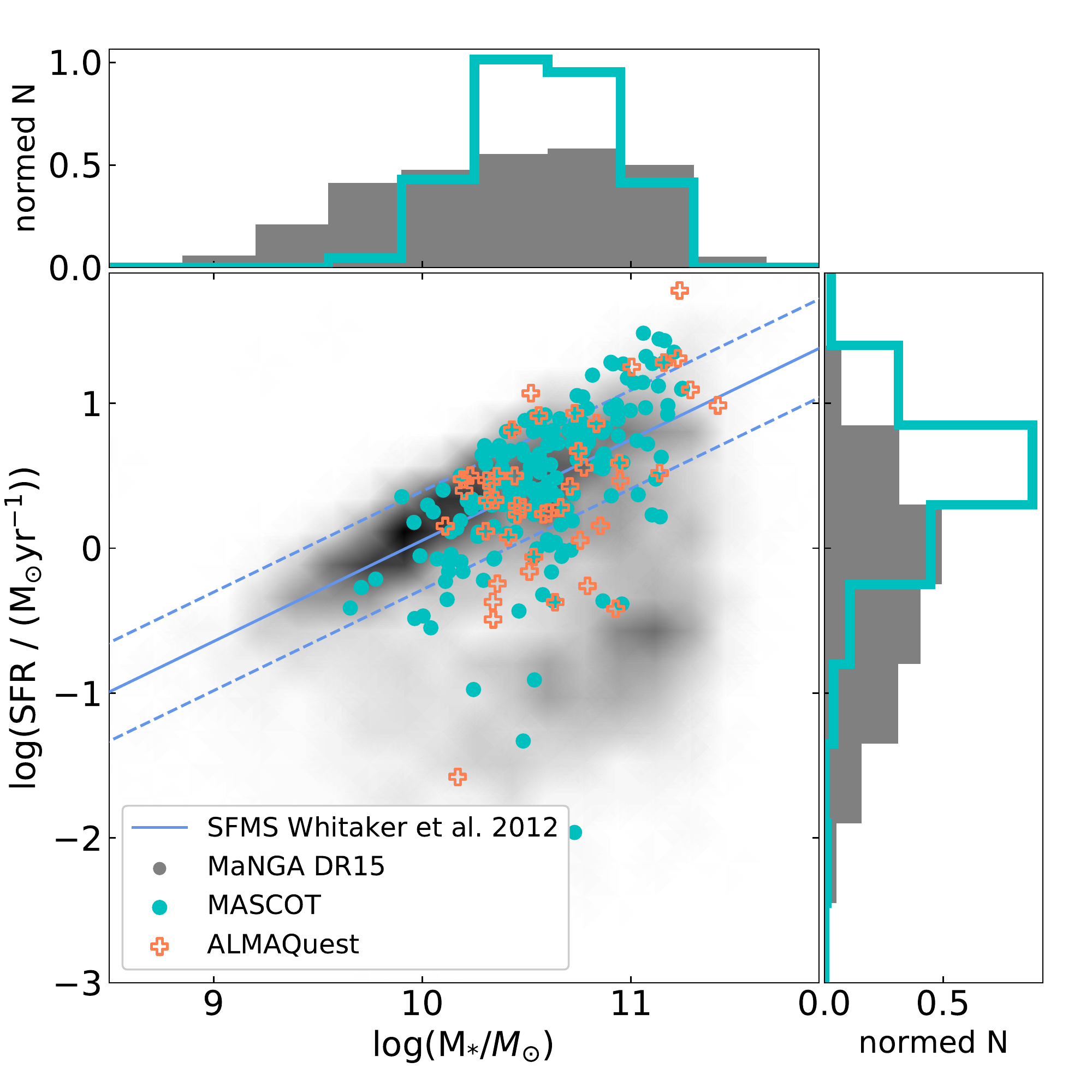}
\caption{Distribution of the MASCOT targets in the SFR-M$_*$ plane (cyan circles). We also show the distribution of the DR15 MaNGA sample (grey density map) and the distribution of targets from the complementary ALMAQuest sample (orange crosses). Additionally, we show the star formation main sequence (SFMS) as derived by \citet{Whitaker_2012} and its associated scatter of $\pm 0.34$ dex (dashed lines). Most MASCOT sources lie above the SFMS owing to the selection process for this first part of the MASCOT survey. The ALMAQuest sources nicely complement the MASCOT sample in sampling sources on and below the SFMS. We also show the normalised distributions of the stellar masses M$_*$ and SFR above and next to the main panel. For the normalisation, each bin displays the bin's raw count divided by the total number of counts and the bin width.}
\label{sfr_mass}
\end{figure}

\begin{figure}
\includegraphics[width=0.45\textwidth, trim = 1cm 0.3cm 2.5cm 1.54cm, clip= true]{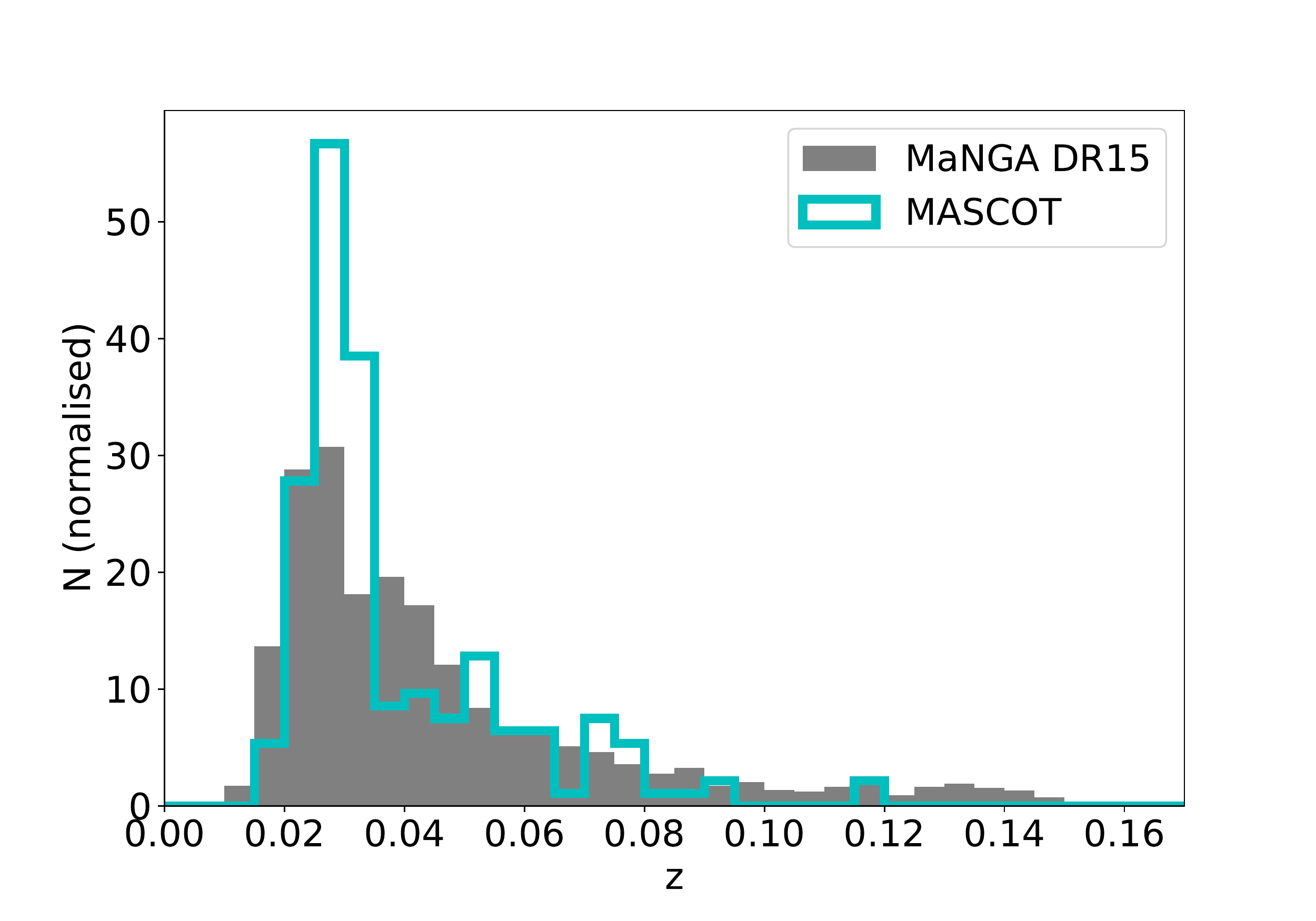}
\caption{Normalised redshift distribution of the MASCOT sources in comparison to the MaNGA sample (DR15). For the normalisation, each bin displays the bin's raw count divided by the total number of counts and the bin width.}
\label{z_dist}
\end{figure}

The number of detections has exceeded our expectations given the previous estimates of the sensitivity of the receiver. Furthermore, the telescope has been upgraded with a new spectrometer (see Section 3.3) since the start of the MASCOT program in 2018. We have therefore adjusted our targeting strategy. To reduce the bias in selecting primarily star-forming galaxies, we have continued the survey drawing from the remaining MaNGA targets with an expected CO(1-0) peak flux density $>15$~mJy and adjusted our target 1-$\sigma_{\rm{rms}}$ sensitivity to 0.25~mK. In Figure \ref{sfr_mass} we show the distribution of the MASCOT sample in the SFR-M$_{*}$ plane in comparison with the full MaNGA sample, where we use the stellar masses and SSP-derived SFR from the Pipe3D catalogue. The current sample of 187 galaxies is located with a median of $\sim$~0.07 dex above the star formation main sequence.

\subsection{Observations}
\label{section: observations}

Observations are carried out with the new 12m millimeter single dish telescope at the Arizona Radio Observatory. At the typical redshifts of the MaNGA sample ($z\sim0.025$), the CO(1-0) line is conveniently redshifted to 112 GHz, where the atmospheric opacity is improved relative to 115 GHz. We used the 3mm receiver on the 12m ARO antenna (equivalent to ALMA Band 3, 84 -- 116 GHz). The first part of the MASCOT survey (MASCOT 1.0) was carried out with the millimetre autocorrelator (MAC) backend on the ARO 12m antenna which provided a bandwidth of 800 MHz (600 MHz usable $\simeq$ 1600 km/s at 112 GHz). In August 2018, the new ARO Wideband Spectrometer (AROWS) backend was successfully commissioned on the 12m. AROWS offers an increased total bandwidth of 4000 MHz, sampling $>$ 5000 km/s around the CO(1-0) line and allowing for an improved baseline subtraction, especially for sources with broad CO(1-0) lines. The second part of the MASCOT survey (MASCOT 2.0) has been carried out with the AROWS backend and so will forthcoming observations. All observations are also still simultaneously recorded with the MAC, as a backup. Both the MAC and the AROWS give a velocity resolution of about $\sim ~ 1$ km/s. Since our sources are `high velocity' sources, i.e. galaxies with $z > 0.02$, we tuned the telescope to the different rest frequencies and set that rest frequency to 0.0 km/s for all sources.

Observations are carried out in fixed observing blocks. The atmospheric conditions varied greatly, with a precipitable water vapor (PWV) ranging between 2 and 20~mm, with a mean of $\sim 5$~mm. Targets for this first data release were chosen and priorized based on their expected peak CO(1-0) flux density as described in Section 3.2 as well as on their observability. For example, we priorised targets with low ($< 2$) airmass and avoided pointing towards the wind direction. We made `real-time' decisions on the targets by performing on-the-fly data reductions of the observations, and observations were stopped once the target was detected at a significance of 5$\sigma$ or when we reached a root mean square ($\sigma_{\rm{rms}}$) of 0.5 mK (0.25 mK, see Section 3.2) per 50 km/s channel, corresponding to $\sim 13$~mJy ($\sim 7$~mJy), -- which ever came first. In good observing conditions and for bright sources, conditions were met within one hour, but typical exposure times ranged between $\sim 2-3$~hours. 

\subsection{Data Reduction}
\label{data_reduction}

The data are reduced with the CLASS software\footnote{\href{http://www.iram.fr/IRAMFR/GILDAS}{http://www.iram.fr/IRAMFR/GILDAS}}. All scans are visually examined and unusually noisy scans, scans with distorted baselines or anomalies are discarded. We then manually set a generous velocity window and estimate the baseline in each scan through a first-order fit to the continuum outside of that velocity window. If the CO line is undetected or very weak, we set the velocity window to [-300, 300]~km/s. The scans are then averaged and saved as a fits file.

These spectra are still in units of the observed source antenna temperature corrected for atmospheric attenuation, radiative loss and rearward scattering and spillover $T_{R}^{*}$  \footnote{see Appendix C of the ARO 12m manual: http://aro.as.arizona.edu/12\_obs\_manual}. We therefore first convert to the main beam temperature $T_{mb}$ using $T_{mb} = T_{R}^{*} / \eta_{m}^{*}$, where $\eta_{m}^{*}$ is the corrected main beam efficiency. The flux density $f_{\nu}$ can then be derived from the Rayleigh-Jeans law such that

\begin{equation}
f_{\nu} = 5.097\times10^{-4}\frac{\rm{BWHM}^2}{\lambda^2} T_{mb}
\end{equation}

where $f_{\nu}$ is in Jy, $\lambda$ is the observing wavelength in cm, and BWHM (beam width at half maximum) is the beam size in arcsec. The beam size at the frequency of our observations is 55 arcsec\footnote{see Table 3.2 of the ARO 12m manual: http://aro.as.arizona.edu/12\_obs\_manual}. The typical uncertainty in the main beam efficiency is $\sim 3-4$\%.

\subsection{Spectral Line Fitting}

\subsubsection{Dynamical spectral binning}

We develop a customised spectral line fitting technique to measure the CO line fluxes and and kinematic parameters. We use the baseline subtracted averaged spectra in units of Jy as described in Section \ref{data_reduction}. Our aim is to dynamically bin the spectra depending on the measured signal-to-noise S/N at a given bin width $dv_{\rm{bin}}$. This procedure utilises non-parametric flux and velocity width measurements which are described in more detail in Section \ref{CO_measurements}.

We begin by fitting a single Gaussian profile to the spectrum at its native spectral resolution (1--2~km/s depending on the redshift of the source) and allow the velocity offset $\Delta v$ to range between [-250, 250]~km/s and the velocity dispersion to range between [50, 500]~km/s. We then measure the cumulative flux 

\begin{equation}
\label{eq:S_calc}
S(v) = \int_{-\infty}^{v} f_{v}(v') dv',
\end{equation}
and the total line flux is given by $S(\infty)$. In practice, we use the interval [-3000, 3000]~km~s$^{-1}$ in the rest-frame of the galaxy for the integration.

We then determine $v_{05}$ such that $S(v_{05}) = 0.05 \cdot S(\infty)$ and $v_{95}$ such that $S(v_{95})~=~0.95~\cdot~S(\infty)$. These are the velocities at which 5\% and 95\% of the total flux $S(\infty)$ are reached, respectively. We use these velocities to determine the width of the fitted line W$_{90}$ using W$_{90} = v_{95} - v_{05}$. We measure the flux $S$ of the line from the data (i.e. not the Gaussian model) within W$_{90}$ and refer to it as S$_{\rm{CO, data}}$ (see Section 3.5.2). For the S$_{\rm{CO, data}}$, we include the flux bins in which $v_{05}$ and $v_{95}$ fall.

We then determine the uncertainty $\epsilon$ on the flux by first determining the root-mean-square (rms) noise of the spectra. We measure the standard deviation of the noise per spectral channel, $
\sigma_{\rm{rms}}$, outside the spectral window of [-250, 250]~km/s (with respect to the 0 km/s rest frequency, see Section 3.3) and then compute 

\begin{equation}
\label{eq:uncertainty}
\epsilon = \frac{  \sigma_{\rm{rms}} W_{90} }{\sqrt{N}},
\end{equation}
where N is the number of channels within W$_{90}$, which can be determined by calculating W$_{90}$ / $dv_{\rm{bin}}$.

If the signal-to-noise ratio, $S/N = S(\infty) / \epsilon$, is determined to be $< 20$, we spectrally bin the spectrum to a resolution of $dv_{\rm{bin}}$~=~10~km/s and repeat the described procedure. We repeat this process of increasing $dv_{\rm{bin}}$ in steps of 10~km/s until the line is measured with a S/N $> 20$ or until we rebin the spectrum to a maximum $dv_{\rm{bin}}$~=~50~km/s. This procedure allows us to retain a better spectral sampling for the bright sources within our sample.

We provide the binned and unbinned spectra as part of the supplementary material to this paper, as well as on the MASCOT website\footnote{\href{https://wwwstaff.ari.uni-heidelberg.de/dwylezalek/mascot.html}{https://wwwstaff.ari.uni-heidelberg.de/dwylezalek/mascot.html}}.

\subsubsection{CO Flux and CO Luminosity Measurements}
\label{CO_measurements}

Once the spectral binning (to a maximum of 50~km/s) has been optimised, we proceed with measuring the CO line flux. A single Gaussian is often insufficient for describing the profile of the CO(1-0) line. This is particularly true when secondary broad components characteristic of potentially outflowing gas or broad or double-peaked CO(1-0) emission in the case of rotating gas disks are contributing significantly to the emission line profile. To evaluate the prevalence of additional kinematic components, we therefore allow for two Gaussian components to be fitted to the CO line. The fitting procedure uses a least squares regression to return best-fit parameters for the single-Gaussian and double-Gaussian model. The $\chi^2$ value is then used to evaluate the goodness of the fit. 

In cases where the spectral shape of an emission line is close to Gaussian, then the calculation of best fit parameters from a single Gaussian fit (i.e. velocity dispersion, full width at half maximum (FWHM), amplitude) are sufficient to describe its kinematic properties. This is not the case when multiple Gaussians are needed to describe the line profile \citep[see also][]{Wylezalek_2020}. We therefore calculate non-parametric values based on the percentages of the total integrated flux \citep[see e.g.][]{liu13b}. We compute the $S(v)$ based on the best-model fit (see equation \ref{eq:S_calc} above). We then compute the line-of-sight velocity $v_{med}$ where $S(v_{med}) = 0.5 \cdot S(\infty)$, i.e. this is the velocity that bisects the total area underneath the emission-line profile. Because the fitting is performed in the rest-frame of the galaxy as determined by its stellar component, $v_{med}$ is measured relative to the rest-frame \citep[see also][]{Wylezalek_2020}. We use the W$_{90}$ parameter to parameterise the velocity width of the CO line. W$_{90}$ refers to the velocity width that encloses 90\% of the total flux. We determine $v_{05}$ and $v_{95}$ and then calculate W$_{90}$ using W$_{90} = v_{95} - v_{05}$. The advantage of using W$_{90}$ over the Gaussian velocity dispersion is that it quasi independent of the underlying model used to fit the profile. We also compute W$_{50} = v_{75} - v_{25}$ for easier comparison with other works and enhancing the legacy value of our work. 

We then compute the CO(1-0) line flux in three different ways: $(i)$ by integrating the spectrum data within $v_{05}$ and $v_{95}$ yielding S$_{\rm{CO, data}}$, $(ii)$ by integrating the model fit within $v_{05}$ and $v_{95}$ yielding S$_{\rm{CO, model}}$, and $(iii)$ by integrating the model fit within [-3000, 3000]~km/s yielding S$_{\rm{CO, tot}}$. 

The measured line fluxes which agree well with one another, generally within 2-10 \%, while the uncertainty on the flux measurements due to the noise in the data ranges between 10-20\%. For the remaining analysis in this paper we therefore use S$_{\rm{CO, model}}$. We refer to this measurement from now on as S$_{\rm{CO}}$ and report it in Table \ref{table_all}. 

We compute the uncertainty $\epsilon$ on the flux measurement following equation \ref{eq:uncertainty} and determine the $S/N$ of the emission line by computing $S/N = S_{\rm{CO}} / \epsilon$. We also compute the signal-to-noise ratio of the emission line peak, $S/N_{\rm{peak}}$, by determining the peak flux density within the velocity window set by $W_{90}$, $f_{\rm{\rm{peak}}}$, and computing $S/N_{\rm{peak}} = f_{\rm{peak}} / \sigma_{\rm{rms}}$. We consider a source to be detected if $S/N > 3$ or $S/N_{\rm{peak}} > 3$. Based on this definition, we detect 162 out of 187 sources. If we had chosen a $S/N$ cut of 4, 132 sources would be considered detected.

For the 162 detected sources, we compute the corresponding CO(1-0) luminosity $L_{\rm{CO}}$ using the equation 

\begin{equation}
L_{\rm{CO}}=4\pi D_{\rm{L}}^2 S_{\rm{CO}}
\end{equation}
where $D_{\rm{L}}$ is the luminosity distance. 

In case of a non-detection, we report the 3$\sigma$ upper limits where $\sigma$ is determined following equation \ref{eq:uncertainty} assuming a constant $W_{90}$ of 200 or 300 km~s$^{-1}$ for galaxies with M$^{*}$ lower or greater than $10^{10}$~M$_{\odot}$, respectively \citep[see also][]{Saintonge_2017}.

Table \ref{table_all} reports the CO line luminosities (or upper limits for the non-detections) in units of L$_{\odot}$ as well as K~km~s$^{-1}$~pc$^2$, the kinematic parameters of the line fit (the velocity width W$_{90}$, the line-of-sight velocity $v_{med}$ as well as the velocity width W$_{50}$), the signal-to-noise ratios and $\sigma_{\rm{rms}}$ of the spectrum.

\subsection{H$_2$ masses}

The $^{12}$CO molecule is a commonly used tracer for the cold molecular gas content in galaxies of which most is in the form of H$_2$. Since the H$_2$ molecule lacks a permanent electric dipole moment, cold H$_2$ is not directly observable but the total molecular gas mass can be estimated following the empirical relation

\begin{equation}
\label{equ:mol_mass}
M_{H_{2}} = \alpha_{\rm{CO}} L^{'}_{\rm{CO(1-0)}}
\end{equation}

where M$_{H_{2}}$ is in units of M$_{\odot}$ and L$^{'}_{\rm{CO(1-0)}}$ is in units of K~km~s$^{-1}$~pc$^2$. The $\alpha_{\rm{CO}}$, the CO-to-H$_2$ conversion factor, can be considered a mass to light ratio. The empirical value of 4.36 M$_{\odot}$ (K km s$^{-1} pc^2$)$^{-1}$, referred to as the Galactic conversion factor, has been determined from observations of the Milky Way and nearby star-forming galaxies, the empirical CO(1-0) conversion factors are consistent with a . While this Galactic conversion factor is often applied to other galaxies, as well, it is well known that the conversion factor is dependent on many parameters, mainly star-formation rate and gas phase metallicity. For example, UV radiation from massive stars destroys CO to a cloud depth of a few $A_V$, such that the Galactic conversion factor would underestimate the true molecular hydrogen content \citep[see][for a review]{Bolatto_2013}. 

\citet{Accurso_2017} recently used results from the xCOLDGASS survey \citep{Saintonge_2017} combined with Herschel observations to carry out a thorough Bayesian analysis revealing that only two parameters, gas phase metallicity $\log\rm{(O/H)}$ and offset from the star formation main sequence (SFMS) $\Delta\rm{(MS)}$ are needed to robustly parametrise changes in the $L_{\rm{[CII]}}/L_{\rm{CO(1-0)}}$ ratio. They use their parametrization of that ratio, alongside radiative transfer modelling, to present a novel conversion function for $\alpha_{\rm{CO}}$:

\begin{equation}
\label{equ:alpha}
\begin{split}
\log{ \alpha_{\rm{CO}}}(\pm0.165~\rm{dex}) = 15.623 - 1.732[12 + \log\rm{(O/H)}] \\
+ 0.051 \log{\Delta\rm{(MS)}}
\end{split}
\end{equation}
where the distance off the SFMS is defined as 

\begin{equation}
\Delta\rm{(MS)} = \frac{\rm{sSFR_{measured}}}{\rm{sSFR_{ms}(z, M_*)}}
\end{equation}
and where the analytical definition of the SFMS by \citet{Whitaker_2012} is used:

\begin{equation}
\begin{split}
\log{\rm{(sSFR_{ms}(z,M_*))}}=-1.12+1.14z-0.19z^2-(0.3+0.13z) \\
\times (\log{M_*} -10.5)[\rm{Gyr^{-1}}].
\end{split}
\end{equation}

We make use of the relation presented by \citet{Accurso_2017} to derive a metallicity- and sSFR-dependent $\alpha_{\rm{CO}}$. We use the $\log\rm{(O/H)}$ oxygen abundance at the effective radius derived using the \citet{Maiolino_2008} calibrator provided in the MaNGA Pipe3D catalog (`OH\_Re\_fit\_M08'), as well as the SFR and stellar masses from the MaNGA Pipe3D catalog (`log\_SFR\_ssp', `log\_mass'). In \citet{Accurso_2017}, the gas-phase metallicity from \citet{Pettini_2004} is used in equation \ref{equ:alpha} which has been shown to have a constant offset of 0.05 from the \citet{Maiolino_2008} calibrator in the mass range of our galaxies \citep{Sanchez_2017}. We therefore correct the Pipe3D $\log\rm{(O/H)}$ oxygen abundance by $+0.05$ before calculating $\alpha_{\rm{CO}}$. For 27 galaxies the $\log\rm{(O/H)}$ abundance is not reported in the Pipe3D catalog due to non sufficient data quality. We assign a $\alpha_{\rm{CO}}$ to those galaxies corresponding to the median value of 2.68 M$_{\odot}$ (K km s$^{-1} pc^2$)$^{-1}$ of our remaining galaxy sample. 

We then calculate the molecular gas masses following equation~\ref{equ:mol_mass}. For sources undetected in the MASCOT CO(1-0) observations we report upper limits based on the 3$\sigma$ upper limits on their CO(1-0) luminosity (see Table \ref{table_all}).

\subsection{Crossmatch with xCOLDGASS and ALLSMOG}

The xCOLDGASS survey is a large legacy survey providing a census of molecular gas in the local Univers by having obtained CO(1-0) observations of 532 galaxies with the IRAM-30m telescope \citep{Saintonge_2017}. The sample was mass-selected in the redshift interval 0.01 < z < 0.05 from the SDSS spectroscopic sample. We therefore cross-match the MASCOT catalog with the xCOLDGASS catalog to identify any overlapping observations. 

\begin{figure}
\includegraphics[width=0.47\textwidth, trim = 16cm 1.5cm 11cm 1cm, clip= true]{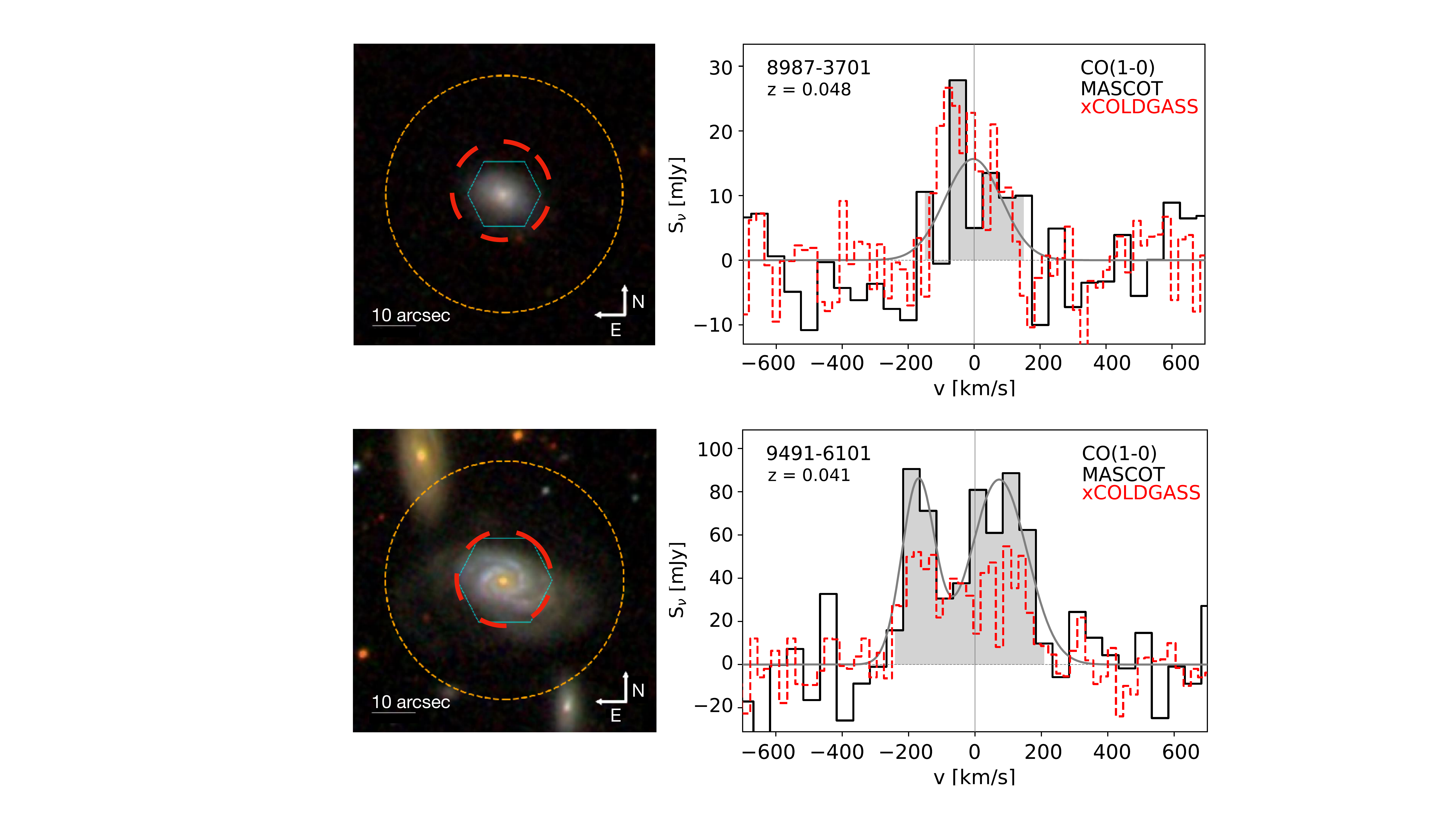}
\caption{Comparison of xCOLDGASS and MASCOT observations for two targets the surveys have in common. We show the SDSS three color image, showing the MaNGA footprint (purple hexagon), the ARO beam (orange dashed circle) and the IRAM 30m beam of the xCOLDGASS observations (red dashed circle). The spectra on the right show the MASCOT CO(1-0) line spectrum (solid black line) and the xCOLDGASS spectrum (red dashed line). The area of the MASCOT spectrum between $v_{05}$ and $v_{95}$ is coloured in light grey.}
\label{xcoldgass_comparison}
\end{figure}

There are three sources observed both within the xCOLDGASS and MASCOT surveys. Figure \ref{xcoldgass_comparison} shows a comparison of two spectra obtained within both projects. The IRAM beam size at the frequency of the CO(1-0) observations is 22'' in contrast to the ARO beam size of 55'' (see Figure \ref{xcoldgass_comparison}). For MASCOT source 8987-3701 (xCOLDGASS ID: 32619) the IRAM beam extents beyond 3~R$_{\rm{eff}}$ of the galaxy (R$_{\rm{eff}}$ = 3.1''), while the effective radius of MASCOT source 9491-6101 (xCOLDGASS ID: 31775) is R$_{\rm{eff}} = 5.6$'' and therefore only covered by the IRAM beam out to $\sim$ 2~R$_{\rm{eff}}$. This difference is reflected in the respective spectra obtained by MASCOT and xCOLDGASS (Figure \ref{xcoldgass_comparison}). While the spectra for 8987-3701 are very similar and their measured CO luminosities are in good agreement (within 15\% and within uncertainties), the xCOLDGASS observations of 9491-6101 seem to miss a significant amount of flux (CO luminosities are within 40\% and are not within uncertainties), even though \citet{Saintonge_2017} do apply an aperture correction to their CO flux measurements. 

The median r-band effective radius of the xCOLDGASS sample is 4.7'', very close to median r-band effective radius of the MaNGA survey of 5.4''. That means that 9491-6101 (xCOLDGASS ID: 31775) is not an outlier with an unusually large size in the xCOLDGASS survey but representative of the entire xCOLDGASS sample. This implies that despite aperture corrections, smaller beam molecular gas surveys may miss a significant amount of molecular gas. Therefore, CO surveys with beam sizes covering galaxies out to several effective radii such as the MASCOT survey are better suited for assessing the total molecular gas content of these low redshift galaxies. 

We also cross-match the MASCOT sample with the ALLSMOG survey \citep{Cicone_2017} but do not find any overlap in sources.

\subsection{ALMaQUEST Survey}

The ALMaQUEST (ALMA-MaNGA QUEnching and STar formation) survey is a program with spatially-resolved 12CO(1-0) measurements obtained with the Atacama Large Millimeter Array (ALMA) for 46 galaxies from MaNGA DR15 \citep{Lin_2020}. The aim of the ALMaQUEST survey is to investigate the dependence of star formation activity on the cold molecular gas content at kpc scales in nearby galaxies. While this survey spatially resolves the CO(1-0) line on scales matching the MaNGA resolution, its field-of-view is $\sim$ 50'', very similar to the MASCOT observations. The ALMaQUEST and MASCOT surveys have four sources in common (ID: 8952-6104, 8950-12705, 8450-6102 and 8655-3701) whose derived molecular gar masses agree well: $\log(M_{\rm{H_2}, MASCOT} / M_{\odot}$) = 9.2 / 9.2 / 9.2 / 10.2, $\log(M_{\rm{H_2, ALMaQUEST}} / M_{\odot}$) = 9.1 / 9.4 / 9.2 / 10.4). The ALMaQUEST Survey is thus well complementary to the MASCOT survey (see also Figure \ref{sfr_mass}). 


\section{Results}

\subsection{Global Relations}

\begin{figure*}
\includegraphics[width=0.48\textwidth, trim = 1cm 1cm 2cm 1.54cm, clip= true ]{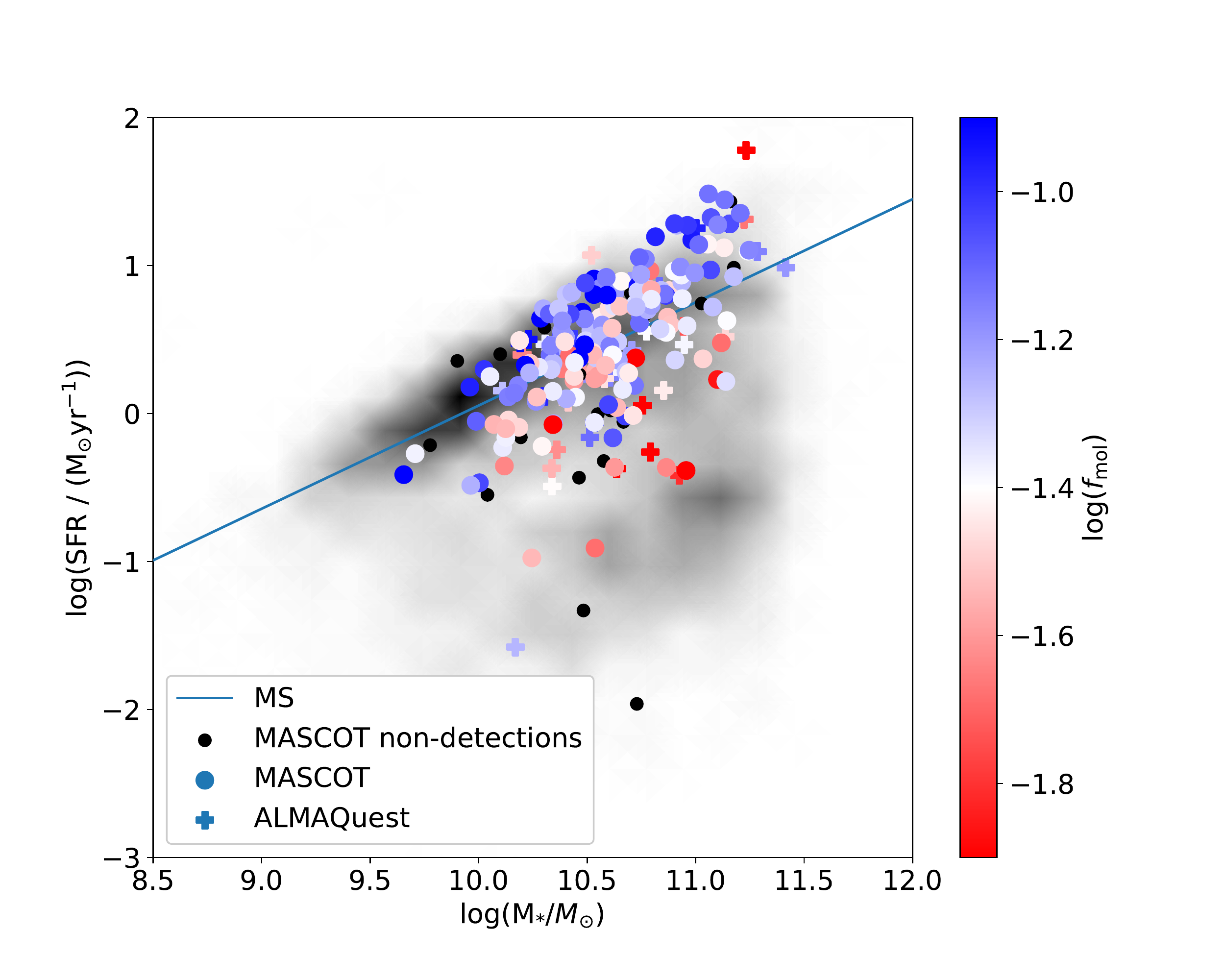}
\includegraphics[width=0.48\textwidth, trim = 1cm 1cm 2cm 1.54cm, clip= true ]{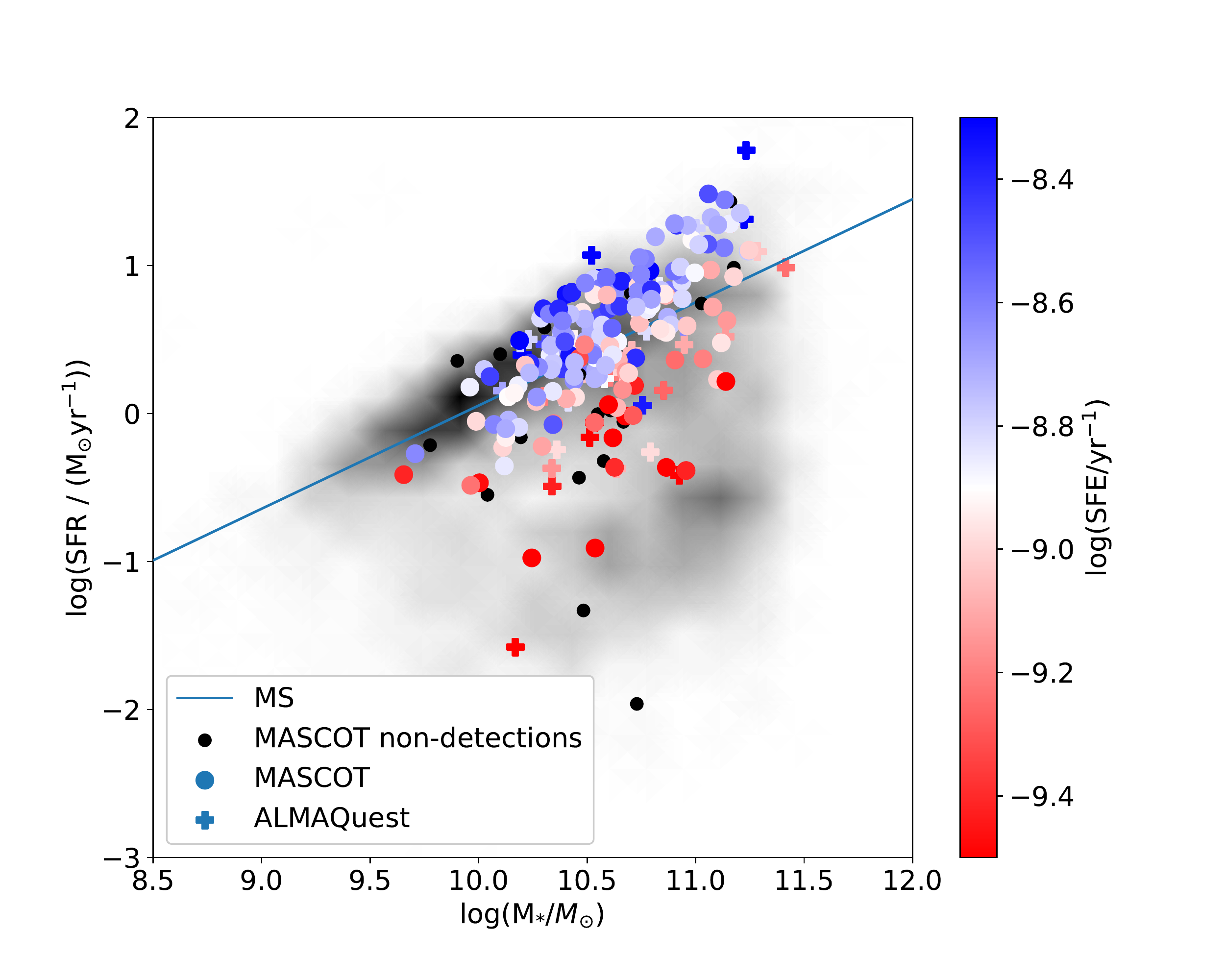}
\includegraphics[width=0.48\textwidth, trim = 1cm 1cm 2cm 1.54cm, clip= true]{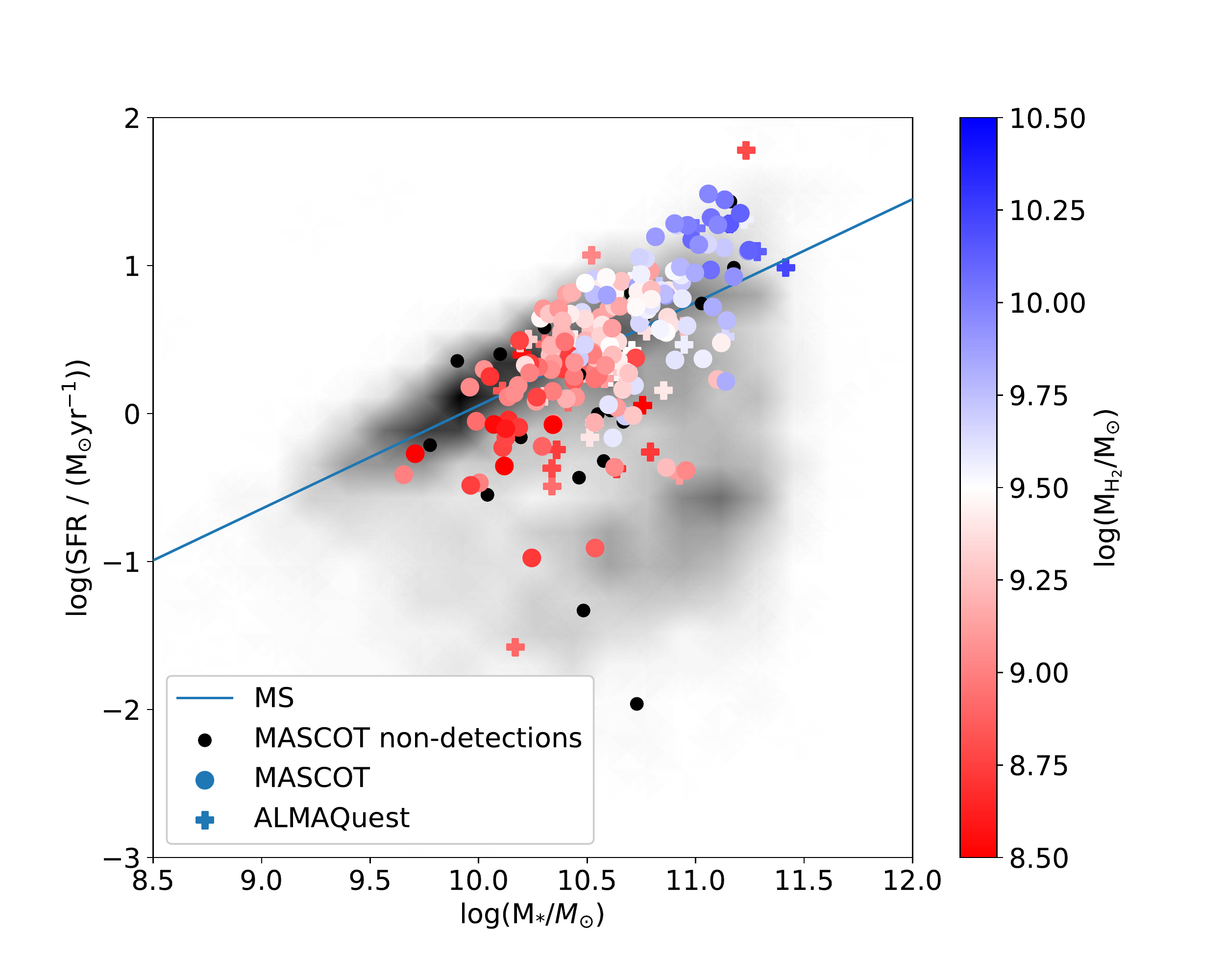}
\caption{We show the distribution of sources in the SFR-M$_{*}$ diagrams, colour-coded by the molecular gas mass fraction f$\rm{_{mol} = M_{H_2} / M_{*}}$ (\textbf{upper left panel}), the star formation efficiency $\rm{SFE = SFR / M_{H_2}}$ (\textbf{upper right panel}), and the molecular gas mass M$_{H_2}$ (\textbf{lower panel}). For reference, we also show the distribution of the DR15 MaNGA sample in the SFR-M$_{*}$ plane, as well as the star formation main sequence SFMS as derived by \citet{Whitaker_2012}. SFRs and stellar masses are both taken from the MaNGA-Pipe3D catalog for all sources, while the molecular gas masses are derived from the 55$\arcsec$ ARO beam for the MASCOT sources and from the MaNGA footprint for the ALMAQuest sources.}
\label{sfr_mstar_mascot}
\end{figure*}

In Figure \ref{sfr_mstar_mascot} we show the distribution of galaxies in the SFR--M$_{*}$ plane. We complement the MASCOT sample with results from the ALMAQuest survey which mainly complements the MASCOT survey in the green valley. While we show the MASCOT non-detections in Figure \ref{sfr_mstar_mascot} for completeness, the following analysis is limited to the 162 MASCOT-detected sources for which we can measure $\rm{M_{H_2}}$. Together with the 43 additional detected sources from the ALMAQuest survey, the following analysis focuses on a total sample size of 205 sources. 

For consistency, and similar to Figure \ref{sfr_mass}, for both the MASCOT and ALMAQuest samples, we use the SFR and M$_{*}$ measurements provided by the Pipe3D catalog. We note that this differs to what is presented and shown in Table 3 and Figure 3 in the ALMAQuest paper by \citet{Lin_2020} where the SFR is computed based on considering star-forming spaxels (as per BPT classification) within 1.5 R$_{\rm{eff}}$. 

We also remind the reader that the ALMAQuest molecular gas mass measurements were measured by summing the CO(1-0) flux over the areas within the MaNGA bundles while the MASCOT beam covers generally a larger area (beam size: 55 arcsec). This implies that the ALMAQuest molecular gas fractions may be lower when directly compared to the MASCOT molecular gas fractions. In the case of an underestimation of the CO(1-0) luminosity by 60\% (as is the case for source 9491-6101 in comparison with the xCOLDGASS survey, see Section 3.7), $\log(\rm{f_{mol}})$ would be underestimated by 0.3. 

The upper left panel of Figure \ref{sfr_mstar_mascot} is color-coded by the molecular gas fraction defined as f$\rm{_{mol} = M_{H_2} / M_{*}}$. For the MASCOT sources, this fraction is directly computed using the $\rm{M_{H_2}}$ measurements described in Section 3.6 and M$_{*}$ from the Pipe3D catalog. We measure a median molecular gas mass fraction for the MASCOT survey of f$\rm{_{mol}} = 0.052$. For the ALMAQuest sources, we use the molecular gas masses reported in Table 3 in \citet{Lin_2020} and the M$_{*}$ from the Pipe3D catalog to compute f$\rm{_{mol}}$. 

The left panel of Figure \ref{sfr_mstar_mascot} illustrates that f$\rm{_{mol}}$ is roughly constant along the SFMS and sharply drops below the SFMS. We observe a significant correlation with specific star formation rate ($\rm{sSFR = SFR /M_{*} }$) with a Spearman rank correlation coefficient of $r_{\rm{corr}}$=0.45 and a $p-$value of $1\times10^{-11}$. The correlation with SFR alone is more moderate with $r_{\rm{corr}} = 0.3$ and a $p-$value of  $1\times10^{-6}$, while we do not observe a significant correlation with stellar mass ($r_{\rm{corr}}$ = 0, $p-$value of $0.5$).

Similar trends are observed when we investigate how the molecular star formation efficiency $\rm{SFE = SFR / M_{H_2}}$ varies across the SFMS. The MASCOT median star formation efficiency is $\rm{\log(SFE/yr^{-1}) = -8.7}$. The inverse of the SFE is the `depletion time' $\tau_{\rm{dep}}$, indicating how much time is necessary to convert all the available molecular gas into stars at the current star formation rate. We find that the depletion time is roughly constant along the SFMS, as well, with depletion times dropping sharply below the SFMS. The correlation between both SFE and sSFR and SFE and SFR are significant with $r_{\rm{corr}}$= 0.68, $p-$value of $10^{-29}$ and $r_{\rm{corr}} = 0.47$, $p-$value of $10^{-13}$, respectively. SFE seems to be independent of stellar mass, though ($r_{\rm{corr}}$ = 0, $p-$value of $0.2$).

In contrast to the first two panels of Figure \ref{sfr_mstar_mascot}, the lower panel shows that the molecular mass $\rm{M_{H_2}}$ varies strongly along the SFMS with significant correlations of $\rm{M_{H_2}}$ with both SFR and $\rm{M_{*}}$ ($\sim $ $r_{\rm{corr}}$ = 0.7 / 0.7, $p-$value = $10^{-31}$ / $p-$value = $10^{-36}$, respectively). The median molecular gas mass in the MASCOT sample is $\rm{\log(M_{H_2}/M_{\odot})} = 9.3$.

The observed relations are similar and consistent with other studies in the literature. For example, \citet{Colombo_2020} carried out a large CO(1-0) survey of 472 galaxies selected from the CALIFA survey with the APEX telescope and the CARMA array. The CALIFA-CO sample is of lower redshift than the MaNGA-CO observations presented here. Furthermore, their CO observations are measured from within $\sim 1 \rm{R_{eff}}$ and therefore only measure the centrally available molecular gas. Despite these differences the observed relations between SFR, $\rm{M_{H_2}}$, f$\rm{_{mol}}$, SFE and $\rm{M_{H_2}}$ are very similar to what we observe here. Similar conclusions were also made by \citet{Saintonge_2017} with the xCOLDGASS survey. These observations imply that it is both the molecular gas mass fraction and the star formation efficiency that determines a galaxy's position in the SFR--M$_{*}$ plane.

\begin{figure}
\includegraphics[trim = 0.7cm 0.8cm 1cm 1.5cm, clip = true, width=0.48\textwidth, ]{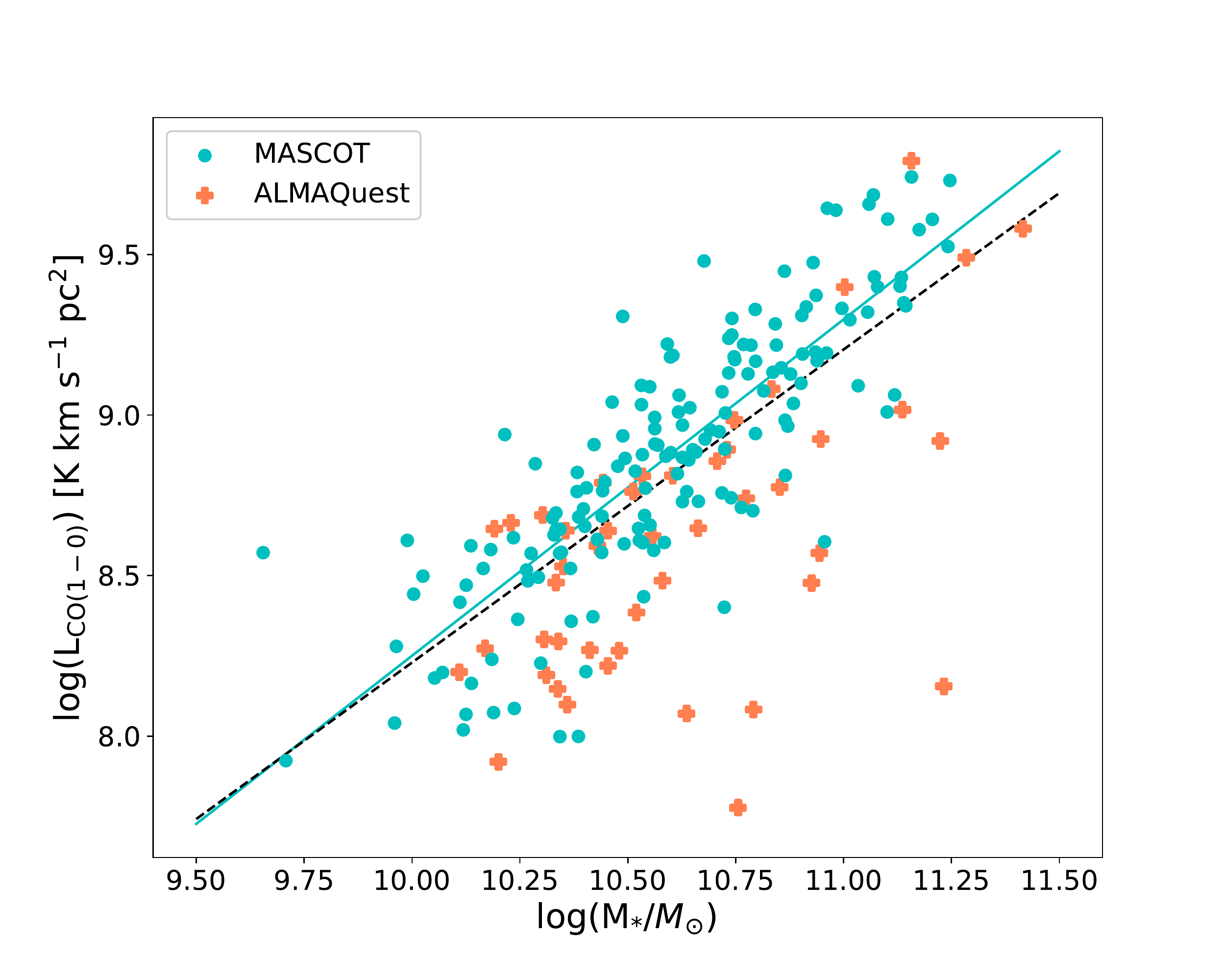}
\caption{CO(1-0) line luminosity as a function of M$_*$. The cyan solid line indicates the best fit  linear relation obtained for the MASCOT data only ($y = -2.21 + 1.05x$), while the black dashed line indicates the best fit relation for the combined MASCOT and ALMAQuest samples ($y = -1.51 + 0.97x$).  }
\label{L_CO_Mass}
\end{figure}

We note that since we are using a prescription for $\alpha_{\rm{CO}}$ that is dependent on $\Delta\rm{(MS)}$, the derived quantities f$\rm{_{mol}}$ and SFE are dependent on M$_{*}$ and SFR in a non-trivial way. We therefore re-peat the above analysis for a constant $\alpha_{\rm{CO}}$ and do not find significant changes in the trends reported in this Section. But even without the added dependency of $\alpha_{\rm{CO}}$ on $\Delta\rm{(MS)}$, derived quantities such as $\rm{SFE = SFR / M_{H_2}}$ and sSFR = $\rm{sSFR = SFR /M_{*} }$ are not independent and therefore some form of correlation is expected. \citet{Cicone_2017} found that L$_{\rm{CO}}$ (from which M$_{H_2}$ is derived) is strongly correlated with M$_*$ and suggested that that this relation may be so tight and linear because the luminosity of optically thick low-$J$ CO transitions is an excellent tracer of the dynamical mass in star-forming galaxies, assuming that in this class of objects the bulk of CO probes molecular clouds in virial motions. We test this correlation with our observations and find a similarly tight correlation as \citet{Cicone_2017} ($r_{\rm{corr}}$ = 0.84, $p-$value = $10^{-45}$, see Figure \ref{L_CO_Mass}). We note the slight offset between the ALMAQuest data points and the MASCOT data points in Figure \ref{L_CO_Mass}. As stated earlier, this is very likely due to the fact that the ALMAQuest molecular gas mass measurements were measured by summing the CO(1-0) flux over the areas within the MaNGA bundles while the MASCOT beam covers generally a larger area. Therefore, the total ALMAQuest CO measurements tend to be slightly underestimated, similar to our findings with the xCOLDGASS galaxies (see Section 3.7). Nevertheless, the best-fit regression parameters and correlation coefficients presented in Figure \ref{L_CO_Mass} are in very good agreement with the \citet{Cicone_2017} results.

\subsection{Resolved Relations}

While the above section reassuringly confirms previous results, the main aim and true strength of the MASCOT survey is to investigate how the molecular gas mass content relates to spatially resolved galaxy parameters. While an in-depth analysis of the resolved relations will be presented in upcoming papers in this series (Bertemes et al. in prep.), we present here an example of how the MASCOT survey is and will be used in investigating global CO measurements together with the spatially resolved nature of the MaNGA observations. 

\begin{figure}
\includegraphics[width=0.48\textwidth, ]{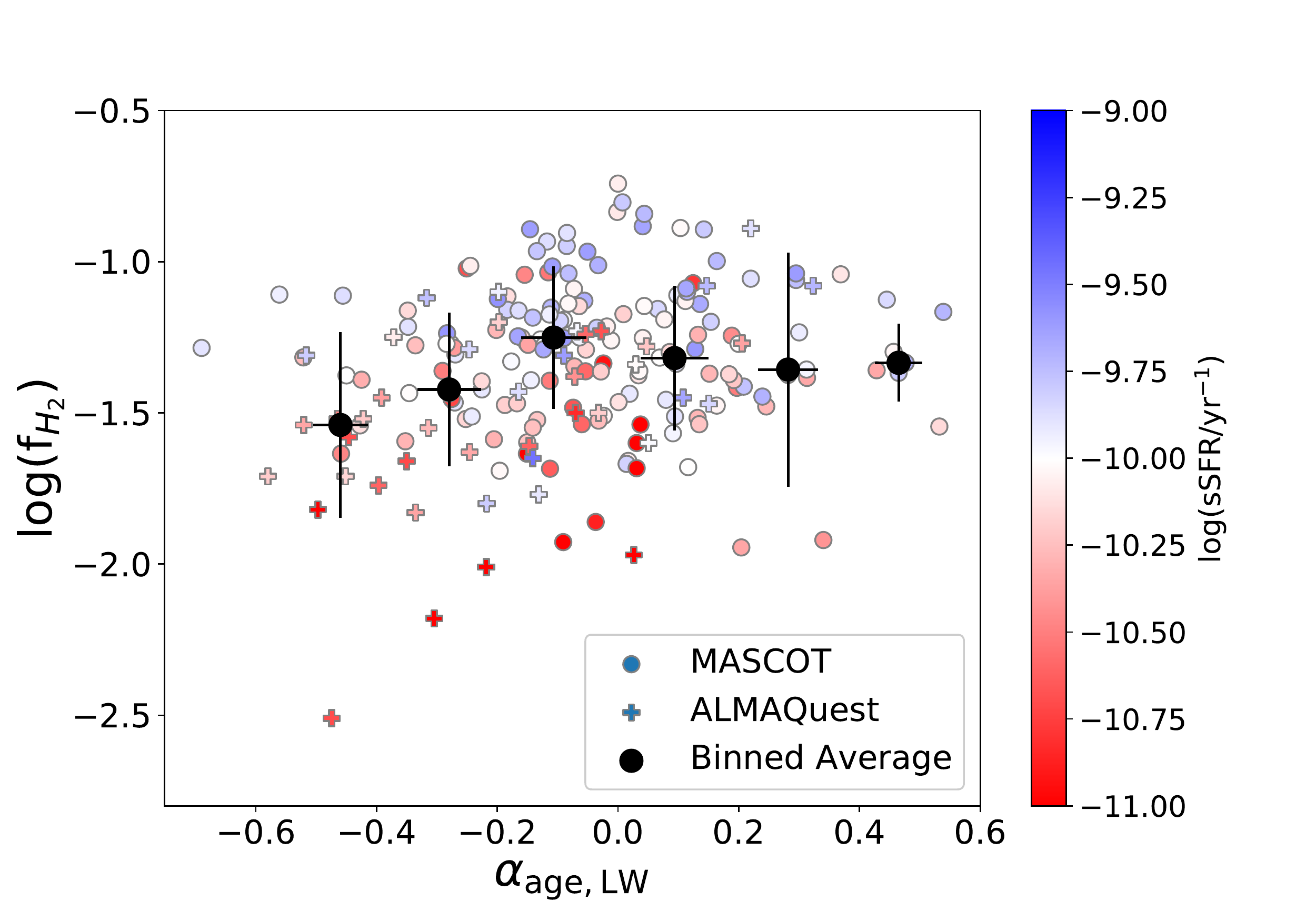}
\caption{The molecular gas mass fraction f$\rm{_{mol}}$ as a function of stellar age gradient $\alpha_{\rm{age,LW}}$. Galaxies with lower f$\rm{_{mol}}$ tend to show older stellar populations close to the galactic center, while the opposite is true for galaxies with higher f$\rm{_{mol}}$, potentially a signature for inside-out quenching being the dominant quenching mechanism in the MASCOT-MaNGA galaxies.}
\label{age_slope}
\end{figure}

In Figure \ref{age_slope} we show how the molecular gas mass fraction, f$\rm{_{mol}}$, relates to the stellar age gradient $\alpha_{\rm{age,LW}}$\footnote{Although we use $\alpha$ to refer to the CO-to-H$_2$ conversion factor, we also use $\alpha$ -- albeit with a different index -- to refer to the stellar age gradient to be consistent with the notation in the Pipe3D catalog.}. The stellar age gradient describes the slope of the gradient of the luminosity-weighted log-age of the stellar population within a galactocentric distance of 0.5 to 2.0 R$_{\rm{eff}}$). This measurement is provided as part of the Pipe3D catalog. While the Pipe3D catalog also reports mass-weighted ages, luminosity-weighted ages are more sensitive to small fractions of recent generations of stars (which contribute significantly in luminosity but not in mass), the mass-weighted age is more representative of the average epoch when the bulk of the stars in a galaxy formed \citep[see e.g.][]{Pasquali_2010}. Since we are investigating potential quenching, or recent star formation triggering effects, the luminosity weighted ages are more meaningful for this analysis. When $\alpha$ is positive, the stellar populations are getting older as we move away from the galactic center. When $\alpha$ is negative, it indicates that the stellar populations are getting younger as we move away from the galactic center. Despite the large scatter, we observe a weak positive correlation between $\alpha_{\rm{age,LW}}$ and f$\rm{_{mol}}$ ($r_{\rm{corr}}$ = 0.2, $p-$value=$4\times10^{-3}$). The correlation is less weak when only considering galaxies with a negative $\alpha_{\rm{age,LW}}$ ($r_{\rm{corr}}$ = 0.3, $p-$value=$2\times10^{-4}$). This means that galaxies with high molecular gas mass fractions tend to host younger stellar populations in their centres than they do at larger distances, while galaxies with low molecular gas mass fractions tend to be centrally quenched. This observation suggests that star formation quenching in these galaxies happens primarily inside-out and that a lower H$_2$ gas fraction seems to be a precursor of quenching. 

Evidence for inside-out quenching being the dominant quenching mode has been suggested by other works \citep{Lin_2017, Bluck_2020, Breda_2020, Brownson_2020}. Such analyses have been in particular possible due to the emergence of large IFU galaxy surveys. For example, \citep{Lin_2019a} classify MaNGA galaxies into inside-out and outside-in quenching types. They base their classification on the strength and spatial distribution of quenched areas which are defined by two non-parametric parameters, quiescence and its concentration, traced by regions with low EW(H$\alpha$) \citep[for details and the definitions of these parameters see][]{Lin_2019a}. Additionally, they classify the galaxies based on their environment into satellites and centrals. They find that the fraction of inside-out quenching is systematically greater than that of outside-in quenching, suggesting that inside-out quenching is the dominant quenching mode in all environments. 

AGN feedback or morphological quenching are potential mechanisms that may suppress the star formation and that may drive the features of the inside-out quenching but which one dominates under what circumstances is still an open point of debate \citep{Lin_2019a, Lin_2017, Ellison_2021c}. The current size of the MASCOT sample and the parameter space it covers does not allow us yet to investigate this question in detail. But as the MASCOT sample continues to grow, we will address these topics in the forthcoming papers in this series.

\section{Conclusion}

In this paper we have presented the first data release of the MASCOT survey which provides CO(1-0) measurements for galaxies that have already been observed as part of the SDSS-IV MaNGA survey. Observations are carried out with the 12m antenna at the Arizona Radio Observatory. We summarise the main points of this paper in the following:

\begin{enumerate}
\item This first data release presents CO(1-0) observations for 187 MaNGA galaxies. These galaxies are mostly located on and above the star formation main sequence and upcoming MASCOT observations are targeting primarily MaNGA galaxies on and below the star formation main sequence.
\item We develop a customised spectral line fitting technique to allow for a dynamic spectral binning of the spectra (from native binning up to spectral bins of 50 km/s) depending on the signal-to-noise ratio of the detection. To measure CO line fluxes, we allow secondary Gaussian components to account for broad, asymmetric line profiles and / or double-Gaussian features. We use non-parametric measurements to compute the flux and kinematics (velocity shift and velocity width) of the CO(1-0) emission line. 
\item We use a metallicity- and sSFR-dependent $\alpha_{\rm{CO}}$ conversion factor to compute the total molecular gas mass within the beam size of the observations, 55 $\arcsec$. 
\item The molecular gas mass fraction f$\rm{_{mol} = M_{H_2} / M_{*}}$ varies strongly across the star formation main sequence and correlates most strongly with sSFR (r$_{corr} = 0.45$) and SFR  (r$_{corr} = 0.3$). Similar trends are observed for the star formation efficiency $\rm{SFE = SFR / M_{H_2}}$. SFE correlates strongly with sSFR (r$_{corr} = 0.68$) and SFR  (r$_{corr} = 0.47$) but is mostly independent of stellar mass. These observations imply that it is both the molecular gas mass fraction and the star formation efficiency that determines a galaxy’s position the SFR-M$_{*}$ plane, confirming previous results. 
\item When investigating the MASCOT CO measurements in the context of the spatially resolved information from the supporting MaNGA data, we find that f$\rm{_{mol}}$ weakly correlates with the luminosity-weighted stellar age gradient $\alpha_{\rm{age,LW}}$. Galaxies with lower f$\rm{_{mol}}$ tend to show older stellar populations close to the galactic center, while the opposite is true for galaxies with higher f$\rm{_{mol}}$, potentially a signature for inside-out quenching being the dominant quenching mechanism in the MASCOT-MaNGA galaxies. 
\end{enumerate}
The MASCOT survey is a large program and this data release only represents the first sets of observations of $\sim 800$ hours of observing time out of 1400 hours total. Upcoming data releases will present results for galaxies primarily on and below the star formation main sequence and the upcoming papers in this series (Bertemes et al. in prep., Wylezalek et al. in prep.) are investigating in depth the relations between molecular gas mass properties and spatially resolved diagnostics from the MaNGA observations. With the current survey efficiency, we expect the final MASCOT sample to consist of a total$\sim$ 250-300 galaxies.

\section*{Acknowledgements}

The entire MASCOT team would like to warmly thank the staff at the Arizona Radio Observatory, in particular the operators of the 12m Telescope, Clayton, Kevin, Mike and Robert, for their continued support and help with the observations. 
 
DW is supported by through the Emmy Noether Programme of the German Research Foundation. SC acknowledges the financial support from the State Agency for Research of the Spanish MCIU through the \lq\lq Center of Excellence Severo Ochoa\rq\rq \ award to the Instituto de Astrof{\'i}sica de Andaluc{\'i}a (SEV-2017-0709). MA acknowledges support from FONDECYT grant 1211951, ``CONICYT + PCI + INSTITUTO MAX PLANCK DE ASTRONOMIA MPG190030'' and ``CONICYT+PCI+REDES 190194''.

This research made use of Marvin, a core Python package and web framework for MaNGA data, developed by Brian Cherinka, José Sánchez-Gallego, Brett Andrews, and Joel Brownstein \citep{marvin_2018}. \href{http://sdss-marvin.readthedocs.io/en/stable/}{http://sdss-marvin.readthedocs.io/en/stable/}

This project makes use of the MaNGA-Pipe3D dataproducts. We thank the IA-UNAM MaNGA team for creating this catalogue, and the ConaCyt-180125 project for supporting them.

Funding for the Sloan Digital Sky 
Survey IV has been provided by the 
Alfred P. Sloan Foundation, the U.S. 
Department of Energy Office of 
Science, and the Participating 
Institutions. 

SDSS-IV acknowledges support and 
resources from the Center for High 
Performance Computing  at the 
University of Utah. The SDSS 
website is www.sdss.org.

SDSS-IV is managed by the 
Astrophysical Research Consortium 
for the Participating Institutions 
of the SDSS Collaboration including 
the Brazilian Participation Group, 
the Carnegie Institution for Science, 
Carnegie Mellon University, Center for 
Astrophysics | Harvard \& 
Smithsonian, the Chilean Participation 
Group, the French Participation Group, 
Instituto de Astrof\'isica de 
Canarias, The Johns Hopkins 
University, Kavli Institute for the 
Physics and Mathematics of the 
Universe (IPMU) / University of 
Tokyo, the Korean Participation Group, 
Lawrence Berkeley National Laboratory, 
Leibniz Institut f\"ur Astrophysik 
Potsdam (AIP),  Max-Planck-Institut 
f\"ur Astronomie (MPIA Heidelberg), 
Max-Planck-Institut f\"ur 
Astrophysik (MPA Garching), 
Max-Planck-Institut f\"ur 
Extraterrestrische Physik (MPE), 
National Astronomical Observatories of 
China, New Mexico State University, 
New York University, University of 
Notre Dame, Observat\'ario 
Nacional / MCTI, The Ohio State 
University, Pennsylvania State 
University, Shanghai 
Astronomical Observatory, United 
Kingdom Participation Group, 
Universidad Nacional Aut\'onoma 
de M\'exico, University of Arizona, 
University of Colorado Boulder, 
University of Oxford, University of 
Portsmouth, University of Utah, 
University of Virginia, University 
of Washington, University of 
Wisconsin, Vanderbilt University, 
and Yale University.

\section*{Data Availability}

The data underlying this article are available in the article and in its supplementary material. We also publish the data on the MASCOT website \href{https://wwwstaff.ari.uni-heidelberg.de/dwylezalek/mascot.html}{https://wwwstaff.ari.uni-heidelberg.de/dwylezalek/mascot.html}.

\bibliographystyle{mnras}
\bibliography{./master_2021.bib} 

\begin{thebibliography}{}
\makeatletter
\relax
\def\mn@urlcharsother{\let\do\@makeother \do\$\do\&\do\#\do\^\do\_\do\%\do\~}
\def\mn@doi{\begingroup\mn@urlcharsother \@ifnextchar [ {\mn@doi@}
  {\mn@doi@[]}}
\def\mn@doi@[#1]#2{\def\@tempa{#1}\ifx\@tempa\@empty \href
  {http://dx.doi.org/#2} {doi:#2}\else \href {http://dx.doi.org/#2} {#1}\fi
  \endgroup}
\def\mn@eprint#1#2{\mn@eprint@#1:#2::\@nil}
\def\mn@eprint@arXiv#1{\href {http://arxiv.org/abs/#1} {{\tt arXiv:#1}}}
\def\mn@eprint@dblp#1{\href {http://dblp.uni-trier.de/rec/bibtex/#1.xml}
  {dblp:#1}}
\def\mn@eprint@#1:#2:#3:#4\@nil{\def\@tempa {#1}\def\@tempb {#2}\def\@tempc
  {#3}\ifx \@tempc \@empty \let \@tempc \@tempb \let \@tempb \@tempa \fi \ifx
  \@tempb \@empty \def\@tempb {arXiv}\fi \@ifundefined
  {mn@eprint@\@tempb}{\@tempb:\@tempc}{\expandafter \expandafter \csname
  mn@eprint@\@tempb\endcsname \expandafter{\@tempc}}}

\bibitem[\protect\citeauthoryear{{Abolfathi} et~al.,}{{Abolfathi}
  et~al.}{2018}]{dr14_2017}
{Abolfathi} B.,  et~al., 2018, \mn@doi [The Astrophysical Journal Supplement
  Series] {10.3847/1538-4365/aa9e8a}, \href
  {https://ui.adsabs.harvard.edu/#abs/2018ApJS..235...42A} {235, 42}

\bibitem[\protect\citeauthoryear{{Accurso} et~al.,}{{Accurso}
  et~al.}{2017}]{Accurso_2017}
{Accurso} G.,  et~al., 2017, \mn@doi [\mnras] {10.1093/mnras/stx1556}, \href
  {https://ui.adsabs.harvard.edu/abs/2017MNRAS.470.4750A} {470, 4750}

\bibitem[\protect\citeauthoryear{{Aguado} et~al.,}{{Aguado}
  et~al.}{2019a}]{dr15_2019}
{Aguado} D.~S.,  et~al., 2019a, \mn@doi [\apjs] {10.3847/1538-4365/aaf651},
  \href {https://ui.adsabs.harvard.edu/abs/2019ApJS..240...23A} {240, 23}

\bibitem[\protect\citeauthoryear{{Aguado} et~al.,}{{Aguado}
  et~al.}{2019b}]{Aguado_2019}
{Aguado} D.~S.,  et~al., 2019b, \mn@doi [\apjs] {10.3847/1538-4365/aaf651},
  \href {https://ui.adsabs.harvard.edu/abs/2019ApJS..240...23A} {240, 23}

\bibitem[\protect\citeauthoryear{{Barrera-Ballesteros}
  et~al.,}{{Barrera-Ballesteros} et~al.}{2020}]{Barrera_2020}
{Barrera-Ballesteros} J.~K.,  et~al., 2020, \mn@doi [\mnras]
  {10.1093/mnras/stz3553}, \href
  {https://ui.adsabs.harvard.edu/abs/2020MNRAS.492.2651B} {492, 2651}

\bibitem[\protect\citeauthoryear{{Barrera-Ballesteros}
  et~al.,}{{Barrera-Ballesteros} et~al.}{2021}]{Barrera_2021}
{Barrera-Ballesteros} J.~K.,  et~al., 2021, \mn@doi [\mnras]
  {10.1093/mnras/stab755}, \href
  {https://ui.adsabs.harvard.edu/abs/2021MNRAS.tmp..743B} {}

\bibitem[\protect\citeauthoryear{{Belfiore} et~al.,}{{Belfiore}
  et~al.}{2019}]{Belfiore_2019}
{Belfiore} F.,  et~al., 2019, \mn@doi [\aj] {10.3847/1538-3881/ab3e4e}, \href
  {https://ui.adsabs.harvard.edu/abs/2019AJ....158..160B} {158, 160}

\bibitem[\protect\citeauthoryear{{Blanton}, {Kazin}, {Muna}, {Weaver}  \&
  {Price-Whelan}}{{Blanton} et~al.}{2011}]{Blanton_2011}
{Blanton} M.~R.,  {Kazin} E.,  {Muna} D.,  {Weaver} B.~A.,   {Price-Whelan} A.,
   2011, \mn@doi [\aj] {10.1088/0004-6256/142/1/31}, \href
  {https://ui.adsabs.harvard.edu/abs/2011AJ....142...31B} {142, 31}

\bibitem[\protect\citeauthoryear{{Blanton} et~al.,}{{Blanton}
  et~al.}{2017}]{Blanton_2017}
{Blanton} M.~R.,  et~al., 2017, \mn@doi [\aj] {10.3847/1538-3881/aa7567}, \href
  {https://ui.adsabs.harvard.edu/#abs/2017AJ....154...28B} {154, 28}

\bibitem[\protect\citeauthoryear{{Bluck} et~al.,}{{Bluck}
  et~al.}{2020}]{Bluck_2020}
{Bluck} A. F.~L.,  et~al., 2020, \mn@doi [\mnras] {10.1093/mnras/staa2806},
  \href {https://ui.adsabs.harvard.edu/abs/2020MNRAS.499..230B} {499, 230}

\bibitem[\protect\citeauthoryear{{Bolatto}, {Wolfire}  \& {Leroy}}{{Bolatto}
  et~al.}{2013}]{Bolatto_2013}
{Bolatto} A.~D.,  {Wolfire} M.,   {Leroy} A.~K.,  2013, \mn@doi [\araa]
  {10.1146/annurev-astro-082812-140944}, \href
  {https://ui.adsabs.harvard.edu/abs/2013ARA&A..51..207B} {51, 207}

\bibitem[\protect\citeauthoryear{{Bolatto} et~al.,}{{Bolatto}
  et~al.}{2017}]{Bolatto_2017}
{Bolatto} A.~D.,  et~al., 2017, \mn@doi [\apj] {10.3847/1538-4357/aa86aa},
  \href {https://ui.adsabs.harvard.edu/abs/2017ApJ...846..159B} {846, 159}

\bibitem[\protect\citeauthoryear{{Bothwell} et~al.,}{{Bothwell}
  et~al.}{2014}]{Bothwell_2014}
{Bothwell} M.~S.,  et~al., 2014, \mn@doi [\mnras] {10.1093/mnras/stu1936},
  \href {https://ui.adsabs.harvard.edu/abs/2014MNRAS.445.2599B} {445, 2599}

\bibitem[\protect\citeauthoryear{{Bouch{\'e}} et~al.,}{{Bouch{\'e}}
  et~al.}{2010}]{Bouche_2010}
{Bouch{\'e}} N.,  et~al., 2010, \mn@doi [\apj] {10.1088/0004-637X/718/2/1001},
  \href {https://ui.adsabs.harvard.edu/abs/2010ApJ...718.1001B} {718, 1001}

\bibitem[\protect\citeauthoryear{{Breda} et~al.,}{{Breda}
  et~al.}{2020}]{Breda_2020}
{Breda} I.,  et~al., 2020, \mn@doi [\aap] {10.1051/0004-6361/201937193}, \href
  {https://ui.adsabs.harvard.edu/abs/2020A&A...635A.177B} {635, A177}

\bibitem[\protect\citeauthoryear{{Brownson}, {Belfiore}, {Maiolino}, {Lin}  \&
  {Carniani}}{{Brownson} et~al.}{2020}]{Brownson_2020}
{Brownson} S.,  {Belfiore} F.,  {Maiolino} R.,  {Lin} L.,   {Carniani} S.,
  2020, \mn@doi [\mnras] {10.1093/mnrasl/slaa128}, \href
  {https://ui.adsabs.harvard.edu/abs/2020MNRAS.498L..66B} {498, L66}

\bibitem[\protect\citeauthoryear{{Bundy} et~al.,}{{Bundy}
  et~al.}{2015}]{Bundy_2015}
{Bundy} K.,  et~al., 2015, \mn@doi [\apj] {10.1088/0004-637X/798/1/7}, \href
  {http://adsabs.harvard.edu/abs/2015ApJ...798....7B} {798, 7}

\bibitem[\protect\citeauthoryear{{Cappellari}}{{Cappellari}}{2017}]{Cappellari_2017}
{Cappellari} M.,  2017, \mn@doi [\mnras] {10.1093/mnras/stw3020}, \href
  {http://adsabs.harvard.edu/abs/2017MNRAS.466..798C} {466, 798}

\bibitem[\protect\citeauthoryear{{Cappellari} \& {Emsellem}}{{Cappellari} \&
  {Emsellem}}{2004}]{Cappellari_2004}
{Cappellari} M.,  {Emsellem} E.,  2004, \mn@doi [\pasp] {10.1086/381875}, \href
  {http://adsabs.harvard.edu/abs/2004PASP..116..138C} {116, 138}

\bibitem[\protect\citeauthoryear{{Cappellari} et~al.,}{{Cappellari}
  et~al.}{2011}]{Cappellari_2011}
{Cappellari} M.,  et~al., 2011, \mn@doi [\mnras]
  {10.1111/j.1365-2966.2010.18174.x}, \href
  {https://ui.adsabs.harvard.edu/abs/2011MNRAS.413..813C} {413, 813}

\bibitem[\protect\citeauthoryear{Cherinka, S{\'a}nchez-Gallego, Andrews  \&
  Brownstein}{Cherinka et~al.}{2018}]{marvin_2018}
Cherinka B.,  S{\'a}nchez-Gallego J.,  Andrews B.,   Brownstein J.,  2018,
  sdss/marvin: Marvin Beta, \mn@doi{10.5281/zenodo.596700}, \url
  {https://doi.org/10.5281/zenodo.596700}

\bibitem[\protect\citeauthoryear{{Cherinka} et~al.,}{{Cherinka}
  et~al.}{2019}]{Cherinka_2019}
{Cherinka} B.,  et~al., 2019, \mn@doi [\aj] {10.3847/1538-3881/ab2634}, \href
  {https://ui.adsabs.harvard.edu/abs/2019AJ....158...74C} {158, 74}

\bibitem[\protect\citeauthoryear{{Cicone} et~al.,}{{Cicone}
  et~al.}{2017}]{Cicone_2017}
{Cicone} C.,  et~al., 2017, \mn@doi [\aap] {10.1051/0004-6361/201730605}, \href
  {https://ui.adsabs.harvard.edu/abs/2017A&A...604A..53C} {604, A53}

\bibitem[\protect\citeauthoryear{{Colombo} et~al.,}{{Colombo}
  et~al.}{2018}]{Colombo_2018}
{Colombo} D.,  et~al., 2018, \mn@doi [\mnras] {10.1093/mnras/stx3233}, \href
  {https://ui.adsabs.harvard.edu/abs/2018MNRAS.475.1791C} {475, 1791}

\bibitem[\protect\citeauthoryear{{Colombo} et~al.,}{{Colombo}
  et~al.}{2020}]{Colombo_2020}
{Colombo} D.,  et~al., 2020, arXiv e-prints, \href
  {https://ui.adsabs.harvard.edu/abs/2020arXiv200908383C} {p. arXiv:2009.08383}

\bibitem[\protect\citeauthoryear{{Combes}, {Young}  \& {Bureau}}{{Combes}
  et~al.}{2007}]{Combes_2007}
{Combes} F.,  {Young} L.~M.,   {Bureau} M.,  2007, \mn@doi [\mnras]
  {10.1111/j.1365-2966.2007.11759.x}, \href
  {https://ui.adsabs.harvard.edu/abs/2007MNRAS.377.1795C} {377, 1795}

\bibitem[\protect\citeauthoryear{{Croom} et~al.,}{{Croom}
  et~al.}{2012}]{Croom_2012}
{Croom} S.~M.,  et~al., 2012, \mn@doi [\mnras]
  {10.1111/j.1365-2966.2011.20365.x}, \href
  {https://ui.adsabs.harvard.edu/abs/2012MNRAS.421..872C} {421, 872}

\bibitem[\protect\citeauthoryear{{Dey} et~al.,}{{Dey} et~al.}{2019}]{Dey_2019}
{Dey} B.,  et~al., 2019, \mn@doi [\mnras] {10.1093/mnras/stz1777}, \href
  {https://ui.adsabs.harvard.edu/abs/2019MNRAS.488.1926D} {488, 1926}

\bibitem[\protect\citeauthoryear{{Drory} et~al.,}{{Drory}
  et~al.}{2015}]{Drory_2015}
{Drory} N.,  et~al., 2015, \mn@doi [\aj] {10.1088/0004-6256/149/2/77}, \href
  {http://adsabs.harvard.edu/abs/2015AJ....149...77D} {149, 77}

\bibitem[\protect\citeauthoryear{{Ellison}, {Thorp}, {Pan}, {Lin}, {Scudder},
  {Bluck}, {S{\'a}nchez}  \& {Sargent}}{{Ellison}
  et~al.}{2020a}]{Ellison_2020a}
{Ellison} S.~L.,  {Thorp} M.~D.,  {Pan} H.-A.,  {Lin} L.,  {Scudder} J.~M.,
  {Bluck} A. F.~L.,  {S{\'a}nchez} S.~F.,   {Sargent} M.,  2020a, \mn@doi
  [\mnras] {10.1093/mnras/staa001}, \href
  {https://ui.adsabs.harvard.edu/abs/2020MNRAS.492.6027E} {492, 6027}

\bibitem[\protect\citeauthoryear{{Ellison} et~al.,}{{Ellison}
  et~al.}{2020b}]{Ellison_2020b}
{Ellison} S.~L.,  et~al., 2020b, \mn@doi [\mnras] {10.1093/mnrasl/slz179},
  \href {https://ui.adsabs.harvard.edu/abs/2020MNRAS.493L..39E} {493, L39}

\bibitem[\protect\citeauthoryear{{Ellison}, {Lin}, {Thorp}, {Pan}, {Scudder},
  {S{\'a}nchez}, {Bluck}  \& {Maiolino}}{{Ellison}
  et~al.}{2021a}]{Ellison_2021a}
{Ellison} S.~L.,  {Lin} L.,  {Thorp} M.~D.,  {Pan} H.-A.,  {Scudder} J.~M.,
  {S{\'a}nchez} S.~F.,  {Bluck} A. F.~L.,   {Maiolino} R.,  2021a, \mn@doi
  [\mnras] {10.1093/mnras/staa3822}, \href
  {https://ui.adsabs.harvard.edu/abs/2021MNRAS.501.4777E} {501, 4777}

\bibitem[\protect\citeauthoryear{{Ellison}, {Lin}, {Thorp}, {Pan},
  {S{\'a}nchez}, {Bluck}  \& {Belfiore}}{{Ellison}
  et~al.}{2021b}]{Ellison_2021b}
{Ellison} S.~L.,  {Lin} L.,  {Thorp} M.~D.,  {Pan} H.-A.,  {S{\'a}nchez} S.~F.,
   {Bluck} A. F.~L.,   {Belfiore} F.,  2021b, \mn@doi [\mnras]
  {10.1093/mnrasl/slaa199}, \href
  {https://ui.adsabs.harvard.edu/abs/2021MNRAS.502L...6E} {502, L6}

\bibitem[\protect\citeauthoryear{{Ellison} et~al.,}{{Ellison}
  et~al.}{2021c}]{Ellison_2021c}
{Ellison} S.~L.,  et~al., 2021c, \mn@doi [\mnras] {10.1093/mnrasl/slab047},
  \href {https://ui.adsabs.harvard.edu/abs/2021MNRAS.505L..46E} {505, L46}

\bibitem[\protect\citeauthoryear{{Gunn} et~al.,}{{Gunn}
  et~al.}{2006}]{Gunn_2006}
{Gunn} J.~E.,  et~al., 2006, \mn@doi [\aj] {10.1086/500975}, \href
  {https://ui.adsabs.harvard.edu/#abs/2006AJ....131.2332G} {131, 2332}

\bibitem[\protect\citeauthoryear{{G{\"u}sten}, {Nyman}, {Schilke}, {Menten},
  {Cesarsky}  \& {Booth}}{{G{\"u}sten} et~al.}{2006}]{Guesten_2006}
{G{\"u}sten} R.,  {Nyman} L.~{\r{A}}.,  {Schilke} P.,  {Menten} K.,  {Cesarsky}
  C.,   {Booth} R.,  2006, \mn@doi [\aap] {10.1051/0004-6361:20065420}, \href
  {https://ui.adsabs.harvard.edu/abs/2006A&A...454L..13G} {454, L13}

\bibitem[\protect\citeauthoryear{{Harrison}}{{Harrison}}{2017}]{harr17}
{Harrison} C.~M.,  2017, \mn@doi [Nature Astronomy] {10.1038/s41550-017-0165},
  \href {http://adsabs.harvard.edu/abs/2017NatAs...1E.165H} {1, 0165}

\bibitem[\protect\citeauthoryear{{Kennicutt}}{{Kennicutt}}{1998a}]{kenn98}
{Kennicutt} Jr. R.~C.,  1998a, \mn@doi [\araa]
  {10.1146/annurev.astro.36.1.189}, \href
  {http://adsabs.harvard.edu/abs/1998ARA%26A..36..189K} {36, 189}

\bibitem[\protect\citeauthoryear{{Kennicutt}}{{Kennicutt}}{1998b}]{kenn98b}
{Kennicutt} Jr. R.~C.,  1998b, \mn@doi [\apj] {10.1086/305588}, \href
  {http://adsabs.harvard.edu/abs/1998ApJ...498..541K} {498, 541}

\bibitem[\protect\citeauthoryear{{Law} et~al.,}{{Law} et~al.}{2015}]{Law_2015}
{Law} D.~R.,  et~al., 2015, \mn@doi [\aj] {10.1088/0004-6256/150/1/19}, \href
  {http://adsabs.harvard.edu/abs/2015AJ....150...19L} {150, 19}

\bibitem[\protect\citeauthoryear{{Law} et~al.,}{{Law} et~al.}{2016}]{Law_2016}
{Law} D.~R.,  et~al., 2016, \mn@doi [\aj] {10.3847/0004-6256/152/4/83}, \href
  {http://adsabs.harvard.edu/abs/2016AJ....152...83L} {152, 83}

\bibitem[\protect\citeauthoryear{{Law} et~al.,}{{Law} et~al.}{2021}]{Law_2021}
{Law} D.~R.,  et~al., 2021, \mn@doi [\aj] {10.3847/1538-3881/abcaa2}, \href
  {https://ui.adsabs.harvard.edu/abs/2021AJ....161...52L} {161, 52}

\bibitem[\protect\citeauthoryear{{Leung} et~al.,}{{Leung}
  et~al.}{2018}]{Leung_2018}
{Leung} G. Y.~C.,  et~al., 2018, \mn@doi [\mnras] {10.1093/mnras/sty288}, \href
  {https://ui.adsabs.harvard.edu/abs/2018MNRAS.477..254L} {477, 254}

\bibitem[\protect\citeauthoryear{{Levy} et~al.,}{{Levy}
  et~al.}{2018}]{Levy_2018}
{Levy} R.~C.,  et~al., 2018, \mn@doi [\apj] {10.3847/1538-4357/aac2e5}, \href
  {https://ui.adsabs.harvard.edu/abs/2018ApJ...860...92L} {860, 92}

\bibitem[\protect\citeauthoryear{{Levy} et~al.,}{{Levy}
  et~al.}{2019}]{Levy_2019}
{Levy} R.~C.,  et~al., 2019, \mn@doi [\apj] {10.3847/1538-4357/ab2ed4}, \href
  {https://ui.adsabs.harvard.edu/abs/2019ApJ...882...84L} {882, 84}

\bibitem[\protect\citeauthoryear{{Lilly}, {Carollo}, {Pipino}, {Renzini}  \&
  {Peng}}{{Lilly} et~al.}{2013}]{Lilly_2013}
{Lilly} S.~J.,  {Carollo} C.~M.,  {Pipino} A.,  {Renzini} A.,   {Peng} Y.,
  2013, \mn@doi [\apj] {10.1088/0004-637X/772/2/119}, \href
  {https://ui.adsabs.harvard.edu/abs/2013ApJ...772..119L} {772, 119}

\bibitem[\protect\citeauthoryear{{Lin} et~al.,}{{Lin} et~al.}{2017}]{Lin_2017}
{Lin} L.,  et~al., 2017, \mn@doi [\apj] {10.3847/1538-4357/aa96ae}, \href
  {https://ui.adsabs.harvard.edu/abs/2017ApJ...851...18L} {851, 18}

\bibitem[\protect\citeauthoryear{{Lin} et~al.,}{{Lin}
  et~al.}{2019a}]{Lin_2019a}
{Lin} L.,  et~al., 2019a, \mn@doi [\apj] {10.3847/1538-4357/aafa84}, \href
  {https://ui.adsabs.harvard.edu/abs/2019ApJ...872...50L} {872, 50}

\bibitem[\protect\citeauthoryear{{Lin} et~al.,}{{Lin} et~al.}{2019b}]{Lin_2019}
{Lin} L.,  et~al., 2019b, \mn@doi [\apjl] {10.3847/2041-8213/ab4815}, \href
  {https://ui.adsabs.harvard.edu/abs/2019ApJ...884L..33L} {884, L33}

\bibitem[\protect\citeauthoryear{{Lin} et~al.,}{{Lin} et~al.}{2020}]{Lin_2020}
{Lin} L.,  et~al., 2020, arXiv e-prints, \href
  {https://ui.adsabs.harvard.edu/abs/2020arXiv201001751L} {p. arXiv:2010.01751}

\bibitem[\protect\citeauthoryear{{Liu}, {Zakamska}, {Greene}, {Nesvadba}  \&
  {Liu}}{{Liu} et~al.}{2013}]{liu13b}
{Liu} G.,  {Zakamska} N.~L.,  {Greene} J.~E.,  {Nesvadba} N.~P.~H.,   {Liu} X.,
   2013, \mn@doi [\mnras] {10.1093/mnras/stt1755}, \href
  {http://adsabs.harvard.edu/abs/2013MNRAS.436.2576L} {436, 2576}

\bibitem[\protect\citeauthoryear{{Maiolino} et~al.,}{{Maiolino}
  et~al.}{2008}]{Maiolino_2008}
{Maiolino} R.,  et~al., 2008, \mn@doi [\aap] {10.1051/0004-6361:200809678},
  \href {https://ui.adsabs.harvard.edu/abs/2008A&A...488..463M} {488, 463}

\bibitem[\protect\citeauthoryear{{Masters} et~al.,}{{Masters}
  et~al.}{2019}]{Masters_2019}
{Masters} K.~L.,  et~al., 2019, \mn@doi [\mnras] {10.1093/mnras/stz1889}, \href
  {https://ui.adsabs.harvard.edu/abs/2019MNRAS.488.3396M} {488, 3396}

\bibitem[\protect\citeauthoryear{{Pasquali}, {Gallazzi}, {Fontanot}, {van den
  Bosch}, {De Lucia}, {Mo}  \& {Yang}}{{Pasquali} et~al.}{2010}]{Pasquali_2010}
{Pasquali} A.,  {Gallazzi} A.,  {Fontanot} F.,  {van den Bosch} F.~C.,  {De
  Lucia} G.,  {Mo} H.~J.,   {Yang} X.,  2010, \mn@doi [\mnras]
  {10.1111/j.1365-2966.2010.17074.x}, \href
  {https://ui.adsabs.harvard.edu/abs/2010MNRAS.407..937P} {407, 937}

\bibitem[\protect\citeauthoryear{{P{\'e}roux} \& {Howk}}{{P{\'e}roux} \&
  {Howk}}{2020}]{Peroux_2020}
{P{\'e}roux} C.,  {Howk} J.~C.,  2020, \mn@doi [\araa]
  {10.1146/annurev-astro-021820-120014}, \href
  {https://ui.adsabs.harvard.edu/abs/2020ARA&A..58..363P} {58, 363}

\bibitem[\protect\citeauthoryear{{Pettini} \& {Pagel}}{{Pettini} \&
  {Pagel}}{2004}]{Pettini_2004}
{Pettini} M.,  {Pagel} B. E.~J.,  2004, \mn@doi [\mnras]
  {10.1111/j.1365-2966.2004.07591.x}, \href
  {https://ui.adsabs.harvard.edu/abs/2004MNRAS.348L..59P} {348, L59}

\bibitem[\protect\citeauthoryear{{Rupke}}{{Rupke}}{2018}]{Rupke_2018}
{Rupke} D.,  2018, \mn@doi [Galaxies] {10.3390/galaxies6040138}, \href
  {https://ui.adsabs.harvard.edu/abs/2018Galax...6..138R} {6, 138}

\bibitem[\protect\citeauthoryear{{Saintonge} et~al.,}{{Saintonge}
  et~al.}{2011}]{Saintonge_2011}
{Saintonge} A.,  et~al., 2011, \mn@doi [\mnras]
  {10.1111/j.1365-2966.2011.18677.x}, \href
  {https://ui.adsabs.harvard.edu/abs/2011MNRAS.415...32S} {415, 32}

\bibitem[\protect\citeauthoryear{{Saintonge} et~al.,}{{Saintonge}
  et~al.}{2016}]{Saintonge_2016}
{Saintonge} A.,  et~al., 2016, \mn@doi [\mnras] {10.1093/mnras/stw1715}, \href
  {https://ui.adsabs.harvard.edu/abs/2016MNRAS.462.1749S} {462, 1749}

\bibitem[\protect\citeauthoryear{{Saintonge} et~al.,}{{Saintonge}
  et~al.}{2017}]{Saintonge_2017}
{Saintonge} A.,  et~al., 2017, \mn@doi [\apjs] {10.3847/1538-4365/aa97e0},
  \href {https://ui.adsabs.harvard.edu/abs/2017ApJS..233...22S} {233, 22}

\bibitem[\protect\citeauthoryear{{Salpeter}}{{Salpeter}}{1955}]{Salpeter_1955}
{Salpeter} E.~E.,  1955, \mn@doi [\apj] {10.1086/145971}, \href
  {https://ui.adsabs.harvard.edu/abs/1955ApJ...121..161S} {121, 161}

\bibitem[\protect\citeauthoryear{{S{\'a}nchez} et~al.,}{{S{\'a}nchez}
  et~al.}{2012}]{Sanchez_2012}
{S{\'a}nchez} S.~F.,  et~al., 2012, \mn@doi [\aap]
  {10.1051/0004-6361/201117353}, \href
  {https://ui.adsabs.harvard.edu/abs/2012A&A...538A...8S} {538, A8}

\bibitem[\protect\citeauthoryear{{S{\'a}nchez} et~al.,}{{S{\'a}nchez}
  et~al.}{2016}]{Sanchez_2016}
{S{\'a}nchez} S.~F.,  et~al., 2016, \rmxaa, \href
  {https://ui.adsabs.harvard.edu/abs/2016RMxAA..52...21S} {52, 21}

\bibitem[\protect\citeauthoryear{{Sanchez} et~al.,}{{Sanchez}
  et~al.}{2017a}]{Sanchez18}
{Sanchez} S.~F.,  et~al., 2017a, preprint, \href
  {http://adsabs.harvard.edu/abs/2017arXiv170905438S} {} (\mn@eprint {arXiv}
  {1709.05438})

\bibitem[\protect\citeauthoryear{{S{\'a}nchez} et~al.,}{{S{\'a}nchez}
  et~al.}{2017b}]{Sanchez_2017}
{S{\'a}nchez} S.~F.,  et~al., 2017b, \mn@doi [\mnras] {10.1093/mnras/stx808},
  \href {https://ui.adsabs.harvard.edu/abs/2017MNRAS.469.2121S} {469, 2121}

\bibitem[\protect\citeauthoryear{{Sanders} \& {Mirabel}}{{Sanders} \&
  {Mirabel}}{1985}]{Sanders_1985}
{Sanders} D.~B.,  {Mirabel} I.~F.,  1985, \mn@doi [\apjl] {10.1086/184561},
  \href {https://ui.adsabs.harvard.edu/abs/1985ApJ...298L..31S} {298, L31}

\bibitem[\protect\citeauthoryear{{Smee} et~al.,}{{Smee} et~al.}{2013}]{smee13}
{Smee} S.~A.,  et~al., 2013, \mn@doi [\aj] {10.1088/0004-6256/146/2/32}, \href
  {http://adsabs.harvard.edu/abs/2013AJ....146...32S} {146, 32}

\bibitem[\protect\citeauthoryear{{Solomon}, {Downes}, {Radford}  \&
  {Barrett}}{{Solomon} et~al.}{1997}]{Solomon_1997}
{Solomon} P.~M.,  {Downes} D.,  {Radford} S.~J.~E.,   {Barrett} J.~W.,  1997,
  \mn@doi [\apj] {10.1086/303765}, \href
  {https://ui.adsabs.harvard.edu/abs/1997ApJ...478..144S} {478, 144}

\bibitem[\protect\citeauthoryear{{Tumlinson}, {Peeples}  \& {Werk}}{{Tumlinson}
  et~al.}{2017}]{Tumlinson_2017}
{Tumlinson} J.,  {Peeples} M.~S.,   {Werk} J.~K.,  2017, \mn@doi [\araa]
  {10.1146/annurev-astro-091916-055240}, \href
  {https://ui.adsabs.harvard.edu/abs/2017ARA&A..55..389T} {55, 389}

\bibitem[\protect\citeauthoryear{{Utomo} et~al.,}{{Utomo}
  et~al.}{2017}]{Utomo_2017}
{Utomo} D.,  et~al., 2017, \mn@doi [\apj] {10.3847/1538-4357/aa88c0}, \href
  {https://ui.adsabs.harvard.edu/abs/2017ApJ...849...26U} {849, 26}

\bibitem[\protect\citeauthoryear{{Wake} et~al.,}{{Wake}
  et~al.}{2017}]{Wake_2017}
{Wake} D.~A.,  et~al., 2017, \mn@doi [\aj] {10.3847/1538-3881/aa7ecc}, \href
  {https://ui.adsabs.harvard.edu/#abs/2017AJ....154...86W} {154, 86}

\bibitem[\protect\citeauthoryear{{Westfall} et~al.,}{{Westfall}
  et~al.}{2019}]{Westfall_2019}
{Westfall} K.~B.,  et~al., 2019, \mn@doi [\aj] {10.3847/1538-3881/ab44a2},
  \href {https://ui.adsabs.harvard.edu/abs/2019AJ....158..231W} {158, 231}

\bibitem[\protect\citeauthoryear{{Whitaker}, {van Dokkum}, {Brammer}  \&
  {Franx}}{{Whitaker} et~al.}{2012}]{Whitaker_2012}
{Whitaker} K.~E.,  {van Dokkum} P.~G.,  {Brammer} G.,   {Franx} M.,  2012,
  \mn@doi [\apjl] {10.1088/2041-8205/754/2/L29}, \href
  {https://ui.adsabs.harvard.edu/abs/2012ApJ...754L..29W} {754, L29}

\bibitem[\protect\citeauthoryear{{Wylezalek}, {Zakamska}, {Greene}, {Riffel},
  {Drory}, {Andrews}, {Merloni}  \& {Thomas}}{{Wylezalek}
  et~al.}{2018}]{wyle18}
{Wylezalek} D.,  {Zakamska} N.~L.,  {Greene} J.~E.,  {Riffel} R.~A.,  {Drory}
  N.,  {Andrews} B.~H.,  {Merloni} A.,   {Thomas} D.,  2018, \mn@doi [\mnras]
  {10.1093/mnras/stx2784}, \href
  {https://ui.adsabs.harvard.edu/#abs/2018MNRAS.474.1499W} {474, 1499}

\bibitem[\protect\citeauthoryear{{Wylezalek}, {Flores}, {Zakamska}, {Greene}
  \& {Riffel}}{{Wylezalek} et~al.}{2020}]{Wylezalek_2020}
{Wylezalek} D.,  {Flores} A.~M.,  {Zakamska} N.~L.,  {Greene} J.~E.,   {Riffel}
  R.~A.,  2020, \mn@doi [\mnras] {10.1093/mnras/staa062}, \href
  {https://ui.adsabs.harvard.edu/abs/2020MNRAS.492.4680W} {492, 4680}

\bibitem[\protect\citeauthoryear{{Yan} et~al.,}{{Yan}
  et~al.}{2016a}]{Yan_2016b}
{Yan} R.,  et~al., 2016a, \mn@doi [\aj] {10.3847/0004-6256/151/1/8}, \href
  {http://adsabs.harvard.edu/abs/2016AJ....151....8Y} {151, 8}

\bibitem[\protect\citeauthoryear{{Yan} et~al.,}{{Yan}
  et~al.}{2016b}]{Yan_2016a}
{Yan} R.,  et~al., 2016b, \mn@doi [\aj] {10.3847/0004-6256/152/6/197}, \href
  {http://adsabs.harvard.edu/abs/2016AJ....152..197Y} {152, 197}

\bibitem[\protect\citeauthoryear{{Young} et~al.,}{{Young}
  et~al.}{2011}]{Young_2011}
{Young} L.~M.,  et~al., 2011, \mn@doi [\mnras]
  {10.1111/j.1365-2966.2011.18561.x}, \href
  {https://ui.adsabs.harvard.edu/abs/2011MNRAS.414..940Y} {414, 940}

\makeatother
\end{thebibliography}

\begin{landscape}
\begin{table}
\caption{The MASCOT sample. In addition to the main galaxy parameters such as MaNGA ID, RA., Dec. and z, we list the SFR and stellar mass as reported in the MaNGA Pipe 3D Value Added Catalog. In column (7) we report the spectral binning $dv$ of the MASCOT spectra used in the analysis (see Section 3.5.1). In column (8), (9) and (10) we report the CO(1-0) line flux and luminosity measured as described in Section 3.5.2 and column (11) indicates if a one-Gaussian or two-Gaussian model was used to describe the line profile.  In column (12) and (13) we report the non-parameteric velocity shift and velocity width of the CO(1-0) line (see Section 3.5.2). Column (14) and column (15) report the measured signal-to-noise ratios (see Section 3.5.2). Column (16) indicates whether the source was formally detected in the MASCOT survey (flag$_{\rm{CO}}$=1) or whether the reported molecular gas mass (column 20) is an upper limit (flag$_{\rm{CO}}$=2). In column (17) and column (18) we furthermore report the rms noise of the spectra in $dv$ in two different units and column (19) lists the metallicity and sSFR-dependent $\alpha_{\rm{CO}}$ conversion factor (see Section 3.6). The full table is available as supplementary material.}
\label{table_all}
\scriptsize
\begin{tabular}{lcccccccccccccccccccc}
\hline
MaNGA ID & R.A. & Dec. & z & $\log(\rm{M_{*}})$ & $\log(\rm{SFR})$ & $dv$ & S$_{\rm{CO}}$ & L$_{\rm{CO}}$ & L$_{\rm{CO}}$ & N$_{\rm{Gauss}}$ & v$_{\rm{med}}$ & W$_{90}$ & W$_{50}$ & S/N & S/N$_{\rm{peak}}$  & flag$_{\rm{CO}}$ & $\sigma_{\rm{rms, \lambda}}$ & $\sigma_{\rm{rms,\nu}}$ & $\alpha_{\rm{CO}}$ & $\log(\rm{M_{H_{2}}})$ \\
 & deg. & deg. &  &  $\log(\rm{M}_{\odot})$ & $\log(\rm{M}_{\odot}/yr)$ & km/s & erg s$^{-1}$ cm$^{-2}$ & L$_{\odot}$ & K km s$^{-1}$ pc$^2$ &  & km/s & km/s & km/s &  &  &  & erg s$^{-1}$ cm$^{-2}$ \AA$^{-1}$ & mJy  & M$_{\odot} $(K km s$^{-1}$ pc$^2$)$^{-1}$ &  \\
\hline
10001-3702 & 132.91276  &  57.10742 & 0.026 & 10.11 & -0.229 & 50 & 3.25e-17 & 1.28e+04 & 2.61e+08 & 1 & 1 & 218 & 90 & 4.9 & 4.0 & 1 & 6.52e-22 & 15.5 & 2.26 & 8.77 \\
7443-12703 & 229.52557  &  42.74584 & 0.040 & 11.07 & 1.324 & 1 & 1.32e-16 & 1.32e+05 & 2.69e+09 & 2 & 48 & 317 & 100 & 32.1 & 6.6 & 1 & 2.40e-21 & 58.6 & 4.17 & 10.05 \\
7443-12704 & 232.46105  &  42.62896 & 0.019 & 10.00 & -0.468 & 50 & 6.49e-17 & 1.36e+04 & 2.77e+08 & 1 & 39 & 273 & 112 & 5.2 & 3.2 & 1 & 1.11e-21 & 25.8 & 3.62 & 9.00 \\
7815-3702  & 317.90320  &  11.49694 & 0.029 & 10.47 & 0.263 & 50 & 4.79e-17 & 2.50e+04 & 5.11e+08 & -- & -- & -- & -- & 2.8 & 2.7 & 2 & 5.96e-22 & 14.2 & 1.99 & 9.01 \\
7815-6104  & 319.19309  &  11.04374 & 0.081 & 11.18 & 0.985 & 50 & 6.11e-17 & 2.59e+05 & 5.29e+09 & -- & -- & -- & -- & 0.4 & 0.7 & 2 & 7.24e-22 & 19.2 & 8.56 & 10.66 \\
7958-6101  & 257.38368  &  34.42703 & 0.024 & 10.44 & 0.223 & 50 & 5.34e-17 & 1.83e+04 & 3.73e+08 & 2 & 78 & 284 & 153 & 6.4 & 4.8 & 1 & 6.71e-22 & 15.8 & 1.91 & 8.85 \\
7968-3701  & 322.21331  &  -1.07011 & 0.052 & 10.68 & -0.017 & 50 & 8.90e-17 & 1.48e+05 & 3.02e+09 & 1 & 183 & 165 & 68 & 10.7 & 11.1 & 1 & 7.86e-22 & 19.5 & 1.65 & 9.70 \\
7990-1902  & 264.52248  &  57.11816 & 0.030 & 10.74 & 0.738 & 50 & 4.95e-17 & 2.71e+04 & 5.53e+08 & 2 & 65 & 405 & 248 & 7.2 & 5.2 & 1 & 5.32e-22 & 12.7 & 2.12 & 9.07 \\
7990-3703  & 262.09934  &  57.54541 & 0.029 & 10.40 & 0.805 & 50 & 1.58e-17 & 7.79e+03 & 1.59e+08 & 1 & 24 & 165 & 68 & 2.4 & 3.0 & 1 & 7.10e-22 & 16.9 & 8.62 & 9.14 \\
7990-6104  & 261.60761  &  58.58885 & 0.026 & 10.29 & 0.644 & 50 & 8.36e-17 & 3.46e+04 & 7.05e+08 & 2 & 4 & 312 & 135 & 6.6 & 4.8 & 1 & 9.79e-22 & 23.2 & 4.12 & 9.46 \\
7991-1901  & 258.54847  &  57.97609 & 0.093 & 10.43 & 1.280 & 50 & 2.44e-17 & 1.39e+05 & 2.83e+09 & -- & -- & -- & -- & -0.5 & 0.7 & 2 & 2.86e-22 & 7.7 & 2.68 & 9.88 \\
7992-6103  & 252.61893  &  64.02066 & 0.092 & 11.03 & 0.370 & 50 & 1.07e-17 & 6.05e+04 & 1.23e+09 & 1 & 63 & 412 & 169 & 3.4 & 2.2 & 1 & 2.45e-22 & 6.6 & 2.98 & 9.57 \\
7992-6104  & 255.27948  &  64.67687 & 0.027 & 10.45 & 0.111 & 50 & 6.84e-17 & 3.04e+04 & 6.19e+08 & 1 & 39 & 228 & 93 & 9.4 & 6.5 & 1 & 6.73e-22 & 15.9 & 1.94 & 9.08 \\
8077-12702 &  40.96904  &   0.15264 & 0.027 & 10.14 & -0.044 & 50 & 1.59e-17 & 7.16e+03 & 1.46e+08 & 1 & -33 & 473 & 194 & 2.8 & 2.1 & 1 & 5.00e-22 & 11.9 & 3.27 & 8.68 \\
8077-12704 &  41.28056  &   0.95016 & 0.025 & 10.12 & -0.354 & 50 & 1.42e-17 & 5.13e+03 & 1.05e+08 & 1 & 44 & 213 & 88 & 3.1 & 2.8 & 1 & 4.91e-22 & 11.6 & 2.98 & 8.49 \\
8077-6102  &  40.55686  &   0.38294 & 0.022 & 9.96 & -0.485 & 50 & 3.21e-17 & 9.34e+03 & 1.91e+08 & 1 & -84 & 400 & 165 & 3.0 & 2.1 & 1 & 7.20e-22 & 16.9 & 2.91 & 8.74 \\
8080-12702 &  47.99941  &  -1.16166 & 0.027 & 10.72 & 0.190 & 50 & 1.34e-16 & 5.79e+04 & 1.18e+09 & 1 & 35 & 590 & 242 & 8.1 & 4.0 & 1 & 1.13e-21 & 26.8 & 3.52 & 9.62 \\
8082-12702 &  48.43841  &  -0.24146 & 0.026 & 11.10 & 0.230 & 50 & 1.23e-16 & 5.01e+04 & 1.02e+09 & 2 & -32 & 447 & 233 & 6.7 & 4.4 & 1 & 1.30e-21 & 30.8 & 1.72 & 9.25 \\
8082-9102  &  50.17993  &  -1.00229 & 0.036 & 10.74 & 0.611 & 50 & 1.23e-16 & 9.80e+04 & 2.00e+09 & 1 & -11 & 454 & 187 & 7.5 & 4.3 & 1 & 1.13e-21 & 27.3 & 2.29 & 9.66 \\
8084-12703 &  51.16991  &  -0.68147 & 0.039 & 10.87 & 0.651 & 50 & 4.82e-17 & 4.53e+04 & 9.24e+08 & 2 & 12 & 264 & 136 & 9.3 & 6.3 & 1 & 4.21e-22 & 10.3 & 2.48 & 9.36 \\
8134-1901  & 113.40018 &  45.94337 & 0.077 & 10.98 & 1.174 & 50 & 5.59e-17 & 2.13e+05 & 4.34e+09 & 2 & 44 & 528 & 261 & 7.1 & 4.2 & 1 & 5.17e-22 & 13.5 & 2.81 & 10.09 \\
8135-1902  & 114.09638  &  39.43827 & 0.118 & 11.16 & 1.433 & 50 & 2.55e-17 & 2.43e+05 & 4.96e+09 & -- & -- & -- & -- & 1.4 & 1.1 & 2 & 2.92e-22 & 8.2 & 2.68 & 10.12 \\
8138-12703 & 116.38968  &  45.77232 & 0.032 & 10.96 & -0.385 & 50 & 3.24e-17 & 1.98e+04 & 4.03e+08 & 1 & 7 & 624 & 256 & 4.8 & 2.5 & 1 & 4.42e-22 & 10.6 & 2.68 & 9.03 \\
8146-1901  & 117.05386  &  28.22509 & 0.027 & 10.38 & 0.522 & 50 & 4.80e-17 & 2.12e+04 & 4.33e+08 & -- & -- & -- & -- & 2.2 & 1.4 & 2 & 5.99e-22 & 14.3 & 2.39 & 9.01 \\
8150-9102  & 149.71387  &  32.07306 & 0.027 & 10.72 & 0.376 & 50 & 2.80e-17 & 1.24e+04 & 2.52e+08 & 1 & -58 & 354 & 144 & 3.7 & 3.1 & 1 & 5.65e-22 & 13.4 & 2.41 & 8.78 \\
8153-1901  &  41.00622  &   0.17629 & 0.023 & 9.99 & -0.052 & 50 & 6.52e-17 & 2.00e+04 & 4.07e+08 & 1 & -17 & 350 & 143 & 5.2 & 3.7 & 1 & 9.82e-22 & 23.1 & 2.12 & 8.94 \\
8155-12701 &  53.17146  &  -1.18362 & 0.030 & 10.64 & 0.361 & 50 & 9.19e-17 & 5.17e+04 & 1.05e+09 & 2 & -11 & 270 & 112 & 9.0 & 6.5 & 1 & 8.28e-22 & 19.8 & 2.64 & 9.44 \\
8158-1901  &  60.85933   &  -5.49184 & 0.038 & 9.78 & -0.213 & 50 & 1.74e-17 & 1.57e+04 & 3.20e+08 & -- & -- & -- & -- & 1.8 & 1.2 & 2 & 3.22e-22 & 7.8 & 6.21 & 9.30 \\
8244-3701  & 132.38439  &  50.87412 & 0.027 & 10.48 & -1.331 & 50 & 4.75e-17 & 2.14e+04 & 4.36e+08 & -- & -- & -- & -- & 0.3 & 1.2 & 2 & 5.92e-22 & 14.1 & 2.68 & 9.07 \\
8249-3704  & 137.87476  &  45.46832 & 0.027 & 10.35 & -0.068 & 50 & 4.22e-17 & 1.83e+04 & 3.74e+08 & 1 & 3 & 958 & 393 & 3.8 & 3.6 & 1 & 4.93e-22 & 11.7 & 3.62 & 9.13 \\
8249-6102  & 137.33592  &  45.06551 & 0.051 & 10.53 & 0.909 & 50 & 3.25e-17 & 5.28e+04 & 1.08e+09 & 1 & -5 & 165 & 68 & 4.4 & 5.9 & 1 & 6.68e-22 & 16.7 & 4.64 & 9.70 \\
8250-6101  & 138.75314  &  42.02438 & 0.028 & 10.56 & 0.849 & 1 & 9.46e-17 & 4.44e+04 & 9.07e+08 & 2 & -27 & 293 & 142 & 24.7 & 3.7 & 1 & 2.51e-21 & 59.9 & 2.95 & 9.43 \\
8252-3701  & 144.84611  &  47.12686 & 0.027 & 10.63 & -0.364 & 50 & 5.87e-17 & 2.63e+04 & 5.37e+08 & 1 & 4 & 764 & 315 & 4.7 & 3.5 & 1 & 6.97e-22 & 16.6 & 2.03 & 9.04 \\
8252-9101  & 144.69238  &  48.56287 & 0.025 & 10.29 & -0.221 & 50 & 4.15e-17 & 1.53e+04 & 3.13e+08 & 2 & 14 & 318 & 182 & 6.7 & 3.6 & 1 & 5.32e-22 & 12.6 & 2.50 & 8.89 \\
8252-9102  & 145.54153  &  48.01286 & 0.056 & 10.81 & 1.195 & 50 & 2.94e-17 & 5.83e+04 & 1.19e+09 & 1 & -17 & 197 & 82 & 5.5 & 4.5 & 1 & 4.83e-22 & 12.2 & 6.64 & 9.90 \\
8253-12704 & 159.15328  &  43.50677 & 0.025 & 10.24 & -0.975 & 50 & 3.11e-17 & 1.13e+04 & 2.31e+08 & 1 & -28 & 305 & 126 & 5.0 & 3.7 & 1 & 5.05e-22 & 12.0 & 2.27 & 8.72 \\
8255-12704 & 165.11773  &  44.26096 & 0.025 & 10.18 & 0.190 & 50 & 4.89e-17 & 1.87e+04 & 3.81e+08 & 1 & 51 & 452 & 186 & 6.7 & 4.8 & 1 & 5.28e-22 & 12.6 & 2.99 & 9.06 \\
8257-12701 & 165.49581  &  45.22802 & 0.020 & 10.63 & 0.813 & 2 & 1.91e-16 & 4.56e+04 & 9.31e+08 & 1 & 27 & 164 & 67 & 29.7 & 5.8 & 1 & 4.45e-21 & 104.3 & 3.07 & 9.46 \\
8257-9102  & 166.76743  &  45.82213 & 0.025 & 10.64 & 0.041 & 50 & 7.41e-17 & 2.83e+04 & 5.78e+08 & 2 & -27 & 476 & 270 & 6.3 & 3.7 & 1 & 8.23e-22 & 19.5 & 2.24 & 9.11 \\
8259-3704  & 179.59330  &  43.81527 & 0.070 & 10.88 & 0.831 & 50 & 2.09e-17 & 6.58e+04 & 1.34e+09 & 1 & -140 & 520 & 213 & 3.0 & 2.6 & 1 & 4.57e-22 & 11.8 & 1.94 & 9.42 \\
8262-3702  & 183.65983  &  43.53621 & 0.024 & 10.39 & 0.302 & 50 & 1.39e-17 & 4.90e+03 & 9.99e+07 & 1 & -69 & 165 & 68 & 2.9 & 2.9 & 1 & 4.73e-22 & 11.2 & 5.45 & 8.74 \\
8262-9102  & 184.55356  &  44.17324 & 0.025 & 10.66 & 0.894 & 1 & 1.04e-16 & 3.76e+04 & 7.67e+08 & 1 & 47 & 282 & 116 & 29.5 & 4.6 & 1 & 2.44e-21 & 57.8 & 2.38 & 9.26 \\
8311-3703  & 205.01217  &  23.14297 & 0.032 & 10.84 & 0.561 & 50 & 1.33e-16 & 8.09e+04 & 1.65e+09 & 2 & -20 & 324 & 193 & 13.0 & 8.0 & 1 & 8.76e-22 & 21.0 & 1.99 & 9.52 \\
8313-12702 & 240.67741  &  41.19726 & 0.033 & 10.73 & 0.861 & 50 & 1.25e-16 & 8.50e+04 & 1.73e+09 & 2 & 2 & 282 & 150 & 9.2 & 5.8 & 1 & 1.17e-21 & 28.1 & 4.13 & 9.85 \\
8313-6102  & 240.91243  &  41.15305 & 0.034 & 10.73 & -1.962 & 50 & 9.99e-17 & 7.10e+04 & 1.45e+09 & -- & -- & -- & -- & 0.5 & 0.9 & 2 & 1.24e-21 & 29.4 & 2.68 & 9.59 \\
8317-6102  & 194.92502  &  43.75317 & 0.058 & 11.13 & 1.120 & 50 & 5.86e-17 & 1.23e+05 & 2.52e+09 & 2 & -16 & 475 & 294 & 8.1 & 4.3 & 1 & 5.17e-22 & 13.1 & 2.05 & 9.71 \\
8318-12702 & 196.44403  &  46.46185 & 0.025 & 10.62 & -0.164 & 50 & 1.52e-16 & 5.65e+04 & 1.15e+09 & 2 & -56 & 361 & 210 & 8.7 & 5.6 & 1 & 1.37e-21 & 32.4 & 3.34 & 9.59 \\
8318-9101  & 196.08628  &  45.05665 & 0.028 & 10.30 & 0.582 & 50 & 2.89e-17 & 1.40e+04 & 2.85e+08 & -- & -- & -- & -- & 1.9 & 1.5 & 2 & 5.40e-22 & 12.8 & 6.90 & 9.29 \\
8319-9101  & 202.53348  &  48.89326 & 0.071 & 11.06 & 1.144 & 50 & 3.19e-17 & 1.02e+05 & 2.09e+09 & 1 & 9 & 178 & 74 & 6.8 & 7.4 & 1 & 4.13e-22 & 10.7 & 2.14 & 9.65 \\
8320-9101  & 206.31385  &  23.31651 & 0.030 & 10.48 & 0.684 & 50 & 6.36e-17 & 3.40e+04 & 6.93e+08 & 1 & 29 & 242 & 99 & 5.4 & 4.4 & 1 & 1.06e-21 & 25.3 & 6.34 & 9.64 \\
8322-1901  & 198.78424  &  30.40377 & 0.023 & 10.38 & 0.458 & 50 & 1.01e-16 & 3.25e+04 & 6.63e+08 & 1 & 150 & 657 & 270 & 5.9 & 4.0 & 1 & 9.67e-22 & 22.8 & 2.26 & 9.18 \\
8322-3701  & 199.06648  &  30.26453& 0.049 & 11.01 & 1.142 & 50 & 6.44e-17 & 9.71e+04 & 1.98e+09 & 2 & 9 & 527 & 270 & 4.2 & 2.7 & 1 & 9.45e-22 & 23.2 & 4.37 & 9.94 \\
8326-6102  & 215.01790  &  47.12133 & 0.070 & 11.03 & 0.744 & 50 & 3.72e-17 & 1.18e+05 & 2.41e+09 & -- & -- & -- & -- & 2.2 & 2.7 & 2 & 4.45e-22 & 11.5 & 3.65 & 9.95 \\
\hline 
\end{tabular}
\end{table}
\end{landscape}

\begin{landscape}
\begin{table}
\scriptsize
\begin{tabular}{lcccccccccccccccccccc}
\hline
MaNGA ID & R.A. & Dec. & z & $\log(\rm{M})$ & $\log(\rm{SFR})$ & $dv$ & S$_{\rm{CO}}$ & L$_{\rm{CO}}$ & L$_{\rm{CO}}$ & N$_{\rm{Gauss}}$ & v$_{\rm{med}}$ & W$_{90}$ & W$_{50}$ & S/N & S/N$_{\rm{peak}}$  & flag$_{\rm{CO}}$ & $\sigma_{\rm{rms, \lambda}}$ & $\sigma_{\rm{rms,\nu}}$ & $\alpha_{\rm{CO}}$ & $\log(\rm{M_{H_{2}}})$ \\
 & deg. & deg. &  &  $\log(\rm{M}_{\odot})$ & $\log(\rm{M}_{\odot}/yr)$ & km/s & erg s$^{-1}$ cm$^{-2}$ & L$_{\odot}$ & K km s$^{-1}$ pc$^2$ &  & km/s & km/s & km/s &  &  &  & erg s$^{-1}$ cm$^{-2}$ \AA$^{-1}$ & mJy  & M$_{\odot} $(K km s$^{-1}$ pc$^2$)$^{-1}$ &  \\
\hline
8329-12701 & 213.41682 &  43.86656 & 0.035 & 10.84 & 0.832 & 50 & 1.26e-16 & 9.42e+04 & 1.92e+09 & 1 & 27 & 511 & 210 & 8.7 & 5.0 & 1 & 9.08e-22 & 21.9 & 2.05 & 9.59 \\
8330-12703 & 203.37460 &  40.52967 & 0.027 & 10.42 & 0.394 & 50 & 2.64e-17 & 1.15e+04 & 2.36e+08 & 1 & 13 & 234 & 96 & 4.3 & 3.1 & 1 & 5.72e-22 & 13.7 & 2.31 & 8.74 \\
8331-3703  & 204.15019 &  43.37336 & 0.053 & 10.55 & 0.622 & 50 & 3.47e-17 & 6.00e+04 & 1.22e+09 & 1 & -100 & 235 & 97 & 3.0 & 2.7 & 1 & 1.07e-21 & 26.6 & 2.44 & 9.47 \\
8331-6102  & 205.91886 &  41.69582 & 0.075 & 11.18 & 0.925 & 50 & 5.08e-17 & 1.85e+05 & 3.78e+09 & 2 & 35 & 254 & 127 & 6.5 & 5.0 & 1 & 6.07e-22 & 15.8 & 2.21 & 9.92 \\
8332-3702  & 207.87280 &  43.80642 & 0.033 & 10.56 & 0.650 & 50 & 2.75e-17 & 1.86e+04 & 3.79e+08 & 1 & 27 & 332 & 136 & 4.3 & 2.8 & 1 & 5.45e-22 & 13.1 & 3.64 & 9.14 \\
8332-9102  & 210.24094 &  42.85564 & 0.032 & 10.60 & 0.060 & 50 & 1.17e-16 & 7.43e+04 & 1.52e+09 & 1 & 17 & 460 & 188 & 12.3 & 7.1 & 1 & 6.68e-22 & 16.1 & 2.66 & 9.61 \\
8335-12705 & 216.75986 &  39.95721 & 0.025 & 10.52 & 0.562 & 50 & 8.51e-17 & 3.28e+04 & 6.69e+08 & 1 & 20 & 296 & 121 & 6.5 & 4.8 & 1 & 1.06e-21 & 25.1 & 2.80 & 9.27 \\
8338-12701 & 172.05119 &  21.99684 & 0.021 & 9.96 & 0.178 & 50 & 2.03e-17 & 5.39e+03 & 1.10e+08 & 1 & 3 & 166 & 68 & 3.8 & 4.4 & 1 & 5.36e-22 & 12.7 & 10.07 & 9.04 \\
8341-12704 & 189.21325 &  45.65117 & 0.030 & 10.72 & 0.724 & 1 & 5.03e-17 & 2.81e+04 & 5.72e+08 & 1 & 69 & 164 & 67 & 22.2 & 5.0 & 1 & 1.91e-21 & 45.7 & 2.68 & 9.19 \\
8439-12704 & 144.03108 &  50.43922 & 0.064 & 10.86 & 0.799 & 50 & 5.27e-17 & 1.37e+05 & 2.80e+09 & 1 & 33 & 503 & 206 & 7.4 & 4.1 & 1 & 4.50e-22 & 11.5 & 2.79 & 9.89 \\
8440-6104  & 135.75896 &  40.43398 & 0.029 & 10.67 & -0.056 & 50 & 5.46e-17 & 2.73e+04 & 5.57e+08 & -- & -- & -- & -- & 2.3 & 2.5 & 2 & 6.80e-22 & 16.2 & 2.68 & 9.17 \\
8442-12701 & 197.66878 &  32.22818 & 0.049 & 10.53 & 0.804 & 50 & 4.07e-17 & 6.06e+04 & 1.24e+09 & 1 & 62 & 277 & 113 & 5.1 & 3.9 & 1 & 6.83e-22 & 16.9 & 4.62 & 9.76 \\
8442-6101  & 200.27335 &  31.83129 & 0.061 & 10.79 & 0.777 & 50 & 3.39e-17 & 8.09e+04 & 1.65e+09 & 1 & 20 & 390 & 160 & 4.5 & 3.3 & 1 & 5.57e-22 & 14.1 & 2.09 & 9.54 \\
8446-1901  & 205.75333 &  36.16565 & 0.024 & 10.04 & -0.549 & 50 & 1.28e-16 & 4.28e+04 & 8.74e+08 & -- & -- & -- & -- & 0.5 & 0.5 & 2 & 2.39e-21 & 55.7 & 4.91 & 9.63 \\
8450-6102  & 171.74883 &  21.14167 & 0.042 & 10.43 & 0.818 & 50 & 1.87e-17 & 2.01e+04 & 4.10e+08 & 1 & -3 & 165 & 68 & 3.0 & 4.6 & 1 & 5.53e-22 & 13.5 & 3.88 & 9.20 \\
8452-12705 & 157.93769 &  46.67167 & 0.025 & 10.33 & 0.394 & 50 & 5.55e-17 & 2.08e+04 & 4.24e+08 & 2 & -93 & 277 & 144 & 5.1 & 4.4 & 1 & 8.79e-22 & 20.8 & 4.25 & 9.26 \\
8452-1902  & 157.77930 &  48.01483 & 0.059 & 10.76 & 0.869 & 50 & 1.16e-17 & 2.53e+04 & 5.16e+08 & 1 & 63 & 181 & 75 & 2.8 & 2.8 & 1 & 4.01e-22 & 10.1 & 6.15 & 9.50 \\
8453-3704  & 154.48083 &  46.60328 & 0.030 & 10.18 & -0.093 & 50 & 1.56e-17 & 8.51e+03 & 1.74e+08 & 1 & 39 & 193 & 80 & 3.6 & 3.5 & 1 & 4.43e-22 & 10.7 & 3.03 & 8.72 \\
8454-12702 & 153.04902 &  45.14215 & 0.076 & 11.13 & 1.444 & 50 & 3.47e-17 & 1.31e+05 & 2.68e+09 & 1 & 194 & 609 & 249 & 4.4 & 2.7 & 1 & 4.68e-22 & 12.2 & 4.07 & 10.04 \\
8455-3701  & 157.17945 &  39.83888 & 0.029 & 10.34 & 0.340 & 50 & 4.16e-17 & 2.16e+04 & 4.41e+08 & 1 & -12 & 361 & 148 & 5.1 & 4.3 & 1 & 6.02e-22 & 14.4 & 2.92 & 9.11 \\
8459-1901  & 147.80179 &  44.00930 & 0.016 & 9.71 & -0.271 & 50 & 2.81e-17 & 4.11e+03 & 8.39e+07 & 1 & 24 & 259 & 106 & 4.2 & 3.0 & 1 & 5.76e-22 & 13.4 & 2.68 & 8.35 \\
8462-9101  & 144.52643 &  36.0392  & 0.022 & 10.07 & -0.074 & 50 & 2.55e-17 & 7.74e+03 & 1.58e+08 & 1 & 58 & 229 & 95 & 3.8 & 2.9 & 1 & 6.94e-22 & 16.4 & 2.18 & 8.54 \\
8464-3704  & 188.10783 &  44.18361 & 0.031 & 10.60 & 0.708 & 50 & 6.51e-17 & 3.74e+04 & 7.64e+08 & 1 & 17 & 165 & 68 & 7.2 & 8.0 & 1 & 7.75e-22 & 18.6 & 1.85 & 9.15 \\
8465-9102  & 198.18916 &  46.93498 & 0.028 & 10.40 & 0.100 & 50 & 6.29e-17 & 2.91e+04 & 5.93e+08 & 1 & 54 & 536 & 219 & 5.8 & 3.6 & 1 & 6.83e-22 & 16.2 & 2.60 & 9.19 \\
8466-12702 & 167.96743 &  45.54139 & 0.028 & 10.40 & 0.516 & 50 & 4.51e-17 & 2.21e+04 & 4.50e+08 & 2 & 16 & 312 & 169 & 5.6 & 5.0 & 1 & 6.36e-22 & 15.2 & 3.62 & 9.21 \\
8483-6101  & 246.45691 &  48.10033 & 0.020 & 10.44 & 0.251 & 50 & 1.01e-16 & 2.37e+04 & 4.84e+08 & 1 & 29 & 362 & 149 & 5.2 & 3.9 & 1 & 1.49e-21 & 34.9 & 2.00 & 8.99 \\
8547-6102  & 217.32486 &  52.66555 & 0.030 & 10.54 & 0.293 & 50 & 4.24e-17 & 2.39e+04 & 4.87e+08 & 1 & 33 & 361 & 148 & 4.5 & 2.9 & 1 & 7.02e-22 & 16.8 & 2.07 & 9.00 \\
8548-12704 & 244.07472 &  47.98160 & 0.021 & 10.27 & 0.082 & 50 & 6.34e-17 & 1.61e+04 & 3.29e+08 & 1 & -18 & 267 & 110 & 5.7 & 3.9 & 1 & 1.03e-21 & 24.2 & 4.02 & 9.12 \\
8550-12703 & 247.67443 &  40.52938 & 0.030 & 10.49 & 0.492 & 50 & 6.67e-17 & 3.59e+04 & 7.33e+08 & 1 & 42 & 165 & 67 & 6.5 & 7.8 & 1 & 8.95e-22 & 21.3 & 2.15 & 9.20 \\
8551-12705 & 233.94092 &  44.83479 & 0.030 & 10.59 & 0.920 & 50 & 6.86e-17 & 3.65e+04 & 7.45e+08 & 2 & 9 & 387 & 92 & 6.0 & 6.6 & 1 & 8.16e-22 & 19.4 & 4.06 & 9.48 \\
8552-12701 & 226.43166 &  44.40490 & 0.028 & 10.66 & 0.163 & 50 & 5.44e-17 & 2.64e+04 & 5.39e+08 & 2 & -101 & 484 & 334 & 6.2 & 4.0 & 1 & 6.64e-22 & 15.8 & 3.89 & 9.32 \\
8553-1901  & 233.96834 &  57.90263 & 0.030 & 10.77 & 0.671 & 50 & 4.48e-17 & 2.48e+04 & 5.06e+08 & -- & -- & -- & -- & 2.4 & 1.8 & 2 & 5.57e-22 & 13.3 & 2.68 & 9.13 \\
8553-9102  & 234.54184 &  57.60365 & 0.074 & 11.24 & 1.097 & 50 & 4.69e-17 & 1.64e+05 & 3.35e+09 & 2 & 19 & 427 & 205 & 5.4 & 3.5 & 1 & 5.75e-22 & 14.9 & 2.18 & 9.86 \\
8588-12705 & 250.31287 &  39.75235 & 0.030 & 10.63 & 0.727 & 50 & 6.57e-17 & 3.62e+04 & 7.38e+08 & 1 & -2 & 417 & 172 & 5.4 & 2.8 & 1 & 9.12e-22 & 21.9 & 2.76 & 9.31 \\
8588-6101  & 248.45675 &  39.26320 & 0.032 & 10.88 & 0.599 & 50 & 8.69e-17 & 5.33e+04 & 1.09e+09 & 1 & 36 & 195 & 80 & 9.5 & 8.5 & 1 & 8.77e-22 & 21.0 & 2.21 & 9.38 \\
8592-6101  & 222.67177 &  51.70446 & 0.026 & 10.54 & 0.376 & 50 & 7.13e-17 & 2.90e+04 & 5.92e+08 & 2 & 34 & 337 & 188 & 9.1 & 5.6 & 1 & 6.39e-22 & 15.2 & 3.16 & 9.27 \\
8595-3703  & 221.43798 &  51.58081 & 0.030 & 10.91 & 0.362 & 50 & 1.43e-16 & 7.59e+04 & 1.55e+09 & 1 & 104 & 566 & 232 & 7.5 & 3.8 & 1 & 1.27e-21 & 30.4 & 2.61 & 9.61 \\
8595-6104  & 219.39661 &  51.58313 & 0.044 & 10.87 & -0.364 & 50 & 2.70e-17 & 3.18e+04 & 6.49e+08 & 1 & 4 & 165 & 68 & 5.0 & 5.3 & 1 & 5.10e-22 & 12.6 & 2.68 & 9.24 \\
8600-1901  & 242.58522 &  43.00964 & 0.025 & 10.19 & 0.493 & 50 & 1.52e-17 & 5.81e+03 & 1.18e+08 & 1 & -10 & 166 & 68 & 2.8 & 3.0 & 1 & 5.86e-22 & 13.9 & 4.84 & 8.76 \\
8600-3704  & 245.84527 &  41.65118 & 0.027 & 10.49 & 0.640 & 50 & 9.36e-17 & 4.22e+04 & 8.61e+08 & 2 & 7 & 284 & 151 & 9.5 & 5.6 & 1 & 8.61e-22 & 20.5 & 2.68 & 9.36 \\
8601-12702 & 247.59167 &  40.92424 & 0.031 & 9.65 & -0.413 & 50 & 3.23e-17 & 1.83e+04 & 3.73e+08 & 1 & 26 & 174 & 71 & 5.8 & 4.7 & 1 & 5.87e-22 & 14.0 & 2.68 & 9.00 \\
8603-12704 & 247.89387 &  40.56559 & 0.027 & 10.78 & 0.702 & 50 & 1.46e-16 & 6.59e+04 & 1.34e+09 & 2 & -9 & 355 & 204 & 15.6 & 11.9 & 1 & 7.47e-22 & 17.8 & 2.68 & 9.56 \\
8603-6103  & 247.80036 &  40.42187 & 0.027 & 10.14 & 0.113 & 50 & 4.36e-17 & 1.92e+04 & 3.92e+08 & 1 & 64 & 314 & 128 & 5.5 & 3.9 & 1 & 6.26e-22 & 14.9 & 2.68 & 9.02 \\
8603-6104  & 247.41995 &  40.68695 & 0.031 & 10.55 & -0.004 & 50 & 3.02e-17 & 1.71e+04 & 3.48e+08 & -- & -- & -- & -- & 2.7 & 2.7 & 2 & 3.75e-22 & 9.0 & 3.52 & 9.09 \\
8604-12702 & 246.65449 &  39.12754 & 0.035 & 10.72 & 0.744 & 50 & 5.04e-17 & 3.84e+04 & 7.84e+08 & 1 & 41 & 382 & 157 & 6.0 & 4.1 & 1 & 6.21e-22 & 15.0 & 3.32 & 9.42 \\
8604-9102  & 246.45535 &  40.34520 & 0.029 & 10.49 & 0.882 & 50 & 3.82e-17 & 1.95e+04 & 3.97e+08 & 1 & -13 & 382 & 157 & 6.2 & 4.2 & 1 & 4.71e-22 & 11.2 & 7.86 & 9.49 \\
8615-3701  & 321.00791 &  -0.36632 & 0.062 & 10.86 & 0.808 & 50 & 2.84e-17 & 6.88e+04 & 1.40e+09 & 1 & 108 & 248 & 101 & 3.8 & 3.7 & 1 & 6.28e-22 & 15.9 & 4.09 & 9.76 \\
8618-3704  & 318.86228 &   9.75781 & 0.070 & 10.94 & 0.889 & 50 & 3.65e-17 & 1.16e+05 & 2.36e+09 & 1 & -16 & 203 & 83 & 6.7 & 6.7 & 1 & 5.08e-22 & 13.1 & 2.17 & 9.71 \\
8624-12702 & 260.58171 &  59.07653 & 0.030 & 10.57 & 0.598 & 50 & 7.22e-17 & 3.95e+04 & 8.06e+08 & 1 & 9 & 412 & 168 & 5.2 & 3.3 & 1 & 1.03e-21 & 24.6 & 3.16 & 9.41 \\
8624-12703 & 264.23953 &  59.20028 & 0.031 & 10.56 & 0.405 & 50 & 6.97e-17 & 3.98e+04 & 8.13e+08 & 1 & -2 & 383 & 157 & 3.1 & 1.9 & 1 & 1.69e-21 & 40.0 & 2.67 & 9.34 \\
8626-12701 & 263.52527 &  56.79961 & 0.029 & 10.46 & -0.433 & 50 & 3.98e-17 & 2.06e+04 & 4.21e+08 & -- & -- & -- & -- & 2.1 & 1.6 & 2 & 4.96e-22 & 11.8 & 4.73 & 9.30 \\
8626-12702 & 263.20536 &  56.90127 & 0.029 & 10.64 & 0.485 & 50 & 6.81e-17 & 3.56e+04 & 7.25e+08 & 2 & 19 & 335 & 164 & 7.9 & 5.1 & 1 & 6.79e-22 & 16.3 & 3.19 & 9.36 \\
8626-12704 & 263.75521 &  57.05243 & 0.047 & 10.70 & 0.811 & 50 & 2.95e-17 & 4.09e+04 & 8.34e+08 & -- & -- & -- & -- & 0.2 & 1.1 & 2 & 5.41e-22 & 13.3 & 35.63 & 10.47 \\
8626-3702  & 261.72122 &  56.84880 & 0.028 & 10.56 & 0.525 & 50 & 1.03e-16 & 4.82e+04 & 9.83e+08 & 2 & -19 & 456 & 258 & 10.4 & 4.5 & 1 & 7.22e-22 & 17.2 & 2.16 & 9.33 \\
8626-3703  & 264.66254 &  56.82424 & 0.029 & 10.33 & 0.297 & 50 & 4.11e-17 & 2.15e+04 & 4.39e+08 & 2 & 11 & 254 & 126 & 5.5 & 4.8 & 1 & 5.57e-22 & 13.3 & 2.60 & 9.06 \\
8655-3701  & 356.75182 &  -0.44738 & 0.071 & 11.16 & 1.282 & 50 & 8.30e-17 & 2.70e+05 & 5.51e+09 & 1 & 7 & 335 & 137 & 6.5 & 4.4 & 1 & 1.01e-21 & 26.3 & 2.51 & 10.14 \\
\hline
\end{tabular}
\end{table}
\end{landscape}

\begin{landscape}
\begin{table}
\scriptsize
\begin{tabular}{lcccccccccccccccccccc}
\hline
MaNGA ID & R.A. & Dec. & z & $\log(\rm{M})$ & $\log(\rm{SFR})$ & $dv$ & S$_{\rm{CO}}$ & L$_{\rm{CO}}$ & L$_{\rm{CO}}$ & N$_{\rm{Gauss}}$ & v$_{\rm{med}}$ & W$_{90}$ & W$_{50}$ & S/N & S/N$_{\rm{peak}}$  & flag$_{\rm{CO}}$ & $\sigma_{\rm{rms, \lambda}}$ & $\sigma_{\rm{rms,\nu}}$ & $\alpha_{\rm{CO}}$ & $\log(\rm{M_{H_{2}}})$ \\
 & deg. & deg. &  &  $\log(\rm{M}_{\odot})$ & $\log(\rm{M}_{\odot}/yr)$ & km/s & erg s$^{-1}$ cm$^{-2}$ & L$_{\odot}$ & K km s$^{-1}$ pc$^2$ &  & km/s & km/s & km/s &  &  &  & erg s$^{-1}$ cm$^{-2}$ \AA$^{-1}$ & mJy  & M$_{\odot} $(K km s$^{-1}$ pc$^2$)$^{-1}$ &  \\
\hline
8712-6101  & 118.72692 & 53.84627  & 0.035 & 10.91 & 1.273 & 1 & 1.43e-16 & 1.06e+05 & 2.17e+09 & 1 & 16 & 249 & 103 & 38.2 & 5.8 & 1 & 2.65e-21 & 64.0 & 2.26 & 9.69 \\
8713-6104  & 118.31731 & 39.05012  & 0.041 & 10.93 & 0.936 & 50 & 7.49e-17 & 7.70e+04 & 1.57e+09 & 2 & 26 & 275 & 142 & 7.5 & 5.7 & 1 & 8.01e-22 & 19.6 & 2.41 & 9.58 \\
8713-9102  & 118.85539 & 39.18609  & 0.033 & 10.52 & 0.414 & 50 & 3.25e-17 & 2.17e+04 & 4.44e+08 & 2 & 108 & 509 & 237 & 5.6 & 4.2 & 1 & 3.62e-22 & 8.7 & 2.68 & 9.08 \\
8715-3702  & 119.92067 & 50.83997  & 0.054 & 10.79 & 0.965 & 50 & 1.33e-17 & 2.47e+04 & 5.04e+08 & 1 & -3 & 184 & 76 & 3.2 & 2.5 & 1 & 4.35e-22 & 10.9 & 2.68 & 9.13 \\
8717-3704  & 117.51833 & 34.47931  & 0.029 & 10.60 & 0.456 & 1 & 1.48e-16 & 7.52e+04 & 1.53e+09 & 2 & -37 & 319 & 184 & 38.9 & 6.0 & 1 & 2.46e-21 & 58.8 & 2.01 & 9.49 \\
8727-12705 &  57.10409  & -6.62606 & 0.021 & 10.42 & 0.672 & 50 & 1.51e-16 & 3.97e+04 & 8.09e+08 & 1 & 9 & 165 & 67 & 12.1 & 19.0 & 1 & 1.10e-21 & 25.9 & 3.28 & 9.42 \\
8727-3701  &  54.90496  & -5.63293 & 0.021 & 10.19 & -0.161 & 50 & 1.68e-17 & 4.39e+03 & 8.96e+07 & -- & -- & -- & -- & 1.8 & 1.6 & 2 & 3.16e-22 & 7.4 & 3.11 & 8.44 \\
8932-9102  & 196.65170 & 27.87294  & 0.021 & 10.16 & 0.138 & 50 & 6.23e-17 & 1.63e+04 & 3.33e+08 & 1 & -1 & 428 & 176 & 3.2 & 2.2 & 1 & 1.46e-21 & 34.2 & 3.44 & 9.06 \\
8935-6104  & 195.53290 & 27.64831  & 0.023 & 10.28 & 0.314 & 50 & 5.73e-17 & 1.82e+04 & 3.71e+08 & 1 & -43 & 217 & 89 & 5.6 & 5.0 & 1 & 9.24e-22 & 21.7 & 2.34 & 8.94 \\
8940-1902  & 120.98712 & 25.48072  & 0.073 & 11.10 & 1.276 & 50 & 5.85e-17 & 2.00e+05 & 4.07e+09 & 1 & 10 & 403 & 166 & 7.5 & 4.1 & 1 & 5.46e-22 & 14.2 & 2.31 & 9.97 \\
8944-6103  & 147.65106 & 34.77714  & 0.053 & 10.75 & 0.775 & 50 & 4.30e-17 & 7.44e+04 & 1.52e+09 & 2 & -3 & 338 & 210 & 5.5 & 4.5 & 1 & 5.78e-22 & 14.5 & 2.05 & 9.49 \\
8945-12701 & 171.89808 & 47.37939  & 0.033 & 10.30 & 0.708 & 50 & 1.27e-17 & 8.28e+03 & 1.69e+08 & 1 & 13 & 165 & 67 & 2.3 & 3.4 & 1 & 5.48e-22 & 13.1 & 7.24 & 9.09 \\
8946-6101  & 168.75209 & 46.63452  & 0.054 & 10.73 & 0.824 & 50 & 3.60e-17 & 6.63e+04 & 1.35e+09 & 2 & 26 & 316 & 154 & 8.0 & 5.6 & 1 & 3.55e-22 & 8.9 & 2.08 & 9.45 \\
8947-12701 & 171.02709 & 48.69629  & 0.058 & 11.14 & 0.217 & 50 & 5.24e-17 & 1.10e+05 & 2.24e+09 & 2 & -73 & 611 & 364 & 6.4 & 3.6 & 1 & 5.34e-22 & 13.5 & 2.98 & 9.82 \\
8947-1901  & 171.38369 & 51.18040  & 0.027 & 10.38 & 0.556 & 50 & 6.38e-17 & 2.83e+04 & 5.78e+08 & 1 & 28 & 227 & 93 & 6.4 & 4.8 & 1 & 9.30e-22 & 22.1 & 2.84 & 9.22 \\
8947-9101  & 168.73450 & 50.33487  & 0.047 & 10.90 & 0.963 & 50 & 4.46e-17 & 6.15e+04 & 1.26e+09 & 1 & 18 & 187 & 76 & 5.3 & 5.2 & 1 & 8.15e-22 & 20.1 & 2.68 & 9.53 \\
8950-12705 & 194.73313 & 27.83344  & 0.025 & 10.53 & -0.060 & 50 & 9.74e-17 & 3.69e+04 & 7.53e+08 & 1 & 51 & 534 & 219 & 6.3 & 3.4 & 1 & 1.03e-21 & 24.5 & 2.06 & 9.19 \\
8952-6104  & 204.93397 & 27.77647  & 0.028 & 10.33 & 0.459 & 50 & 5.05e-17 & 2.43e+04 & 4.96e+08 & 1 & 11 & 165 & 67 & 6.6 & 10.1 & 1 & 6.38e-22 & 15.2 & 3.22 & 9.20 \\
8978-3701  & 248.50746 & 41.34794  & 0.028 & 10.34 & 0.150 & 50 & 3.73e-17 & 1.82e+04 & 3.71e+08 & 1 & 29 & 248 & 101 & 5.0 & 3.9 & 1 & 7.14e-22 & 17.0 & 2.68 & 9.00 \\
8979-3704  & 244.42296 & 40.93381  & 0.062 & 10.68 & 0.283 & 50 & 1.69e-17 & 4.13e+04 & 8.42e+08 & 1 & 40 & 269 & 110 & 3.4 & 3.2 & 1 & 5.56e-22 & 14.1 & 3.11 & 9.42 \\
8979-6102  & 241.82338 & 41.40360  & 0.035 & 10.53 & 0.396 & 50 & 2.73e-17 & 2.00e+04 & 4.07e+08 & 2 & 19 & 303 & 169 & 3.8 & 3.0 & 1 & 5.42e-22 & 13.1 & 2.44 & 9.00 \\
8982-9101  & 201.71365 & 26.59123  & 0.023 & 10.24 & 0.340 & 50 & 1.81e-17 & 5.98e+03 & 1.22e+08 & 1 & -84 & 335 & 137 & 2.8 & 2.0 & 1 & 5.68e-22 & 13.4 & 4.49 & 8.74 \\
8987-3701  & 136.24988 & 28.34772  & 0.049 & 10.37 & 0.674 & 50 & 1.11e-17 & 1.63e+04 & 3.33e+08 & 1 & -4 & 281 & 115 & 2.9 & 3.0 & 1 & 3.71e-22 & 9.2 & 3.40 & 9.05 \\
8989-9102  & 177.52435 & 50.46744  & 0.024 & 10.27 & 0.111 & 50 & 4.49e-17 & 1.49e+04 & 3.05e+08 & 1 & 16 & 267 & 110 & 7.2 & 4.6 & 1 & 5.72e-22 & 13.5 & 1.89 & 8.76 \\
8993-12705 & 165.39153 & 45.65386  & 0.029 & 10.46 & 0.374 & 50 & 1.03e-16 & 5.38e+04 & 1.10e+09 & 1 & -13 & 527 & 217 & 8.5 & 6.0 & 1 & 7.84e-22 & 18.7 & 4.54 & 9.70 \\
8993-6104  & 166.53203 & 45.12225  & 0.021 & 8.04 & -1.187 & 50 & 3.25e-17 & 8.75e+03 & 1.79e+08 & -- & -- & -- & -- & 1.9 & 2.7 & 2 & 6.12e-22 & 14.4 & 5.61 & 9.00 \\
8996-3703  & 172.30201 & 53.73396  & 0.027 & 10.44 & 0.345 & 50 & 6.36e-17 & 2.85e+04 & 5.81e+08 & 2 & 17 & 296 & 146 & 8.5 & 5.5 & 1 & 6.50e-22 & 15.5 & 2.09 & 9.08 \\
8997-12704 & 170.16277 & 52.62504  & 0.048 & 11.12 & 0.478 & 50 & 3.89e-17 & 5.66e+04 & 1.16e+09 & 2 & 185 & 704 & 514 & 4.1 & 2.8 & 1 & 5.86e-22 & 14.5 & 2.40 & 9.44 \\
9000-1901  & 171.40065 & 54.38257  & 0.021 & 10.34 & -0.075 & 50 & 1.91e-17 & 4.89e+03 & 9.98e+07 & 1 & -11 & 242 & 99 & 2.9 & 2.5 & 1 & 5.78e-22 & 13.5 & 2.68 & 8.43 \\
9024-1902  & 224.43778 & 33.16563  & 0.030 & 9.90 & 0.356 & 50 & 3.07e-17 & 1.68e+04 & 3.43e+08 & -- & -- & -- & -- & 2.3 & 2.3 & 2 & 5.73e-22 & 13.7 & 2.68 & 8.96 \\
9027-12701 & 243.93592 & 31.96395  & 0.032 & 10.77 & 1.045 & 50 & 1.35e-16 & 8.13e+04 & 1.66e+09 & 2 & 5 & 650 & 211 & 9.6 & 8.0 & 1 & 8.68e-22 & 20.8 & 2.68 & 9.65 \\
9029-12702 & 247.25172 & 41.28426  & 0.032 & 10.80 & 0.731 & 50 & 1.69e-16 & 1.05e+05 & 2.13e+09 & 1 & 22 & 164 & 67 & 13.8 & 14.6 & 1 & 1.16e-21 & 27.8 & 1.90 & 9.61 \\
9029-6102  & 246.39291 & 41.68139  & 0.028 & 10.03 & 0.298 & 50 & 3.28e-17 & 1.54e+04 & 3.15e+08 & 2 & -28 & 403 & 263 & 3.0 & 3.5 & 1 & 7.42e-22 & 17.7 & 3.76 & 9.07 \\
9034-1901  & 226.72354 & 47.05696  & 0.088 & 11.07 & 0.970 & 50 & 4.67e-17 & 2.38e+05 & 4.85e+09 & 2 & 12 & 265 & 114 & 9.3 & 7.0 & 1 & 4.08e-22 & 10.9 & 2.42 & 10.07 \\
9034-3702  & 226.28032 & 46.96365  & 0.038 & 10.18 & 0.500 & 50 & 1.73e-17 & 1.51e+04 & 3.08e+08 & -- & -- & -- & -- & 2.3 & 2.4 & 2 & 3.21e-22 & 7.8 & 6.17 & 9.28 \\
9036-9102  & 240.56931 & 42.91689  & 0.024 & 10.21 & 0.327 & 50 & 1.19e-16 & 4.26e+04 & 8.70e+08 & 2 & 36 & 233 & 108 & 10.2 & 8.0 & 1 & 1.09e-21 & 25.8 & 2.68 & 9.37 \\
9038-3702  & 237.51955 & 42.30863  & 0.021 & 10.05 & 0.250 & 50 & 2.93e-17 & 7.44e+03 & 1.52e+08 & 1 & 9 & 164 & 67 & 3.4 & 5.4 & 1 & 7.21e-22 & 16.9 & 3.33 & 8.70 \\
9041-12701 & 236.94151 & 28.64169  & 0.033 & 10.49 & 0.465 & 2 & 1.48e-16 & 9.94e+04 & 2.03e+09 & 1 & 24 & 193 & 80 & 25.2 & 5.5 & 1 & 3.80e-21 & 91.4 & 2.25 & 9.66 \\
9042-12703 & 235.15267 & 28.51243  & 0.033 & 10.71 & -0.013 & 50 & 6.67e-17 & 4.35e+04 & 8.88e+08 & 1 & 11 & 443 & 181 & 5.2 & 3.4 & 1 & 9.24e-22 & 22.3 & 2.11 & 9.27 \\
9042-6101  & 232.55107 & 27.13411  & 0.033 & 10.54 & -0.909 & 50 & 2.06e-17 & 1.33e+04 & 2.72e+08 & 1 & 173 & 382 & 156 & 3.8 & 1.9 & 1 & 4.49e-22 & 10.8 & 2.68 & 8.86 \\
9044-6101  & 230.68698 & 29.76960  & 0.023 & 10.24 & 0.275 & 50 & 6.46e-17 & 2.03e+04 & 4.15e+08 & 2 & 49 & 386 & 188 & 6.3 & 4.4 & 1 & 7.58e-22 & 17.9 & 2.46 & 9.01 \\
9047-6104  & 248.14087 & 26.38073  & 0.059 & 11.25 & 1.104 & 50 & 1.21e-16 & 2.63e+05 & 5.37e+09 & 2 & -41 & 586 & 294 & 8.6 & 3.9 & 1 & 9.12e-22 & 23.1 & 2.43 & 10.12 \\
9049-6104  & 248.46298 & 25.86004  & 0.051 & 10.80 & 0.837 & 50 & 2.62e-17 & 4.29e+04 & 8.76e+08 & 2 & 43 & 513 & 282 & 3.9 & 2.7 & 1 & 4.37e-22 & 10.9 & 1.99 & 9.24 \\
9050-3704  & 245.99500 & 22.39336  & 0.037 & 10.74 & 1.054 & 50 & 1.04e-16 & 8.70e+04 & 1.77e+09 & 1 & 2 & 233 & 96 & 8.0 & 6.6 & 1 & 1.17e-21 & 28.3 & 2.68 & 9.68 \\
9050-9101  & 245.99843 & 21.79503  & 0.032 & 10.61 & 0.577 & 50 & 5.13e-17 & 3.22e+04 & 6.57e+08 & 1 & -1 & 242 & 99 & 6.7 & 5.9 & 1 & 7.02e-22 & 16.9 & 2.00 & 9.12 \\
9085-12703 & 260.30627 & 29.18589  & 0.047 & 10.61 & 0.021 & 50 & 7.67e-17 & 1.06e+05 & 2.17e+09 & -- & -- & -- & -- & 2.6 & 2.2 & 2 & 9.38e-22 & 23.1 & 1.74 & 9.58 \\
9095-1901  & 242.81005 & 24.22502  & 0.033 & 10.62 & 0.396 & 50 & 7.78e-17 & 5.01e+04 & 1.02e+09 & 1 & 66 & 275 & 113 & 7.5 & 5.1 & 1 & 9.11e-22 & 21.9 & 1.72 & 9.24 \\
9095-9102  & 243.08488 & 23.00202  & 0.032 & 10.65 & 0.725 & 50 & 6.04e-17 & 3.82e+04 & 7.80e+08 & 2 & 43 & 295 & 148 & 6.0 & 3.9 & 1 & 8.30e-22 & 19.9 & 1.82 & 9.15 \\
9185-9101  & 256.21228 & 34.81733  & 0.056 & 11.21 & 1.354 & 50 & 9.97e-17 & 1.99e+05 & 4.07e+09 & 1 & 46 & 285 & 116 & 11.4 & 7.6 & 1 & 7.03e-22 & 17.7 & 3.20 & 10.11 \\
9195-3702  &  27.84278  & 13.06033  & 0.064 & 11.14 & 0.628 & 50 & 4.09e-17 & 1.07e+05 & 2.19e+09 & 2 & -29 & 583 & 337 & 5.8 & 3.3 & 1 & 4.24e-22 & 10.9 & 2.68 & 9.77 \\
9196-6103  & 260.35471 & 54.30492  & 0.030 & 10.55 & 0.255 & 50 & 4.12e-17 & 2.23e+04 & 4.54e+08 & 2 & 60 & 384 & 216 & 4.5 & 3.8 & 1 & 6.79e-22 & 16.2 & 2.05 & 8.97 \\
9196-6104  & 262.31771 & 54.20592  & 0.079 & 10.93 & 0.991 & 50 & 3.60e-17 & 1.46e+05 & 2.98e+09 & 1 & 58 & 300 & 122 & 3.4 & 2.7 & 1 & 8.68e-22 & 22.7 & 2.04 & 9.78 \\
9485-12705 & 121.77993 & 36.23347  & 0.032 & 10.86 & 0.548 & 50 & 7.44e-17 & 4.72e+04 & 9.64e+08 & 1 & -13 & 701 & 287 & 9.4 & 5.2 & 1 & 4.45e-22 & 10.7 & 3.22 & 9.49 \\
9485-1901  & 120.77803 & 37.02355  & 0.071 & 10.96 & 1.272 & 50 & 6.62e-17 & 2.16e+05 & 4.40e+09 & 1 & 6 & 165 & 68 & 9.9 & 12.7 & 1 & 5.49e-22 & 14.2 & 2.24 & 10.00 \\
9485-3701  & 119.53203 & 36.83126  & 0.039 & 10.13 & -0.162 & 50 & 1.53e-17 & 1.45e+04 & 2.95e+08 & 1 & -52 & 353 & 144 & 3.4 & 2.3 & 1 & 3.58e-22 & 8.7 & 2.01 & 8.77 \\
9486-12701 & 121.67826 & 39.09021  & 0.042 & 10.96 & 0.593 & 50 & 7.19e-17 & 7.64e+04 & 1.56e+09 & 2 & 34 & 446 & 262 & 7.3 & 5.7 & 1 & 6.86e-22 & 16.8 & 2.68 & 9.62 \\
9486-12702 & 121.12959 & 40.20604  & 0.040 & 10.84 & 0.571 & 50 & 6.68e-17 & 6.66e+04 & 1.36e+09 & 2 & 1 & 274 & 142 & 13.3 & 10.0 & 1 & 4.08e-22 & 9.9 & 2.56 & 9.54 \\
9487-3702  & 123.33054 & 46.14715  & 0.054 & 10.58 & -0.321 & 50 & 3.17e-17 & 5.76e+04 & 1.18e+09 & -- & -- & -- & -- & 2.0 & 2.5 & 2 & 3.85e-22 & 9.7 & 2.68 & 9.50 \\
9487-3703  & 124.81375 & 45.29875  & 0.043 & 10.33 & 0.673 & 50 & 2.02e-17 & 2.34e+04 & 4.78e+08 & 1 & 7 & 387 & 159 & 4.3 & 2.9 & 1 & 3.43e-22 & 8.4 & 3.93 & 9.27 \\
9487-9102  & 123.82032 & 46.07525  & 0.041 & 11.00 & 0.951 & 50 & 1.02e-16 & 1.05e+05 & 2.15e+09 & 1 & 112 & 317 & 130 & 12.0 & 8.5 & 1 & 6.79e-22 & 16.6 & 3.18 & 9.83 \\
9491-1902  & 120.04224 & 19.23433  & 0.078 & 11.08 & 0.720 & 50 & 3.10e-17 & 1.23e+05 & 2.51e+09 & 2 & -3 & 625 & 360 & 6.4 & 3.6 & 1 & 2.93e-22 & 7.7 & 2.68 & 9.83 \\
\hline
\end{tabular}
\end{table}
\end{landscape}

\begin{landscape}
\begin{table}
\scriptsize
\begin{tabular}{lcccccccccccccccccccc}
\hline
MaNGA ID & R.A. & Dec. & z & $\log(\rm{M})$ & $\log(\rm{SFR})$ & $dv$ & S$_{\rm{CO}}$ & L$_{\rm{CO}}$ & L$_{\rm{CO}}$ & N$_{\rm{Gauss}}$ & v$_{\rm{med}}$ & W$_{90}$ & W$_{50}$ & S/N & S/N$_{\rm{peak}}$  & flag$_{\rm{CO}}$ & $\sigma_{\rm{rms, \lambda}}$ & $\sigma_{\rm{rms,\nu}}$ & $\alpha_{\rm{CO}}$ & $\log(\rm{M_{H_{2}}})$ \\
 & deg. & deg. &  &  $\log(\rm{M}_{\odot})$ & $\log(\rm{M}_{\odot}/yr)$ & km/s & erg s$^{-1}$ cm$^{-2}$ & L$_{\odot}$ & K km s$^{-1}$ pc$^2$ &  & km/s & km/s & km/s &  &  &  & erg s$^{-1}$ cm$^{-2}$ \AA$^{-1}$ & mJy  & M$_{\odot} $(K km s$^{-1}$ pc$^2$)$^{-1}$ &  \\
\hline
9491-6101  & 119.174377409 & 17.99116 & 0.041 & 10.90 & 1.284 & 50 & 9.55e-17 & 1.00e+05 & 2.04e+09 & 2 & 3 & 417 & 244 & 9.1 & 5.0 & 1 & 7.37e-22 & 18.0 & 4.18 & 9.93 \\
9494-3704  & 128.019903751 & 21.62862 & 0.054 & 10.75 & 0.940 & 50 & 4.02e-17 & 7.29e+04 & 1.49e+09 & 1 & 7 & 397 & 163 & 7.4 & 4.1 & 1 & 3.91e-22 & 9.8 & 2.42 & 9.56 \\
9497-12705 & 118.074345115 & 19.59508 & 0.117 & 10.905 & 1.728 & 50 & 3.66e-17 & 3.42e+05 & 6.99e+09 & -- & -- & -- & -- & -1.2 & 0.3 & 2 & 4.19e-22 & 11.8 & 2.68 & 10.27 \\
9500-1901  & 131.725410562 & 25.37008 & 0.051 & 10.86 & 0.638 & 50 & 3.69e-17 & 5.95e+04 & 1.21e+09 & -- & -- & -- & -- & 0.6 & 1.3 & 2 & 4.50e-22 & 11.2 & 2.68 & 9.51 \\
9506-3701  & 133.569904521 & 27.26652 & 0.064 & 11.06 & 1.484 & 50 & 8.59e-17 & 2.22e+05 & 4.54e+09 & 1 & 4 & 188 & 77 & 14.5 & 13.3 & 1 & 5.31e-22 & 13.6 & 2.06 & 9.97 \\
9508-12705 & 127.273132937 & 25.77262 & 0.028 & 10.54 & 0.235 & 50 & 4.19e-17 & 1.96e+04 & 4.00e+08 & 1 & 31 & 552 & 226 & 4.8 & 3.5 & 1 & 5.63e-22 & 13.4 & 2.27 & 8.96 \\
9508-6101  & 127.13101957  & 26.34868 & 0.053 & 10.59 & 0.800 & 50 & 4.63e-17 & 8.15e+04 & 1.66e+09 & 1 & 9 & 486 & 201 & 3.9 & 2.8 & 1 & 6.77e-22 & 17.0 & 4.37 & 9.86 \\
9508-6104  & 127.553016498 & 26.62732 & 0.053 & 10.80 & 0.772 & 50 & 4.10e-17 & 7.21e+04 & 1.47e+09 & 1 & 37 & 301 & 123 & 4.7 & 3.5 & 1 & 7.11e-22 & 17.8 & 1.93 & 9.45 \\
9509-3702  & 122.439749732 & 25.88030 & 0.025 & 10.10 & 0.401 & 50 & 3.56e-17 & 1.35e+04 & 2.75e+08 & -- & -- & -- & -- & 2.7 & 2.2 & 2 & 6.67e-22 & 15.8 & 10.06 & 9.44 \\
9865-9102  & 224.897977138 & 49.32744 & 0.027 & 10.37 & 0.708 & 50 & 2.59e-17 & 1.12e+04 & 2.28e+08 & 1 & -33 & 187 & 76 & 4.5 & 3.7 & 1 & 6.24e-22 & 14.8 & 5.54 & 9.10 \\
9870-3702  & 229.934607505 & 44.32184 & 0.028 & 10.58 & 0.325 & 50 & 4.09e-17 & 1.97e+04 & 4.01e+08 & 1 & 31 & 216 & 89 & 4.0 & 3.3 & 1 & 9.94e-22 & 23.7 & 2.94 & 9.07 \\
9871-6101  & 227.863793703 & 41.08999 & 0.031 & 10.39 & 0.627 & 50 & 4.11e-17 & 2.36e+04 & 4.81e+08 & 2 & 36 & 275 & 145 & 5.1 & 4.5 & 1 & 6.33e-22 & 15.2 & 3.53 & 9.23 \\
9872-3701  & 233.231957146 & 42.43825 & 0.020 & 10.13 & -0.102 & 50 & 2.48e-17 & 5.73e+03 & 1.17e+08 & 1 & -18 & 235 & 97 & 3.3 & 2.8 & 1 & 6.64e-22 & 15.5 & 3.40 & 8.60 \\
9881-12705 & 205.373570269 & 23.16696 & 0.031 & 10.69 & 0.273 & 50 & 7.70e-17 & 4.40e+04 & 8.98e+08 & 2 & 96 & 273 & 143 & 6.3 & 5.1 & 1 & 1.03e-21 & 24.7 & 2.08 & 9.27 \\
9881-3702  & 203.81165409  & 25.04475 & 0.026 & 10.40 & 0.485 & 50 & 6.10e-17 & 2.50e+04 & 5.11e+08 & 2 & 4 & 262 & 134 & 9.5 & 6.2 & 1 & 5.55e-22 & 13.2 & 1.77 & 8.96 \\
9888-12704 & 236.006339163 & 28.27702 & 0.033 & 10.73 & 0.720 & 50 & 7.40e-17 & 4.97e+04 & 1.01e+09 & 2 & 55 & 275 & 143 & 9.7 & 6.0 & 1 & 6.45e-22 & 15.5 & 2.96 & 9.48 \\
9888-12705 & 236.901377571 & 26.06376 & 0.032 & 10.94 & 0.777 & 50 & 1.20e-16 & 7.24e+04 & 1.48e+09 & 1 & -4 & 466 & 191 & 9.1 & 6.1 & 1 & 9.18e-22 & 22.1 & 2.60 & 9.59 \\
\hline
\end{tabular}
\end{table}
\end{landscape} 
\begin{figure*} 
 \centering 
 \includegraphics[width = 0.17\textwidth, trim = 0cm 0cm 0cm 0cm, clip = true]{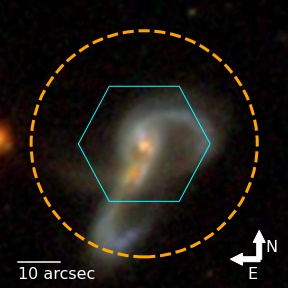}
 \includegraphics[width = 0.29\textwidth, trim = 0cm 0cm 0cm 0cm, clip = true]{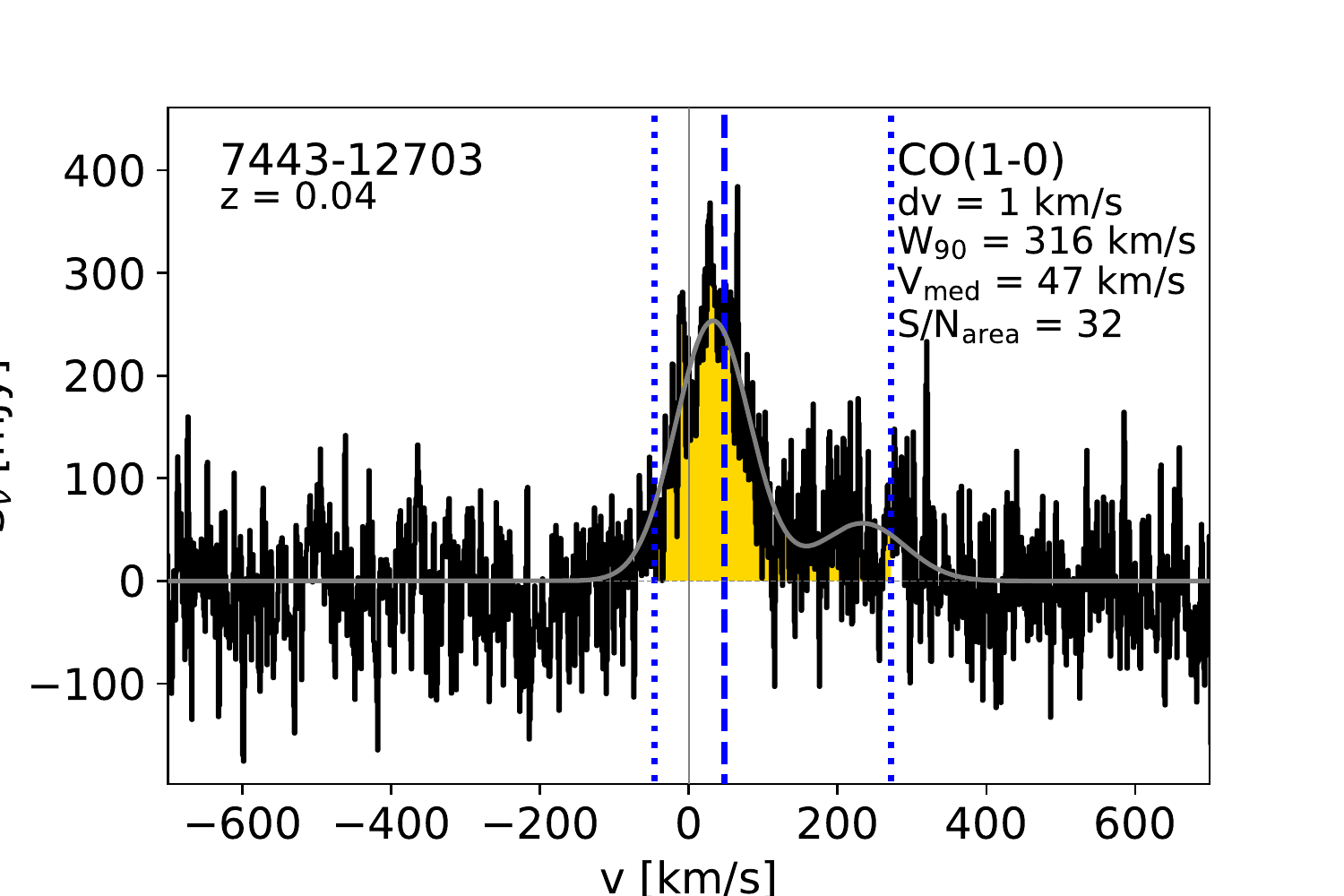} 
 \hspace{0.4cm}
 \includegraphics[width = 0.17\textwidth, trim = 0cm 0cm 0cm 0cm, clip = true]{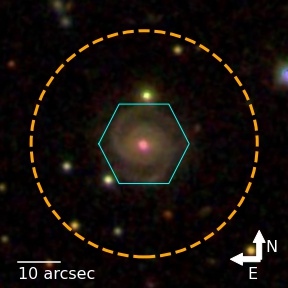}
 \includegraphics[width = 0.29\textwidth, trim = 0cm 0cm 0cm 0cm, clip = true]{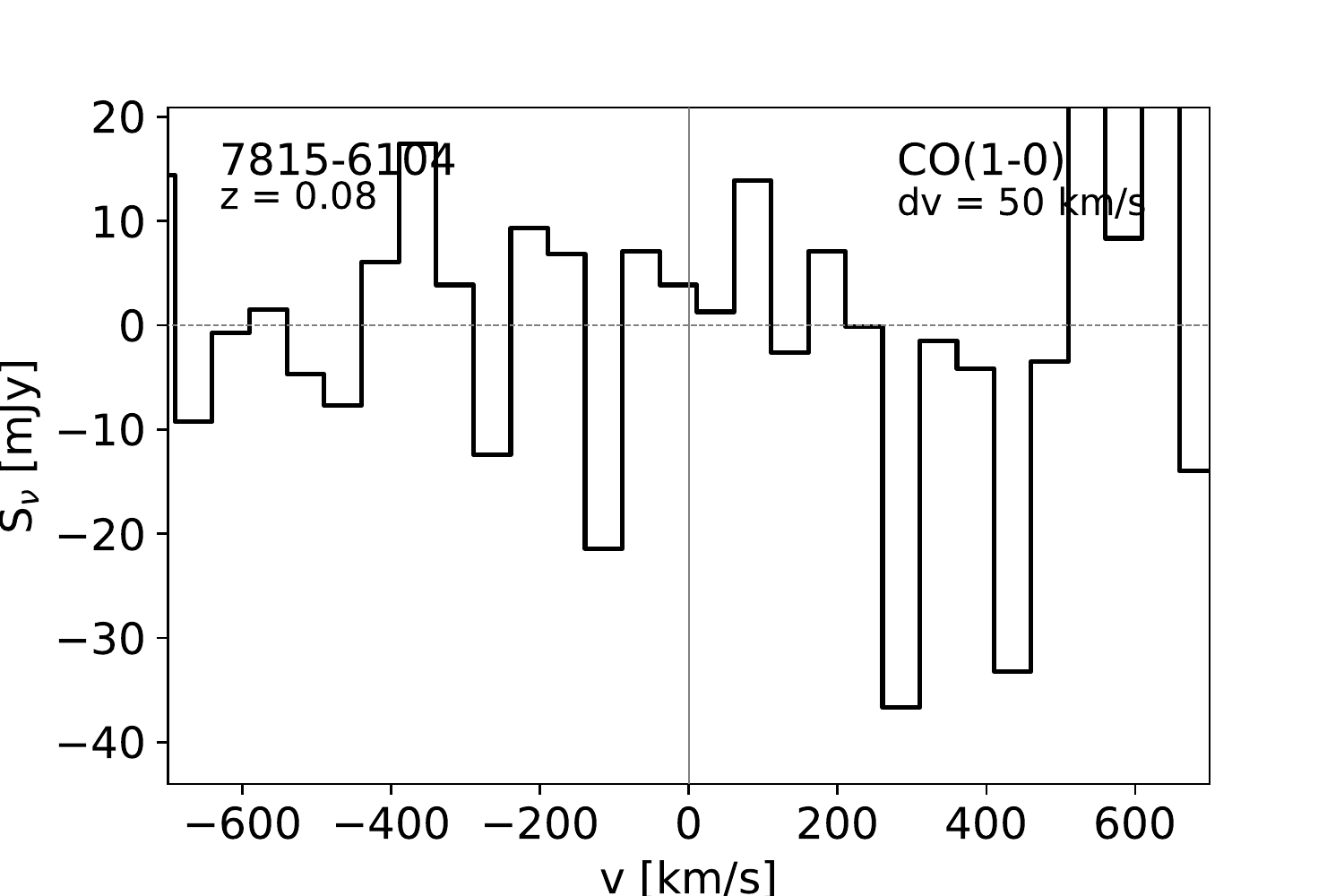} 
\end{figure*} 

\begin{figure*} 
 \centering 
 \includegraphics[width = 0.17\textwidth, trim = 0cm 0cm 0cm 0cm, clip = true]{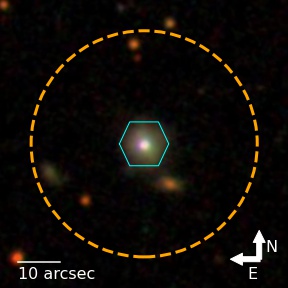}
 \includegraphics[width = 0.29\textwidth, trim = 0cm 0cm 0cm 0cm, clip = true]{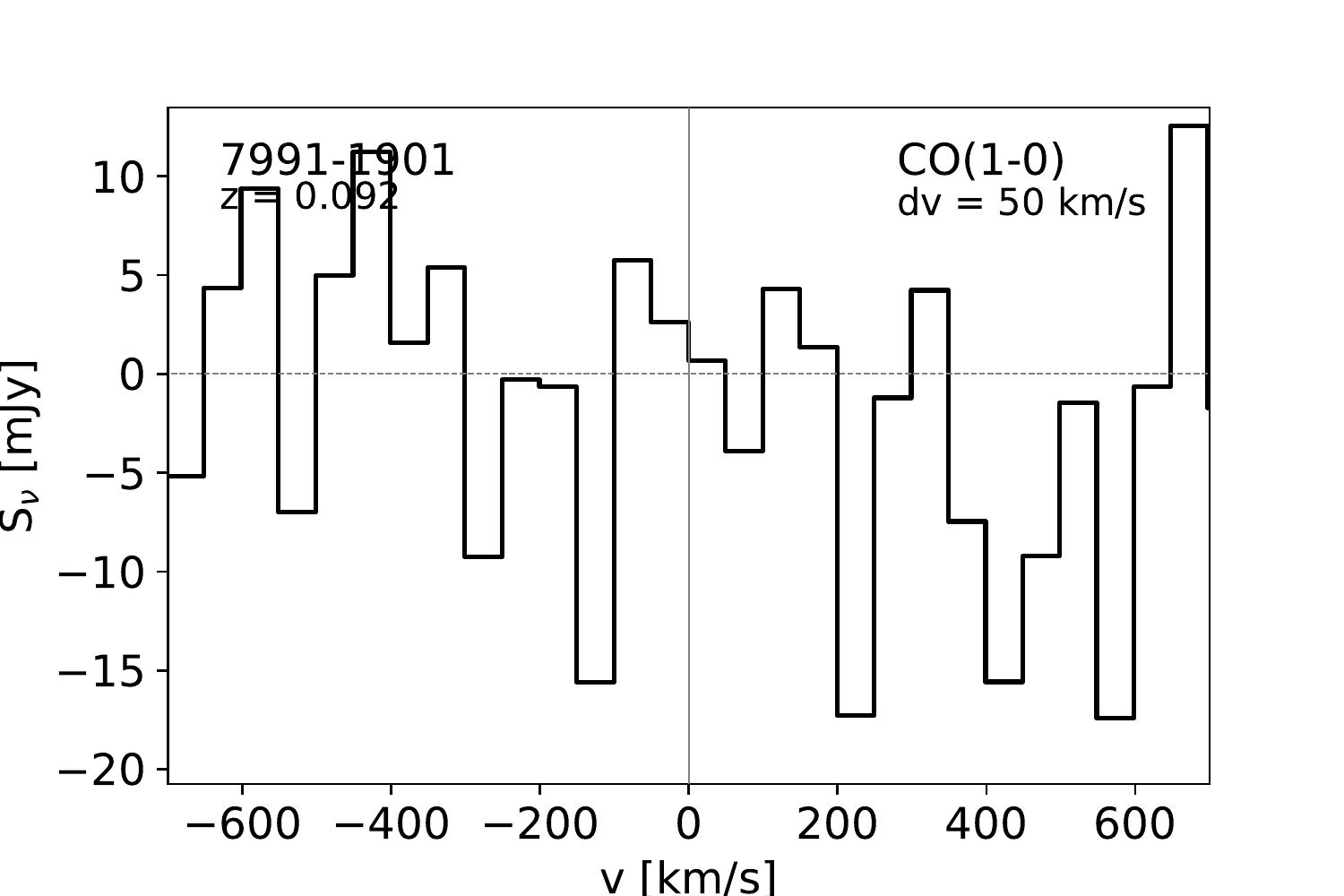} 
 \hspace{0.4cm}
 \centering 
 \includegraphics[width = 0.17\textwidth, trim = 0cm 0cm 0cm 0cm, clip = true]{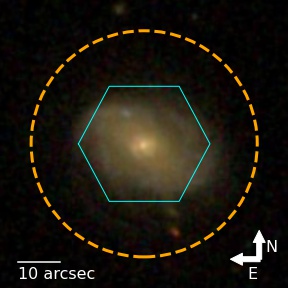}
 \includegraphics[width = 0.29\textwidth, trim = 0cm 0cm 0cm 0cm, clip = true]{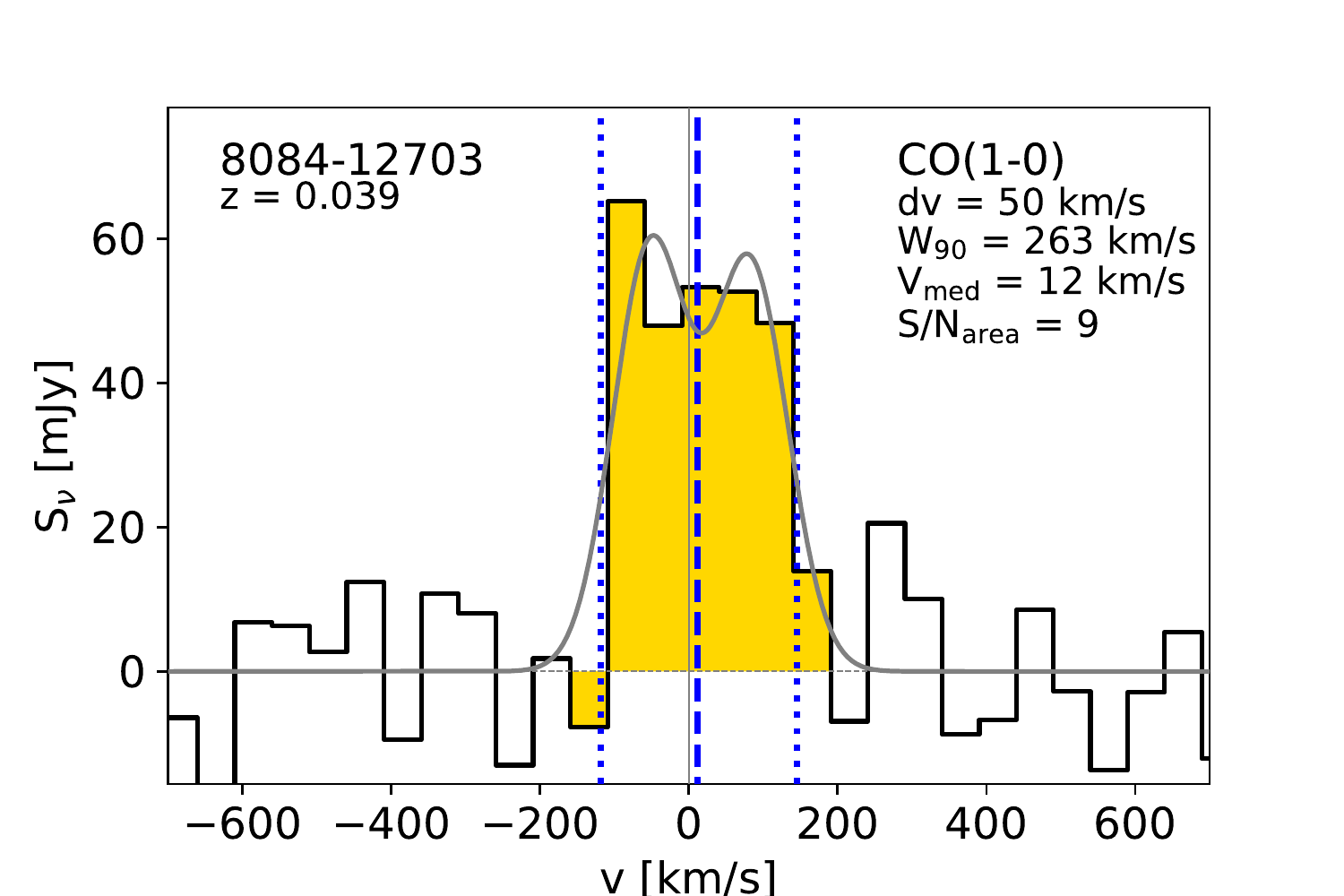} 

\end{figure*}

\begin{figure*} 
 \centering 
 \includegraphics[width = 0.17\textwidth, trim = 0cm 0cm 0cm 0cm, clip = true]{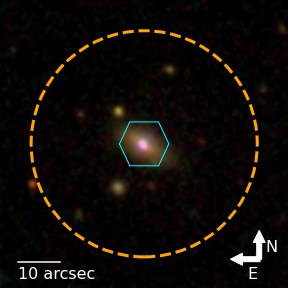}
 \includegraphics[width = 0.29\textwidth, trim = 0cm 0cm 0cm 0cm, clip = true]{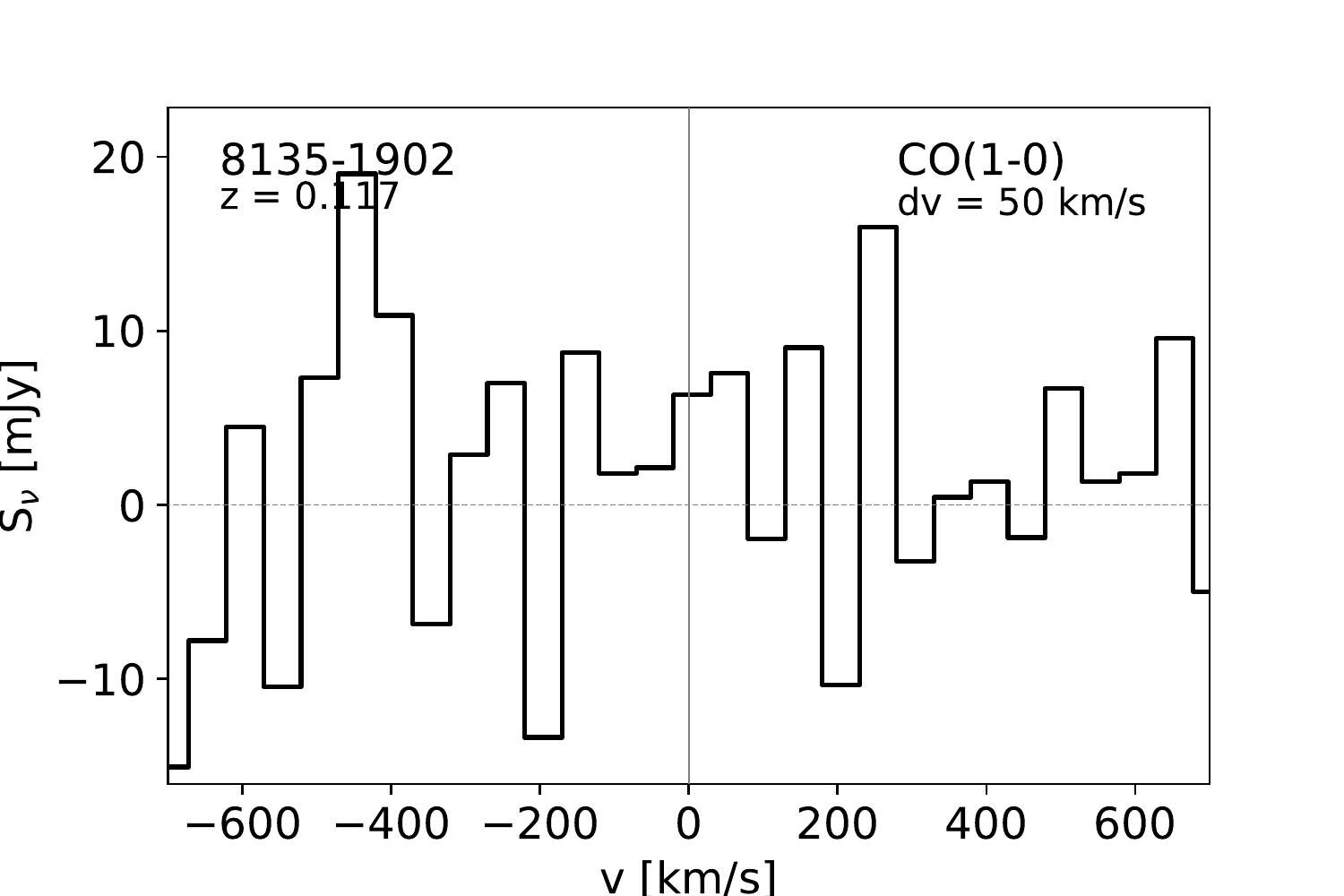} 
 \hspace{0.4cm}
 \centering 
 \includegraphics[width = 0.17\textwidth, trim = 0cm 0cm 0cm 0cm, clip = true]{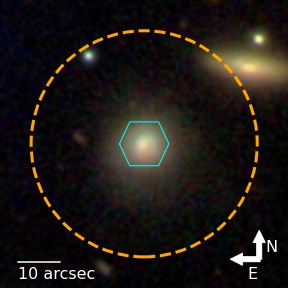}
 \includegraphics[width = 0.29\textwidth, trim = 0cm 0cm 0cm 0cm, clip = true]{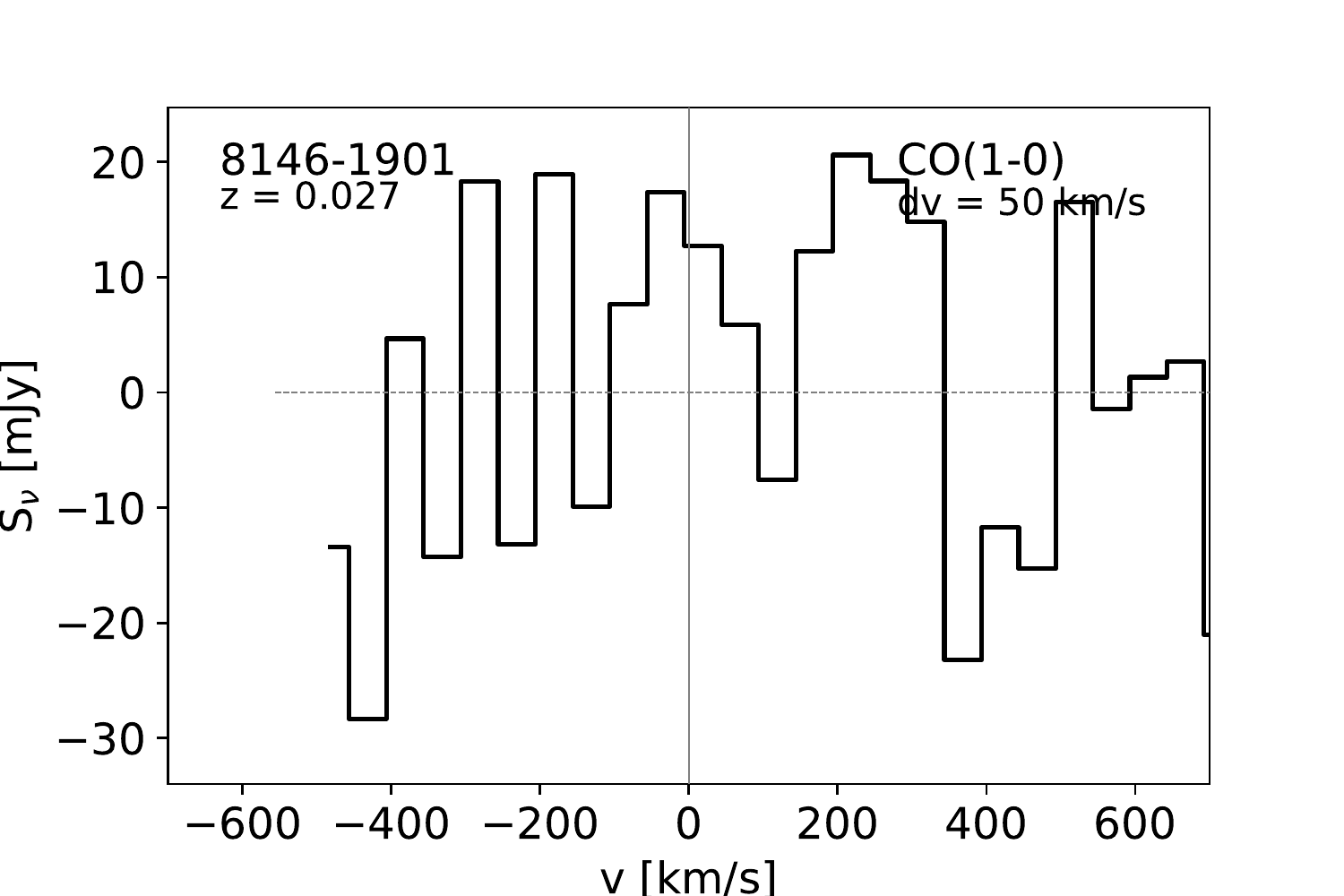} 

\end{figure*}

\begin{figure*} 
 \centering 
 \includegraphics[width = 0.17\textwidth, trim = 0cm 0cm 0cm 0cm, clip = true]{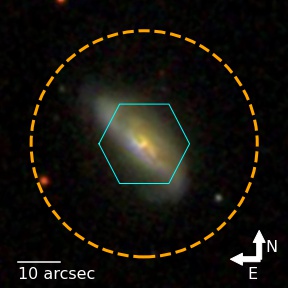}
 \includegraphics[width = 0.29\textwidth, trim = 0cm 0cm 0cm 0cm, clip = true]{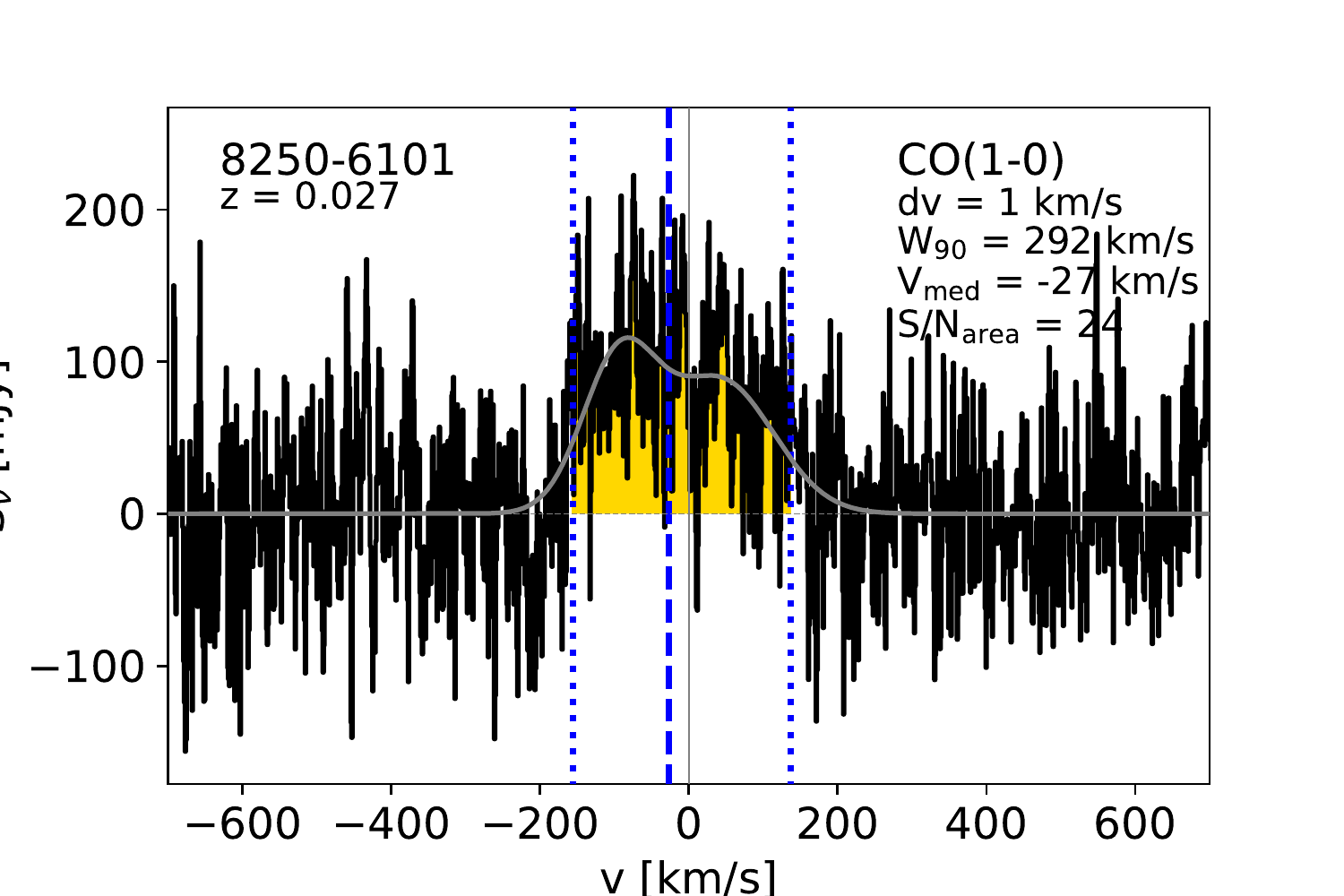} 
 \hspace{0.4cm}
 \centering 
 \includegraphics[width = 0.17\textwidth, trim = 0cm 0cm 0cm 0cm, clip = true]{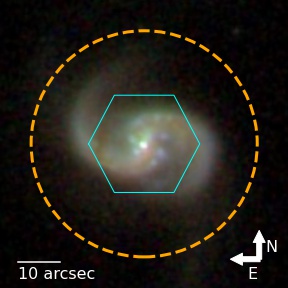}
 \includegraphics[width = 0.29\textwidth, trim = 0cm 0cm 0cm 0cm, clip = true]{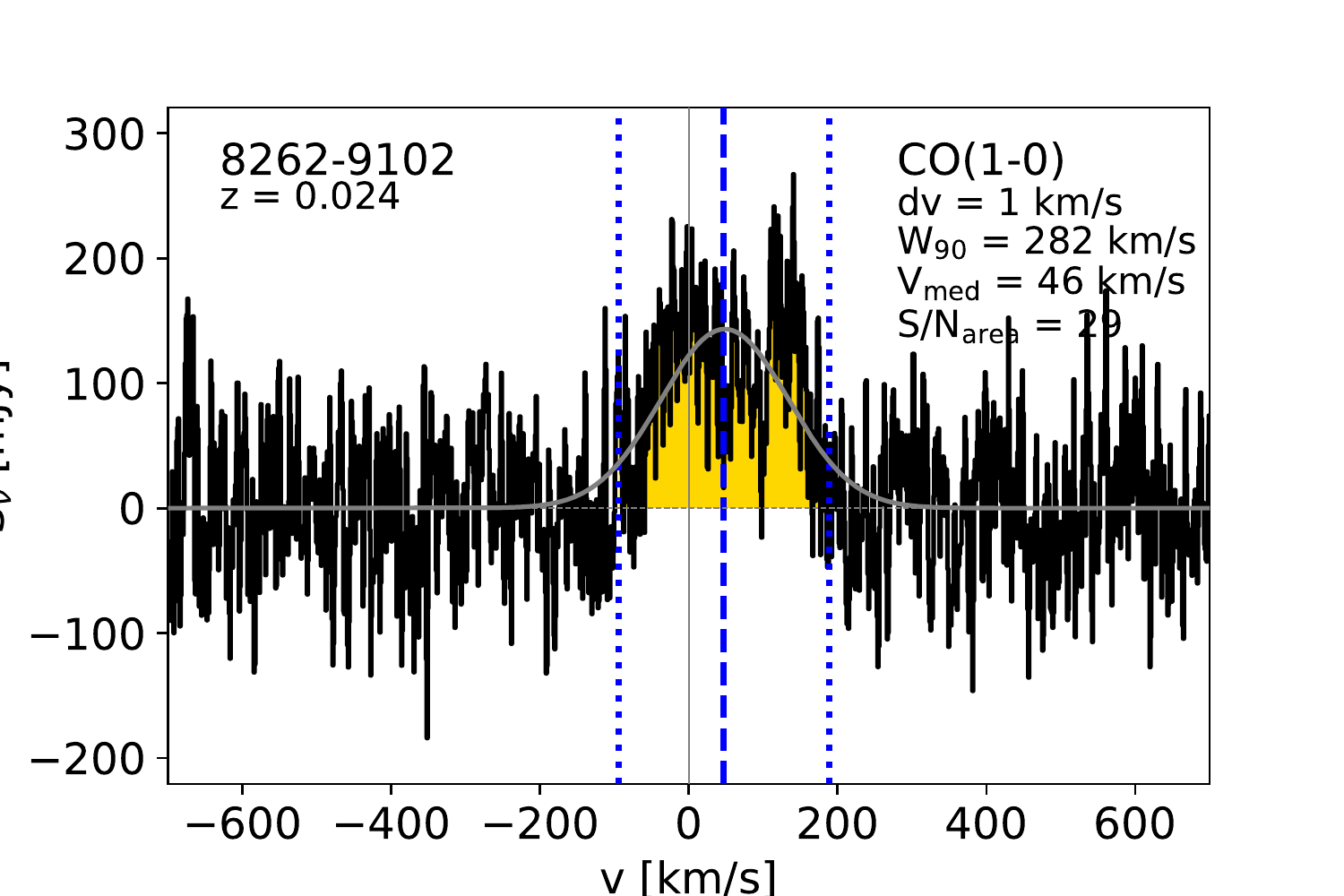} 

\end{figure*}

\begin{figure*} 
 \centering 
 \includegraphics[width = 0.17\textwidth, trim = 0cm 0cm 0cm 0cm, clip = true]{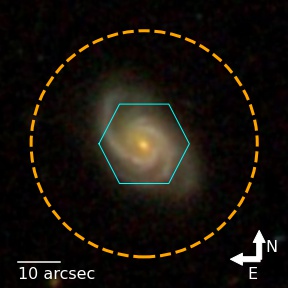}
 \includegraphics[width = 0.29\textwidth, trim = 0cm 0cm 0cm 0cm, clip = true]{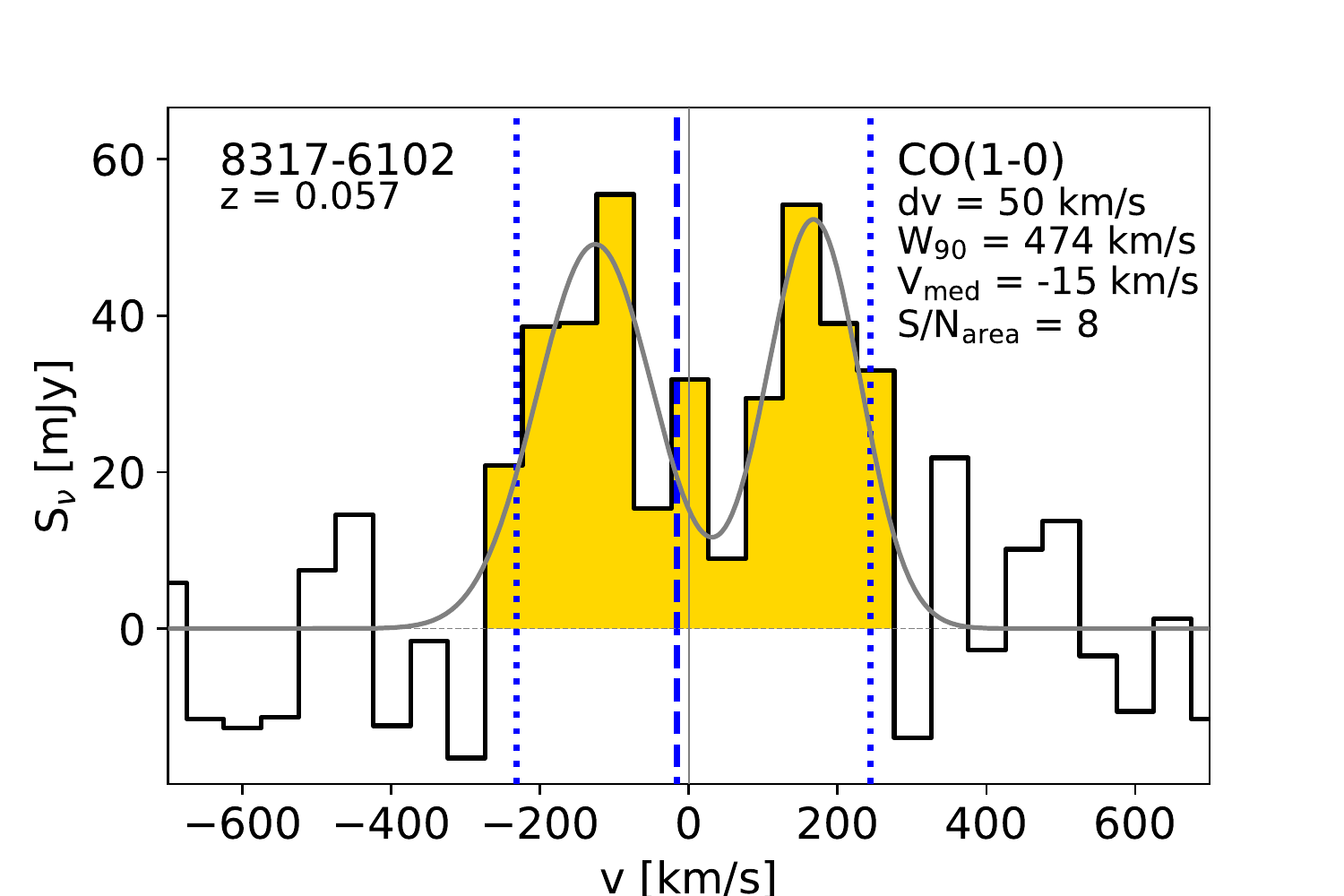} 
 \hspace{0.4cm}
 \centering 
 \includegraphics[width = 0.17\textwidth, trim = 0cm 0cm 0cm 0cm, clip = true]{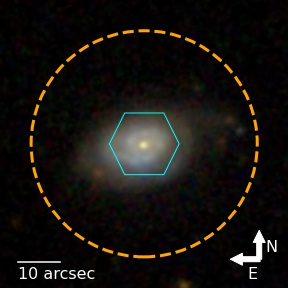}
 \includegraphics[width = 0.29\textwidth, trim = 0cm 0cm 0cm 0cm, clip = true]{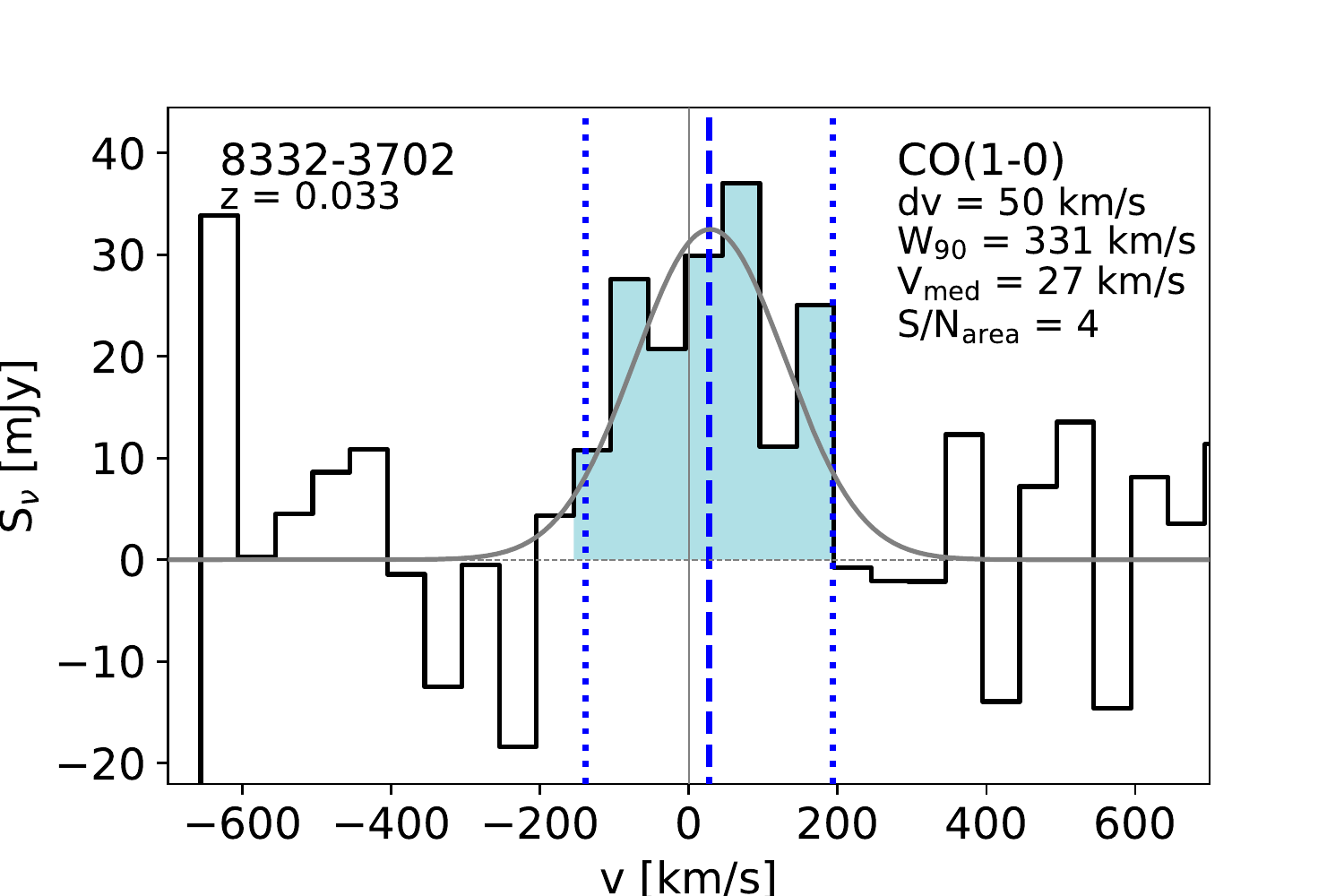} 
\caption{\textbf{Left:} SDSS three color image, also showing the MaNGA footprint (purple hexagon) and the ARO beam (orange dashed circle). \textbf{Right:} the MASCOT CO(1-0) line spectrum (solid black line). In cases where the spectrum was spectrally binned, the spectrum in native spectral resolution is shown in light grey. The final spectral resolution $dv$ is reported in the plot. When the CO(1-0) line is detected, the model fit is shown as the grey solid line and the line-of-sight velocity $v_{med}$ is indicated by the blue dashed line. Both $v_{05}$ and $v_{95}$, which are used for the $W_{90}$ determination, are indicated by the dotted blue lines. For sources detected at a S/N is $> 5$, the part of the spectrum between $v_{05}$ and $v_{95}$ which is used for the data-based line flux measurement S$_{\rm{CO, data}}$ is coloured in yellow. For tentative detections with $ 3 < $ S/N $ < 5$, that part is coloured in light blue.}
\end{figure*}

\begin{figure*} 
 \centering 
 \includegraphics[width = 0.17\textwidth, trim = 0cm 0cm 0cm 0cm, clip = true]{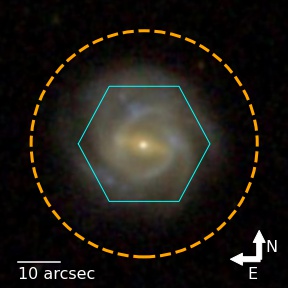}
 \includegraphics[width = 0.29\textwidth, trim = 0cm 0cm 0cm 0cm, clip = true]{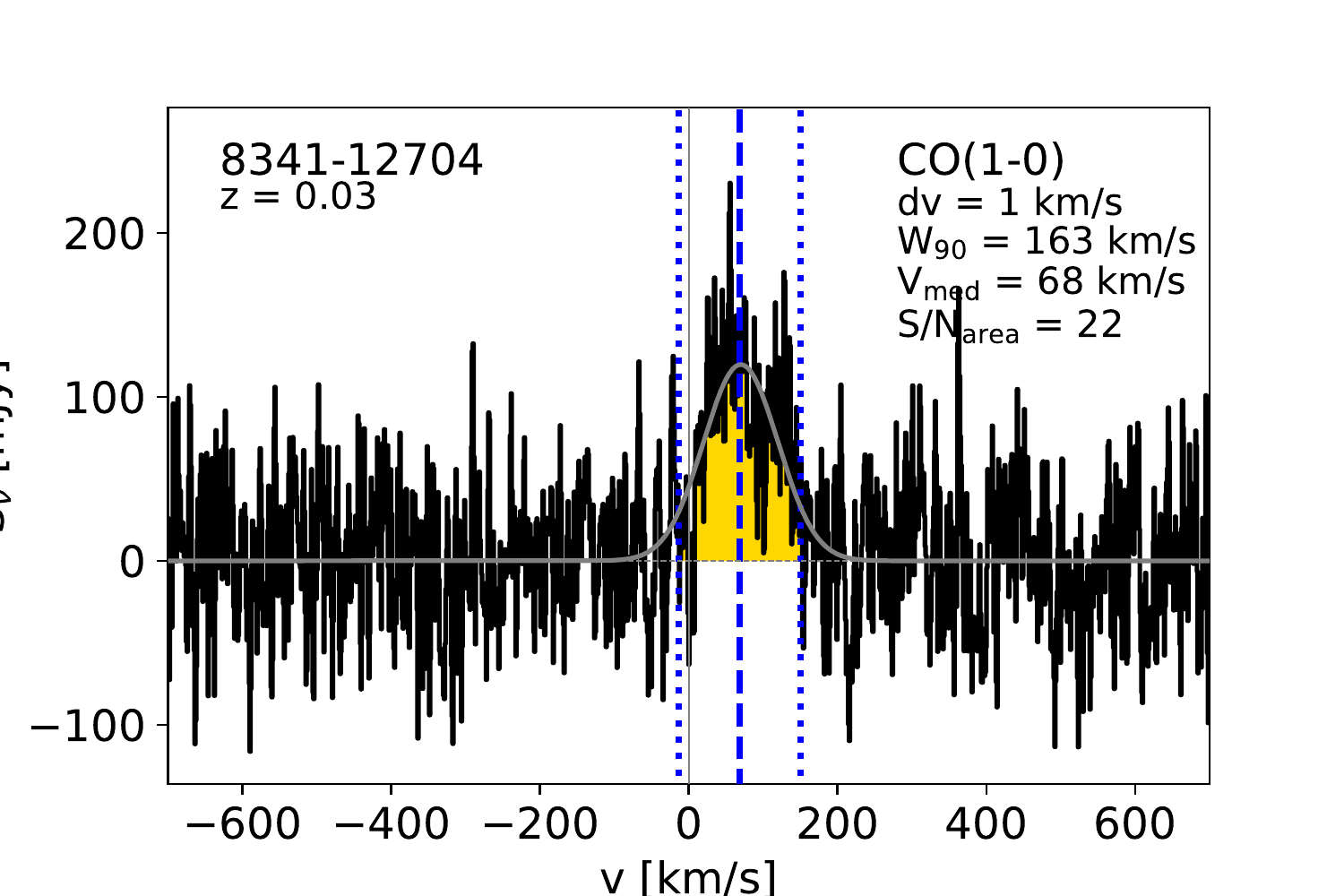} 
 \hspace{0.4cm}
 \centering 
 \includegraphics[width = 0.17\textwidth, trim = 0cm 0cm 0cm 0cm, clip = true]{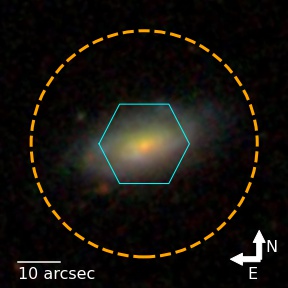}
 \includegraphics[width = 0.29\textwidth, trim = 0cm 0cm 0cm 0cm, clip = true]{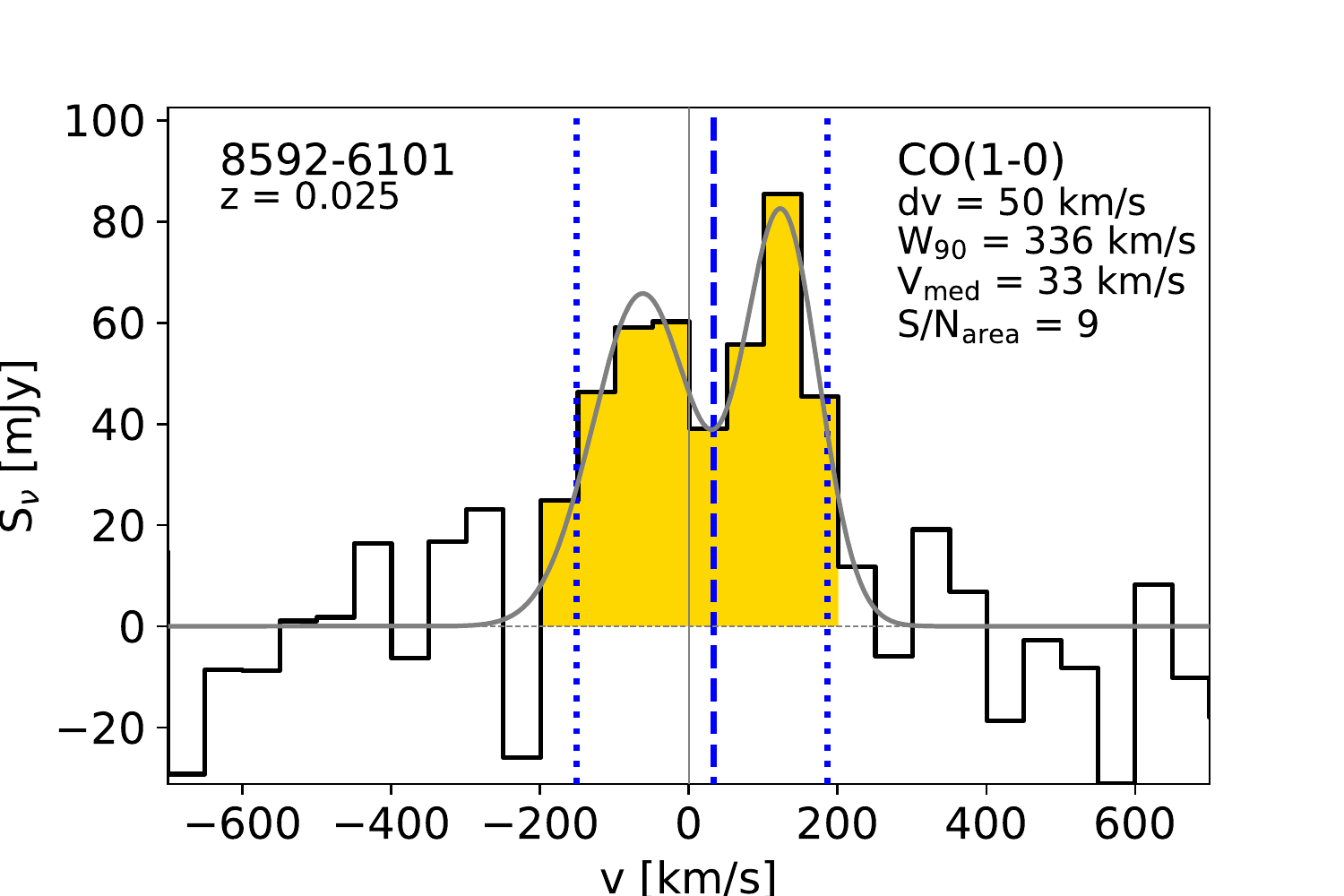} 
\end{figure*}

\begin{figure*} 
 \centering
 \includegraphics[width = 0.17\textwidth, trim = 0cm 0cm 0cm 0cm, clip = true]{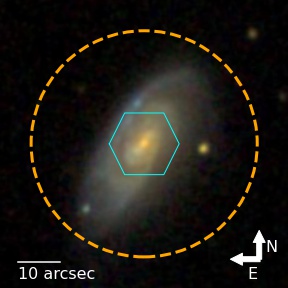}
 \includegraphics[width = 0.29\textwidth, trim = 0cm 0cm 0cm 0cm, clip = true]{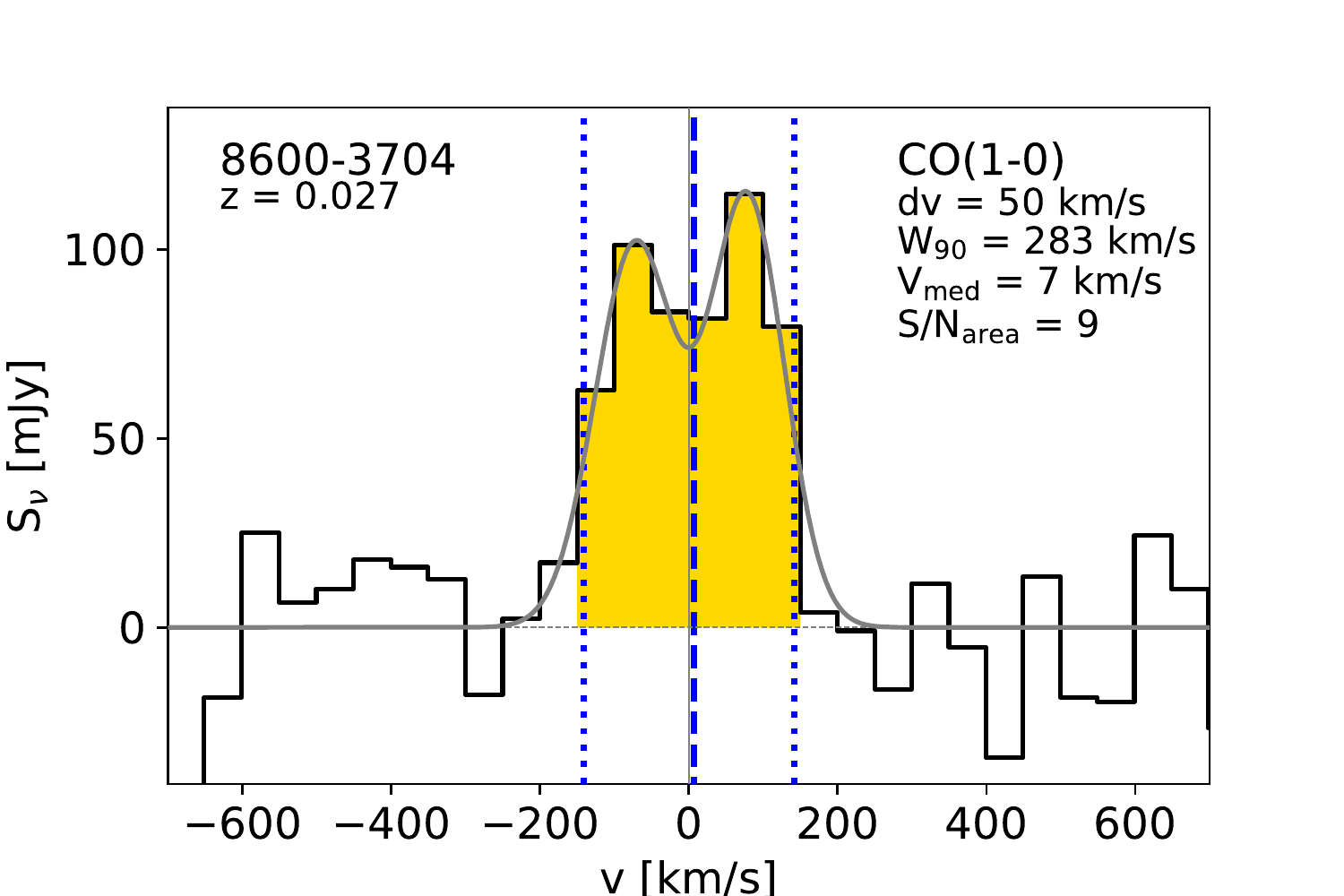} 
 \hspace{0.4cm}
 \centering 
 \includegraphics[width = 0.17\textwidth, trim = 0cm 0cm 0cm 0cm, clip = true]{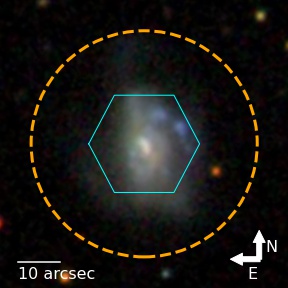}
 \includegraphics[width = 0.29\textwidth, trim = 0cm 0cm 0cm 0cm, clip = true]{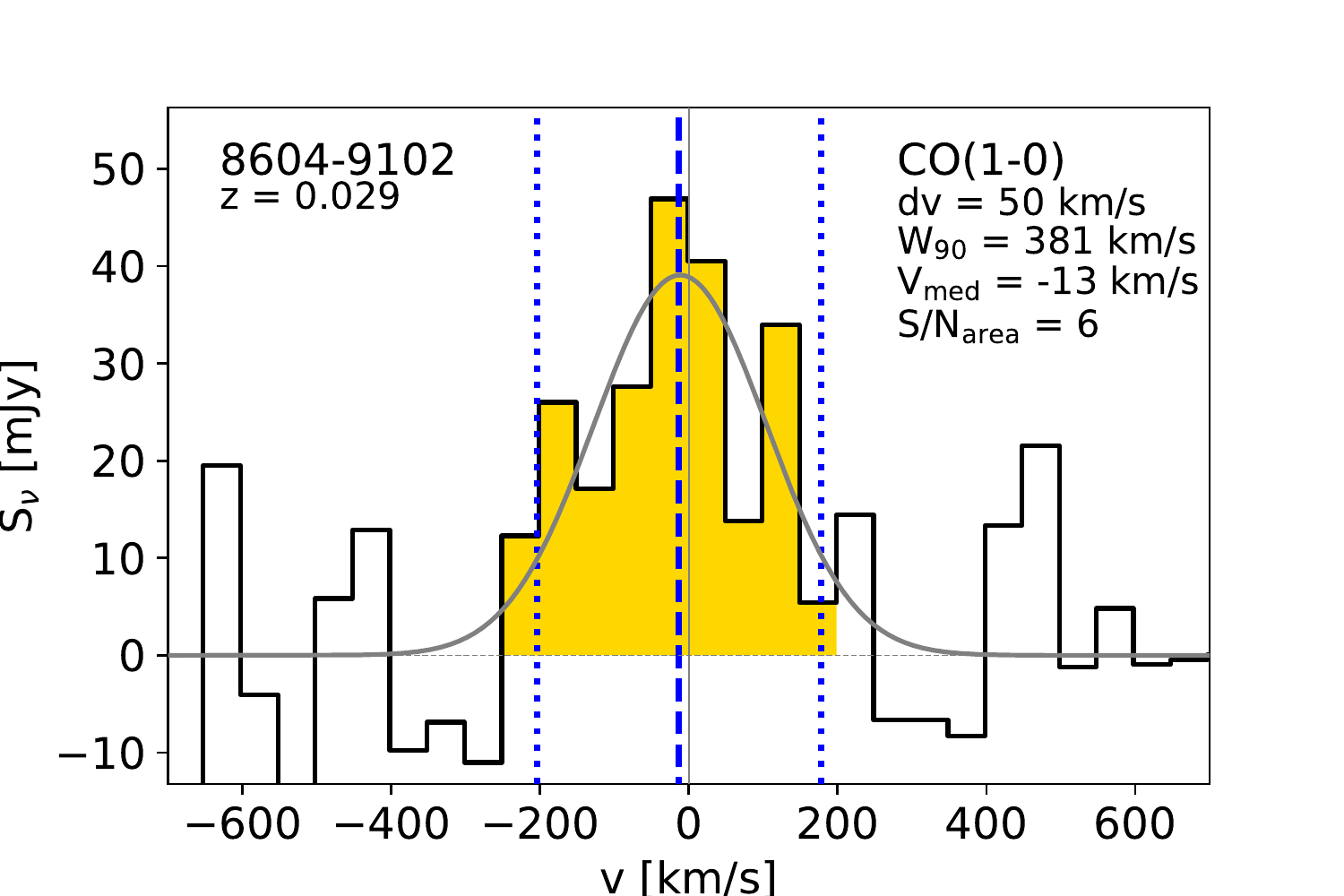} 

\end{figure*}

\begin{figure*} 
 \centering 
 \includegraphics[width = 0.17\textwidth, trim = 0cm 0cm 0cm 0cm, clip = true]{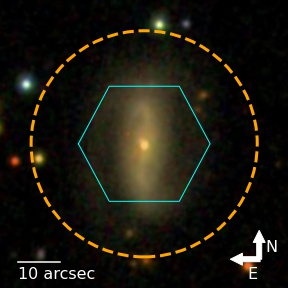}
 \includegraphics[width = 0.29\textwidth, trim = 0cm 0cm 0cm 0cm, clip = true]{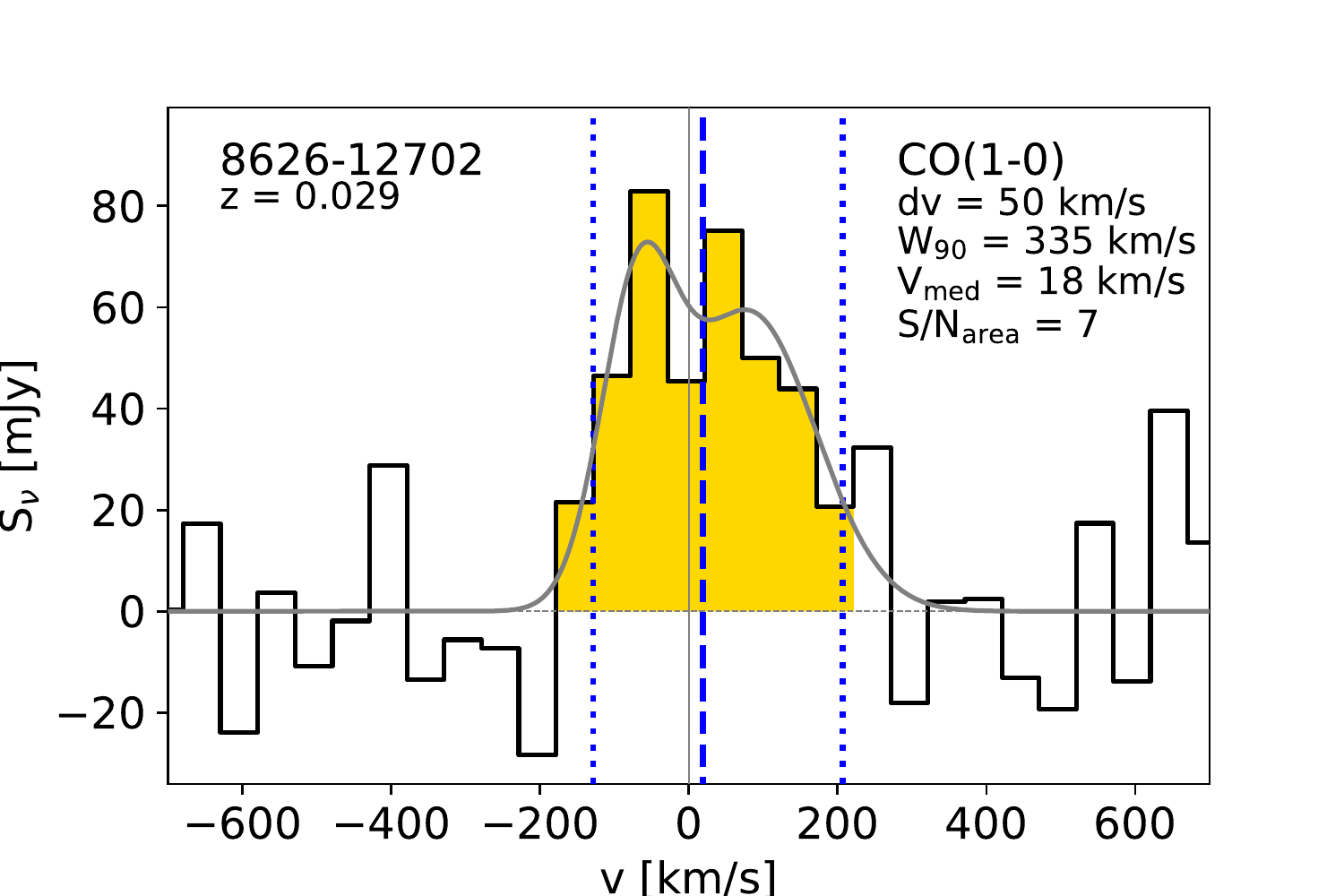} 
 \hspace{0.4cm}
 \centering 
 \includegraphics[width = 0.17\textwidth, trim = 0cm 0cm 0cm 0cm, clip = true]{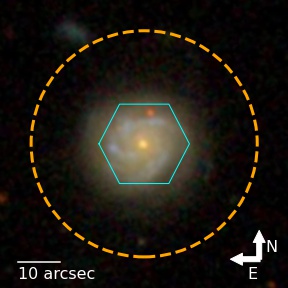}
 \includegraphics[width = 0.29\textwidth, trim = 0cm 0cm 0cm 0cm, clip = true]{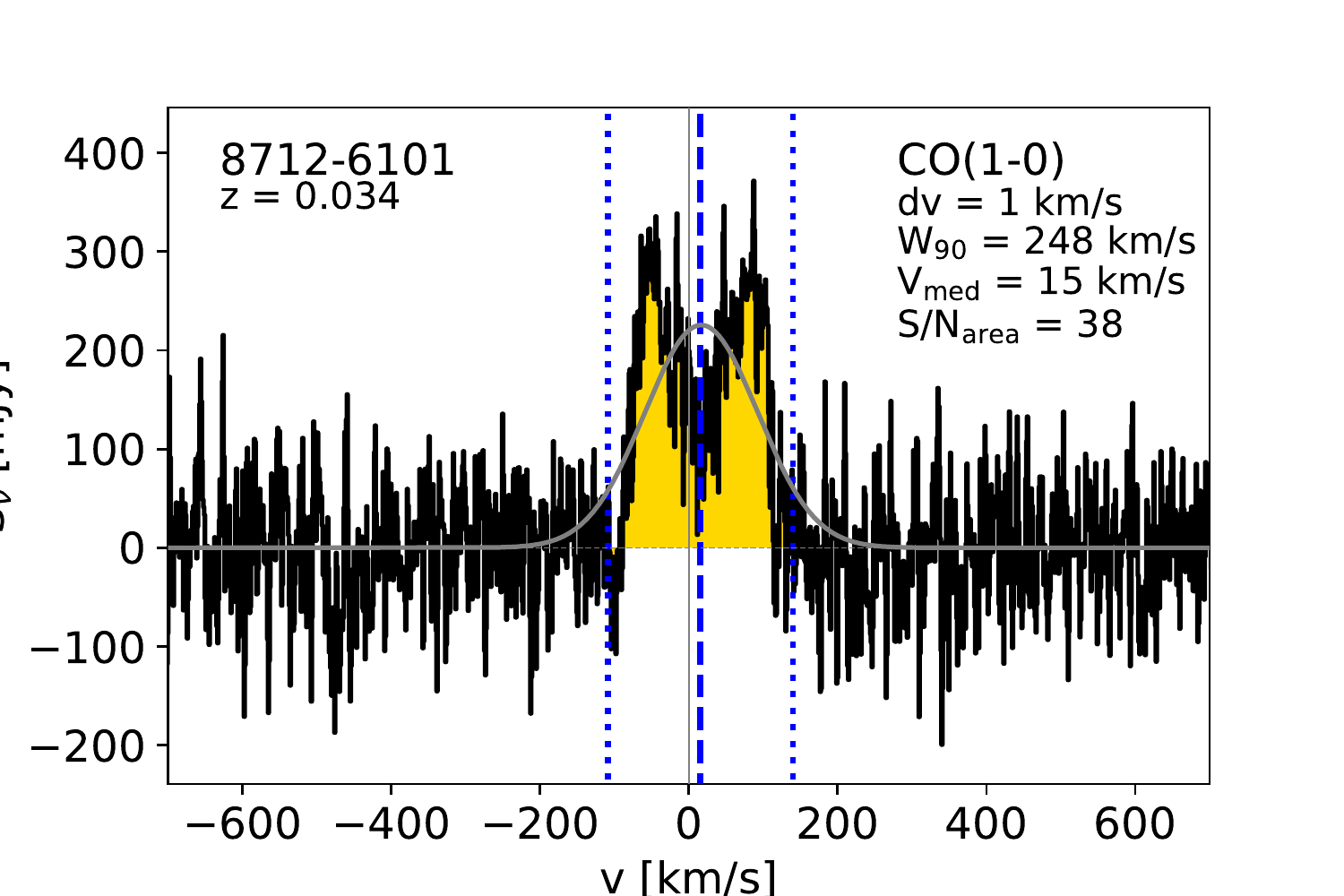} 

\end{figure*}

\begin{figure*} 
 \centering 
 \includegraphics[width = 0.17\textwidth, trim = 0cm 0cm 0cm 0cm, clip = true]{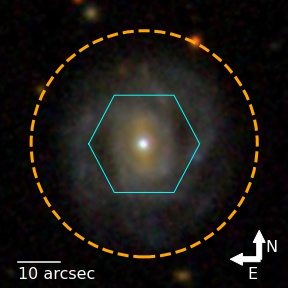}
 \includegraphics[width = 0.29\textwidth, trim = 0cm 0cm 0cm 0cm, clip = true]{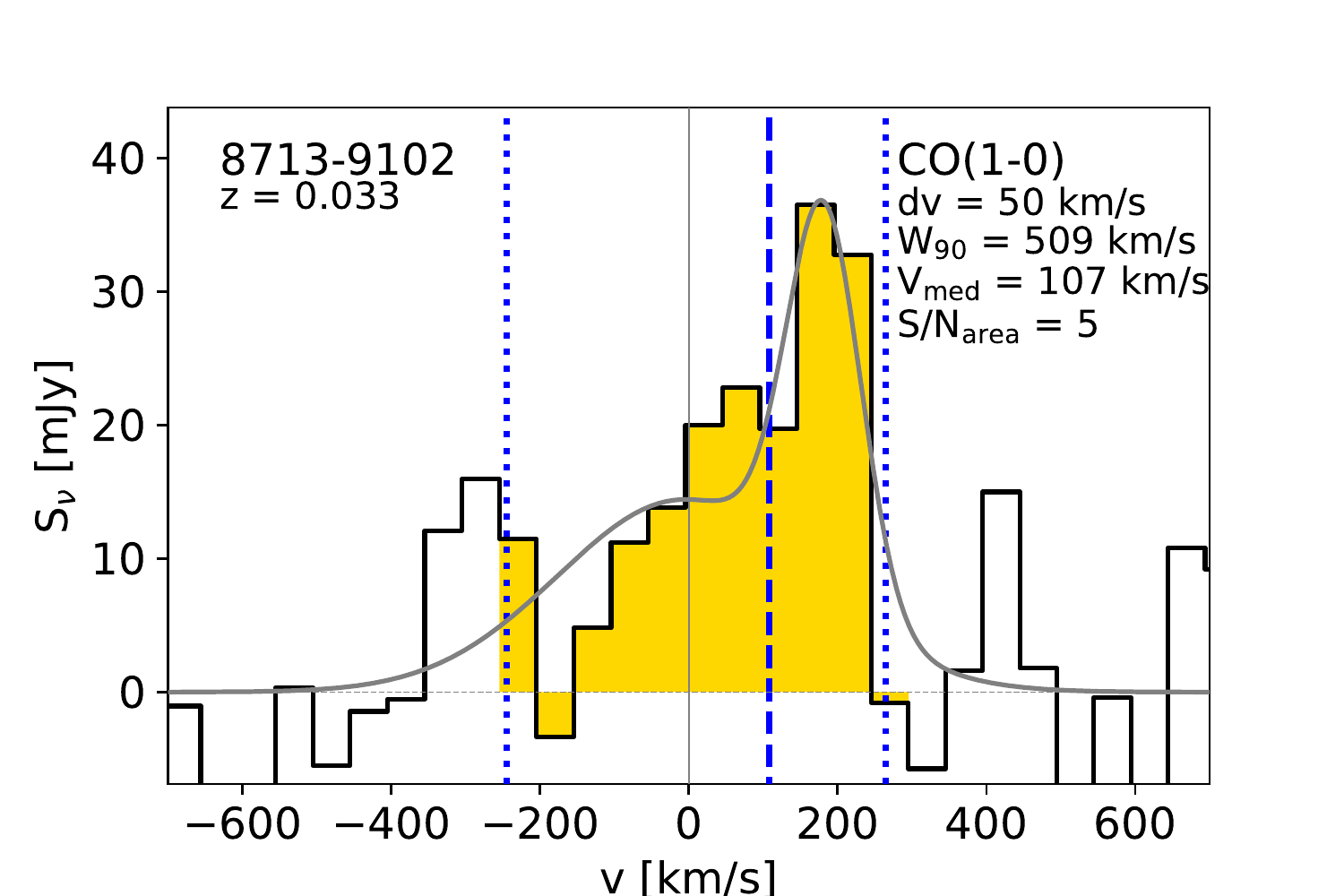} 
 \hspace{0.4cm}
 \centering 
 \includegraphics[width = 0.17\textwidth, trim = 0cm 0cm 0cm 0cm, clip = true]{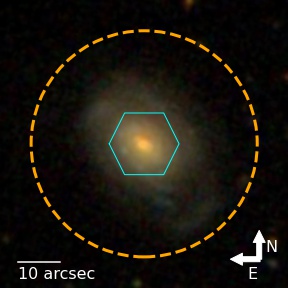}
 \includegraphics[width = 0.29\textwidth, trim = 0cm 0cm 0cm 0cm, clip = true]{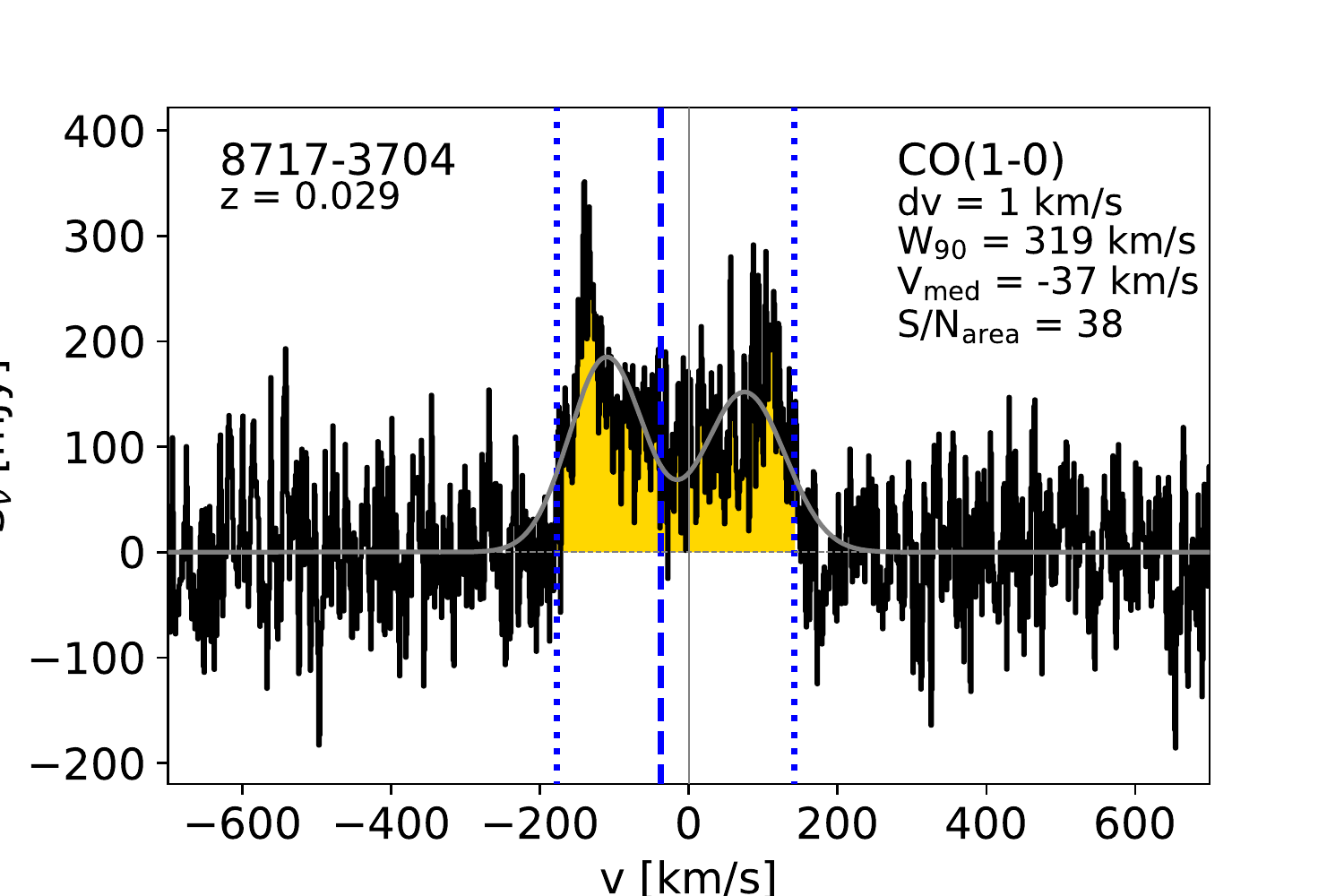} 

\end{figure*}

\begin{figure*} 
 \centering 
 \includegraphics[width = 0.17\textwidth, trim = 0cm 0cm 0cm 0cm, clip = true]{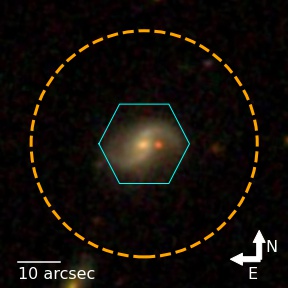}
 \includegraphics[width = 0.29\textwidth, trim = 0cm 0cm 0cm 0cm, clip = true]{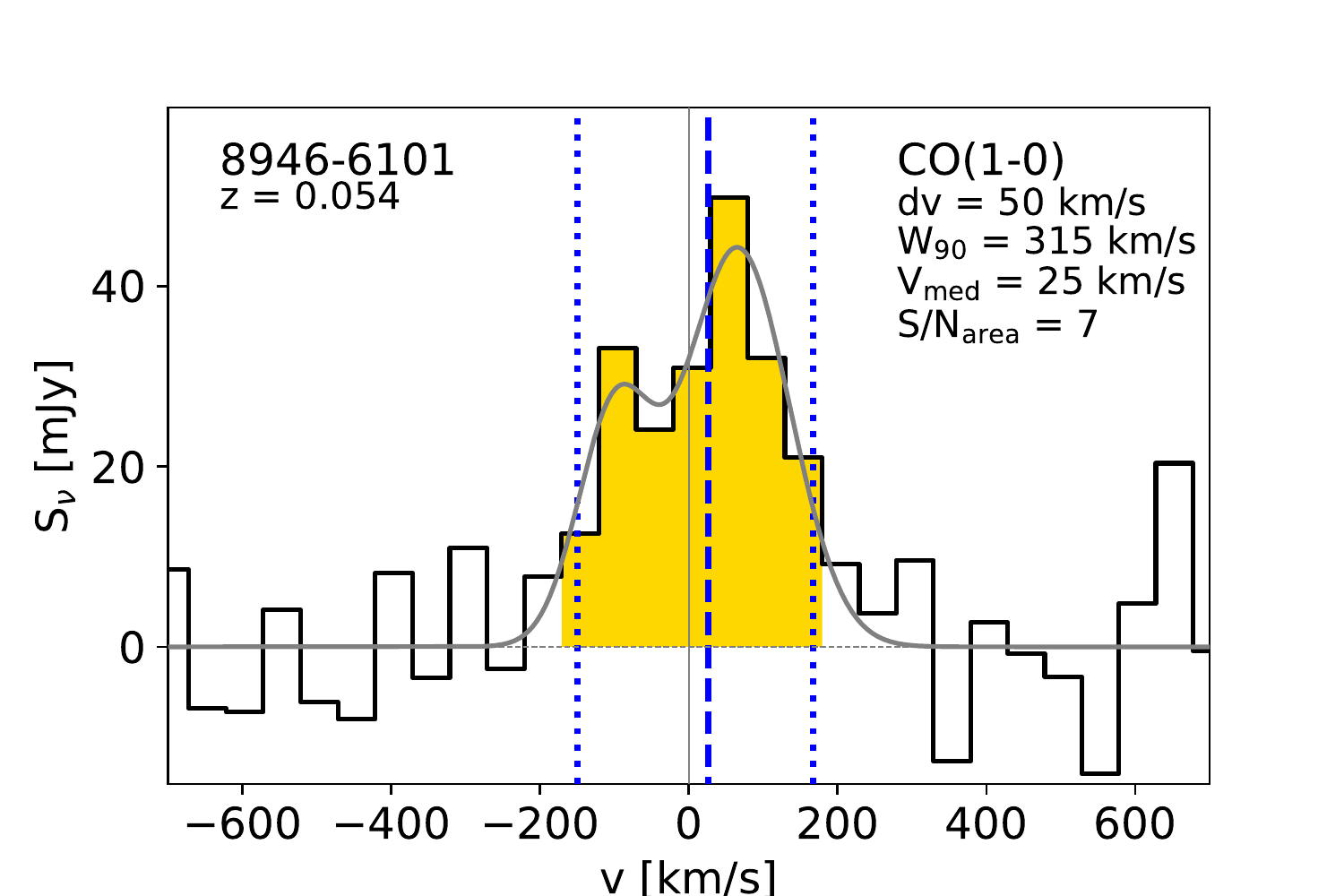} 
 \hspace{0.4cm}
 \centering 
 \includegraphics[width = 0.17\textwidth, trim = 0cm 0cm 0cm 0cm, clip = true]{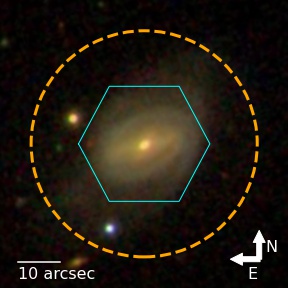}
 \includegraphics[width = 0.29\textwidth, trim = 0cm 0cm 0cm 0cm, clip = true]{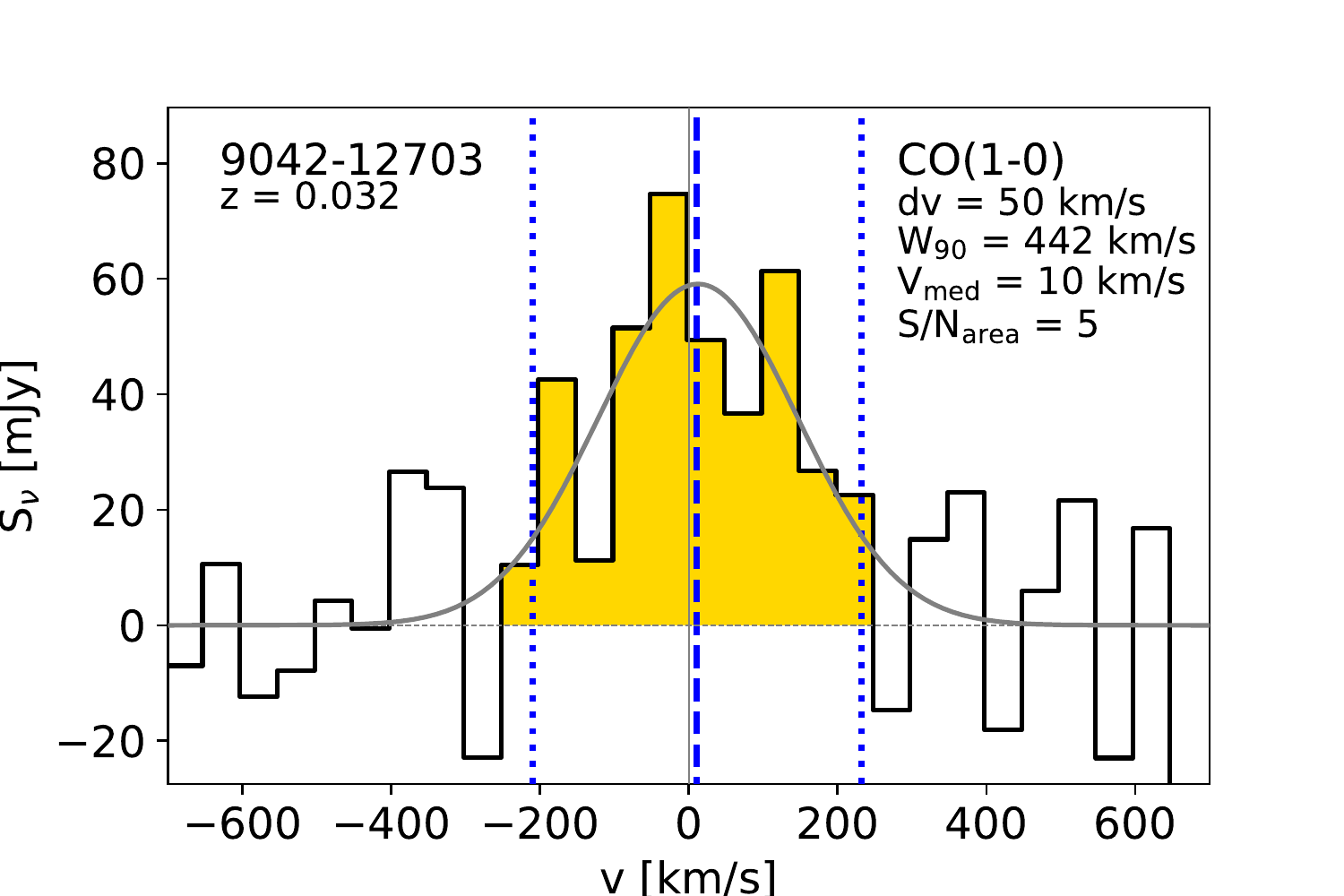} 

\end{figure*}

\begin{figure*} 
   \ContinuedFloat
 \centering 
 \includegraphics[width = 0.17\textwidth, trim = 0cm 0cm 0cm 0cm, clip = true]{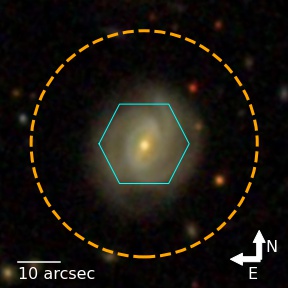}
 \includegraphics[width = 0.29\textwidth, trim = 0cm 0cm 0cm 0cm, clip = true]{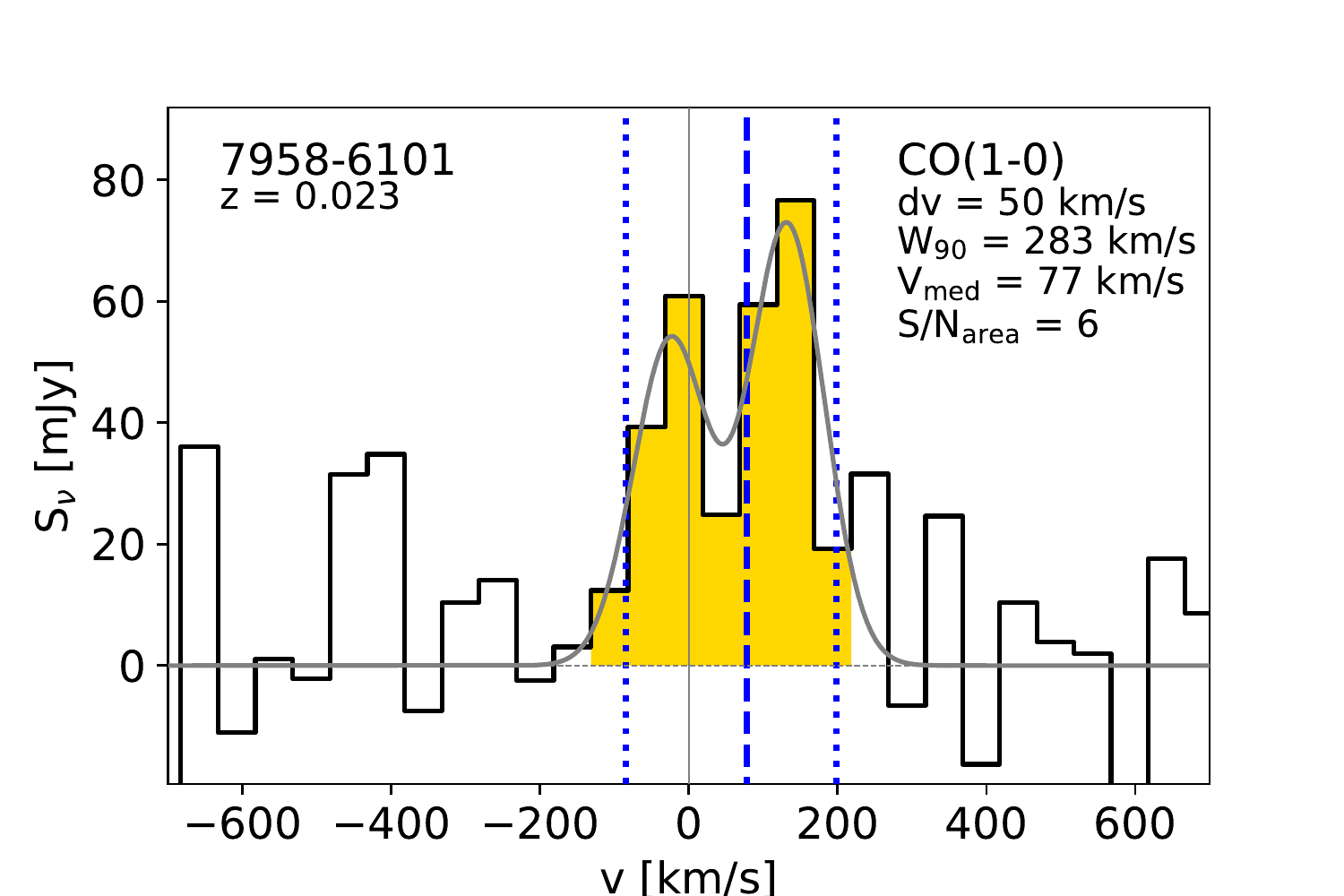} 
 \hspace{0.4cm}
 \centering 
 \includegraphics[width = 0.17\textwidth, trim = 0cm 0cm 0cm 0cm, clip = true]{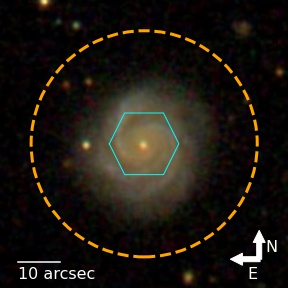}
 \includegraphics[width = 0.29\textwidth, trim = 0cm 0cm 0cm 0cm, clip = true]{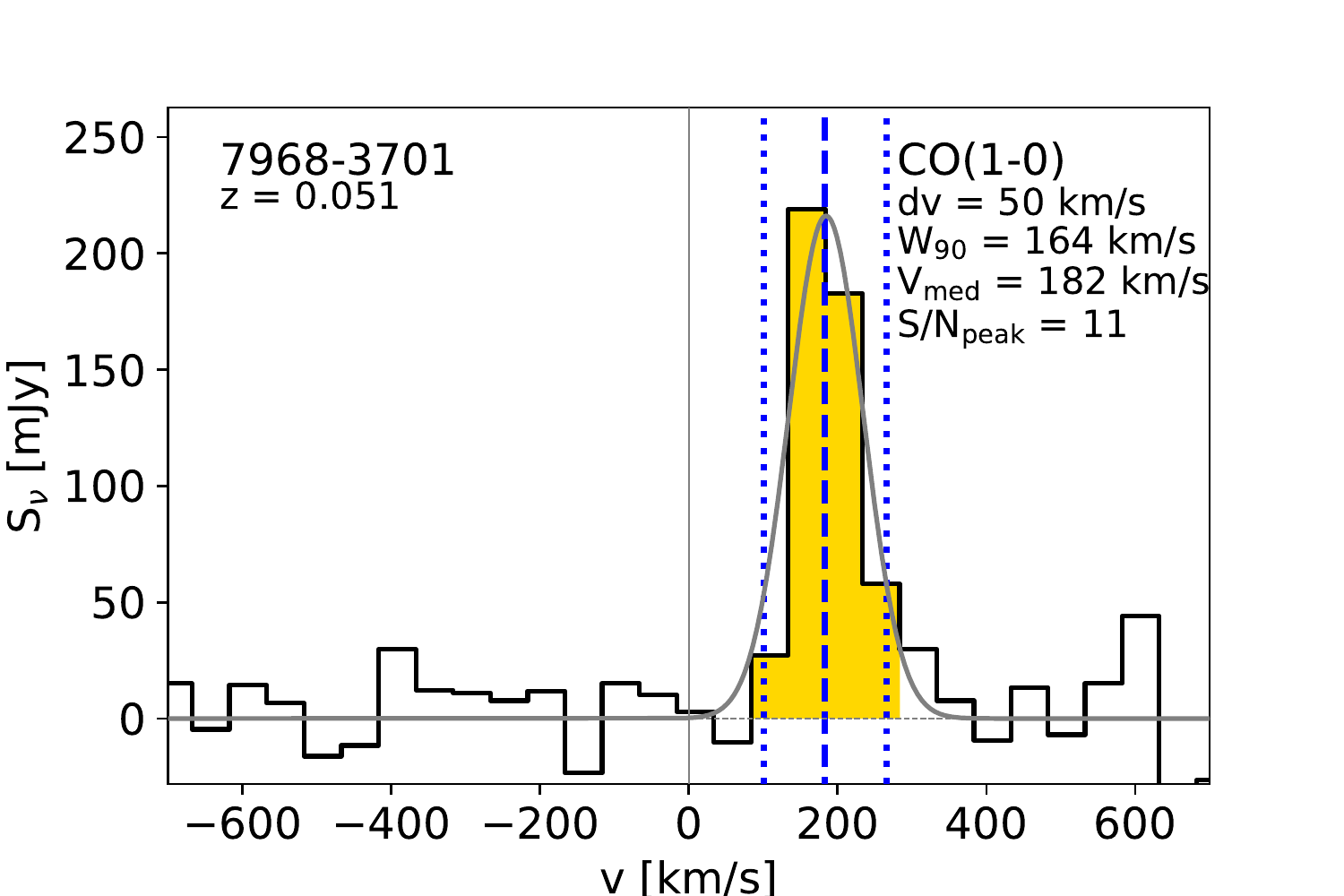} 
     \caption{continued.}
\end{figure*}

\begin{figure*} 
 \centering 
 \includegraphics[width = 0.17\textwidth, trim = 0cm 0cm 0cm 0cm, clip = true]{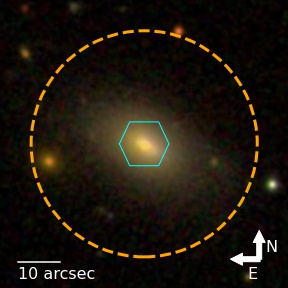}
 \includegraphics[width = 0.29\textwidth, trim = 0cm 0cm 0cm 0cm, clip = true]{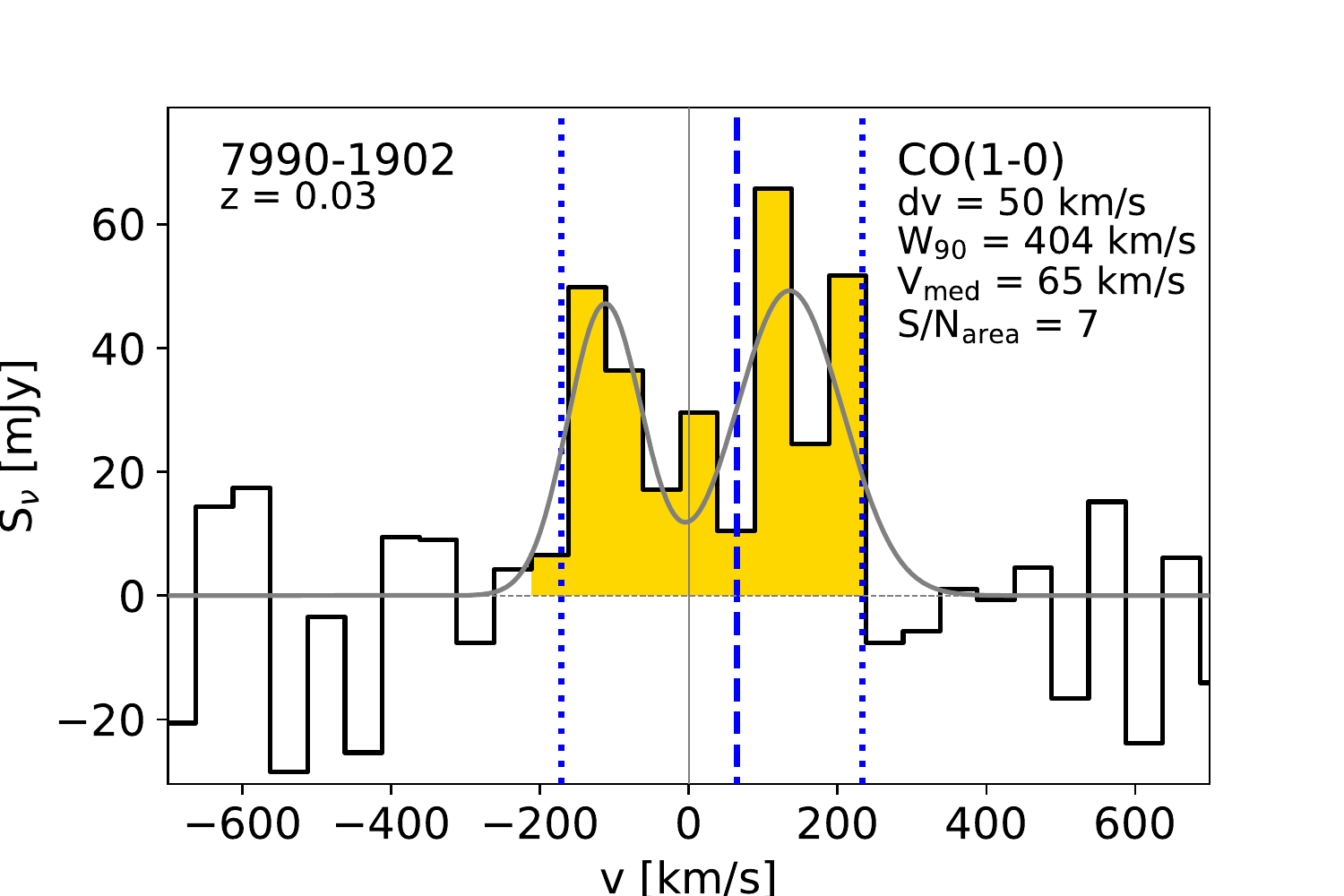} 
 \hspace{0.4cm}
 \centering 
 \includegraphics[width = 0.17\textwidth, trim = 0cm 0cm 0cm 0cm, clip = true]{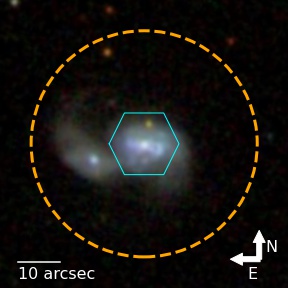}
 \includegraphics[width = 0.29\textwidth, trim = 0cm 0cm 0cm 0cm, clip = true]{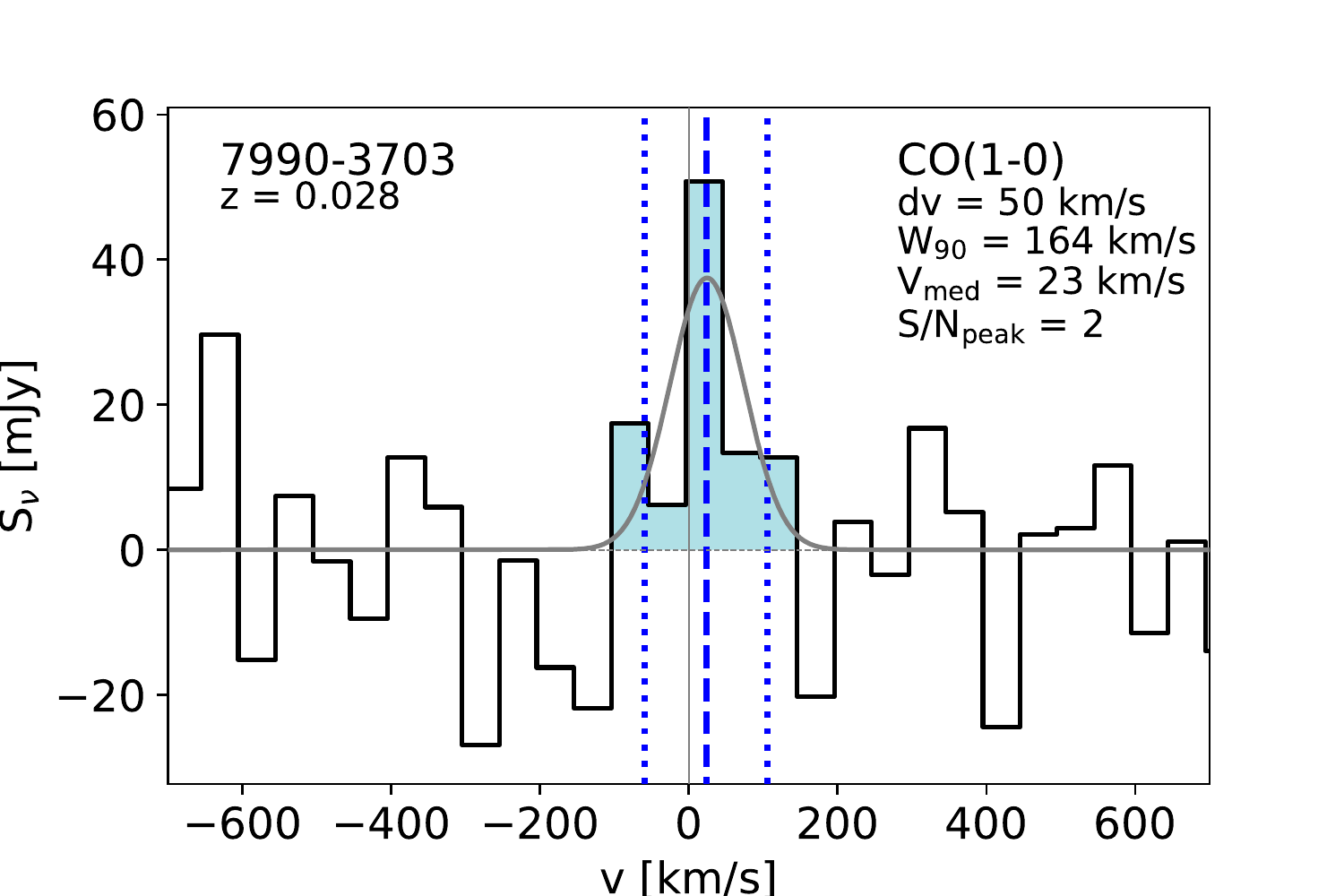} 

\end{figure*}

\begin{figure*} 
 \centering 
 \includegraphics[width = 0.17\textwidth, trim = 0cm 0cm 0cm 0cm, clip = true]{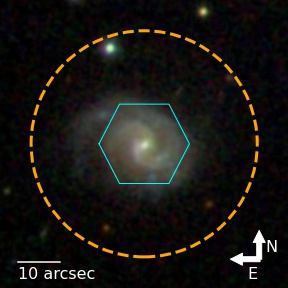}
 \includegraphics[width = 0.29\textwidth, trim = 0cm 0cm 0cm 0cm, clip = true]{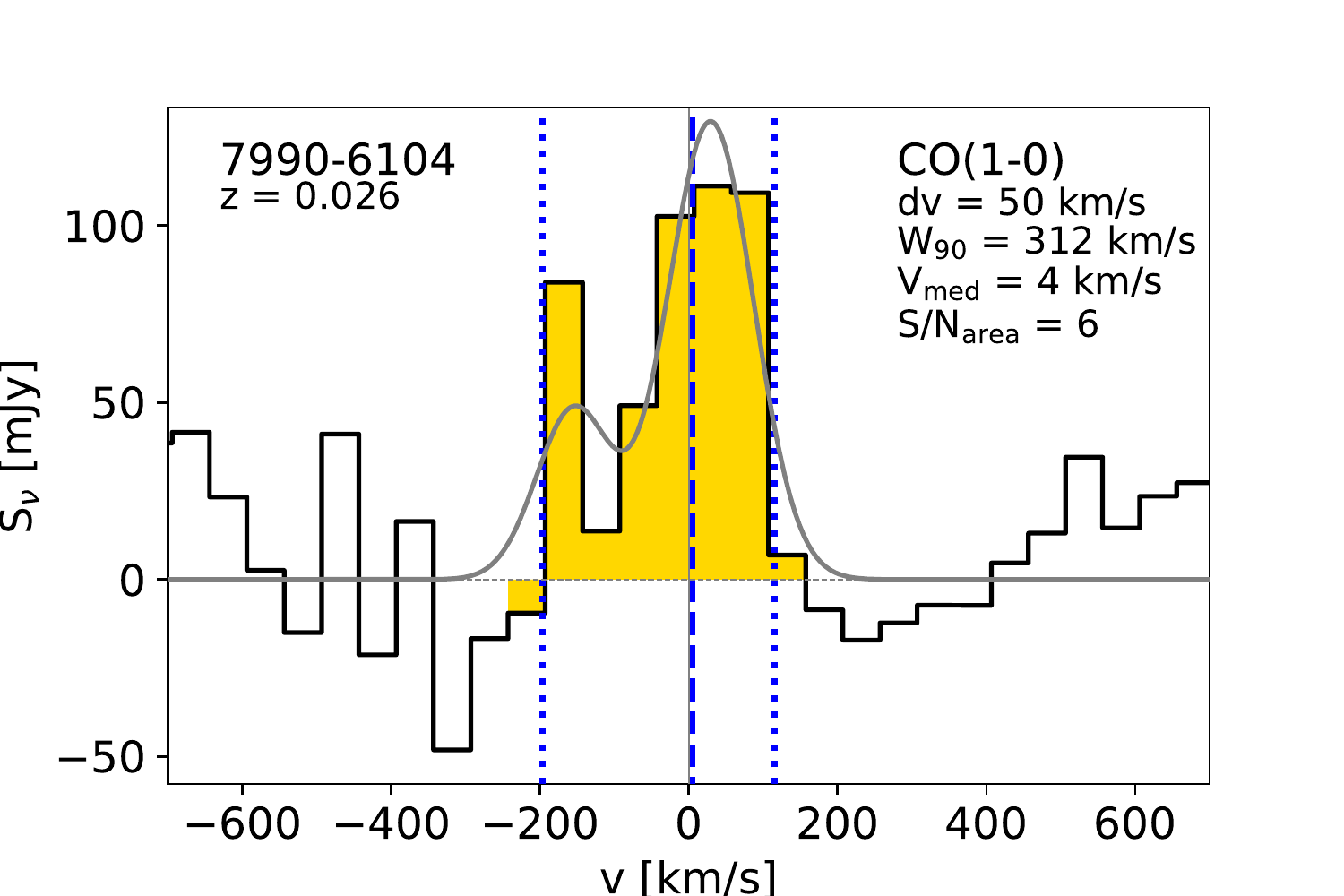} 
 \hspace{0.4cm}
 \centering 
 \includegraphics[width = 0.17\textwidth, trim = 0cm 0cm 0cm 0cm, clip = true]{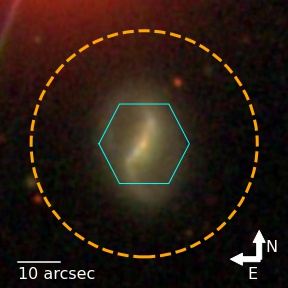}
 \includegraphics[width = 0.29\textwidth, trim = 0cm 0cm 0cm 0cm, clip = true]{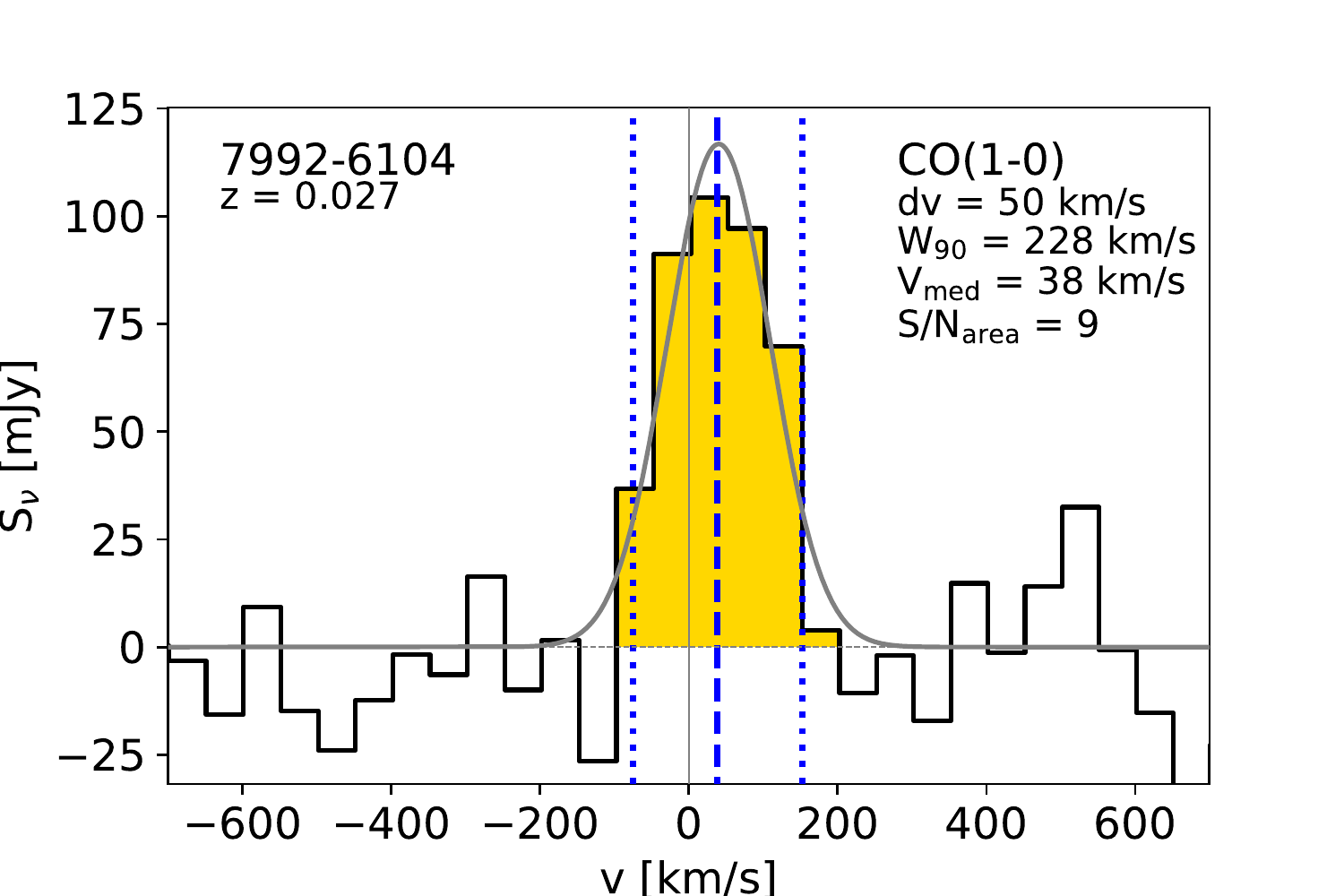} 

\end{figure*}

\begin{figure*} 
 \centering 
 \includegraphics[width = 0.17\textwidth, trim = 0cm 0cm 0cm 0cm, clip = true]{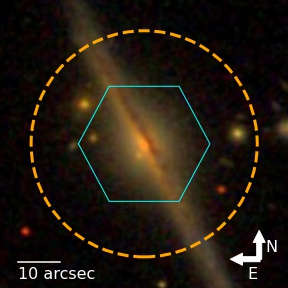}
 \includegraphics[width = 0.29\textwidth, trim = 0cm 0cm 0cm 0cm, clip = true]{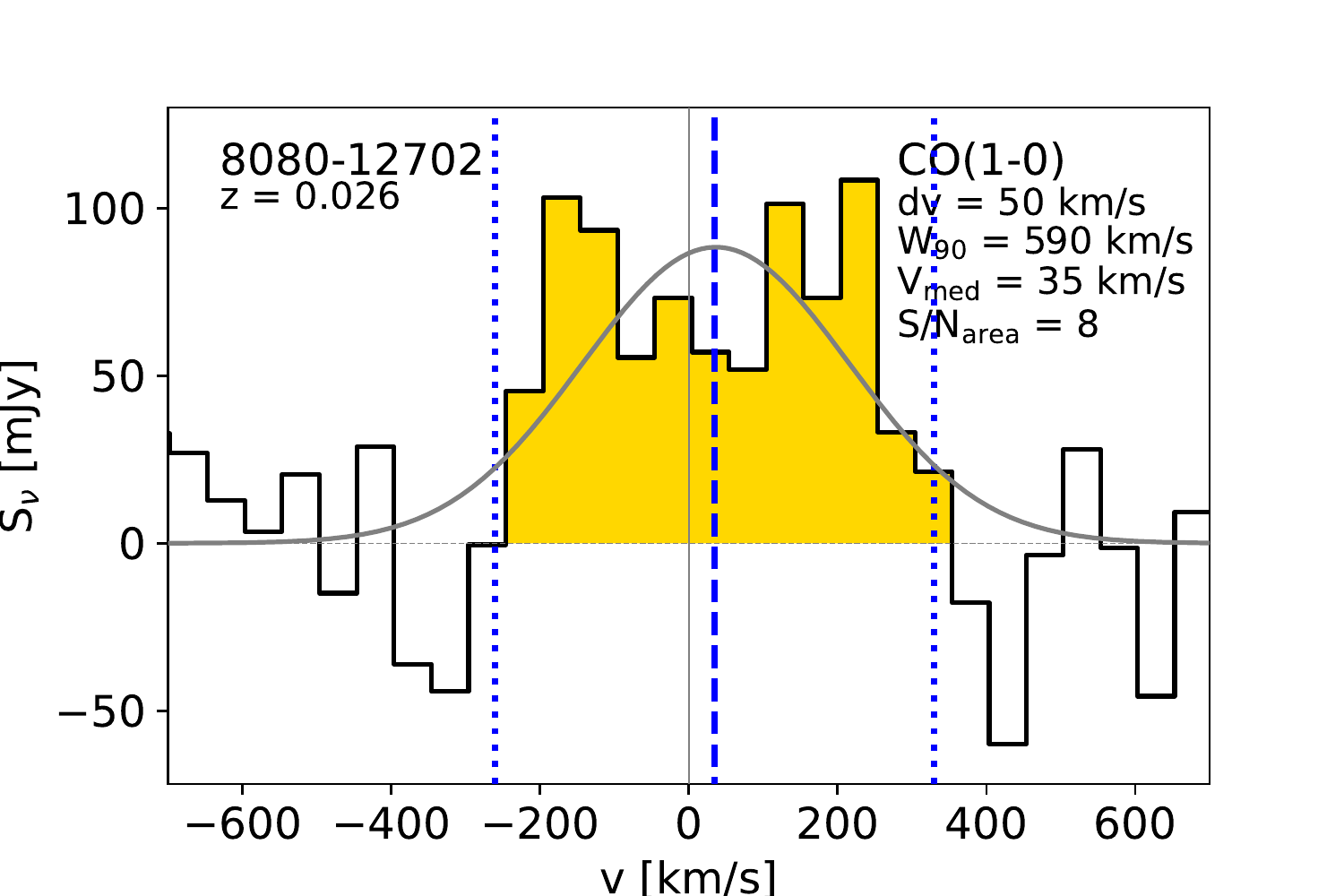} 
 \hspace{0.4cm}
 \centering 
 \includegraphics[width = 0.17\textwidth, trim = 0cm 0cm 0cm 0cm, clip = true]{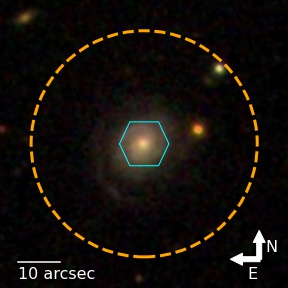}
 \includegraphics[width = 0.29\textwidth, trim = 0cm 0cm 0cm 0cm, clip = true]{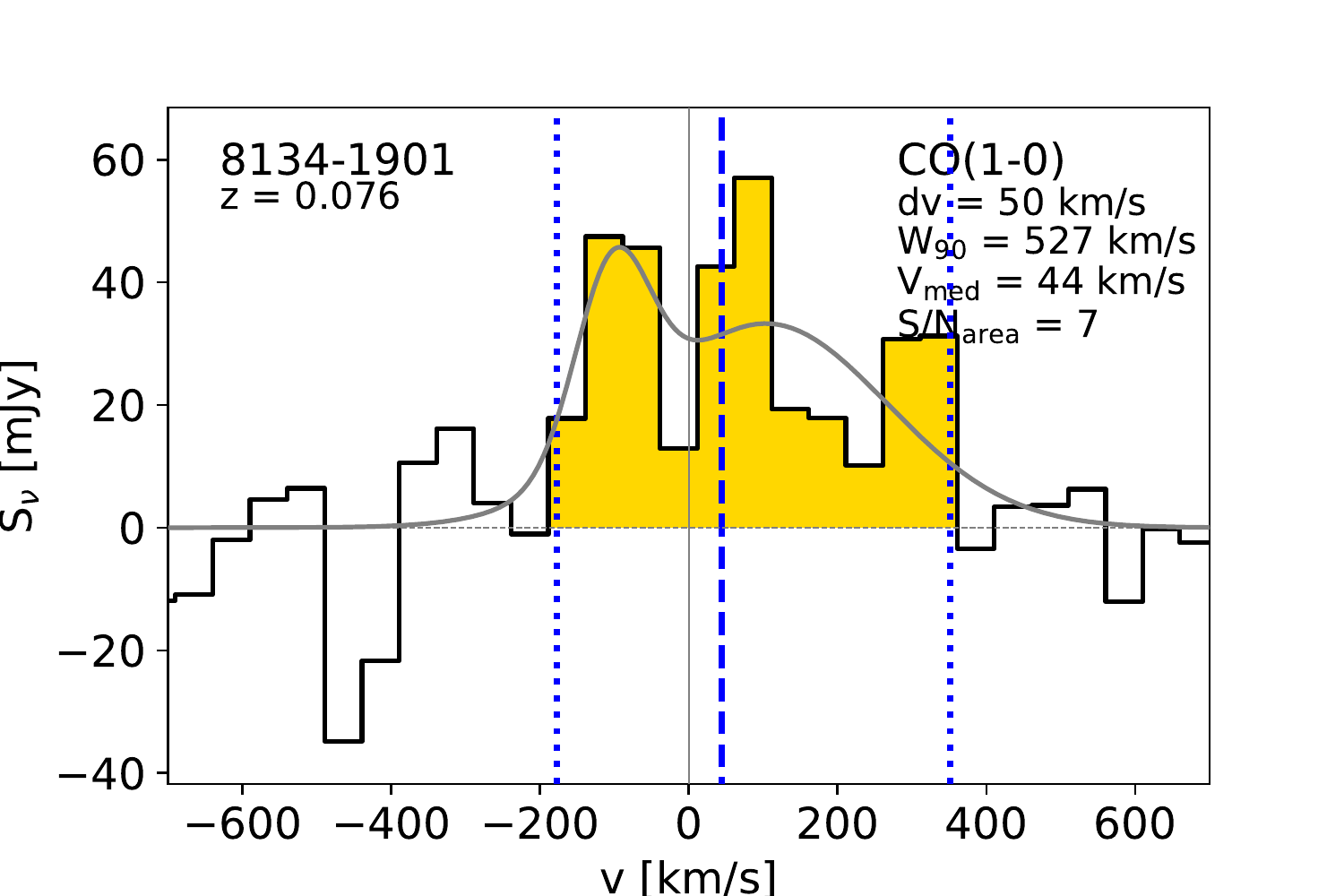} 

\end{figure*}

\begin{figure*} 
 \centering 
 \includegraphics[width = 0.17\textwidth, trim = 0cm 0cm 0cm 0cm, clip = true]{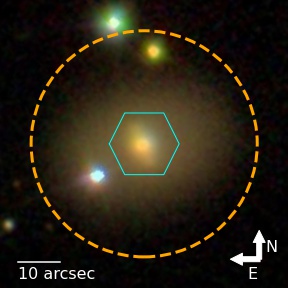}
 \includegraphics[width = 0.29\textwidth, trim = 0cm 0cm 0cm 0cm, clip = true]{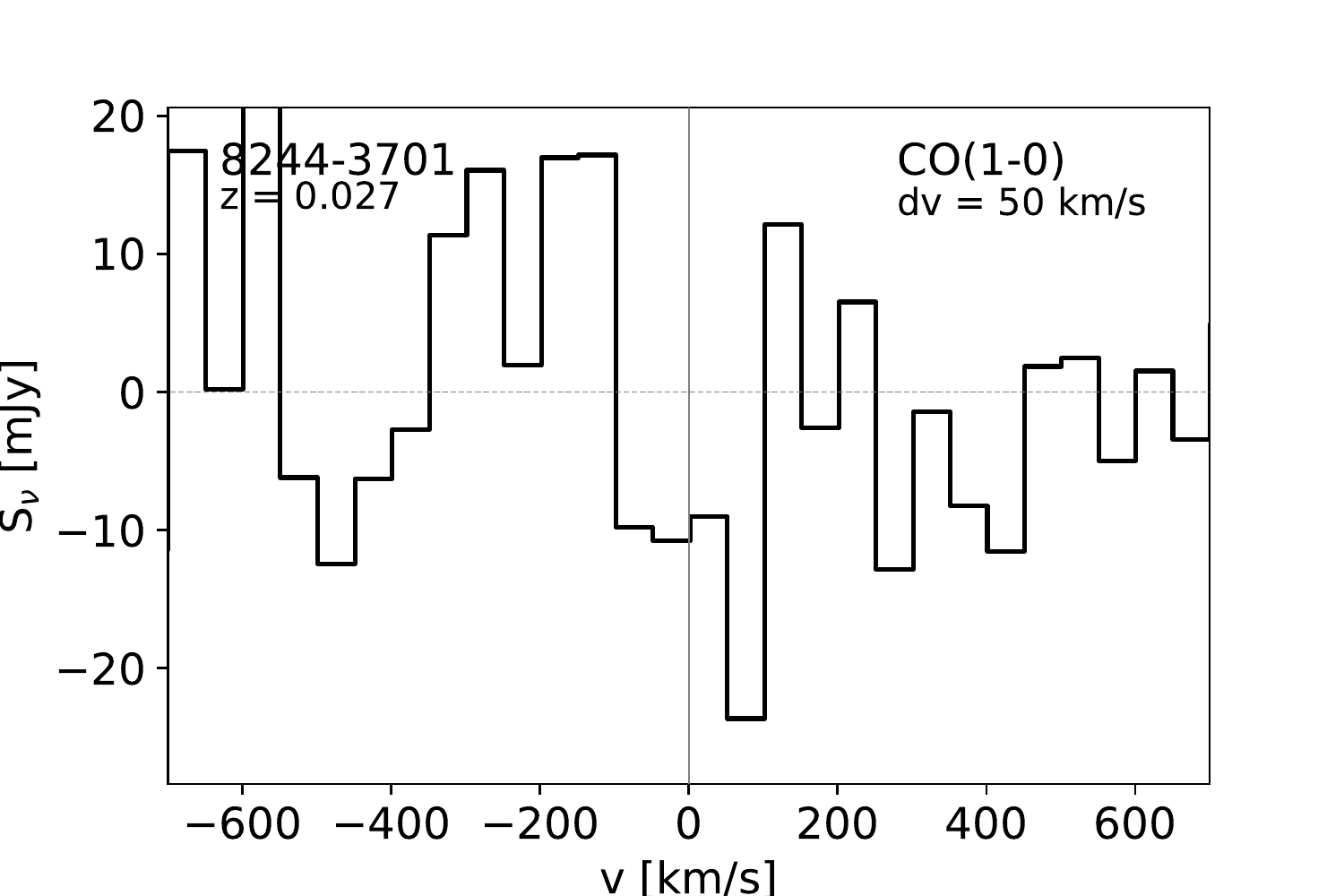} 
 \hspace{0.4cm}
 \centering 
 \includegraphics[width = 0.17\textwidth, trim = 0cm 0cm 0cm 0cm, clip = true]{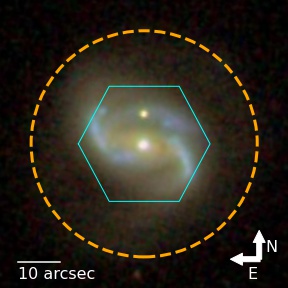}
 \includegraphics[width = 0.29\textwidth, trim = 0cm 0cm 0cm 0cm, clip = true]{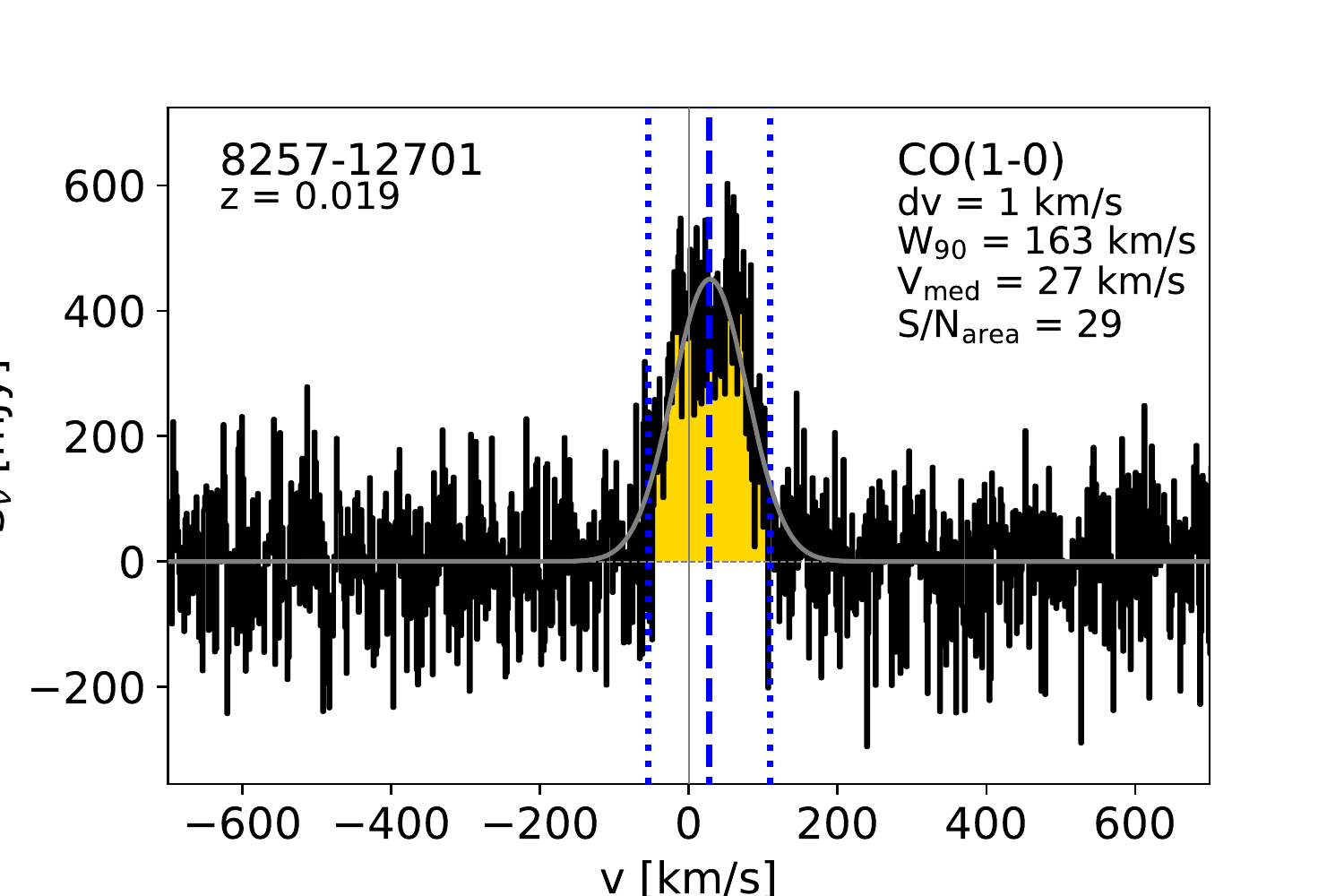} 

\end{figure*}

\begin{figure*} 
 \centering 
 \includegraphics[width = 0.17\textwidth, trim = 0cm 0cm 0cm 0cm, clip = true]{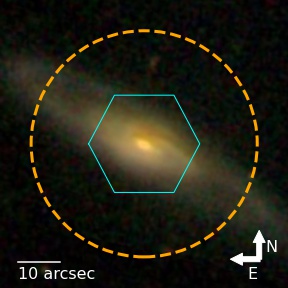}
 \includegraphics[width = 0.29\textwidth, trim = 0cm 0cm 0cm 0cm, clip = true]{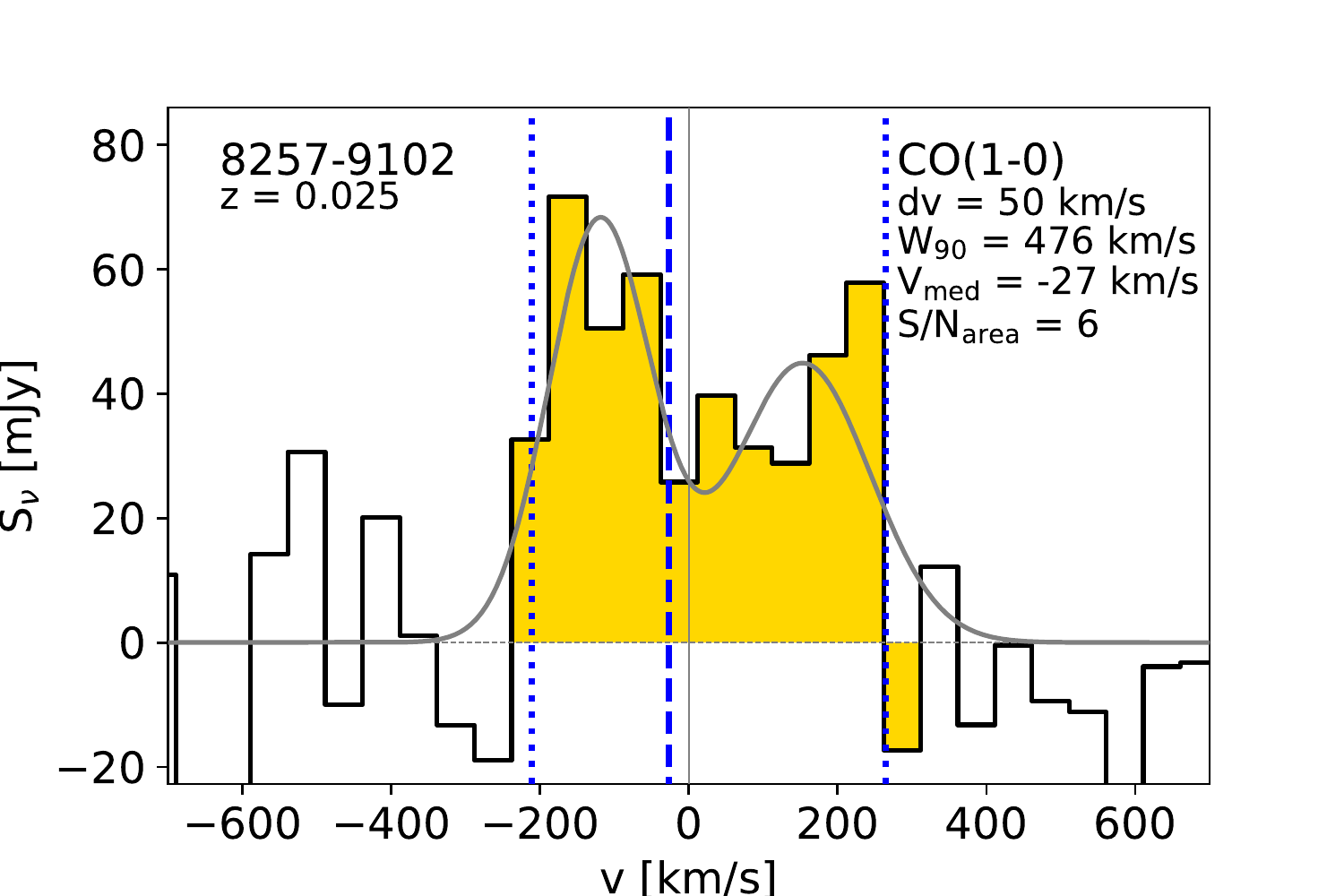} 
 \hspace{0.4cm}
 \centering 
 \includegraphics[width = 0.17\textwidth, trim = 0cm 0cm 0cm 0cm, clip = true]{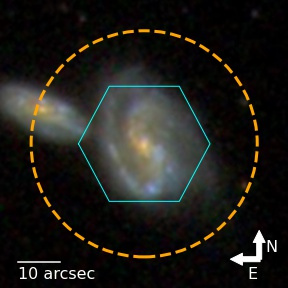}
 \includegraphics[width = 0.29\textwidth, trim = 0cm 0cm 0cm 0cm, clip = true]{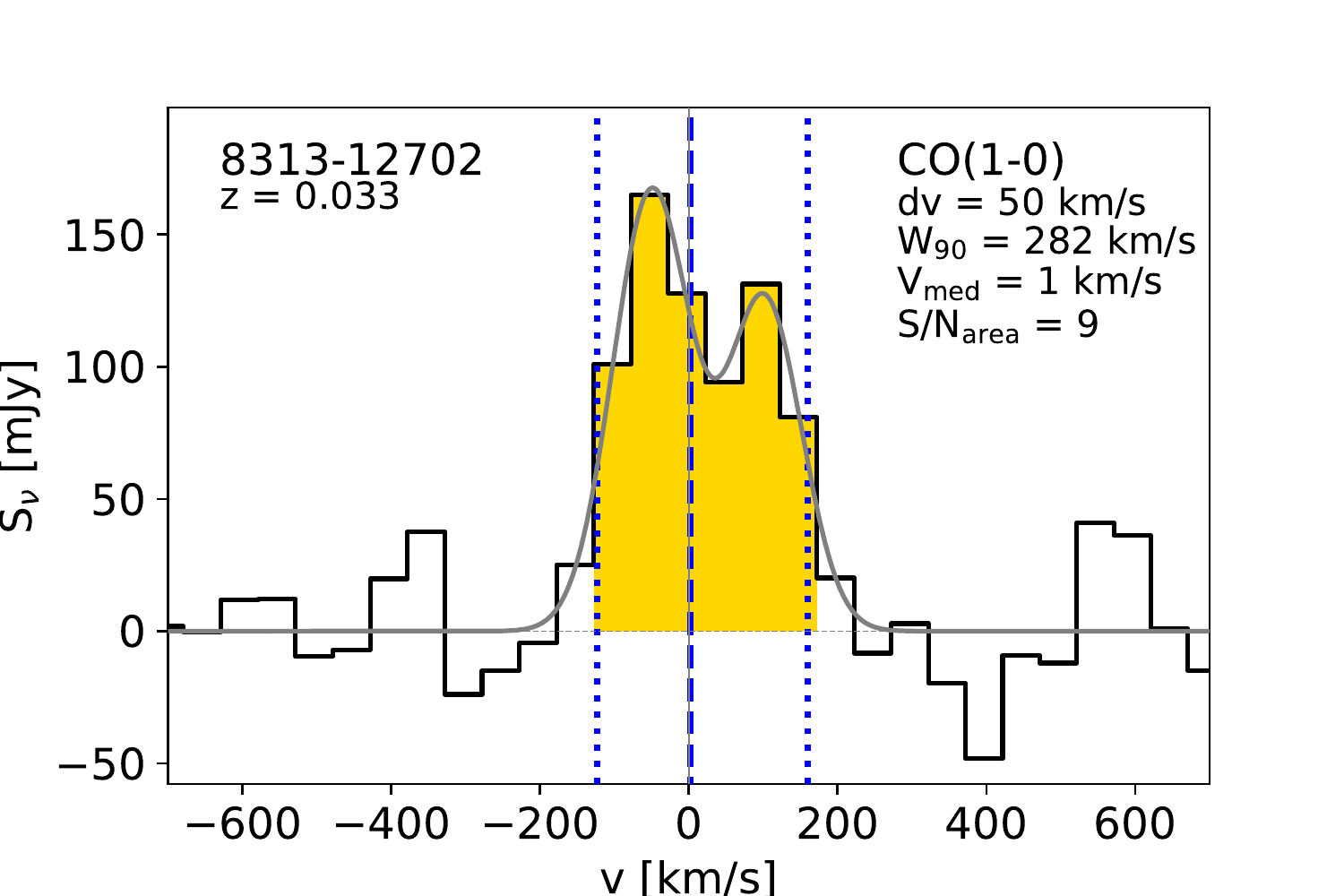} 

\end{figure*}

\begin{figure*} 
   \ContinuedFloat
 \centering 
 \includegraphics[width = 0.17\textwidth, trim = 0cm 0cm 0cm 0cm, clip = true]{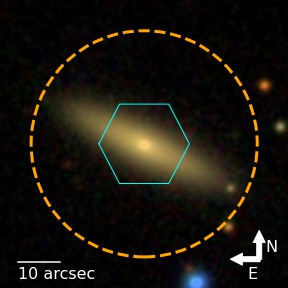}
 \includegraphics[width = 0.29\textwidth, trim = 0cm 0cm 0cm 0cm, clip = true]{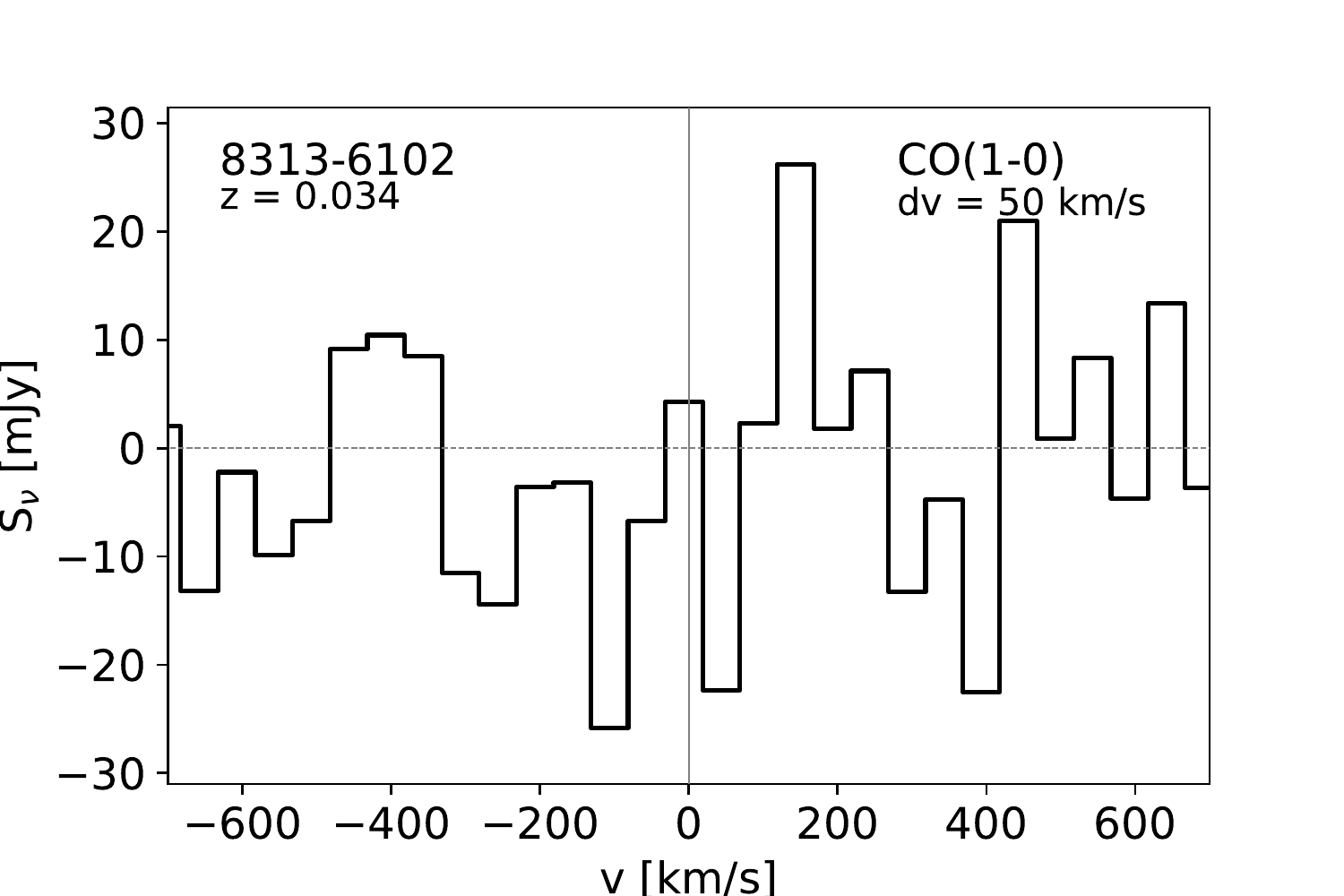} 
 \hspace{0.4cm}
 \centering 
 \includegraphics[width = 0.17\textwidth, trim = 0cm 0cm 0cm 0cm, clip = true]{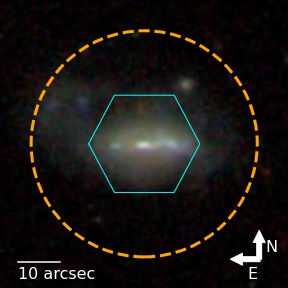}
 \includegraphics[width = 0.29\textwidth, trim = 0cm 0cm 0cm 0cm, clip = true]{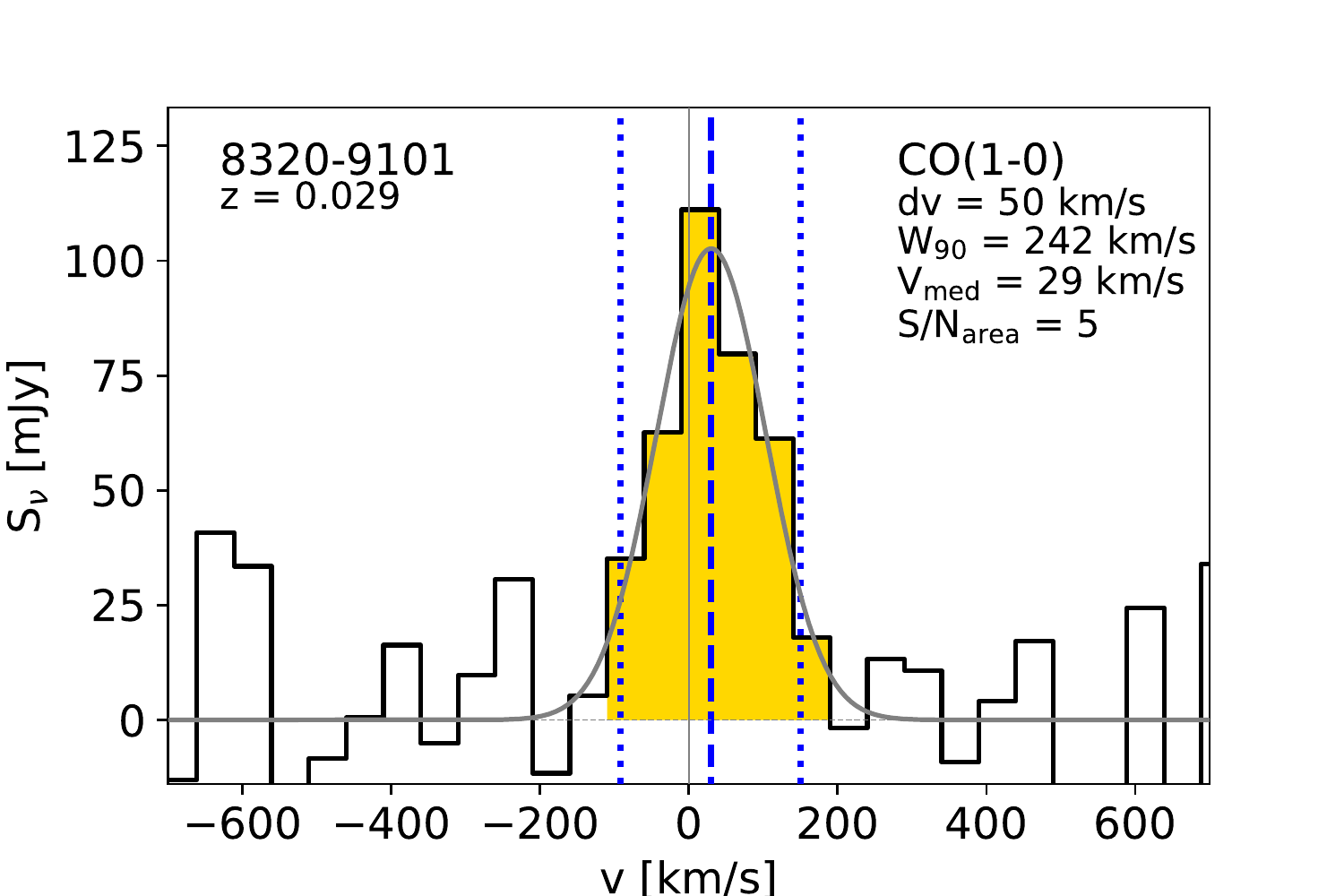} 
\caption{continued.}
\end{figure*}

\begin{figure*} 
 \centering 
 \includegraphics[width = 0.17\textwidth, trim = 0cm 0cm 0cm 0cm, clip = true]{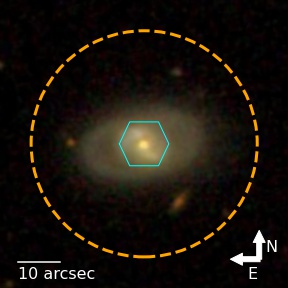}
 \includegraphics[width = 0.29\textwidth, trim = 0cm 0cm 0cm 0cm, clip = true]{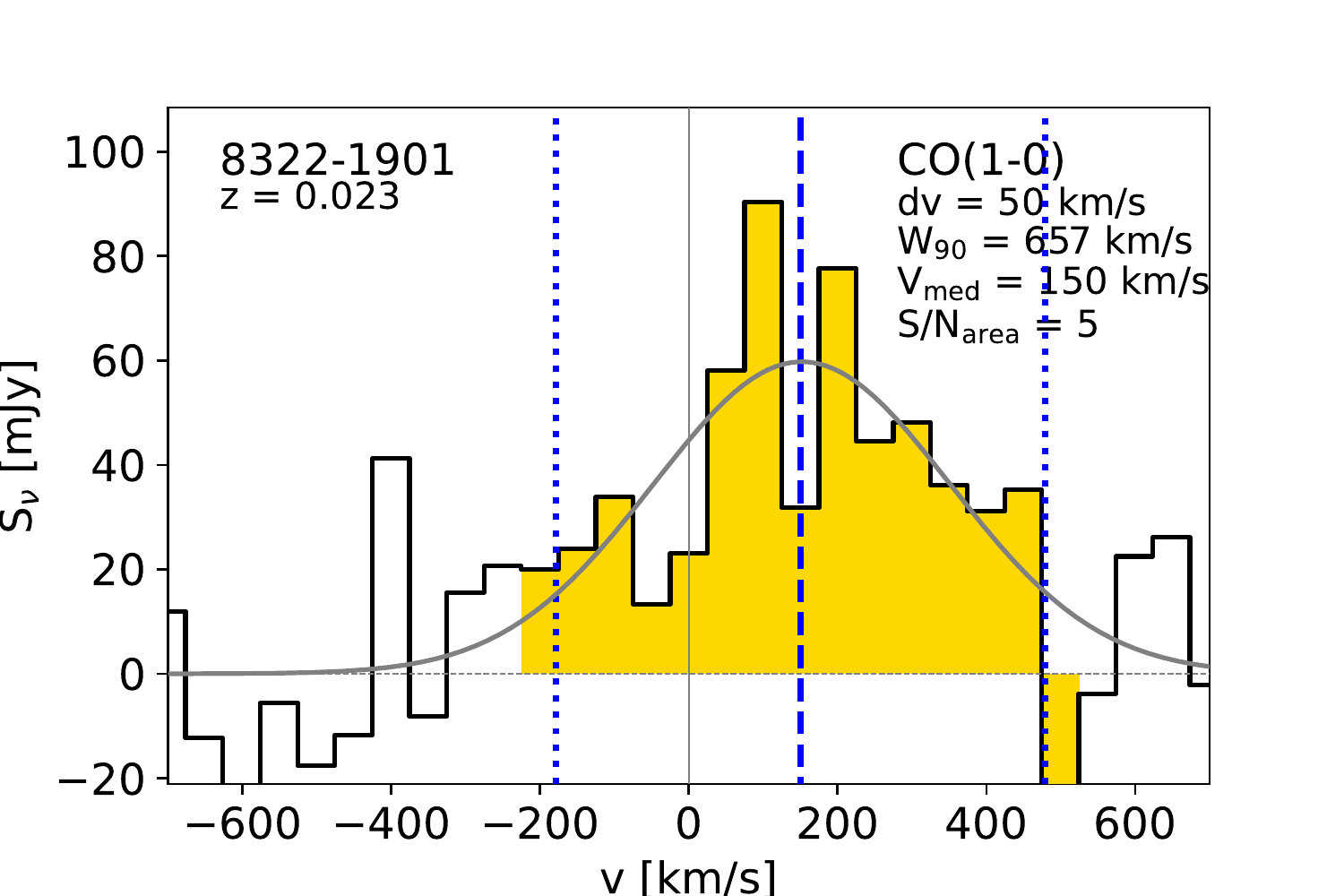} 
 \hspace{0.4cm}
 \centering 
 \includegraphics[width = 0.17\textwidth, trim = 0cm 0cm 0cm 0cm, clip = true]{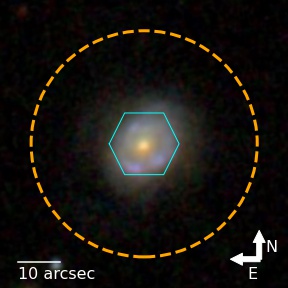}
 \includegraphics[width = 0.29\textwidth, trim = 0cm 0cm 0cm 0cm, clip = true]{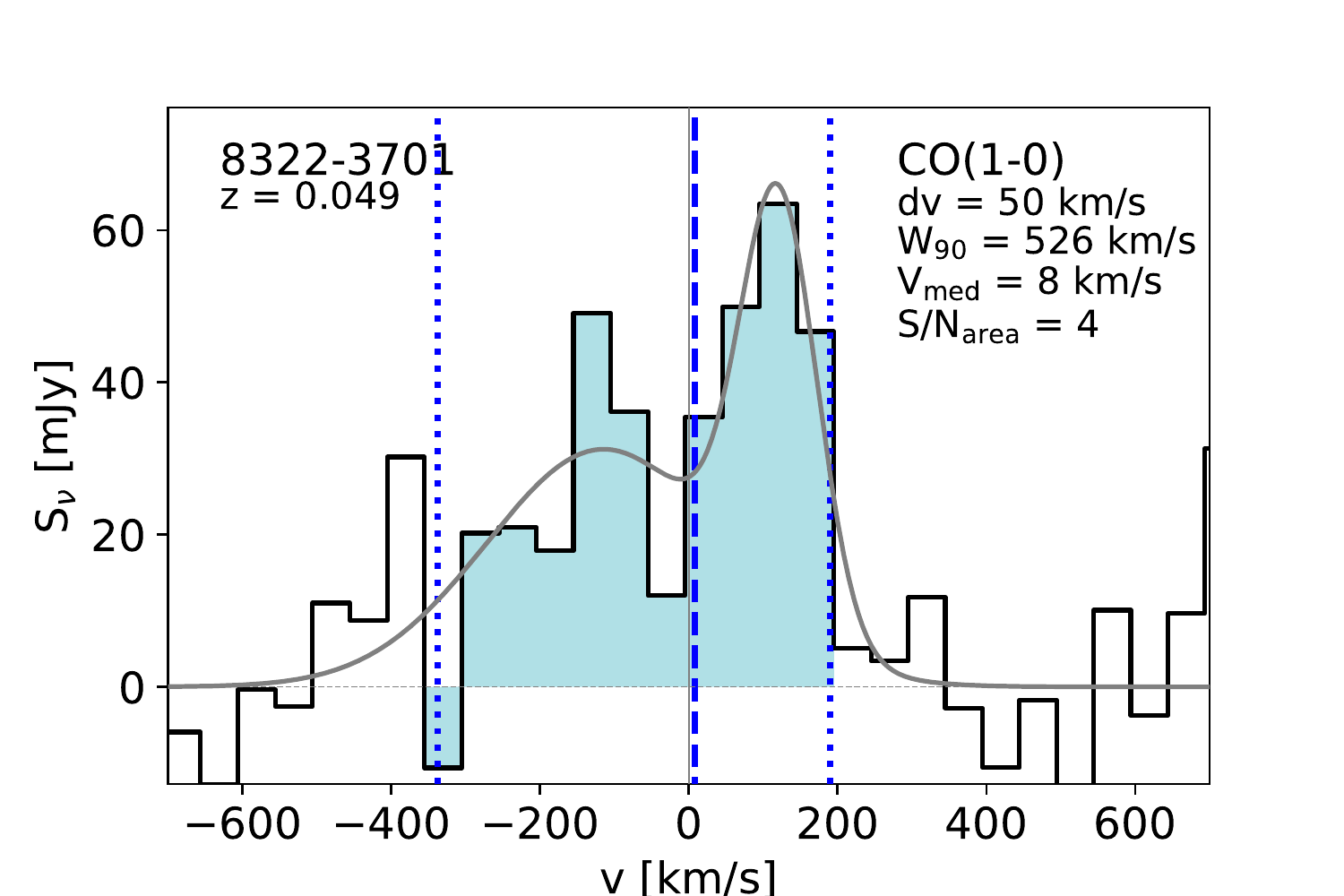} 

\end{figure*}

\begin{figure*} 
 \centering 
 \includegraphics[width = 0.17\textwidth, trim = 0cm 0cm 0cm 0cm, clip = true]{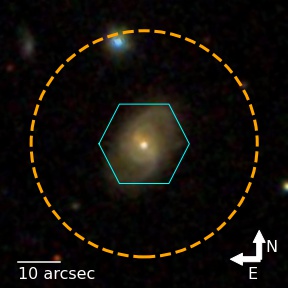}
 \includegraphics[width = 0.29\textwidth, trim = 0cm 0cm 0cm 0cm, clip = true]{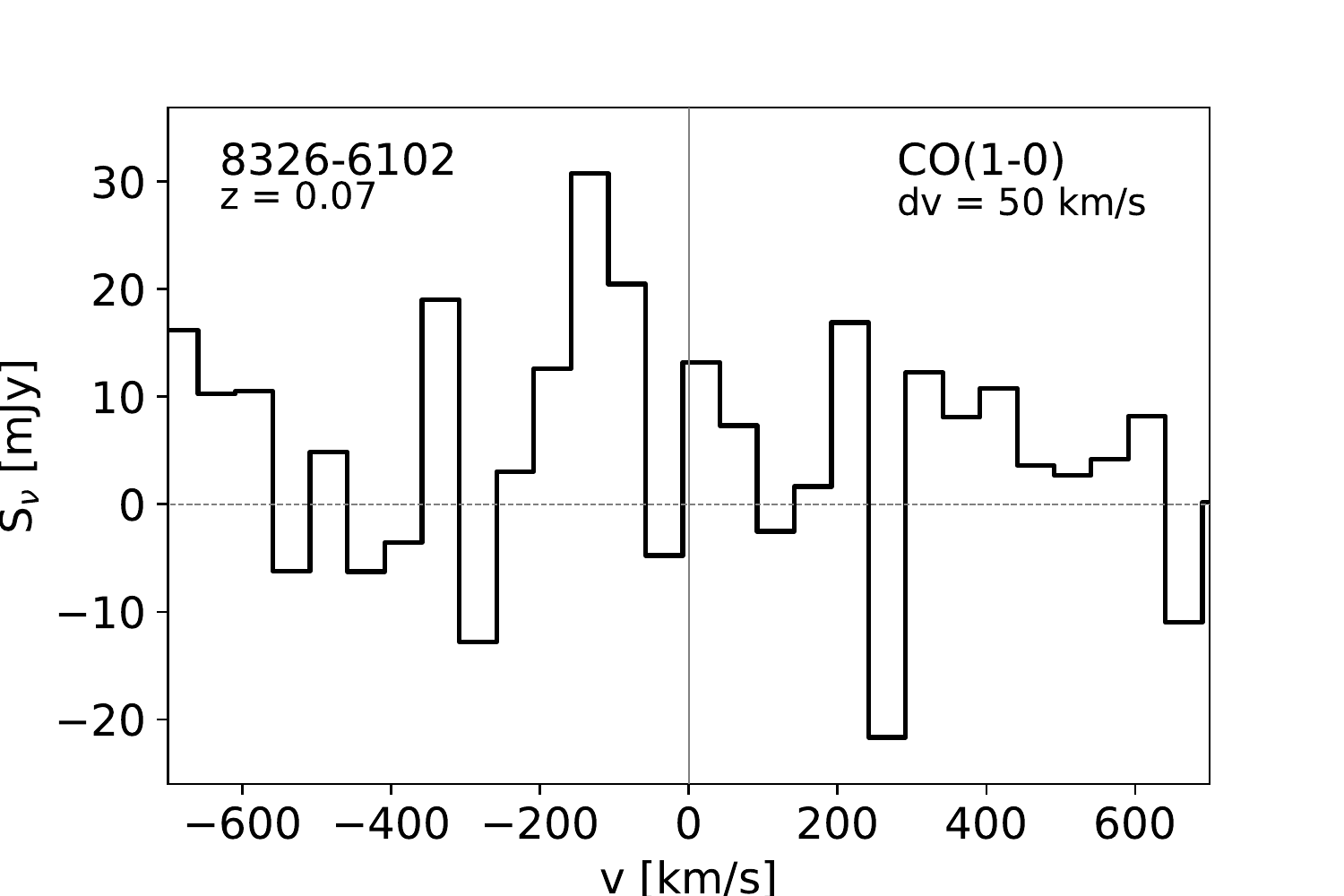} 
 \hspace{0.4cm}
 \centering 
 \includegraphics[width = 0.17\textwidth, trim = 0cm 0cm 0cm 0cm, clip = true]{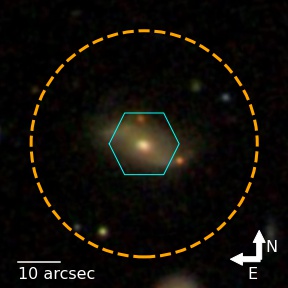}
 \includegraphics[width = 0.29\textwidth, trim = 0cm 0cm 0cm 0cm, clip = true]{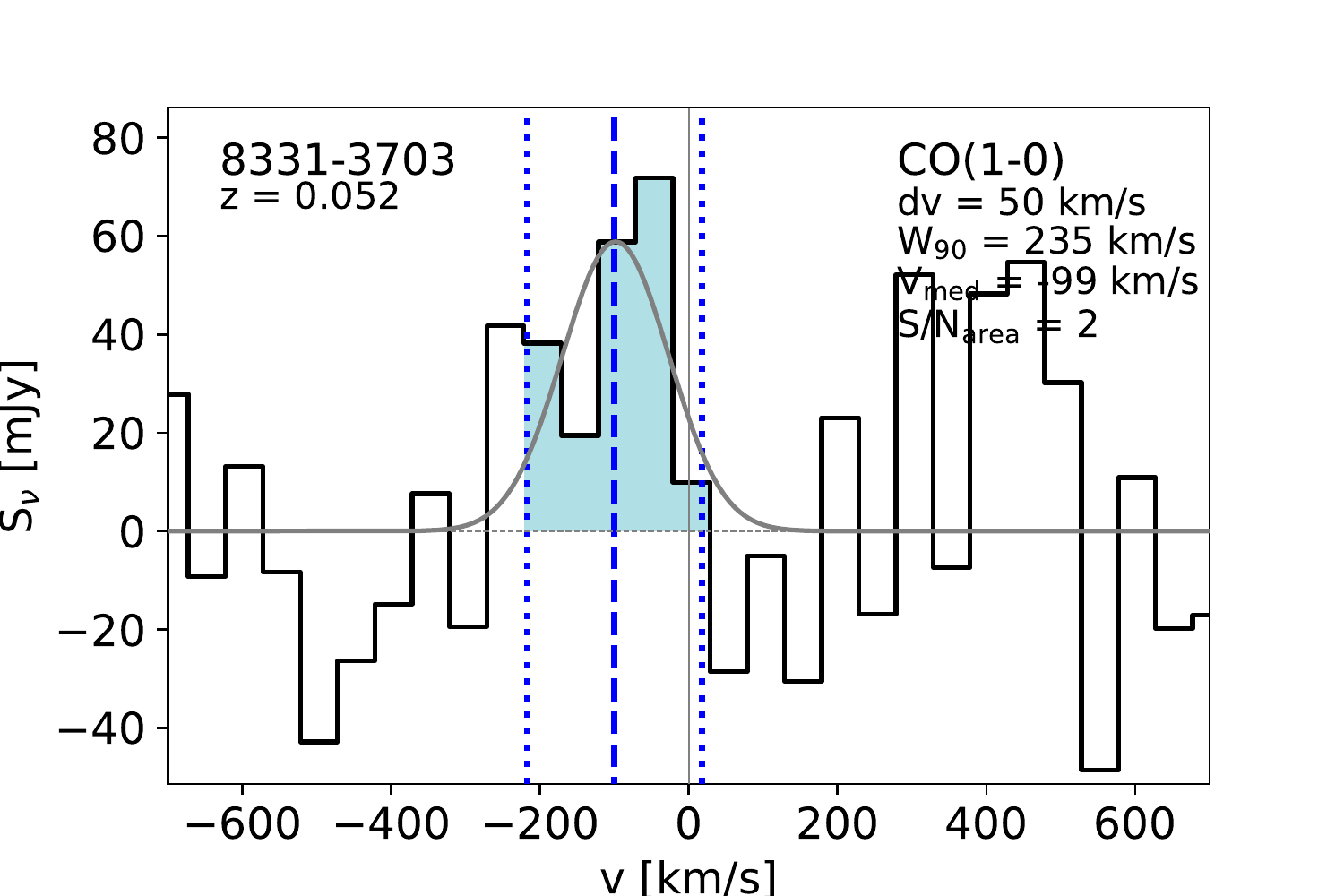} 

\end{figure*}

\begin{figure*} 
 \centering 
 \includegraphics[width = 0.17\textwidth, trim = 0cm 0cm 0cm 0cm, clip = true]{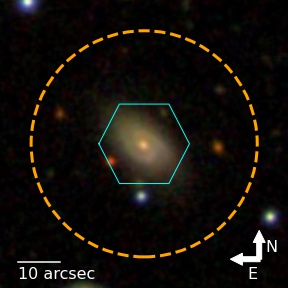}
 \includegraphics[width = 0.29\textwidth, trim = 0cm 0cm 0cm 0cm, clip = true]{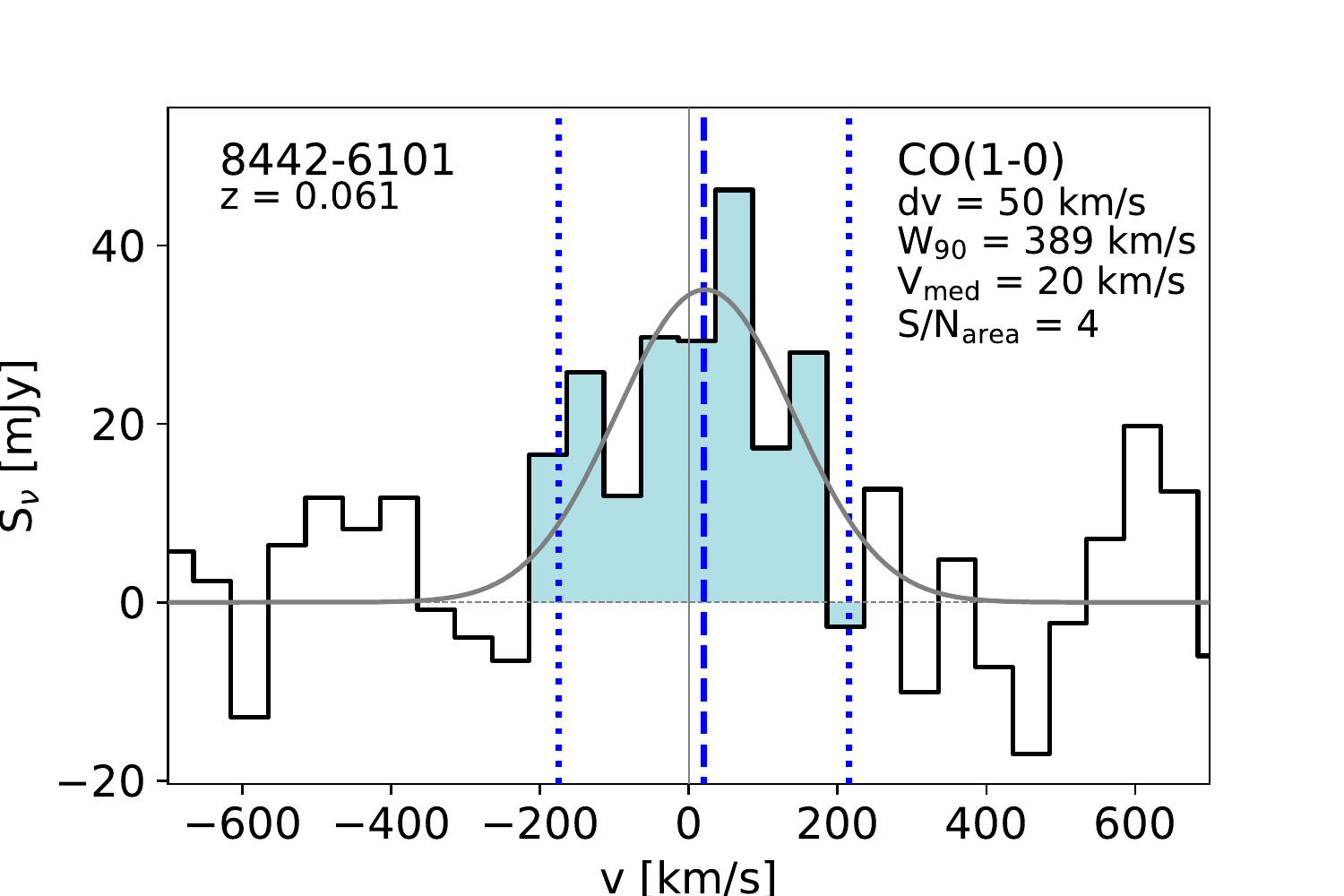} 
 \hspace{0.4cm}
 \centering 
 \includegraphics[width = 0.17\textwidth, trim = 0cm 0cm 0cm 0cm, clip = true]{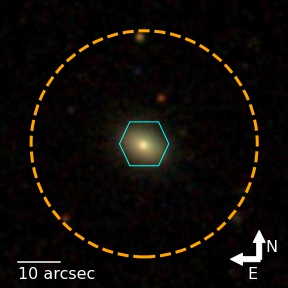}
 \includegraphics[width = 0.29\textwidth, trim = 0cm 0cm 0cm 0cm, clip = true]{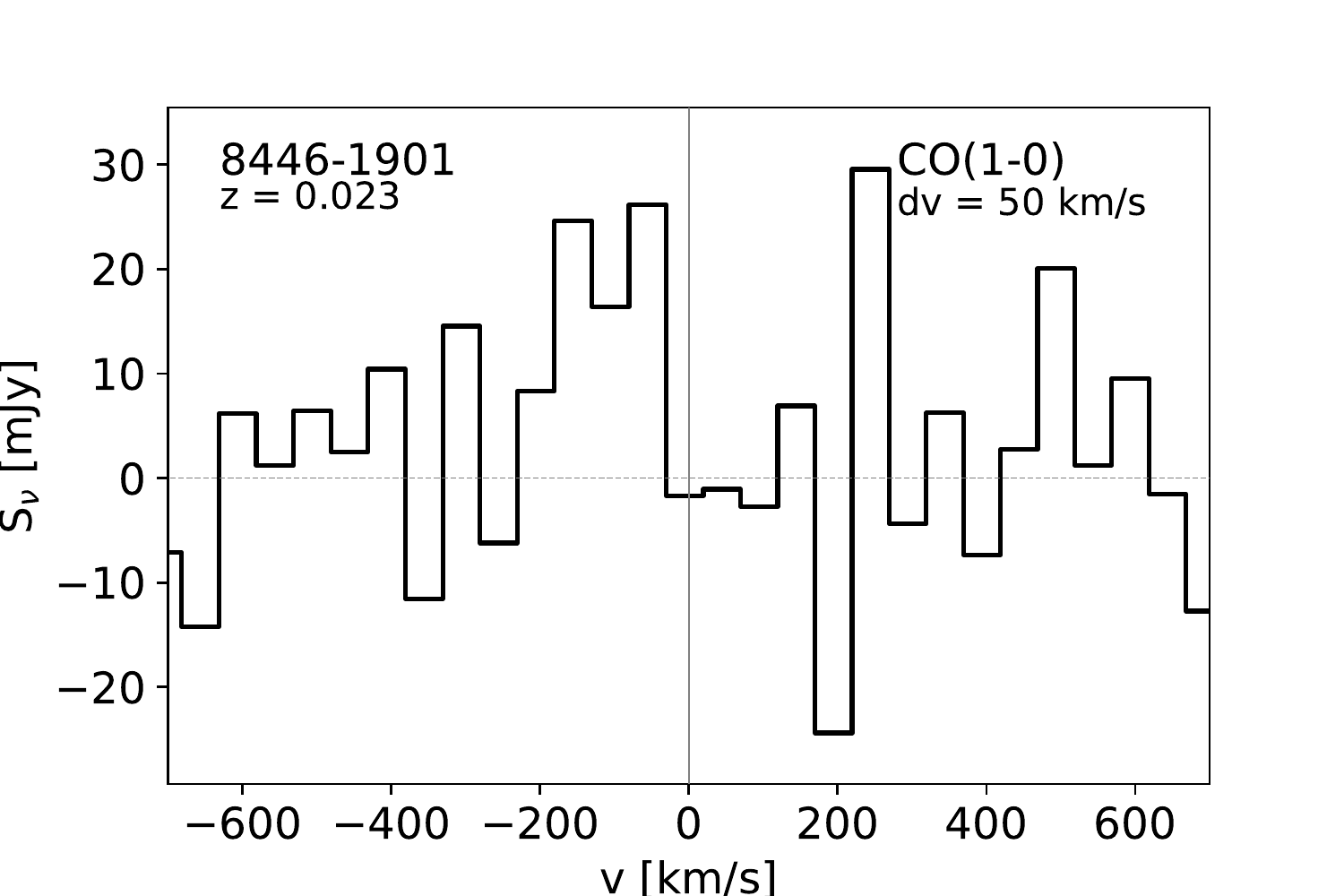} 

\end{figure*}

\begin{figure*} 
 \centering 
 \includegraphics[width = 0.17\textwidth, trim = 0cm 0cm 0cm 0cm, clip = true]{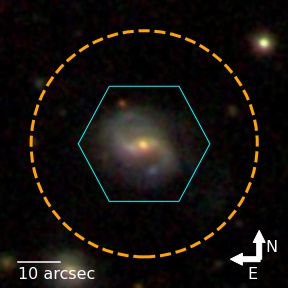}
 \includegraphics[width = 0.29\textwidth, trim = 0cm 0cm 0cm 0cm, clip = true]{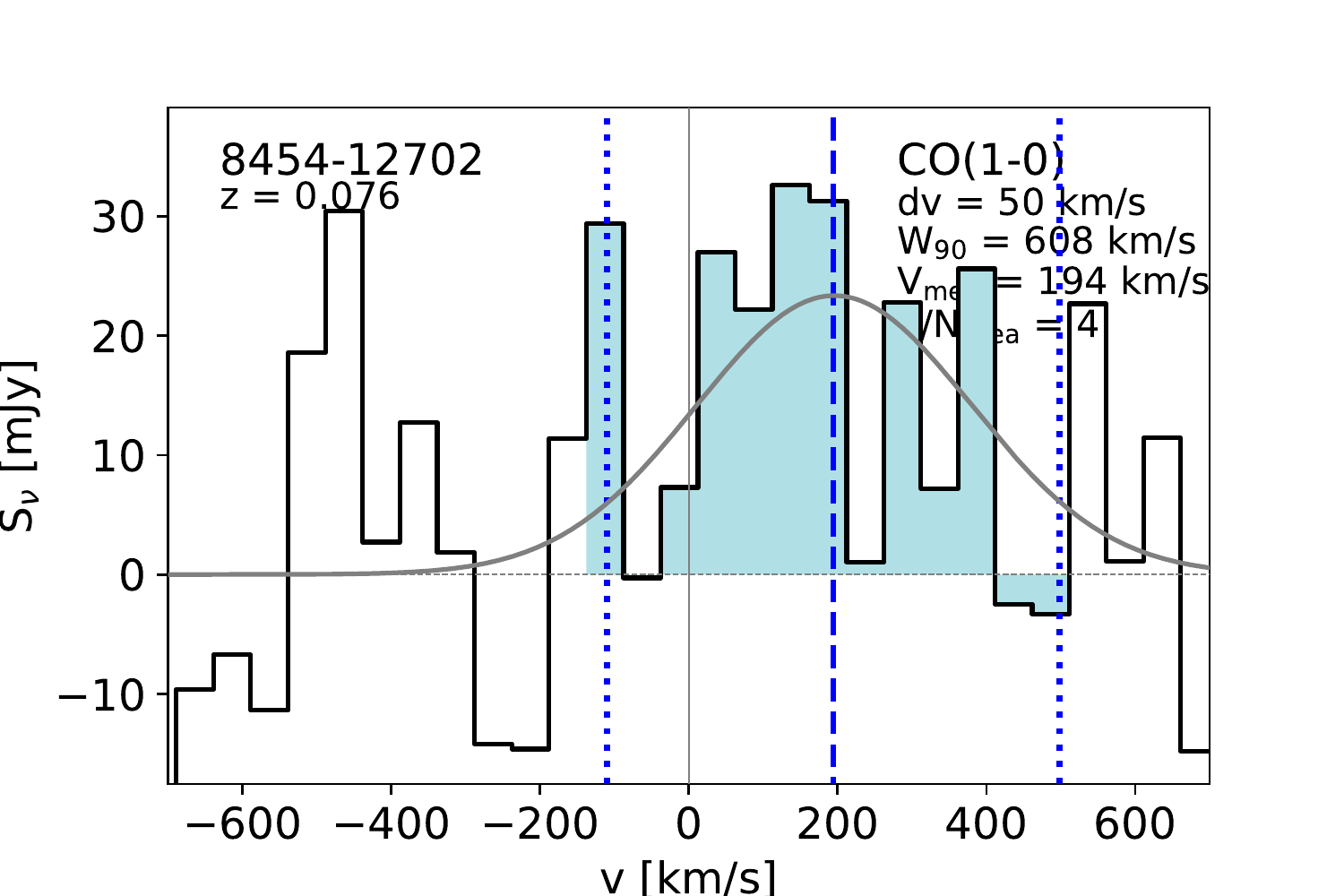} 
 \hspace{0.4cm}
 \centering 
 \includegraphics[width = 0.17\textwidth, trim = 0cm 0cm 0cm 0cm, clip = true]{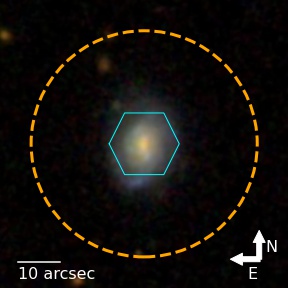}
 \includegraphics[width = 0.29\textwidth, trim = 0cm 0cm 0cm 0cm, clip = true]{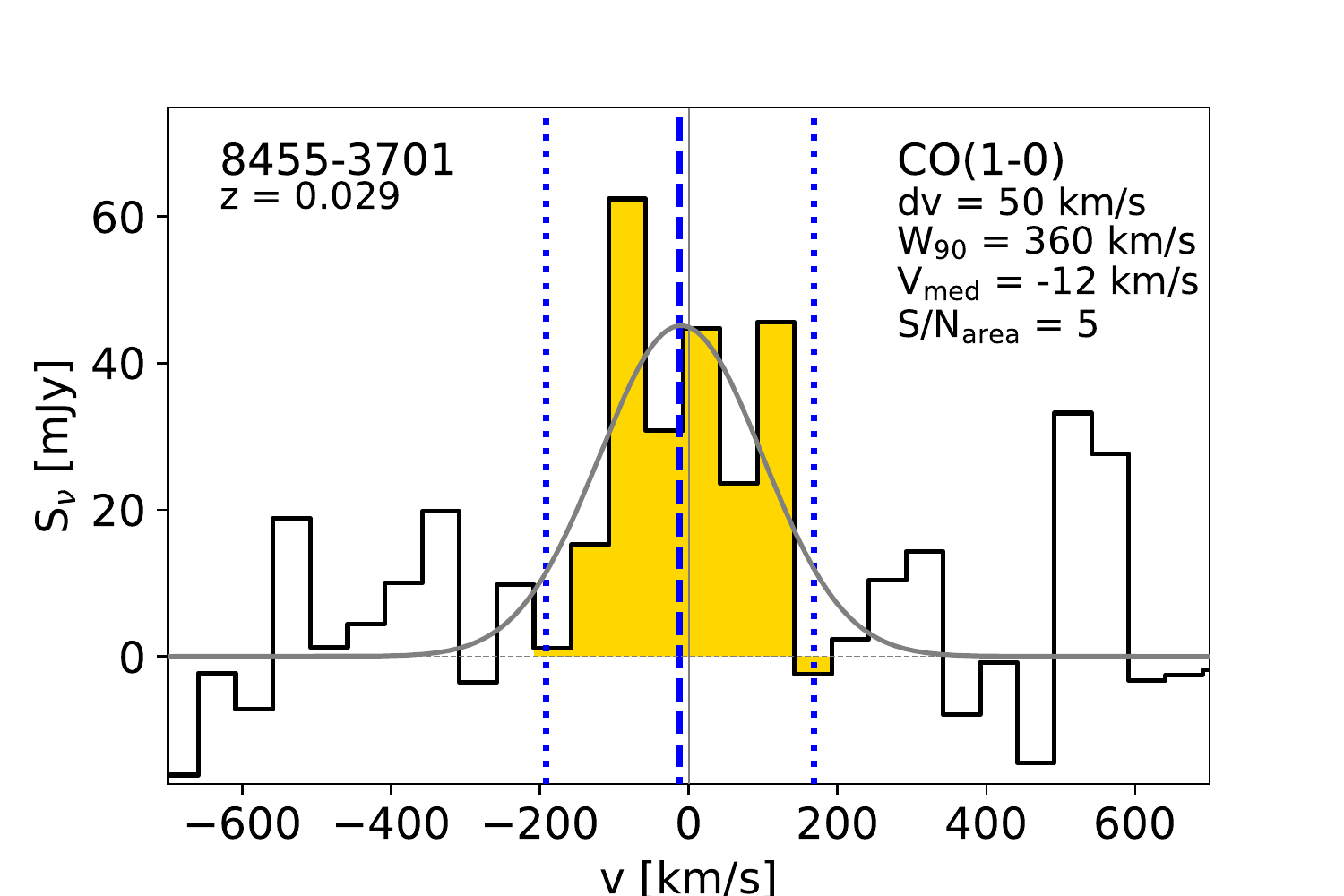} 

\end{figure*}

\begin{figure*} 
 \centering 
 \includegraphics[width = 0.17\textwidth, trim = 0cm 0cm 0cm 0cm, clip = true]{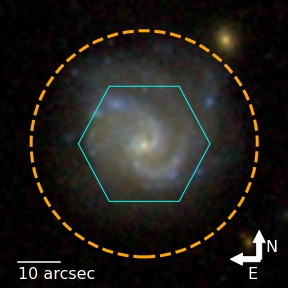}
 \includegraphics[width = 0.29\textwidth, trim = 0cm 0cm 0cm 0cm, clip = true]{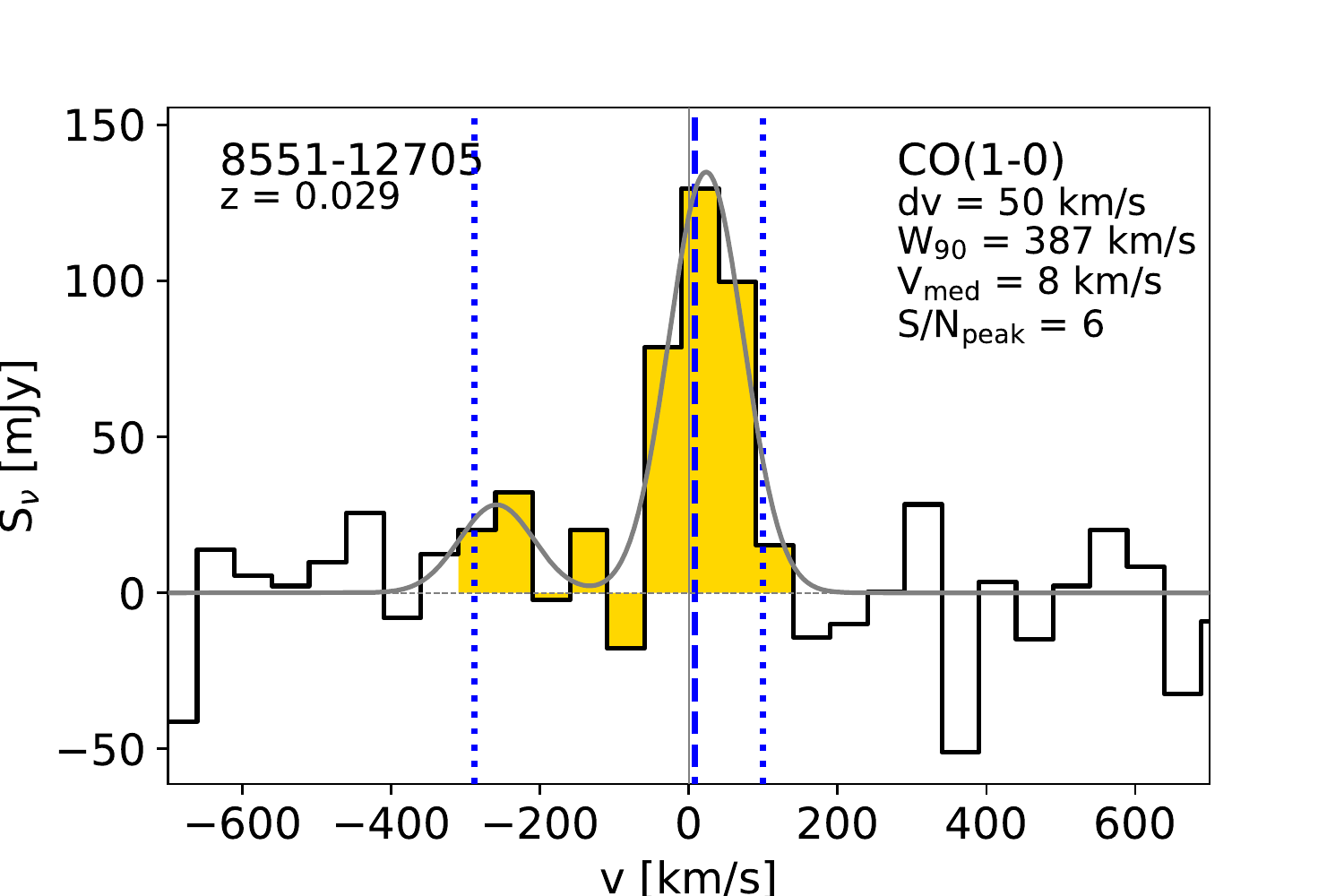} 
 \hspace{0.4cm}
 \centering 
 \includegraphics[width = 0.17\textwidth, trim = 0cm 0cm 0cm 0cm, clip = true]{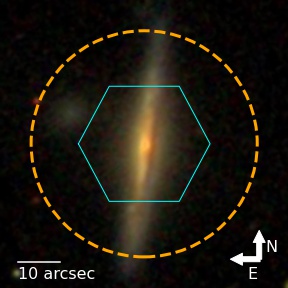}
 \includegraphics[width = 0.29\textwidth, trim = 0cm 0cm 0cm 0cm, clip = true]{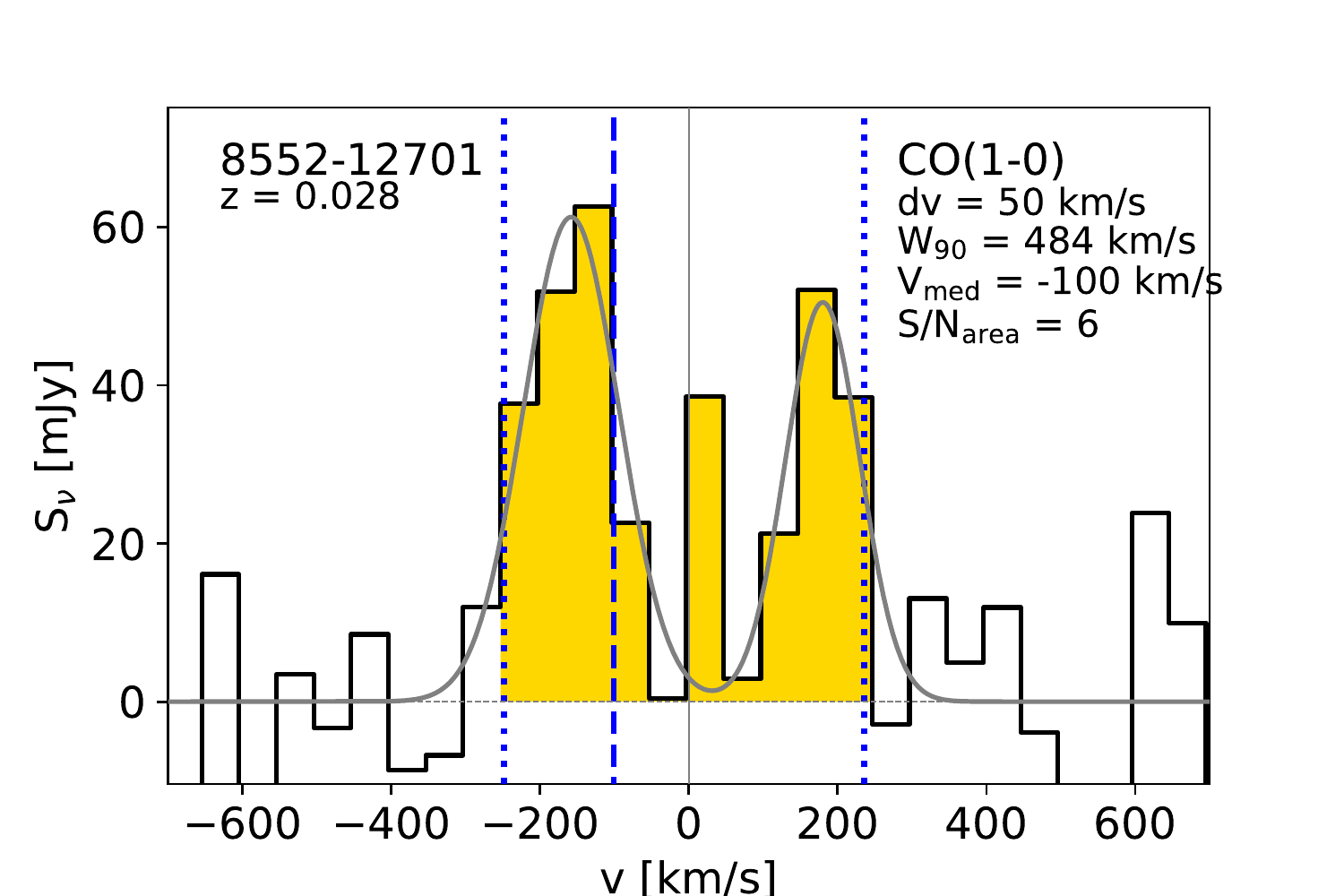} 

\end{figure*}

\begin{figure*} 
   \ContinuedFloat
 \centering 
 \includegraphics[width = 0.17\textwidth, trim = 0cm 0cm 0cm 0cm, clip = true]{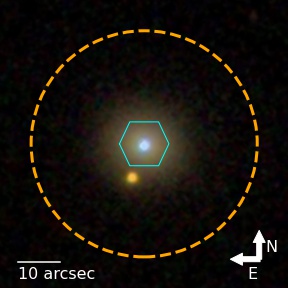}
 \includegraphics[width = 0.29\textwidth, trim = 0cm 0cm 0cm 0cm, clip = true]{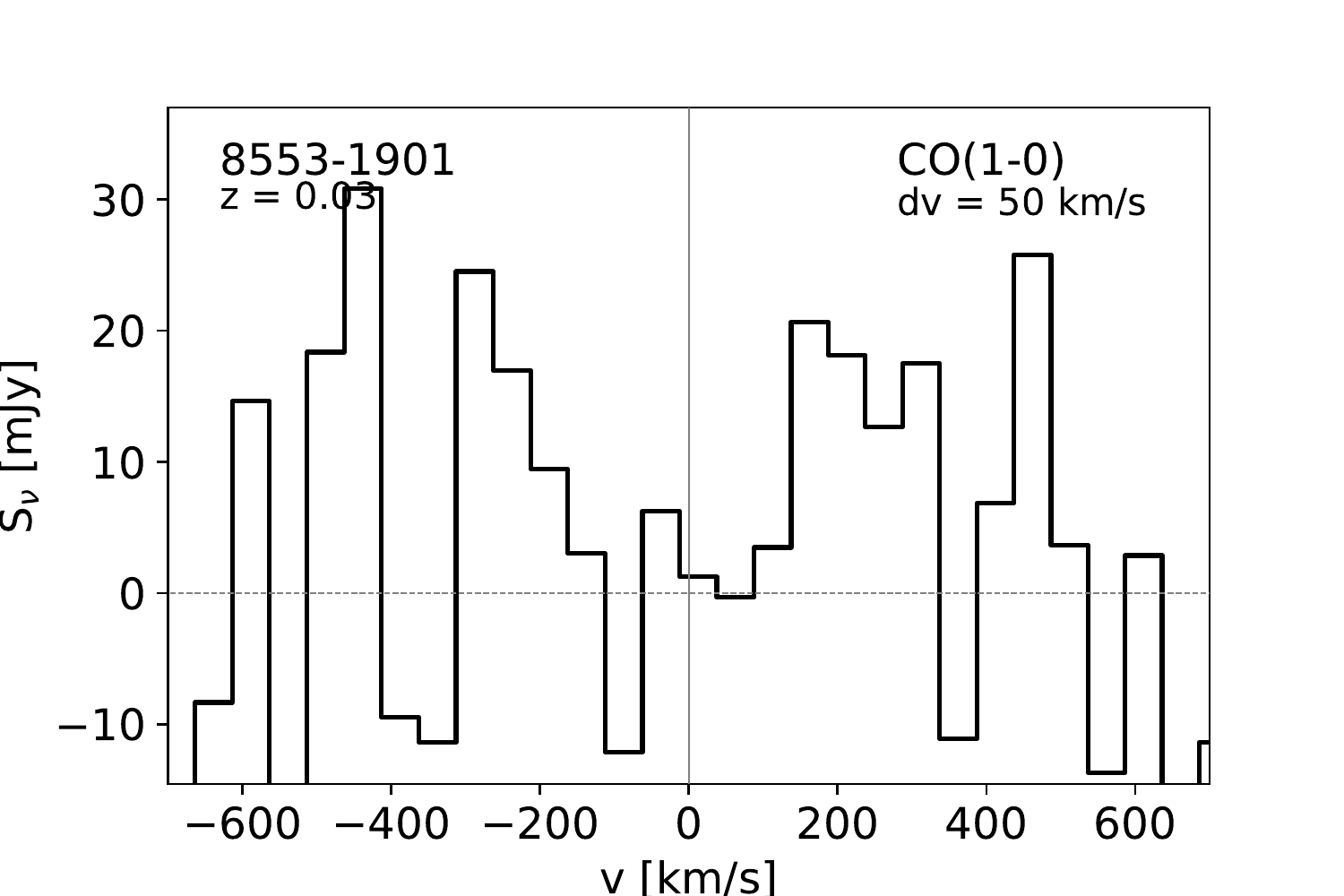} 
 \hspace{0.4cm}
 \centering 
 \includegraphics[width = 0.17\textwidth, trim = 0cm 0cm 0cm 0cm, clip = true]{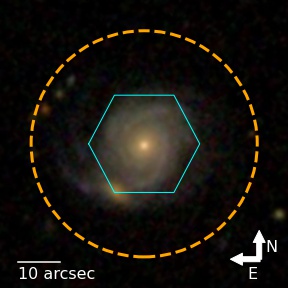}
 \includegraphics[width = 0.29\textwidth, trim = 0cm 0cm 0cm 0cm, clip = true]{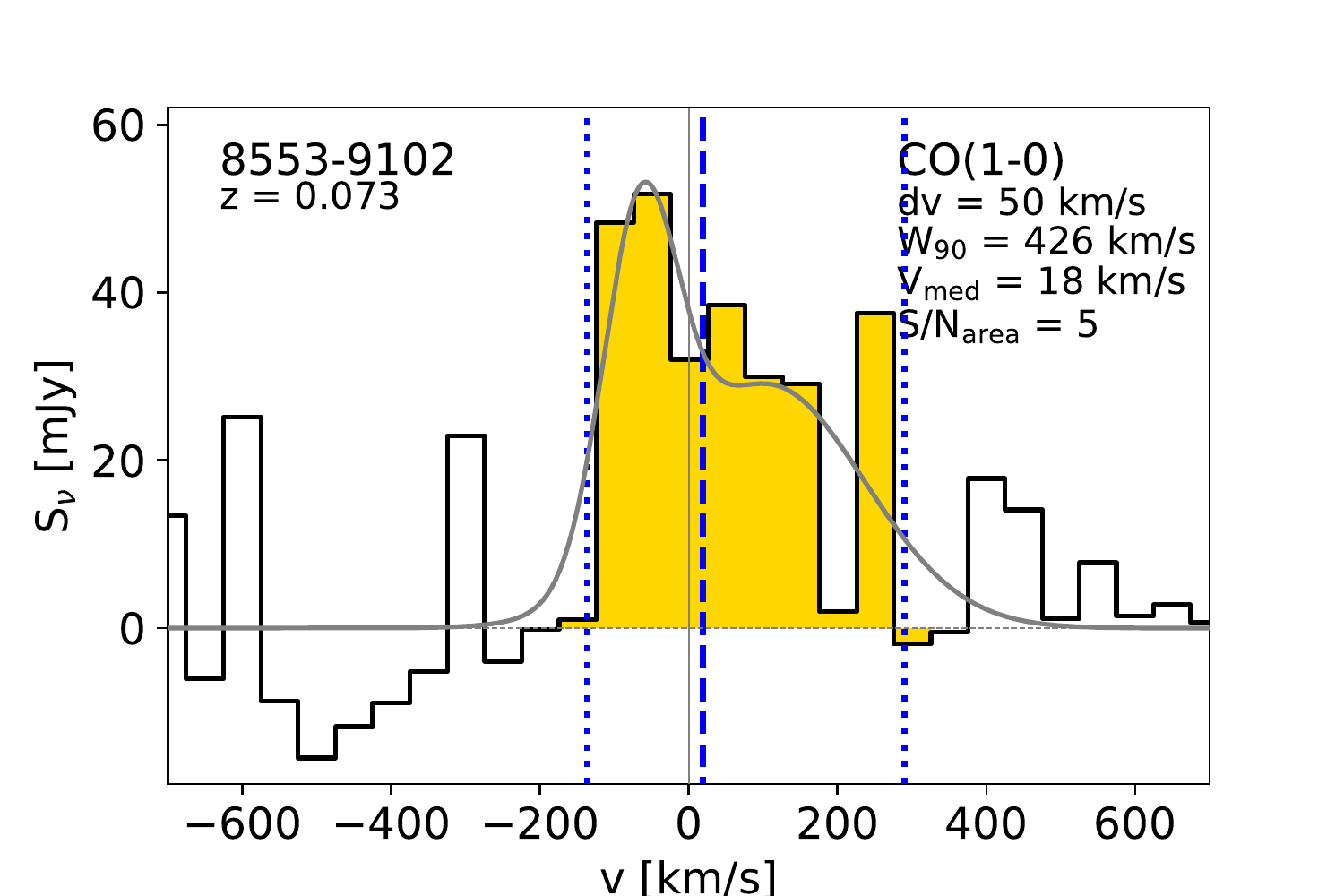} 
\caption{continued.}
\end{figure*}

\begin{figure*} 
 \centering 
 \includegraphics[width = 0.17\textwidth, trim = 0cm 0cm 0cm 0cm, clip = true]{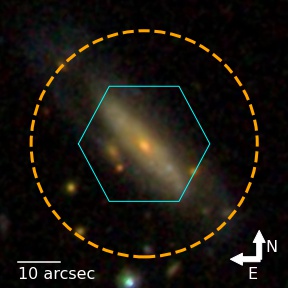}
 \includegraphics[width = 0.29\textwidth, trim = 0cm 0cm 0cm 0cm, clip = true]{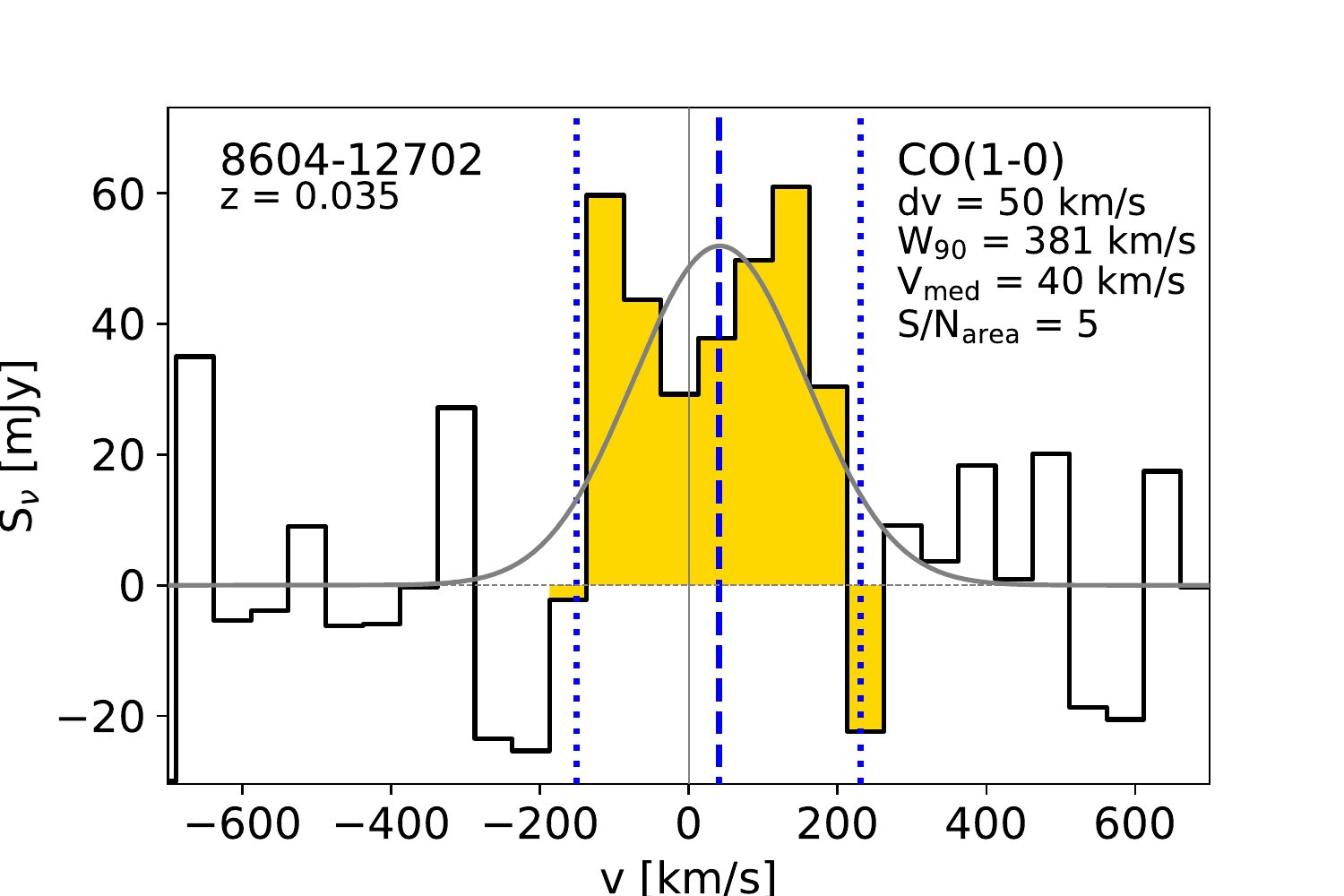} 
 \hspace{0.4cm}
 \centering 
 \includegraphics[width = 0.17\textwidth, trim = 0cm 0cm 0cm 0cm, clip = true]{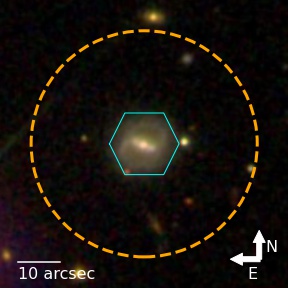}
 \includegraphics[width = 0.29\textwidth, trim = 0cm 0cm 0cm 0cm, clip = true]{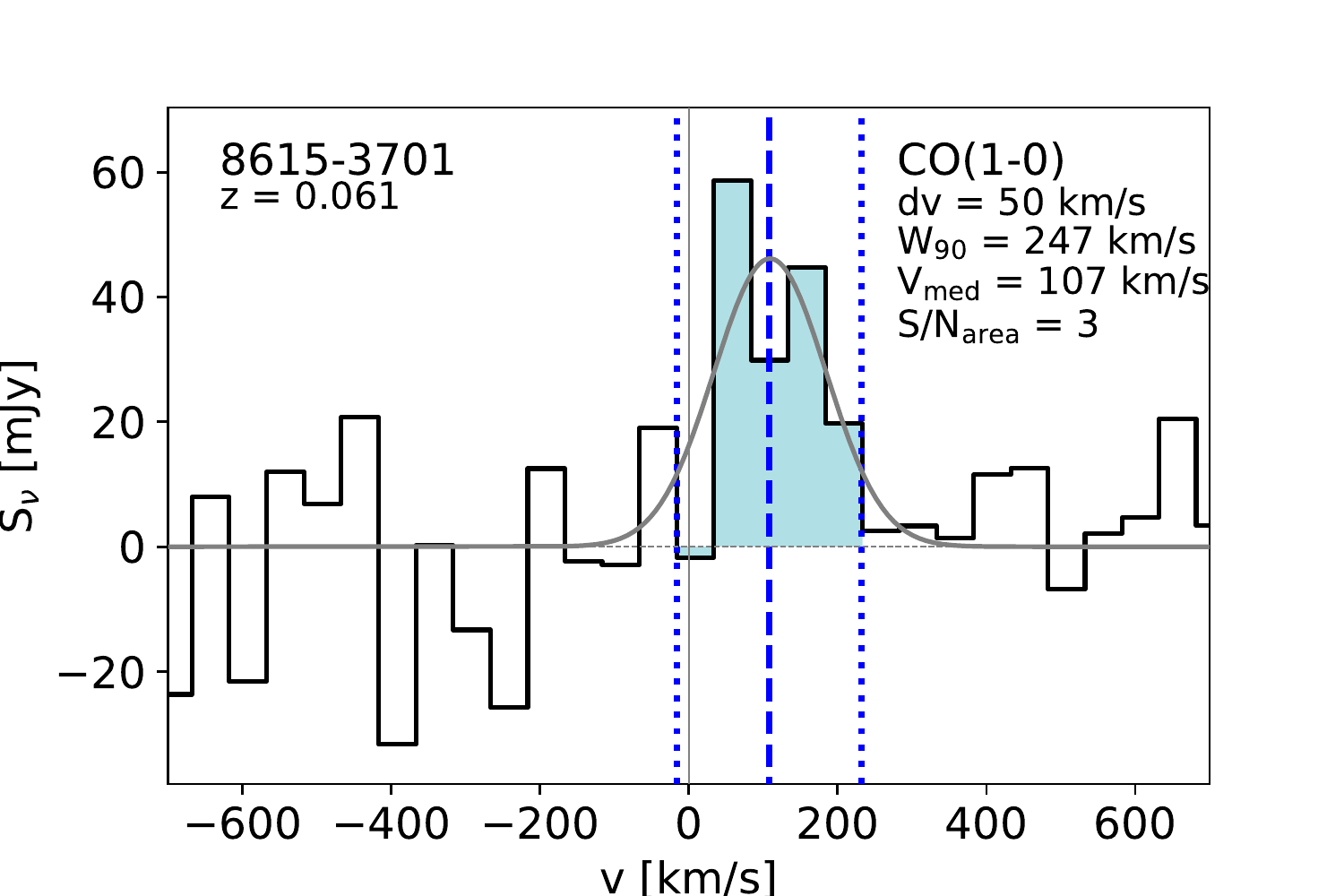} 

\end{figure*}

\begin{figure*} 
 \centering 
 \includegraphics[width = 0.17\textwidth, trim = 0cm 0cm 0cm 0cm, clip = true]{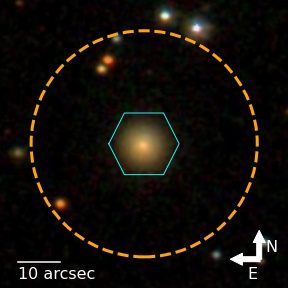}
 \includegraphics[width = 0.29\textwidth, trim = 0cm 0cm 0cm 0cm, clip = true]{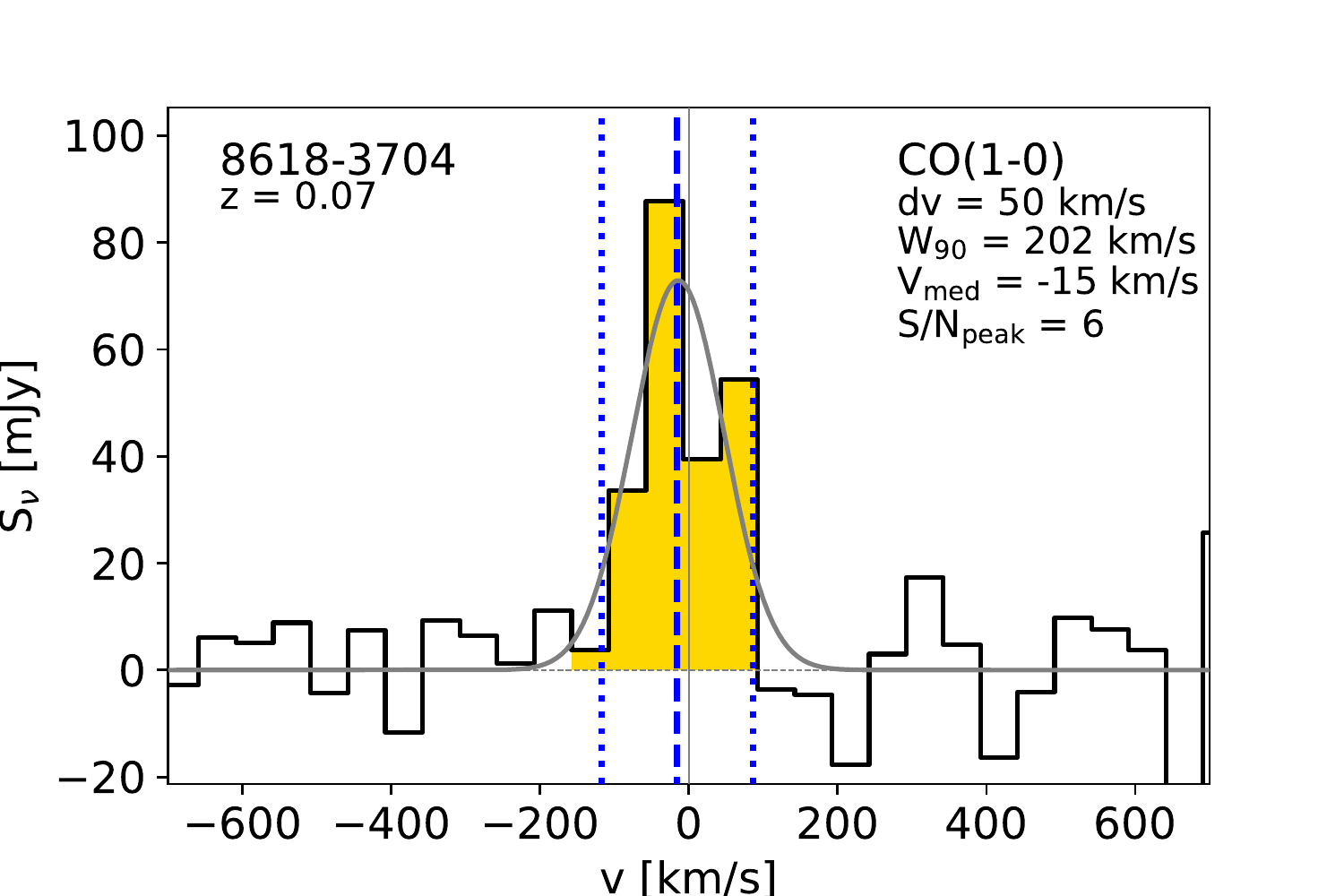} 
 \hspace{0.4cm}
 \centering 
 \includegraphics[width = 0.17\textwidth, trim = 0cm 0cm 0cm 0cm, clip = true]{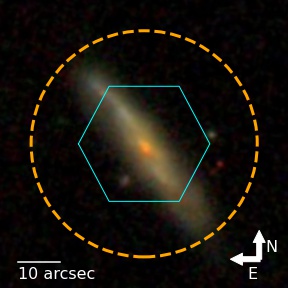}
 \includegraphics[width = 0.29\textwidth, trim = 0cm 0cm 0cm 0cm, clip = true]{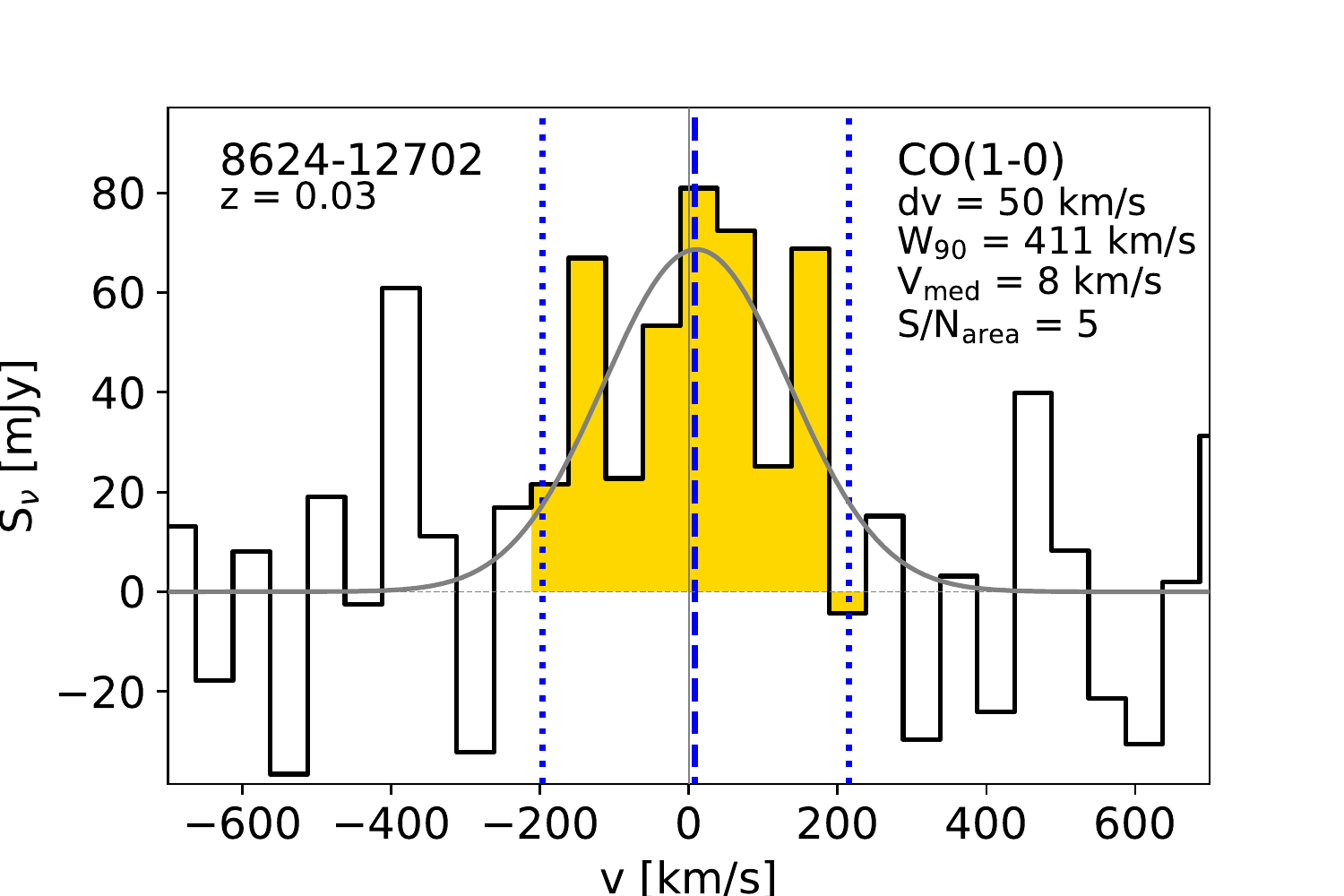} 

\end{figure*}

\begin{figure*} 
 \centering 
 \includegraphics[width = 0.17\textwidth, trim = 0cm 0cm 0cm 0cm, clip = true]{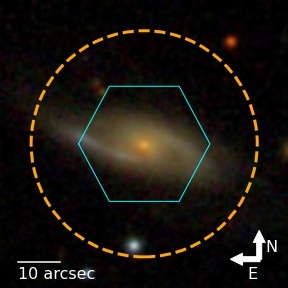}
 \includegraphics[width = 0.29\textwidth, trim = 0cm 0cm 0cm 0cm, clip = true]{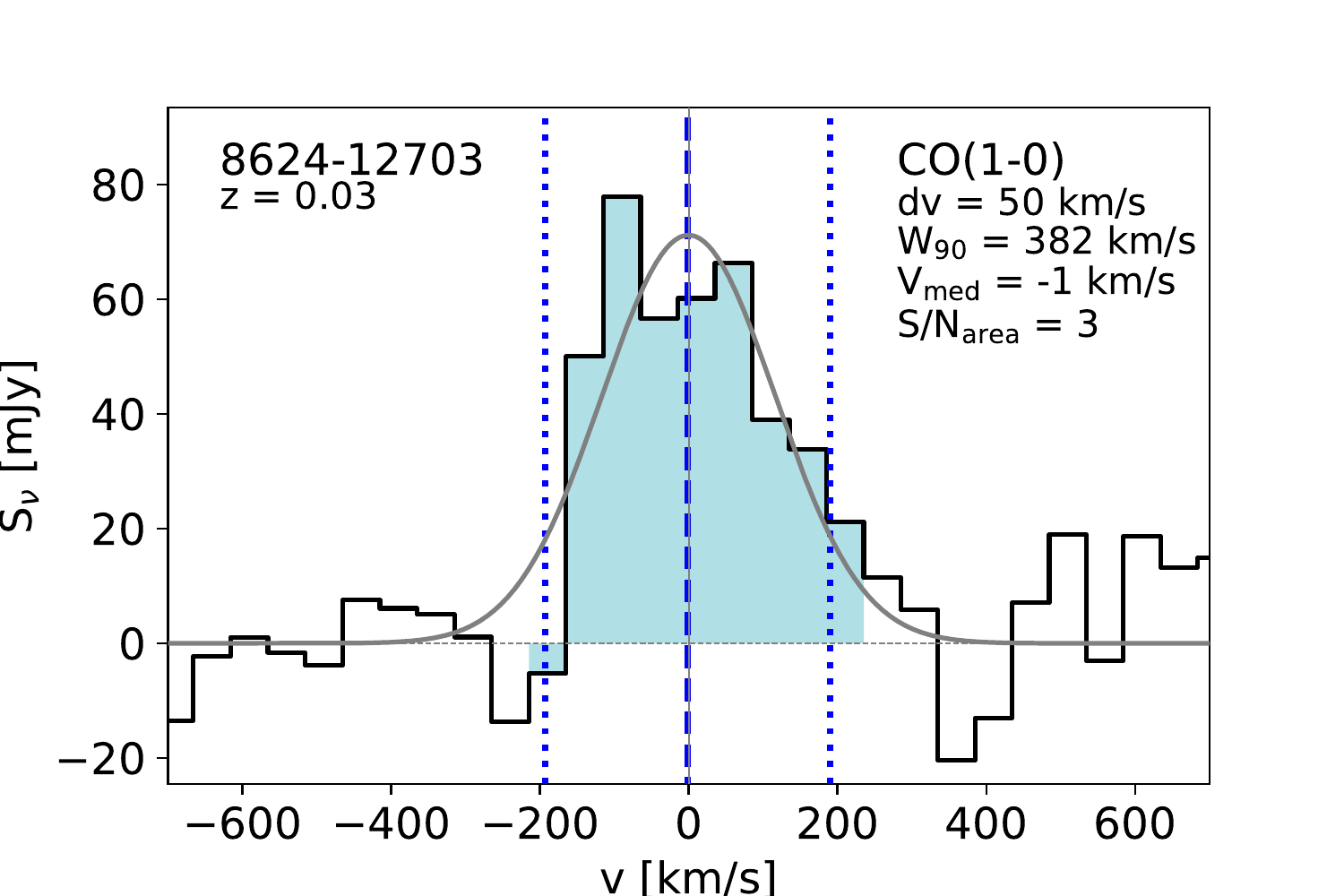} 
 \hspace{0.4cm}
 \centering 
 \includegraphics[width = 0.17\textwidth, trim = 0cm 0cm 0cm 0cm, clip = true]{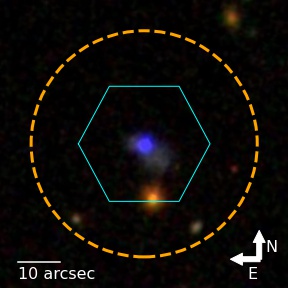}
 \includegraphics[width = 0.29\textwidth, trim = 0cm 0cm 0cm 0cm, clip = true]{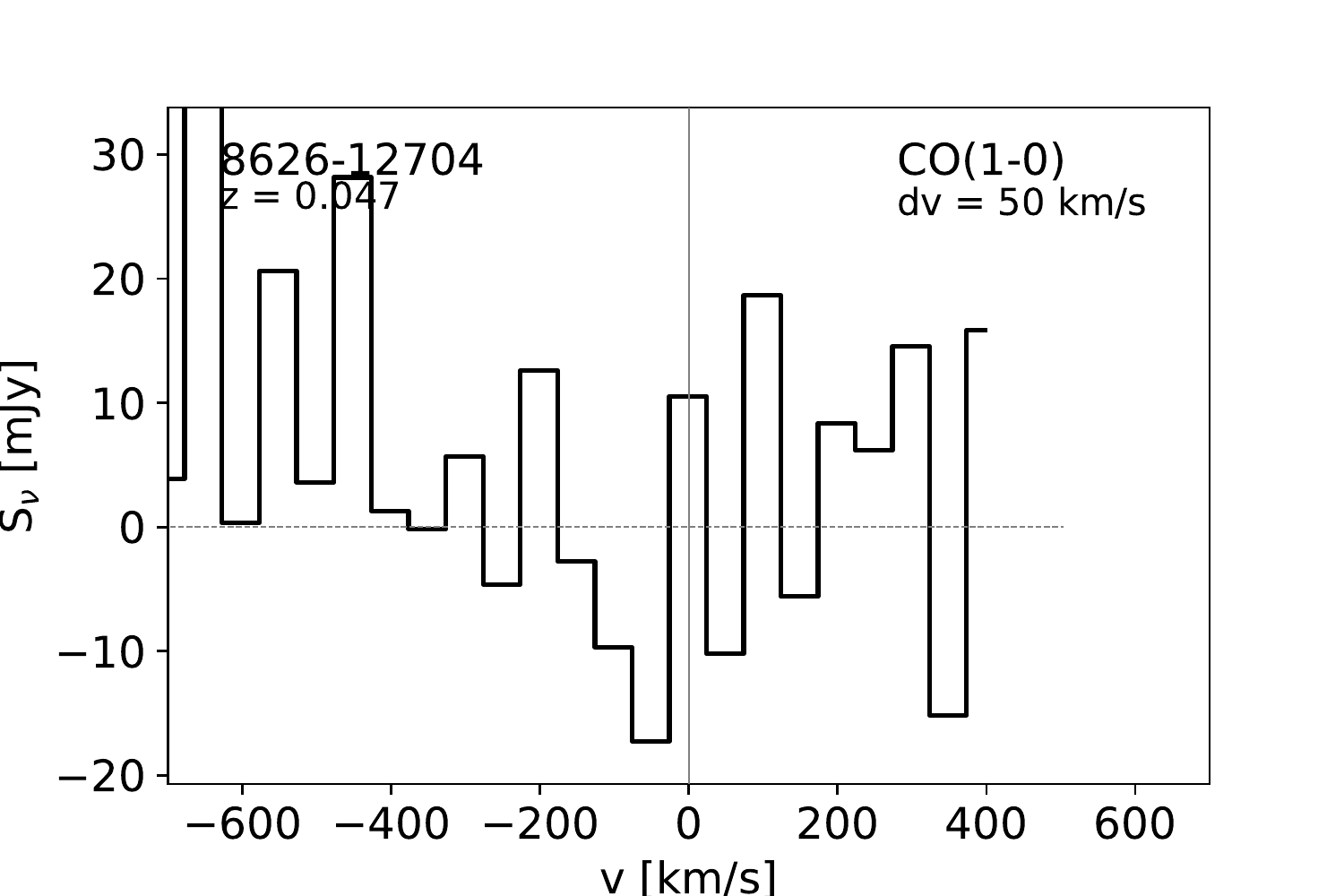} 

\end{figure*}

\begin{figure*} 
 \centering 
 \includegraphics[width = 0.17\textwidth, trim = 0cm 0cm 0cm 0cm, clip = true]{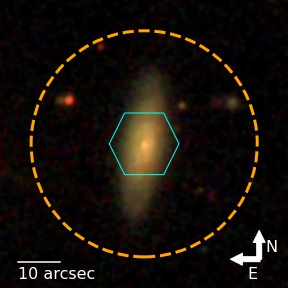}
 \includegraphics[width = 0.29\textwidth, trim = 0cm 0cm 0cm 0cm, clip = true]{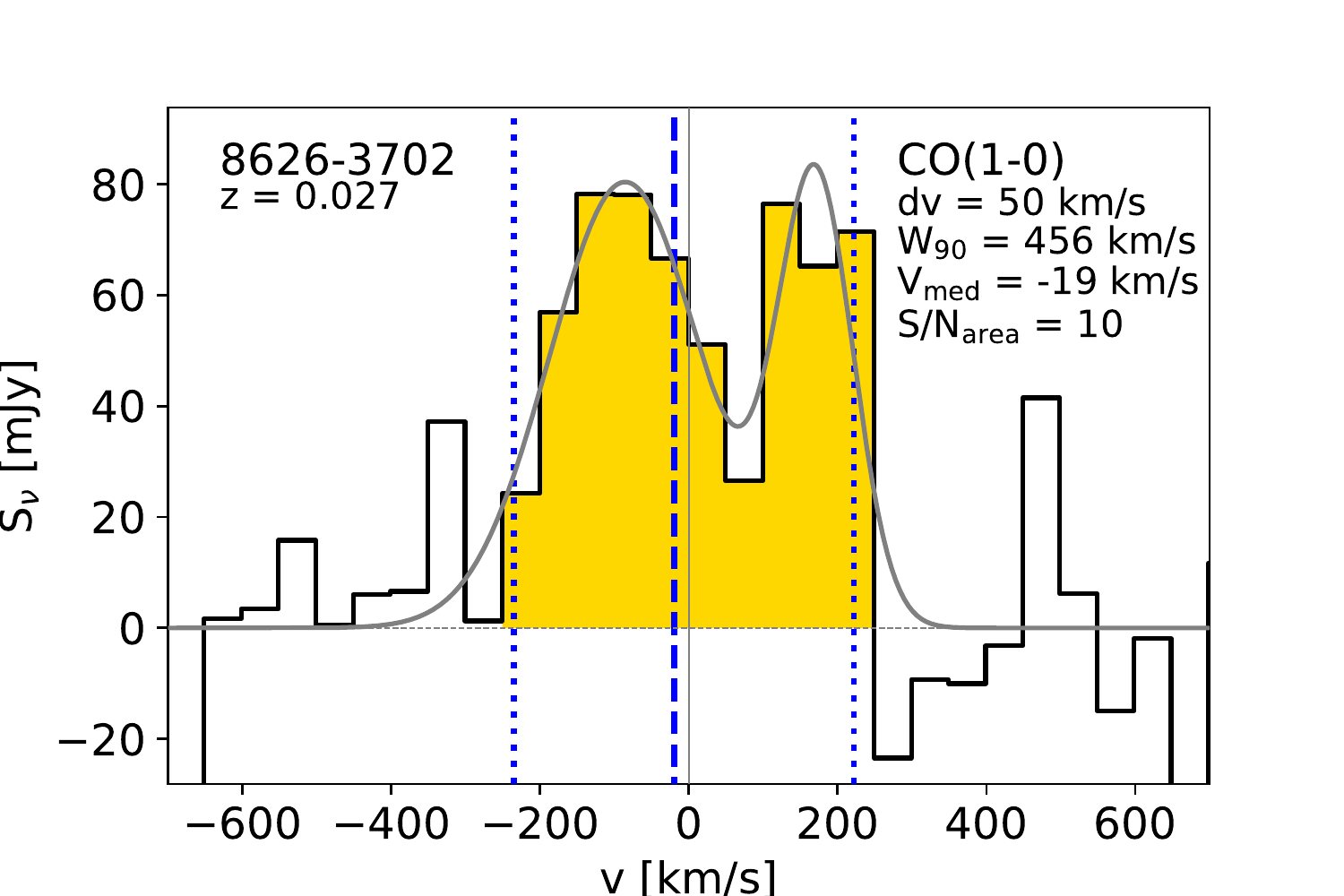} 
 \hspace{0.4cm}
 \centering 
 \includegraphics[width = 0.17\textwidth, trim = 0cm 0cm 0cm 0cm, clip = true]{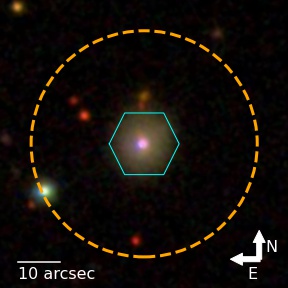}
 \includegraphics[width = 0.29\textwidth, trim = 0cm 0cm 0cm 0cm, clip = true]{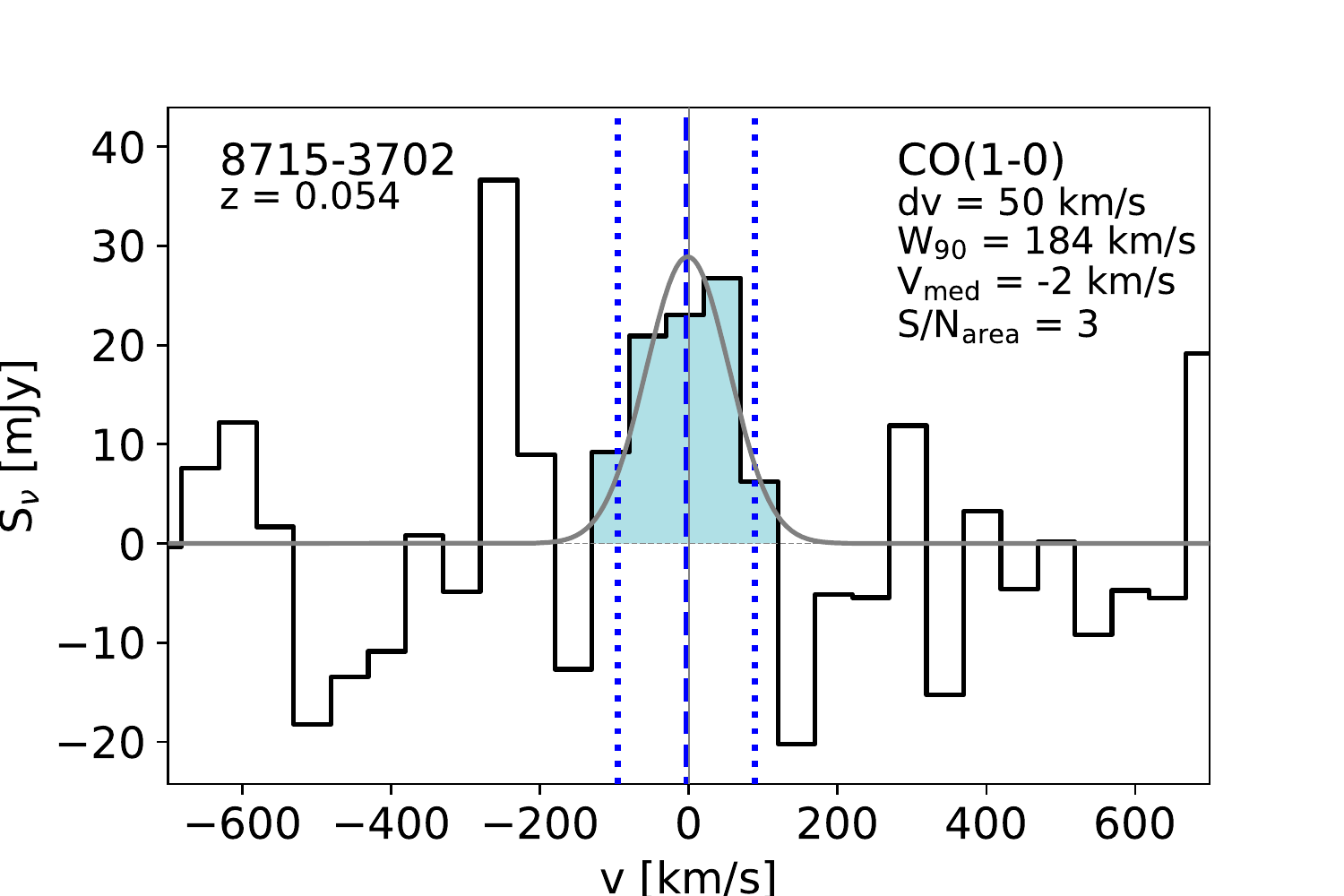} 

\end{figure*}

\begin{figure*} 
 \centering 
 \includegraphics[width = 0.17\textwidth, trim = 0cm 0cm 0cm 0cm, clip = true]{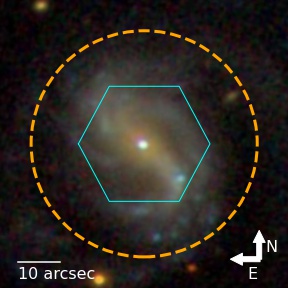}
 \includegraphics[width = 0.29\textwidth, trim = 0cm 0cm 0cm 0cm, clip = true]{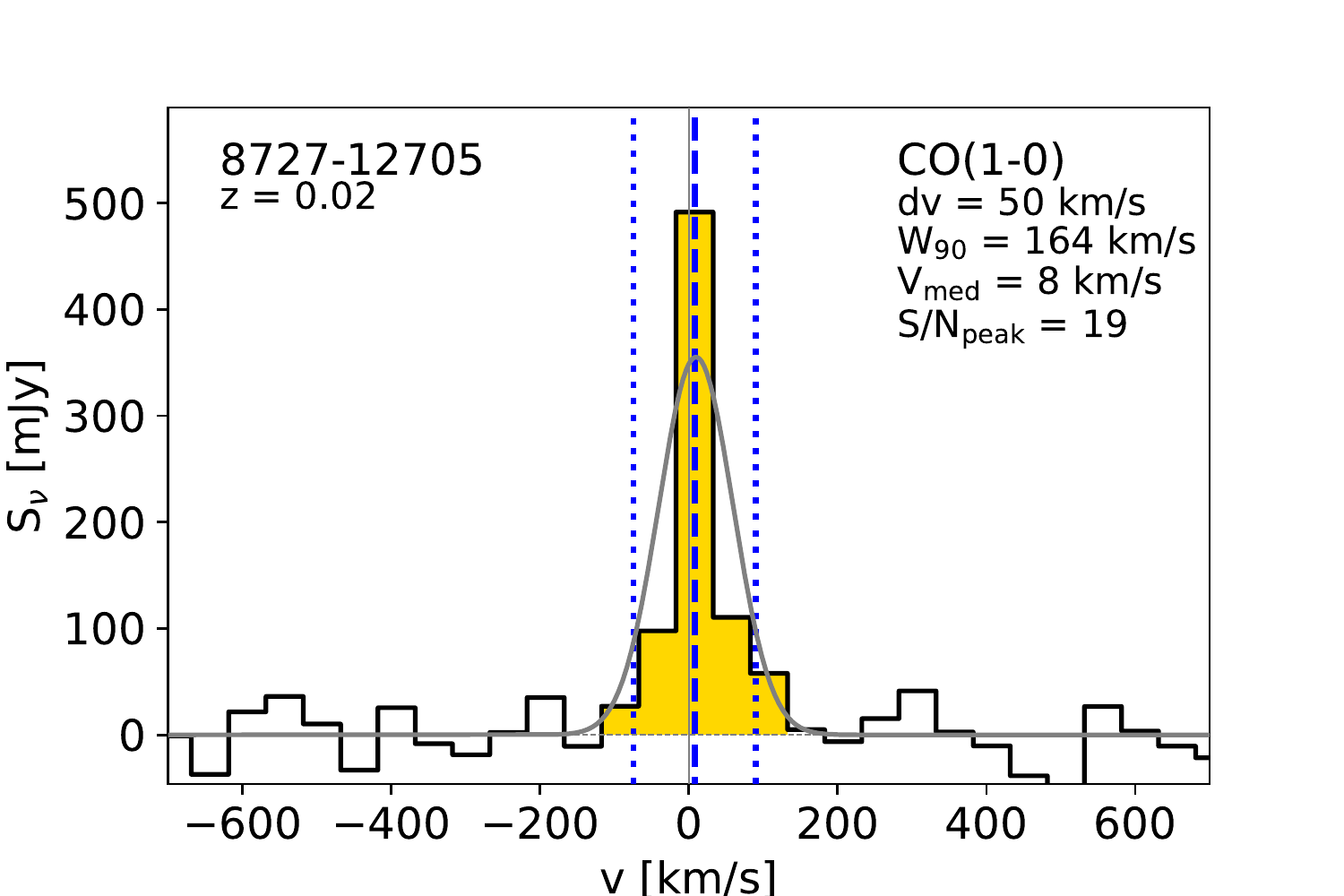} 
 \hspace{0.4cm}
 \centering 
 \includegraphics[width = 0.17\textwidth, trim = 0cm 0cm 0cm 0cm, clip = true]{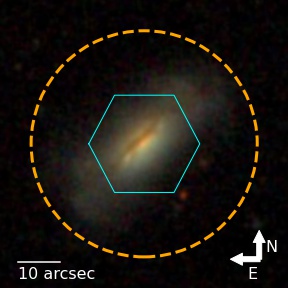}
 \includegraphics[width = 0.29\textwidth, trim = 0cm 0cm 0cm 0cm, clip = true]{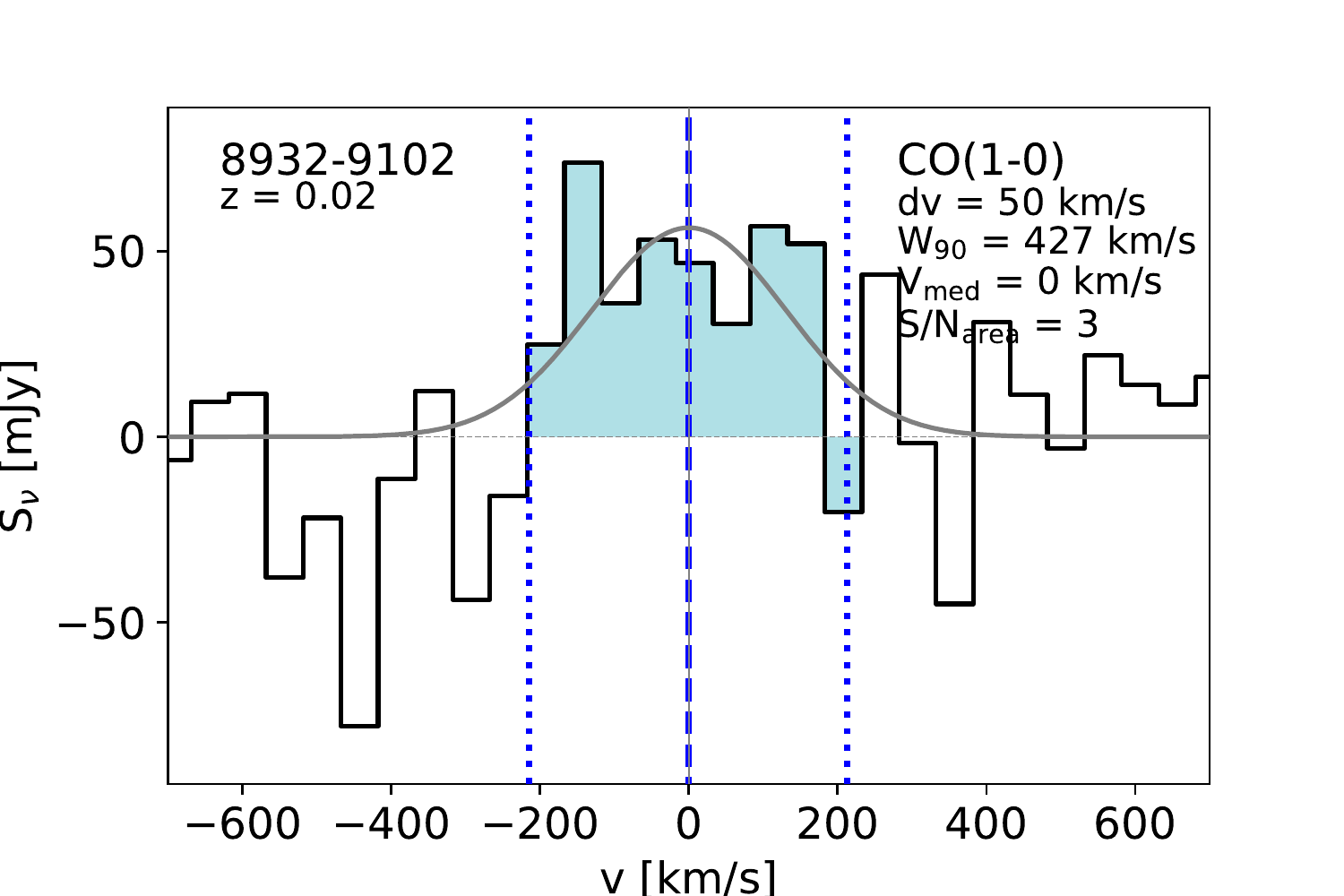} 

\end{figure*}

\begin{figure*} 
   \ContinuedFloat 
 \centering 
 \includegraphics[width = 0.17\textwidth, trim = 0cm 0cm 0cm 0cm, clip = true]{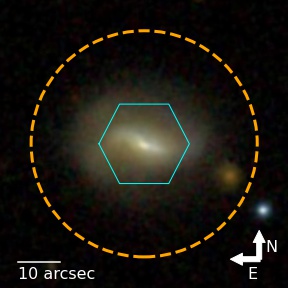}
 \includegraphics[width = 0.29\textwidth, trim = 0cm 0cm 0cm 0cm, clip = true]{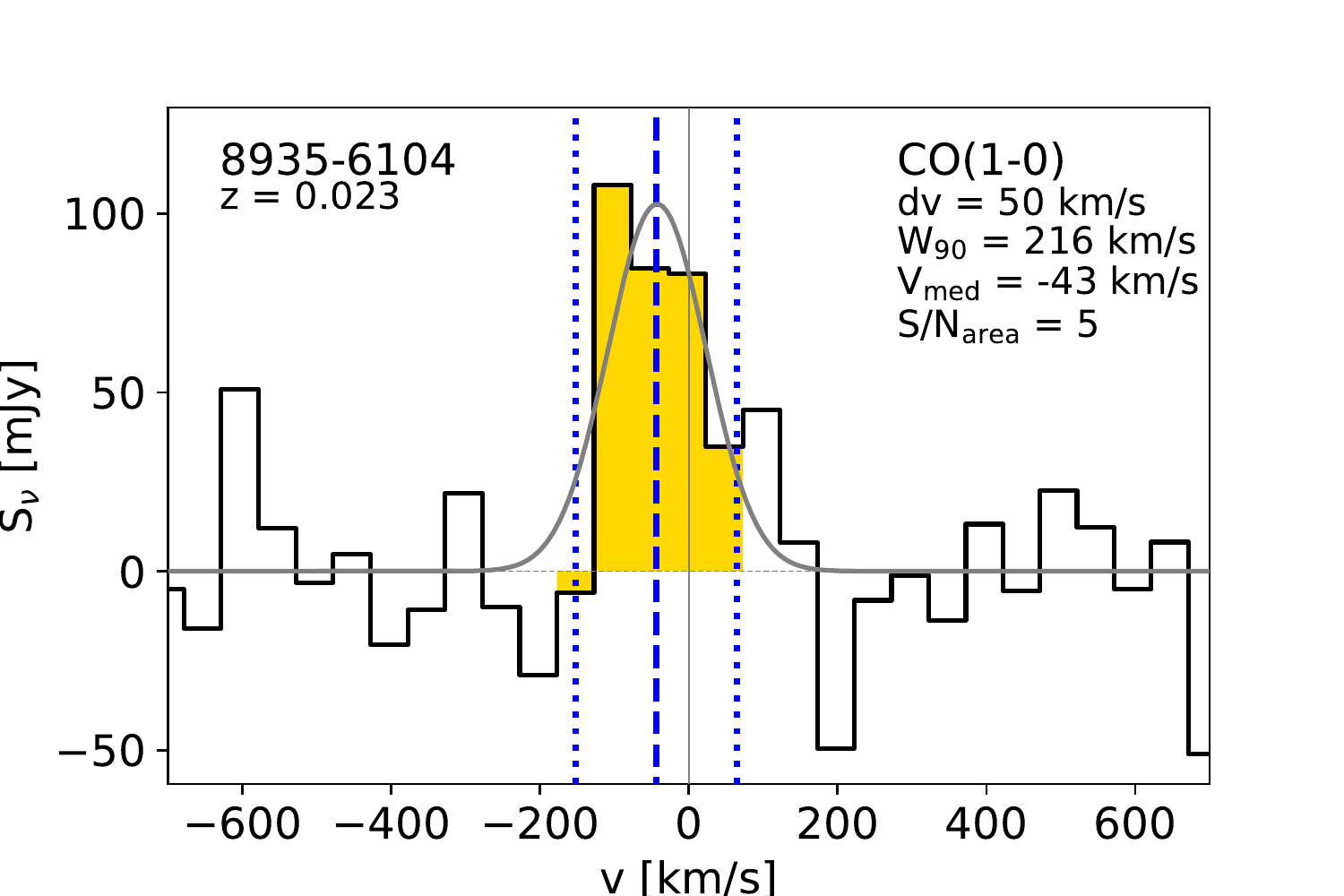} 
 \hspace{0.4cm}
 \centering 
 \includegraphics[width = 0.17\textwidth, trim = 0cm 0cm 0cm 0cm, clip = true]{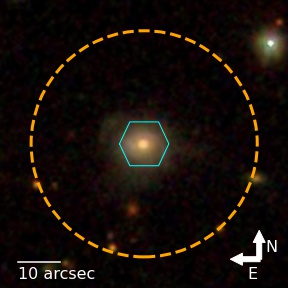}
 \includegraphics[width = 0.29\textwidth, trim = 0cm 0cm 0cm 0cm, clip = true]{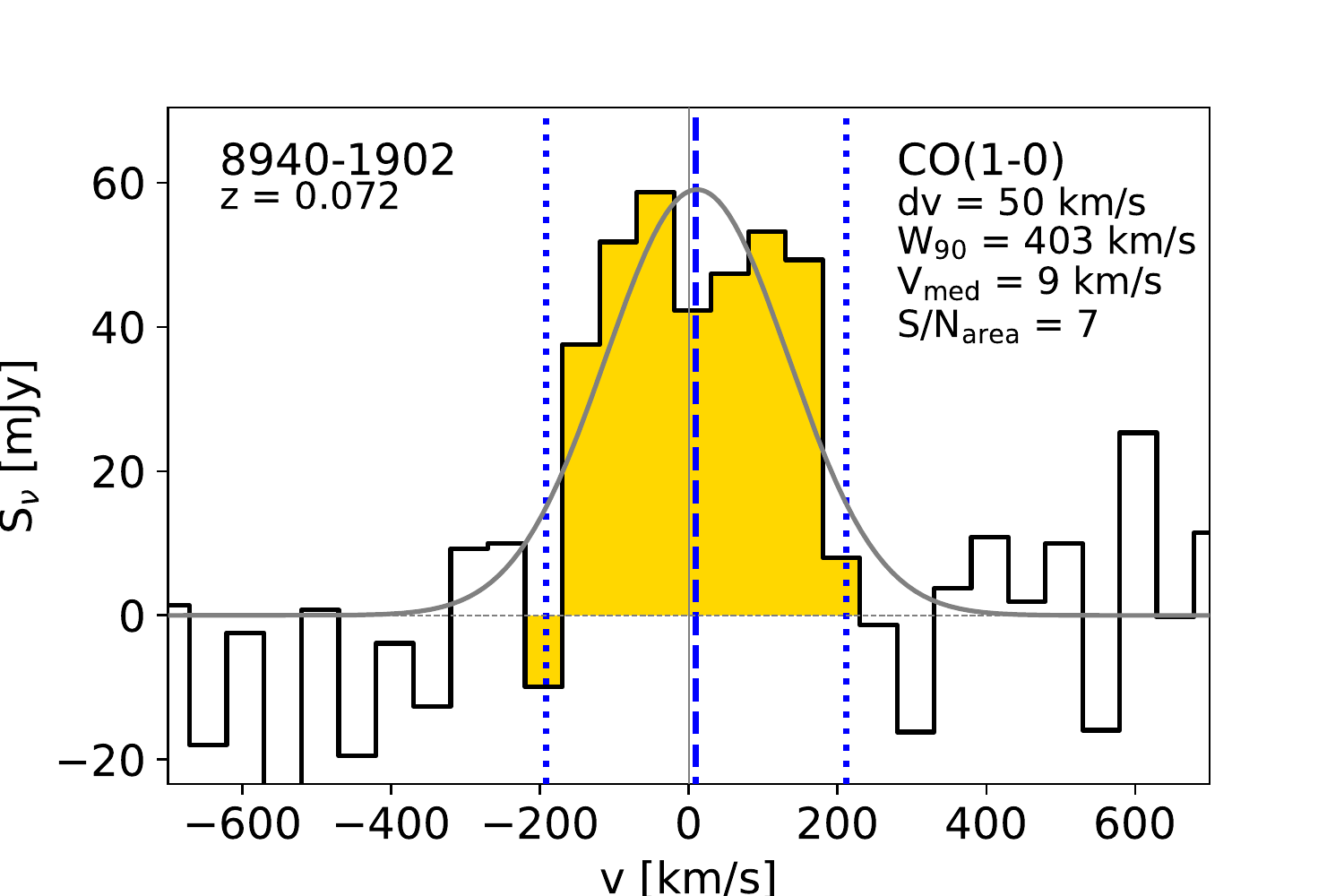} 
 \caption{continued.}
\end{figure*}

\begin{figure*} 
 \centering 
 \includegraphics[width = 0.17\textwidth, trim = 0cm 0cm 0cm 0cm, clip = true]{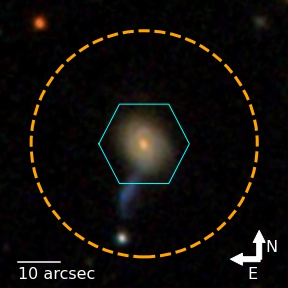}
 \includegraphics[width = 0.29\textwidth, trim = 0cm 0cm 0cm 0cm, clip = true]{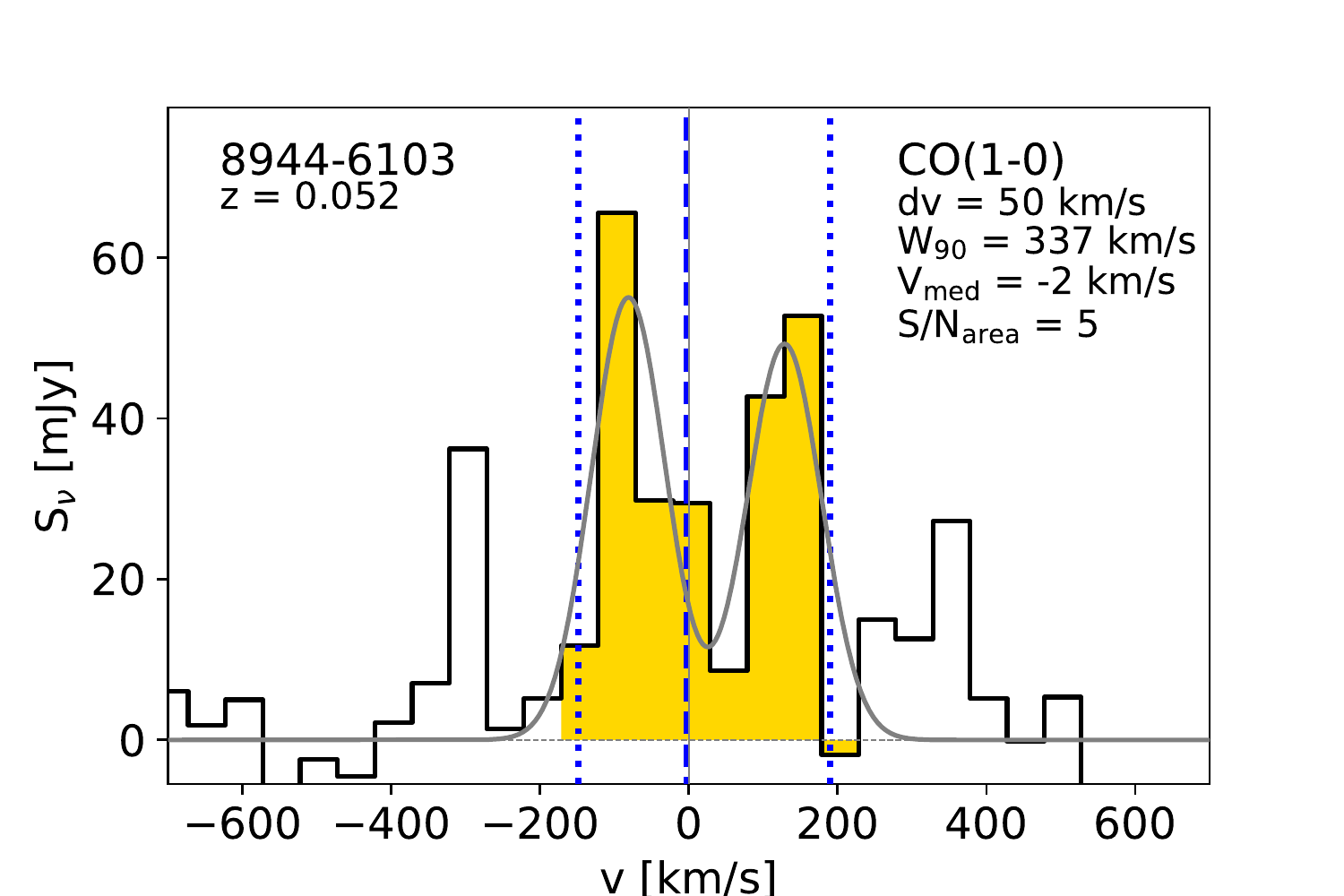} 
 \hspace{0.4cm}
 \centering 
 \includegraphics[width = 0.17\textwidth, trim = 0cm 0cm 0cm 0cm, clip = true]{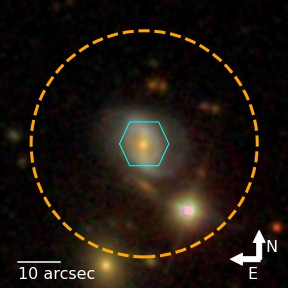}
 \includegraphics[width = 0.29\textwidth, trim = 0cm 0cm 0cm 0cm, clip = true]{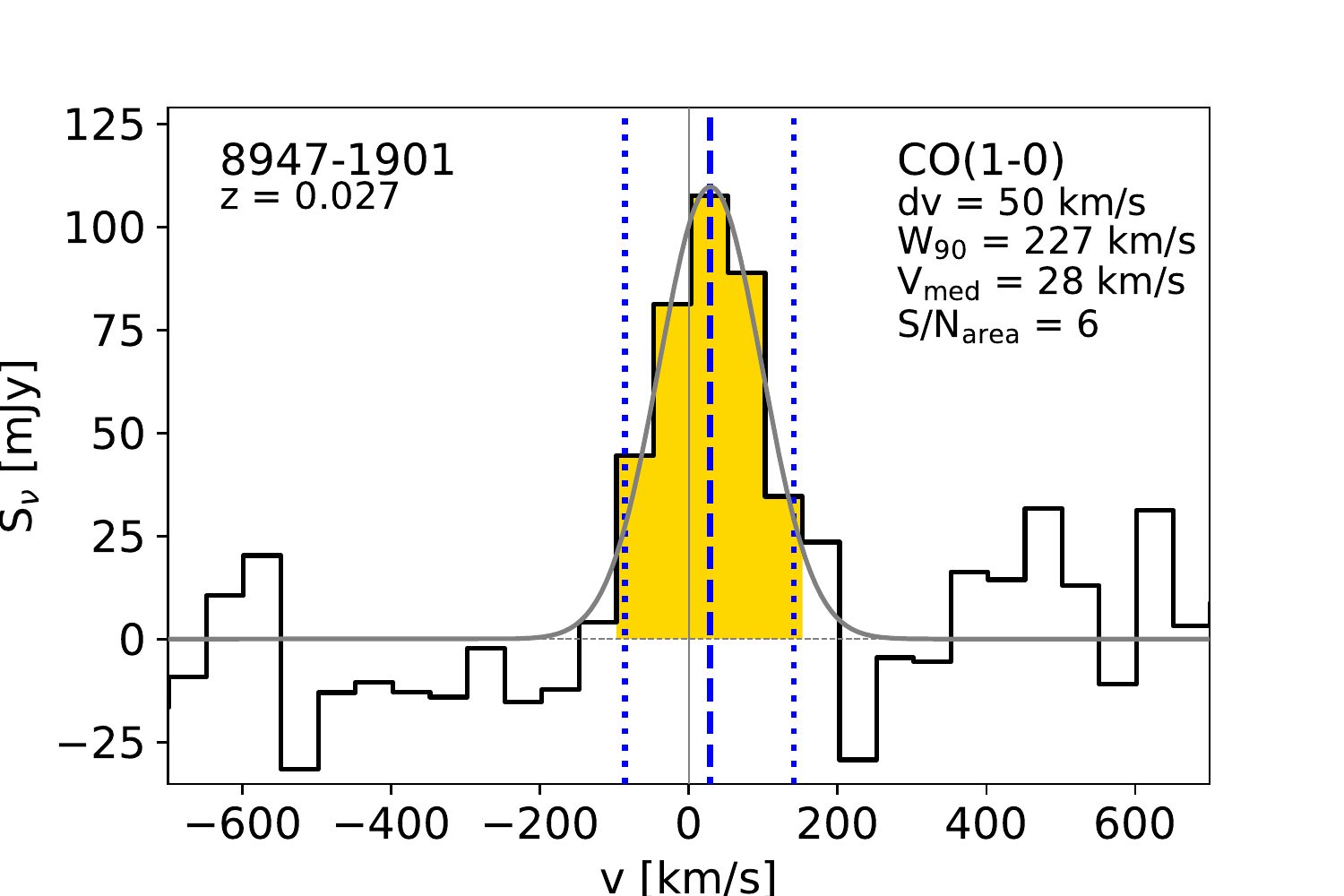} 

\end{figure*}

\begin{figure*} 
 \centering 
 \includegraphics[width = 0.17\textwidth, trim = 0cm 0cm 0cm 0cm, clip = true]{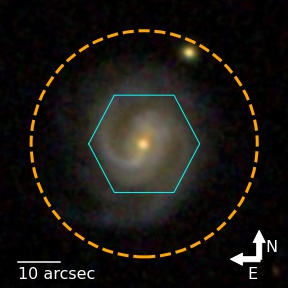}
 \includegraphics[width = 0.29\textwidth, trim = 0cm 0cm 0cm 0cm, clip = true]{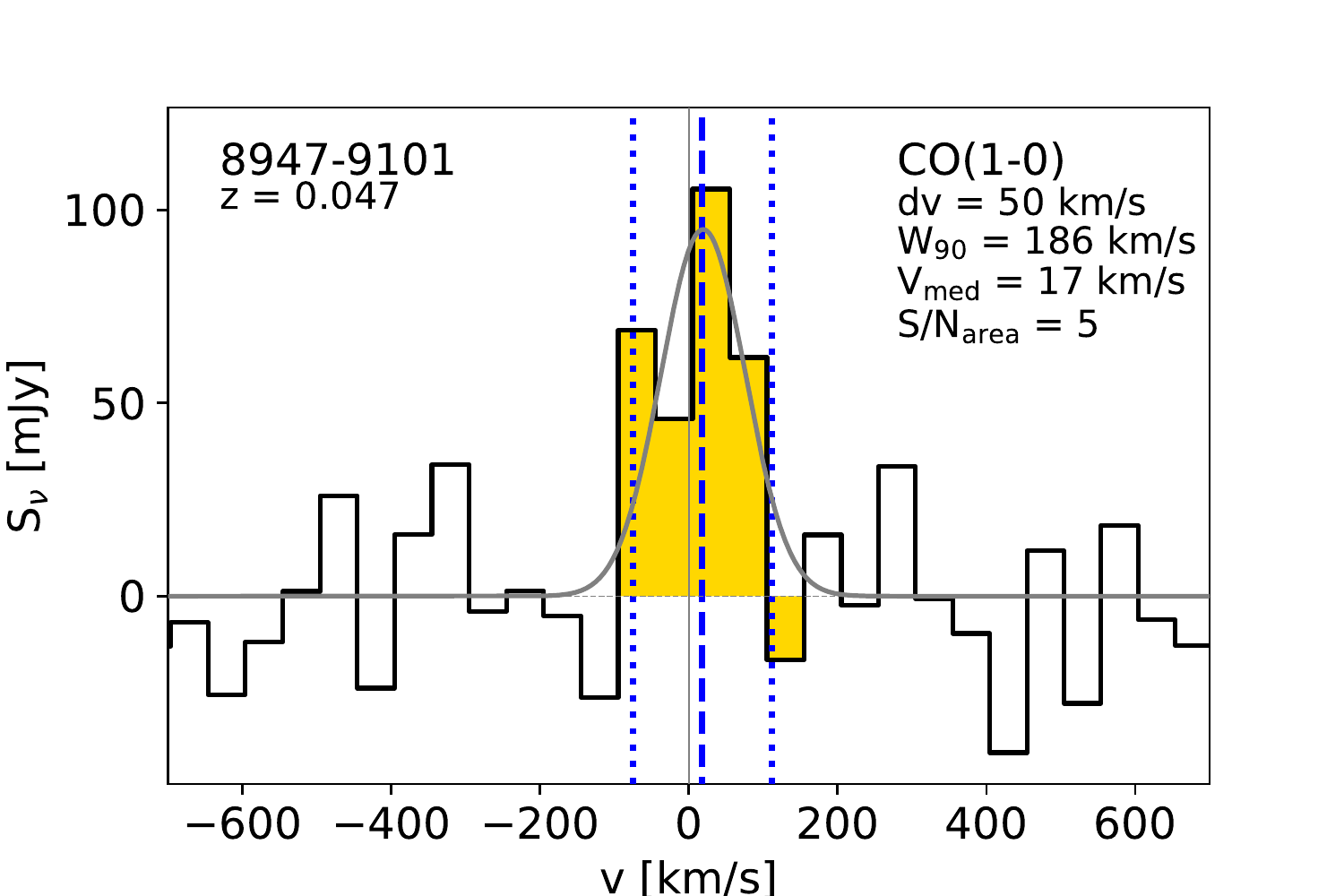} 
 \hspace{0.4cm}
 \centering 
 \includegraphics[width = 0.17\textwidth, trim = 0cm 0cm 0cm 0cm, clip = true]{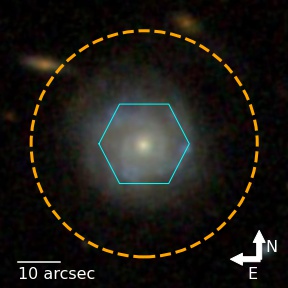}
 \includegraphics[width = 0.29\textwidth, trim = 0cm 0cm 0cm 0cm, clip = true]{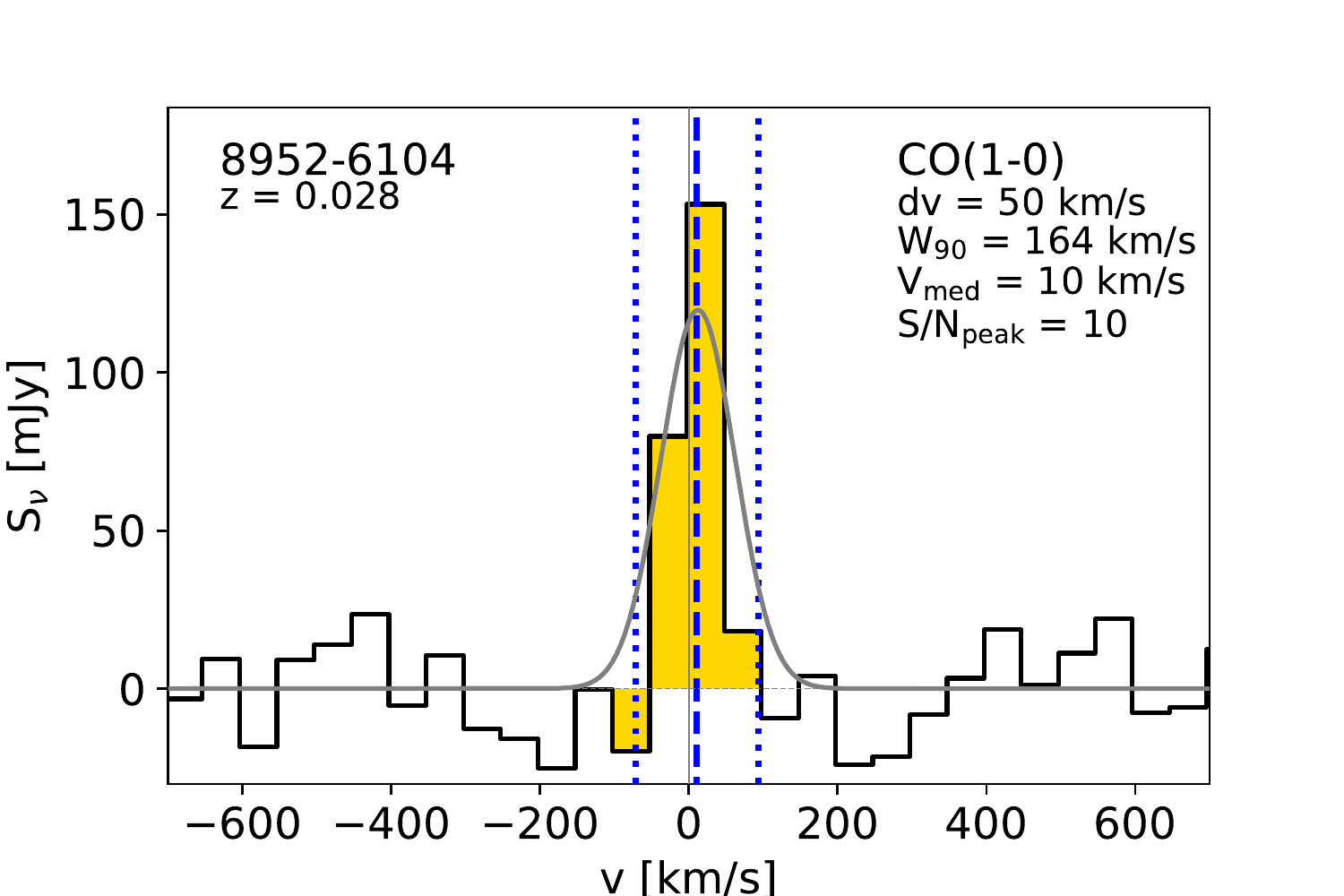} 

\end{figure*}

\begin{figure*} 
 \centering 
 \includegraphics[width = 0.17\textwidth, trim = 0cm 0cm 0cm 0cm, clip = true]{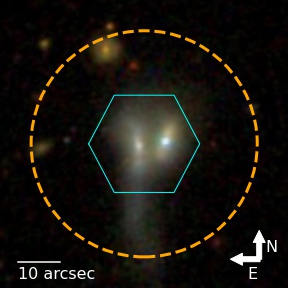}
 \includegraphics[width = 0.29\textwidth, trim = 0cm 0cm 0cm 0cm, clip = true]{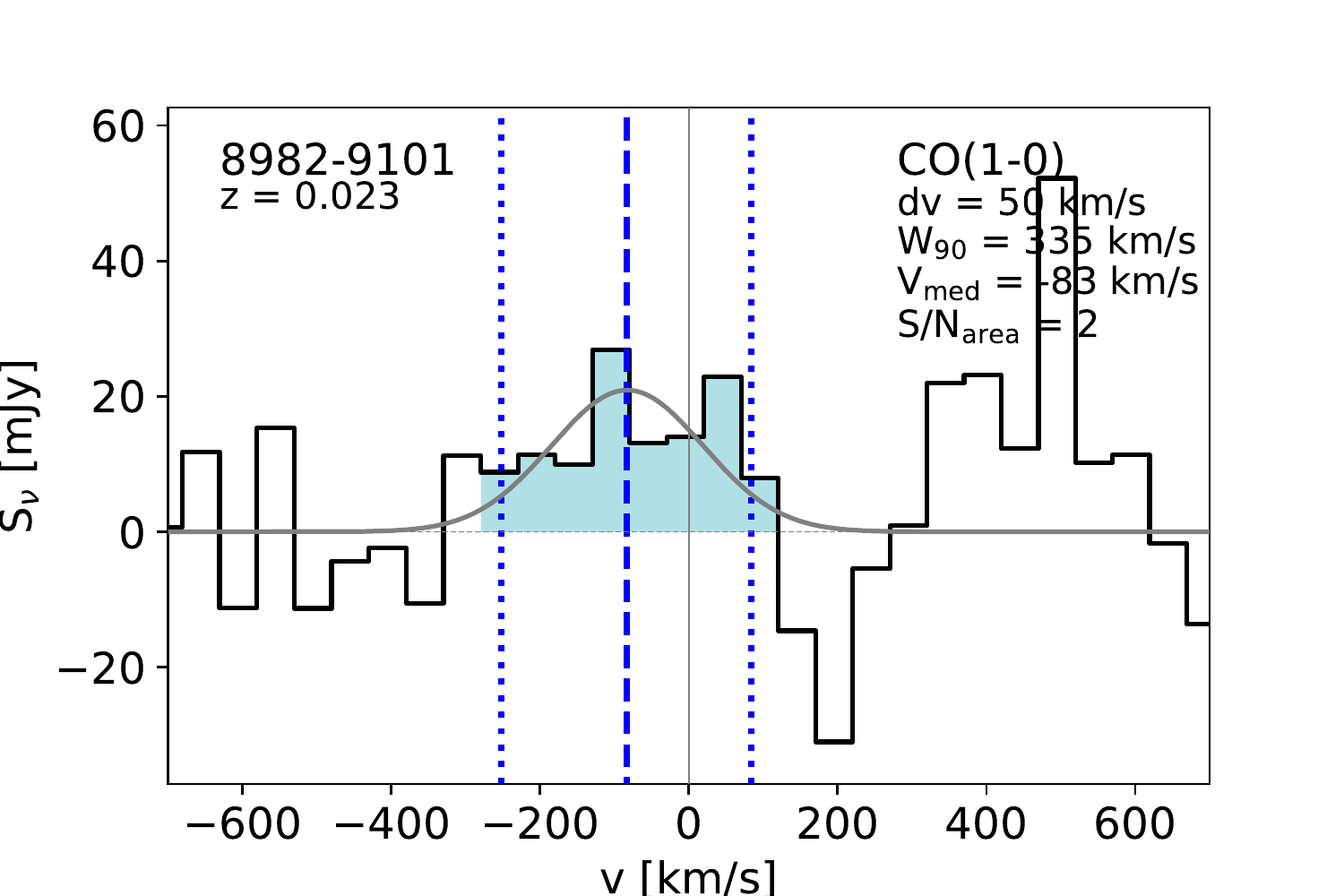} 
 \hspace{0.4cm}
 \centering 
 \includegraphics[width = 0.17\textwidth, trim = 0cm 0cm 0cm 0cm, clip = true]{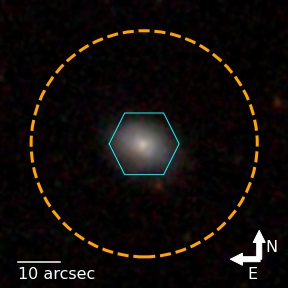}
 \includegraphics[width = 0.29\textwidth, trim = 0cm 0cm 0cm 0cm, clip = true]{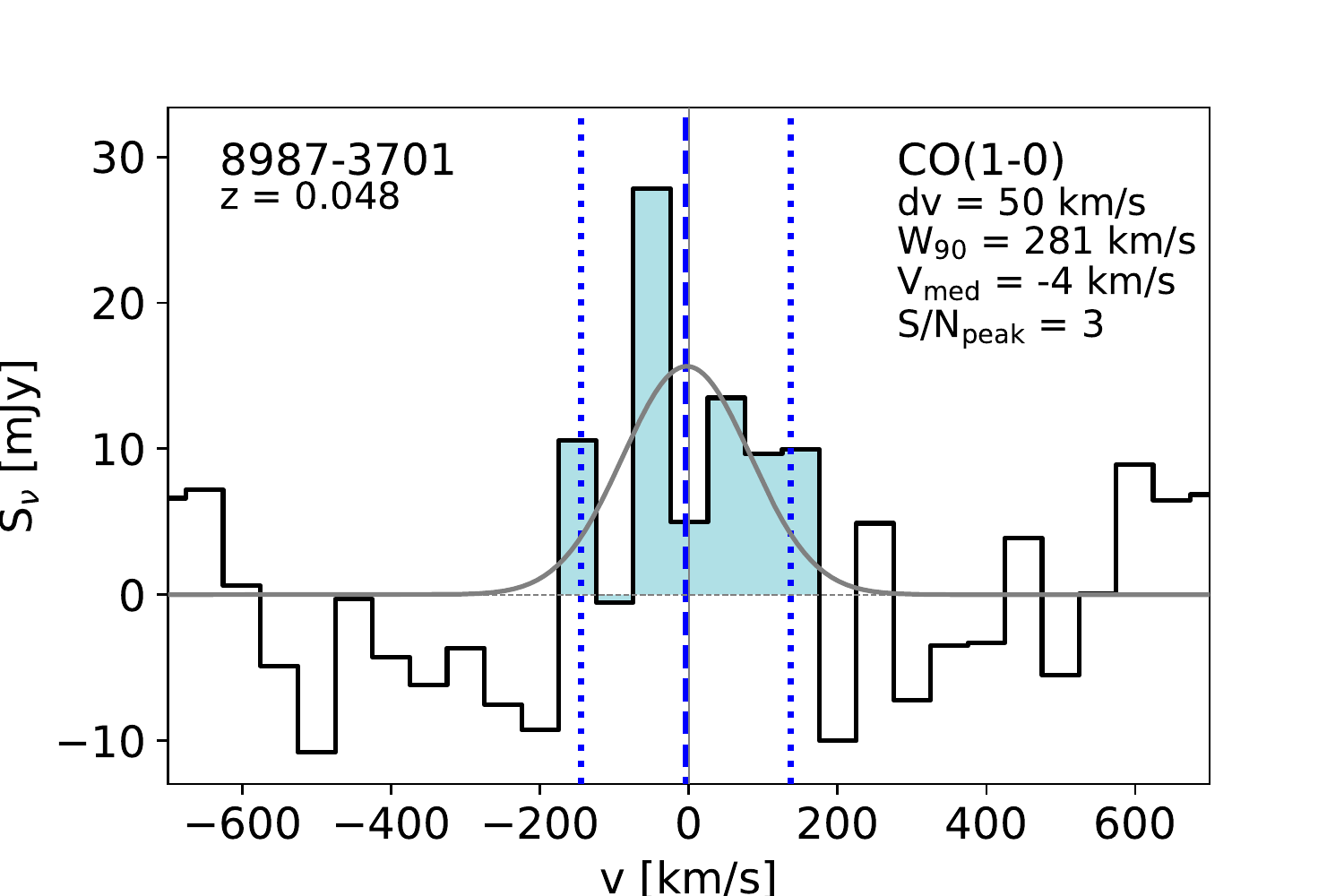} 

\end{figure*}

\begin{figure*} 
 \centering 
 \includegraphics[width = 0.17\textwidth, trim = 0cm 0cm 0cm 0cm, clip = true]{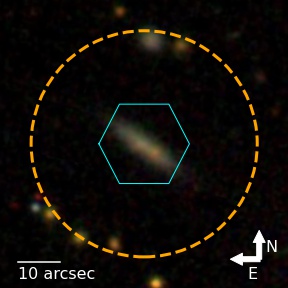}
 \includegraphics[width = 0.29\textwidth, trim = 0cm 0cm 0cm 0cm, clip = true]{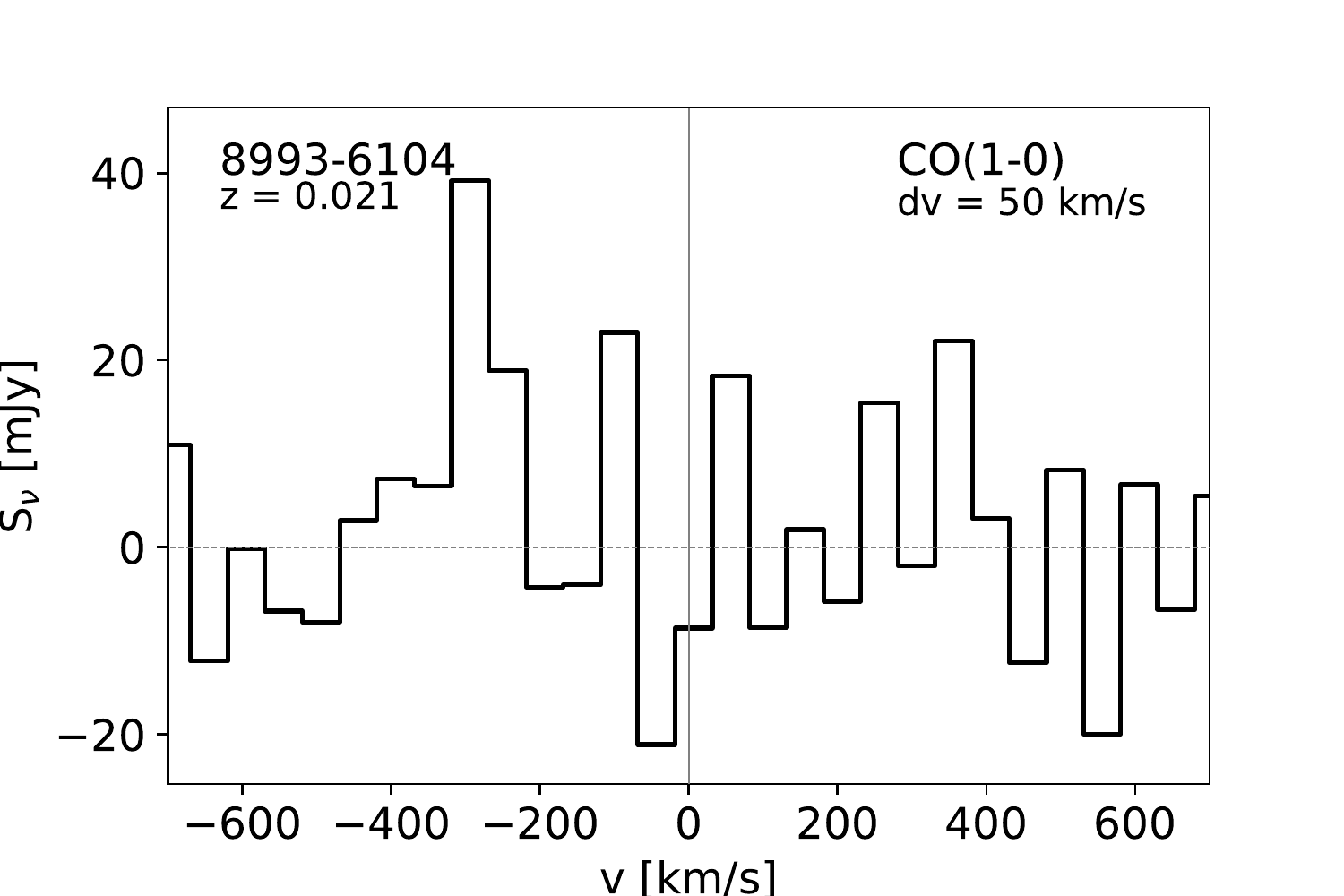} 
 \hspace{0.4cm}
 \centering 
 \includegraphics[width = 0.17\textwidth, trim = 0cm 0cm 0cm 0cm, clip = true]{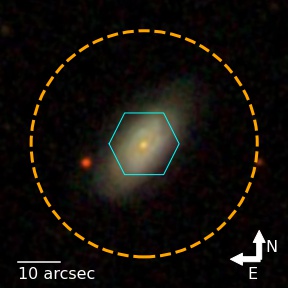}
 \includegraphics[width = 0.29\textwidth, trim = 0cm 0cm 0cm 0cm, clip = true]{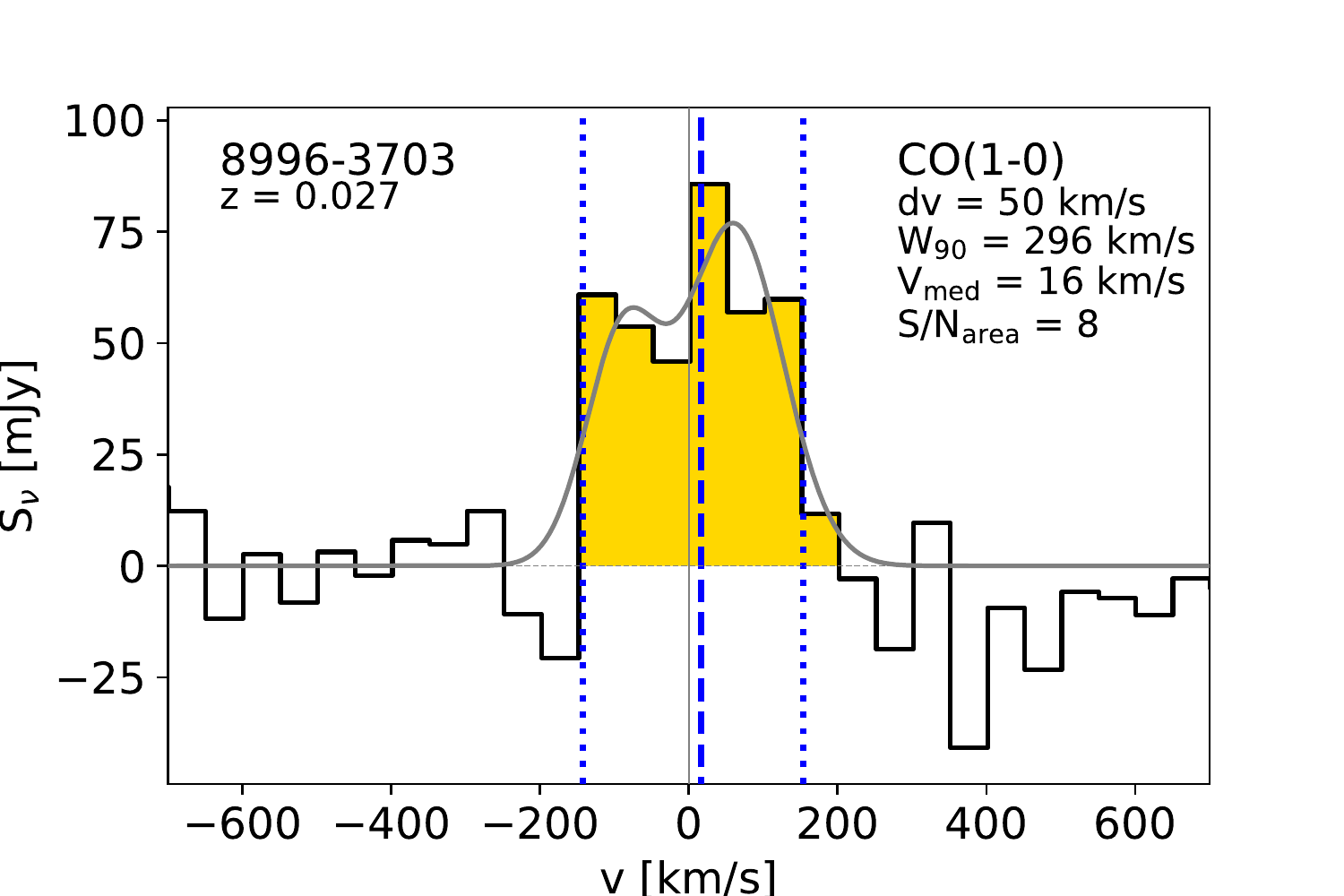} 

\end{figure*}

\begin{figure*} 
 \centering 
 \includegraphics[width = 0.17\textwidth, trim = 0cm 0cm 0cm 0cm, clip = true]{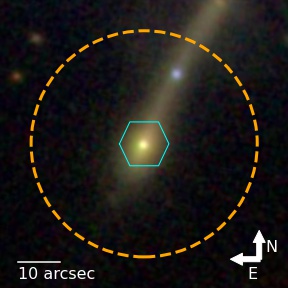}
 \includegraphics[width = 0.29\textwidth, trim = 0cm 0cm 0cm 0cm, clip = true]{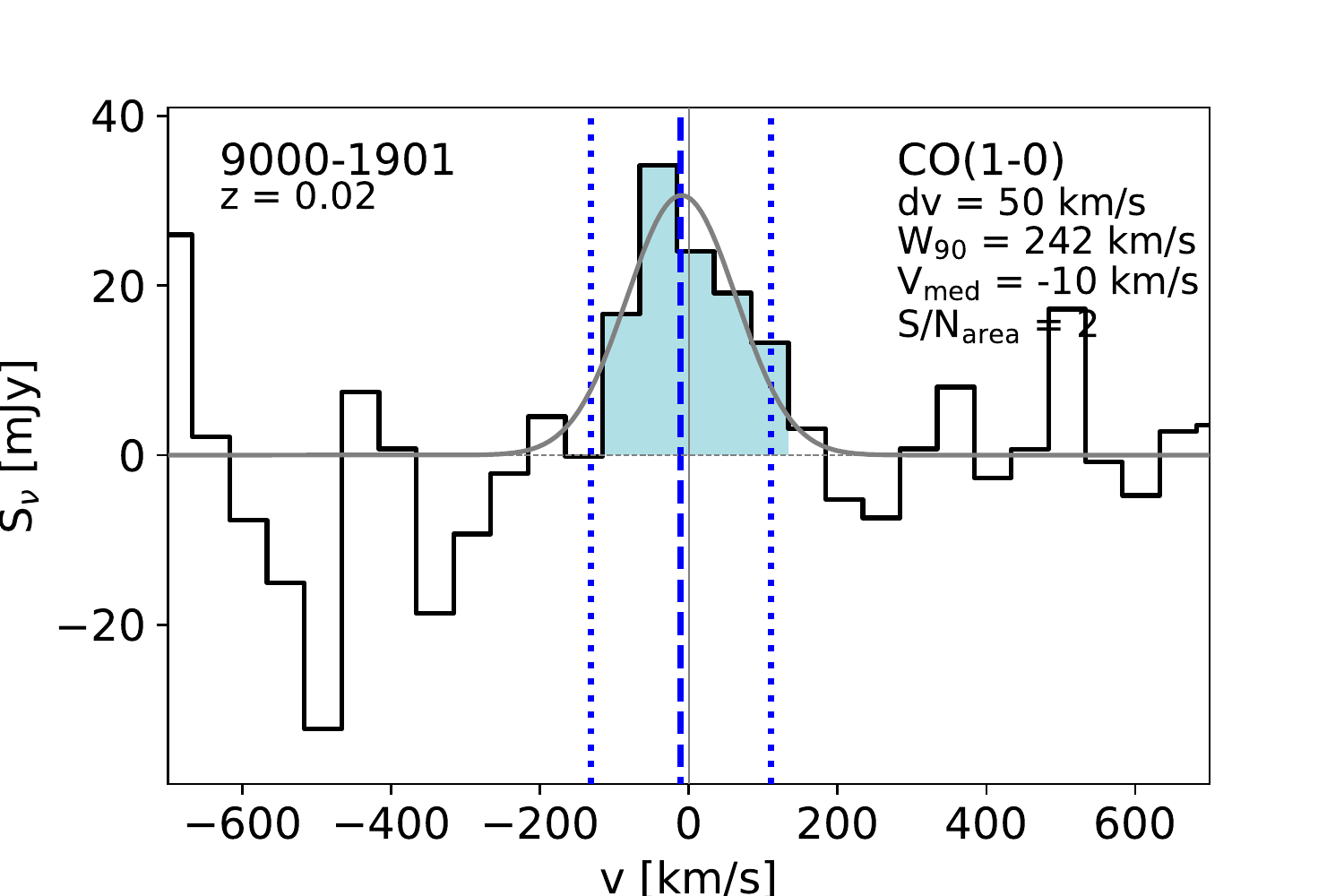} 
 \hspace{0.4cm}
 \centering 
 \includegraphics[width = 0.17\textwidth, trim = 0cm 0cm 0cm 0cm, clip = true]{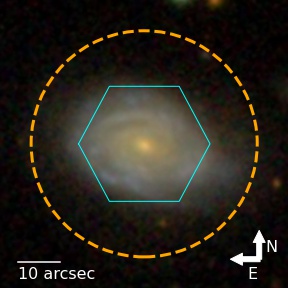}
 \includegraphics[width = 0.29\textwidth, trim = 0cm 0cm 0cm 0cm, clip = true]{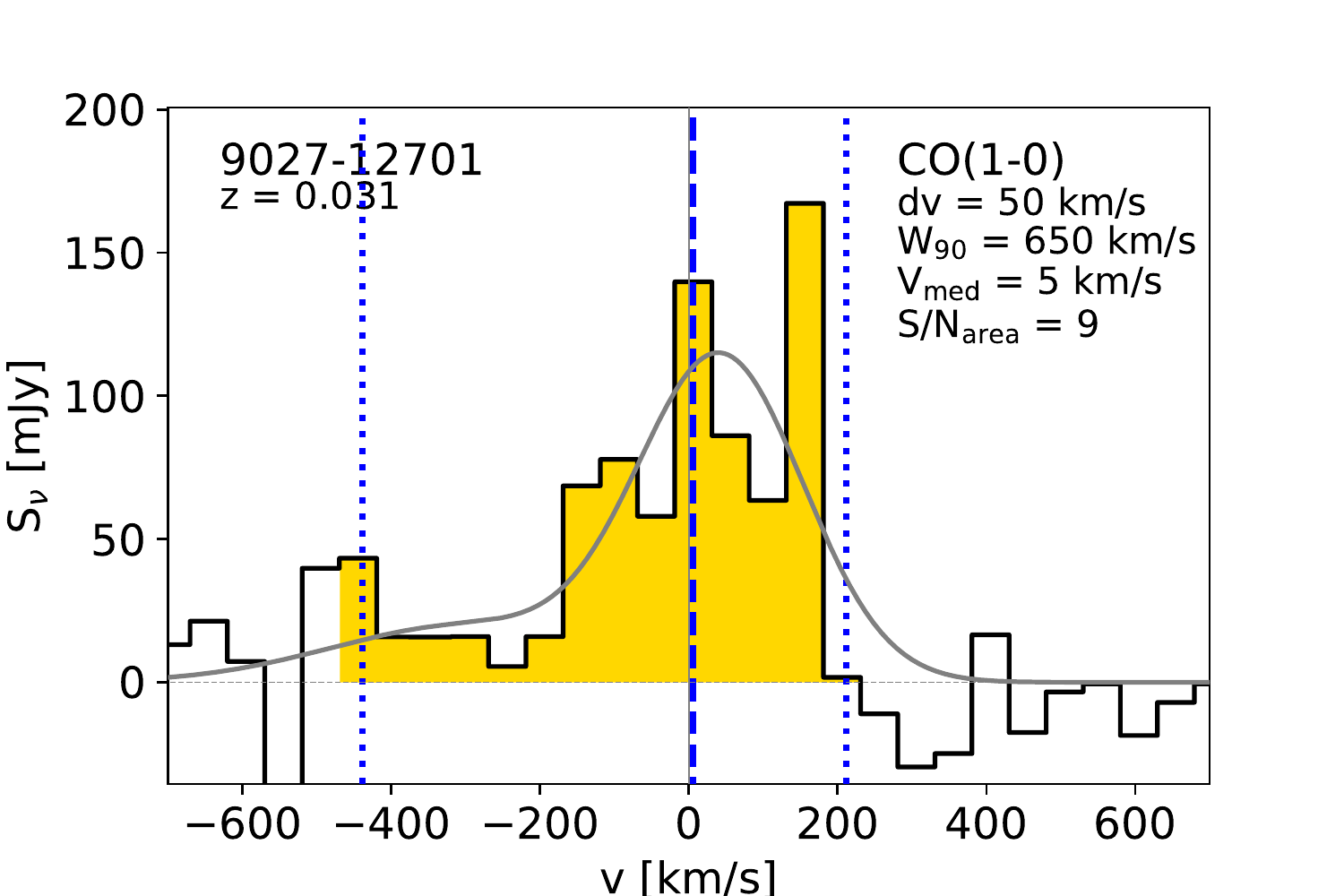} 

\end{figure*}

\begin{figure*} 
   \ContinuedFloat 
 \centering 
 \includegraphics[width = 0.17\textwidth, trim = 0cm 0cm 0cm 0cm, clip = true]{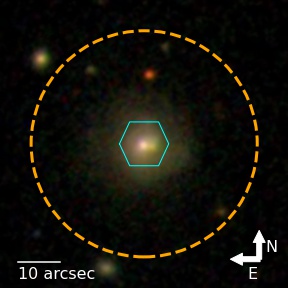}
 \includegraphics[width = 0.29\textwidth, trim = 0cm 0cm 0cm 0cm, clip = true]{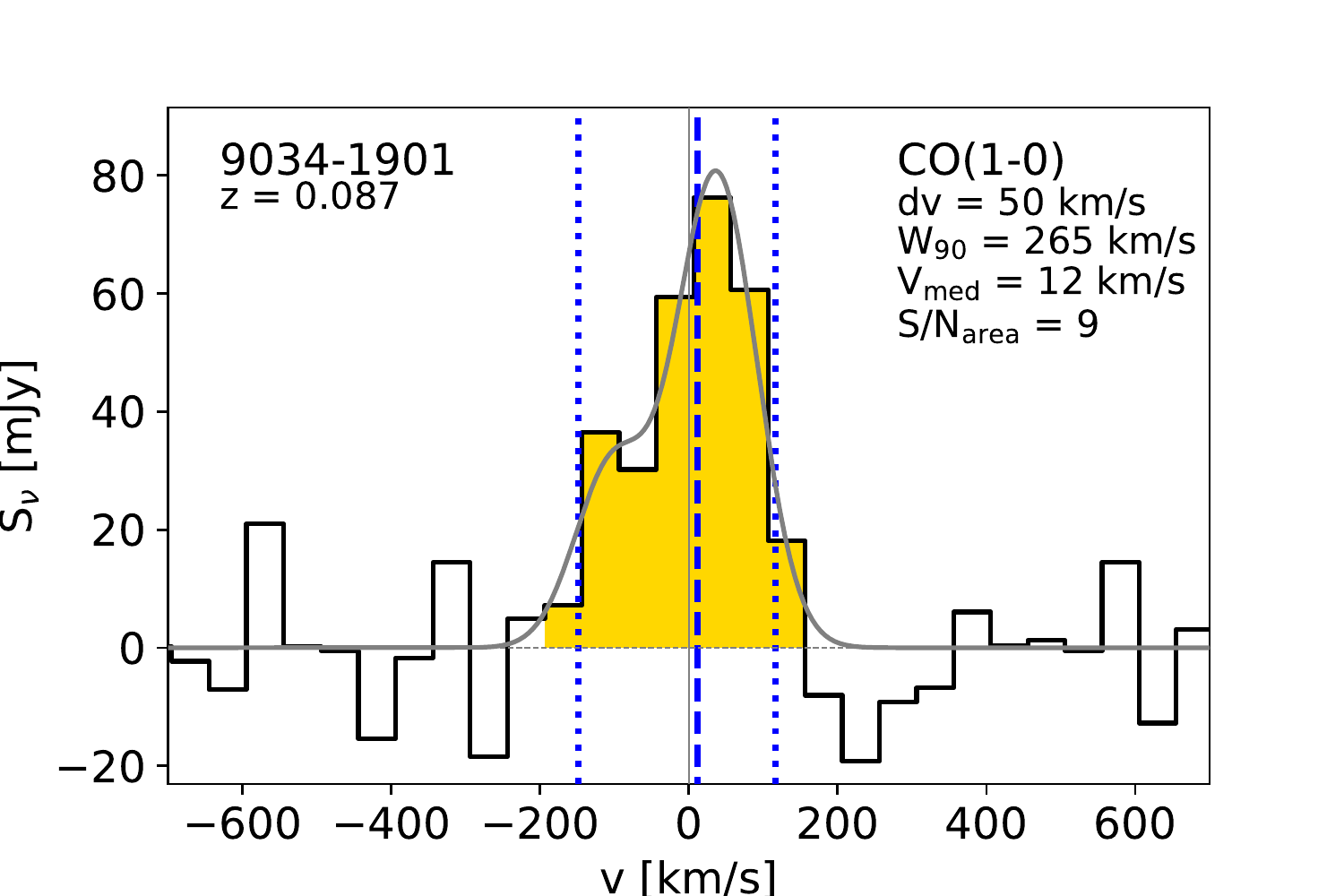} 
 \hspace{0.4cm}
 \centering 
 \includegraphics[width = 0.17\textwidth, trim = 0cm 0cm 0cm 0cm, clip = true]{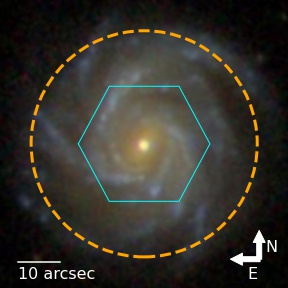}
 \includegraphics[width = 0.29\textwidth, trim = 0cm 0cm 0cm 0cm, clip = true]{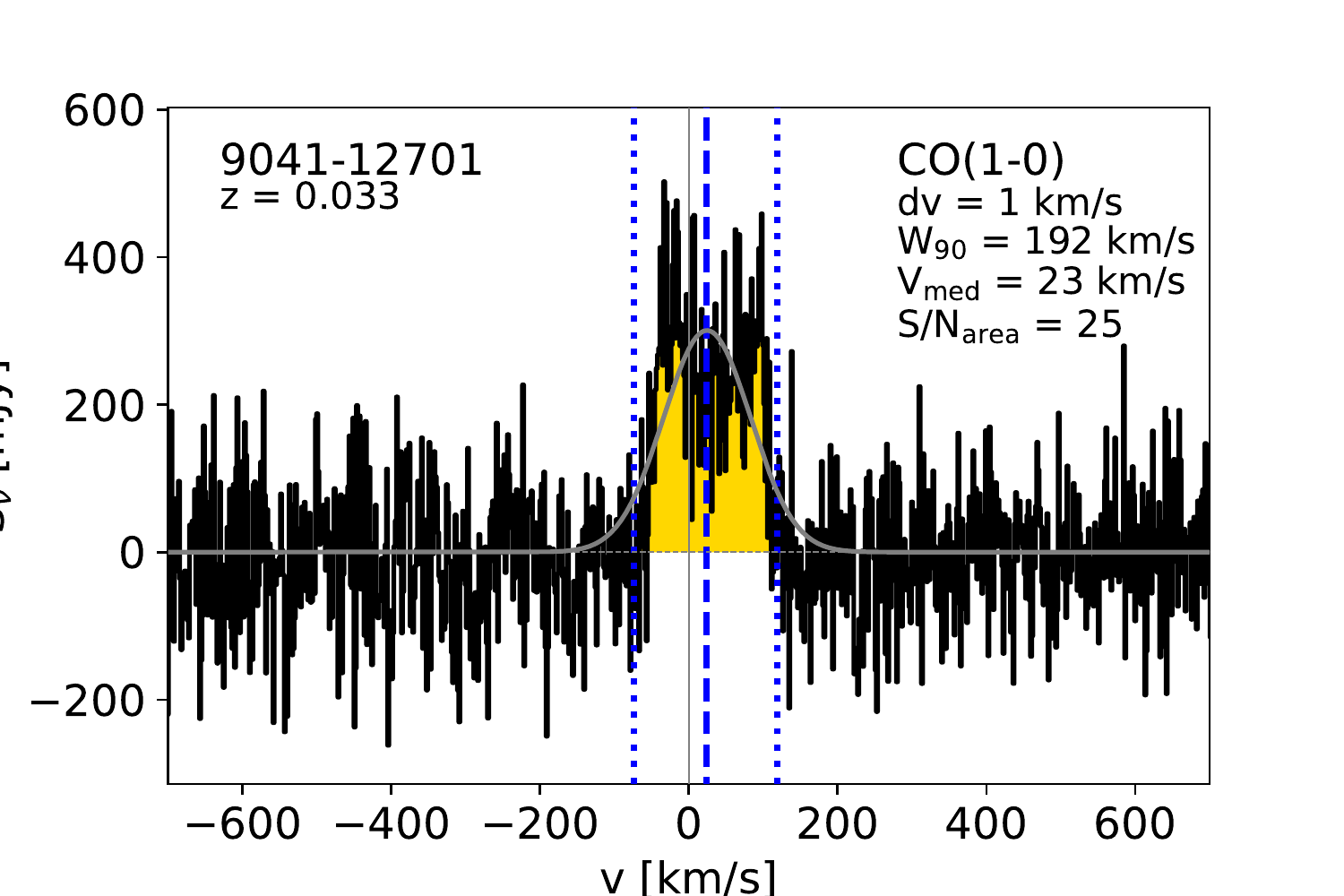} 
\caption{continued.}
\end{figure*}

\begin{figure*} 
 \centering 
 \includegraphics[width = 0.17\textwidth, trim = 0cm 0cm 0cm 0cm, clip = true]{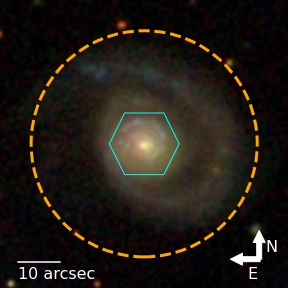}
 \includegraphics[width = 0.29\textwidth, trim = 0cm 0cm 0cm 0cm, clip = true]{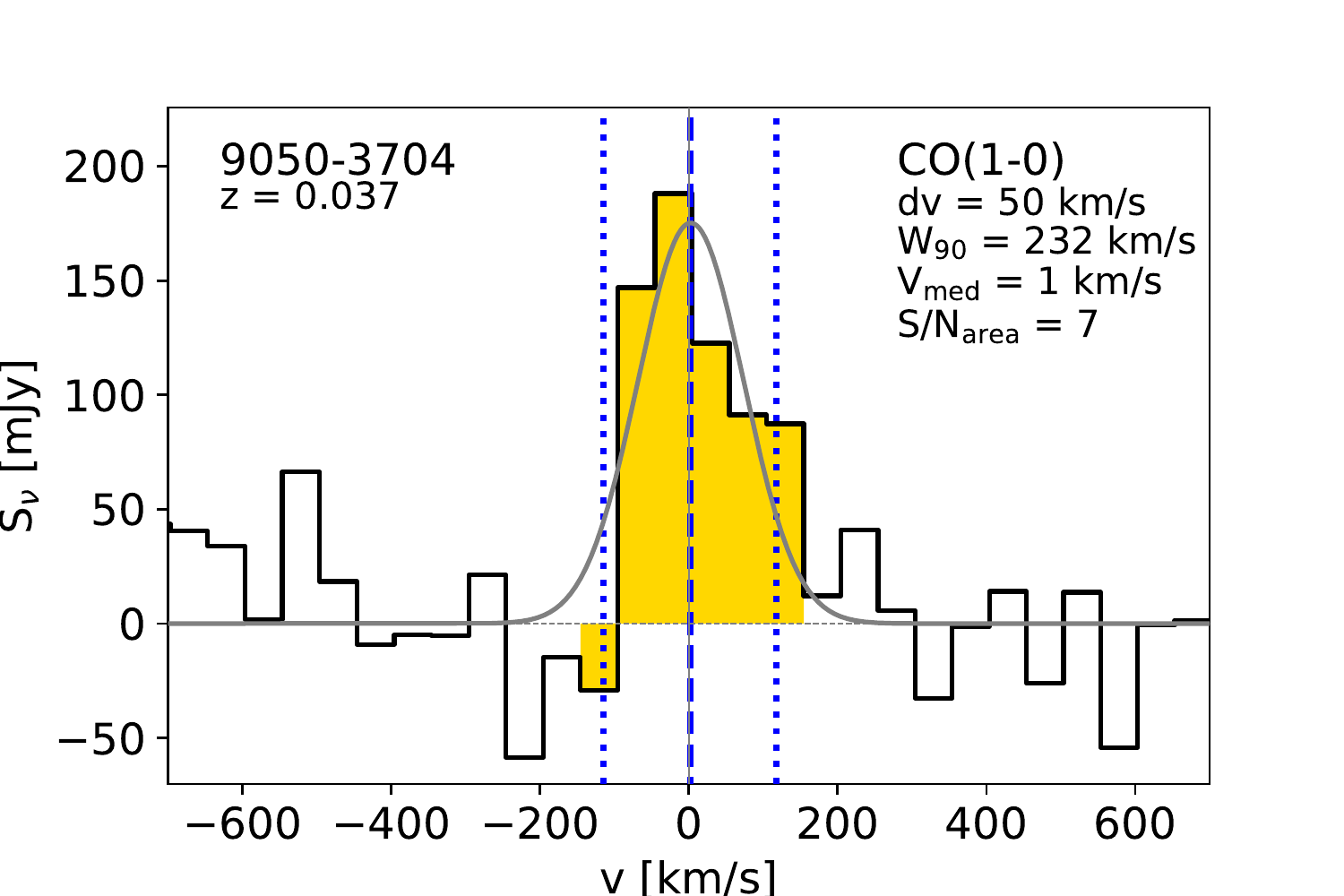} 
 \hspace{0.4cm}
 \centering 
 \includegraphics[width = 0.17\textwidth, trim = 0cm 0cm 0cm 0cm, clip = true]{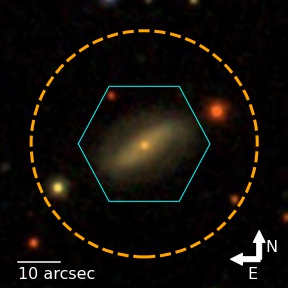}
 \includegraphics[width = 0.29\textwidth, trim = 0cm 0cm 0cm 0cm, clip = true]{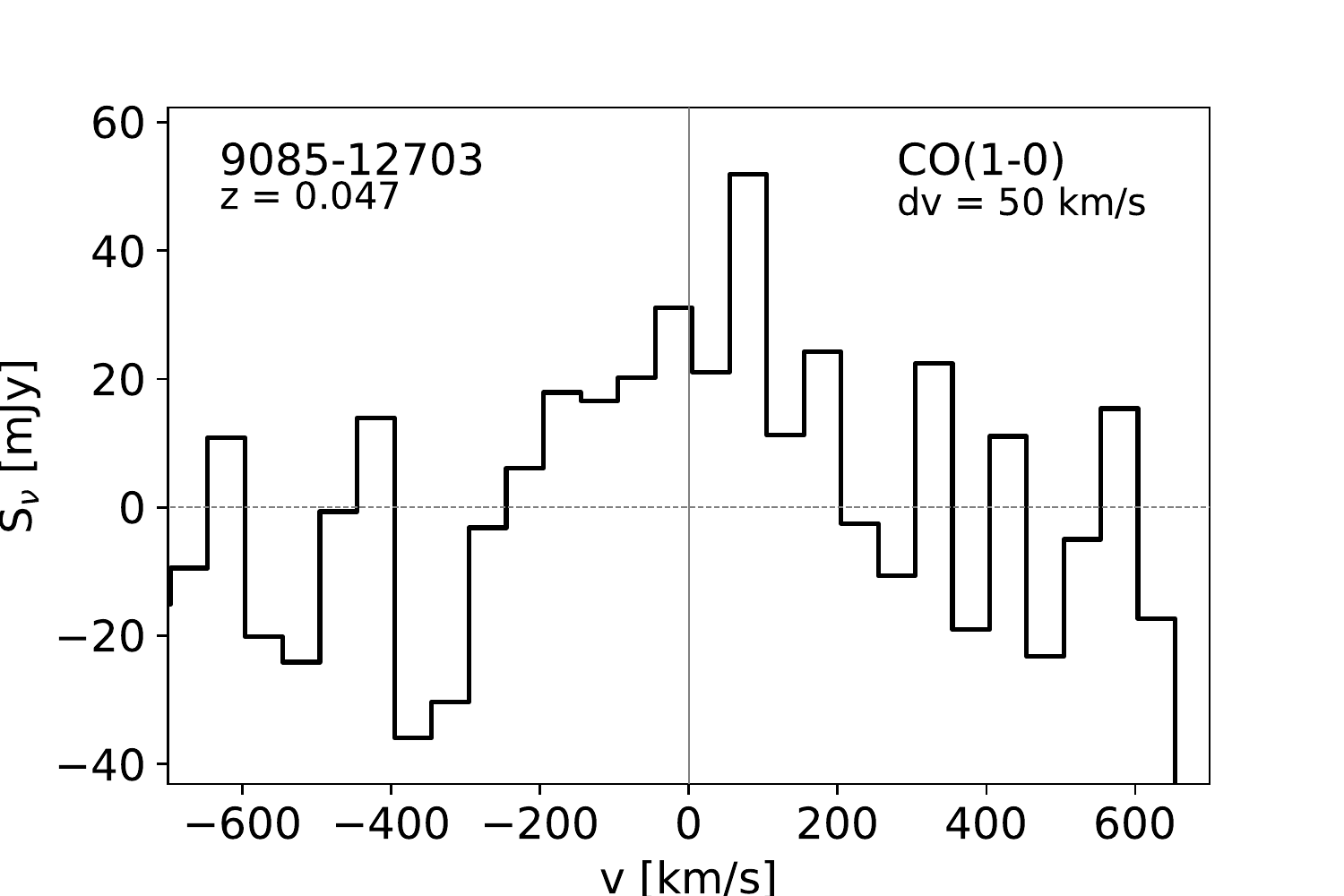} 

\end{figure*}

\begin{figure*} 
 \centering 
 \includegraphics[width = 0.17\textwidth, trim = 0cm 0cm 0cm 0cm, clip = true]{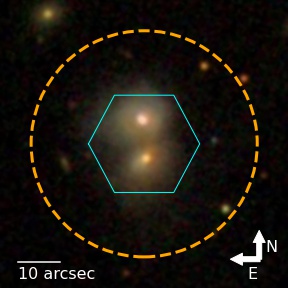}
 \includegraphics[width = 0.29\textwidth, trim = 0cm 0cm 0cm 0cm, clip = true]{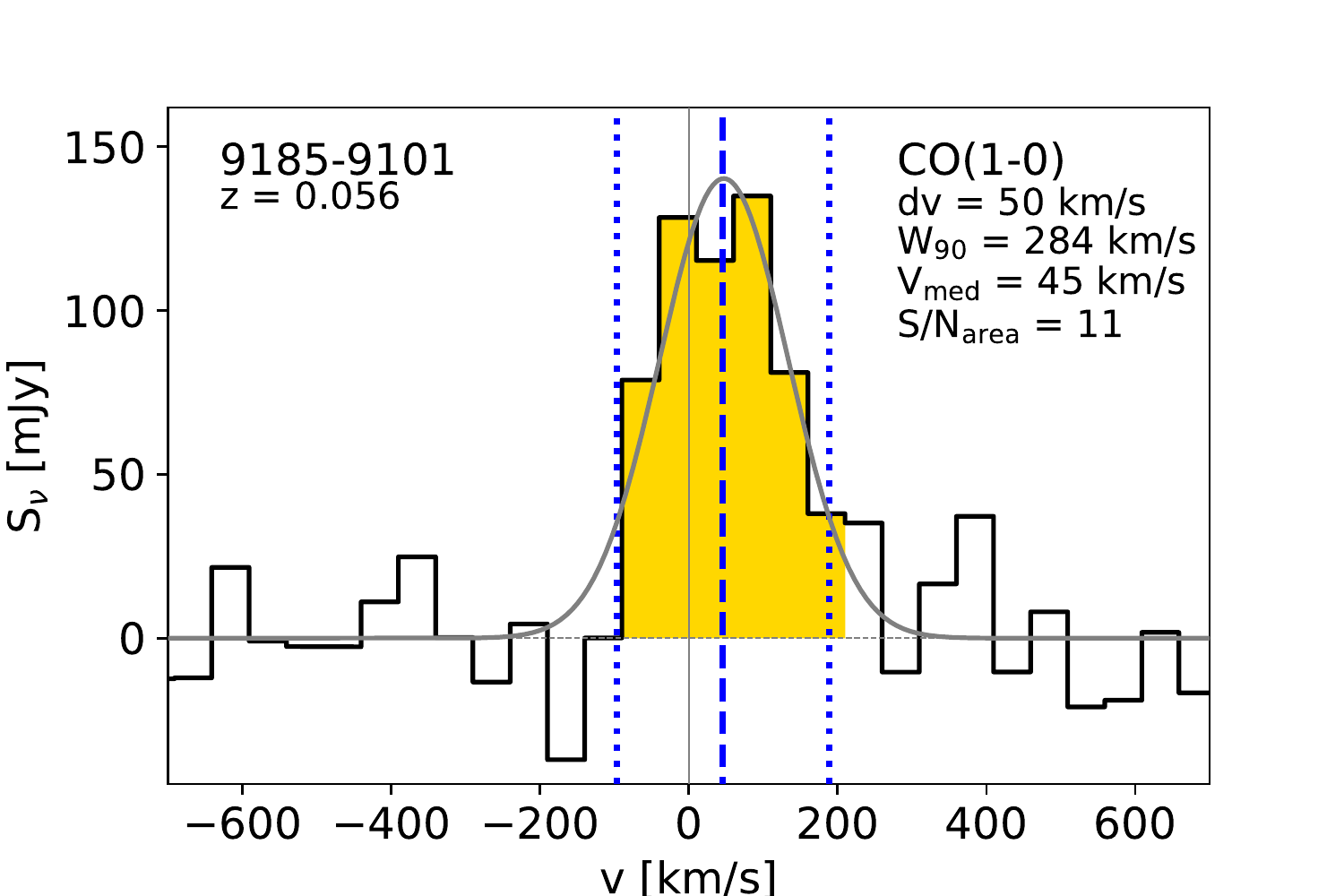} 
 \hspace{0.4cm}
 \centering 
 \includegraphics[width = 0.17\textwidth, trim = 0cm 0cm 0cm 0cm, clip = true]{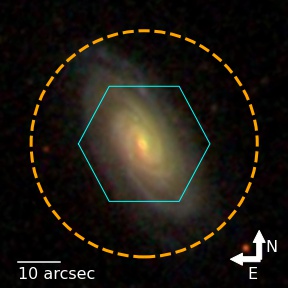}
 \includegraphics[width = 0.29\textwidth, trim = 0cm 0cm 0cm 0cm, clip = true]{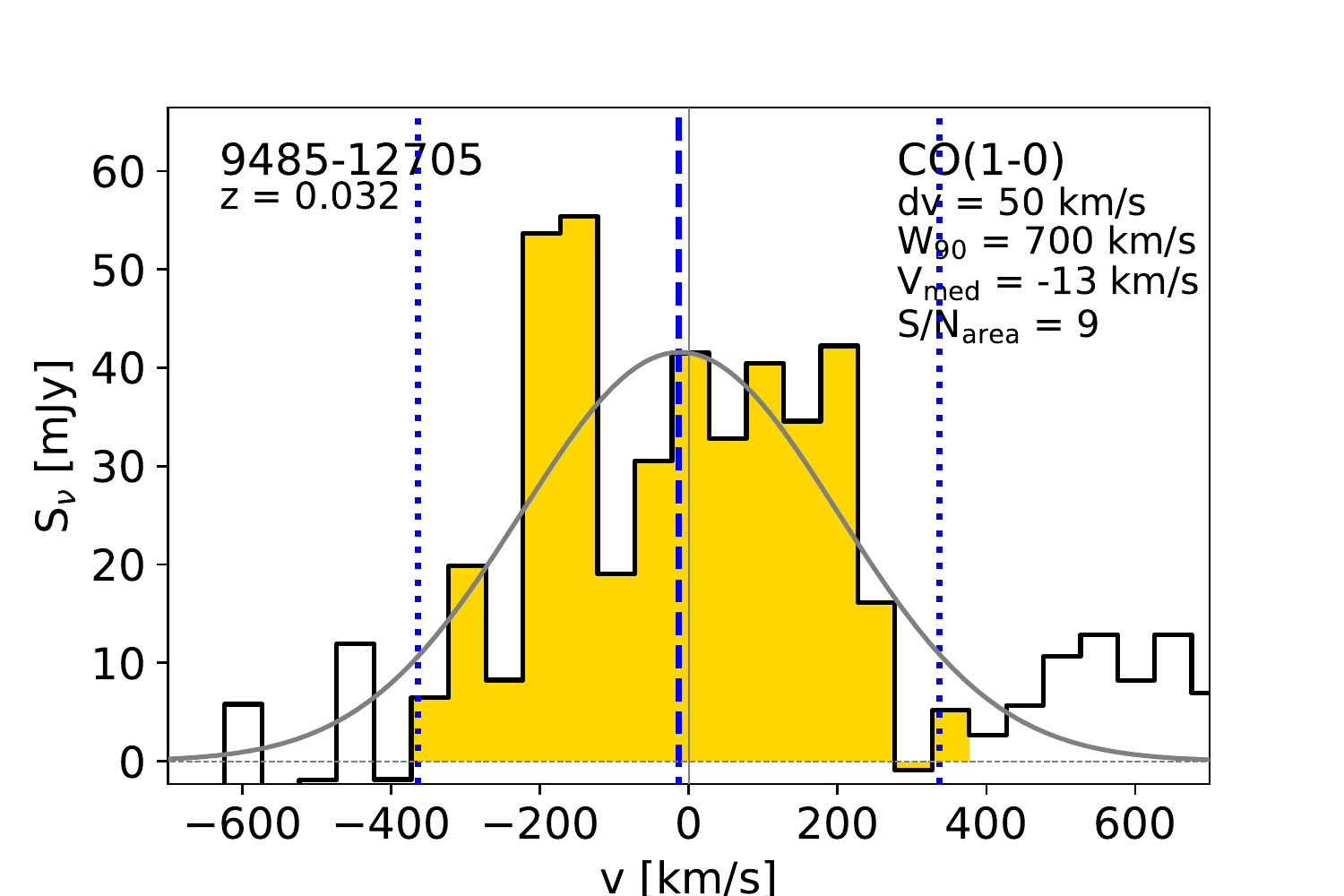} 

\end{figure*}

\begin{figure*} 
 \centering 
 \includegraphics[width = 0.17\textwidth, trim = 0cm 0cm 0cm 0cm, clip = true]{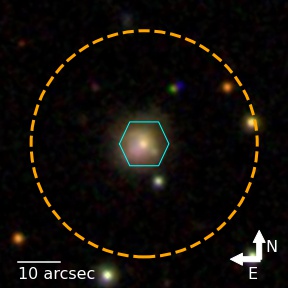}
 \includegraphics[width = 0.29\textwidth, trim = 0cm 0cm 0cm 0cm, clip = true]{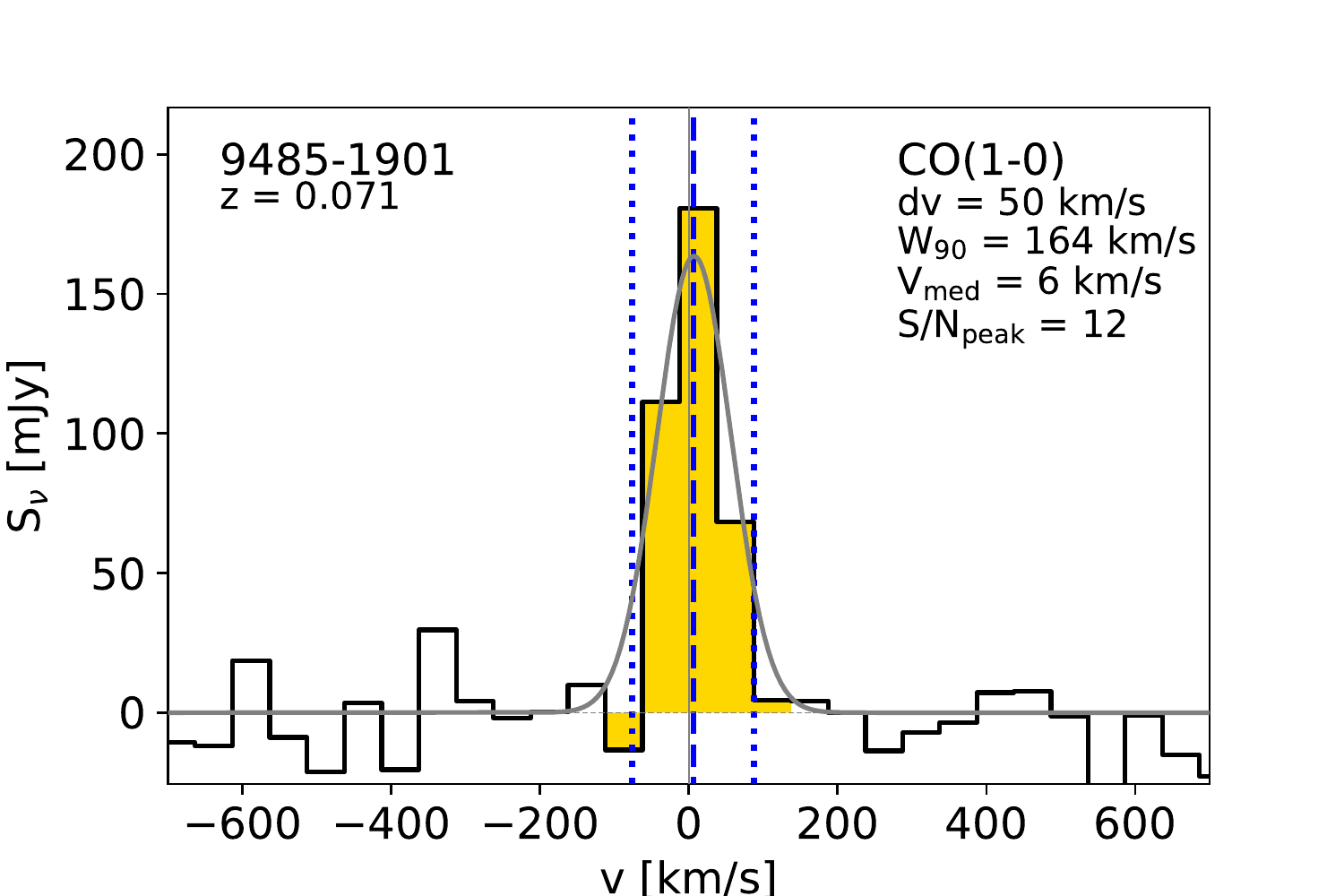} 
 \hspace{0.4cm}
 \centering 
 \includegraphics[width = 0.17\textwidth, trim = 0cm 0cm 0cm 0cm, clip = true]{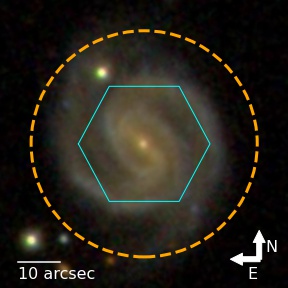}
 \includegraphics[width = 0.29\textwidth, trim = 0cm 0cm 0cm 0cm, clip = true]{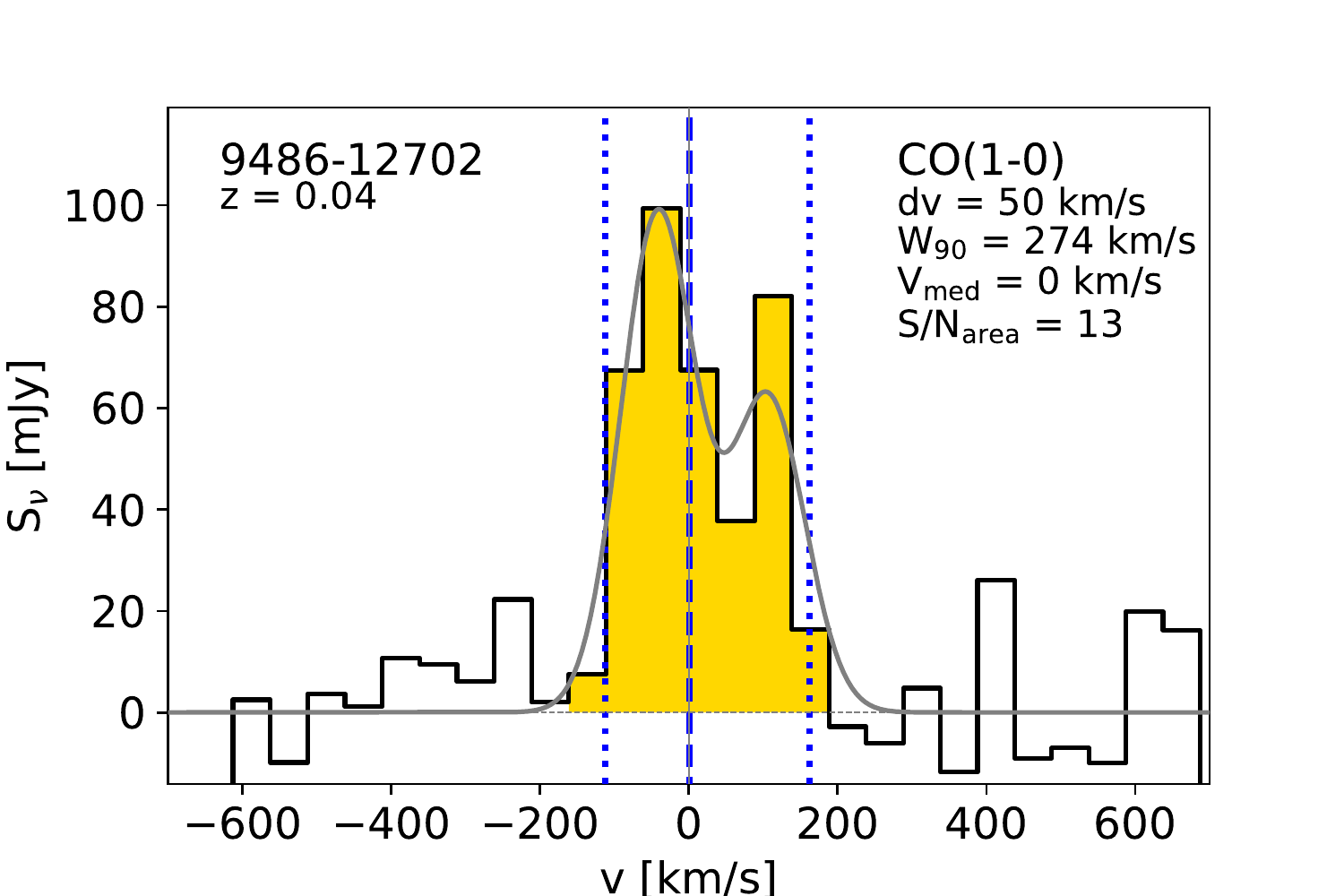} 

\end{figure*}

\begin{figure*} 
 \centering 
 \includegraphics[width = 0.17\textwidth, trim = 0cm 0cm 0cm 0cm, clip = true]{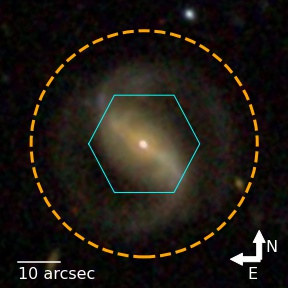}
 \includegraphics[width = 0.29\textwidth, trim = 0cm 0cm 0cm 0cm, clip = true]{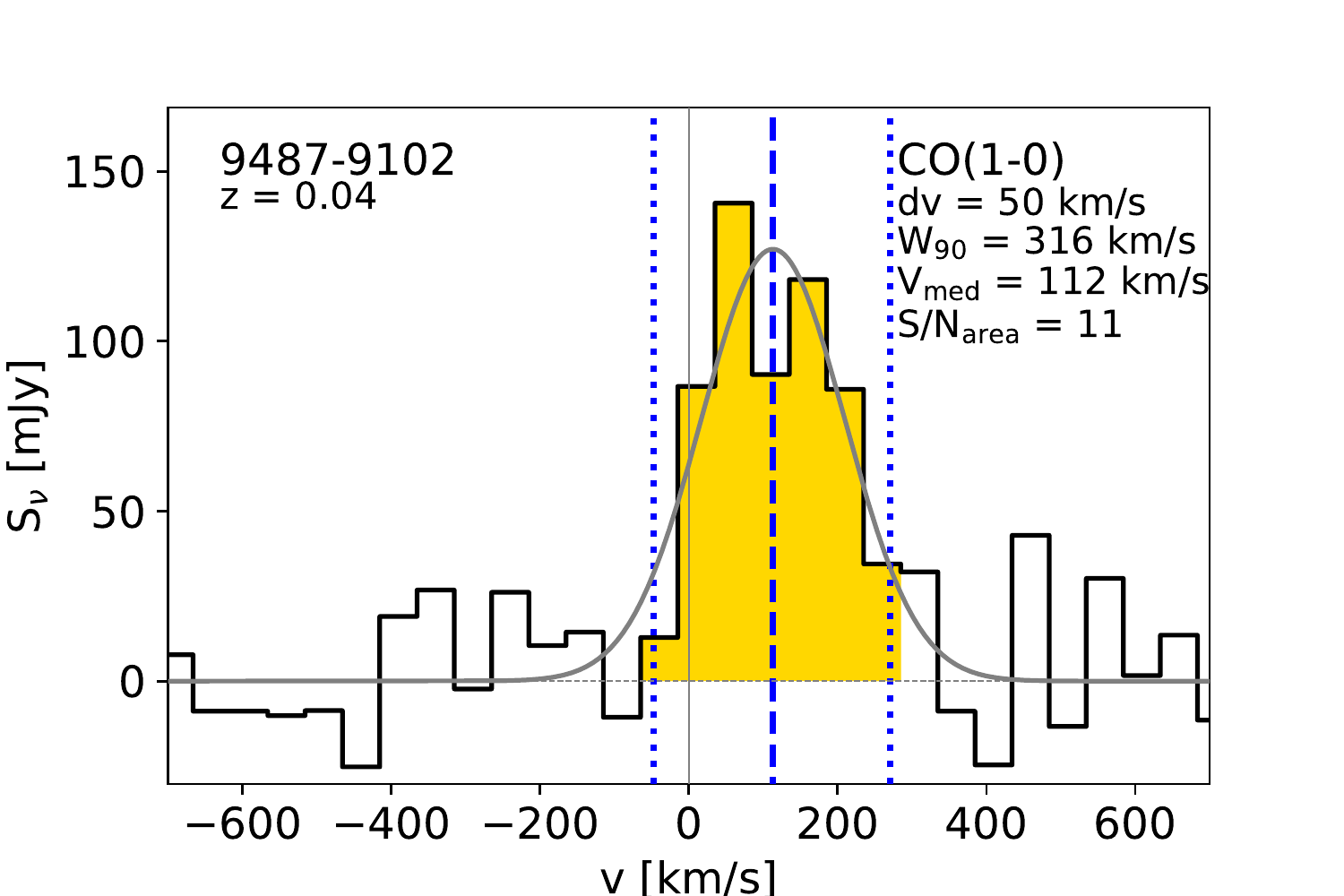} 
 \hspace{0.4cm}
 \centering 
 \includegraphics[width = 0.17\textwidth, trim = 0cm 0cm 0cm 0cm, clip = true]{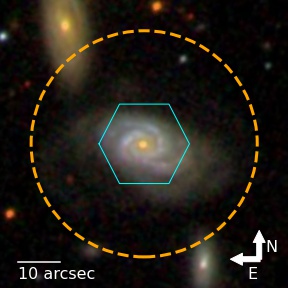}
 \includegraphics[width = 0.29\textwidth, trim = 0cm 0cm 0cm 0cm, clip = true]{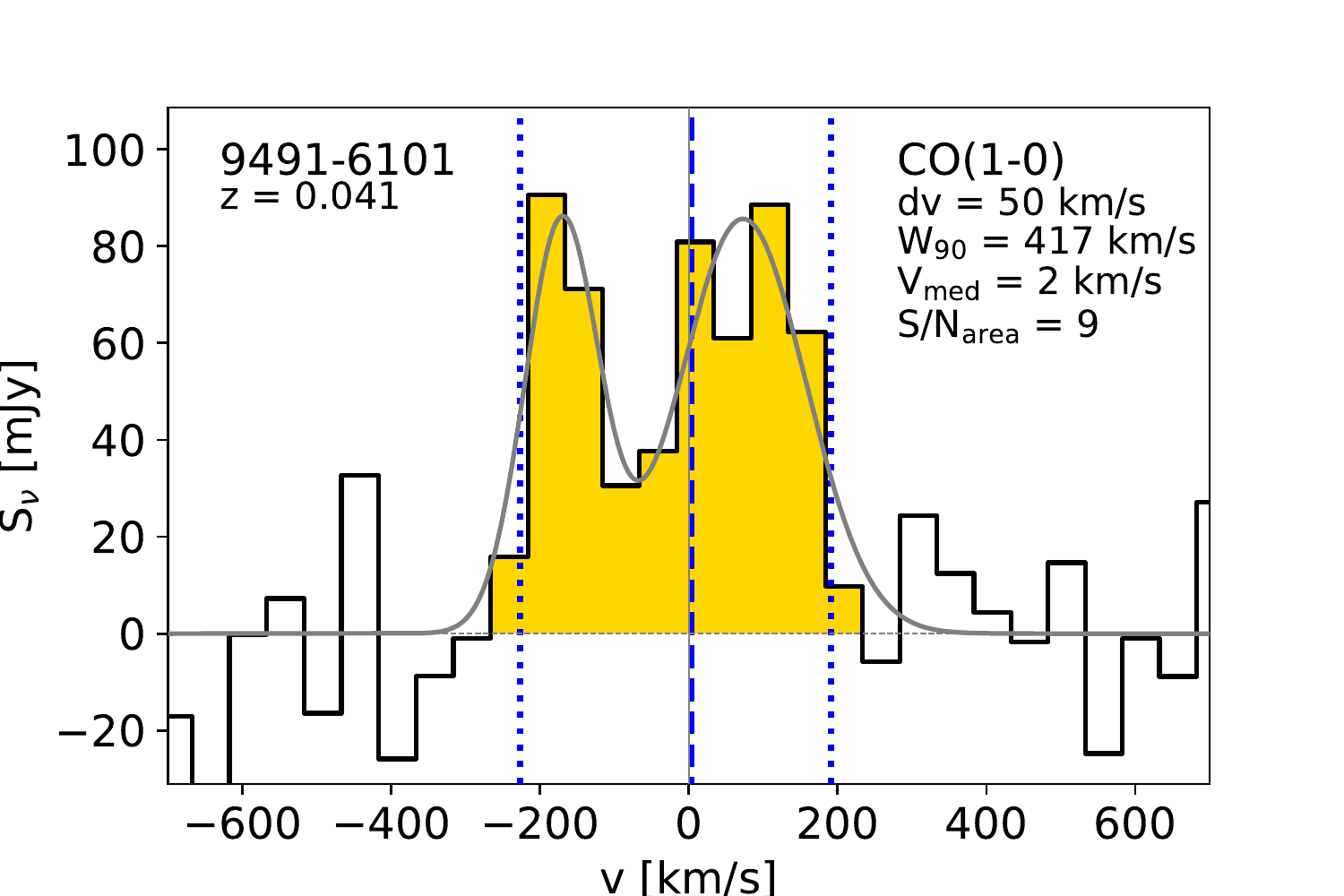} 

\end{figure*}

\begin{figure*} 
 \centering 
 \includegraphics[width = 0.17\textwidth, trim = 0cm 0cm 0cm 0cm, clip = true]{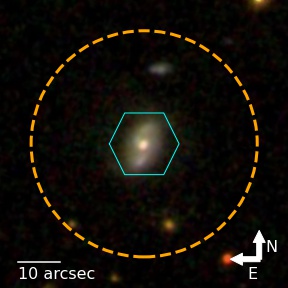}
 \includegraphics[width = 0.29\textwidth, trim = 0cm 0cm 0cm 0cm, clip = true]{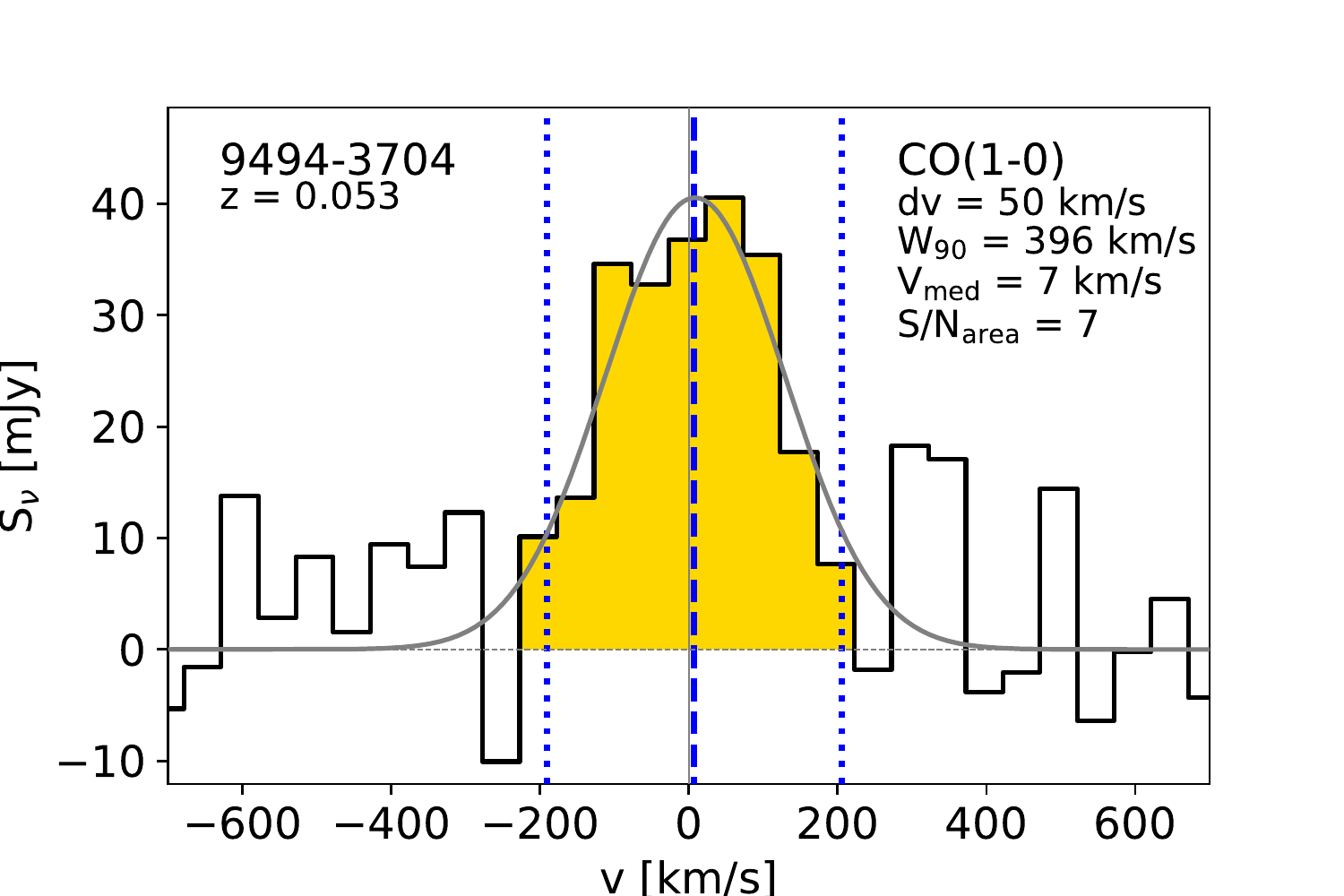} 
 \hspace{0.4cm}
 \centering 
 \includegraphics[width = 0.17\textwidth, trim = 0cm 0cm 0cm 0cm, clip = true]{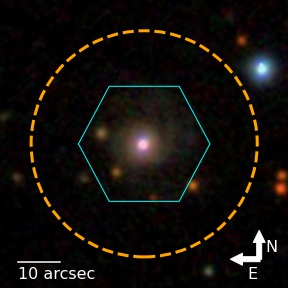}
 \includegraphics[width = 0.29\textwidth, trim = 0cm 0cm 0cm 0cm, clip = true]{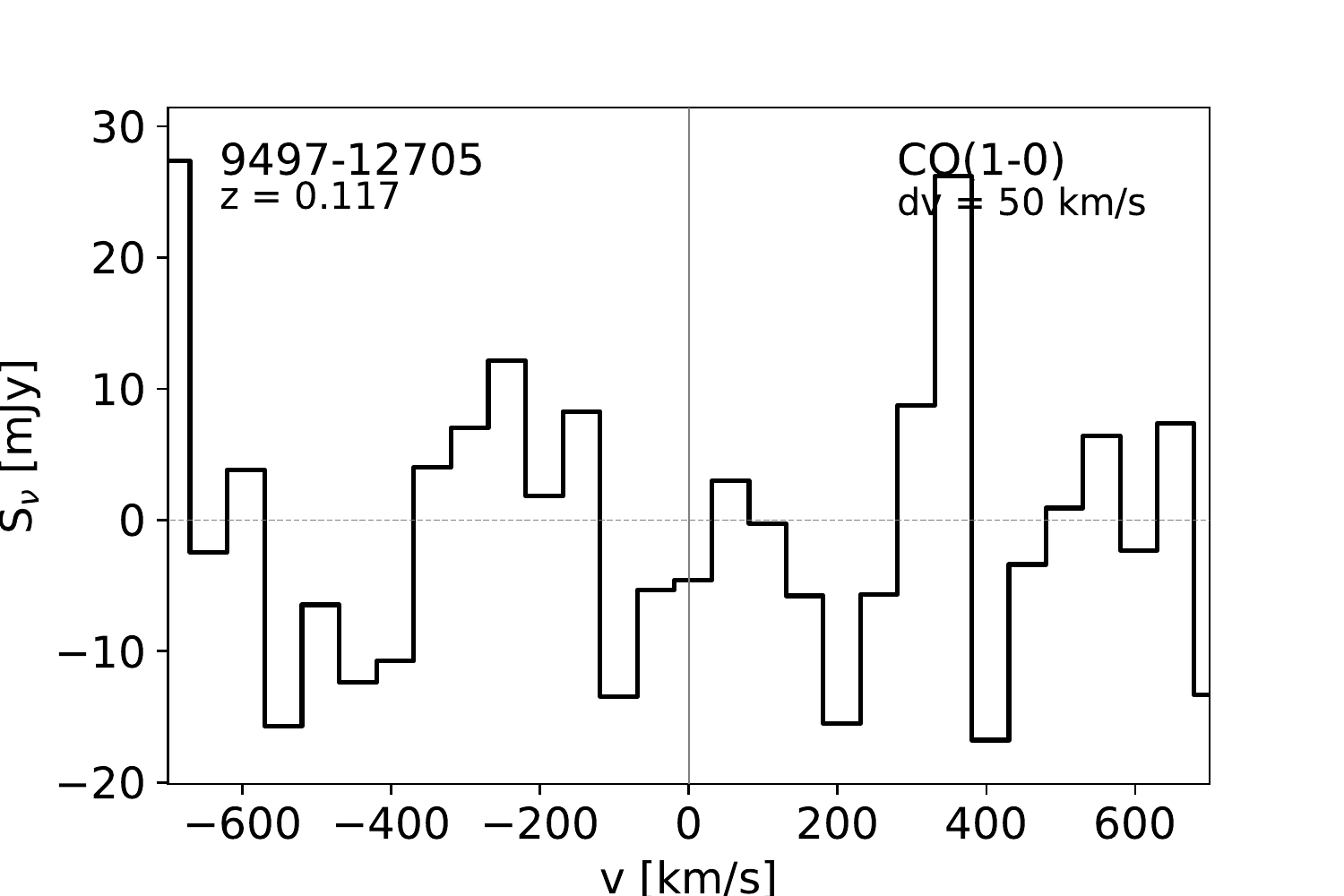} 

\end{figure*}

\begin{figure*} 
   \ContinuedFloat
 \centering 
 \includegraphics[width = 0.17\textwidth, trim = 0cm 0cm 0cm 0cm, clip = true]{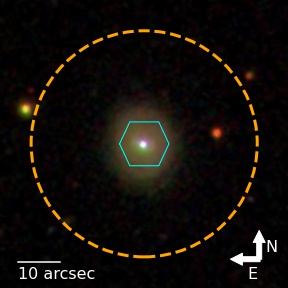}
 \includegraphics[width = 0.29\textwidth, trim = 0cm 0cm 0cm 0cm, clip = true]{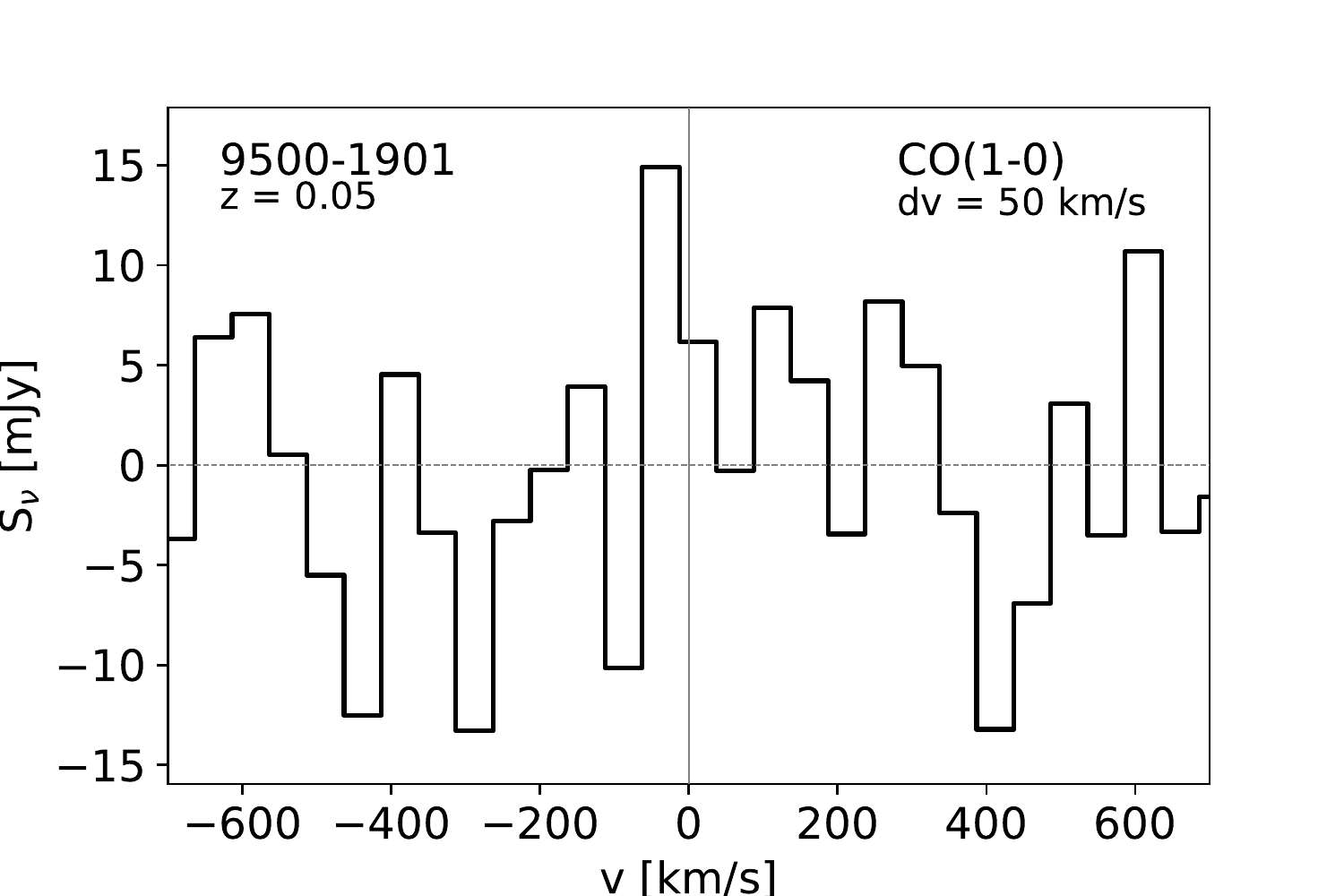} 
 \hspace{0.4cm}
 \centering 
 \includegraphics[width = 0.17\textwidth, trim = 0cm 0cm 0cm 0cm, clip = true]{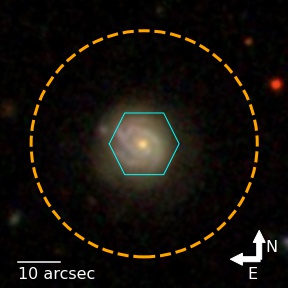}
 \includegraphics[width = 0.29\textwidth, trim = 0cm 0cm 0cm 0cm, clip = true]{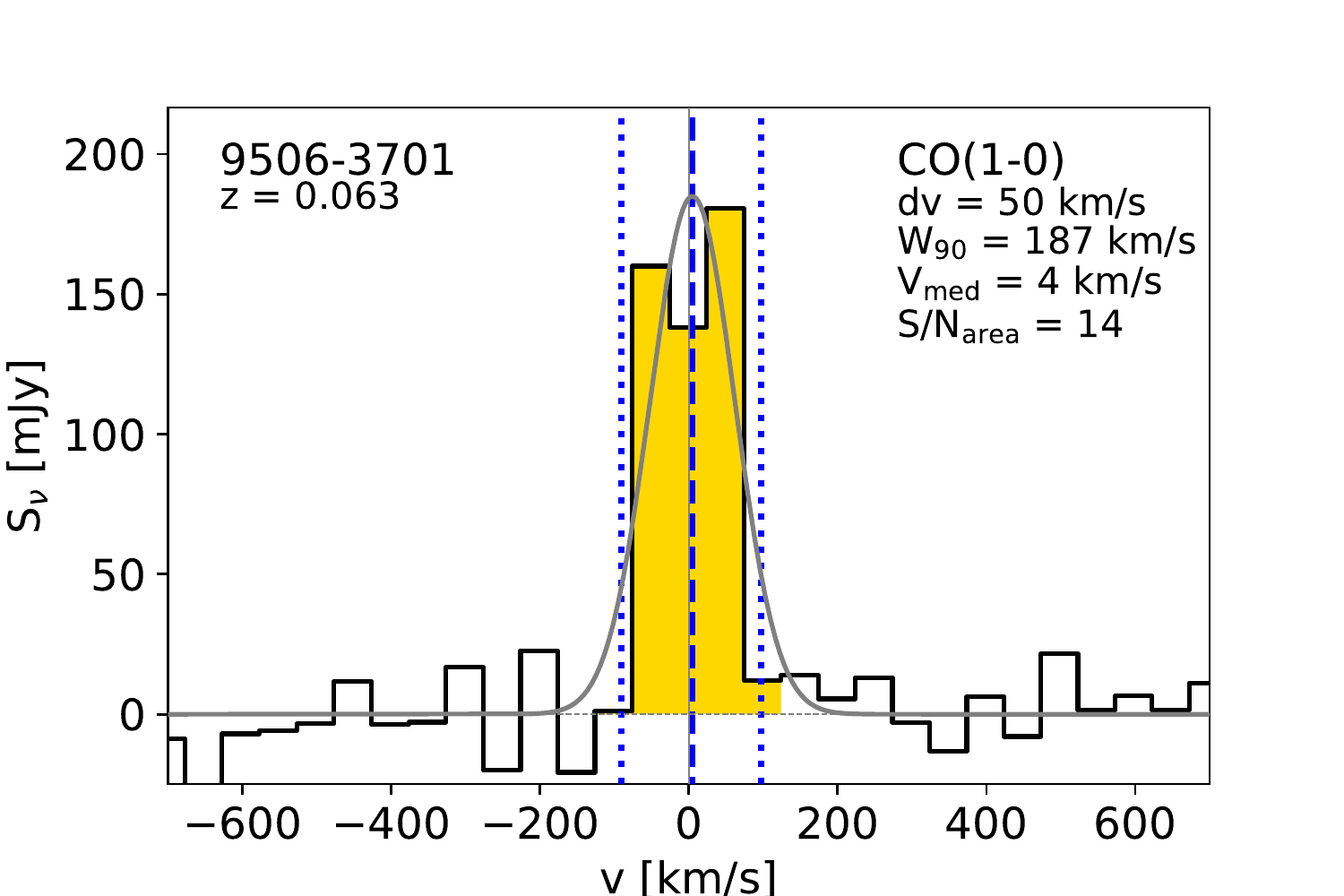} 
\caption{continued.}
\end{figure*}

\begin{figure*} 
 \centering 
 \includegraphics[width = 0.17\textwidth, trim = 0cm 0cm 0cm 0cm, clip = true]{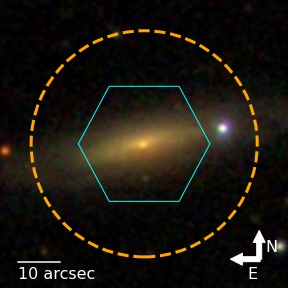}
 \includegraphics[width = 0.29\textwidth, trim = 0cm 0cm 0cm 0cm, clip = true]{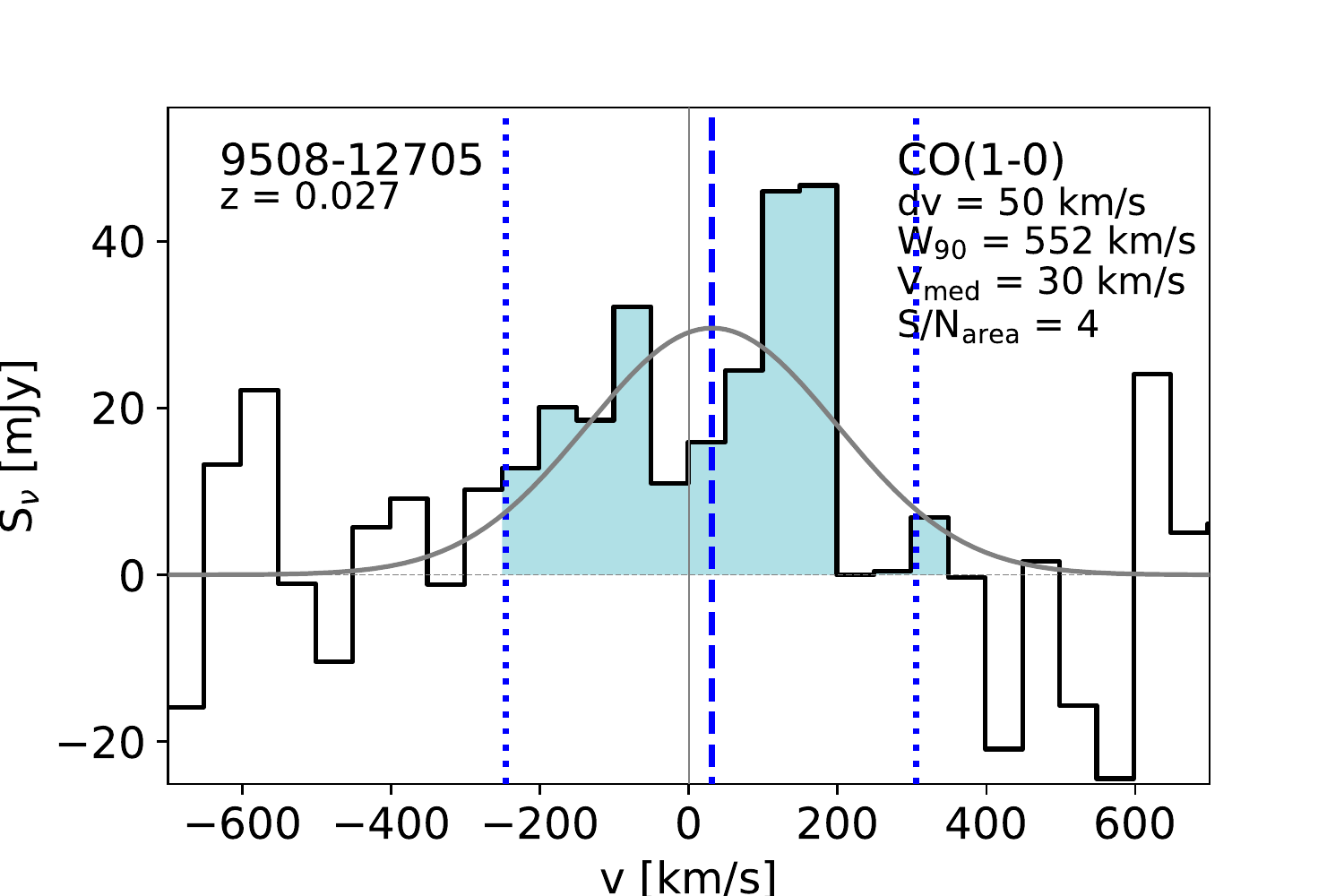} 
 \hspace{0.4cm}
 \centering 
 \includegraphics[width = 0.17\textwidth, trim = 0cm 0cm 0cm 0cm, clip = true]{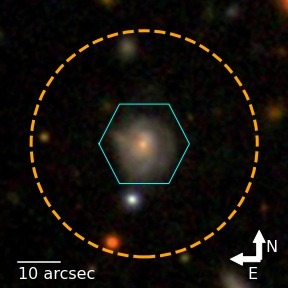}
 \includegraphics[width = 0.29\textwidth, trim = 0cm 0cm 0cm 0cm, clip = true]{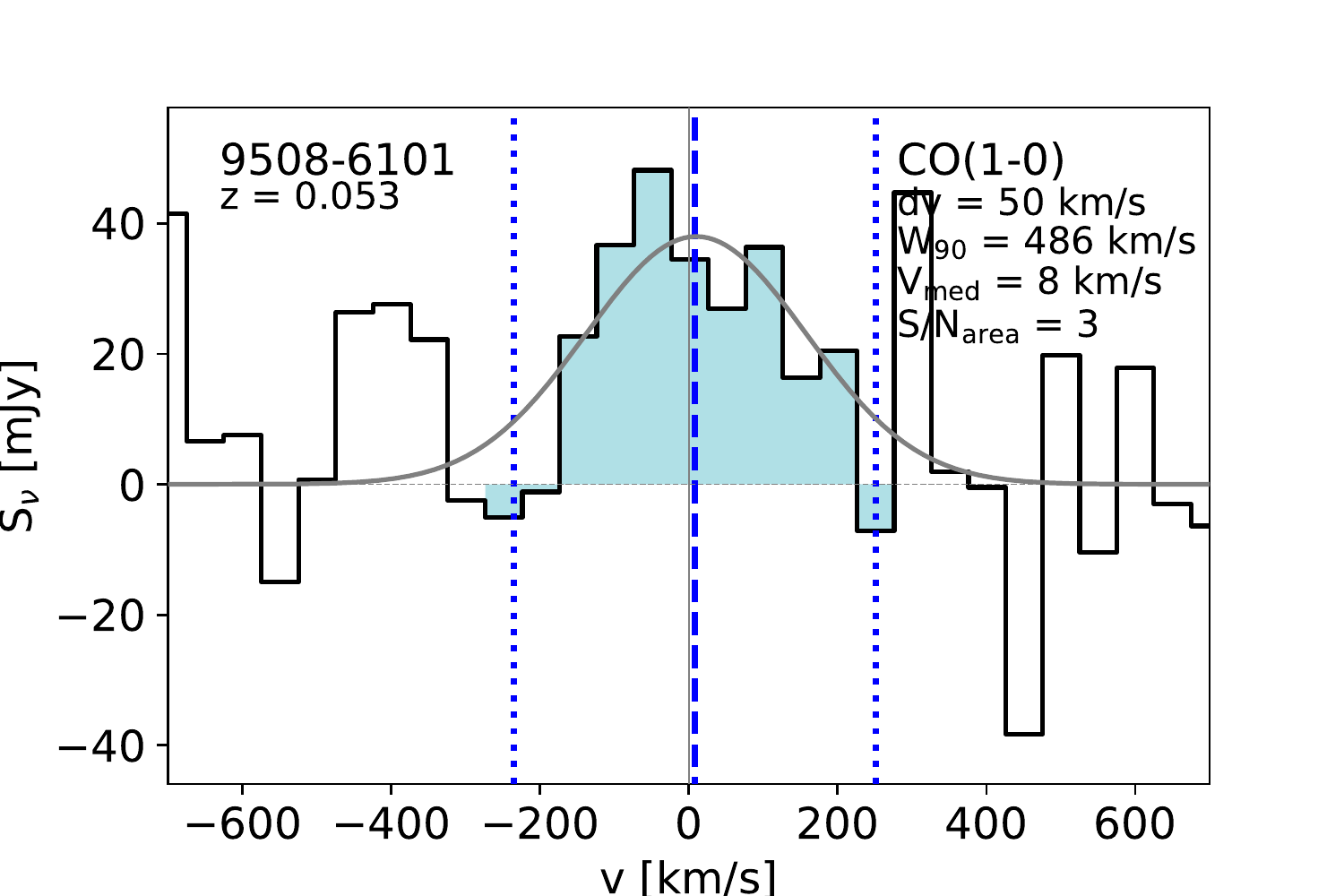} 

\end{figure*}

\begin{figure*} 
 \centering 
 \includegraphics[width = 0.17\textwidth, trim = 0cm 0cm 0cm 0cm, clip = true]{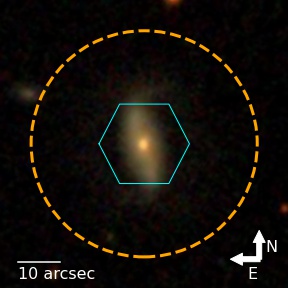}
 \includegraphics[width = 0.29\textwidth, trim = 0cm 0cm 0cm 0cm, clip = true]{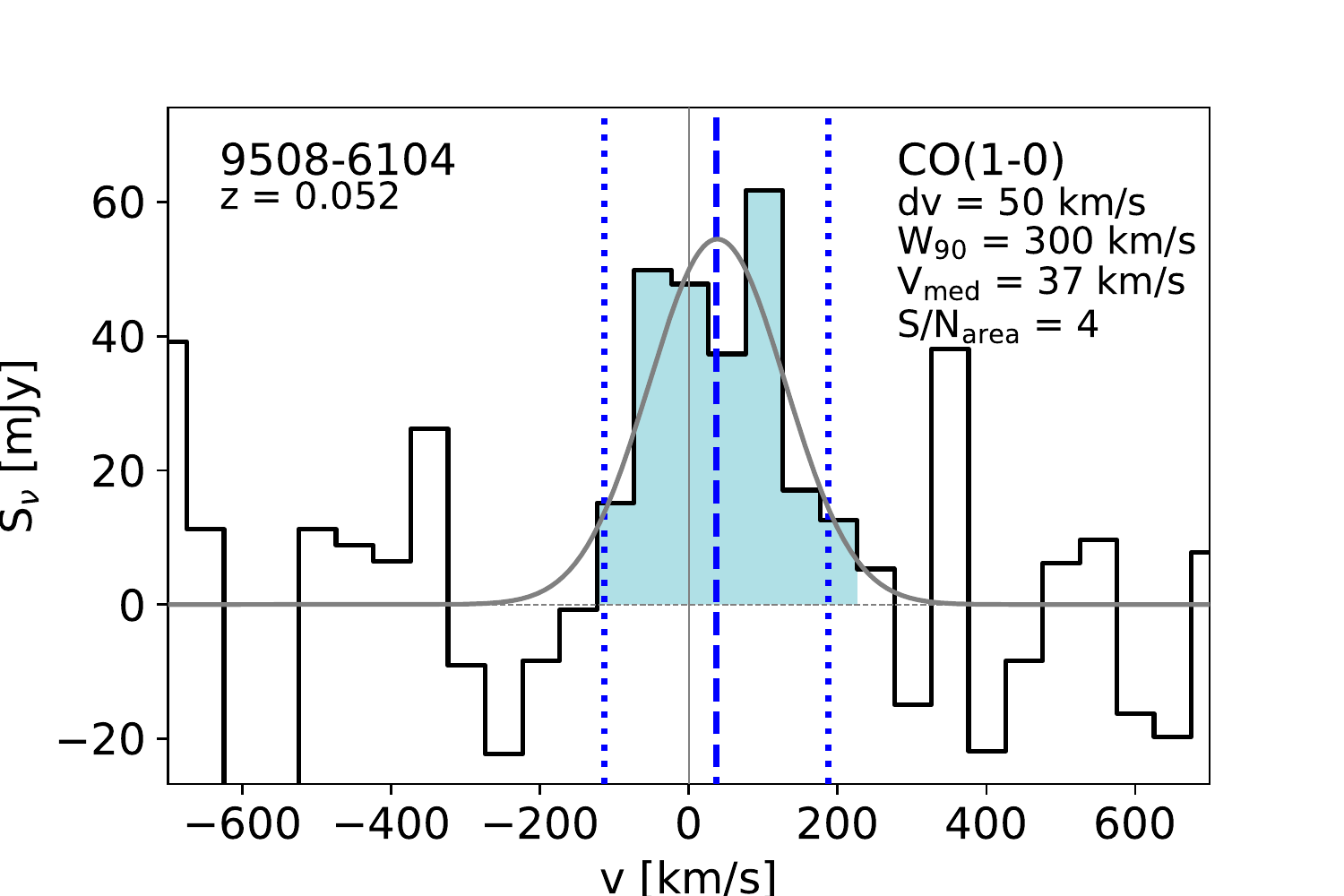} 
 \hspace{0.4cm}
 \centering 
 \includegraphics[width = 0.17\textwidth, trim = 0cm 0cm 0cm 0cm, clip = true]{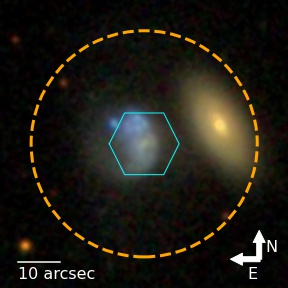}
 \includegraphics[width = 0.29\textwidth, trim = 0cm 0cm 0cm 0cm, clip = true]{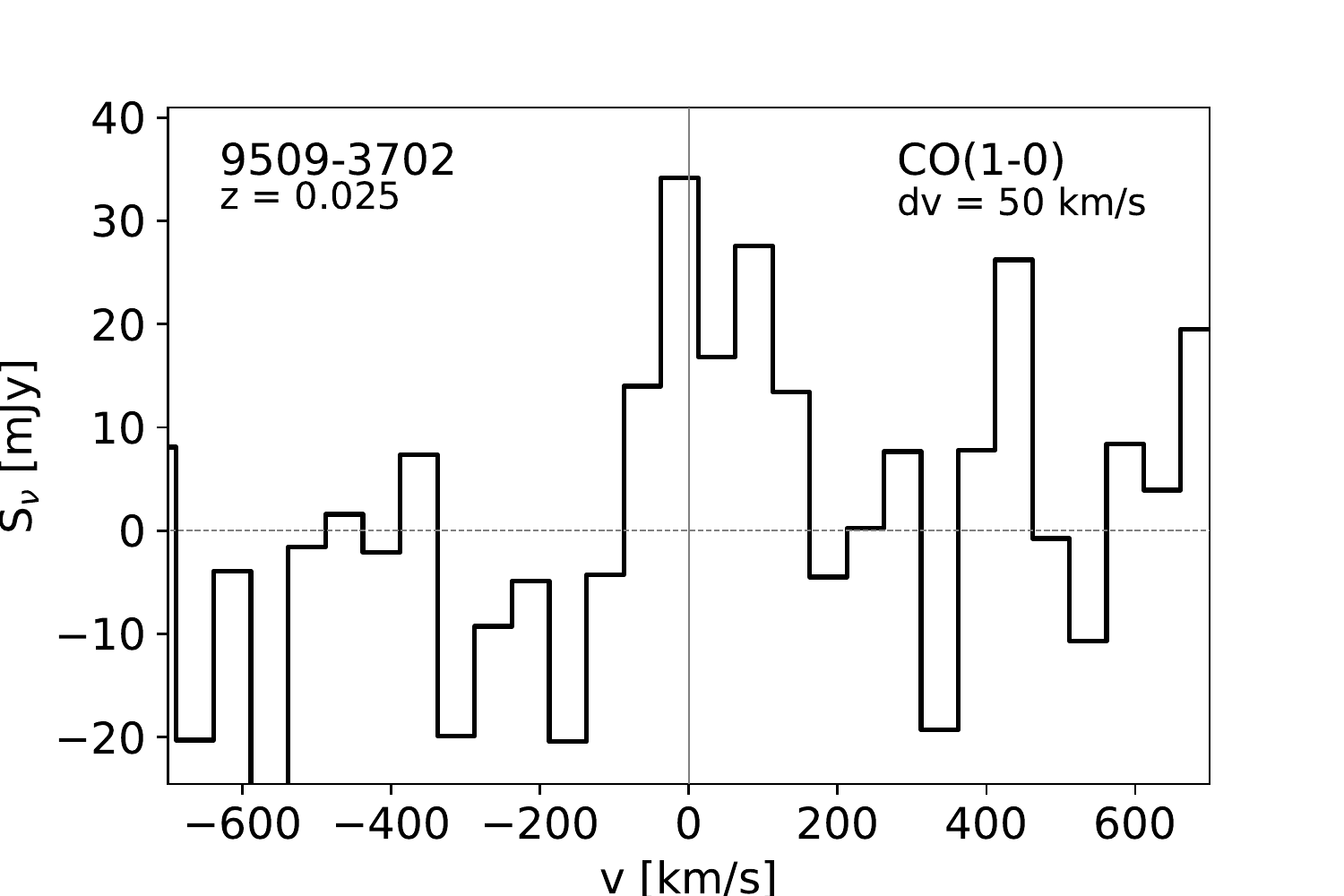} 

\end{figure*}

\begin{figure*} 
 \centering 
 \includegraphics[width = 0.17\textwidth, trim = 0cm 0cm 0cm 0cm, clip = true]{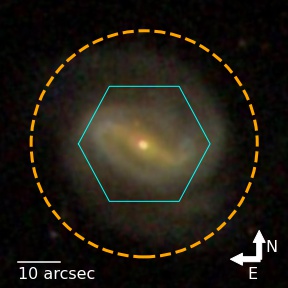}
 \includegraphics[width = 0.29\textwidth, trim = 0cm 0cm 0cm 0cm, clip = true]{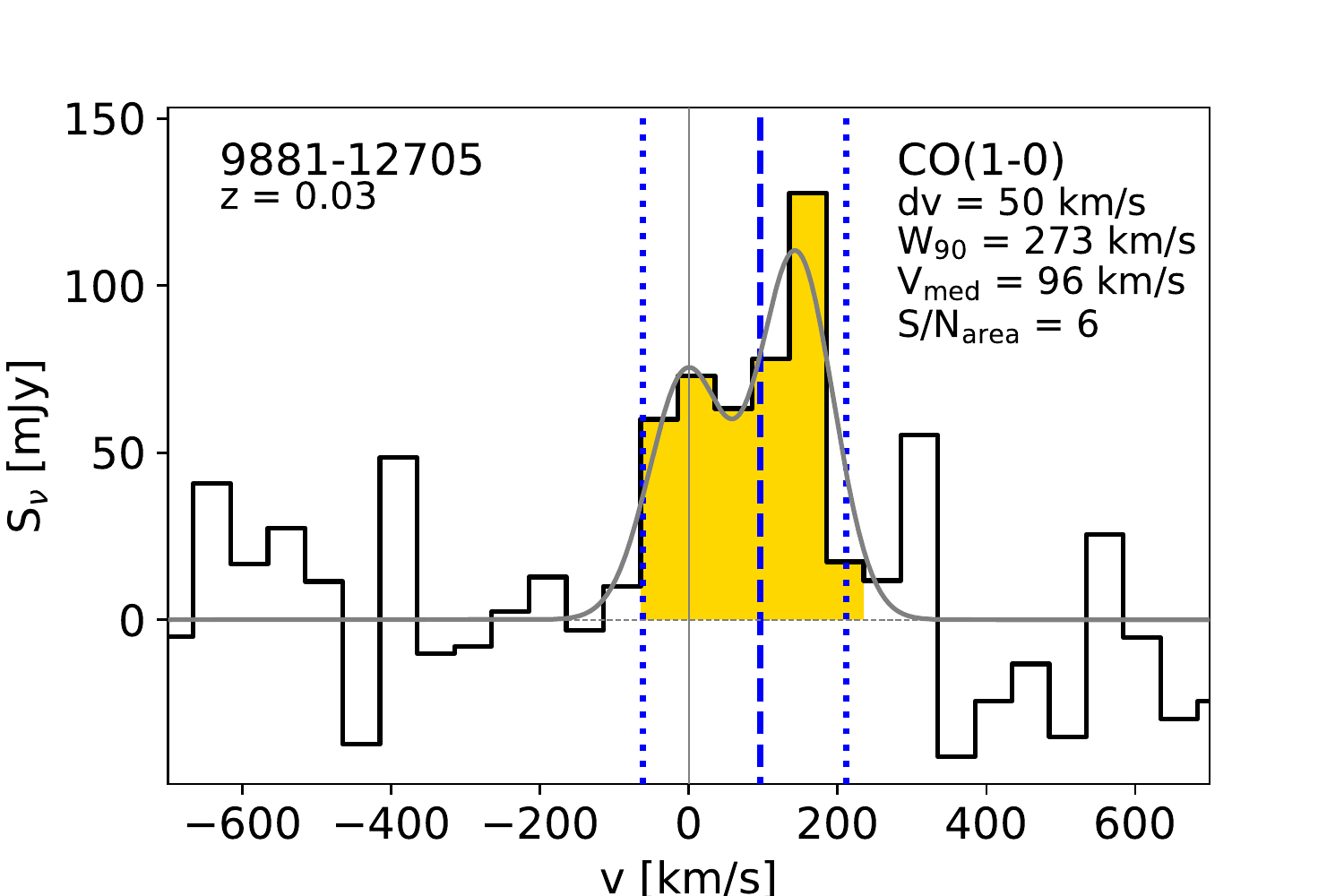} 
 \hspace{0.4cm}
 \centering 
 \includegraphics[width = 0.17\textwidth, trim = 0cm 0cm 0cm 0cm, clip = true]{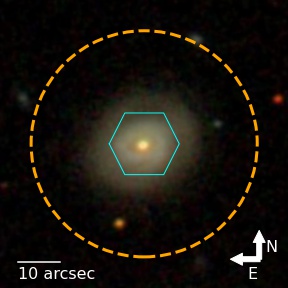}
 \includegraphics[width = 0.29\textwidth, trim = 0cm 0cm 0cm 0cm, clip = true]{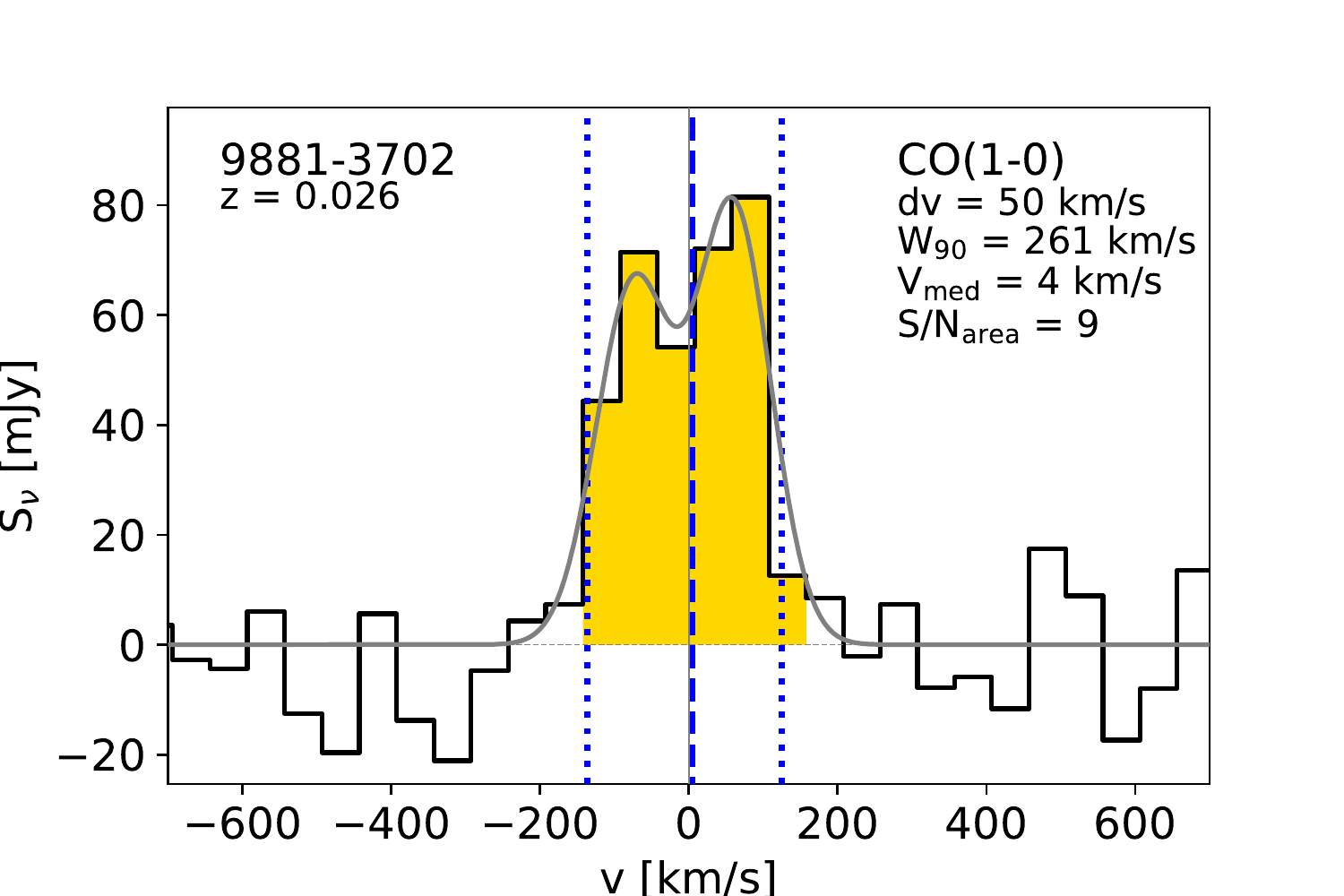} 
\end{figure*}

\begin{figure*} 
 \centering 
 \includegraphics[width = 0.17\textwidth, trim = 0cm 0cm 0cm 0cm, clip = true]{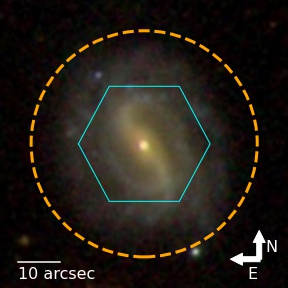}
 \includegraphics[width = 0.29\textwidth, trim = 0cm 0cm 0cm 0cm, clip = true]{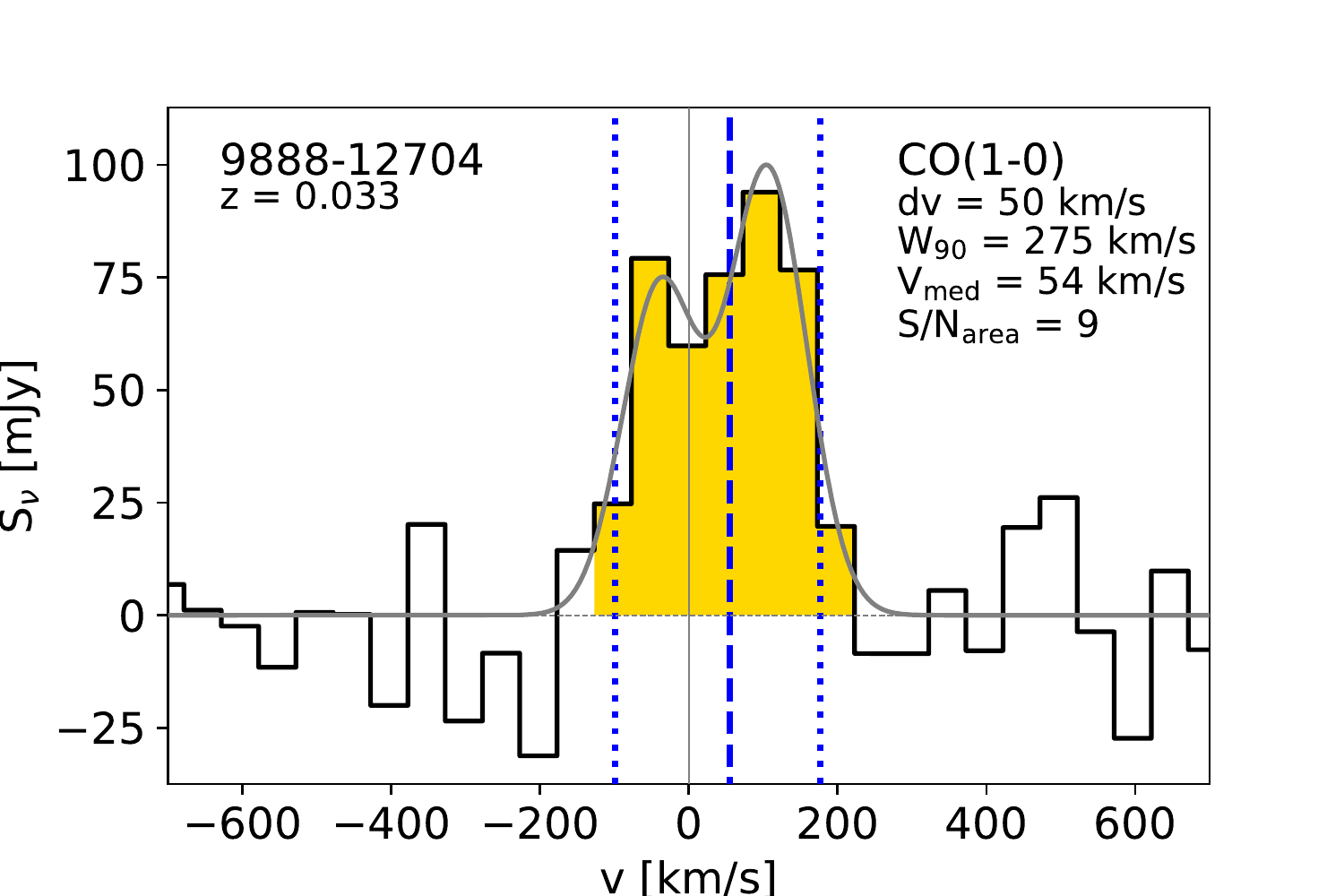} 
 \hspace{0.4cm}
 \centering 
 \includegraphics[width = 0.17\textwidth, trim = 0cm 0cm 0cm 0cm, clip = true]{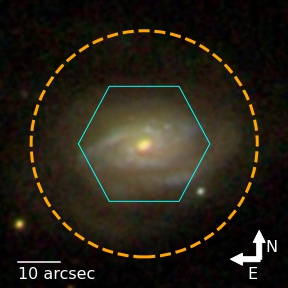}
 \includegraphics[width = 0.29\textwidth, trim = 0cm 0cm 0cm 0cm, clip = true]{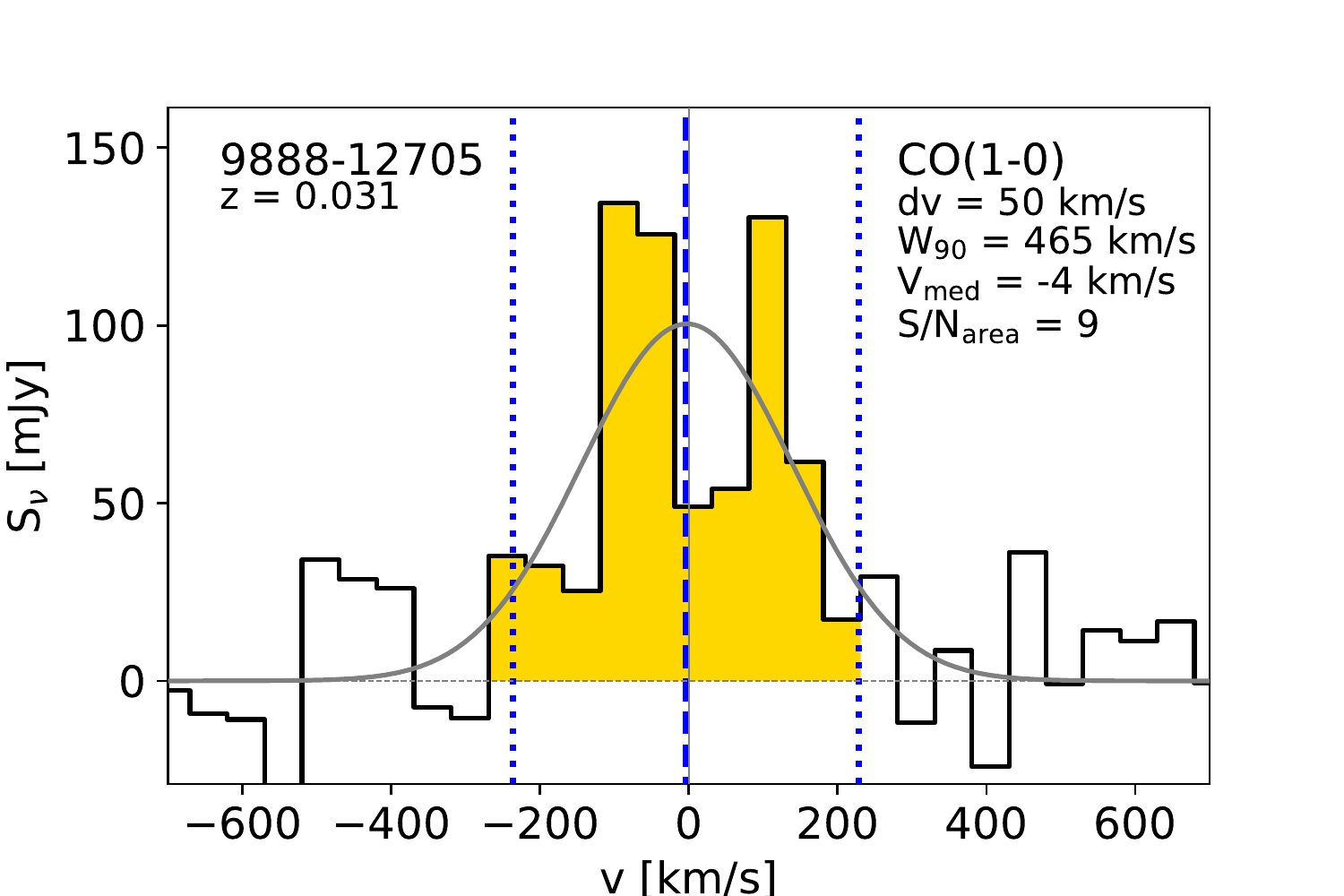} 

\end{figure*}

\begin{figure*} 
 \centering 
 \includegraphics[width = 0.17\textwidth, trim = 0cm 0cm 0cm 0cm, clip = true]{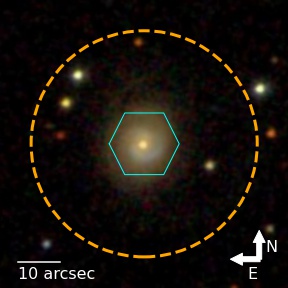}
 \includegraphics[width = 0.29\textwidth, trim = 0cm 0cm 0cm 0cm, clip = true]{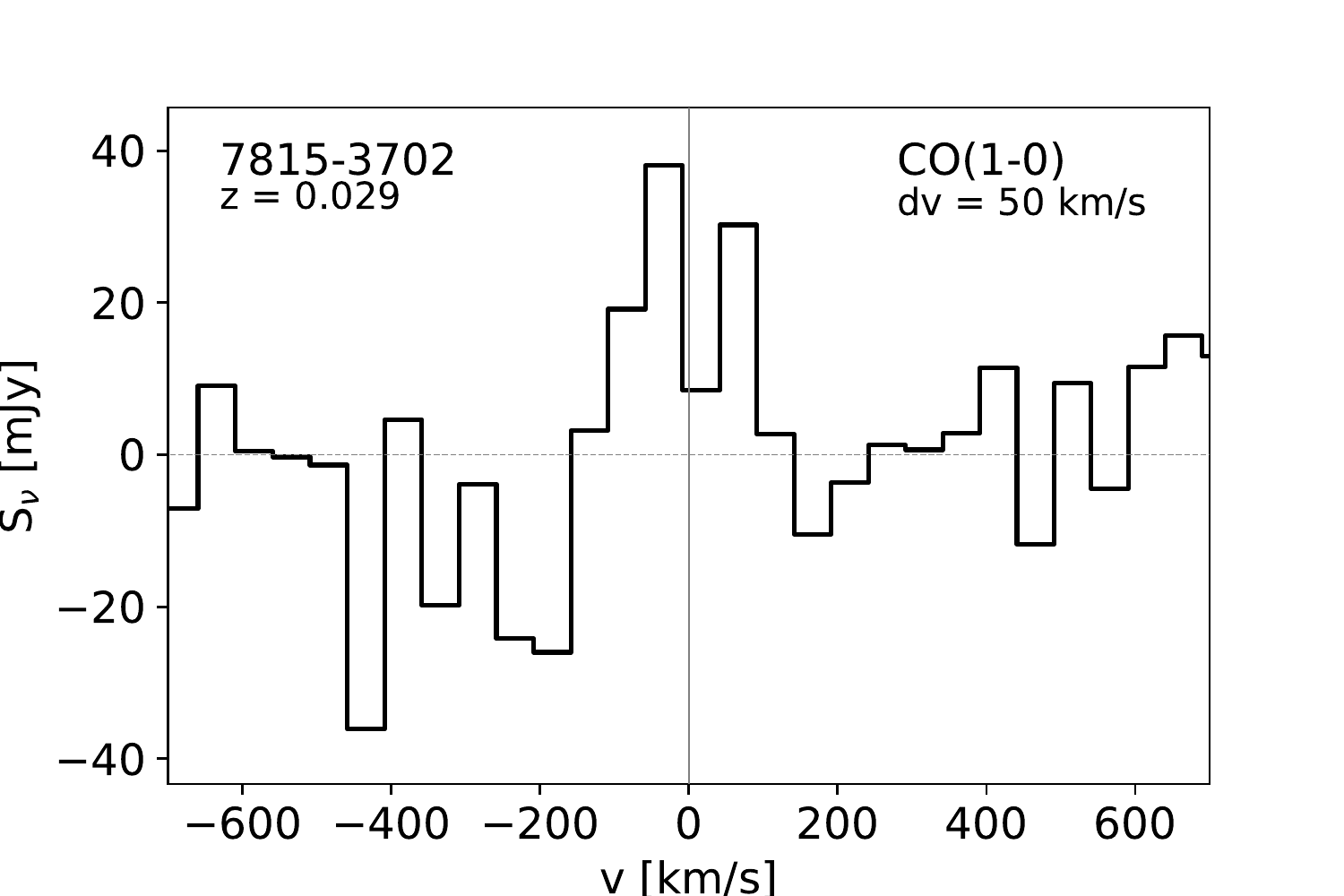} 
  \hspace{0.4cm}
  \centering  \includegraphics[width = 0.17\textwidth, trim = 0cm 0cm 0cm 0cm, clip = true]{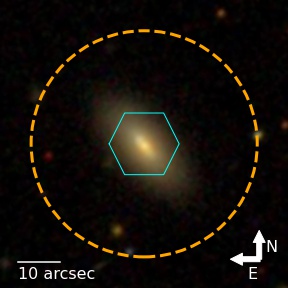} \includegraphics[width = 0.29\textwidth, trim = 0cm 0cm 0cm 0cm, clip = true]{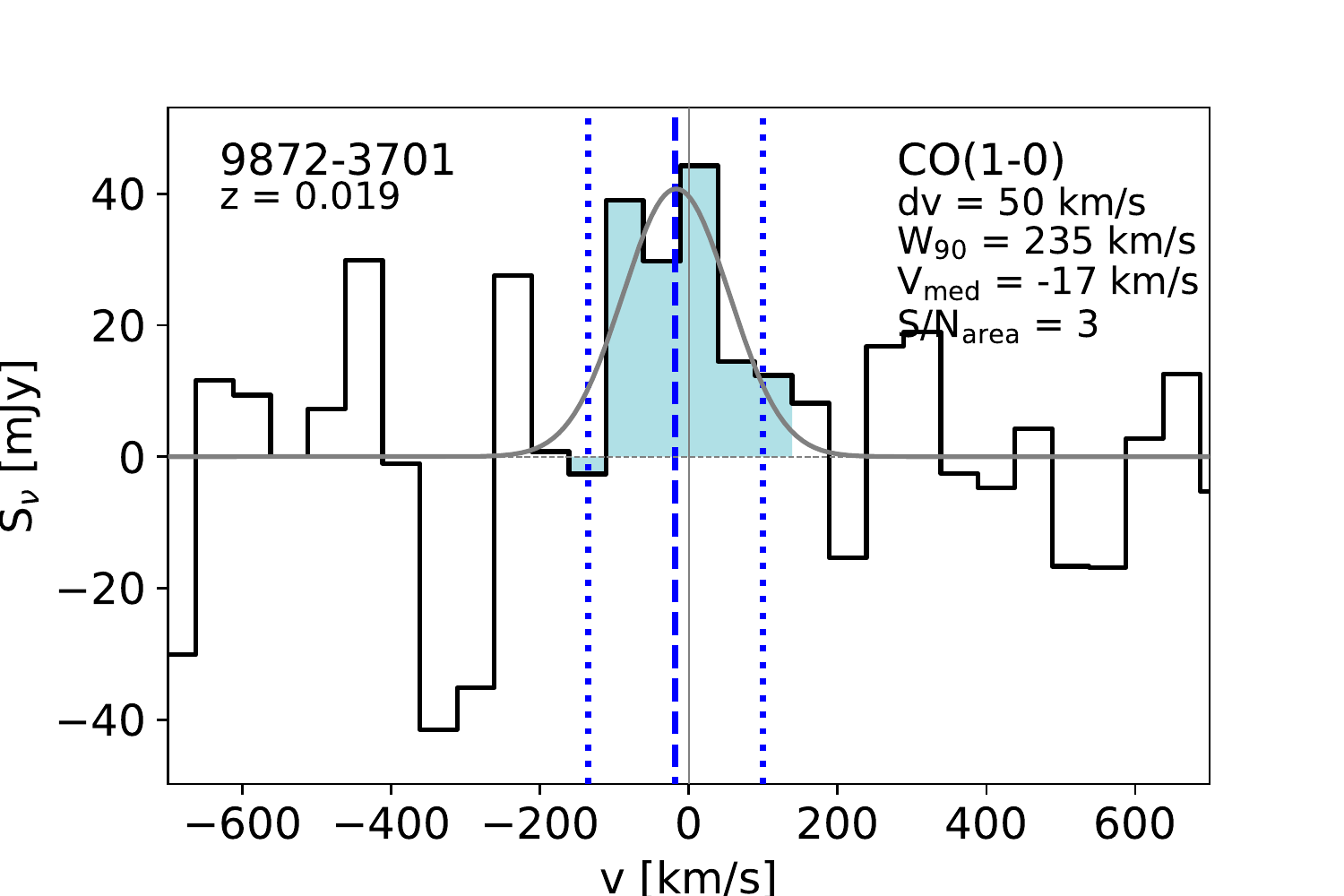}
\end{figure*} 

\begin{figure*} 
   \ContinuedFloat
 \centering  \includegraphics[width = 0.17\textwidth, trim = 0cm 0cm 0cm 0cm, clip = true]{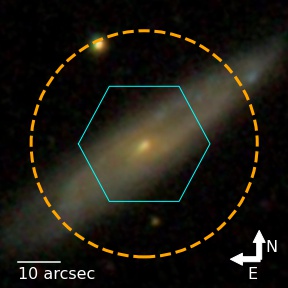} \includegraphics[width = 0.29\textwidth, trim = 0cm 0cm 0cm 0cm, clip = true]{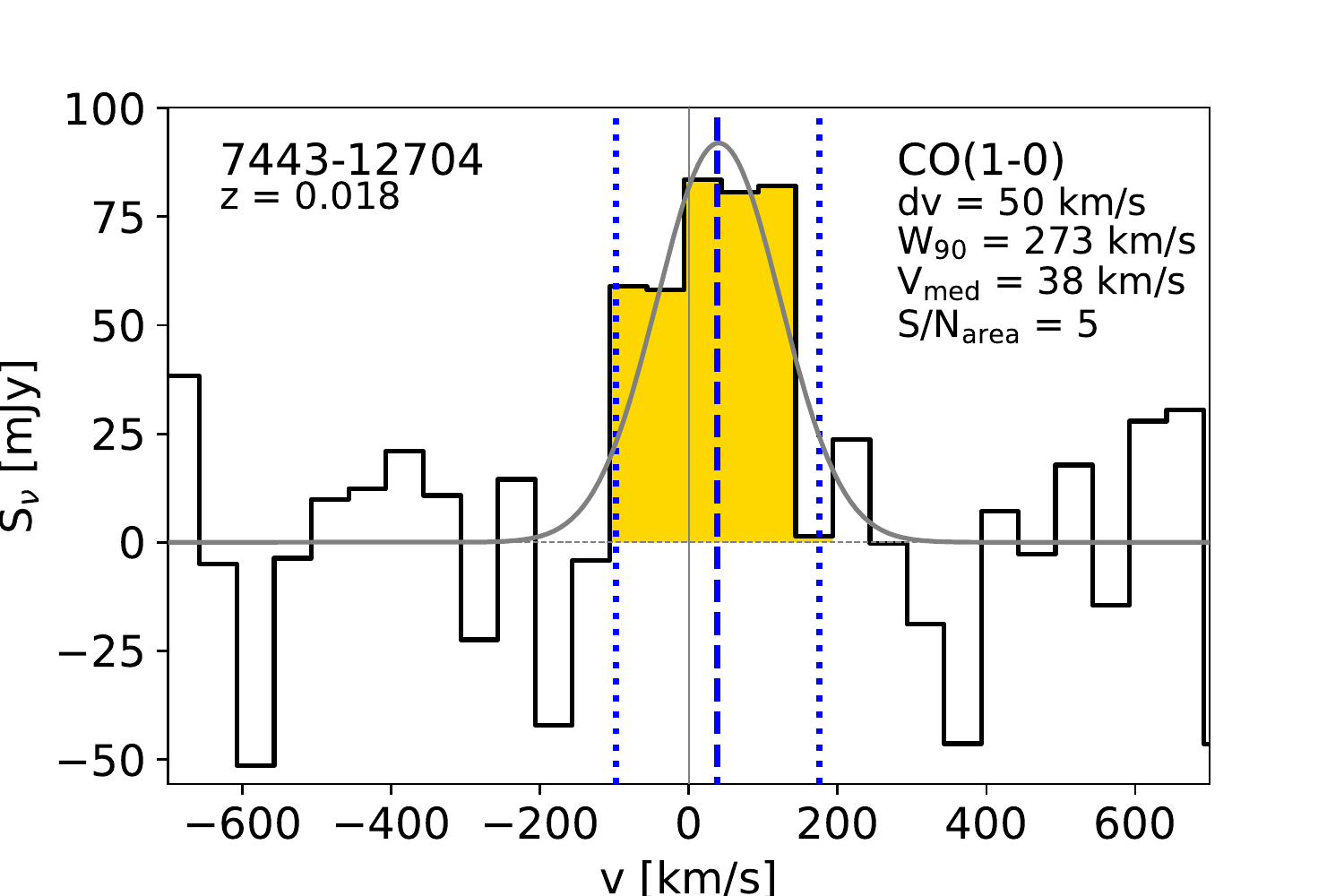}
\hspace{0.4cm}  \centering  \includegraphics[width = 0.17\textwidth, trim = 0cm 0cm 0cm 0cm, clip = true]{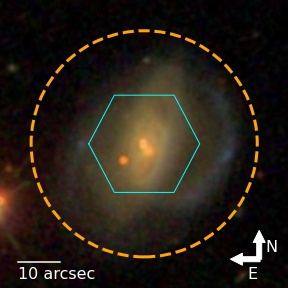} \includegraphics[width = 0.29\textwidth, trim = 0cm 0cm 0cm 0cm, clip = true]{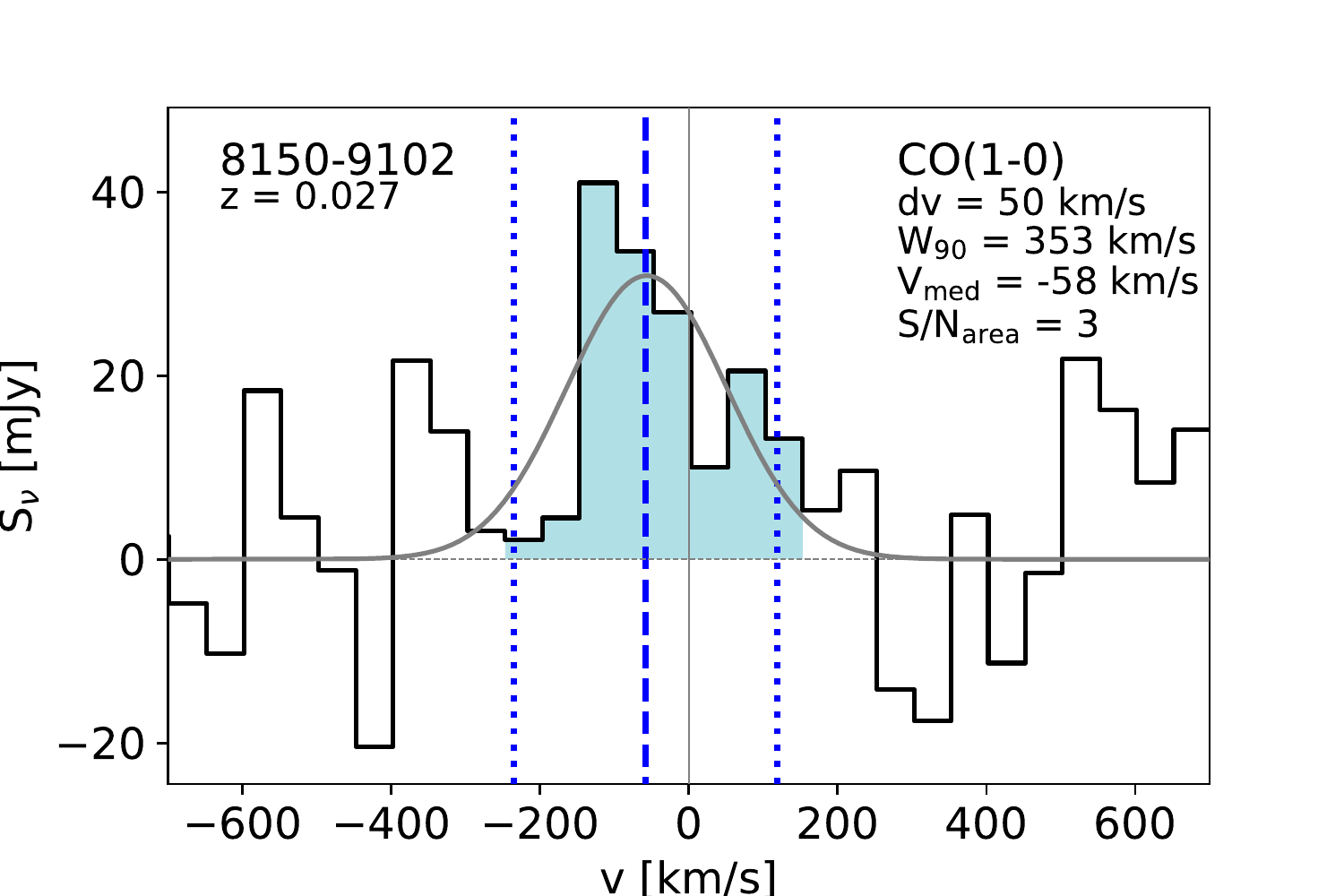}
 \caption{continued.}
\end{figure*}

\begin{figure*}  \centering  \includegraphics[width = 0.17\textwidth, trim = 0cm 0cm 0cm 0cm, clip = true]{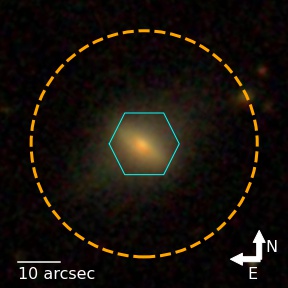} \includegraphics[width = 0.29\textwidth, trim = 0cm 0cm 0cm 0cm, clip = true]{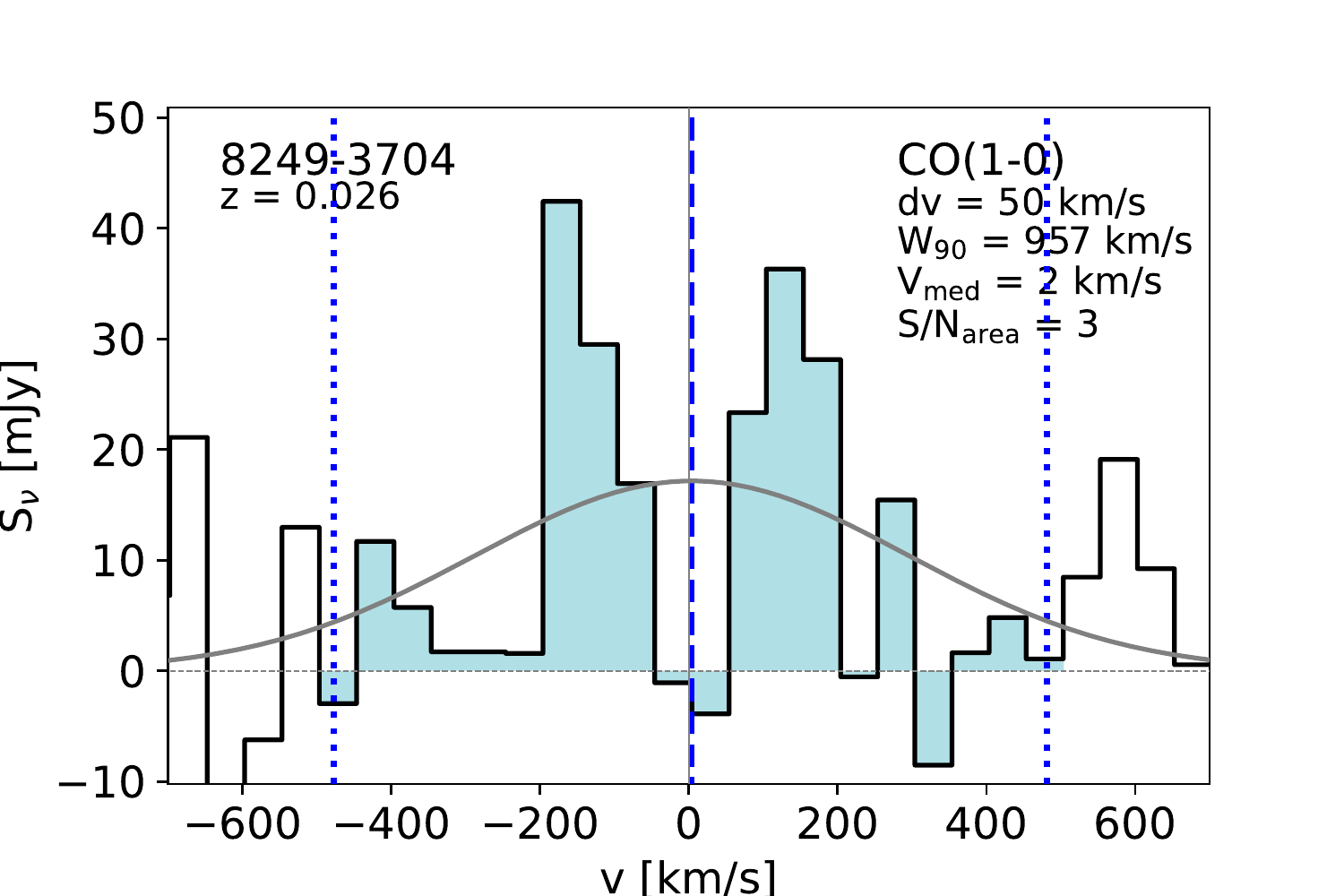}
\hspace{0.4cm}  \centering  \includegraphics[width = 0.17\textwidth, trim = 0cm 0cm 0cm 0cm, clip = true]{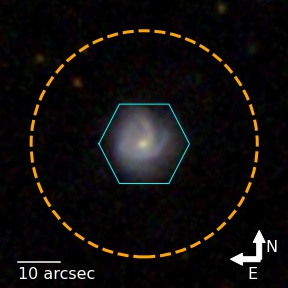} \includegraphics[width = 0.29\textwidth, trim = 0cm 0cm 0cm 0cm, clip = true]{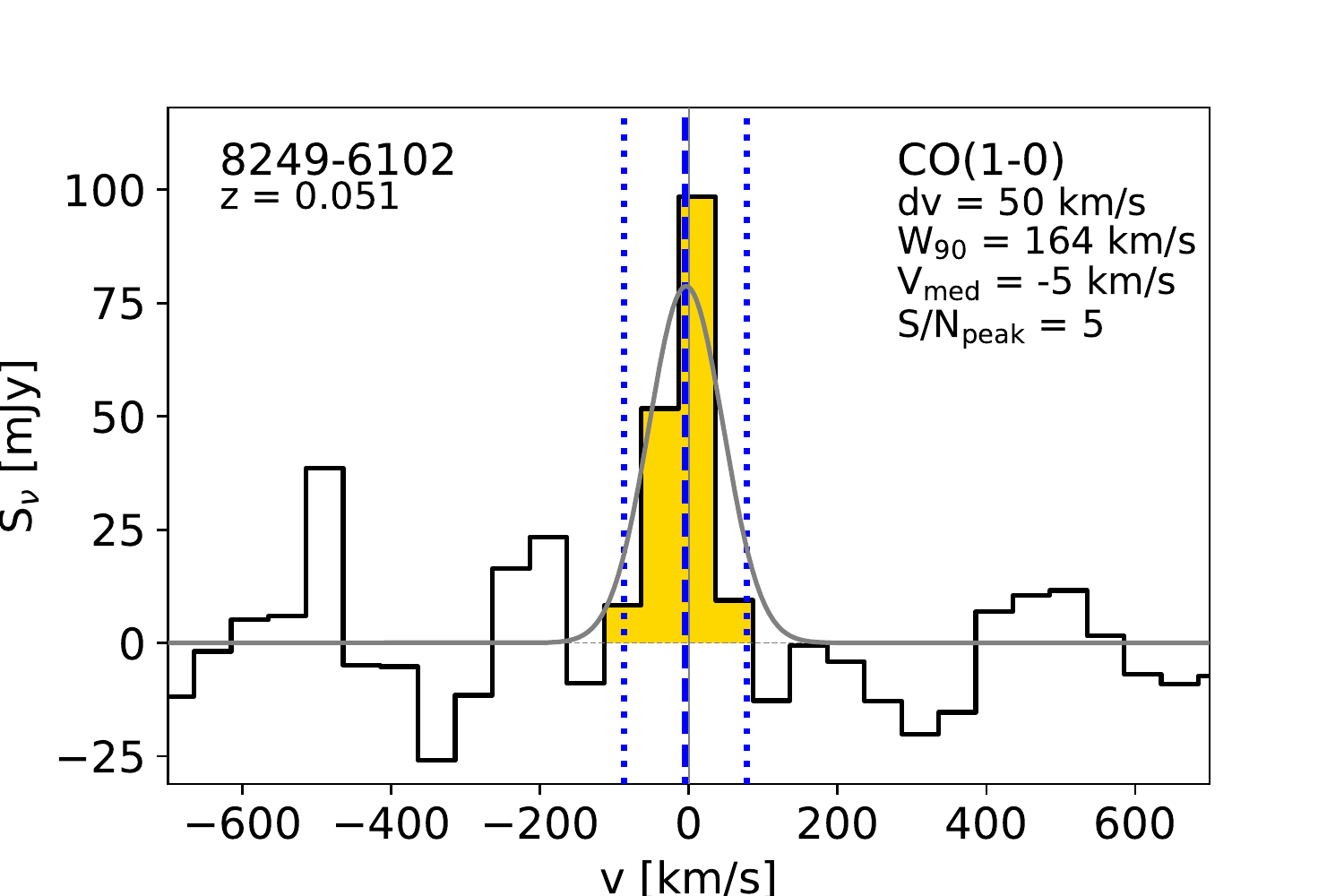}
\end{figure*}

\begin{figure*}  \centering  \includegraphics[width = 0.17\textwidth, trim = 0cm 0cm 0cm 0cm, clip = true]{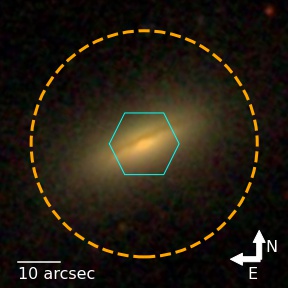} \includegraphics[width = 0.29\textwidth, trim = 0cm 0cm 0cm 0cm, clip = true]{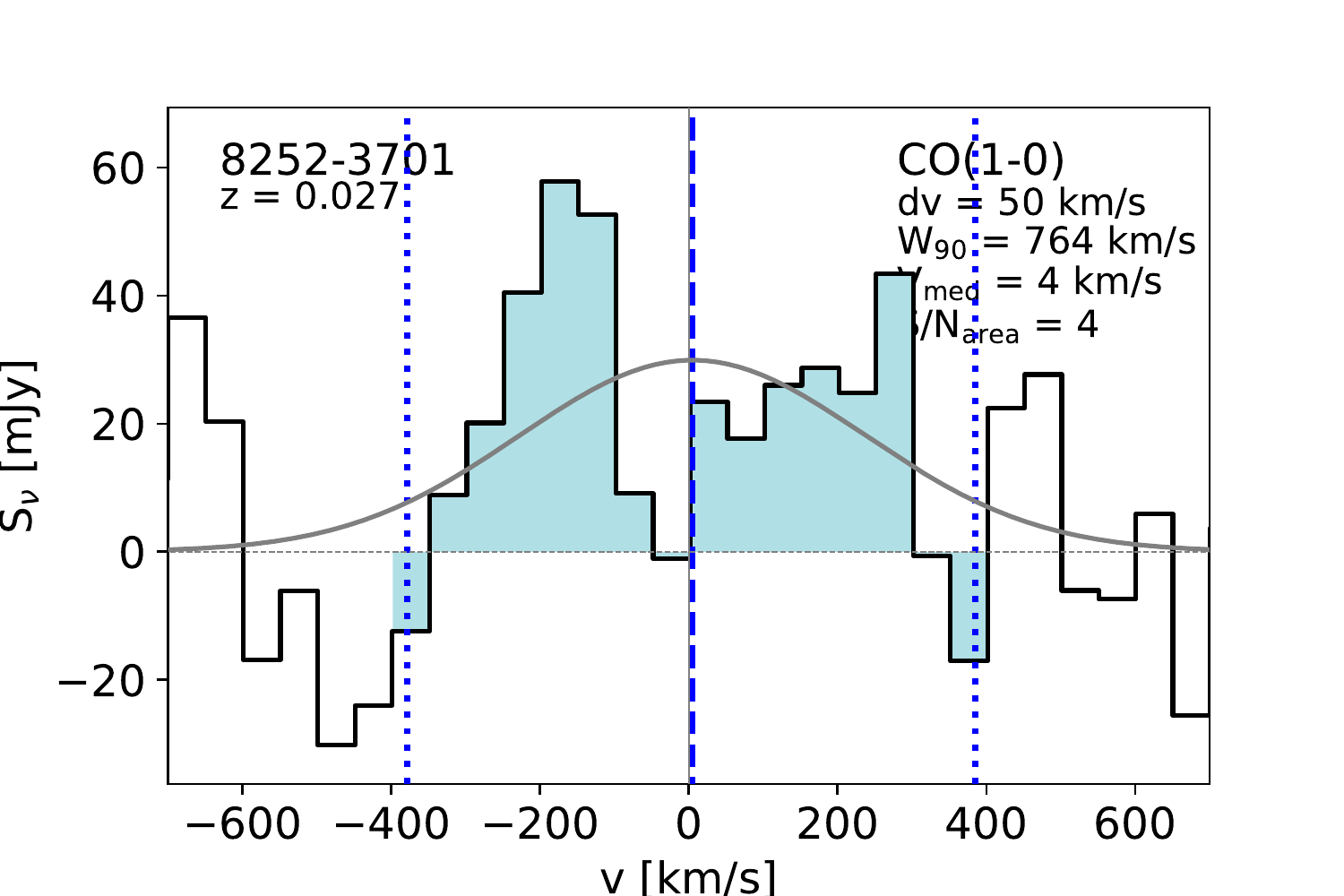}
\hspace{0.4cm}  \centering  \includegraphics[width = 0.17\textwidth, trim = 0cm 0cm 0cm 0cm, clip = true]{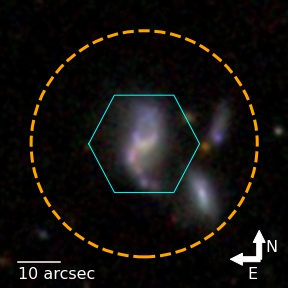} \includegraphics[width = 0.29\textwidth, trim = 0cm 0cm 0cm 0cm, clip = true]{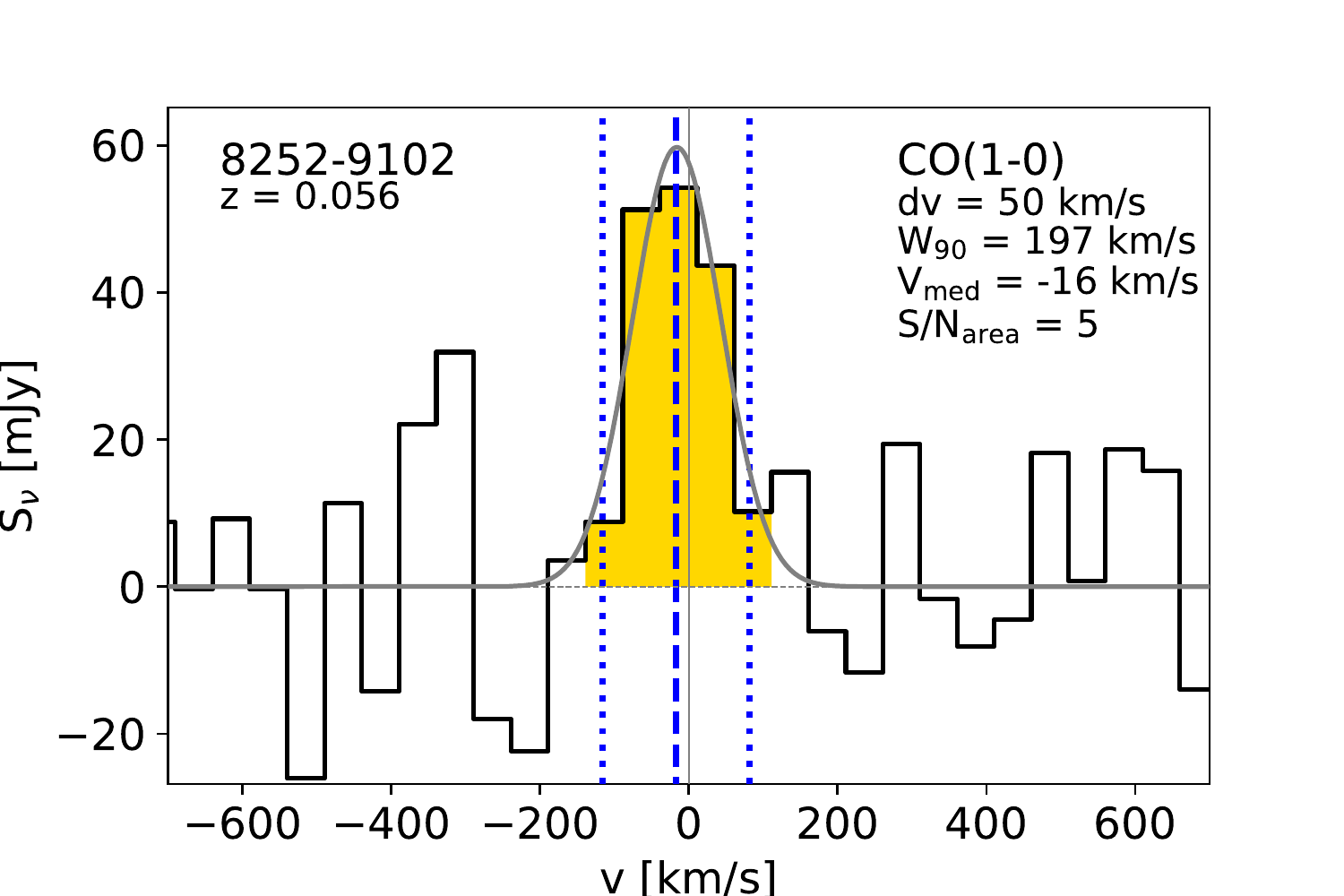}
\end{figure*}

\begin{figure*}  \centering  \includegraphics[width = 0.17\textwidth, trim = 0cm 0cm 0cm 0cm, clip = true]{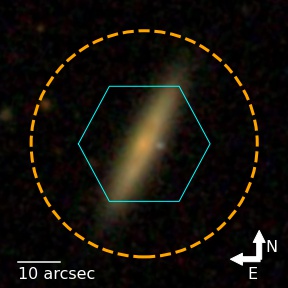} \includegraphics[width = 0.29\textwidth, trim = 0cm 0cm 0cm 0cm, clip = true]{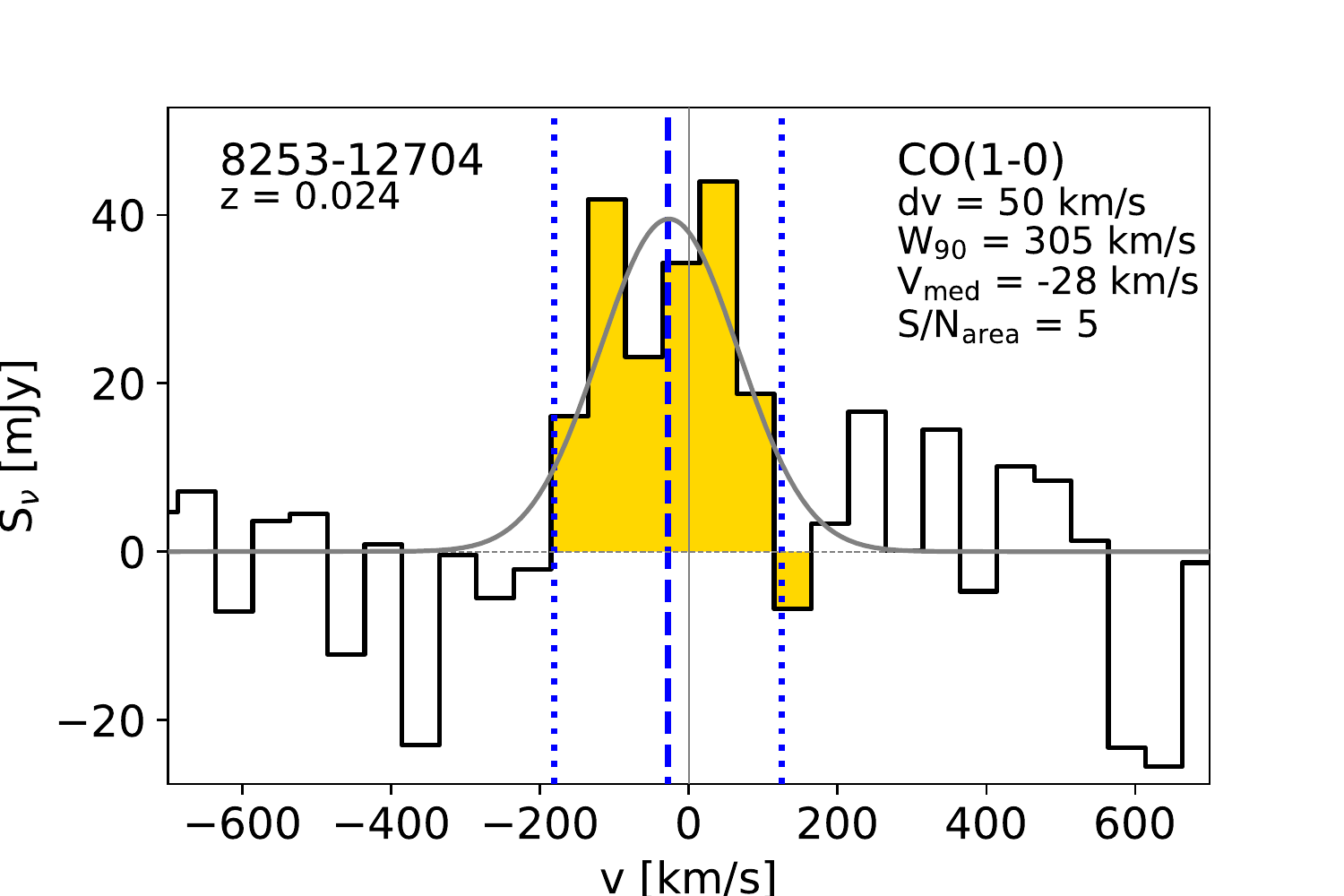}
\hspace{0.4cm}  \centering  \includegraphics[width = 0.17\textwidth, trim = 0cm 0cm 0cm 0cm, clip = true]{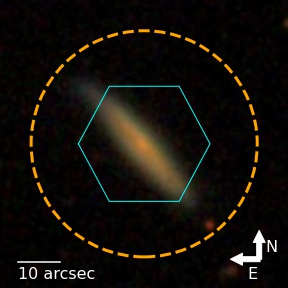} \includegraphics[width = 0.29\textwidth, trim = 0cm 0cm 0cm 0cm, clip = true]{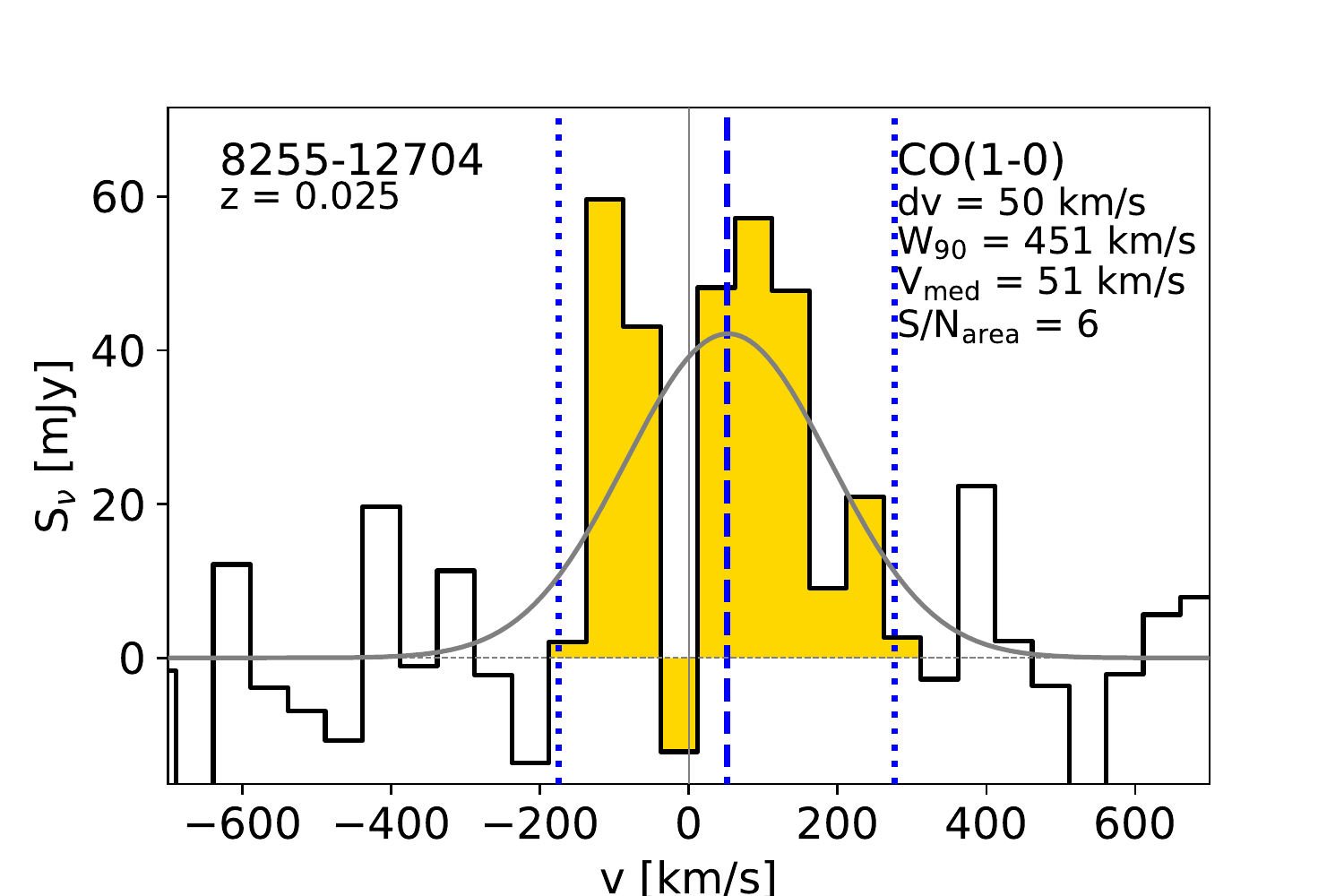}
\end{figure*}

\begin{figure*}  \centering  \includegraphics[width = 0.17\textwidth, trim = 0cm 0cm 0cm 0cm, clip = true]{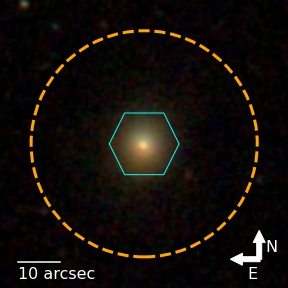} \includegraphics[width = 0.29\textwidth, trim = 0cm 0cm 0cm 0cm, clip = true]{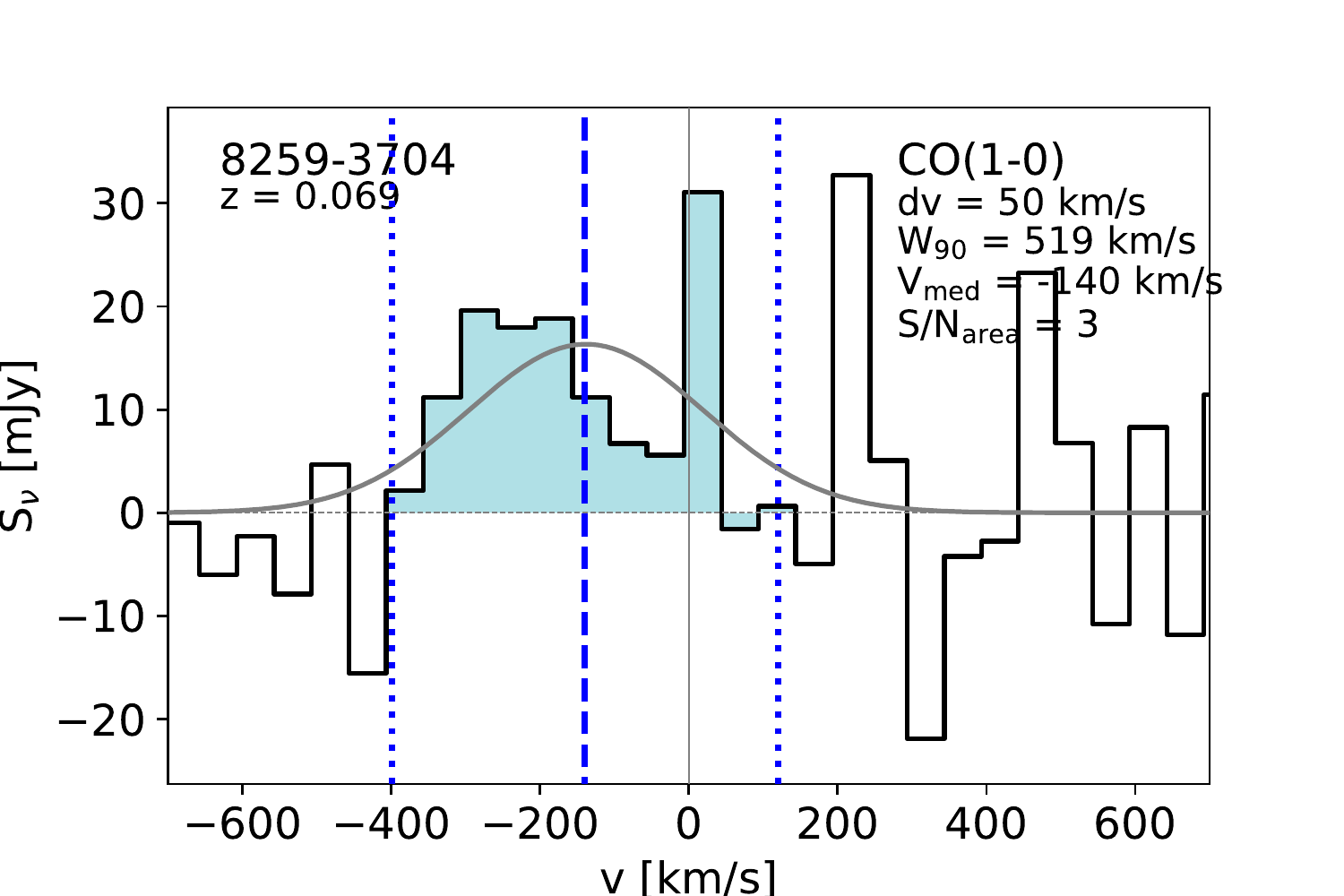}
\hspace{0.4cm}  \centering  \includegraphics[width = 0.17\textwidth, trim = 0cm 0cm 0cm 0cm, clip = true]{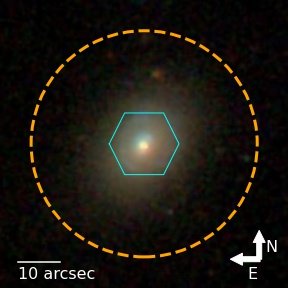} \includegraphics[width = 0.29\textwidth, trim = 0cm 0cm 0cm 0cm, clip = true]{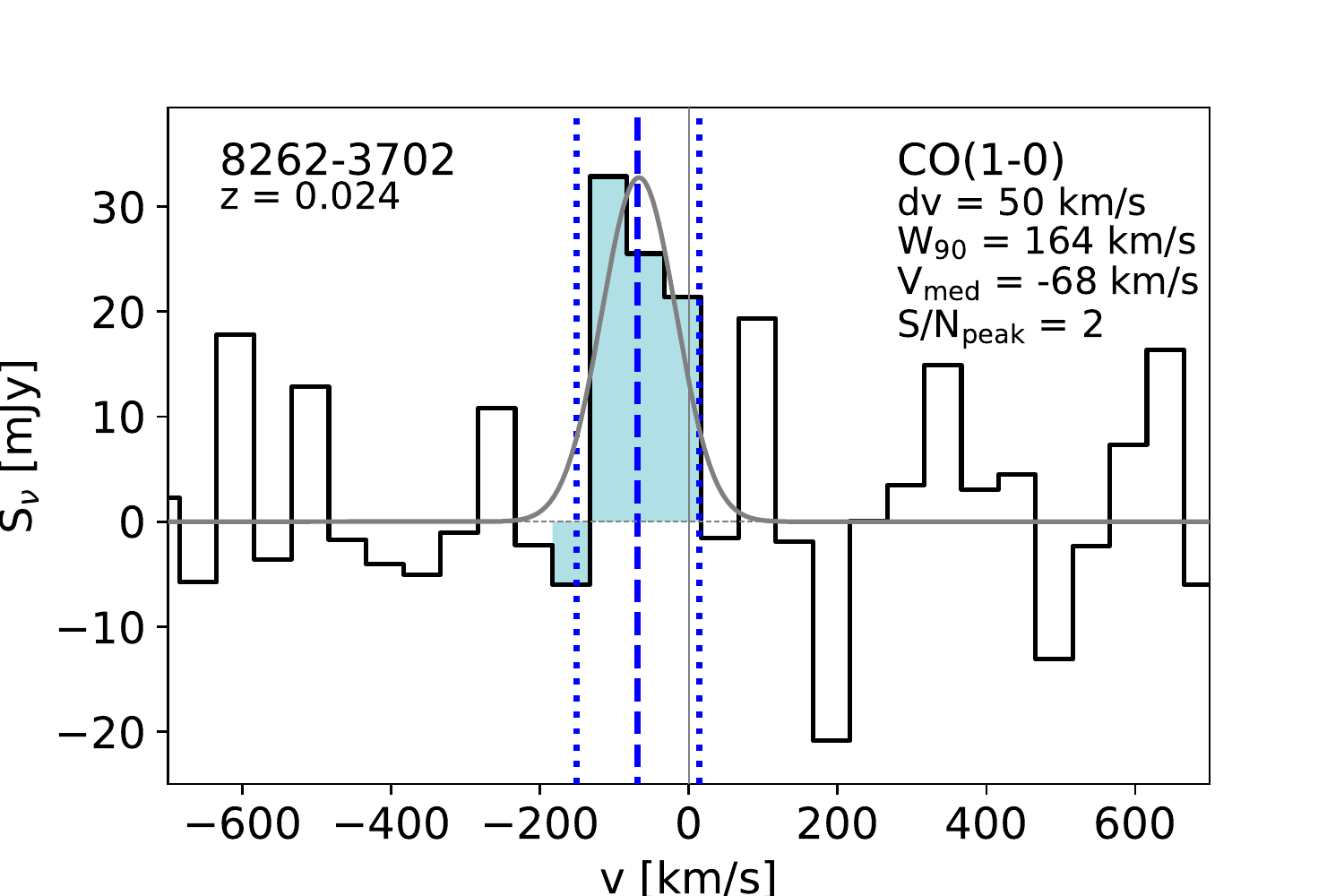}
\end{figure*}

\begin{figure*}  \centering  \includegraphics[width = 0.17\textwidth, trim = 0cm 0cm 0cm 0cm, clip = true]{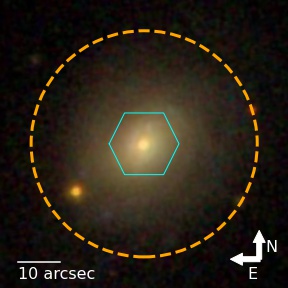} \includegraphics[width = 0.29\textwidth, trim = 0cm 0cm 0cm 0cm, clip = true]{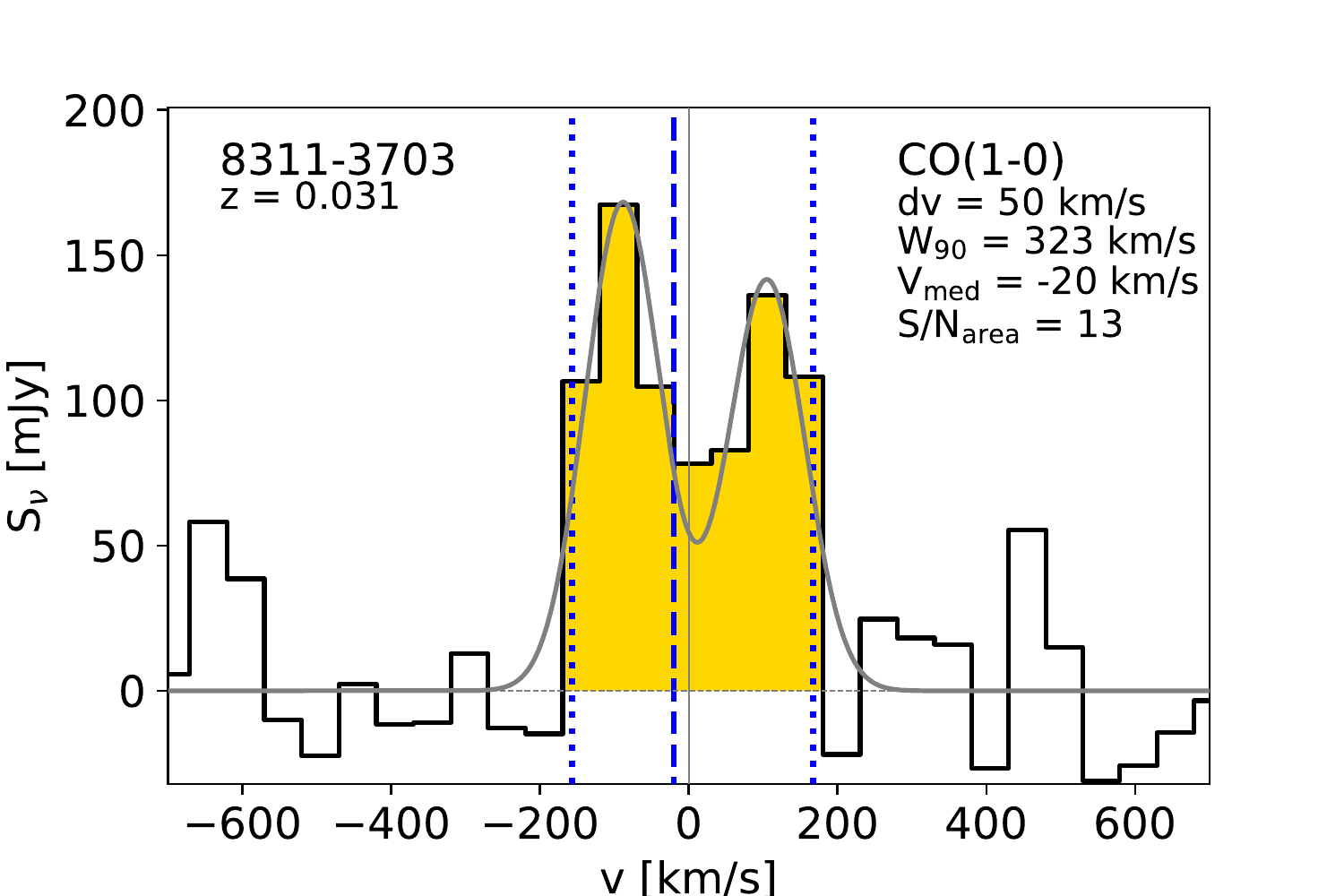}
\hspace{0.4cm}  \centering  \includegraphics[width = 0.17\textwidth, trim = 0cm 0cm 0cm 0cm, clip = true]{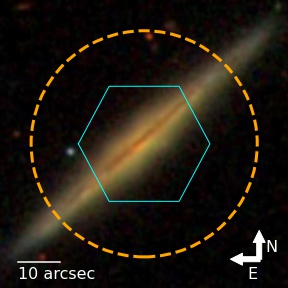} \includegraphics[width = 0.29\textwidth, trim = 0cm 0cm 0cm 0cm, clip = true]{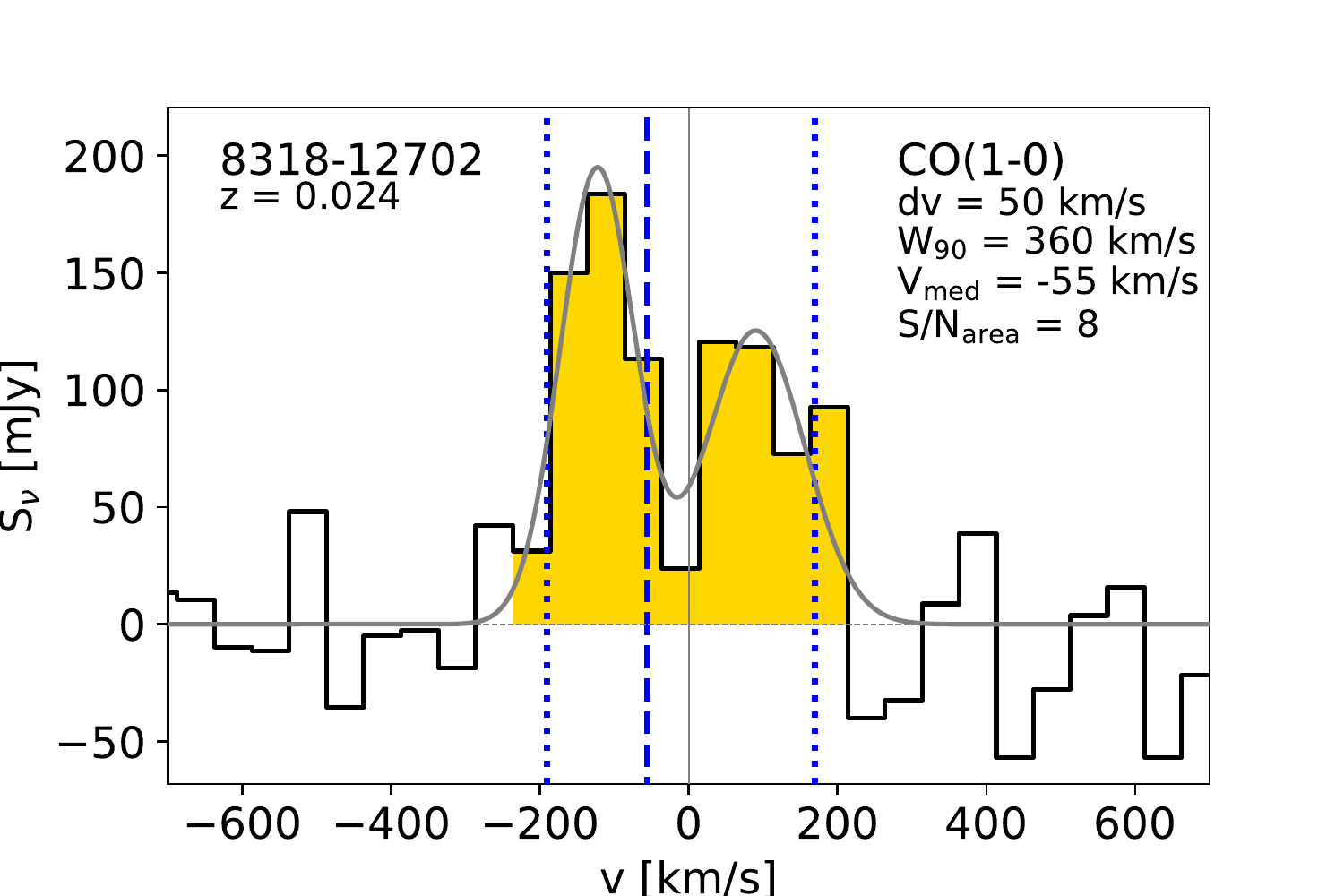}
\end{figure*}

\begin{figure*} 
   \ContinuedFloat
 \centering  \includegraphics[width = 0.17\textwidth, trim = 0cm 0cm 0cm 0cm, clip = true]{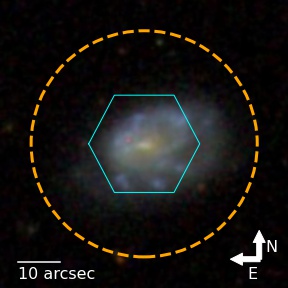} \includegraphics[width = 0.29\textwidth, trim = 0cm 0cm 0cm 0cm, clip = true]{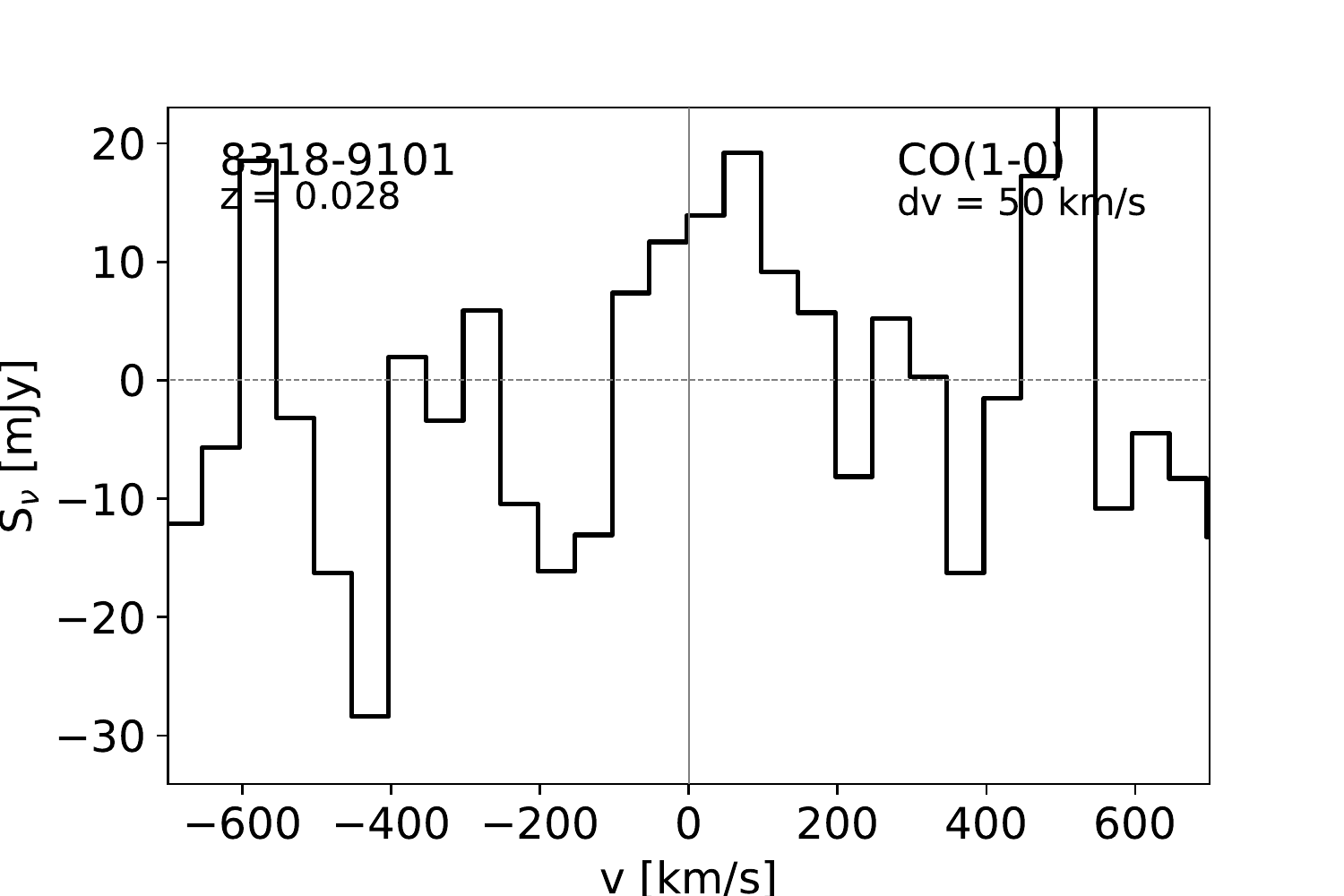}
\hspace{0.4cm}  \centering  \includegraphics[width = 0.17\textwidth, trim = 0cm 0cm 0cm 0cm, clip = true]{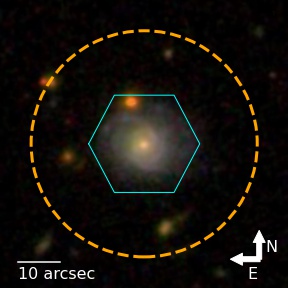} \includegraphics[width = 0.29\textwidth, trim = 0cm 0cm 0cm 0cm, clip = true]{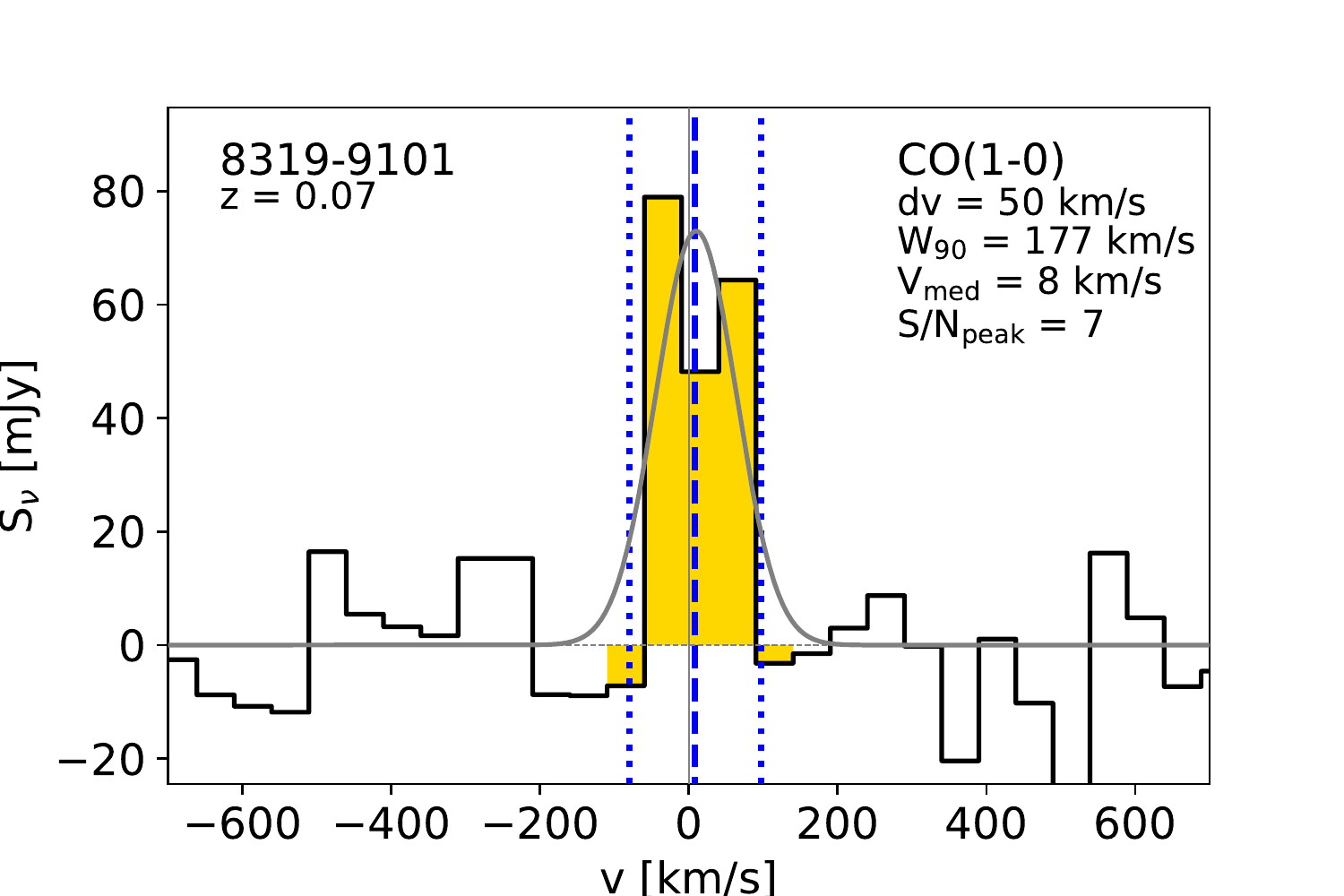}
 \caption{continued.}
\end{figure*}

\begin{figure*}  \centering  \includegraphics[width = 0.17\textwidth, trim = 0cm 0cm 0cm 0cm, clip = true]{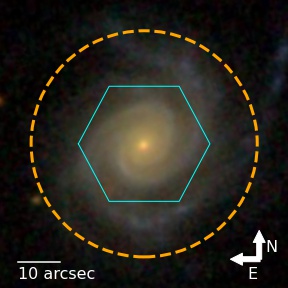} \includegraphics[width = 0.29\textwidth, trim = 0cm 0cm 0cm 0cm, clip = true]{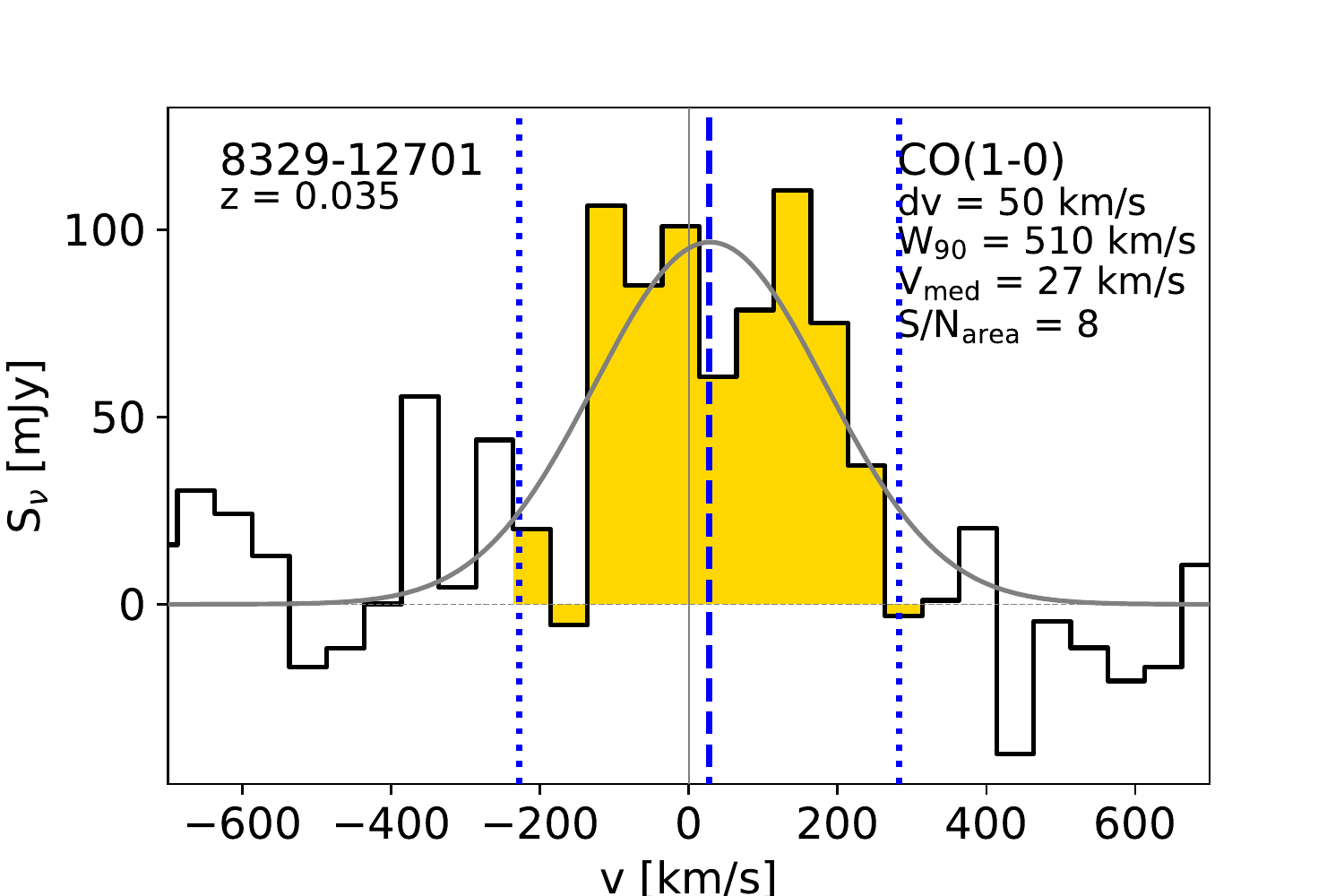}
\hspace{0.4cm}  \centering  \includegraphics[width = 0.17\textwidth, trim = 0cm 0cm 0cm 0cm, clip = true]{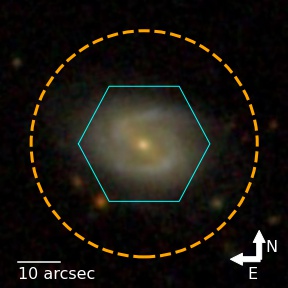} \includegraphics[width = 0.29\textwidth, trim = 0cm 0cm 0cm 0cm, clip = true]{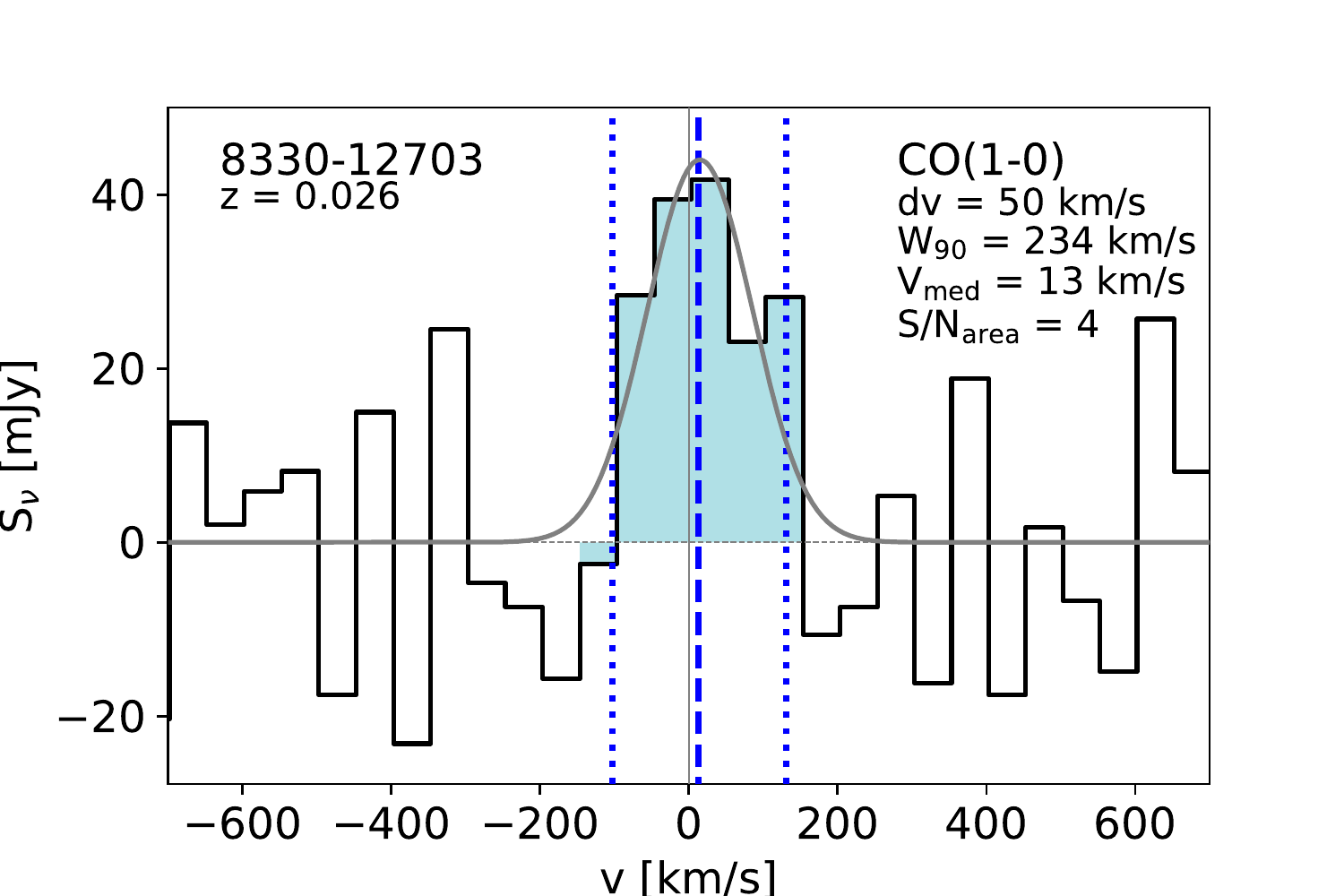}
\end{figure*}

\begin{figure*}  \centering  \includegraphics[width = 0.17\textwidth, trim = 0cm 0cm 0cm 0cm, clip = true]{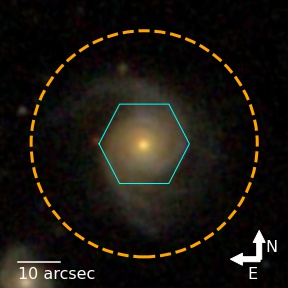} \includegraphics[width = 0.29\textwidth, trim = 0cm 0cm 0cm 0cm, clip = true]{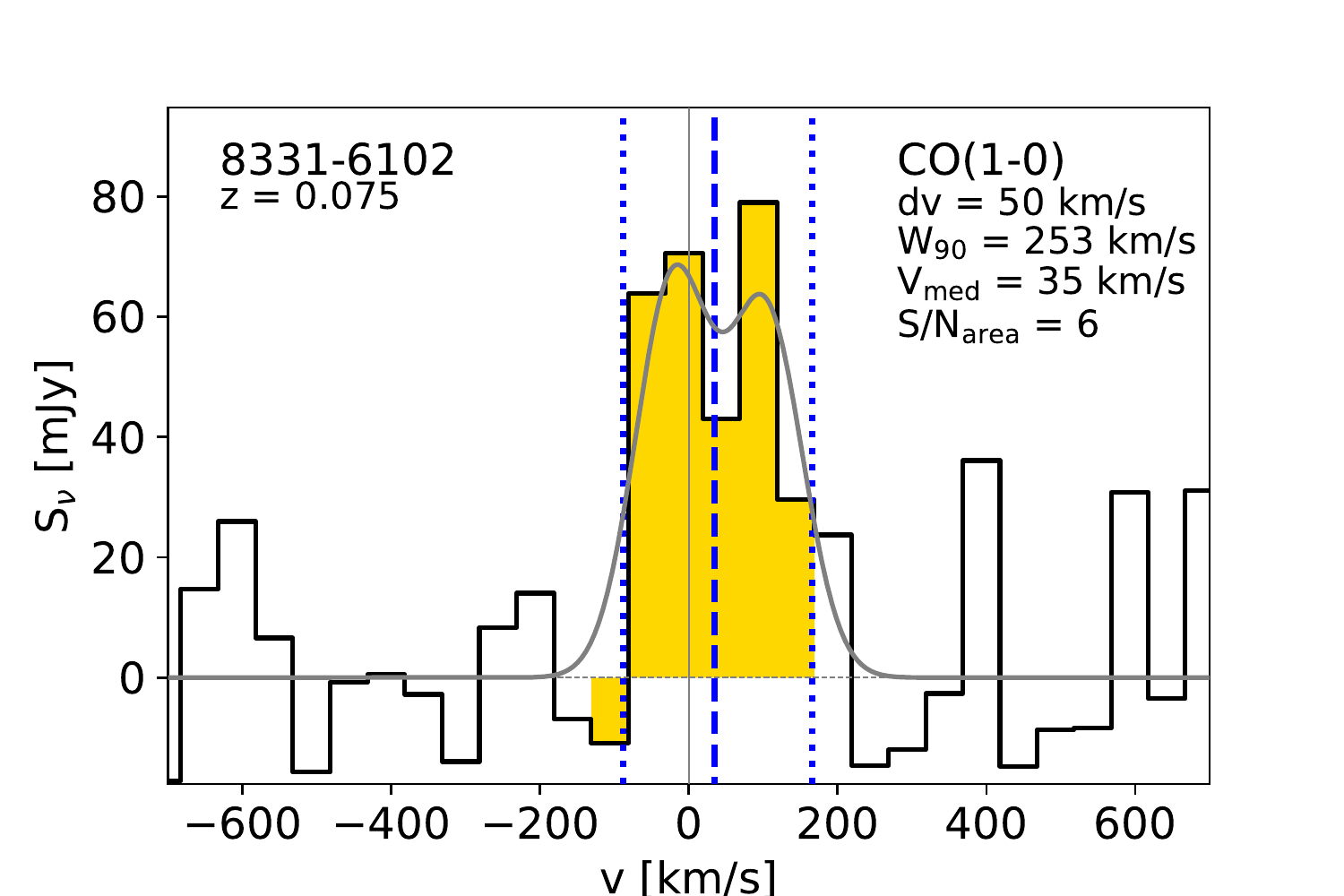}
\hspace{0.4cm}  \centering  \includegraphics[width = 0.17\textwidth, trim = 0cm 0cm 0cm 0cm, clip = true]{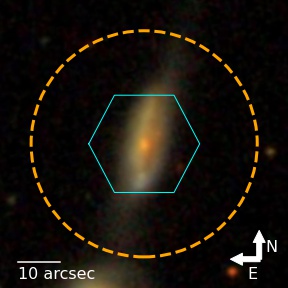} \includegraphics[width = 0.29\textwidth, trim = 0cm 0cm 0cm 0cm, clip = true]{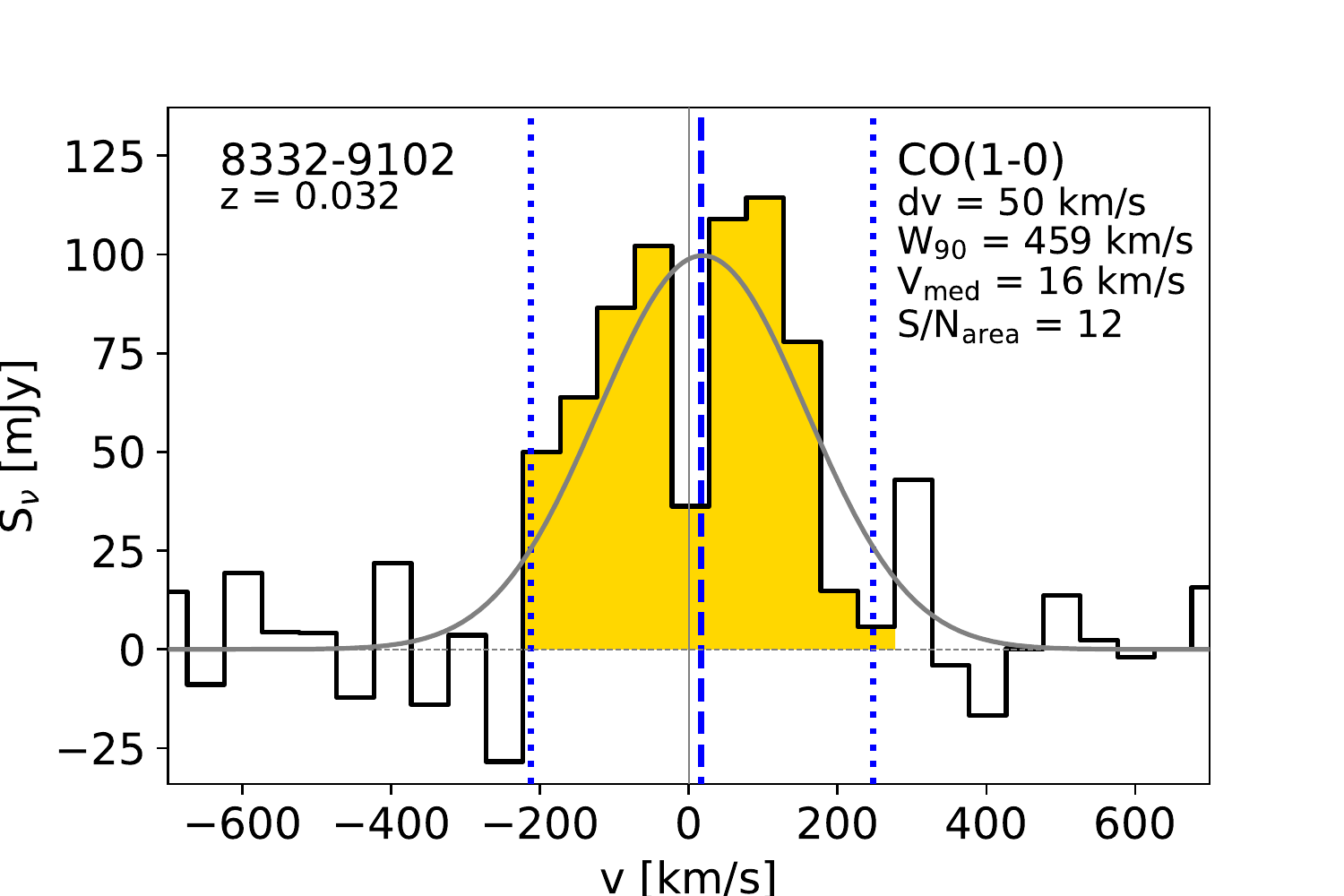}
\end{figure*}

\begin{figure*}  \centering  \includegraphics[width = 0.17\textwidth, trim = 0cm 0cm 0cm 0cm, clip = true]{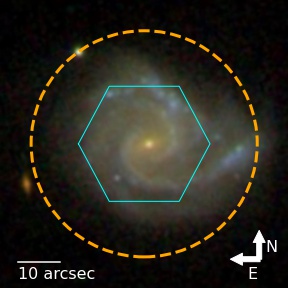} \includegraphics[width = 0.29\textwidth, trim = 0cm 0cm 0cm 0cm, clip = true]{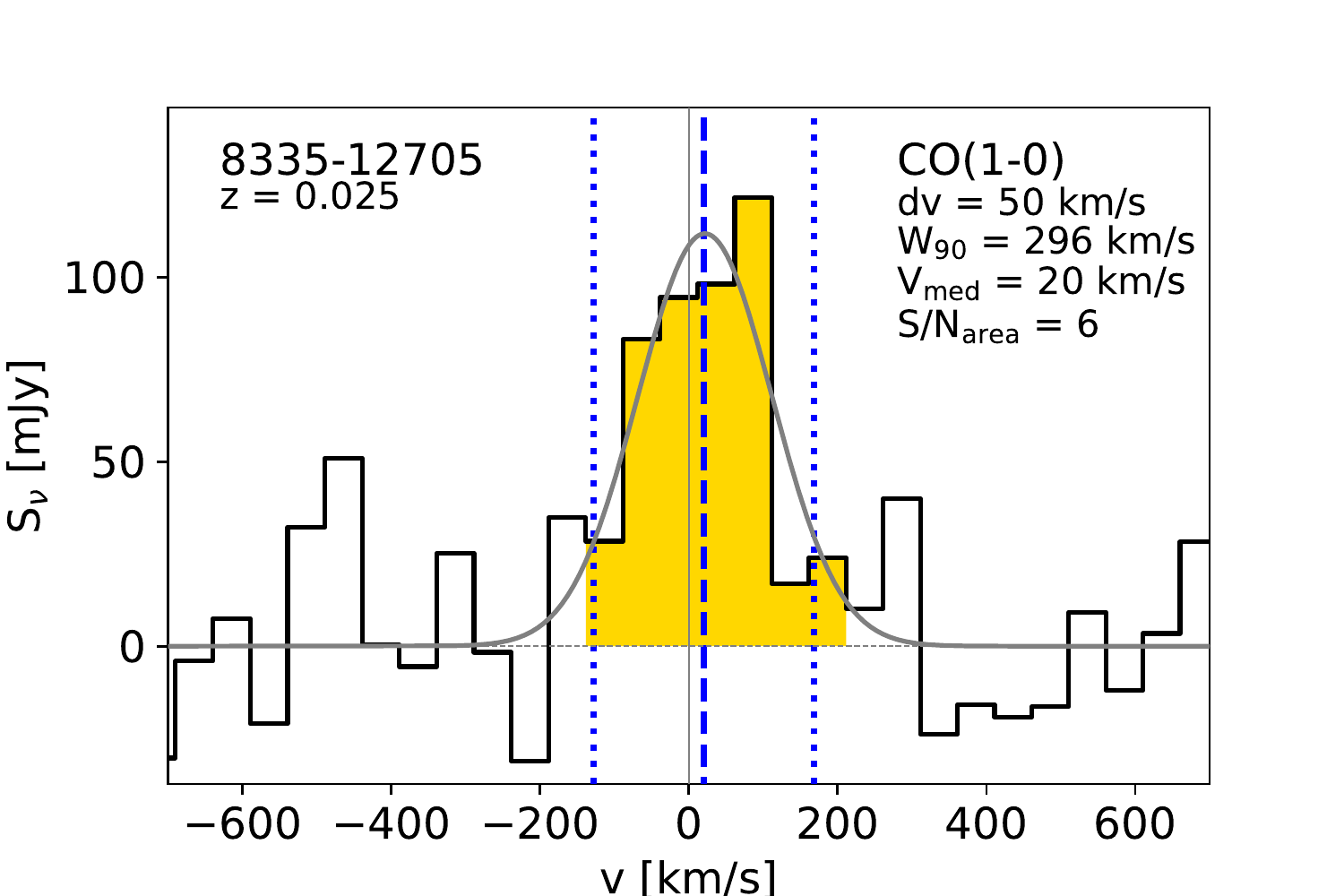}
\hspace{0.4cm}  \centering  \includegraphics[width = 0.17\textwidth, trim = 0cm 0cm 0cm 0cm, clip = true]{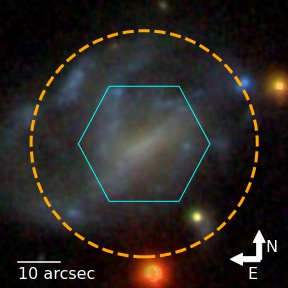} \includegraphics[width = 0.29\textwidth, trim = 0cm 0cm 0cm 0cm, clip = true]{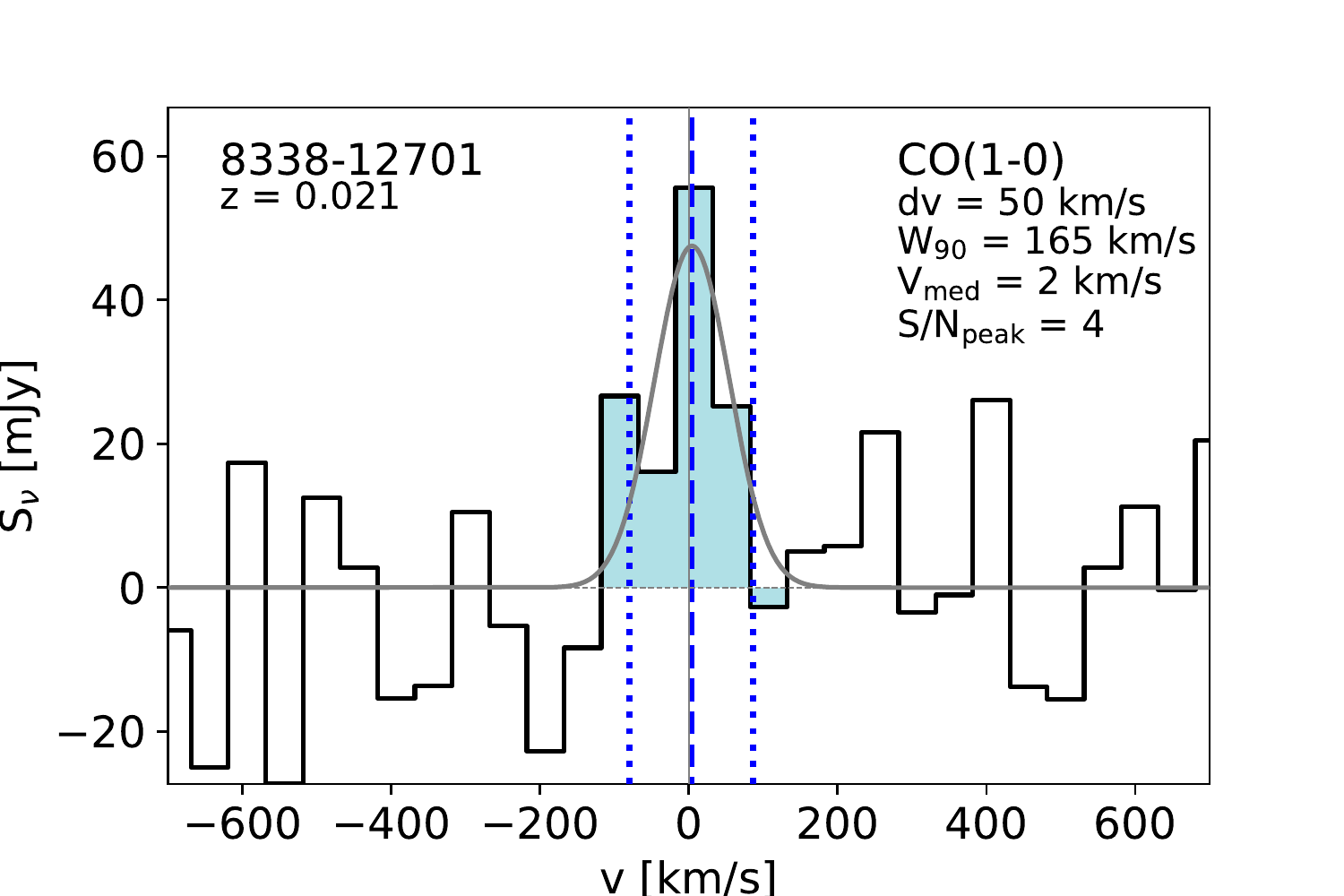}
\end{figure*}

\begin{figure*}  \centering  \includegraphics[width = 0.17\textwidth, trim = 0cm 0cm 0cm 0cm, clip = true]{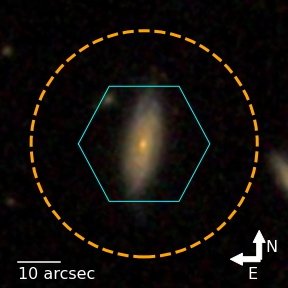} \includegraphics[width = 0.29\textwidth, trim = 0cm 0cm 0cm 0cm, clip = true]{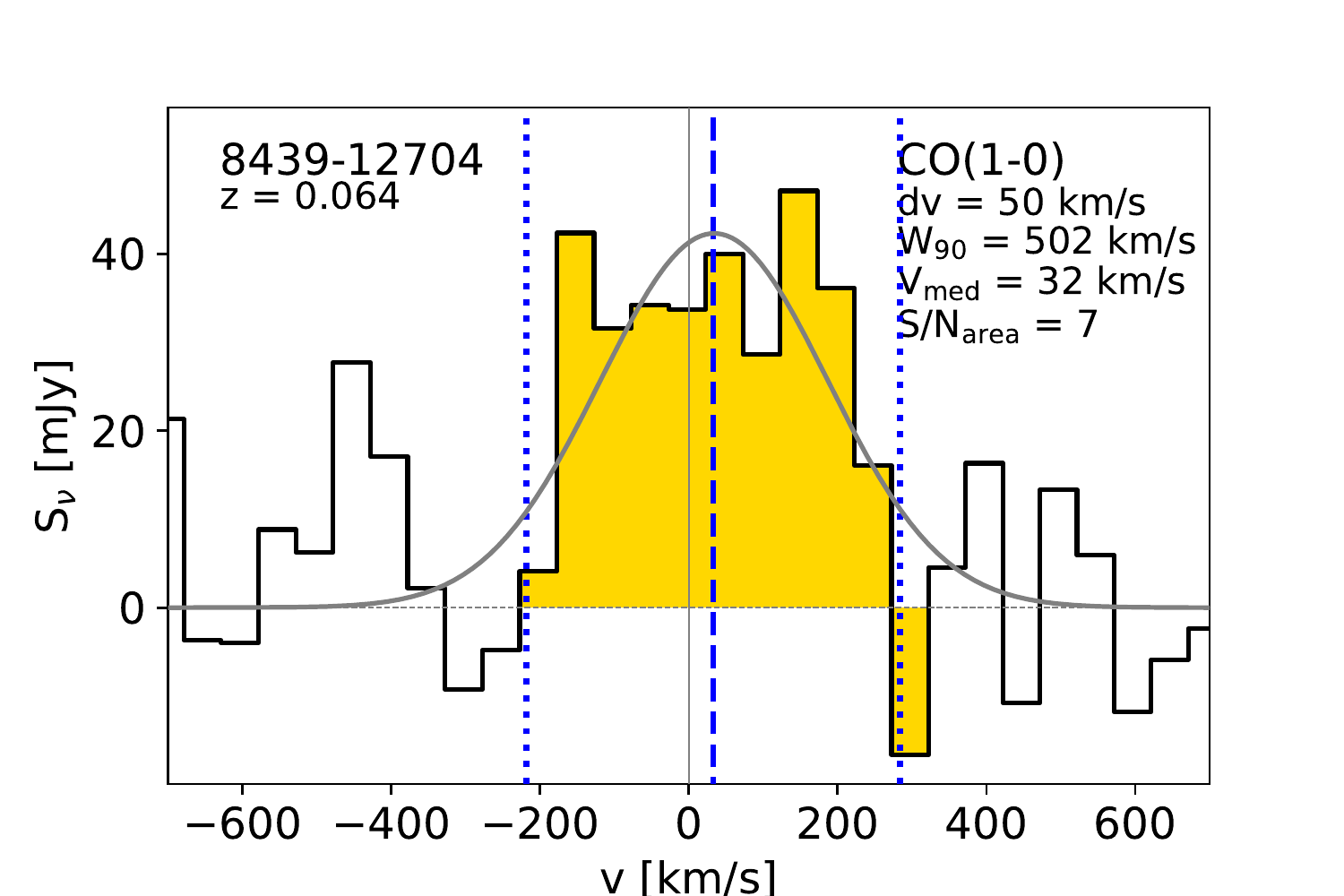}
\hspace{0.4cm}  \centering  \includegraphics[width = 0.17\textwidth, trim = 0cm 0cm 0cm 0cm, clip = true]{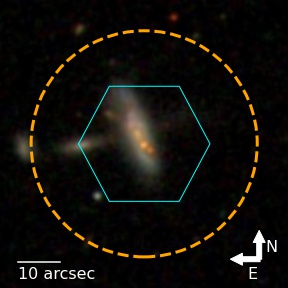} \includegraphics[width = 0.29\textwidth, trim = 0cm 0cm 0cm 0cm, clip = true]{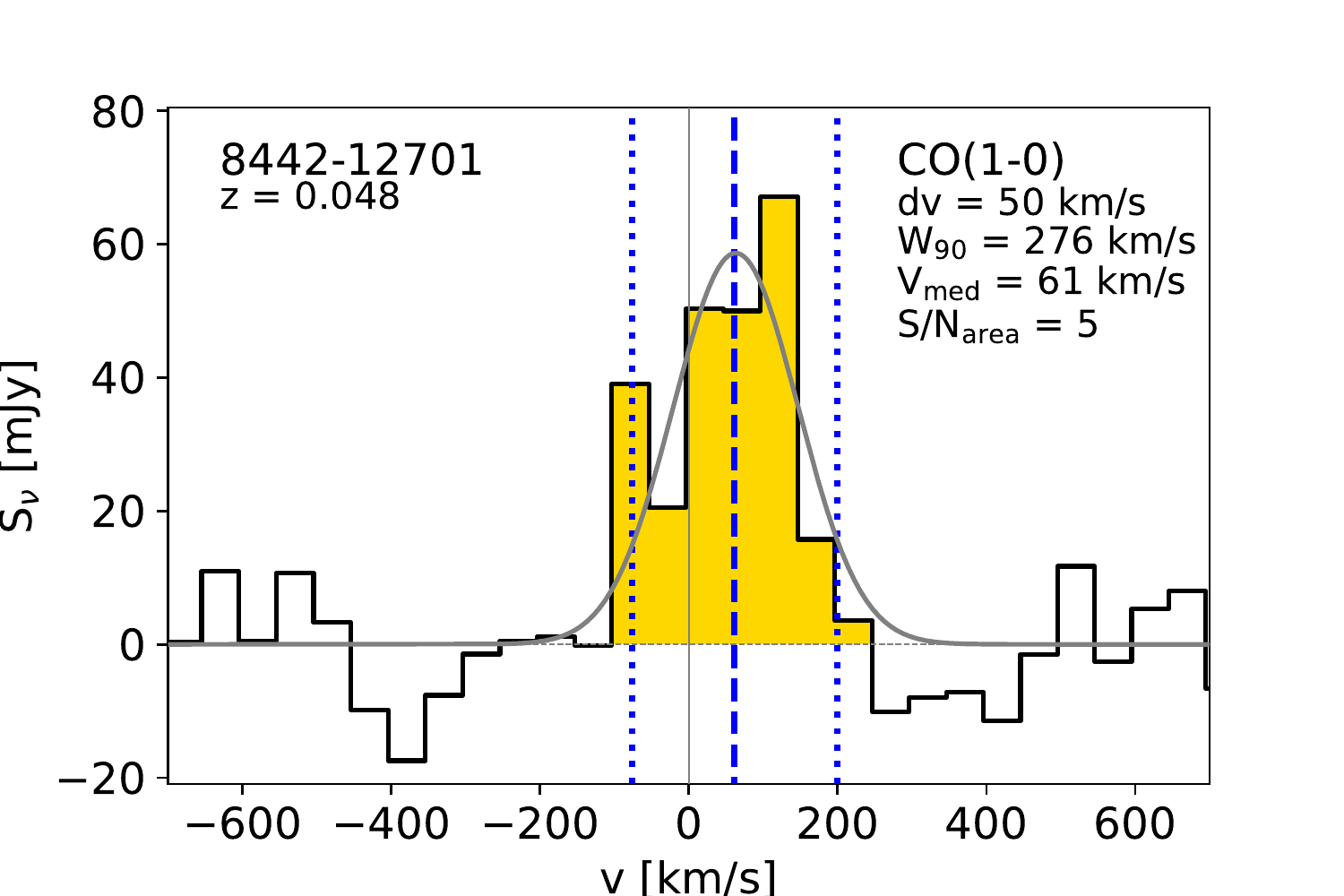}
\end{figure*}

\begin{figure*}  \centering  \includegraphics[width = 0.17\textwidth, trim = 0cm 0cm 0cm 0cm, clip = true]{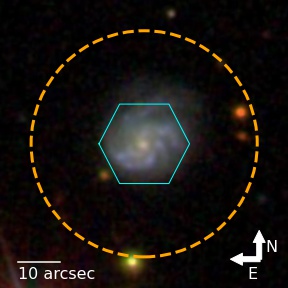} \includegraphics[width = 0.29\textwidth, trim = 0cm 0cm 0cm 0cm, clip = true]{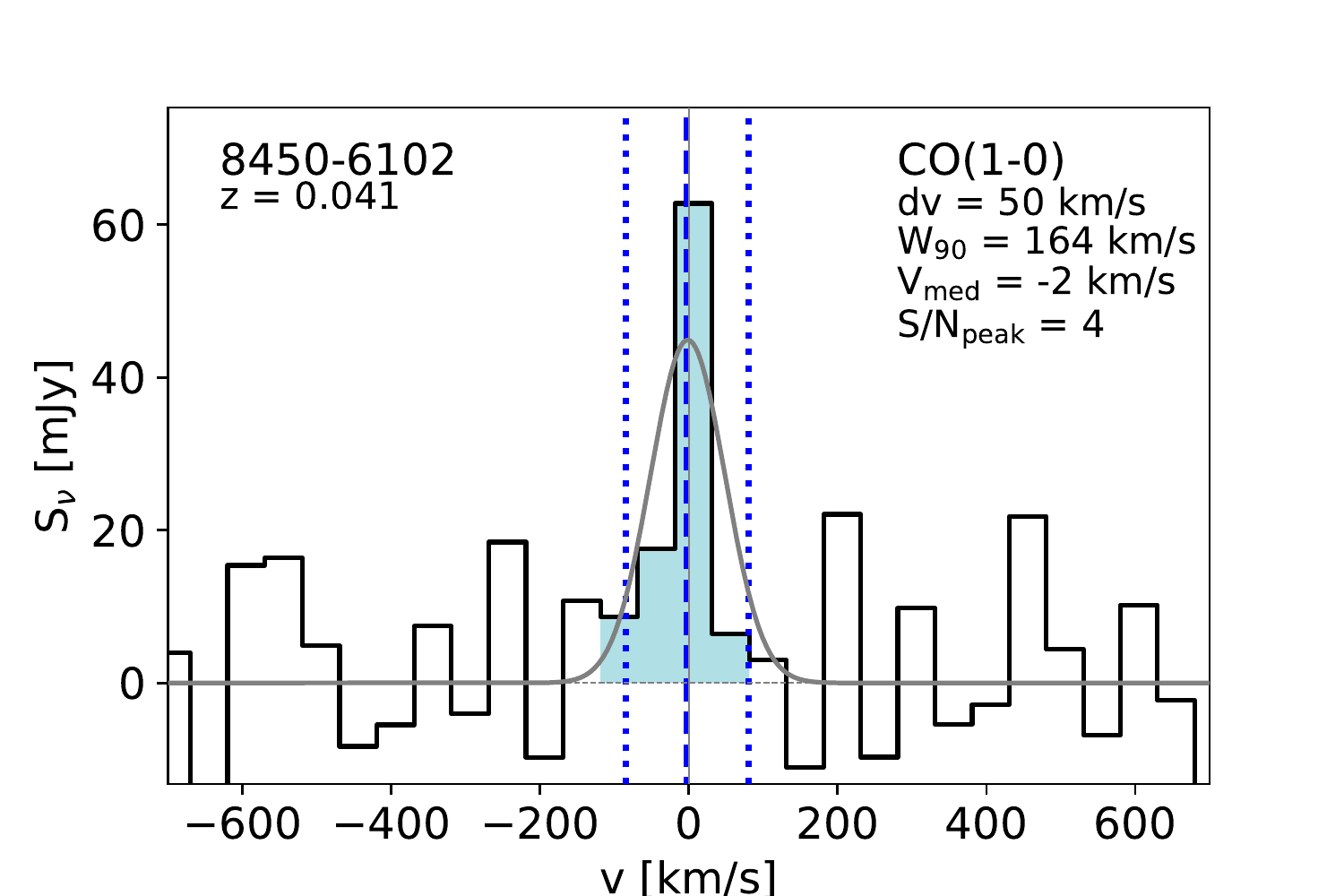}
\hspace{0.4cm}  \centering  \includegraphics[width = 0.17\textwidth, trim = 0cm 0cm 0cm 0cm, clip = true]{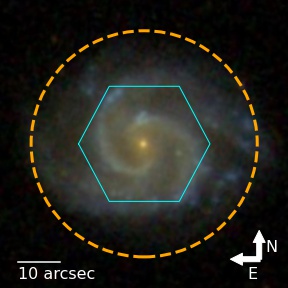} \includegraphics[width = 0.29\textwidth, trim = 0cm 0cm 0cm 0cm, clip = true]{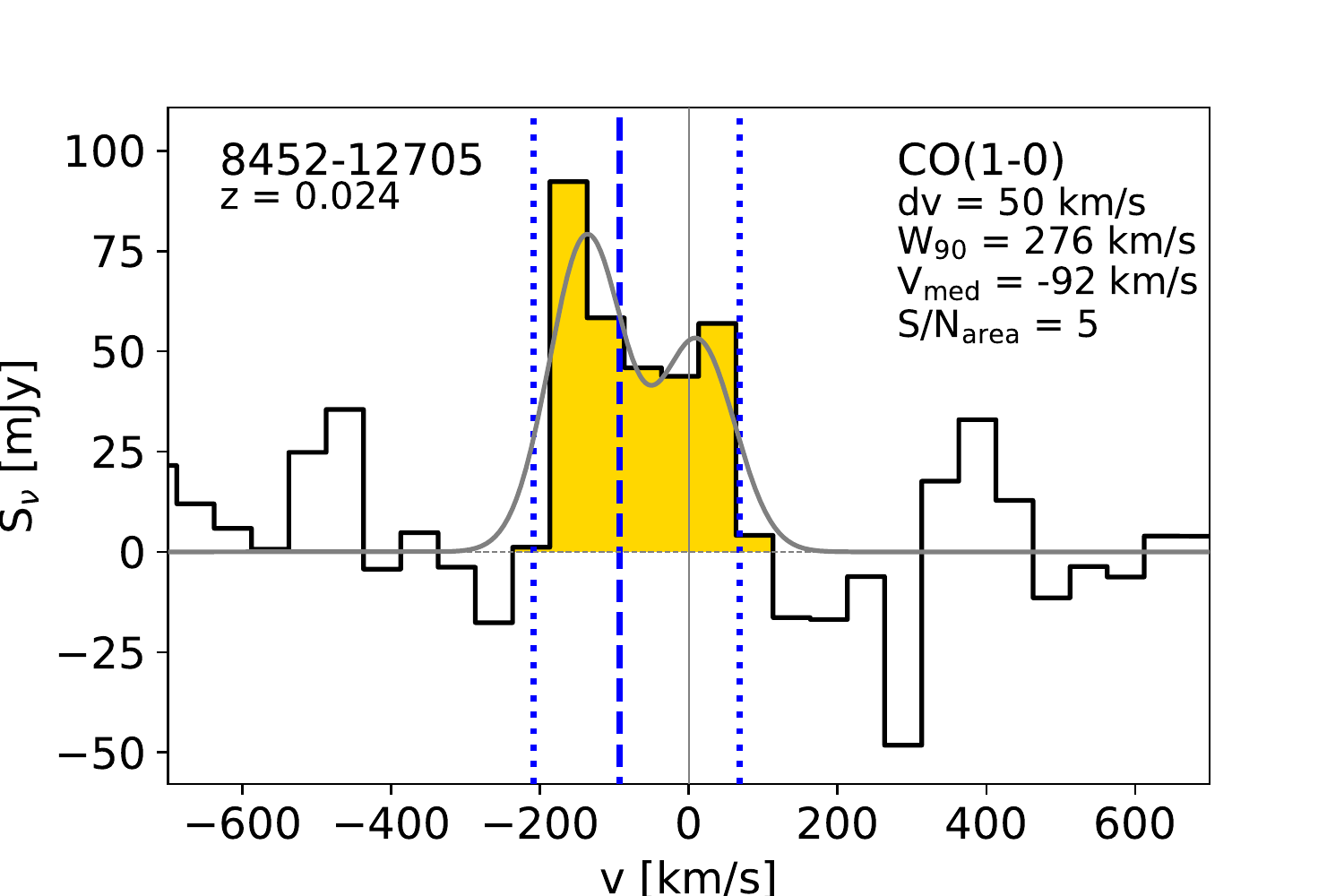}
\end{figure*}

\begin{figure*}  
   \ContinuedFloat
\centering  \includegraphics[width = 0.17\textwidth, trim = 0cm 0cm 0cm 0cm, clip = true]{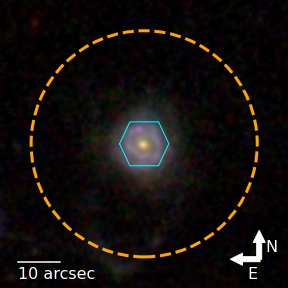} \includegraphics[width = 0.29\textwidth, trim = 0cm 0cm 0cm 0cm, clip = true]{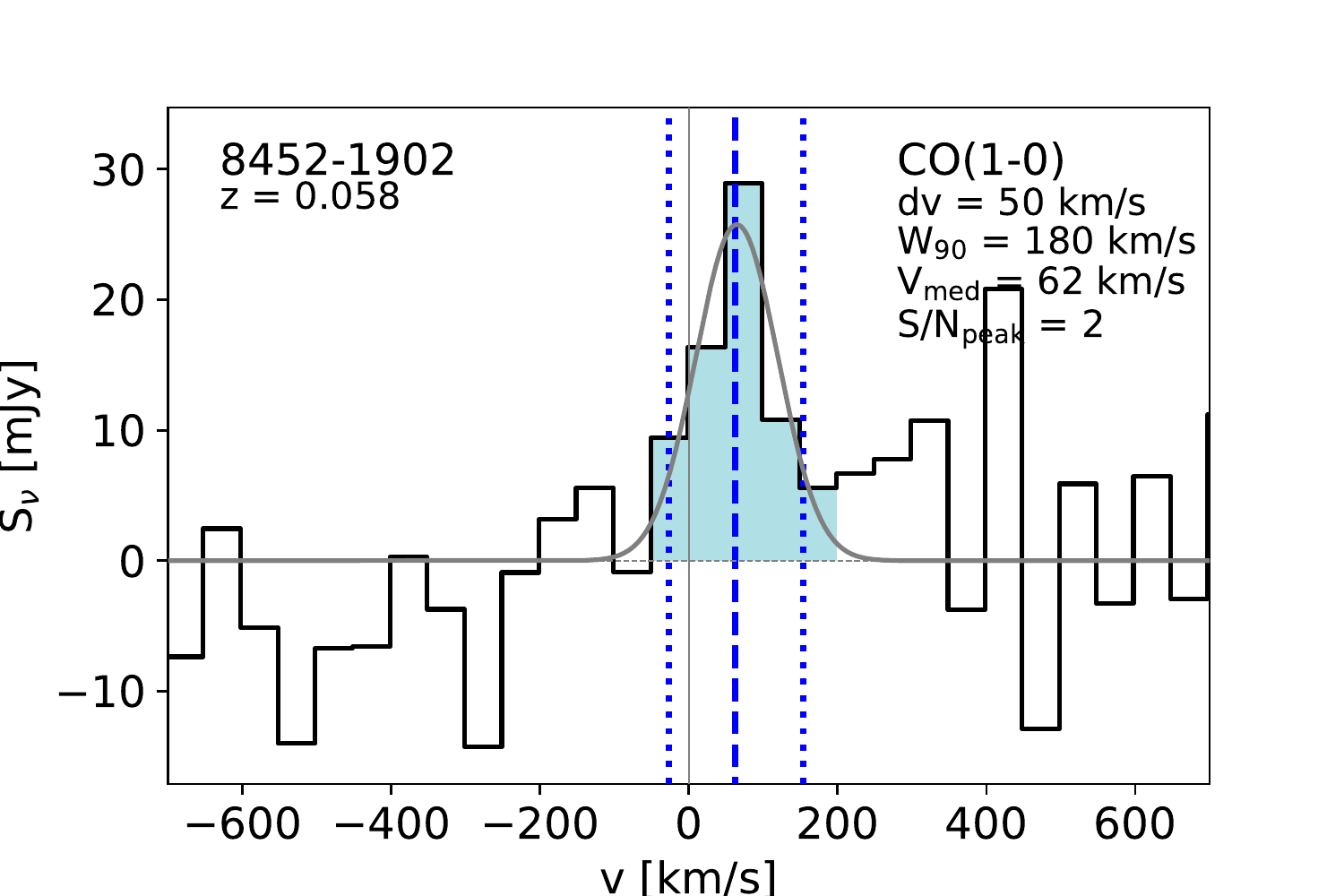}
\hspace{0.4cm}  \centering  \includegraphics[width = 0.17\textwidth, trim = 0cm 0cm 0cm 0cm, clip = true]{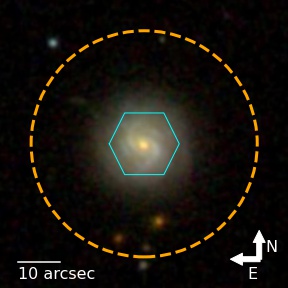} \includegraphics[width = 0.29\textwidth, trim = 0cm 0cm 0cm 0cm, clip = true]{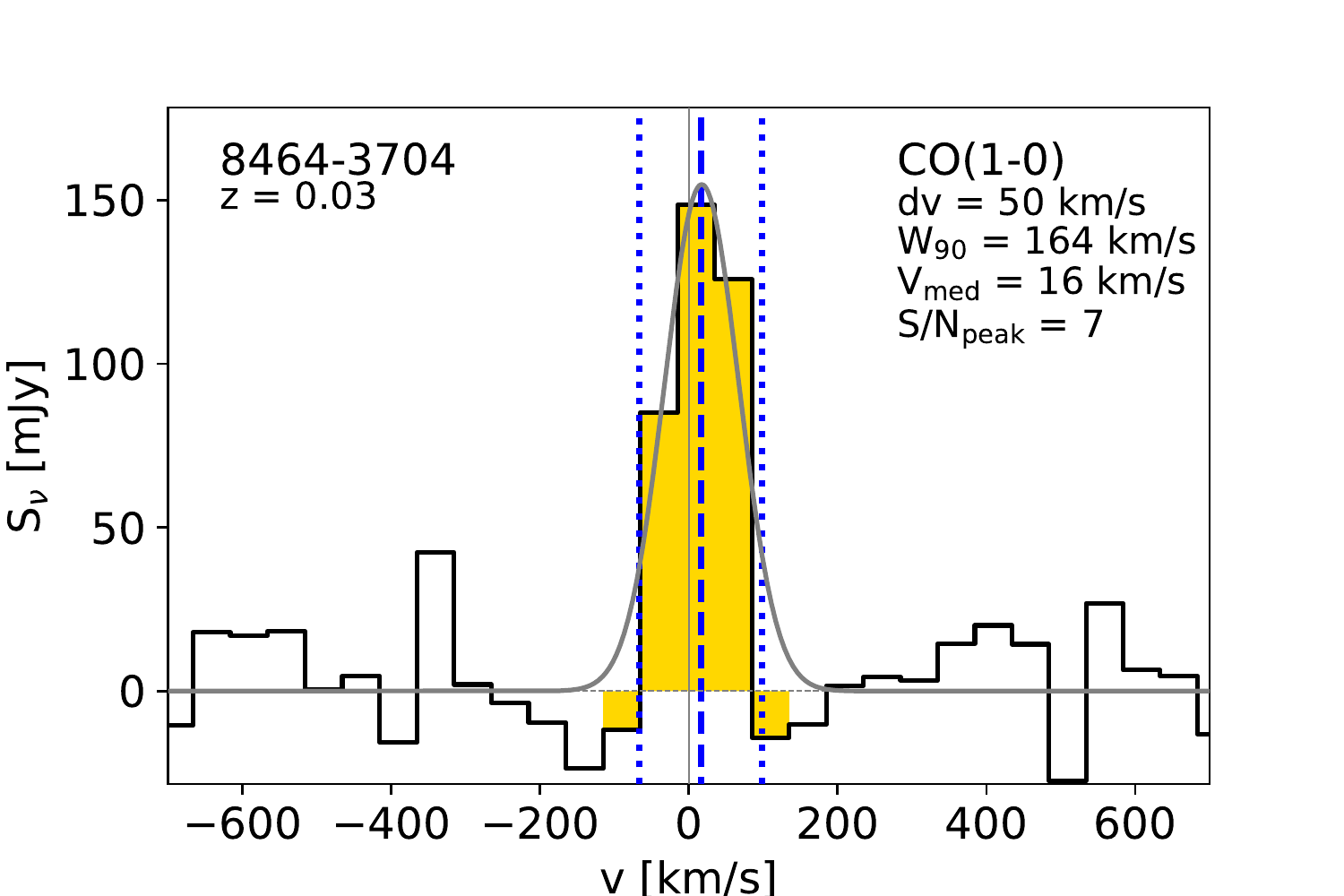}
 \caption{continued.}
\end{figure*}

\begin{figure*}  \centering  \includegraphics[width = 0.17\textwidth, trim = 0cm 0cm 0cm 0cm, clip = true]{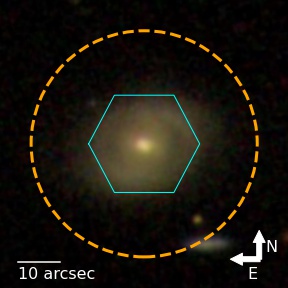} \includegraphics[width = 0.29\textwidth, trim = 0cm 0cm 0cm 0cm, clip = true]{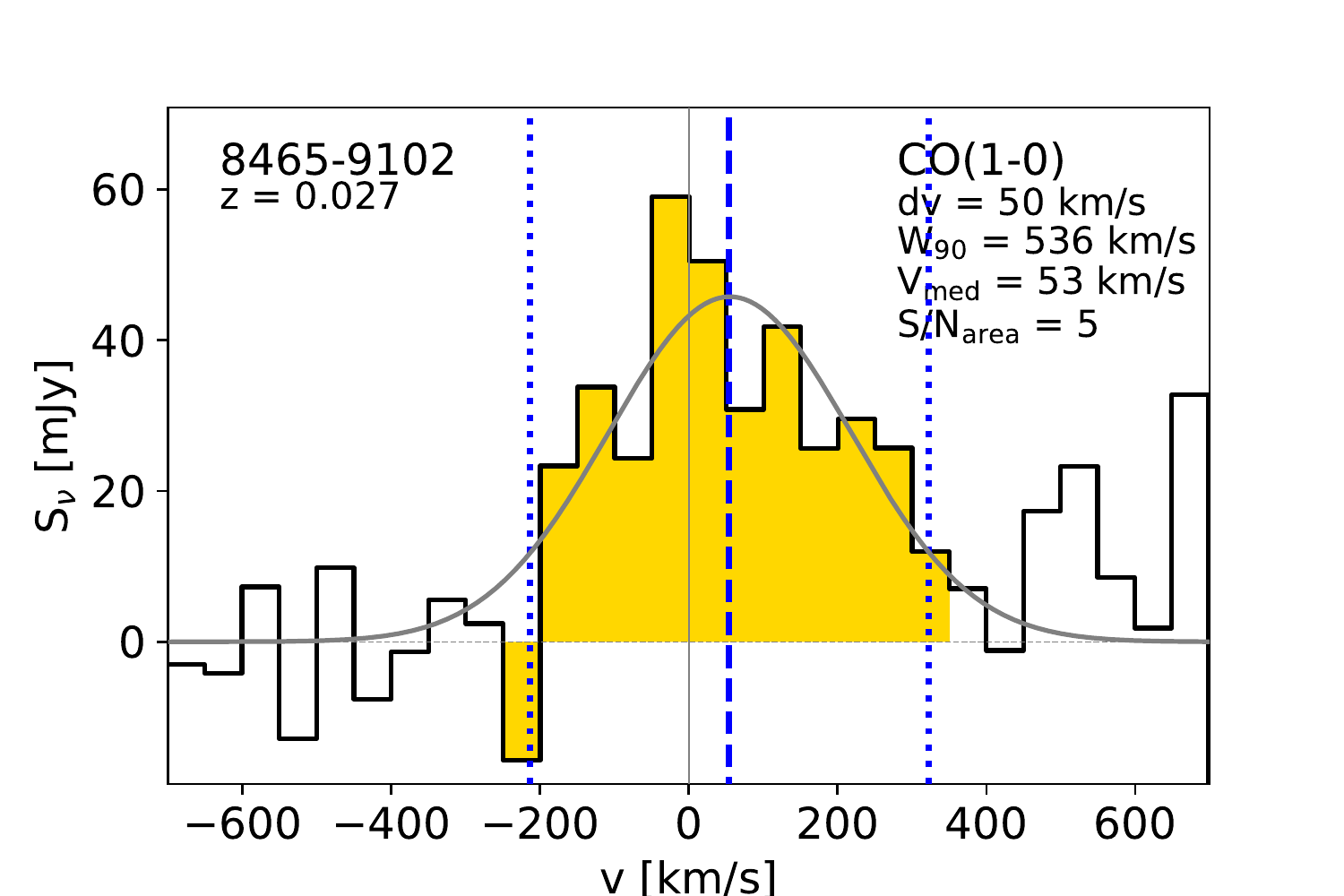}
\hspace{0.4cm}  \centering  \includegraphics[width = 0.17\textwidth, trim = 0cm 0cm 0cm 0cm, clip = true]{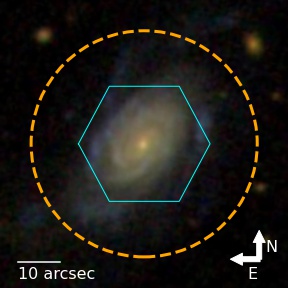} \includegraphics[width = 0.29\textwidth, trim = 0cm 0cm 0cm 0cm, clip = true]{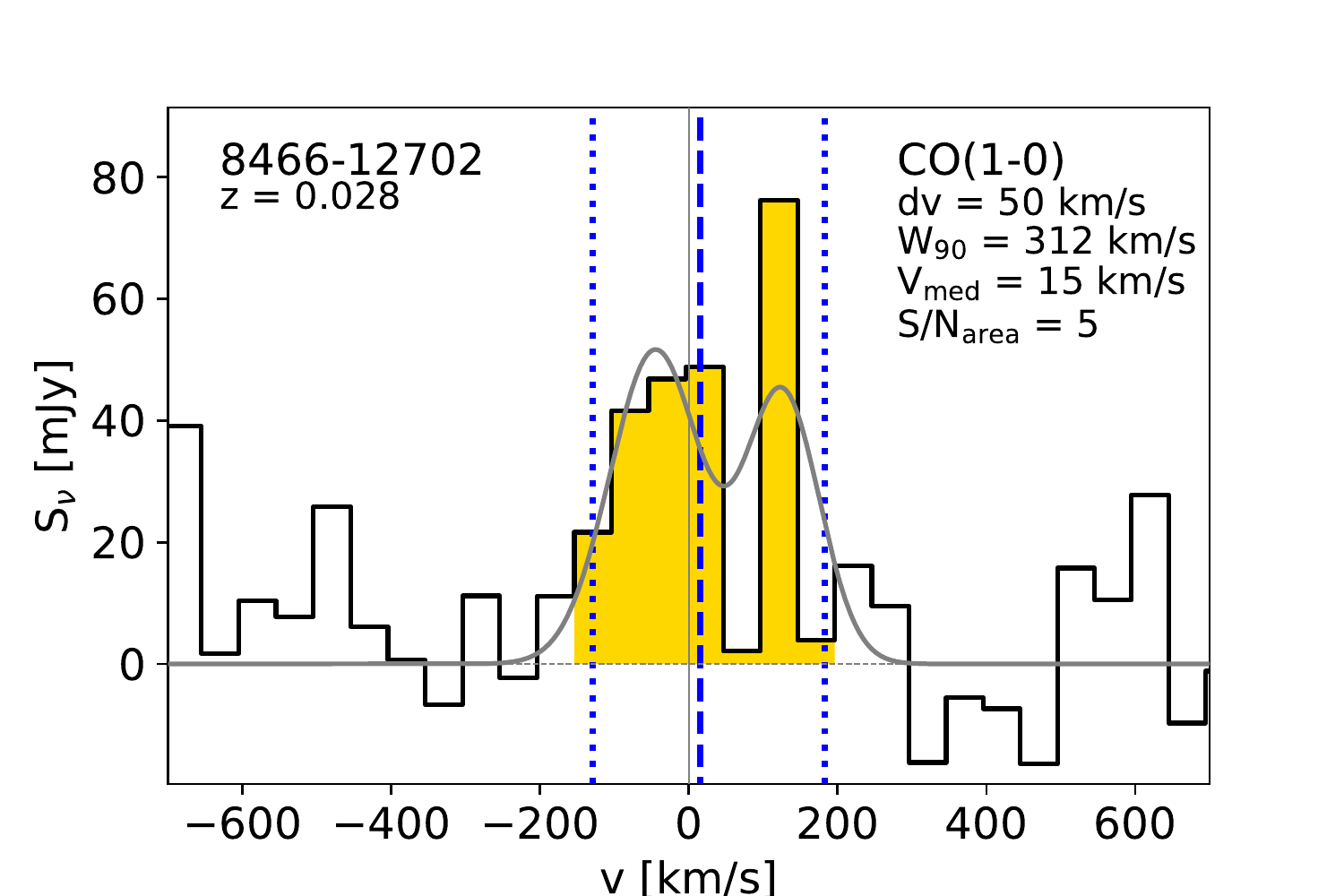}
\end{figure*}

\begin{figure*}  \centering  \includegraphics[width = 0.17\textwidth, trim = 0cm 0cm 0cm 0cm, clip = true]{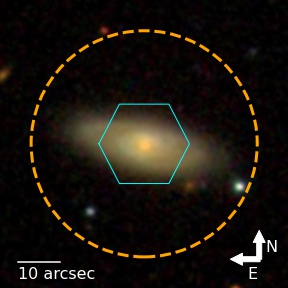} \includegraphics[width = 0.29\textwidth, trim = 0cm 0cm 0cm 0cm, clip = true]{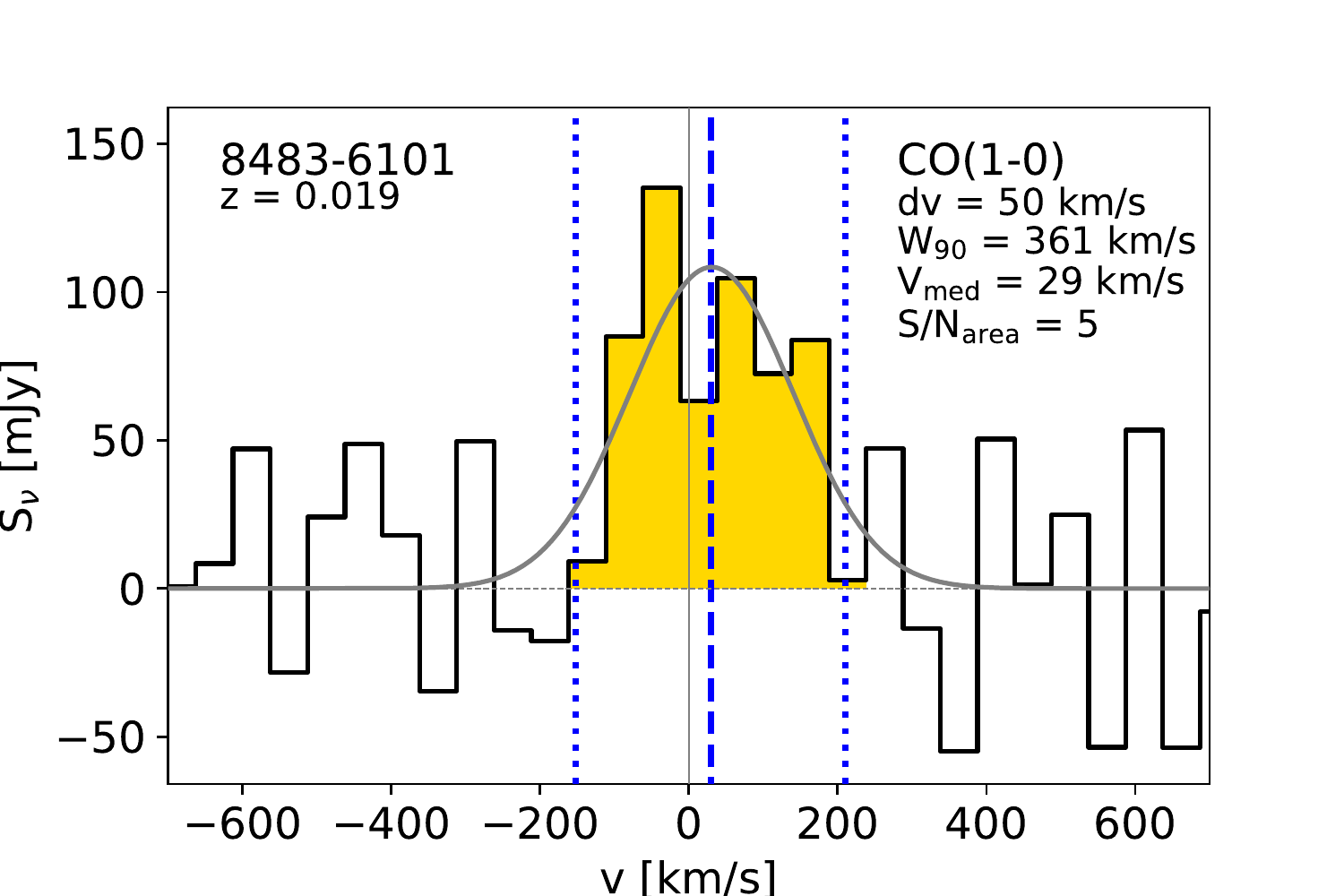}
\hspace{0.4cm}  \centering  \includegraphics[width = 0.17\textwidth, trim = 0cm 0cm 0cm 0cm, clip = true]{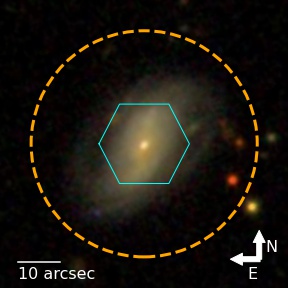} \includegraphics[width = 0.29\textwidth, trim = 0cm 0cm 0cm 0cm, clip = true]{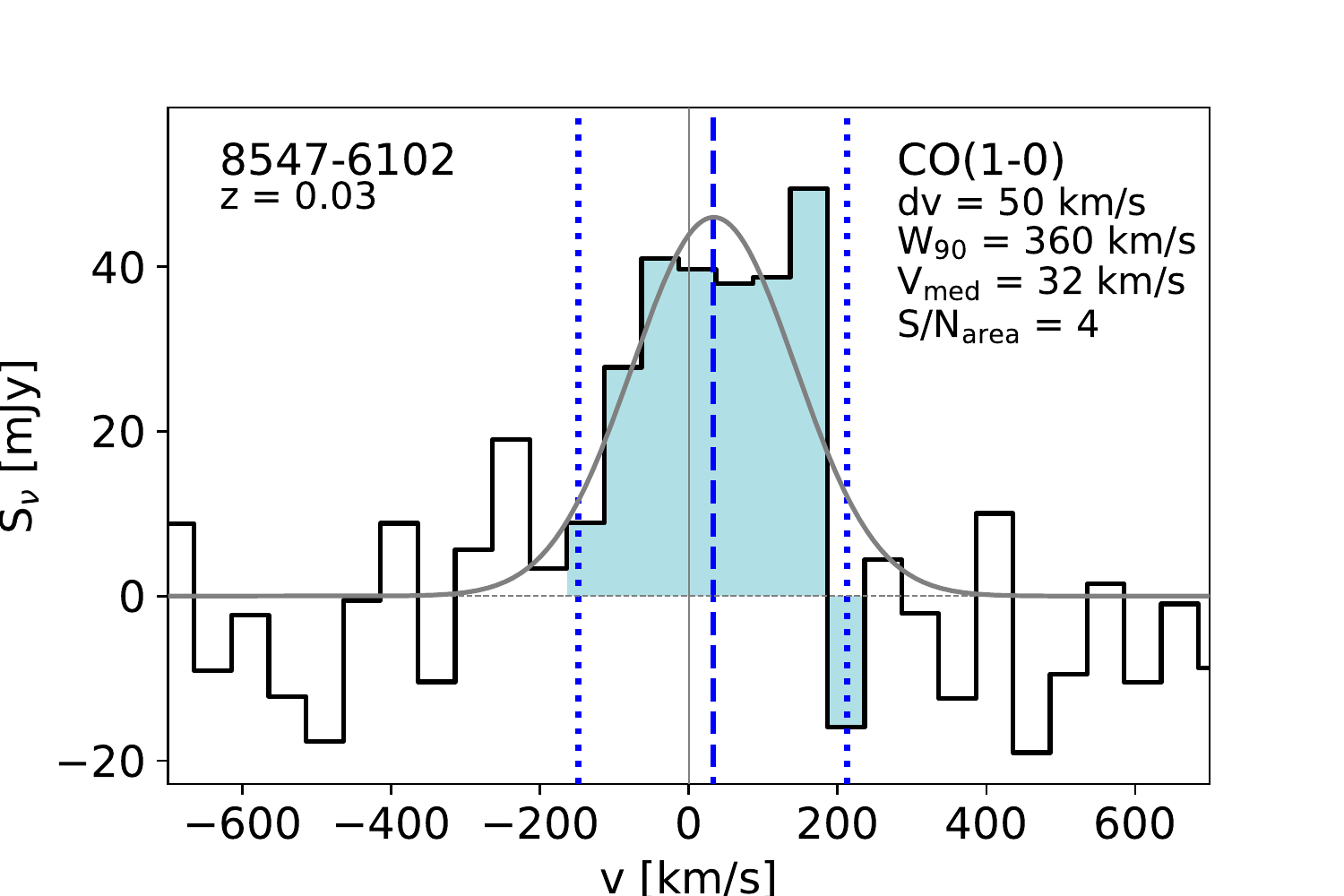}
\end{figure*}

\begin{figure*}  \centering  \includegraphics[width = 0.17\textwidth, trim = 0cm 0cm 0cm 0cm, clip = true]{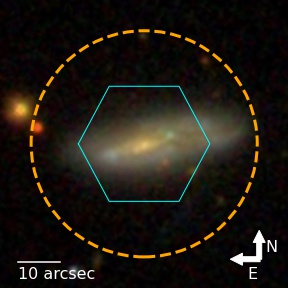} \includegraphics[width = 0.29\textwidth, trim = 0cm 0cm 0cm 0cm, clip = true]{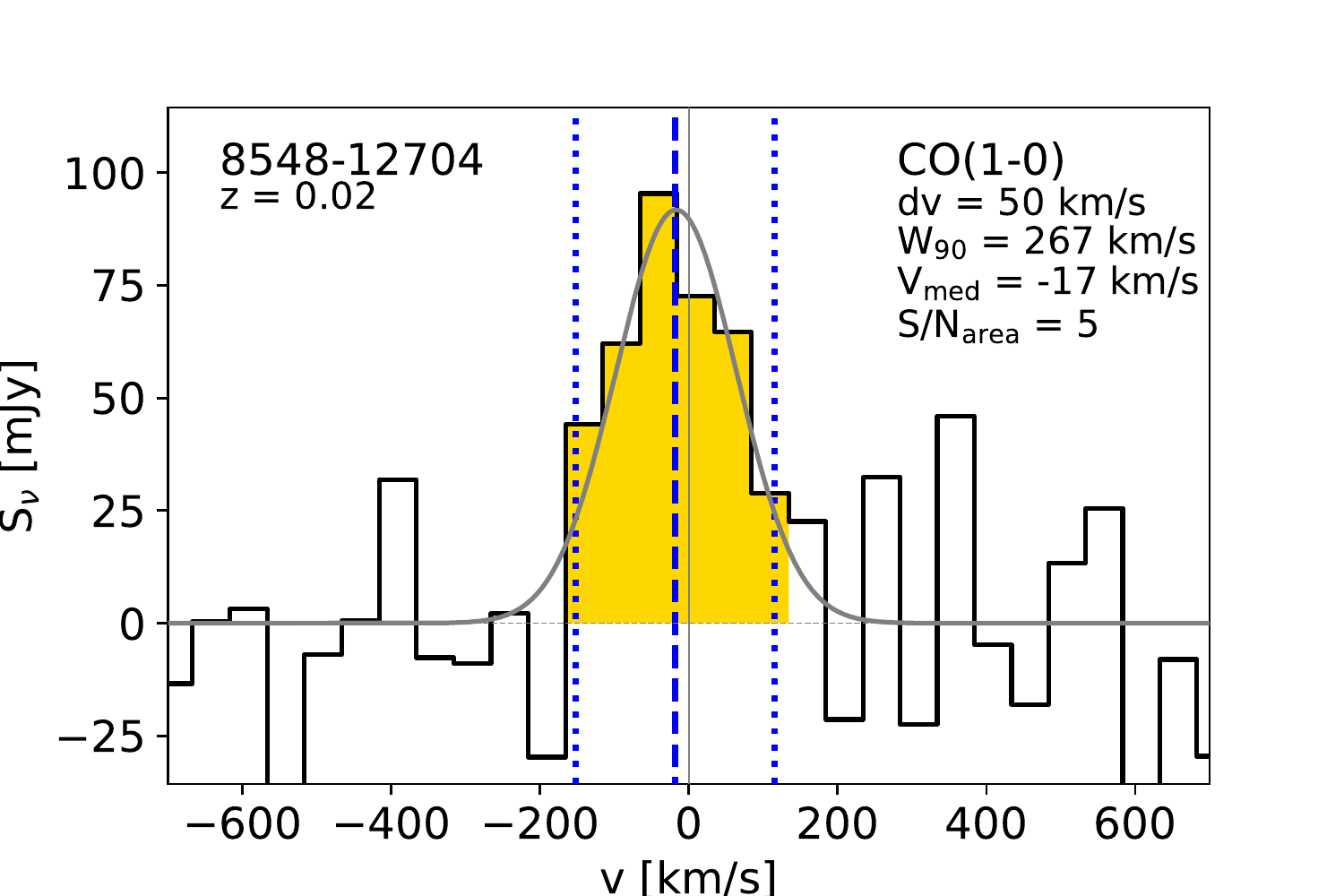}
\hspace{0.4cm}  \centering  \includegraphics[width = 0.17\textwidth, trim = 0cm 0cm 0cm 0cm, clip = true]{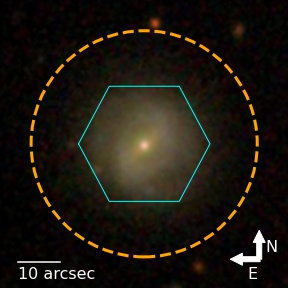} \includegraphics[width = 0.29\textwidth, trim = 0cm 0cm 0cm 0cm, clip = true]{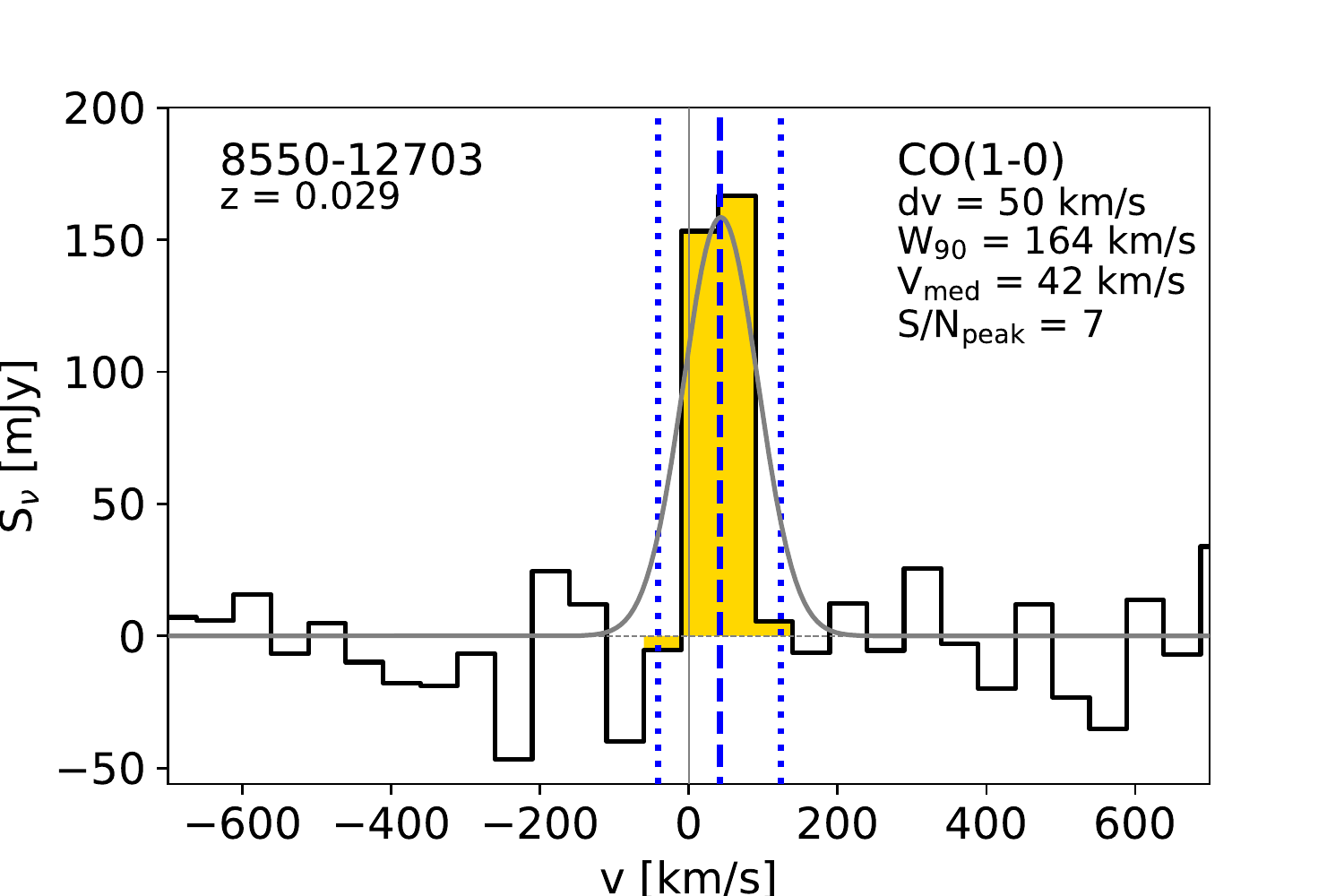}
\end{figure*}

\begin{figure*}  \centering  \includegraphics[width = 0.17\textwidth, trim = 0cm 0cm 0cm 0cm, clip = true]{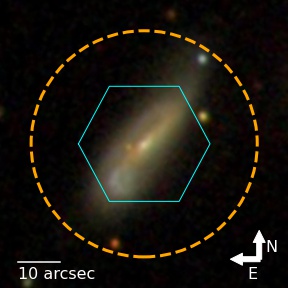} \includegraphics[width = 0.29\textwidth, trim = 0cm 0cm 0cm 0cm, clip = true]{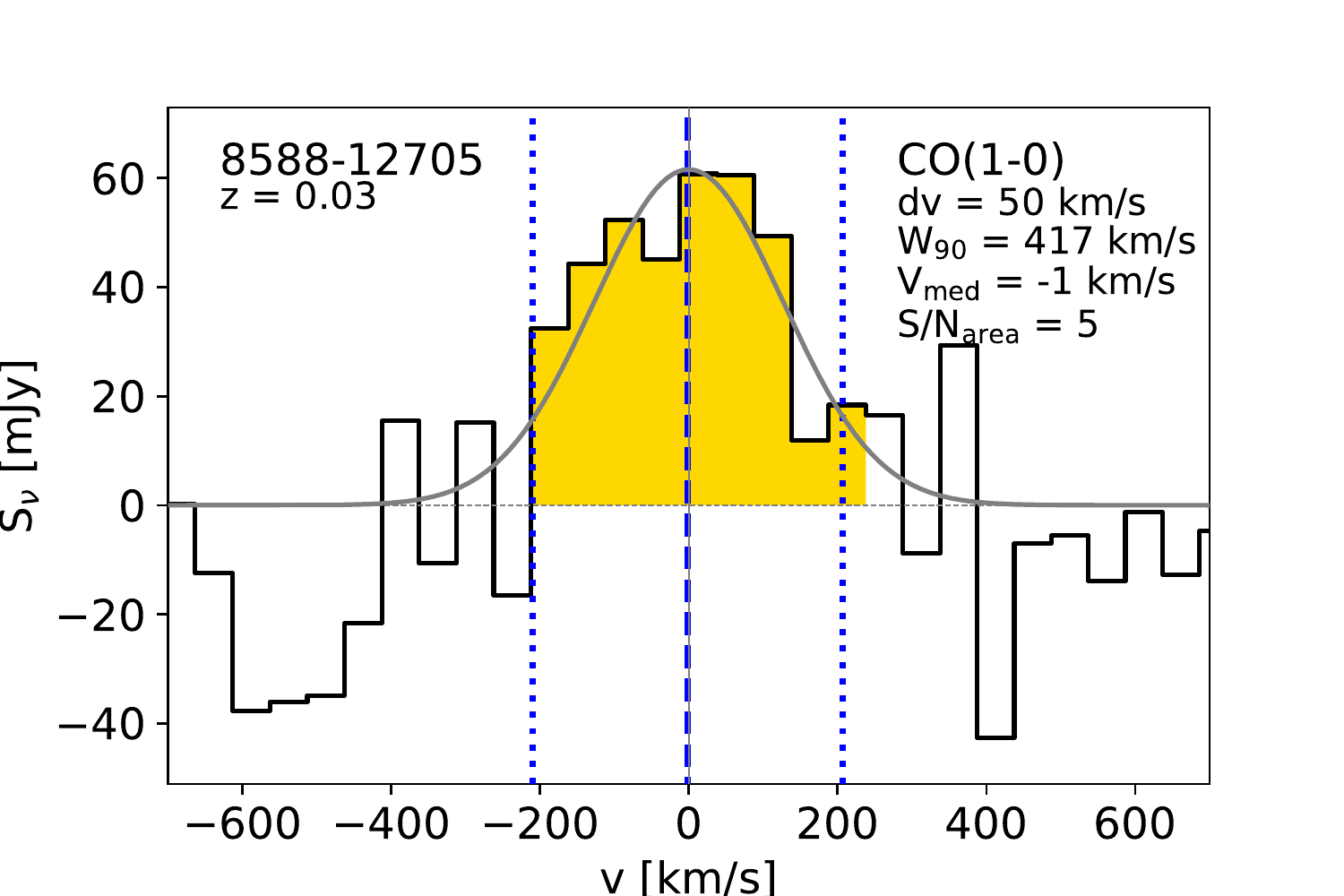}
\hspace{0.4cm}  \centering  \includegraphics[width = 0.17\textwidth, trim = 0cm 0cm 0cm 0cm, clip = true]{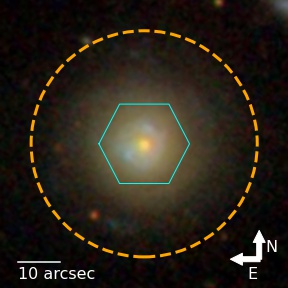} \includegraphics[width = 0.29\textwidth, trim = 0cm 0cm 0cm 0cm, clip = true]{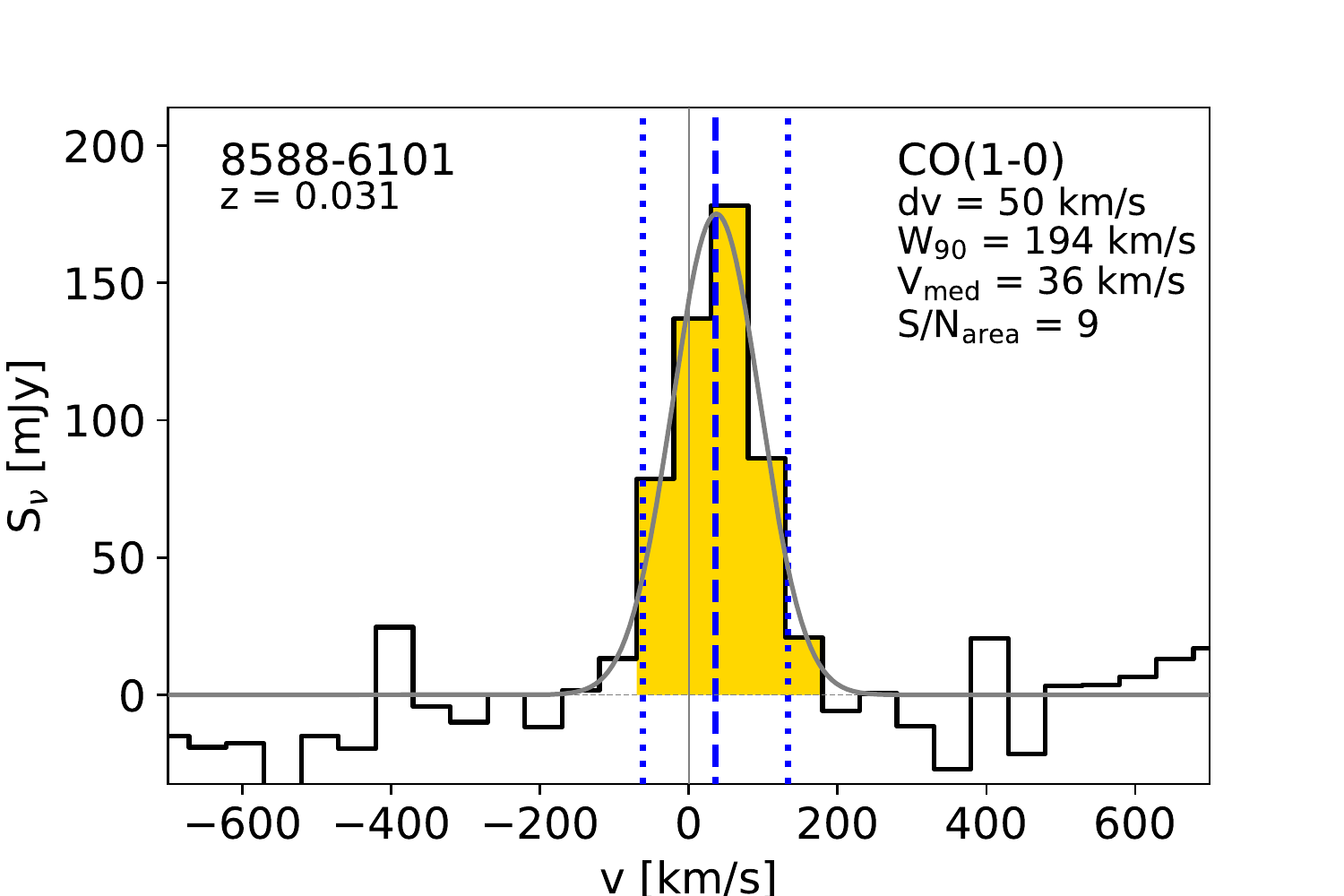}
\end{figure*}

\begin{figure*}  \centering  \includegraphics[width = 0.17\textwidth, trim = 0cm 0cm 0cm 0cm, clip = true]{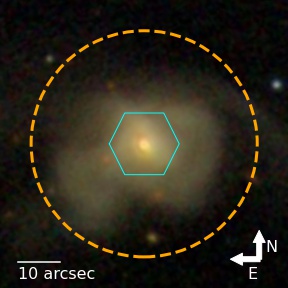} \includegraphics[width = 0.29\textwidth, trim = 0cm 0cm 0cm 0cm, clip = true]{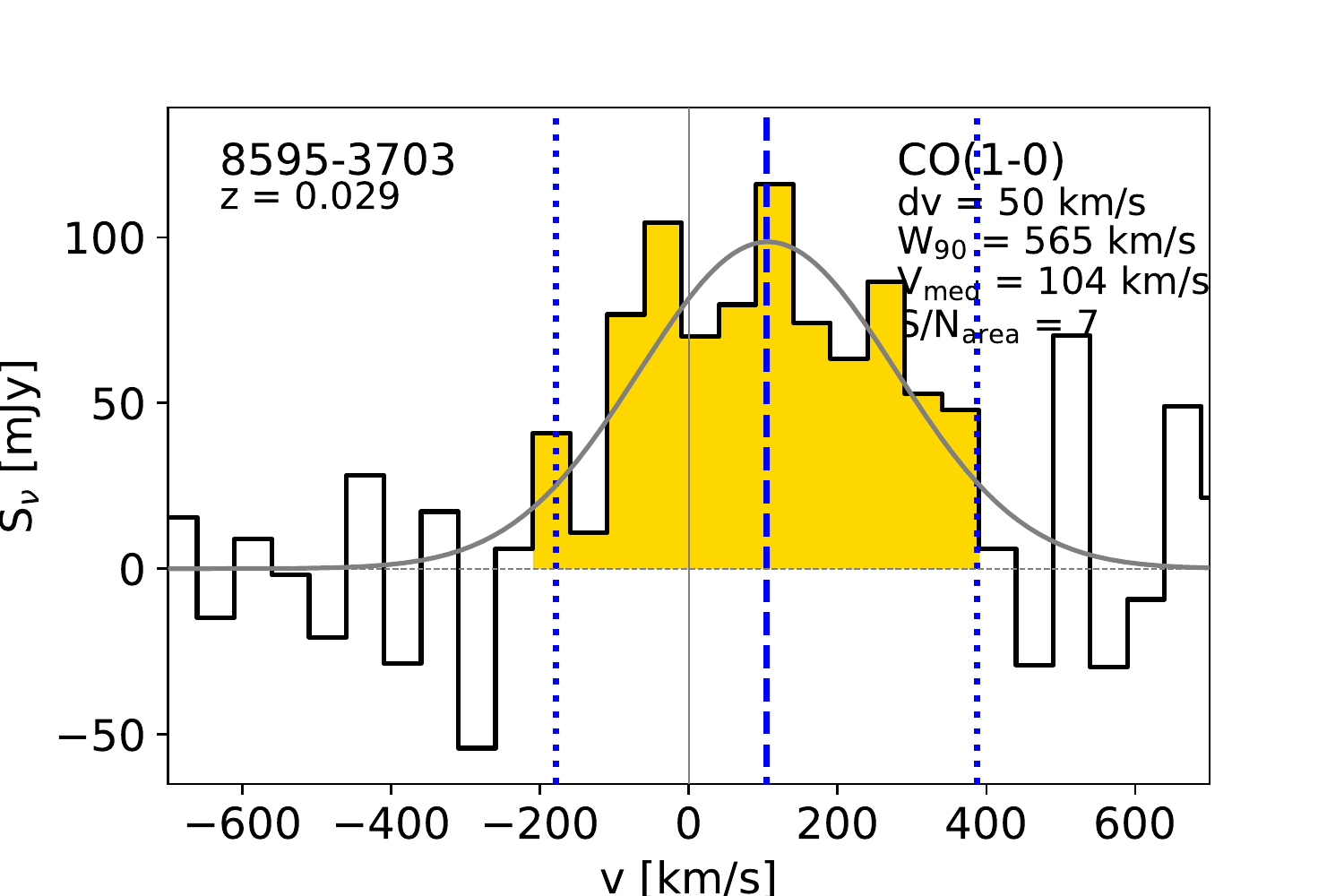}
\hspace{0.4cm}  \centering  \includegraphics[width = 0.17\textwidth, trim = 0cm 0cm 0cm 0cm, clip = true]{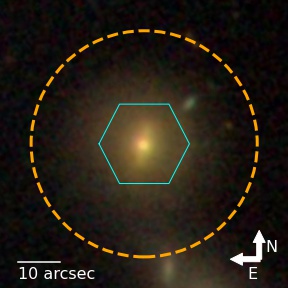} \includegraphics[width = 0.29\textwidth, trim = 0cm 0cm 0cm 0cm, clip = true]{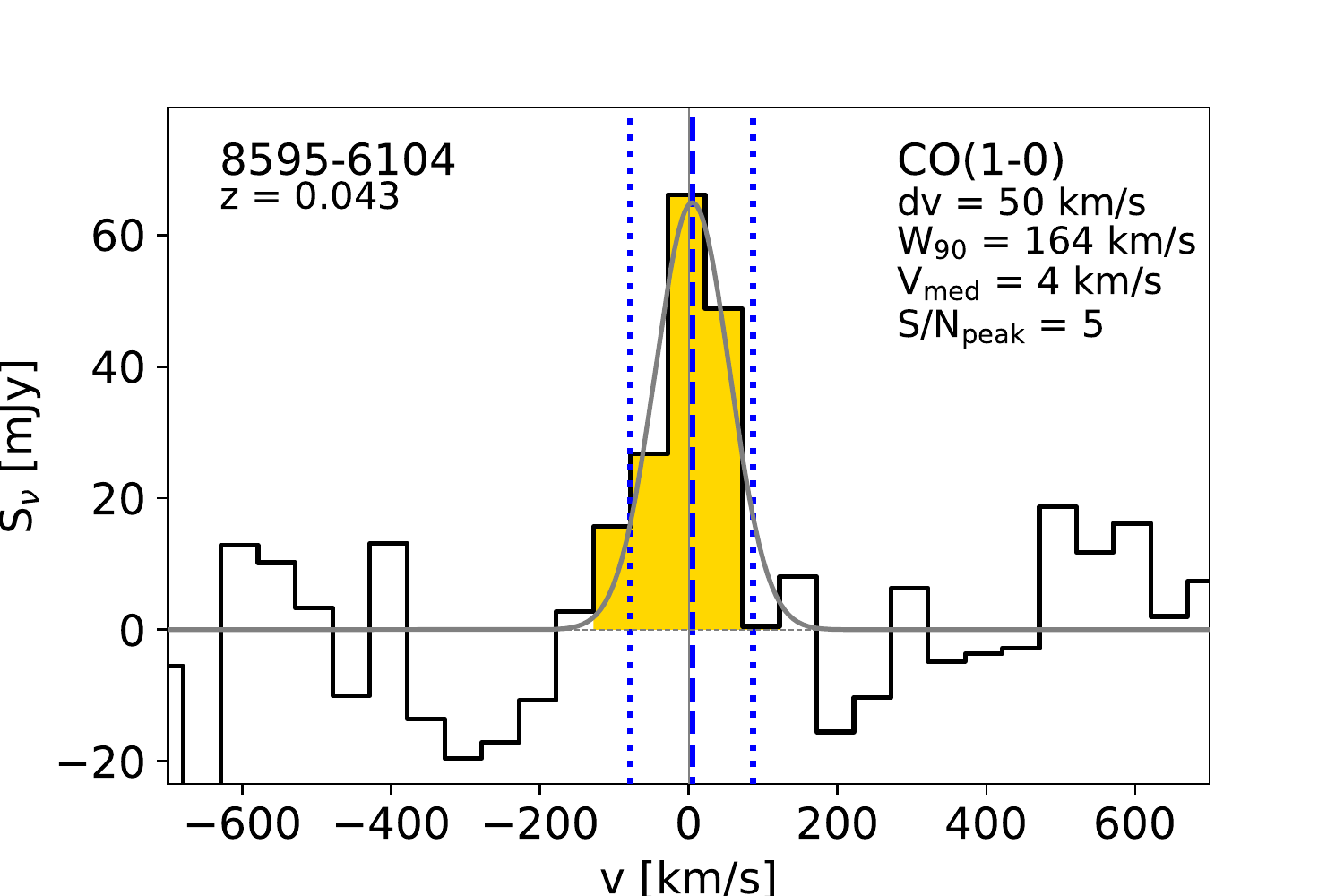}
\end{figure*}

\begin{figure*}
   \ContinuedFloat 
  \centering  \includegraphics[width = 0.17\textwidth, trim = 0cm 0cm 0cm 0cm, clip = true]{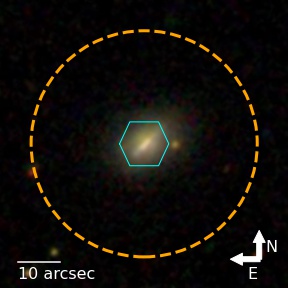} \includegraphics[width = 0.29\textwidth, trim = 0cm 0cm 0cm 0cm, clip = true]{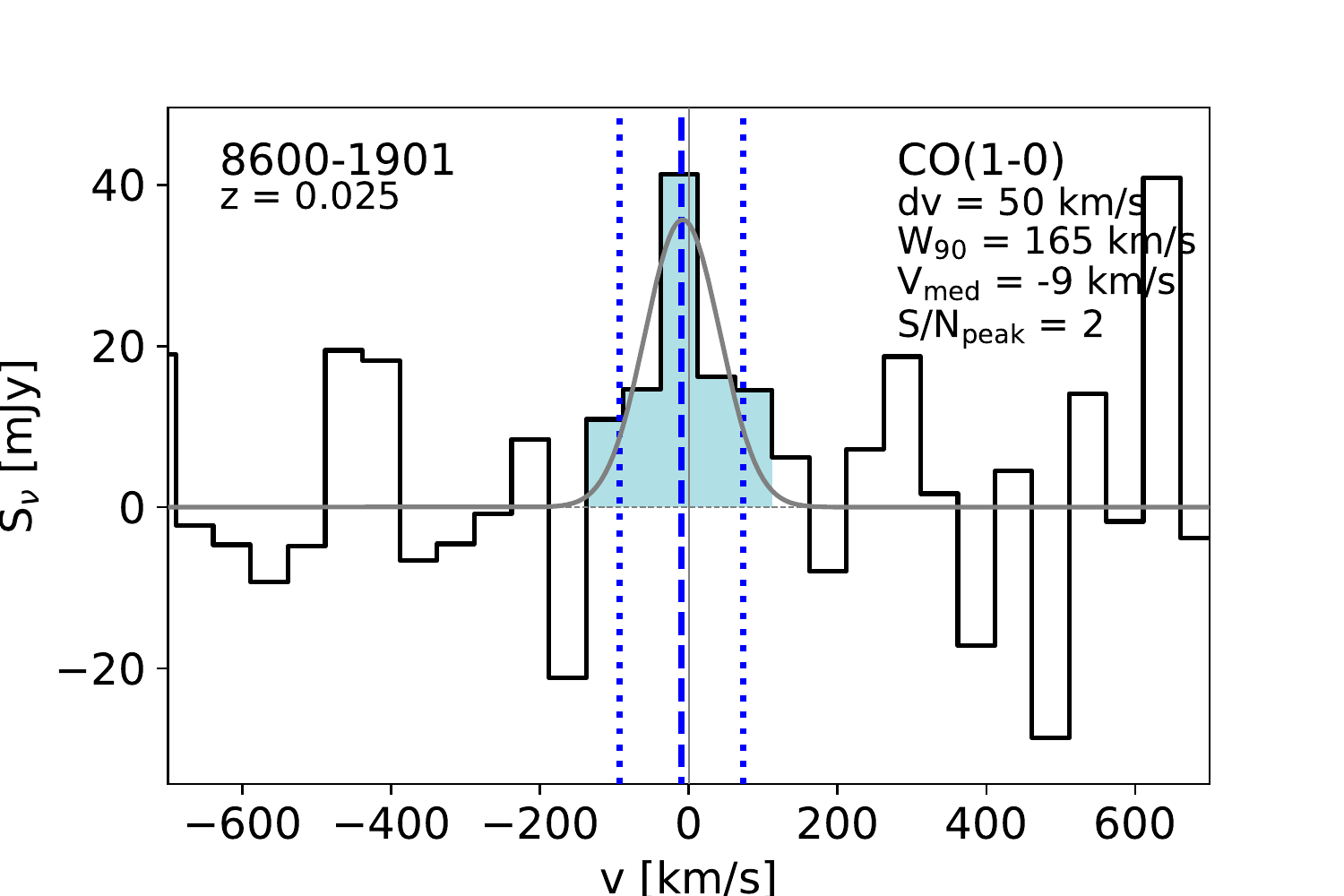}
\hspace{0.4cm}  \centering  \includegraphics[width = 0.17\textwidth, trim = 0cm 0cm 0cm 0cm, clip = true]{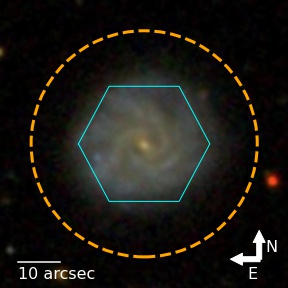} \includegraphics[width = 0.29\textwidth, trim = 0cm 0cm 0cm 0cm, clip = true]{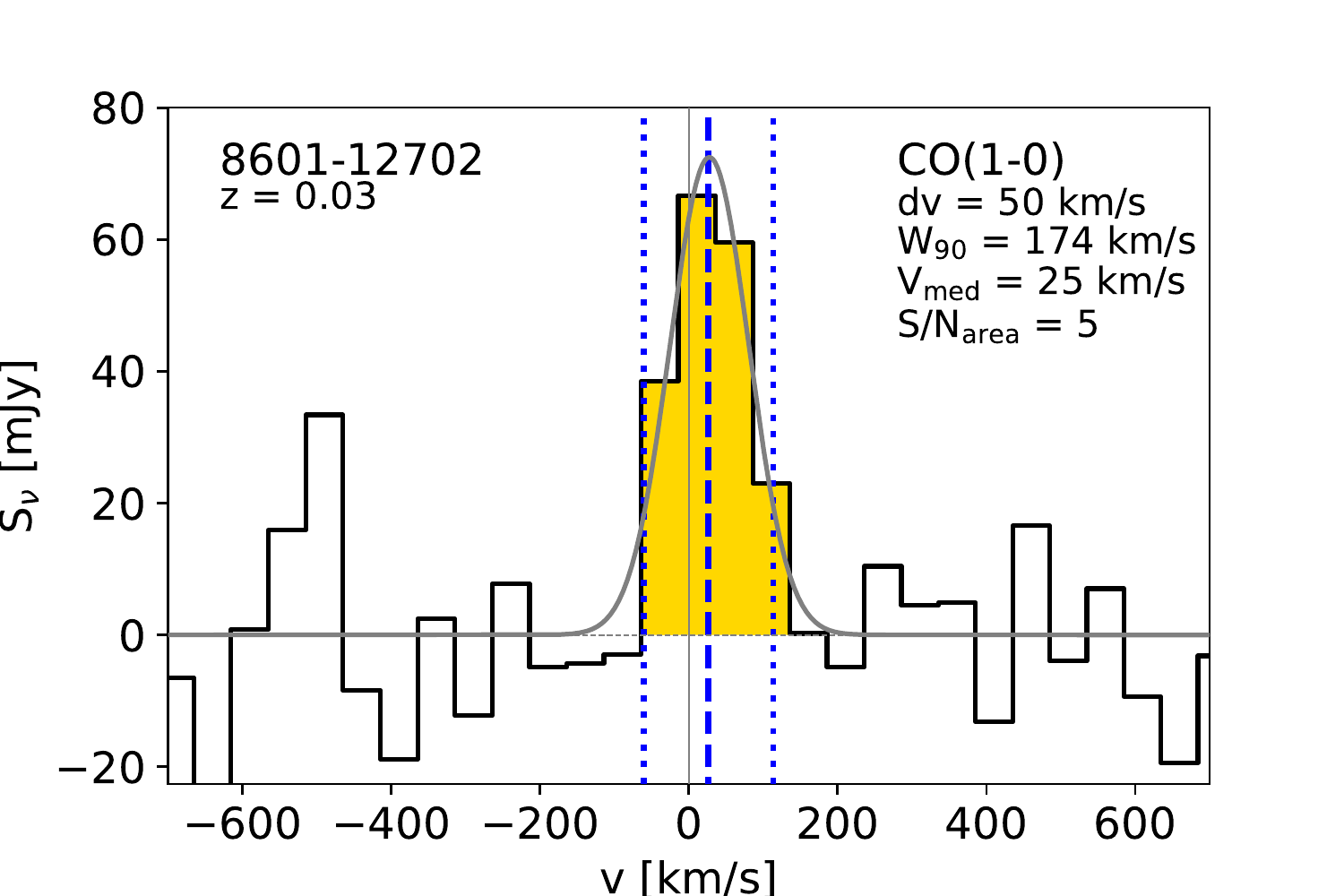}
 \caption{continued.}
\end{figure*}

\begin{figure*}  \centering  \includegraphics[width = 0.17\textwidth, trim = 0cm 0cm 0cm 0cm, clip = true]{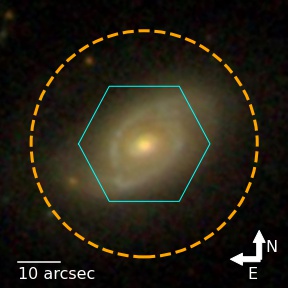} \includegraphics[width = 0.29\textwidth, trim = 0cm 0cm 0cm 0cm, clip = true]{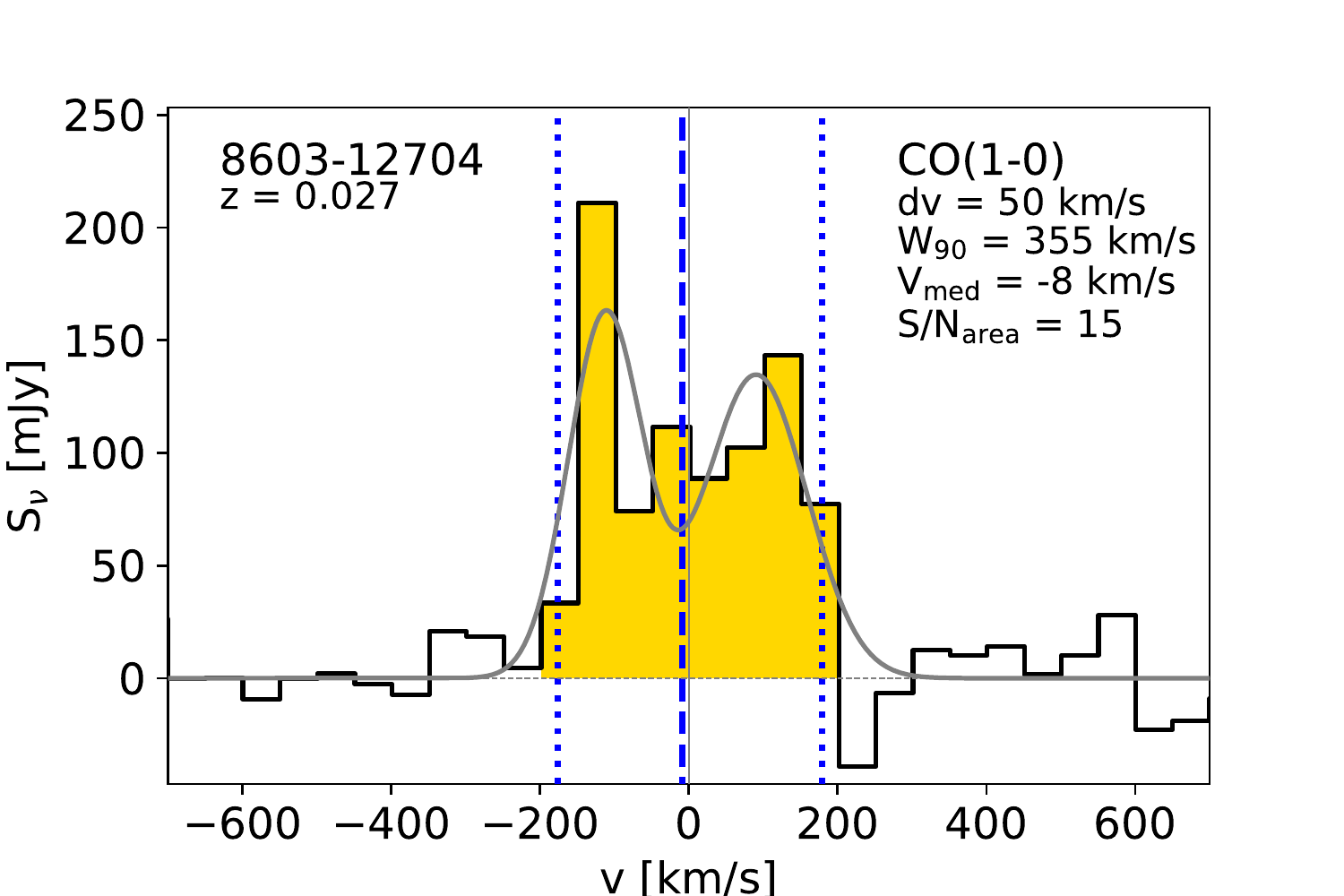}
\hspace{0.4cm}  \centering  \includegraphics[width = 0.17\textwidth, trim = 0cm 0cm 0cm 0cm, clip = true]{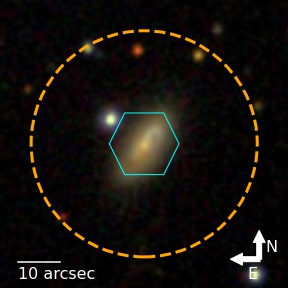} \includegraphics[width = 0.29\textwidth, trim = 0cm 0cm 0cm 0cm, clip = true]{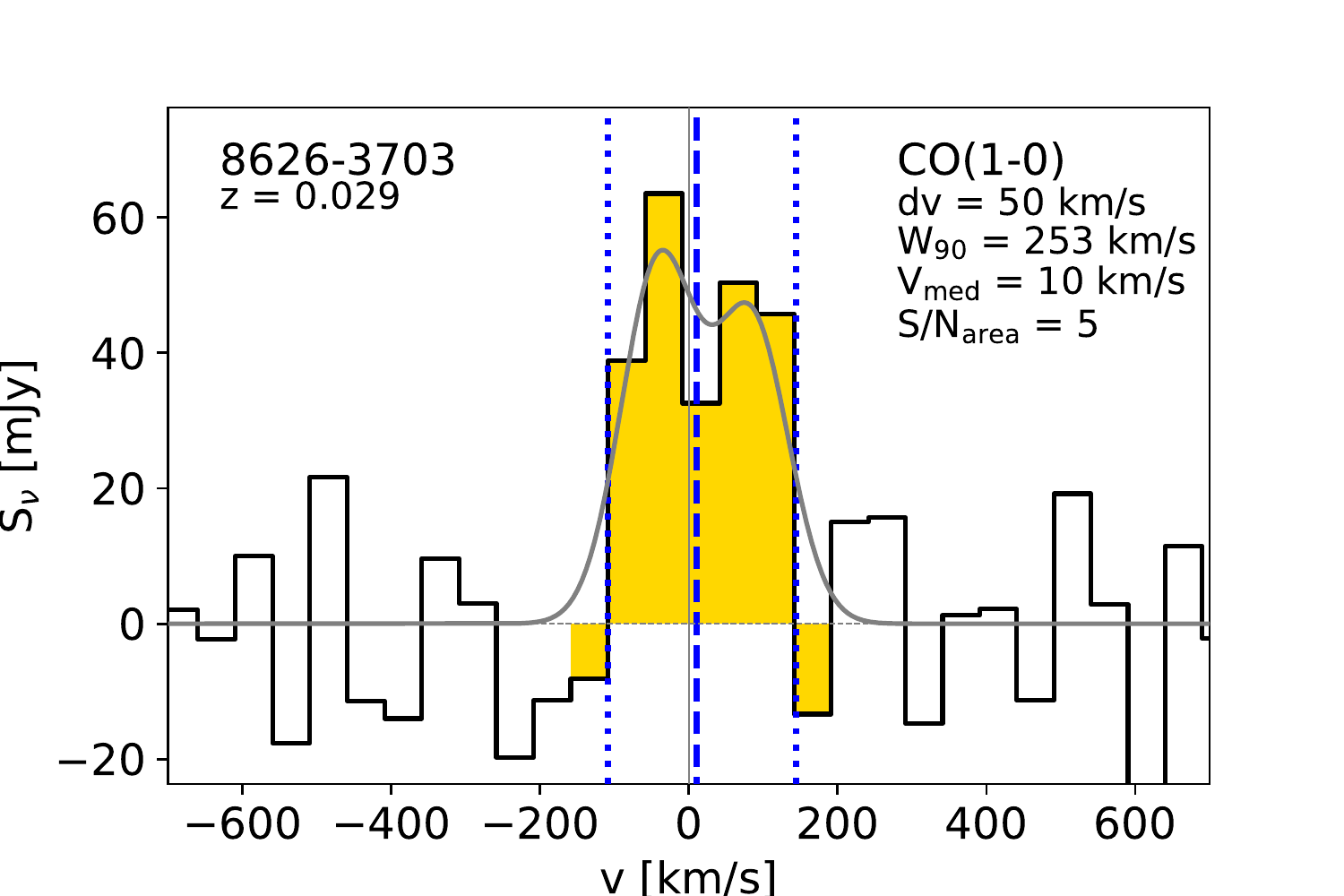}
\end{figure*}

\begin{figure*}  \centering  \includegraphics[width = 0.17\textwidth, trim = 0cm 0cm 0cm 0cm, clip = true]{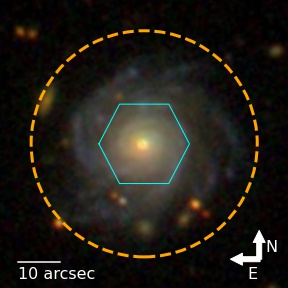} \includegraphics[width = 0.29\textwidth, trim = 0cm 0cm 0cm 0cm, clip = true]{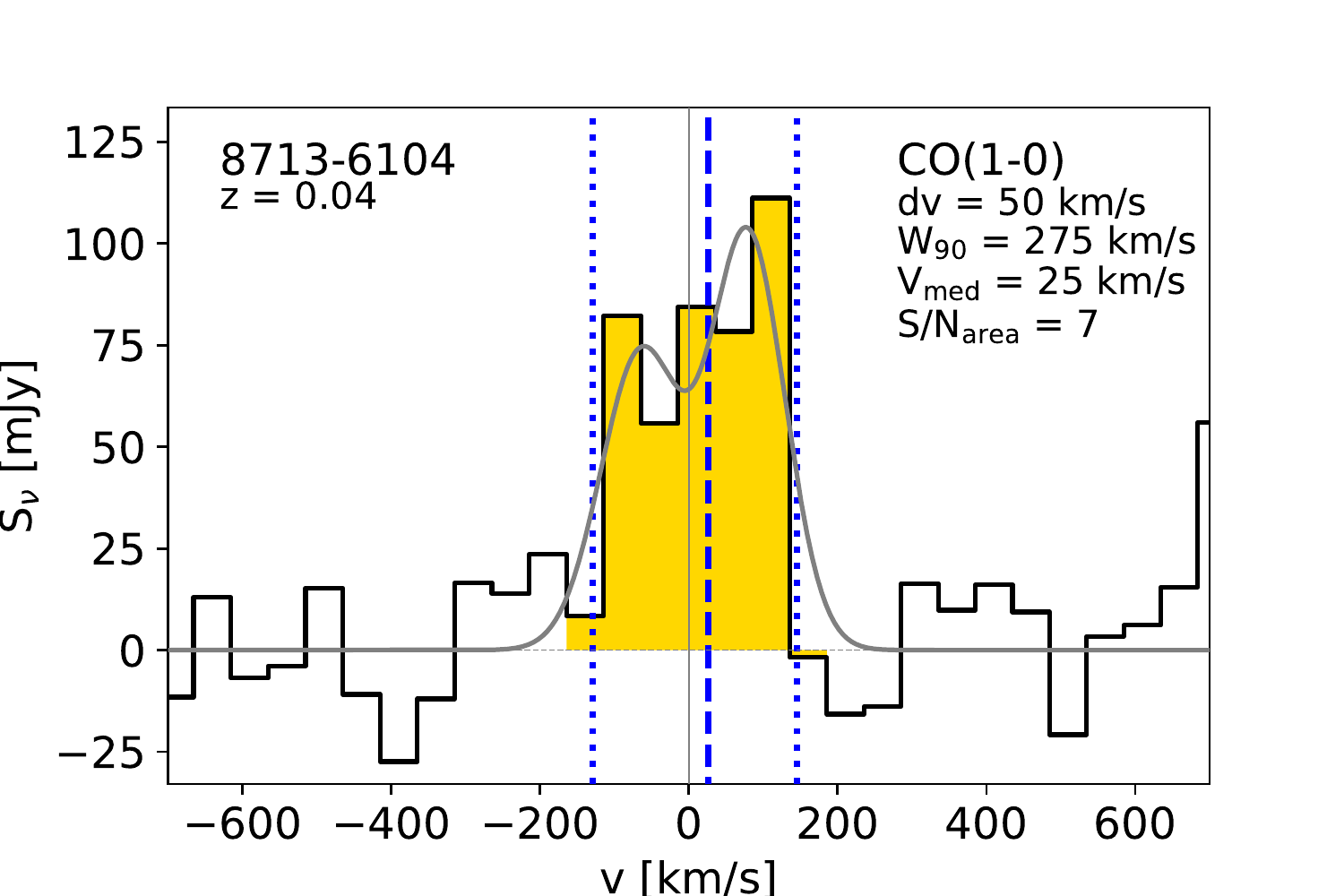}
\hspace{0.4cm}  \centering  \includegraphics[width = 0.17\textwidth, trim = 0cm 0cm 0cm 0cm, clip = true]{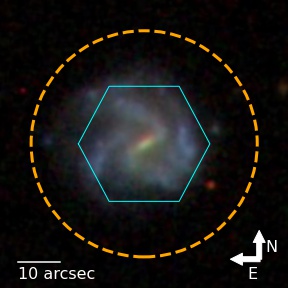} \includegraphics[width = 0.29\textwidth, trim = 0cm 0cm 0cm 0cm, clip = true]{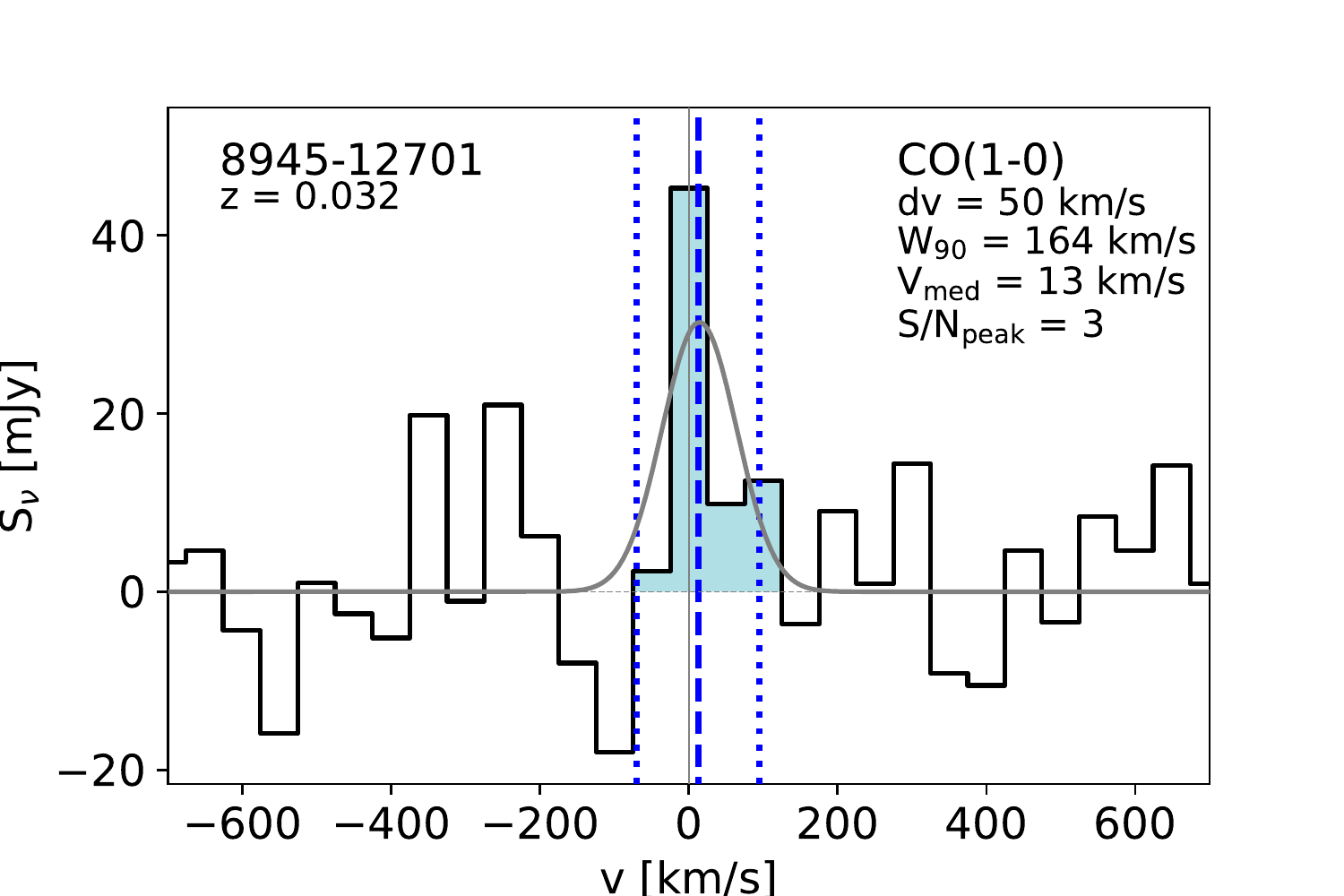}
\end{figure*}

\begin{figure*}  \centering  \includegraphics[width = 0.17\textwidth, trim = 0cm 0cm 0cm 0cm, clip = true]{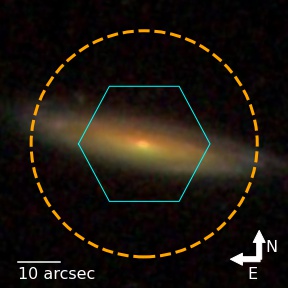} \includegraphics[width = 0.29\textwidth, trim = 0cm 0cm 0cm 0cm, clip = true]{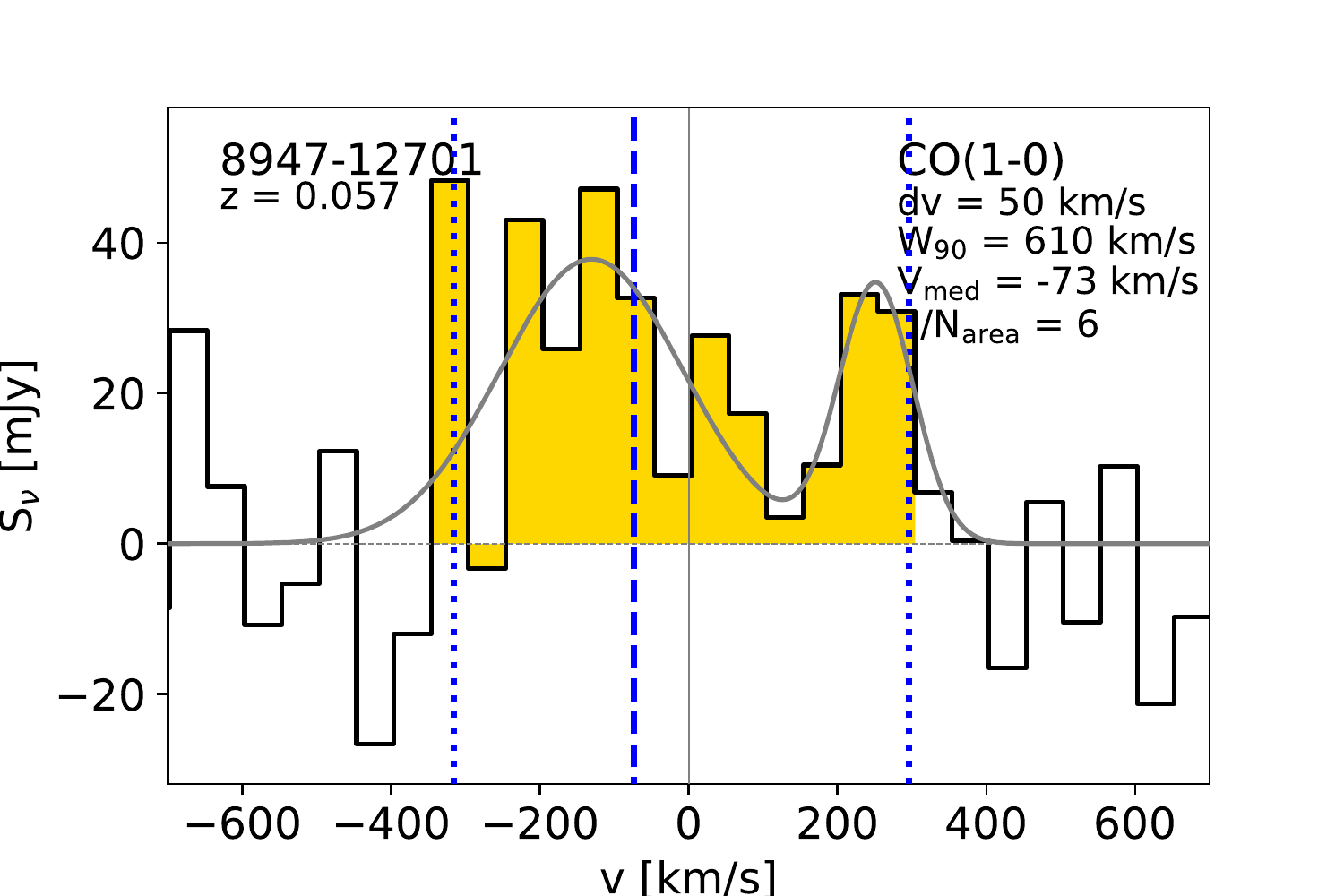}
\hspace{0.4cm}  \centering  \includegraphics[width = 0.17\textwidth, trim = 0cm 0cm 0cm 0cm, clip = true]{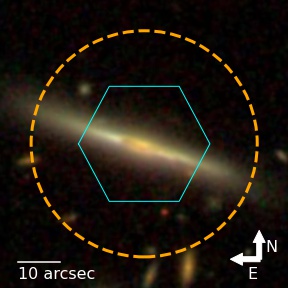} \includegraphics[width = 0.29\textwidth, trim = 0cm 0cm 0cm 0cm, clip = true]{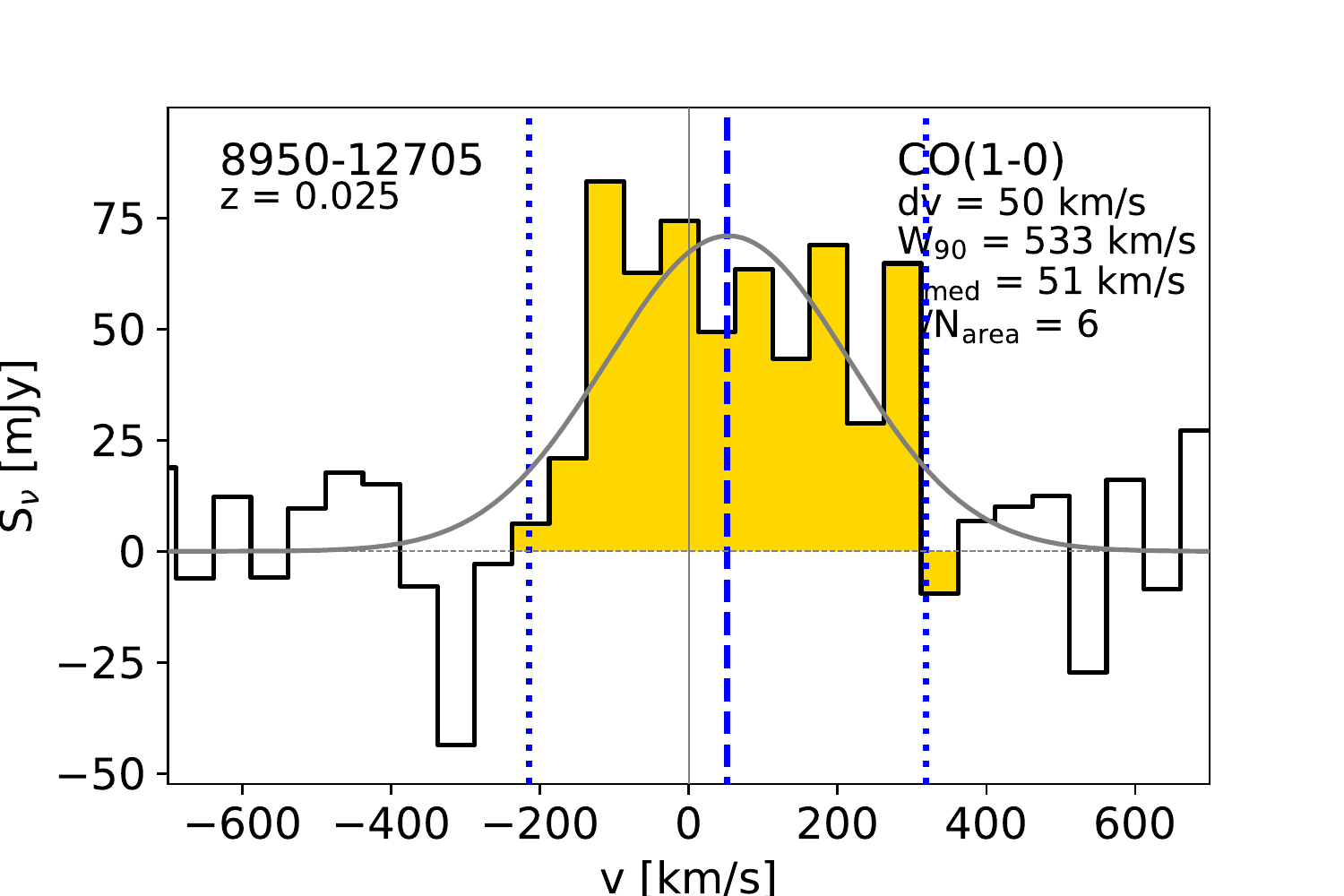}
\end{figure*}

\begin{figure*}  \centering  \includegraphics[width = 0.17\textwidth, trim = 0cm 0cm 0cm 0cm, clip = true]{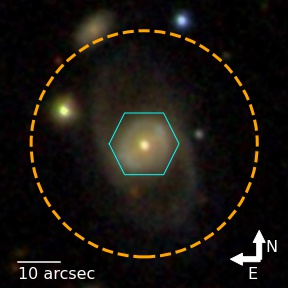} \includegraphics[width = 0.29\textwidth, trim = 0cm 0cm 0cm 0cm, clip = true]{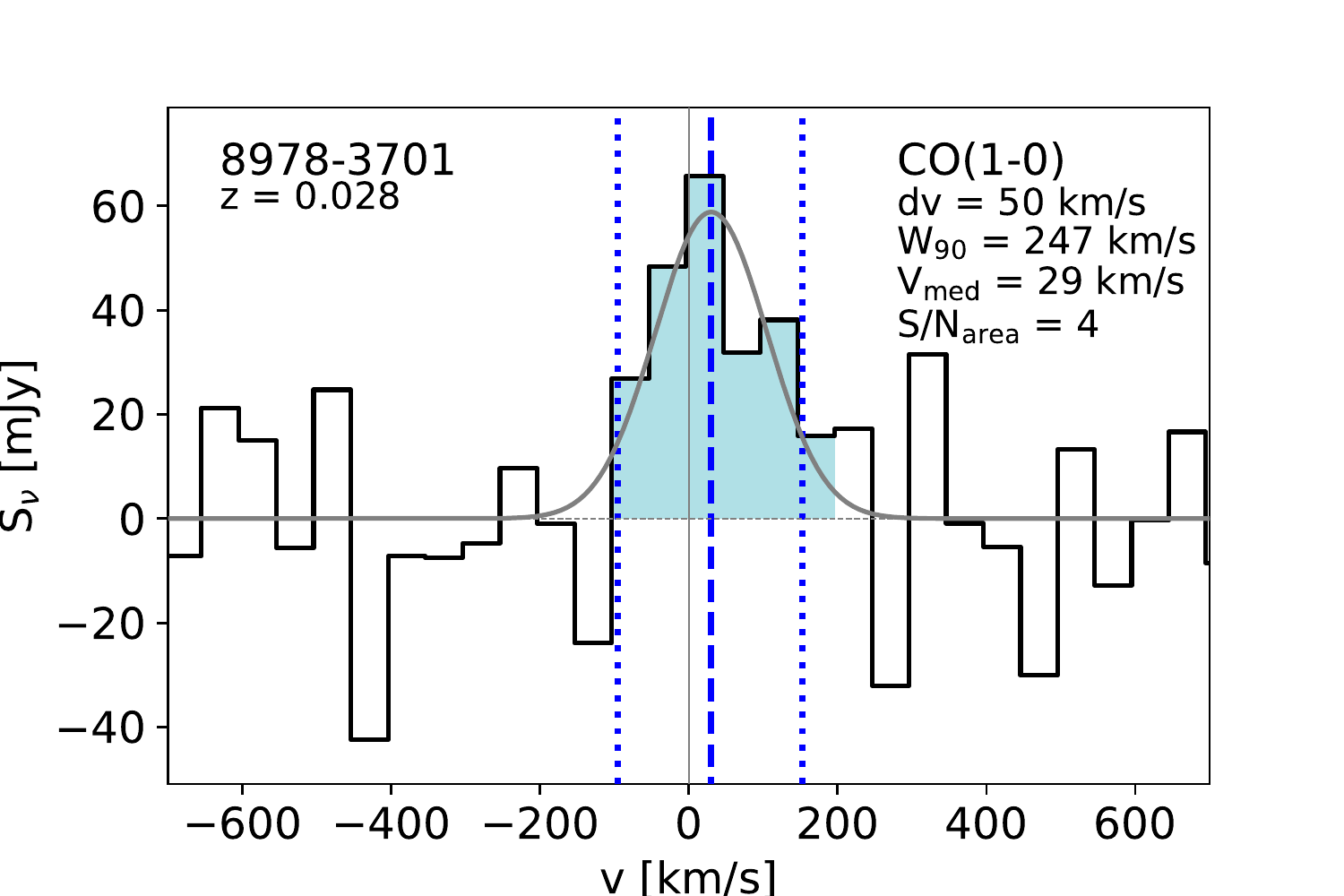}
\hspace{0.4cm}  \centering  \includegraphics[width = 0.17\textwidth, trim = 0cm 0cm 0cm 0cm, clip = true]{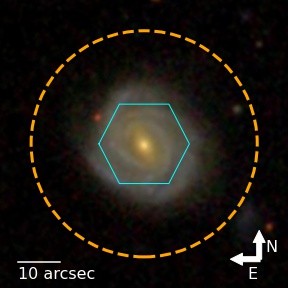} \includegraphics[width = 0.29\textwidth, trim = 0cm 0cm 0cm 0cm, clip = true]{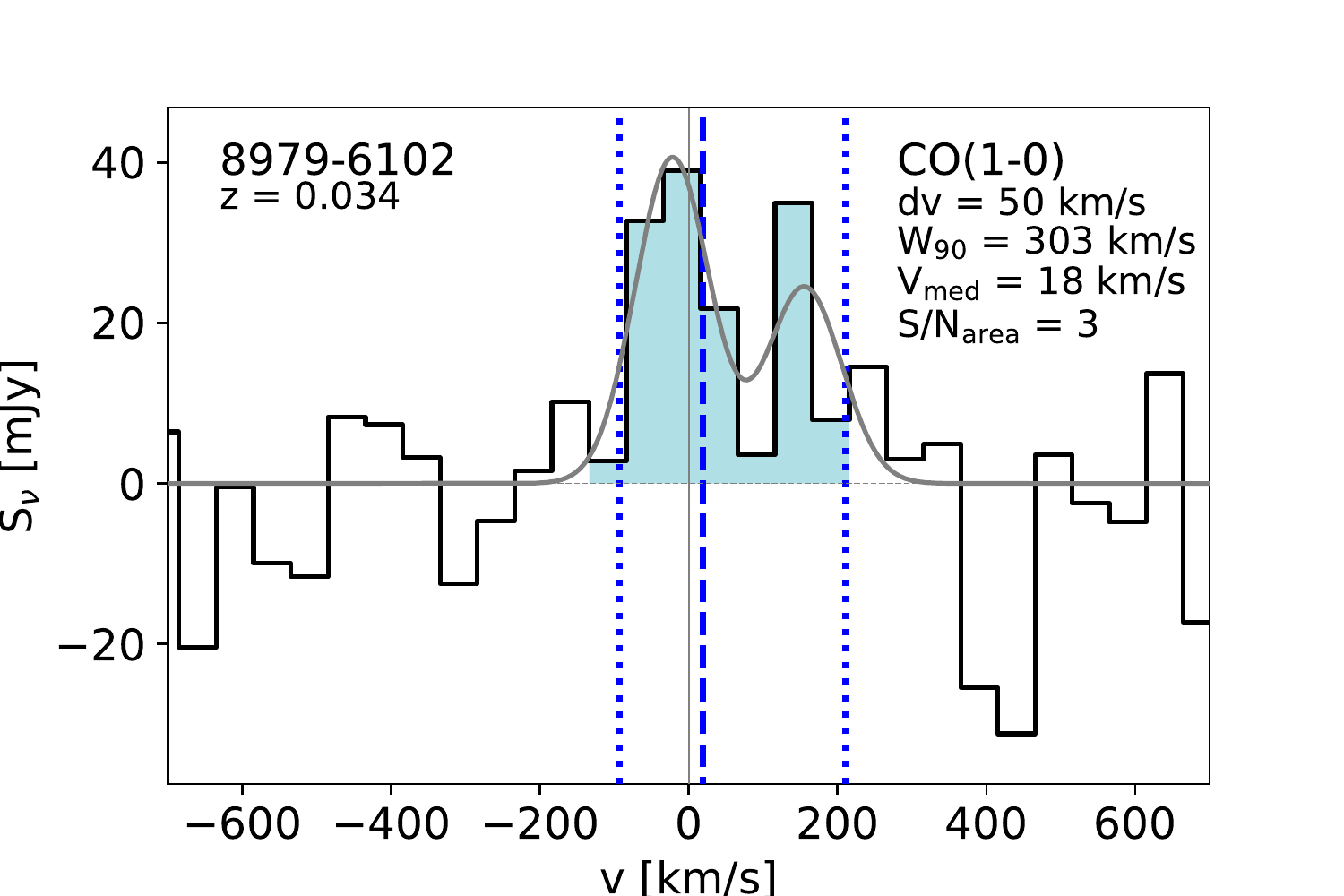}
\end{figure*}

\begin{figure*}  \centering  \includegraphics[width = 0.17\textwidth, trim = 0cm 0cm 0cm 0cm, clip = true]{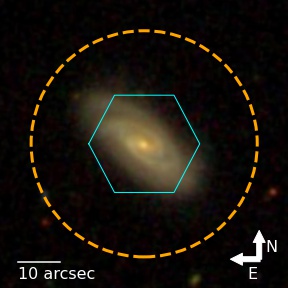} \includegraphics[width = 0.29\textwidth, trim = 0cm 0cm 0cm 0cm, clip = true]{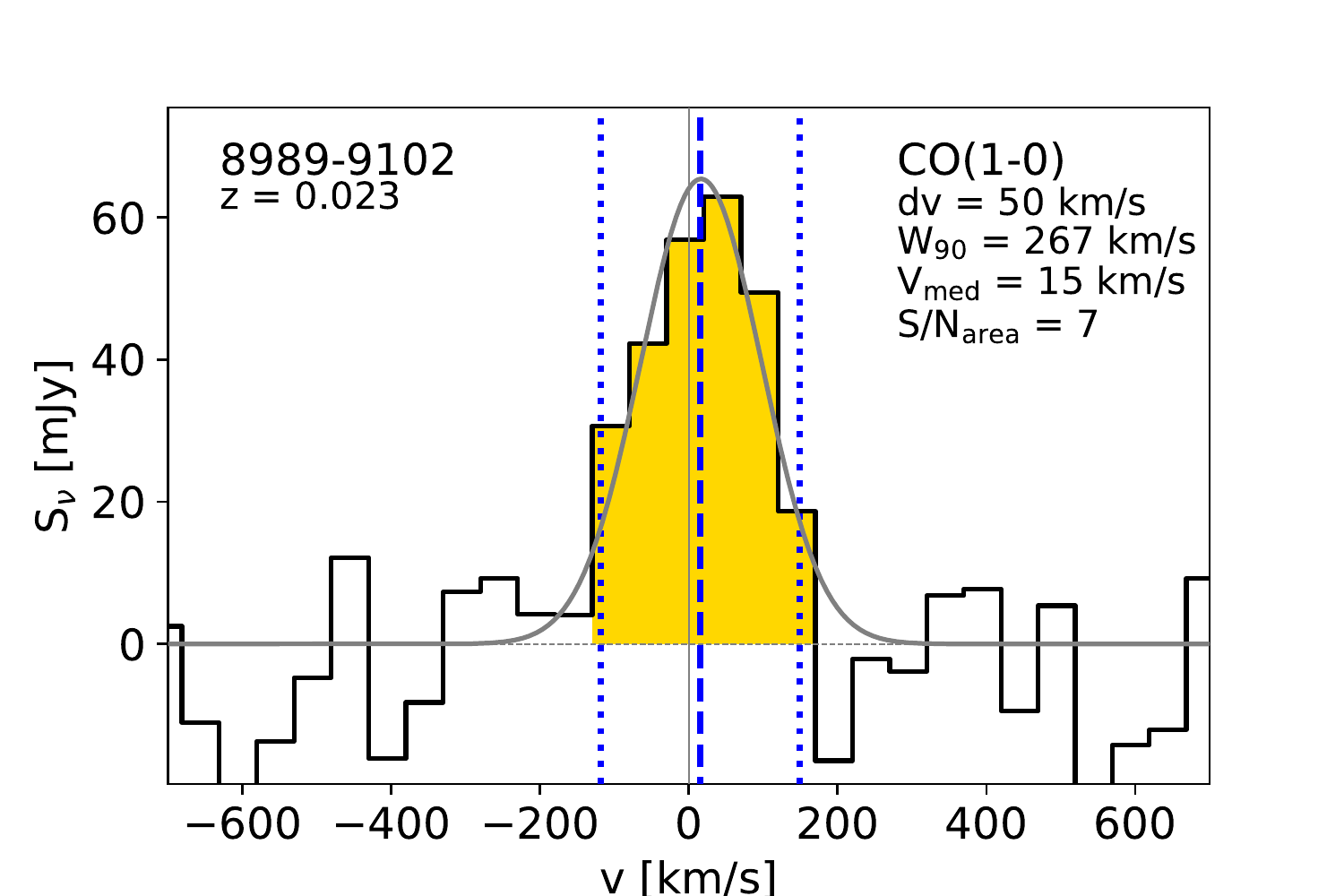}
\hspace{0.4cm}  \centering  \includegraphics[width = 0.17\textwidth, trim = 0cm 0cm 0cm 0cm, clip = true]{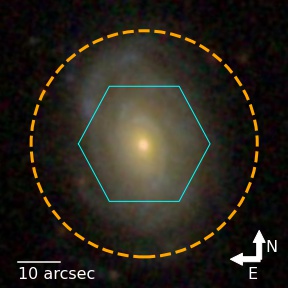} \includegraphics[width = 0.29\textwidth, trim = 0cm 0cm 0cm 0cm, clip = true]{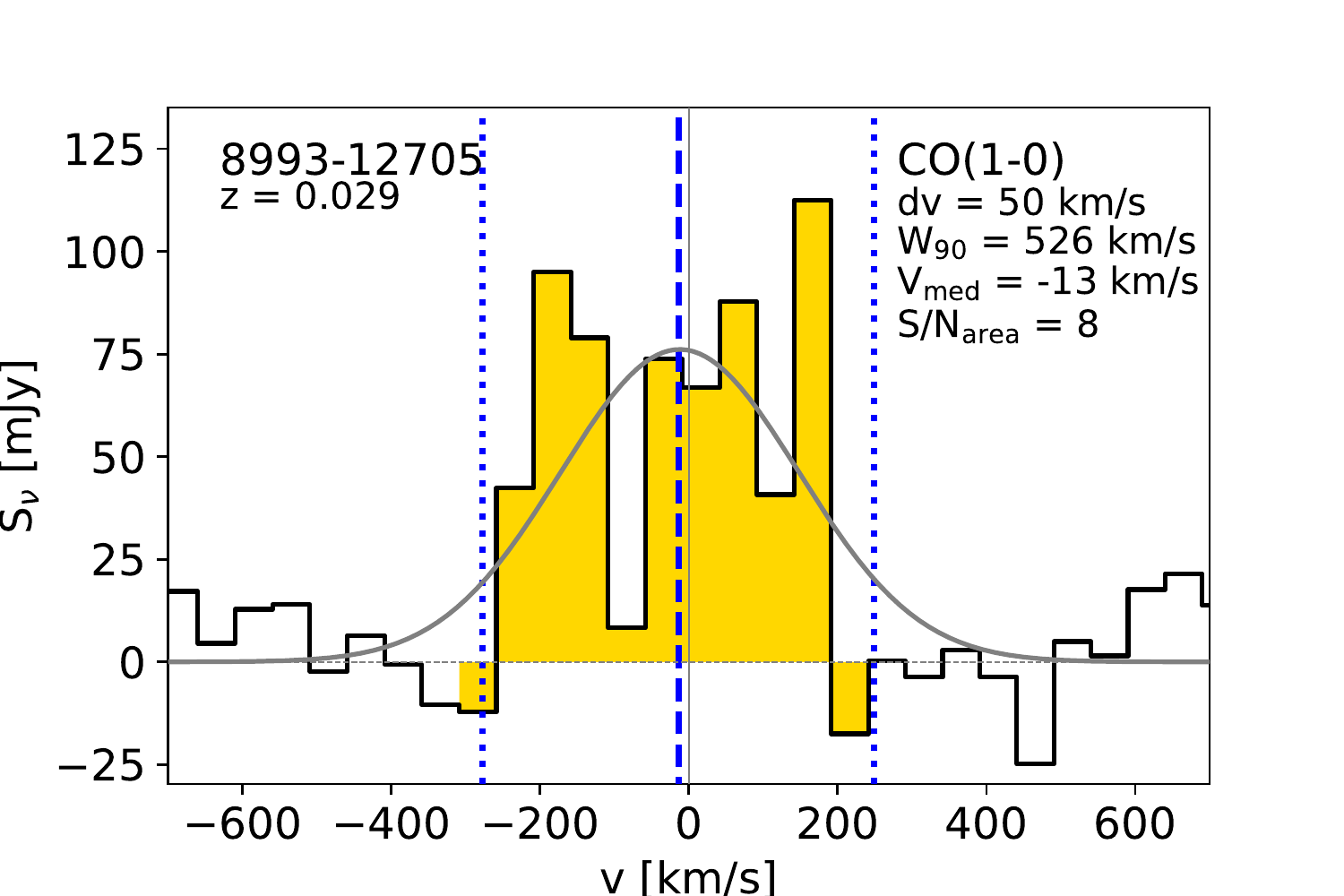}
\end{figure*}

\begin{figure*}
   \ContinuedFloat 
  \centering  \includegraphics[width = 0.17\textwidth, trim = 0cm 0cm 0cm 0cm, clip = true]{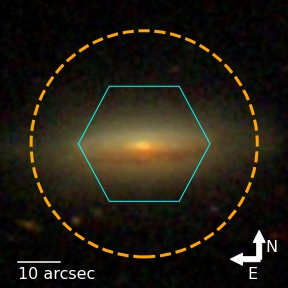} \includegraphics[width = 0.29\textwidth, trim = 0cm 0cm 0cm 0cm, clip = true]{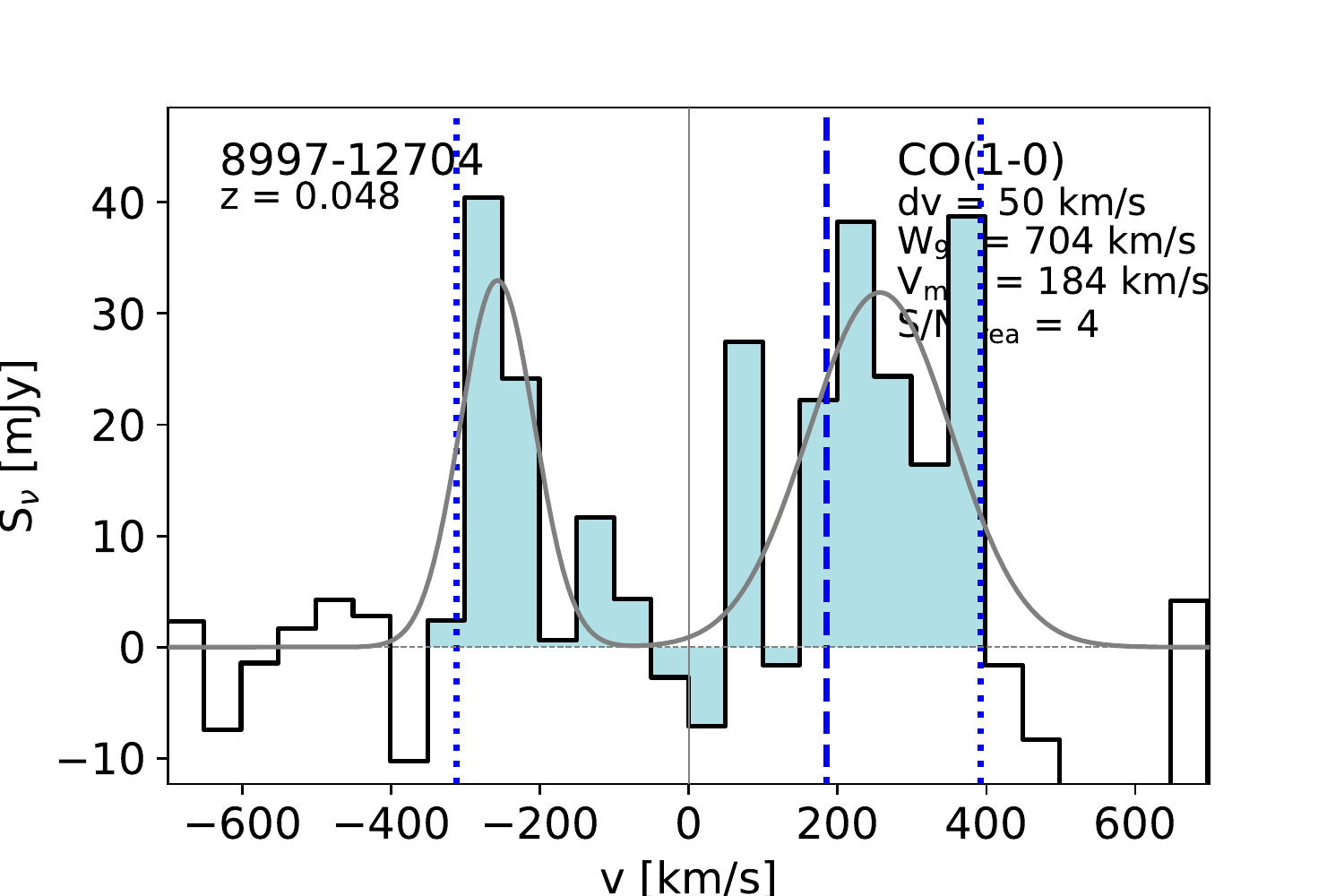}
\hspace{0.4cm}  \centering  \includegraphics[width = 0.17\textwidth, trim = 0cm 0cm 0cm 0cm, clip = true]{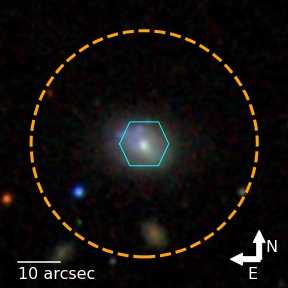} \includegraphics[width = 0.29\textwidth, trim = 0cm 0cm 0cm 0cm, clip = true]{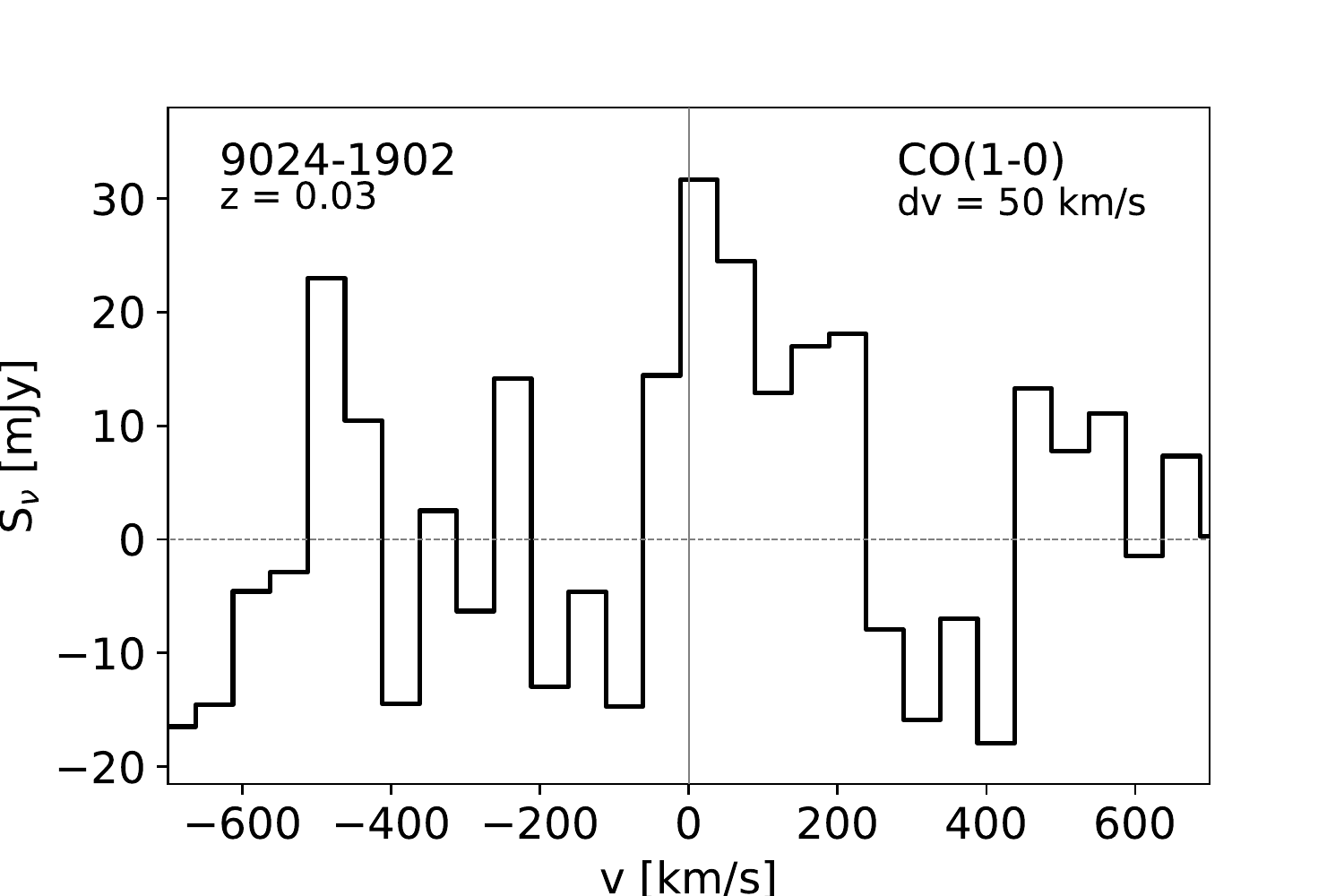}
  \caption{continued.}
\end{figure*}

\begin{figure*}  \centering  \includegraphics[width = 0.17\textwidth, trim = 0cm 0cm 0cm 0cm, clip = true]{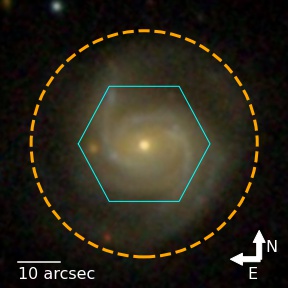} \includegraphics[width = 0.29\textwidth, trim = 0cm 0cm 0cm 0cm, clip = true]{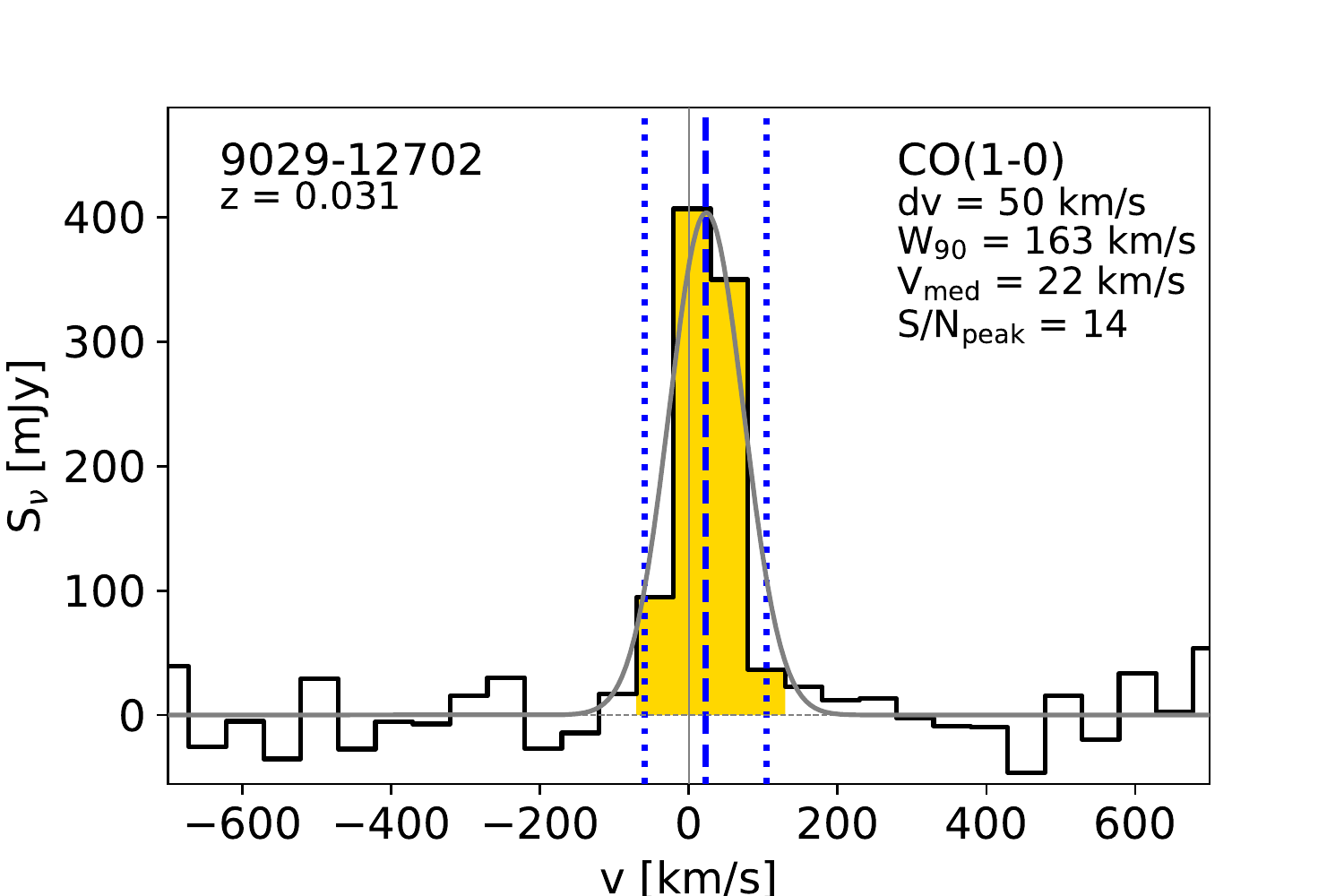}
\hspace{0.4cm}  \centering  \includegraphics[width = 0.17\textwidth, trim = 0cm 0cm 0cm 0cm, clip = true]{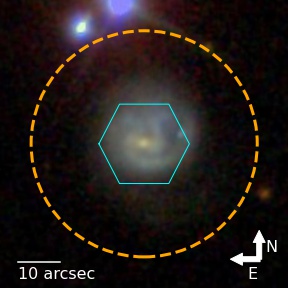} \includegraphics[width = 0.29\textwidth, trim = 0cm 0cm 0cm 0cm, clip = true]{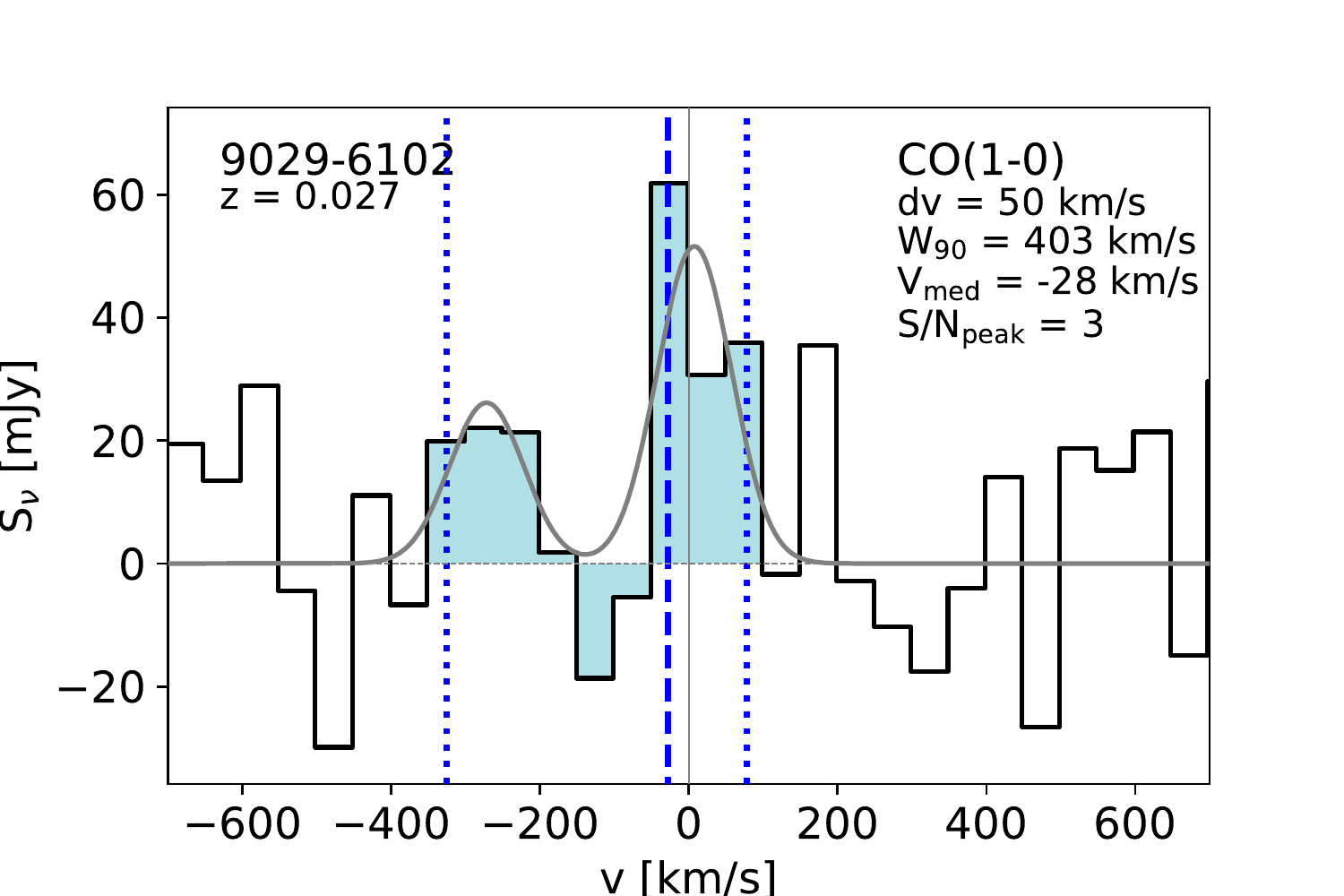}
\end{figure*}

\begin{figure*}  \centering  \includegraphics[width = 0.17\textwidth, trim = 0cm 0cm 0cm 0cm, clip = true]{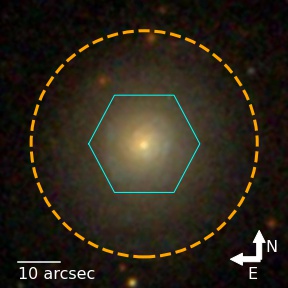} \includegraphics[width = 0.29\textwidth, trim = 0cm 0cm 0cm 0cm, clip = true]{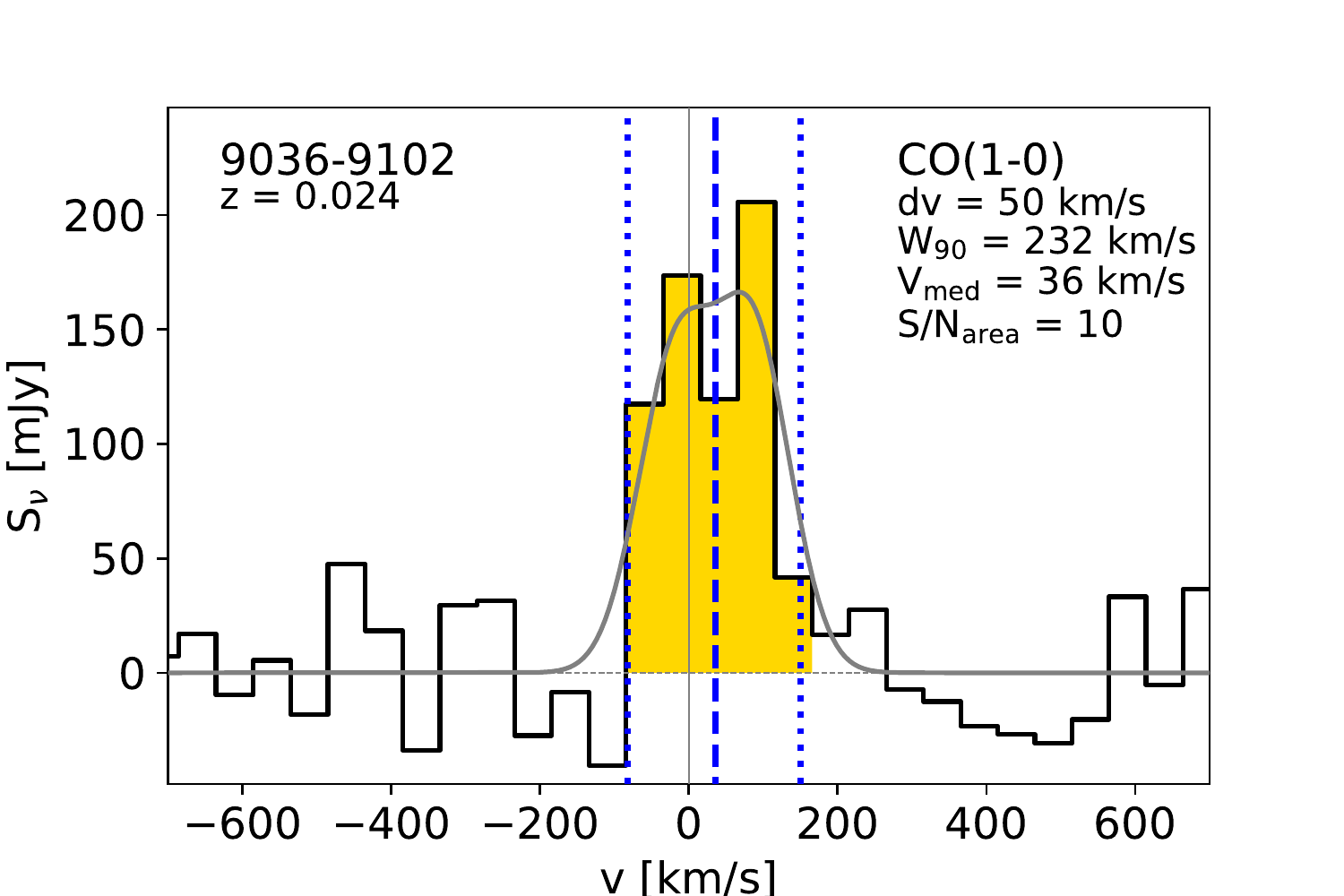}
\hspace{0.4cm}  \centering  \includegraphics[width = 0.17\textwidth, trim = 0cm 0cm 0cm 0cm, clip = true]{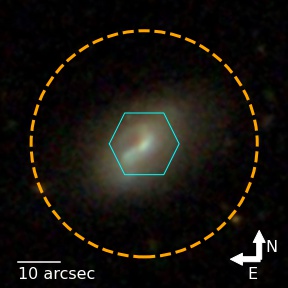} \includegraphics[width = 0.29\textwidth, trim = 0cm 0cm 0cm 0cm, clip = true]{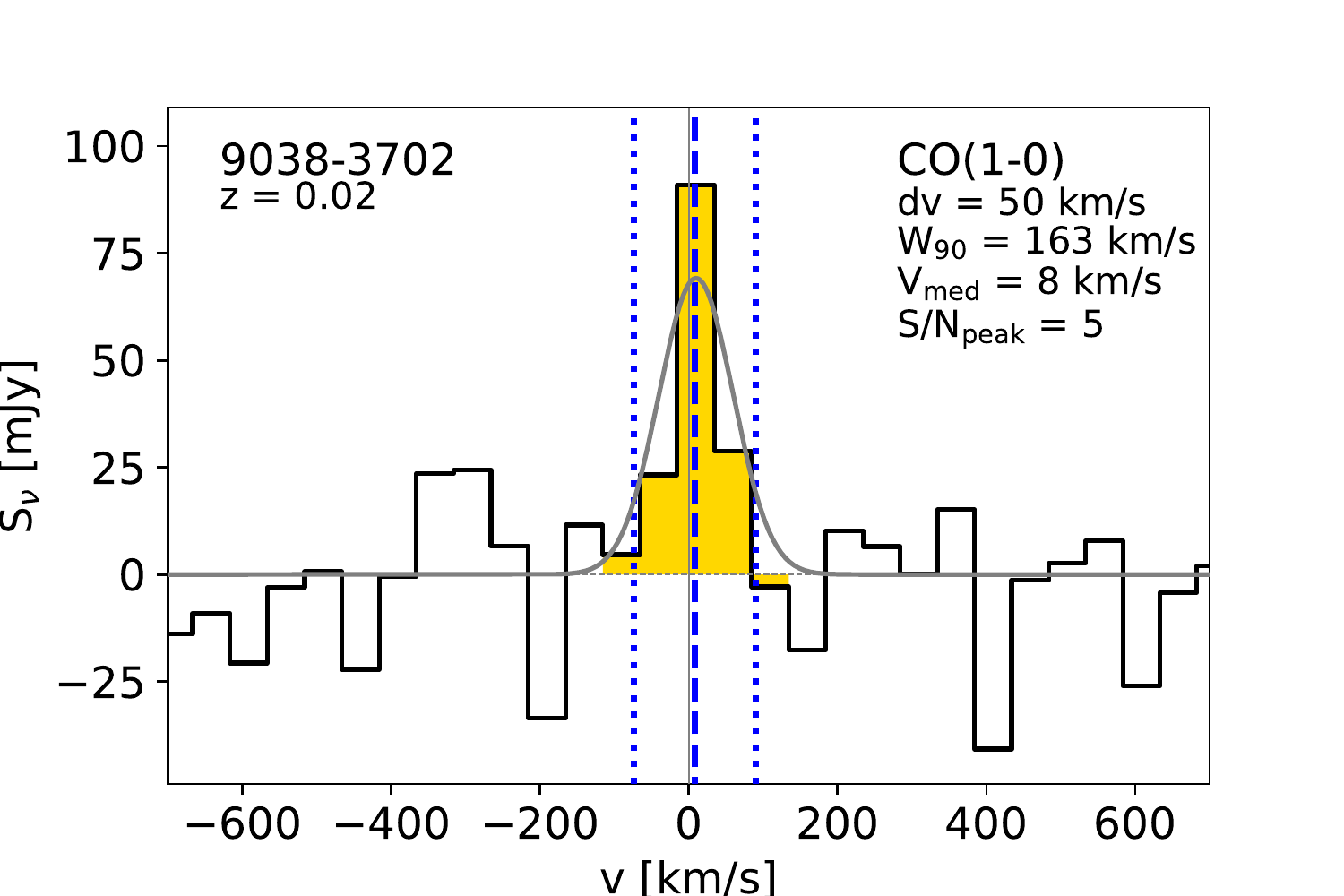}
\end{figure*}

\begin{figure*}  \centering  \includegraphics[width = 0.17\textwidth, trim = 0cm 0cm 0cm 0cm, clip = true]{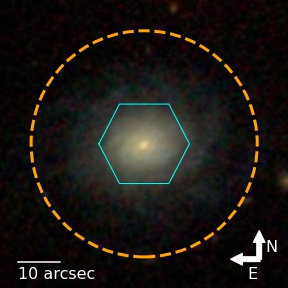} \includegraphics[width = 0.29\textwidth, trim = 0cm 0cm 0cm 0cm, clip = true]{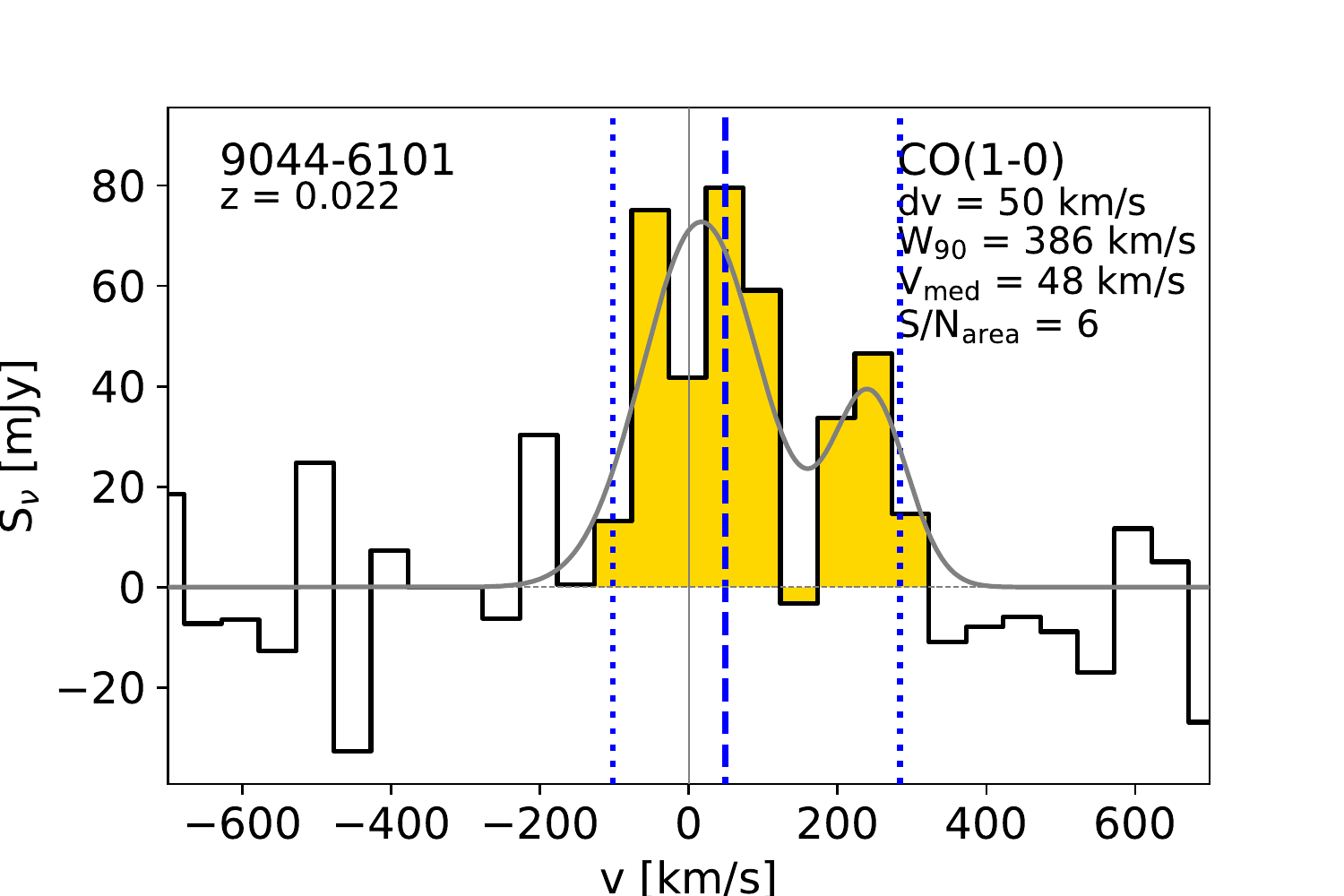}
\hspace{0.4cm}  \centering  \includegraphics[width = 0.17\textwidth, trim = 0cm 0cm 0cm 0cm, clip = true]{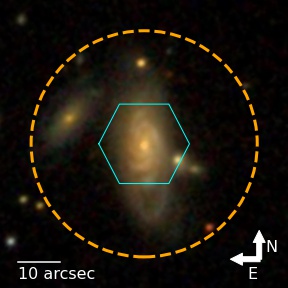} \includegraphics[width = 0.29\textwidth, trim = 0cm 0cm 0cm 0cm, clip = true]{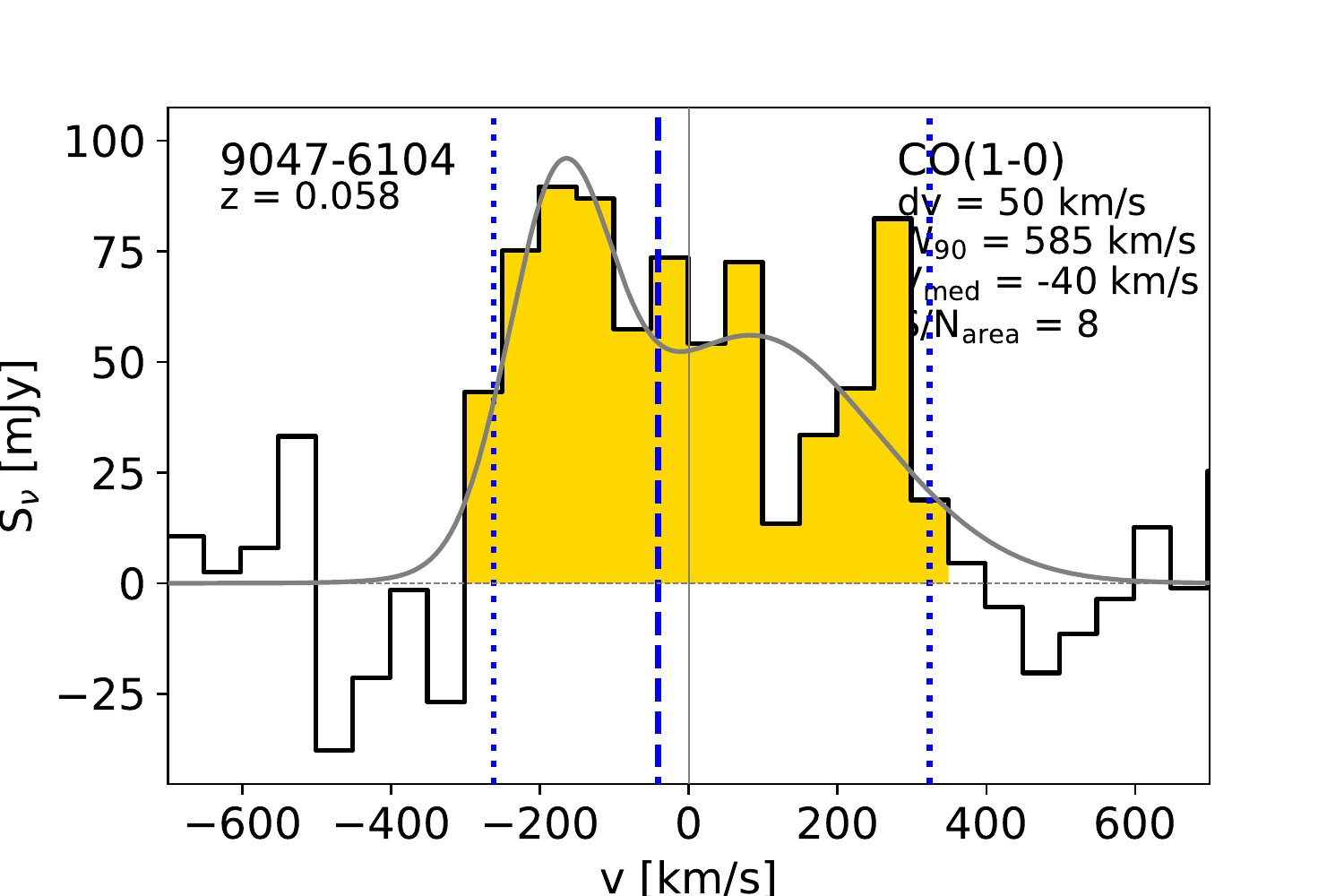}
\end{figure*}

\begin{figure*}  \centering  \includegraphics[width = 0.17\textwidth, trim = 0cm 0cm 0cm 0cm, clip = true]{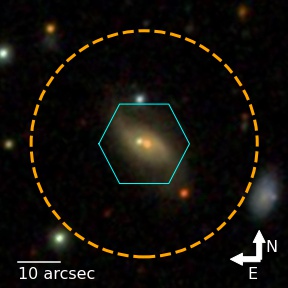} \includegraphics[width = 0.29\textwidth, trim = 0cm 0cm 0cm 0cm, clip = true]{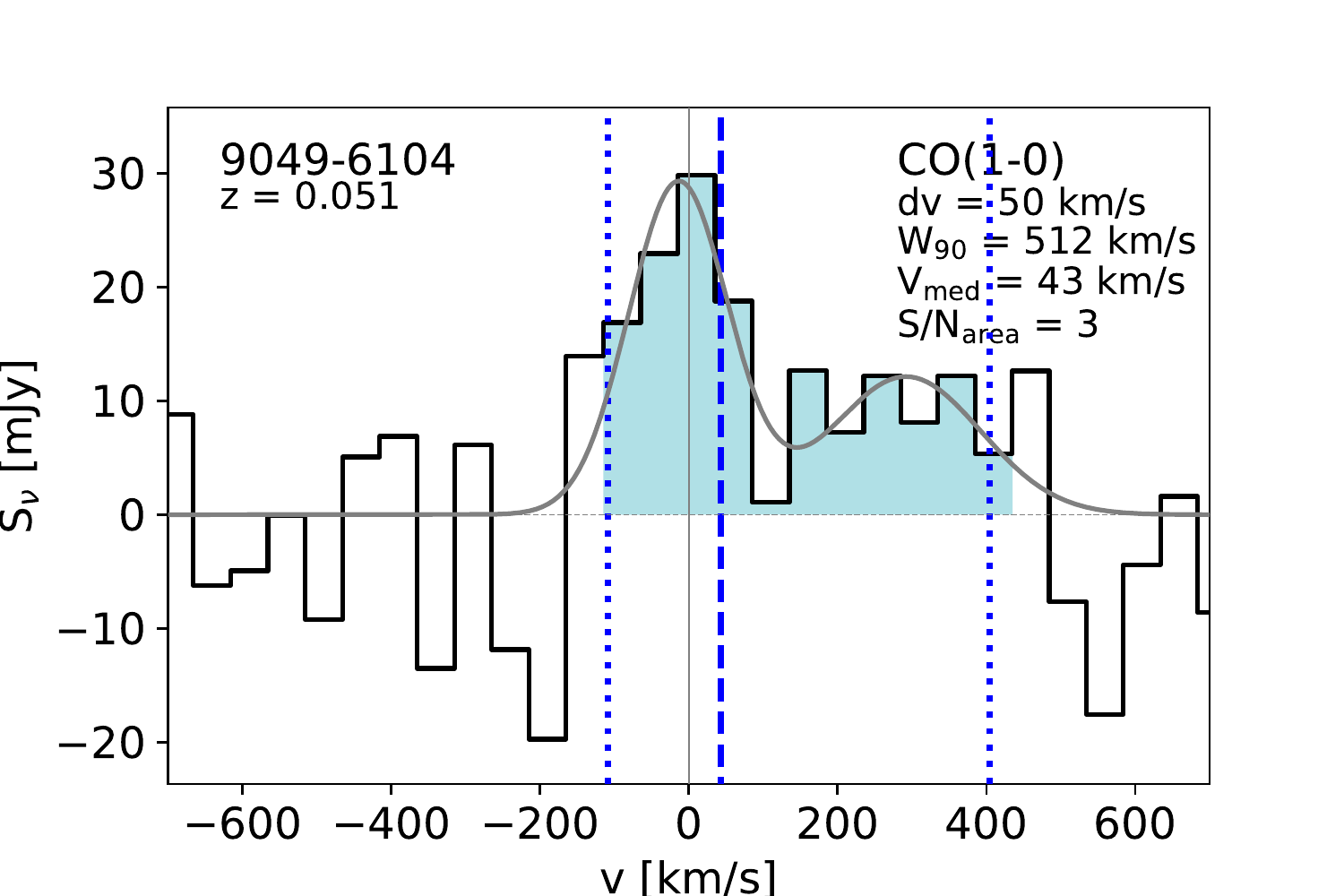}
\hspace{0.4cm}  \centering  \includegraphics[width = 0.17\textwidth, trim = 0cm 0cm 0cm 0cm, clip = true]{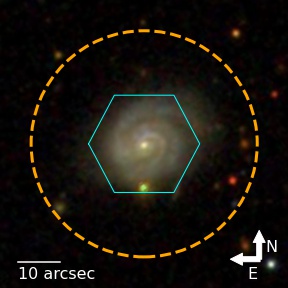} \includegraphics[width = 0.29\textwidth, trim = 0cm 0cm 0cm 0cm, clip = true]{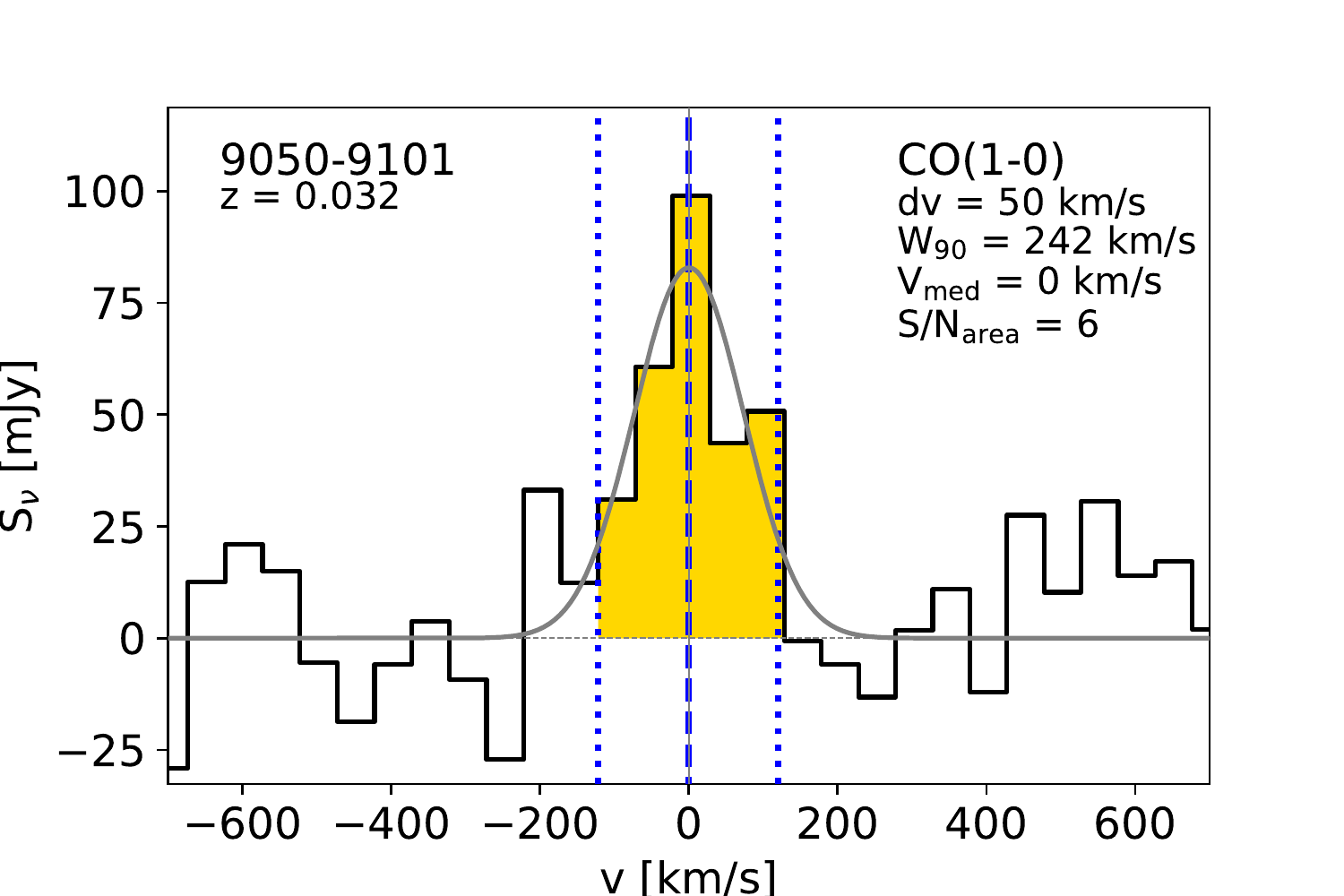}
\end{figure*}

\begin{figure*}  \centering  \includegraphics[width = 0.17\textwidth, trim = 0cm 0cm 0cm 0cm, clip = true]{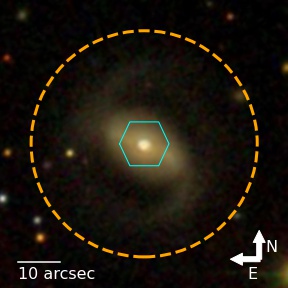} \includegraphics[width = 0.29\textwidth, trim = 0cm 0cm 0cm 0cm, clip = true]{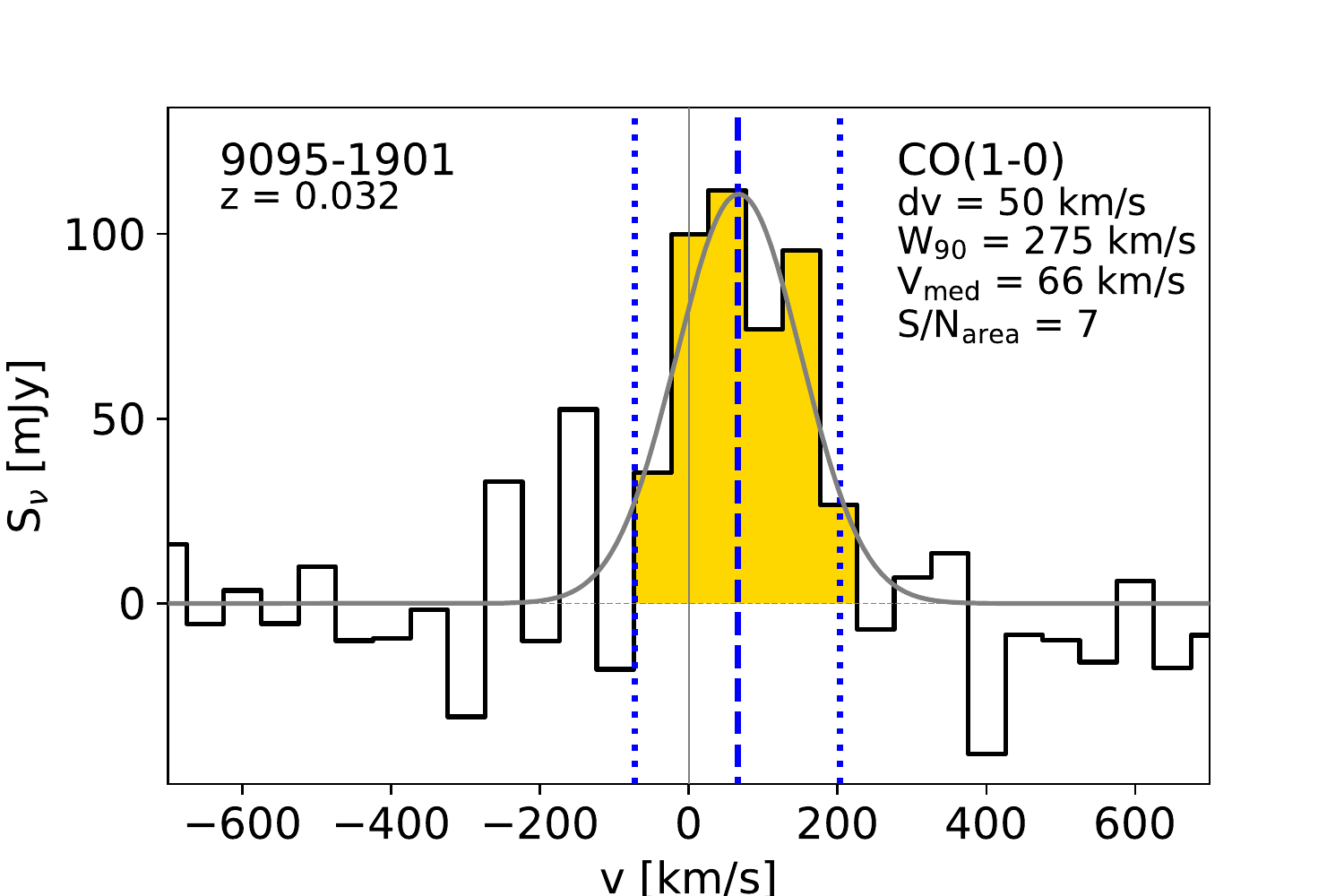}
\hspace{0.4cm}  \centering  \includegraphics[width = 0.17\textwidth, trim = 0cm 0cm 0cm 0cm, clip = true]{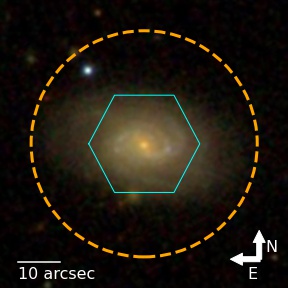} \includegraphics[width = 0.29\textwidth, trim = 0cm 0cm 0cm 0cm, clip = true]{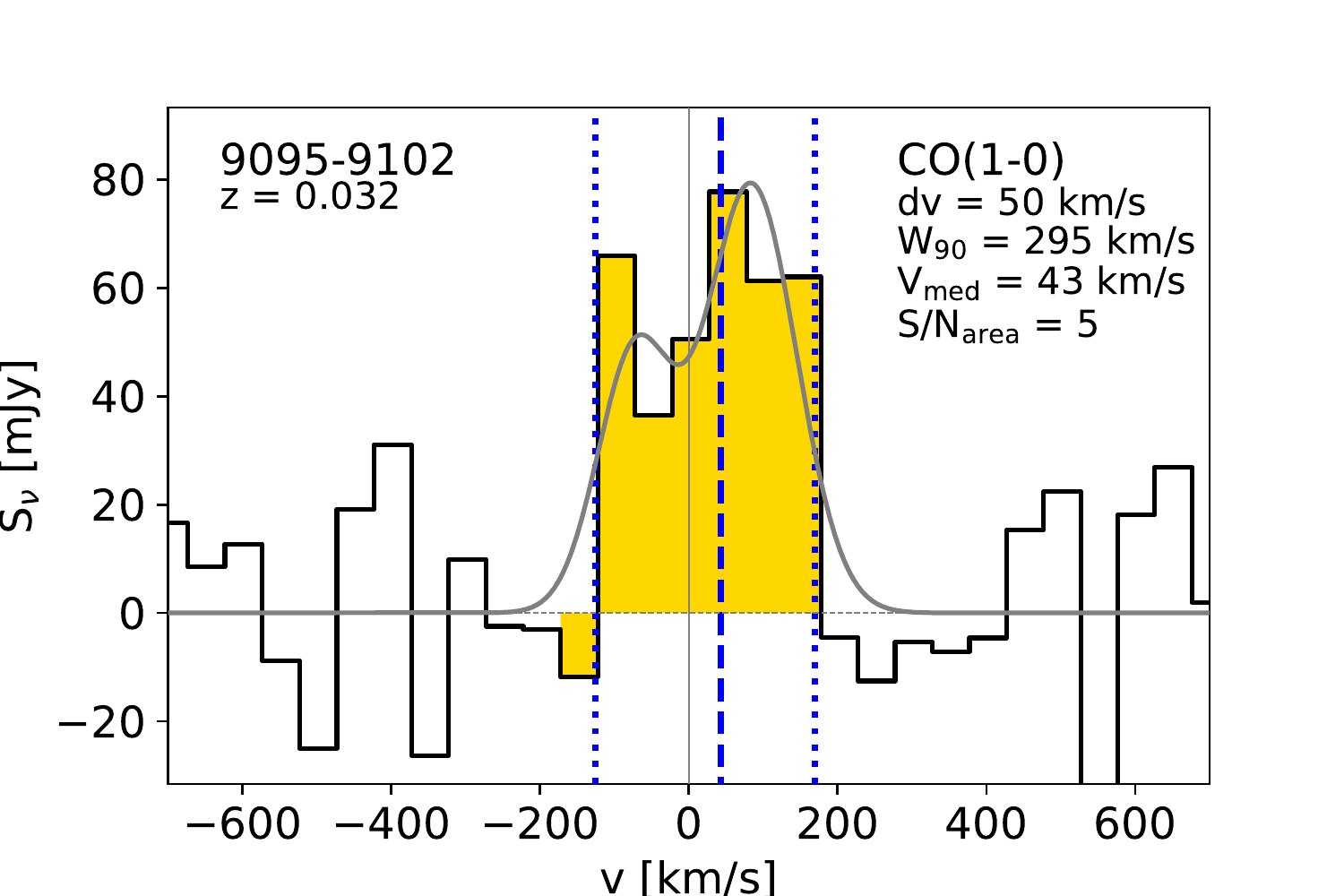}
\end{figure*}

\begin{figure*} 
   \ContinuedFloat 
 \centering  \includegraphics[width = 0.17\textwidth, trim = 0cm 0cm 0cm 0cm, clip = true]{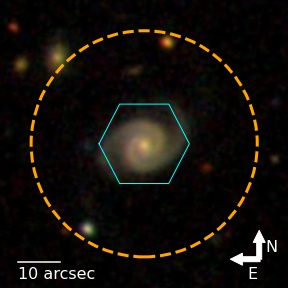} \includegraphics[width = 0.29\textwidth, trim = 0cm 0cm 0cm 0cm, clip = true]{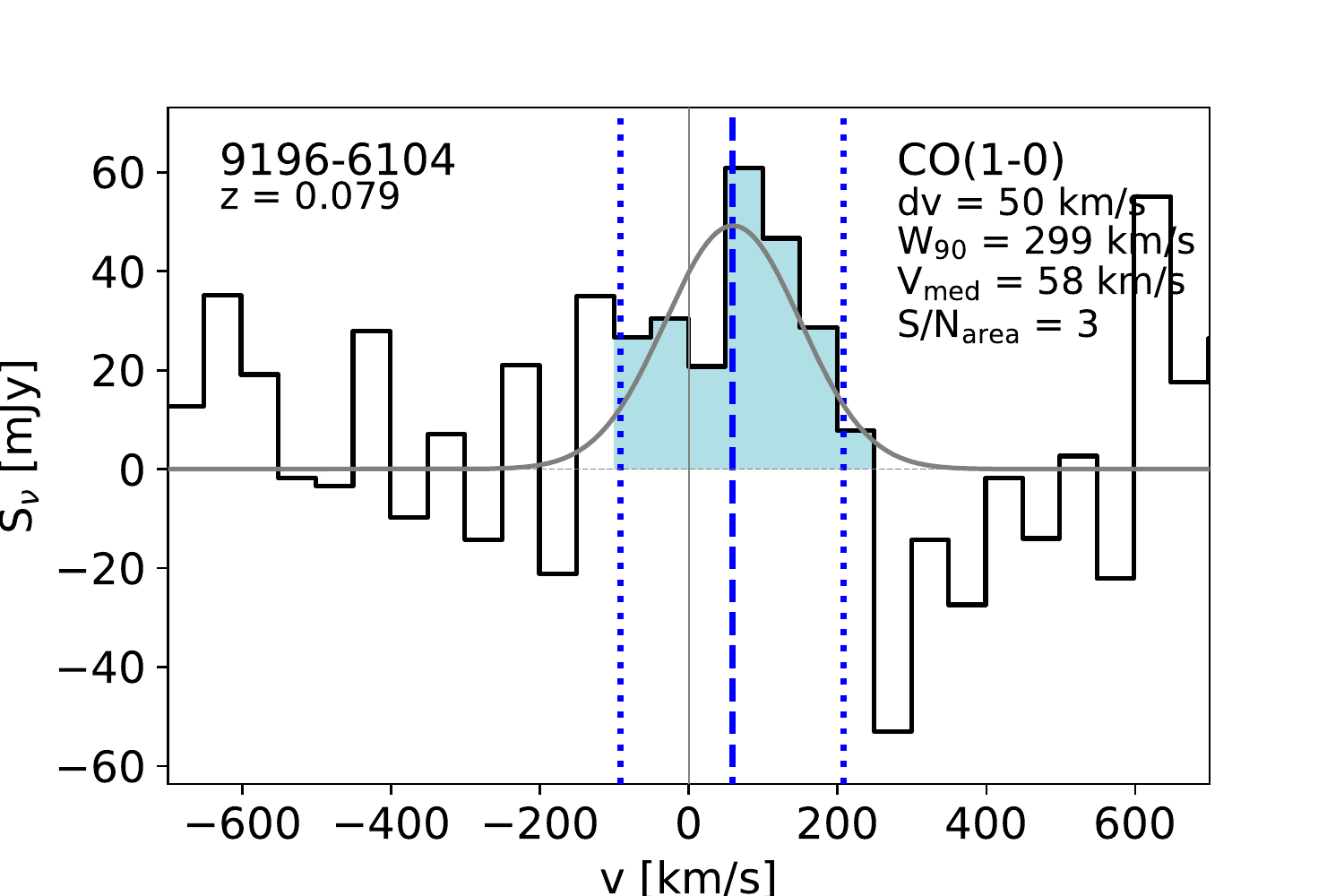}
\hspace{0.4cm}  \centering  \includegraphics[width = 0.17\textwidth, trim = 0cm 0cm 0cm 0cm, clip = true]{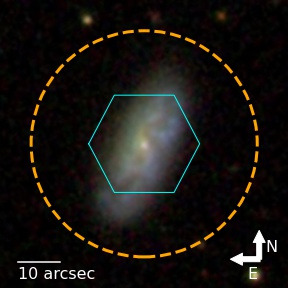} \includegraphics[width = 0.29\textwidth, trim = 0cm 0cm 0cm 0cm, clip = true]{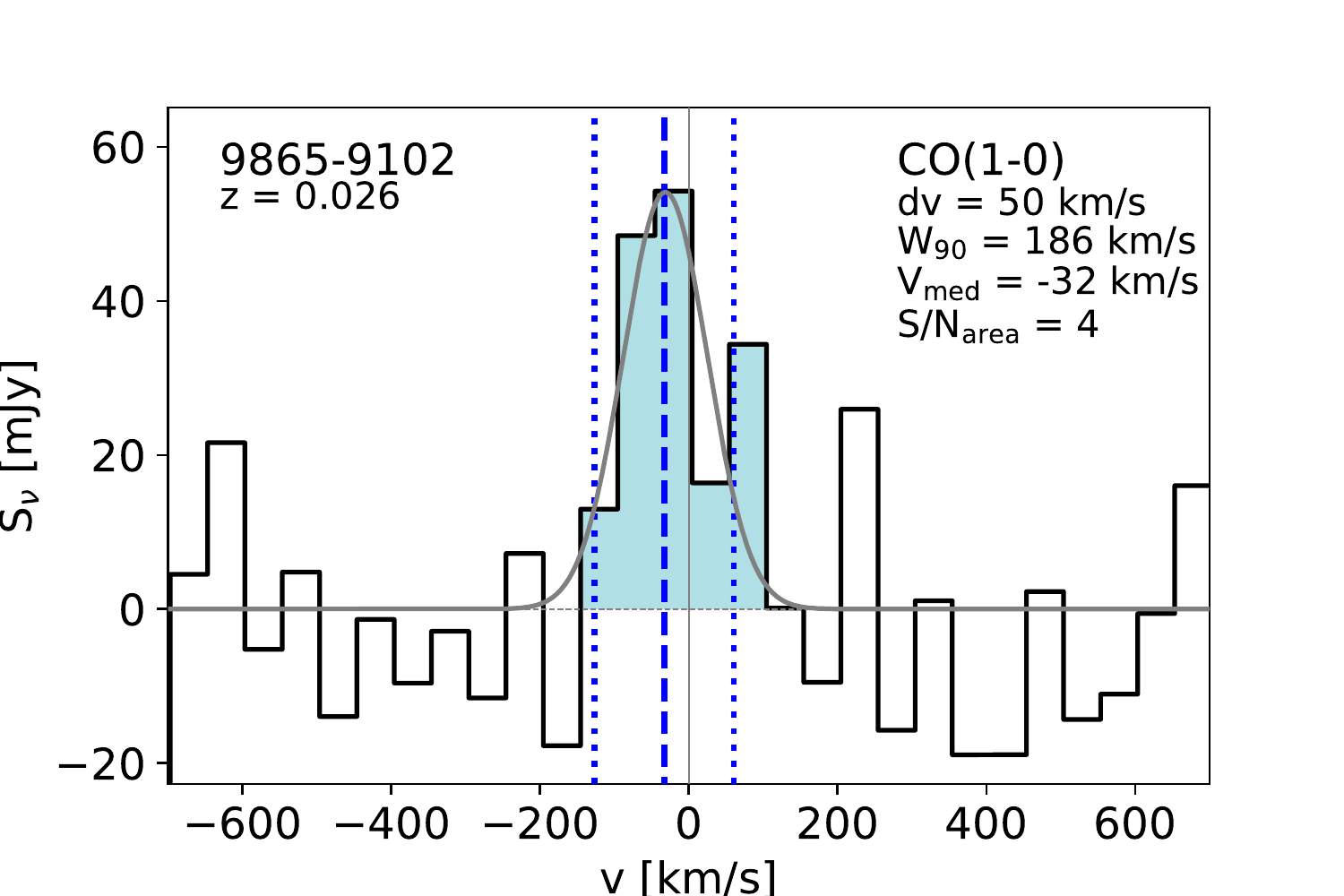}
 \caption{continued.}
\end{figure*}

\begin{figure*}    
 \centering  \includegraphics[width = 0.17\textwidth, trim = 0cm 0cm 0cm 0cm, clip = true]{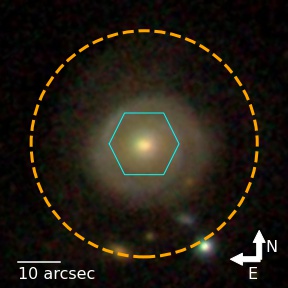} \includegraphics[width = 0.29\textwidth, trim = 0cm 0cm 0cm 0cm, clip = true]{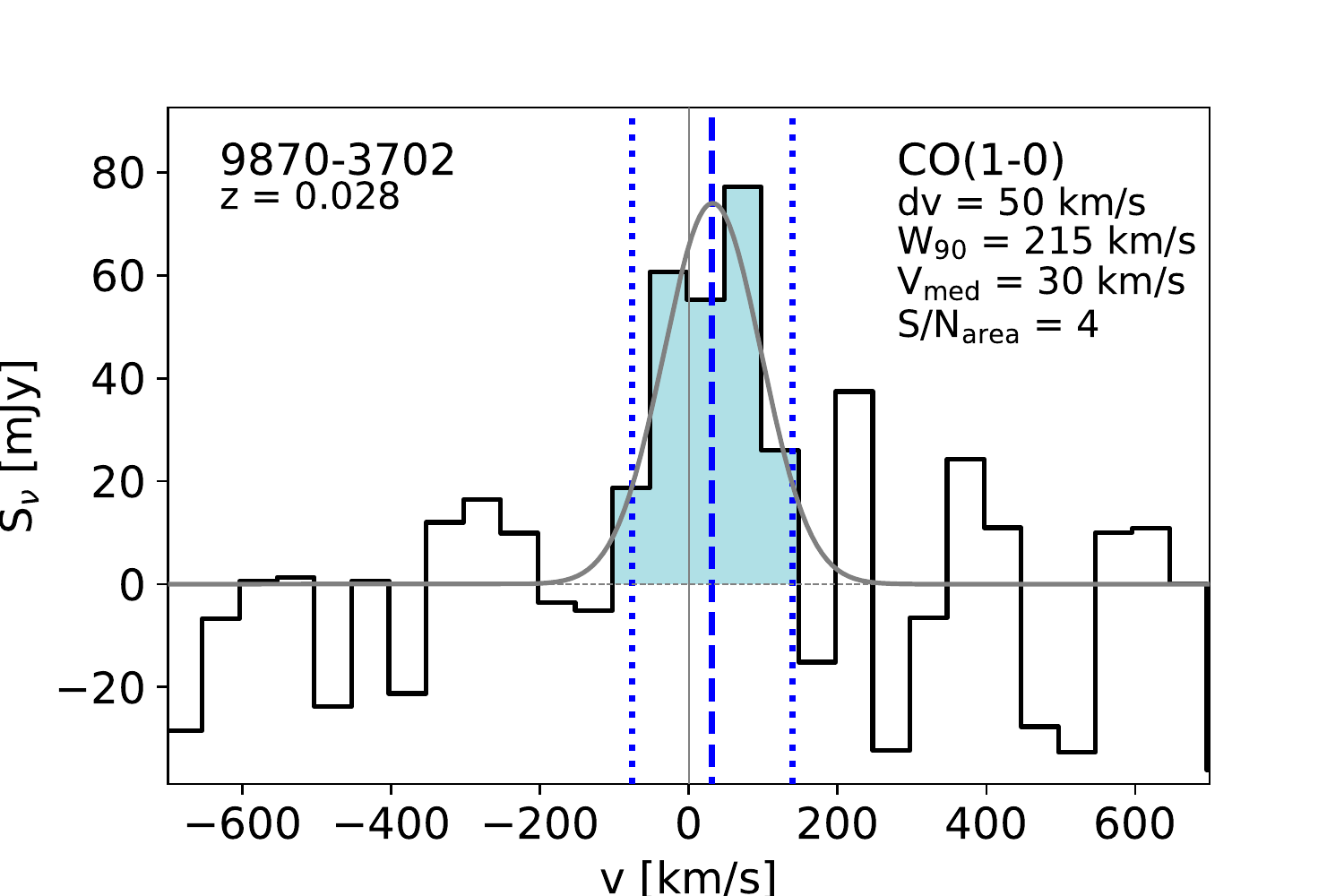}
\hspace{0.4cm}  \centering  \includegraphics[width = 0.17\textwidth, trim = 0cm 0cm 0cm 0cm, clip = true]{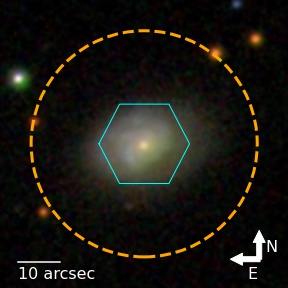} \includegraphics[width = 0.29\textwidth, trim = 0cm 0cm 0cm 0cm, clip = true]{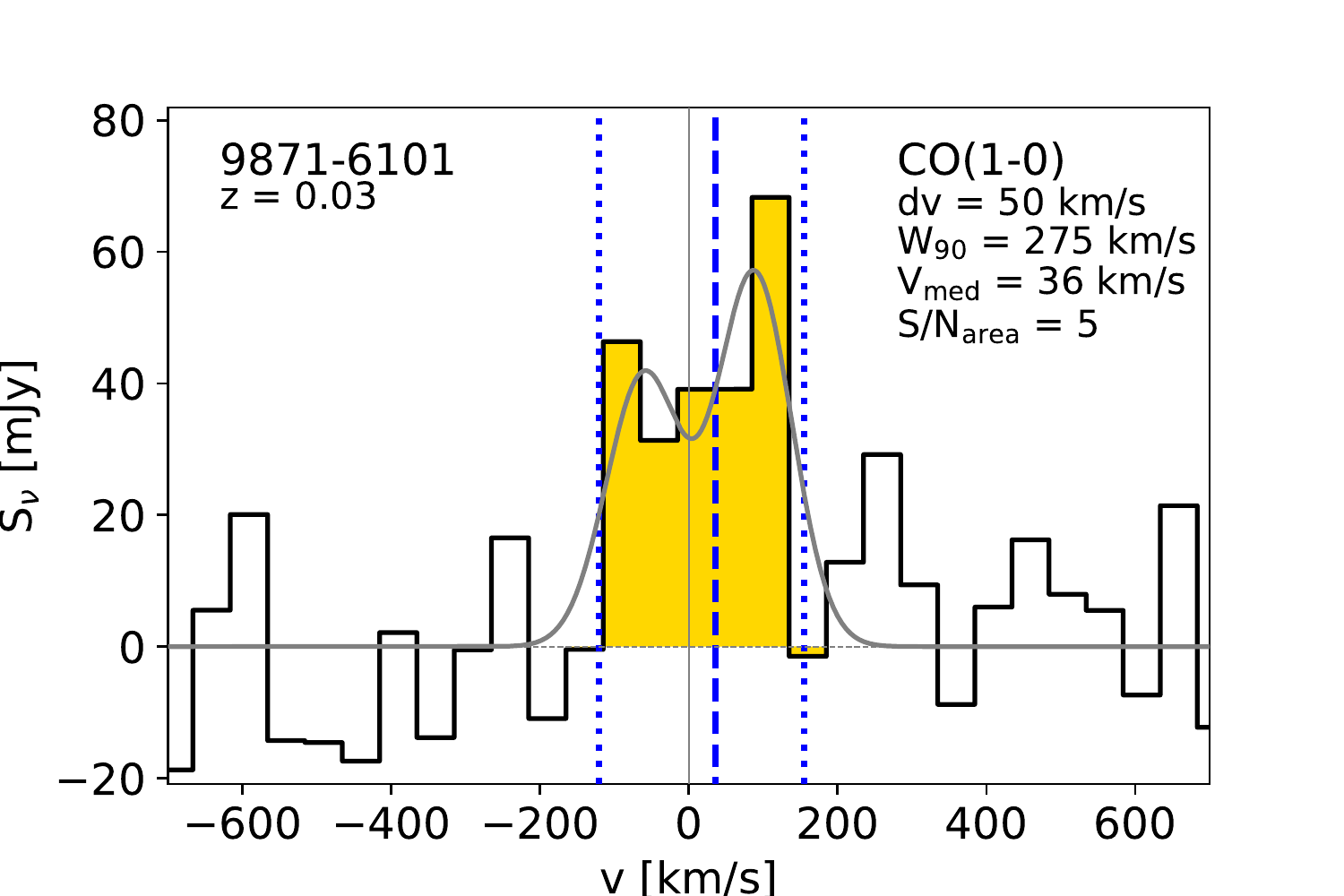}
\end{figure*}

\begin{figure*}    
 \centering  \includegraphics[width = 0.17\textwidth, trim = 0cm 0cm 0cm 0cm, clip = true]{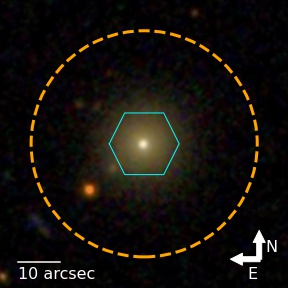} \includegraphics[width = 0.29\textwidth, trim = 0cm 0cm 0cm 0cm, clip = true]{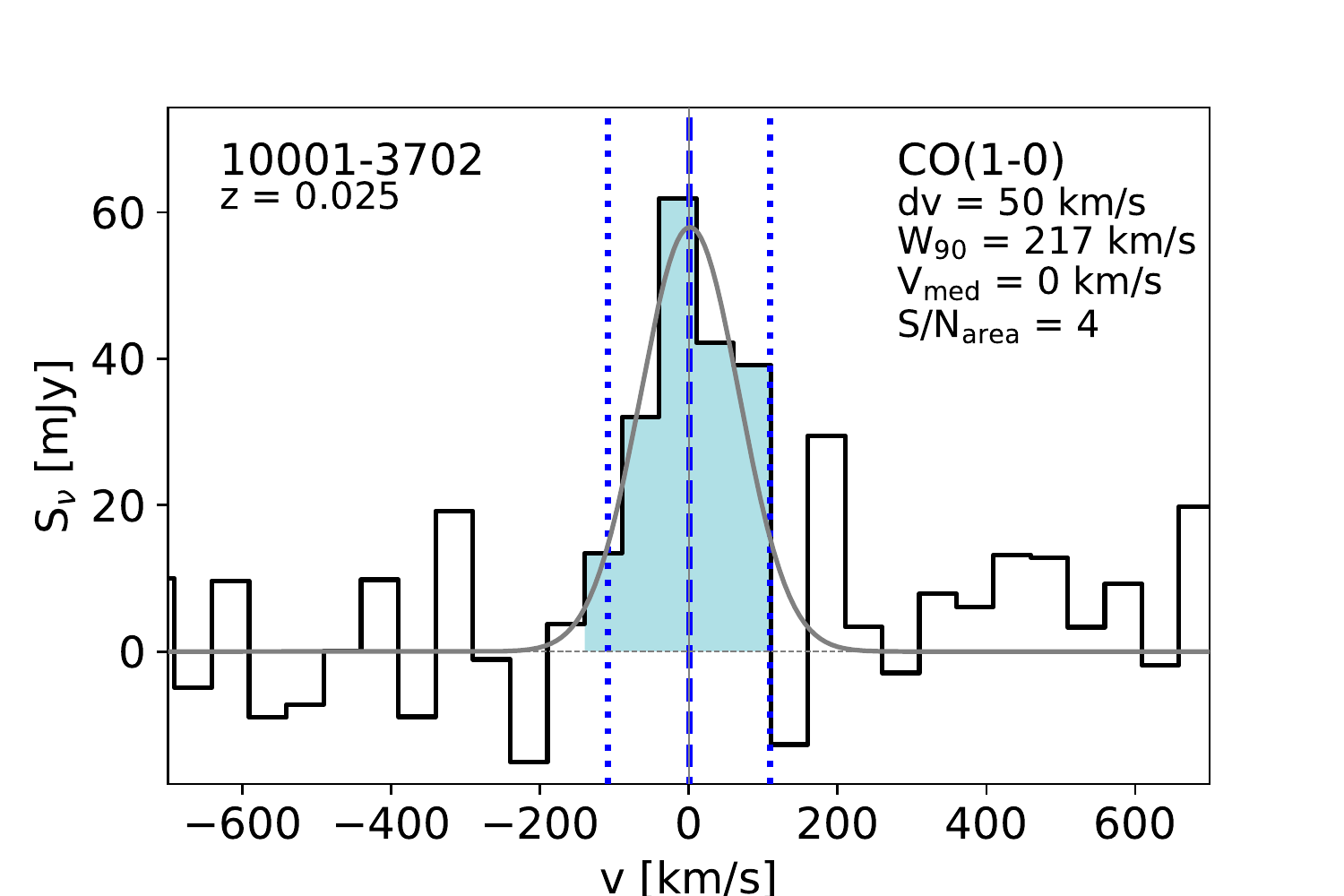}
\hspace{0.4cm}  \centering  \includegraphics[width = 0.17\textwidth, trim = 0cm 0cm 0cm 0cm, clip = true]{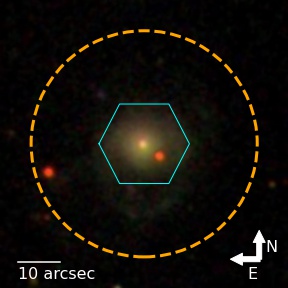} \includegraphics[width = 0.29\textwidth, trim = 0cm 0cm 0cm 0cm, clip = true]{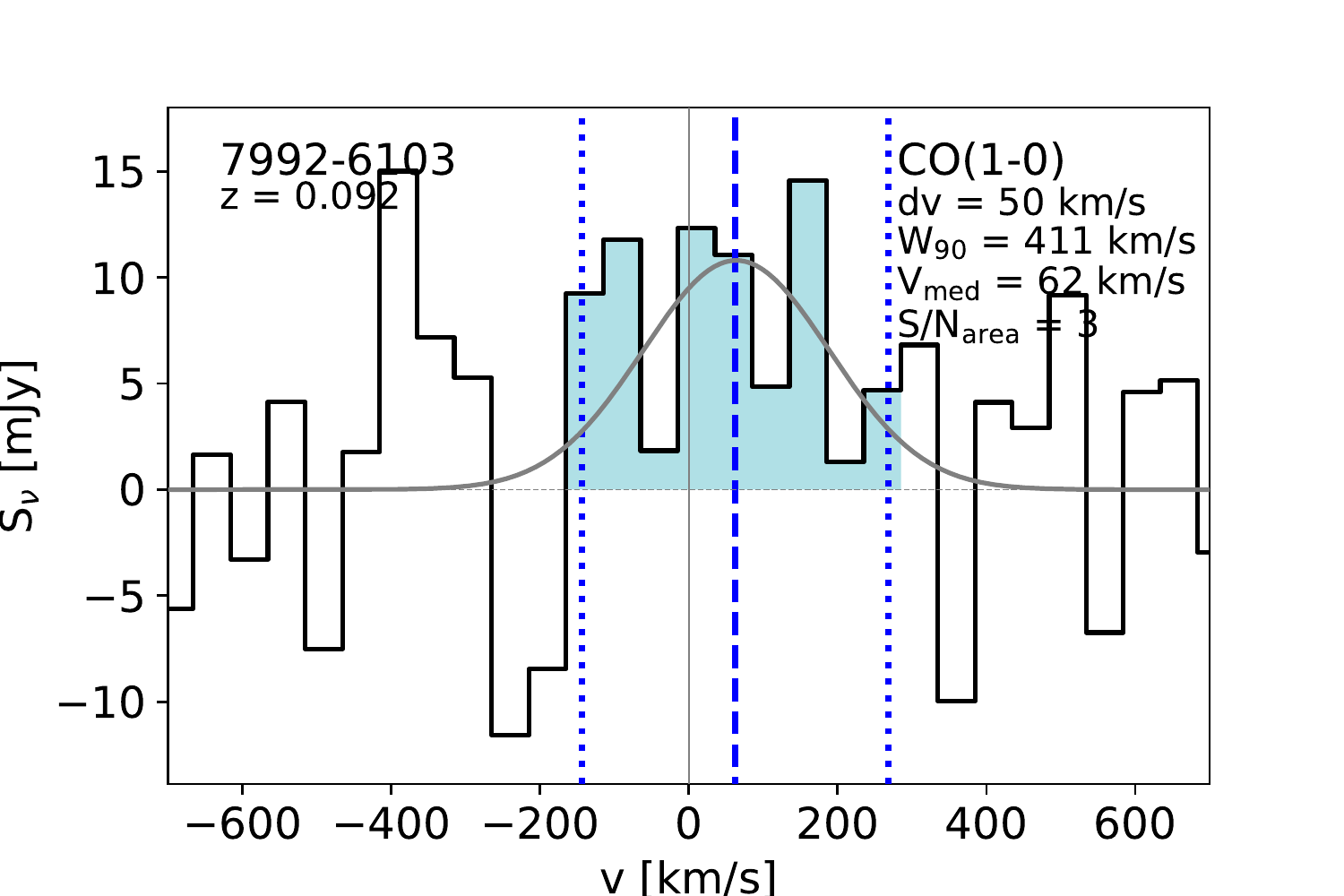}
\end{figure*}

\begin{figure*}    
 \centering  \includegraphics[width = 0.17\textwidth, trim = 0cm 0cm 0cm 0cm, clip = true]{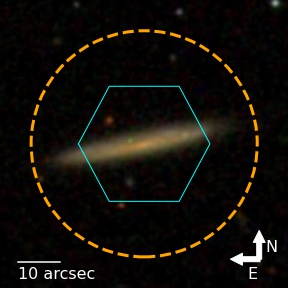} \includegraphics[width = 0.29\textwidth, trim = 0cm 0cm 0cm 0cm, clip = true]{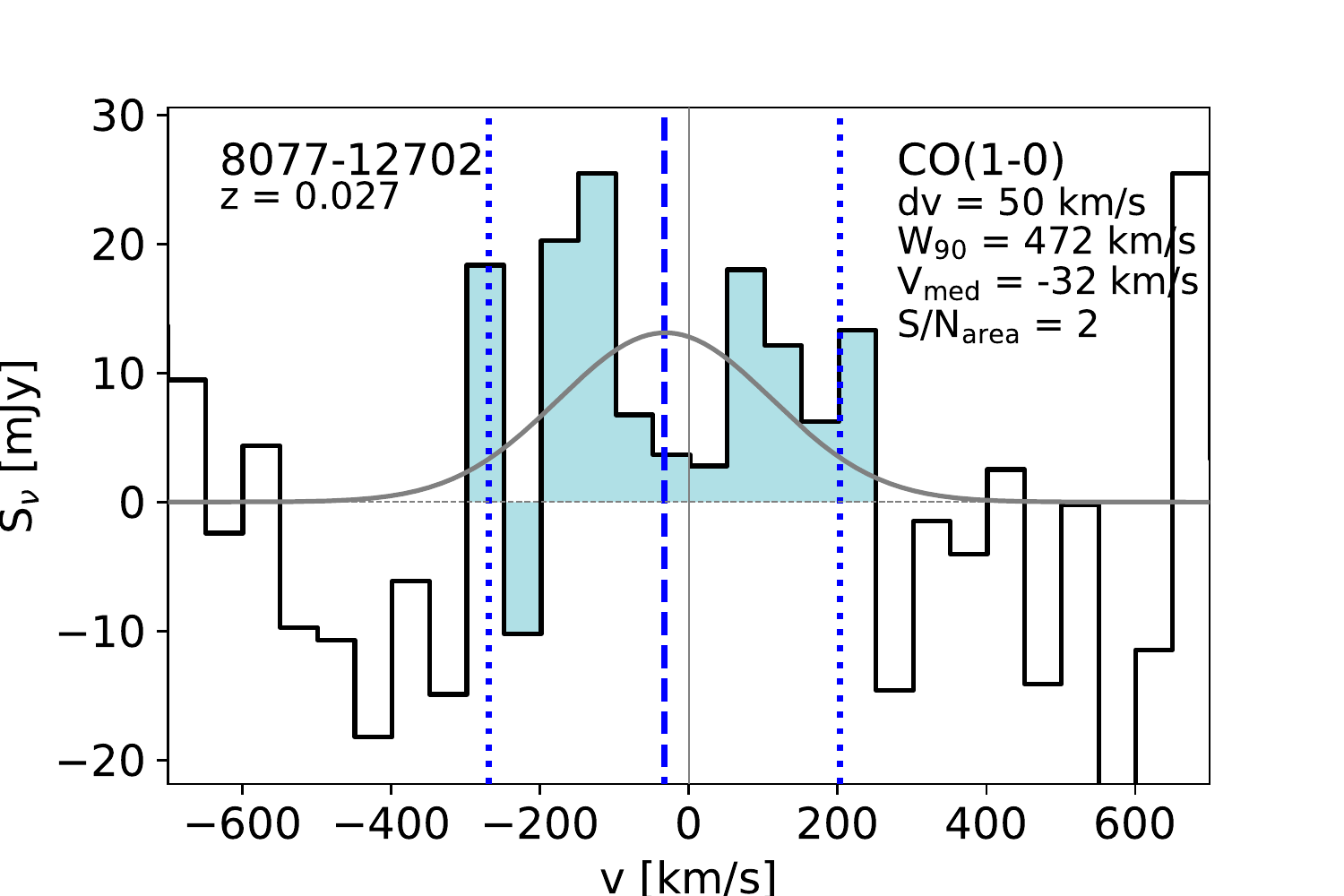}
\hspace{0.4cm}  \centering  \includegraphics[width = 0.17\textwidth, trim = 0cm 0cm 0cm 0cm, clip = true]{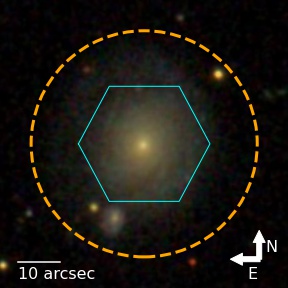} \includegraphics[width = 0.29\textwidth, trim = 0cm 0cm 0cm 0cm, clip = true]{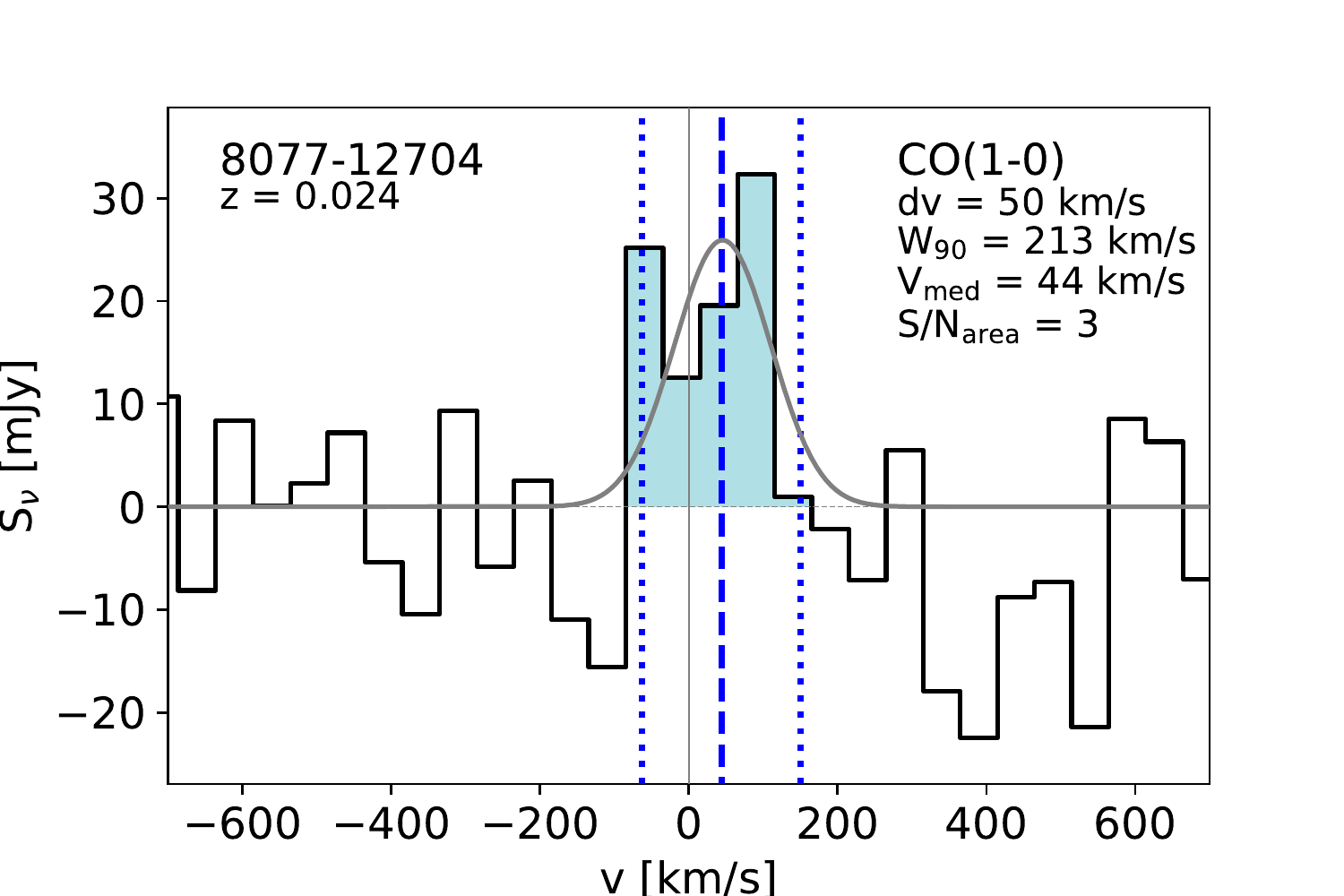}
\end{figure*}

\begin{figure*}    
 \centering  \includegraphics[width = 0.17\textwidth, trim = 0cm 0cm 0cm 0cm, clip = true]{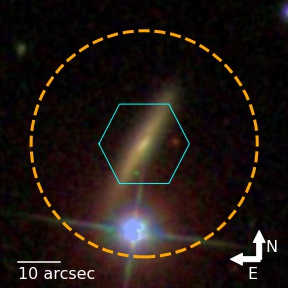} \includegraphics[width = 0.29\textwidth, trim = 0cm 0cm 0cm 0cm, clip = true]{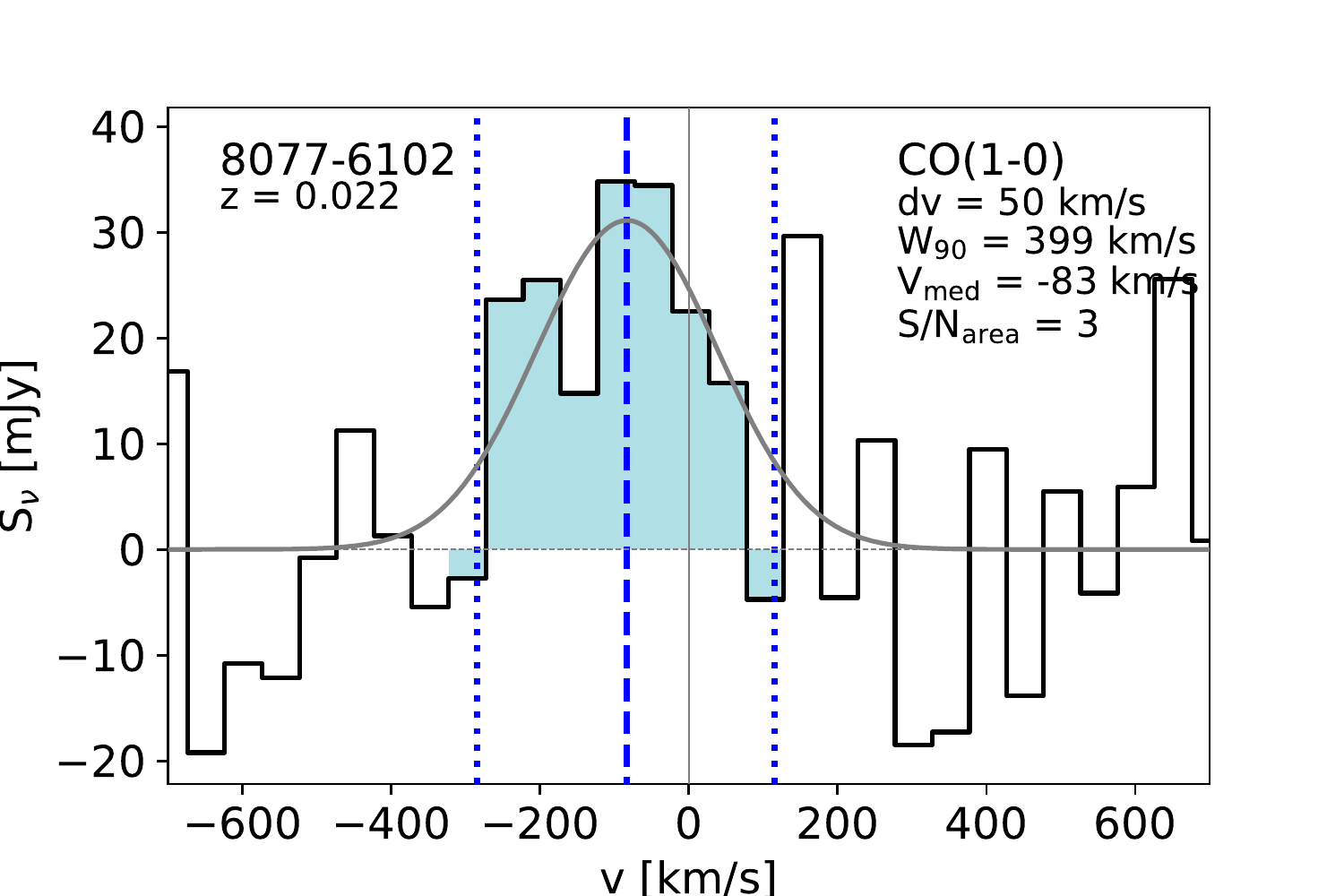}
\hspace{0.4cm}  \centering  \includegraphics[width = 0.17\textwidth, trim = 0cm 0cm 0cm 0cm, clip = true]{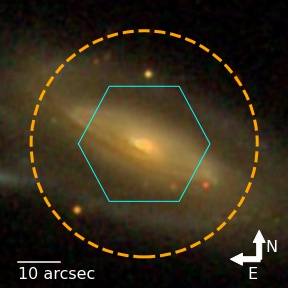} \includegraphics[width = 0.29\textwidth, trim = 0cm 0cm 0cm 0cm, clip = true]{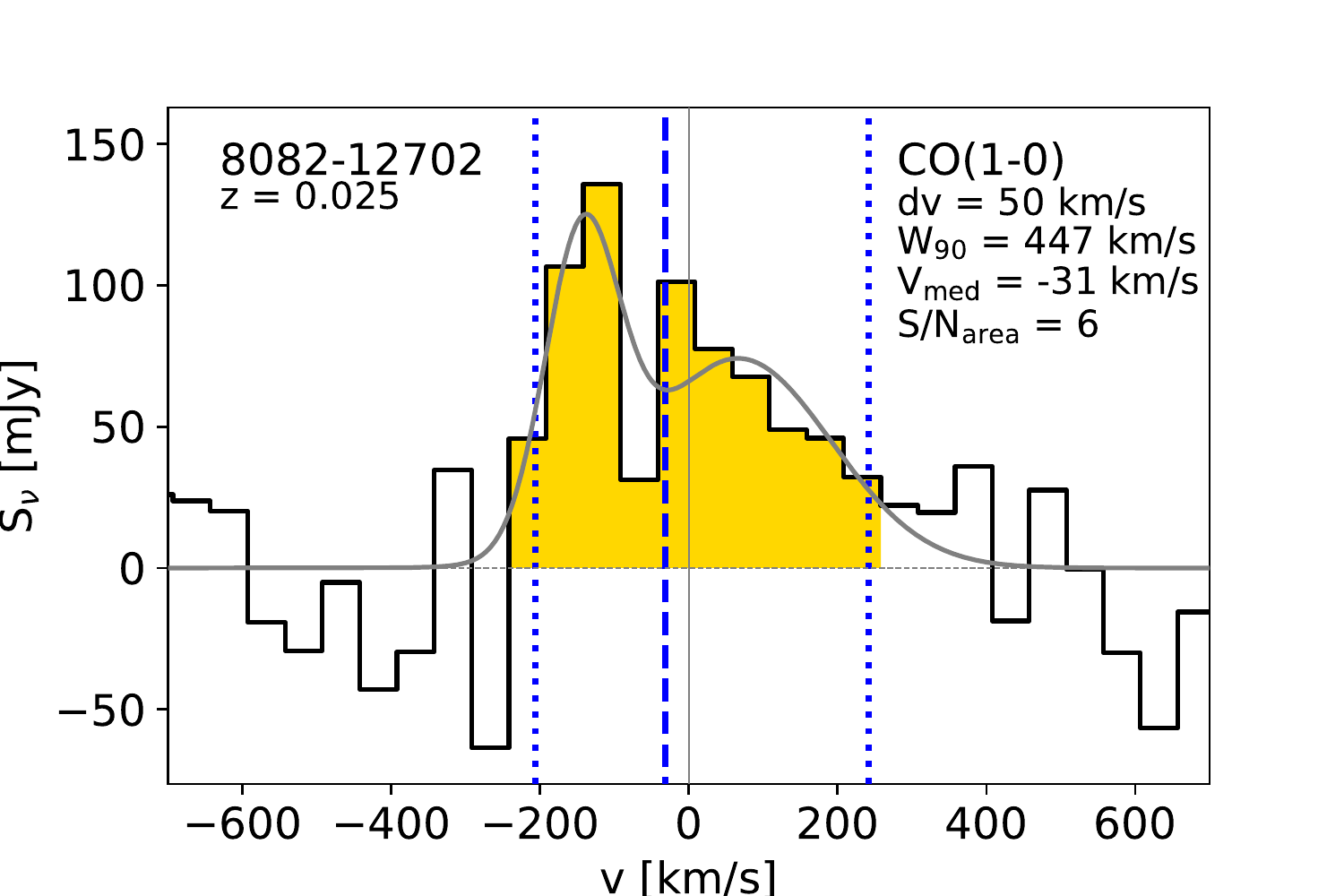}
\end{figure*}

\begin{figure*}    
 \centering  \includegraphics[width = 0.17\textwidth, trim = 0cm 0cm 0cm 0cm, clip = true]{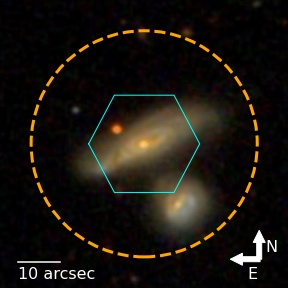} \includegraphics[width = 0.29\textwidth, trim = 0cm 0cm 0cm 0cm, clip = true]{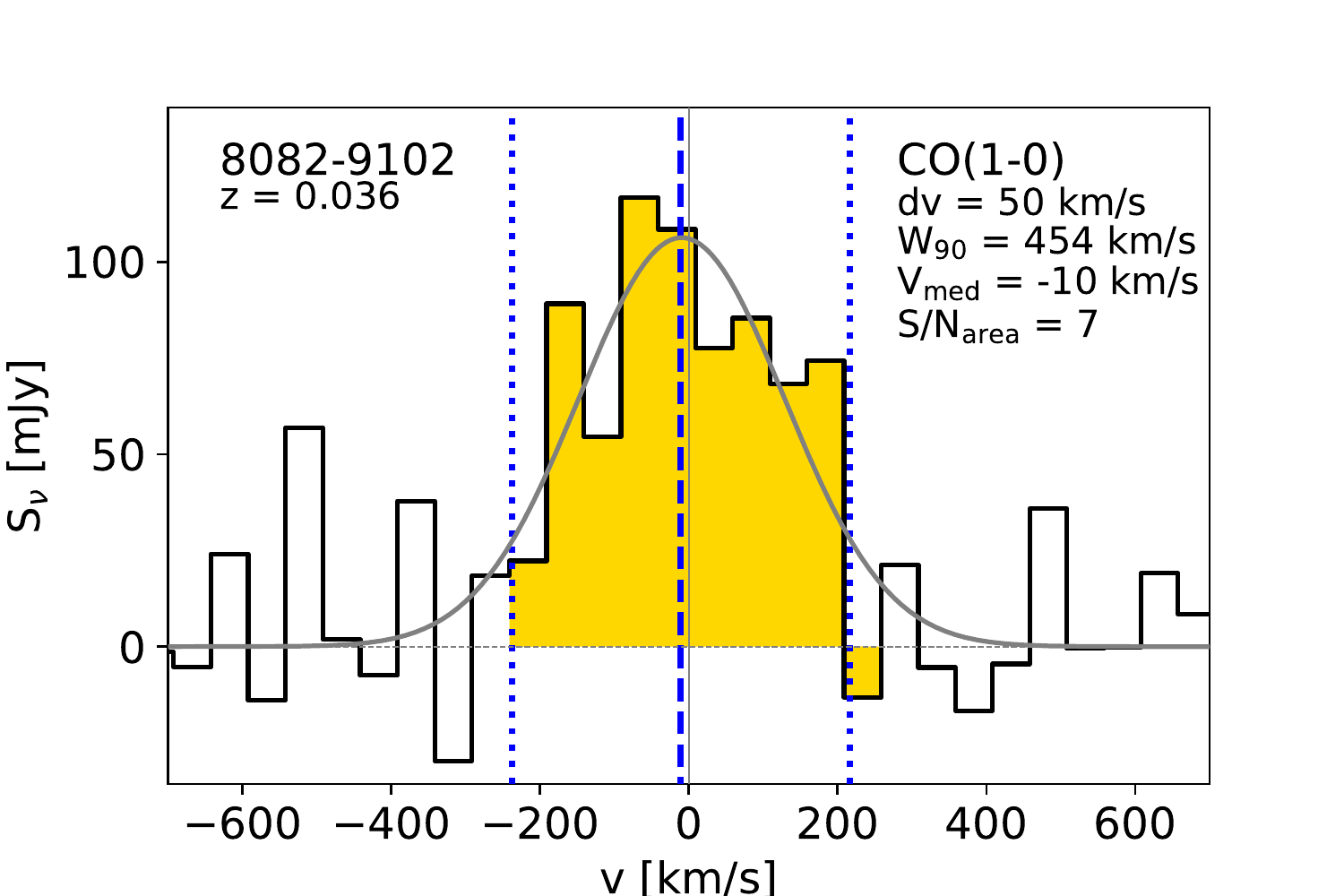}
\hspace{0.4cm}  \centering  \includegraphics[width = 0.17\textwidth, trim = 0cm 0cm 0cm 0cm, clip = true]{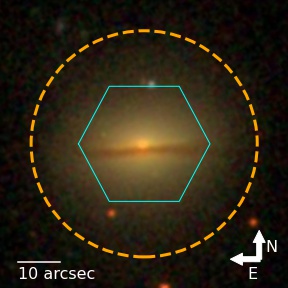} \includegraphics[width = 0.29\textwidth, trim = 0cm 0cm 0cm 0cm, clip = true]{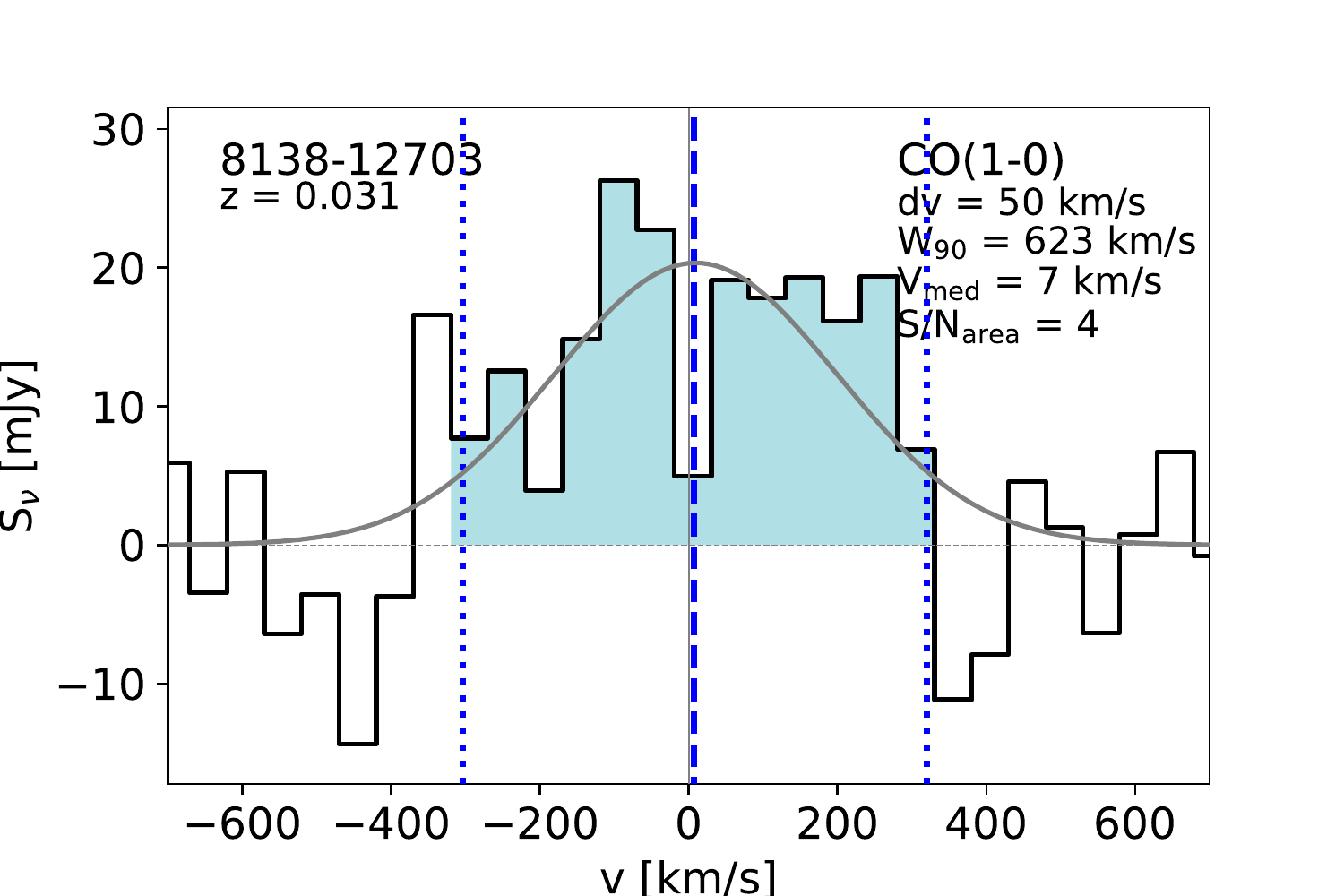}
\end{figure*}

\begin{figure*}
   \ContinuedFloat 
 \centering  \includegraphics[width = 0.17\textwidth, trim = 0cm 0cm 0cm 0cm, clip = true]{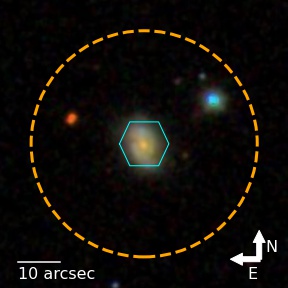} \includegraphics[width = 0.29\textwidth, trim = 0cm 0cm 0cm 0cm, clip = true]{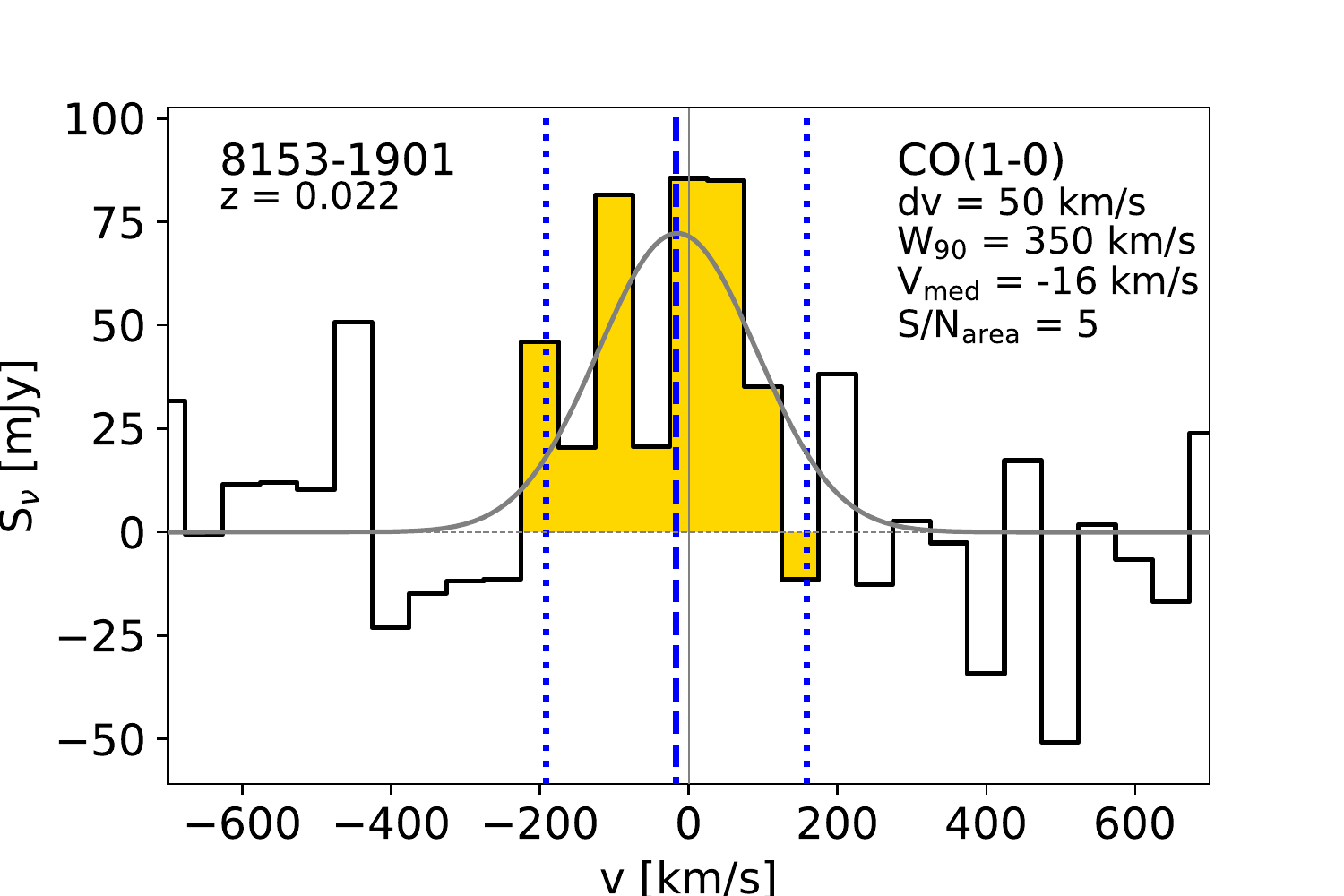}
\hspace{0.4cm}  \centering  \includegraphics[width = 0.17\textwidth, trim = 0cm 0cm 0cm 0cm, clip = true]{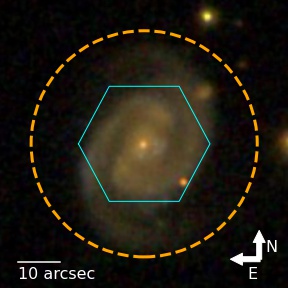} \includegraphics[width = 0.29\textwidth, trim = 0cm 0cm 0cm 0cm, clip = true]{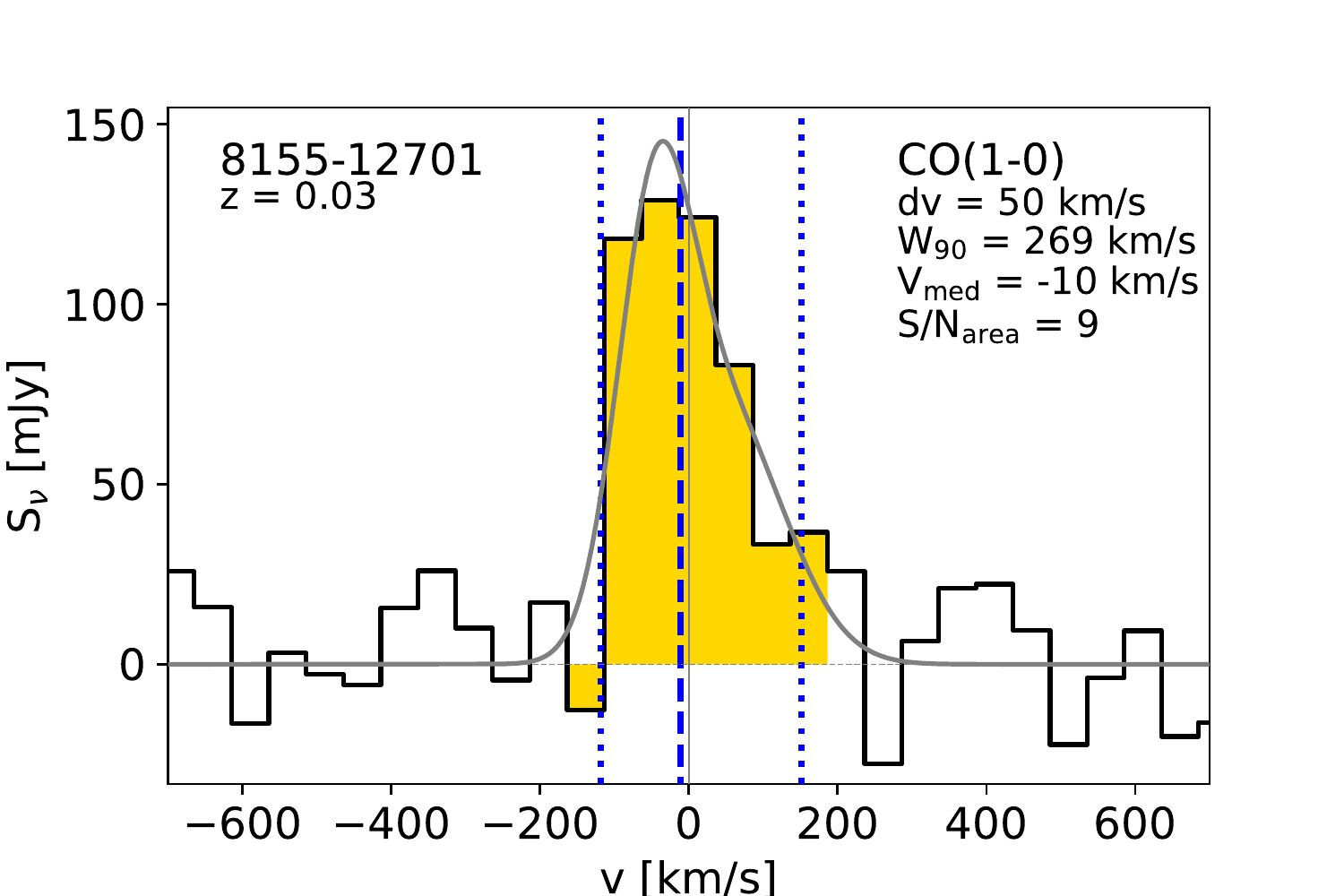}
 \caption{continued.}
\end{figure*}

\begin{figure*}    
 \centering  \includegraphics[width = 0.17\textwidth, trim = 0cm 0cm 0cm 0cm, clip = true]{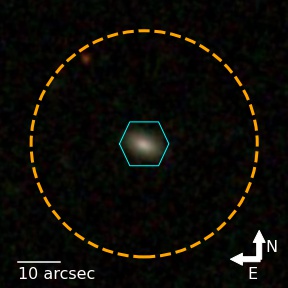} \includegraphics[width = 0.29\textwidth, trim = 0cm 0cm 0cm 0cm, clip = true]{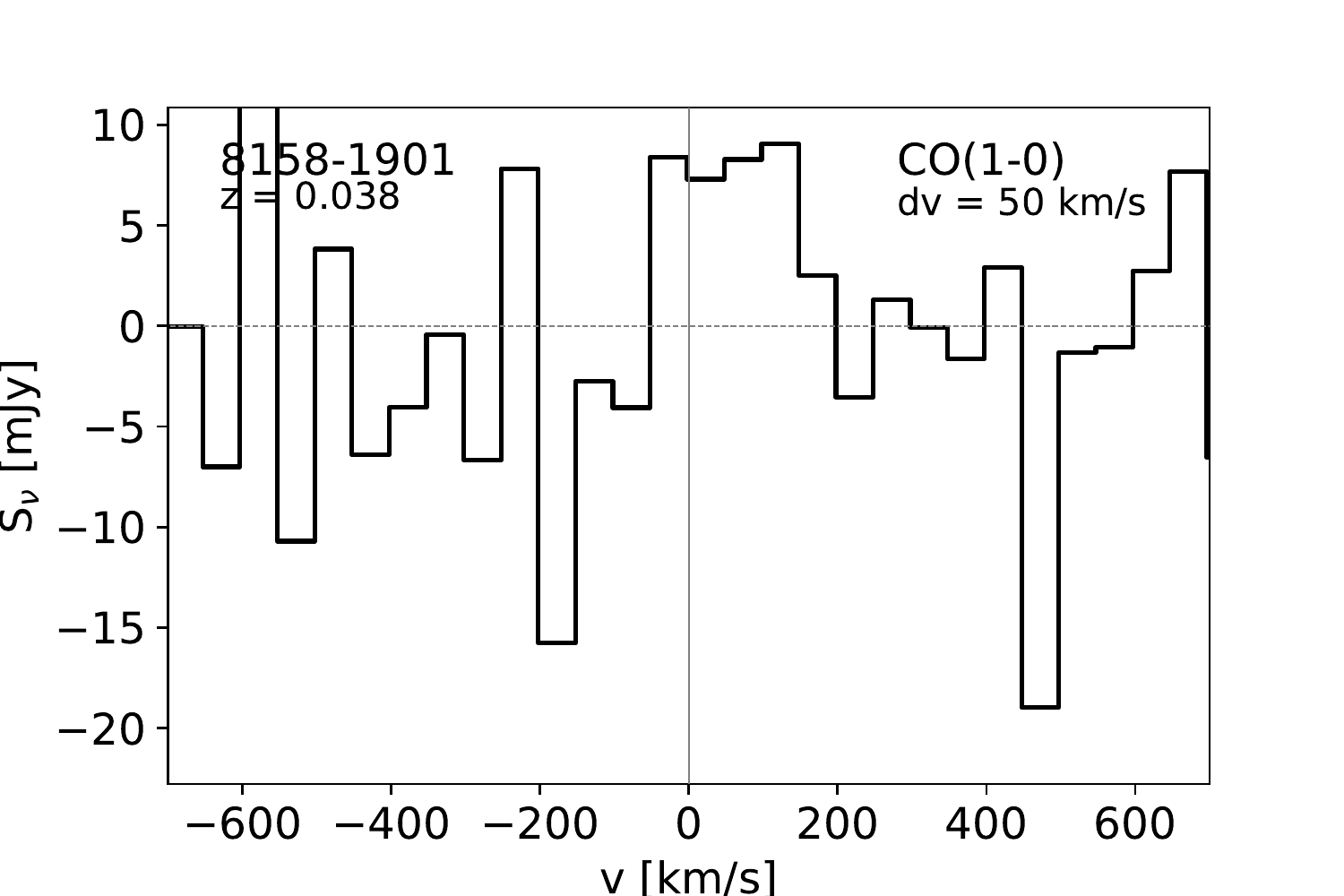}
\hspace{0.4cm}  \centering  \includegraphics[width = 0.17\textwidth, trim = 0cm 0cm 0cm 0cm, clip = true]{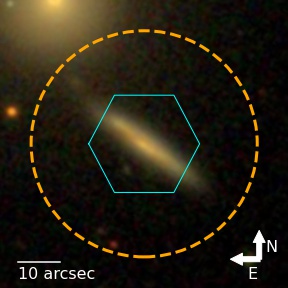} \includegraphics[width = 0.29\textwidth, trim = 0cm 0cm 0cm 0cm, clip = true]{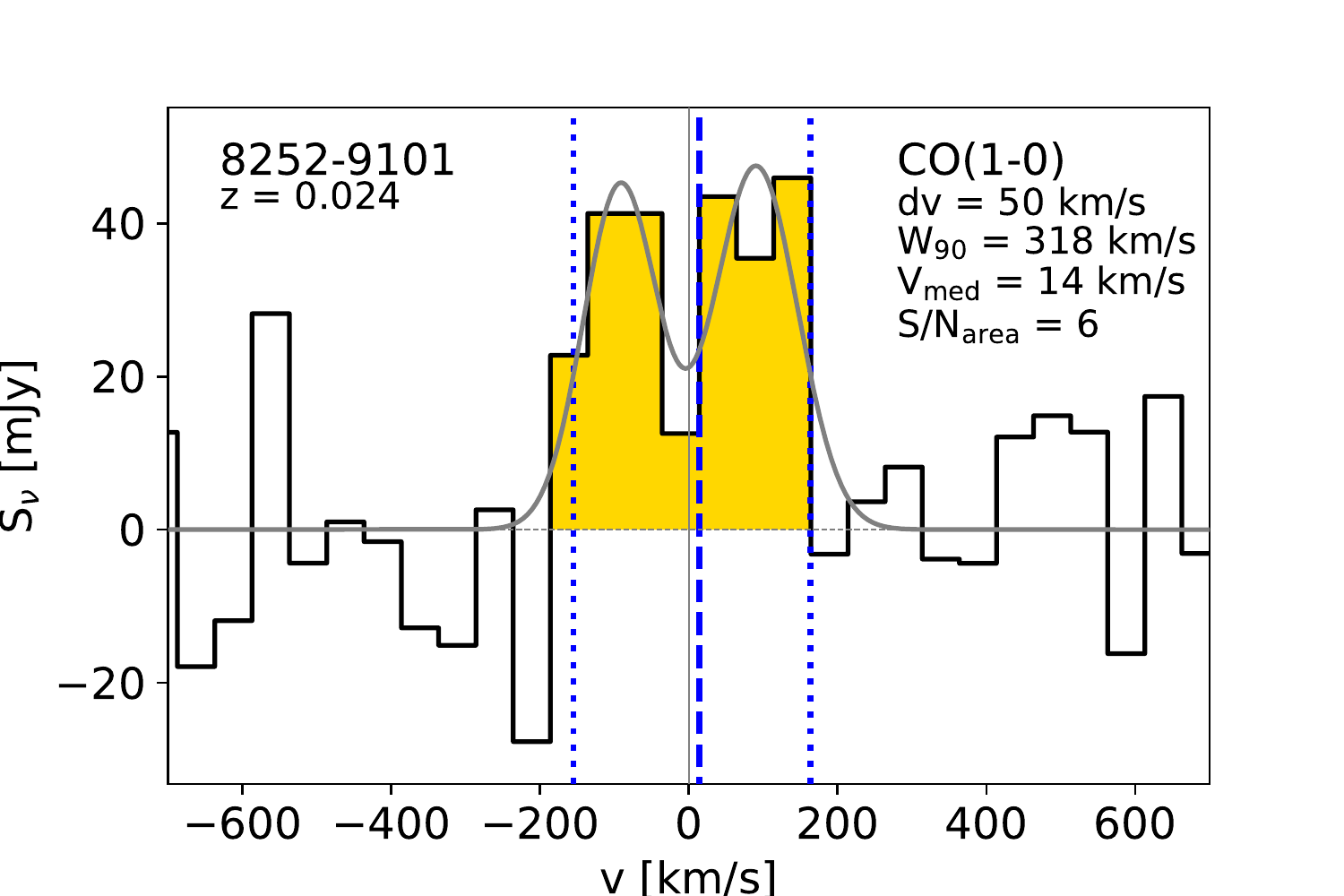}
\end{figure*}

\begin{figure*}    
 \centering  \includegraphics[width = 0.17\textwidth, trim = 0cm 0cm 0cm 0cm, clip = true]{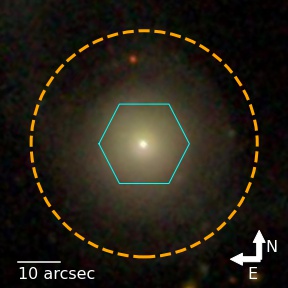} \includegraphics[width = 0.29\textwidth, trim = 0cm 0cm 0cm 0cm, clip = true]{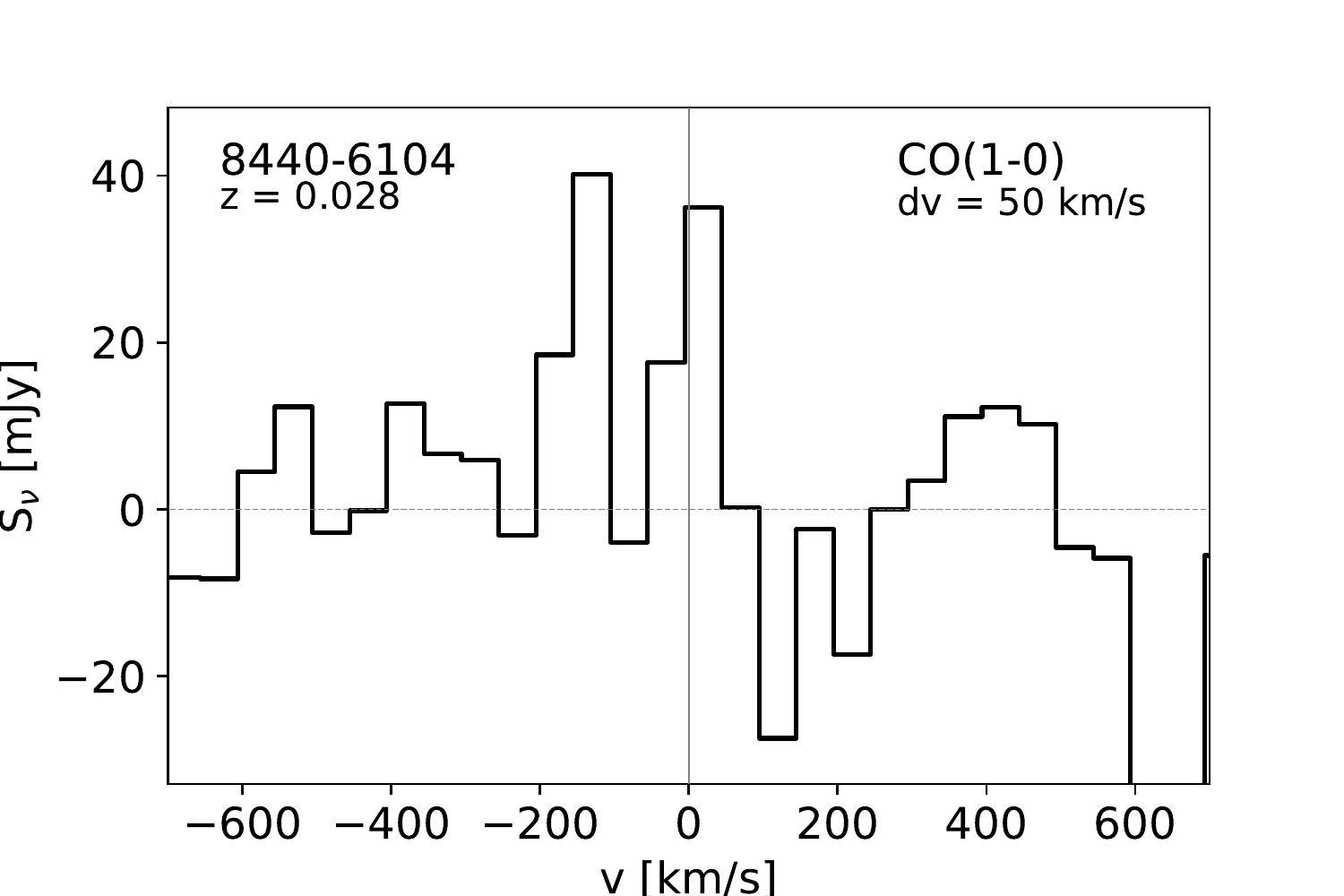}
\hspace{0.4cm}  \centering  \includegraphics[width = 0.17\textwidth, trim = 0cm 0cm 0cm 0cm, clip = true]{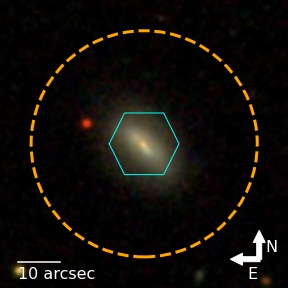} \includegraphics[width = 0.29\textwidth, trim = 0cm 0cm 0cm 0cm, clip = true]{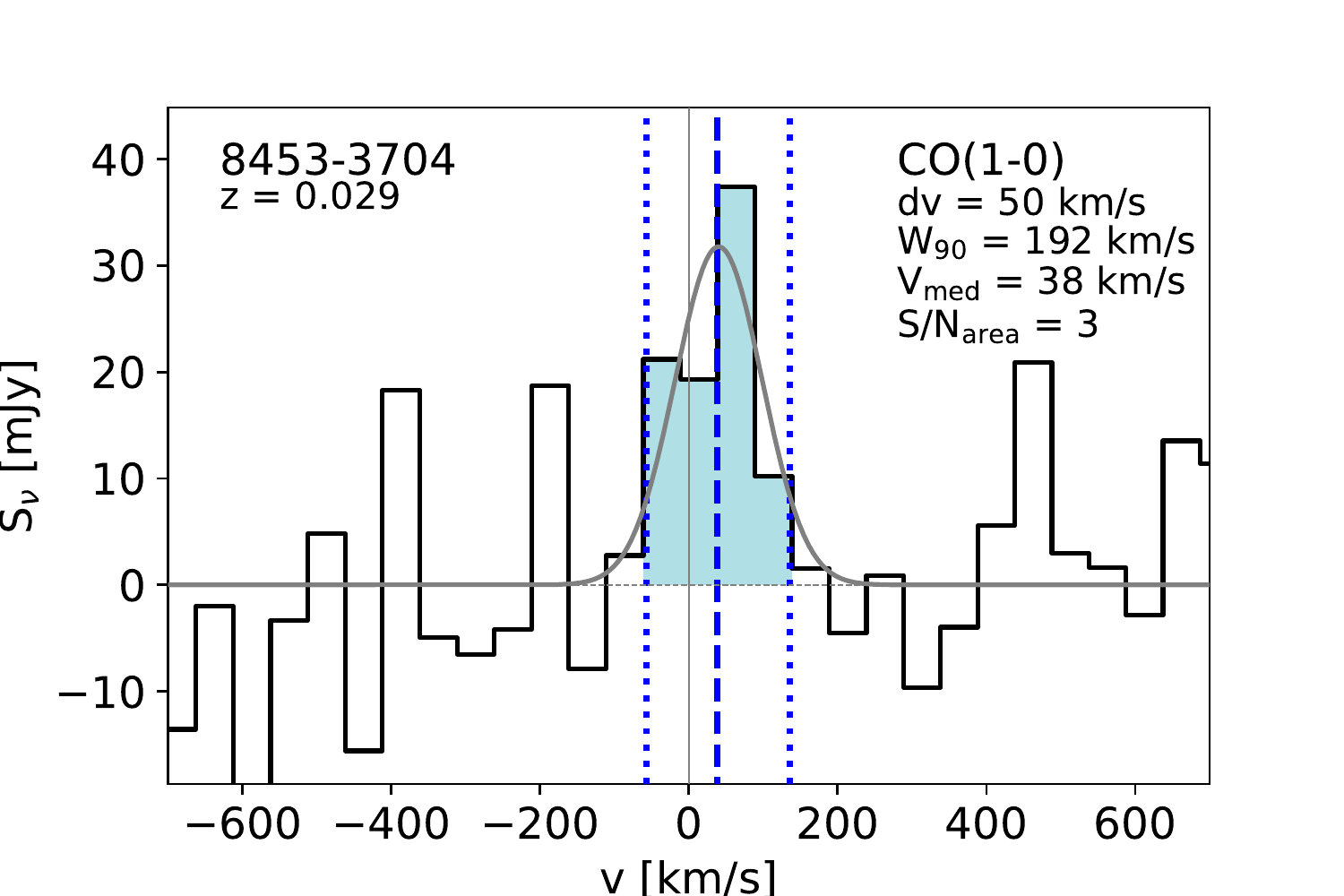}
\end{figure*}

\begin{figure*}    
 \centering  \includegraphics[width = 0.17\textwidth, trim = 0cm 0cm 0cm 0cm, clip = true]{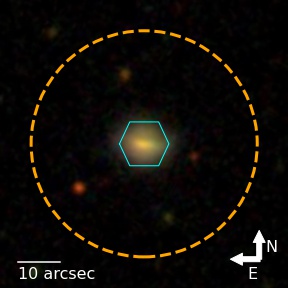} \includegraphics[width = 0.29\textwidth, trim = 0cm 0cm 0cm 0cm, clip = true]{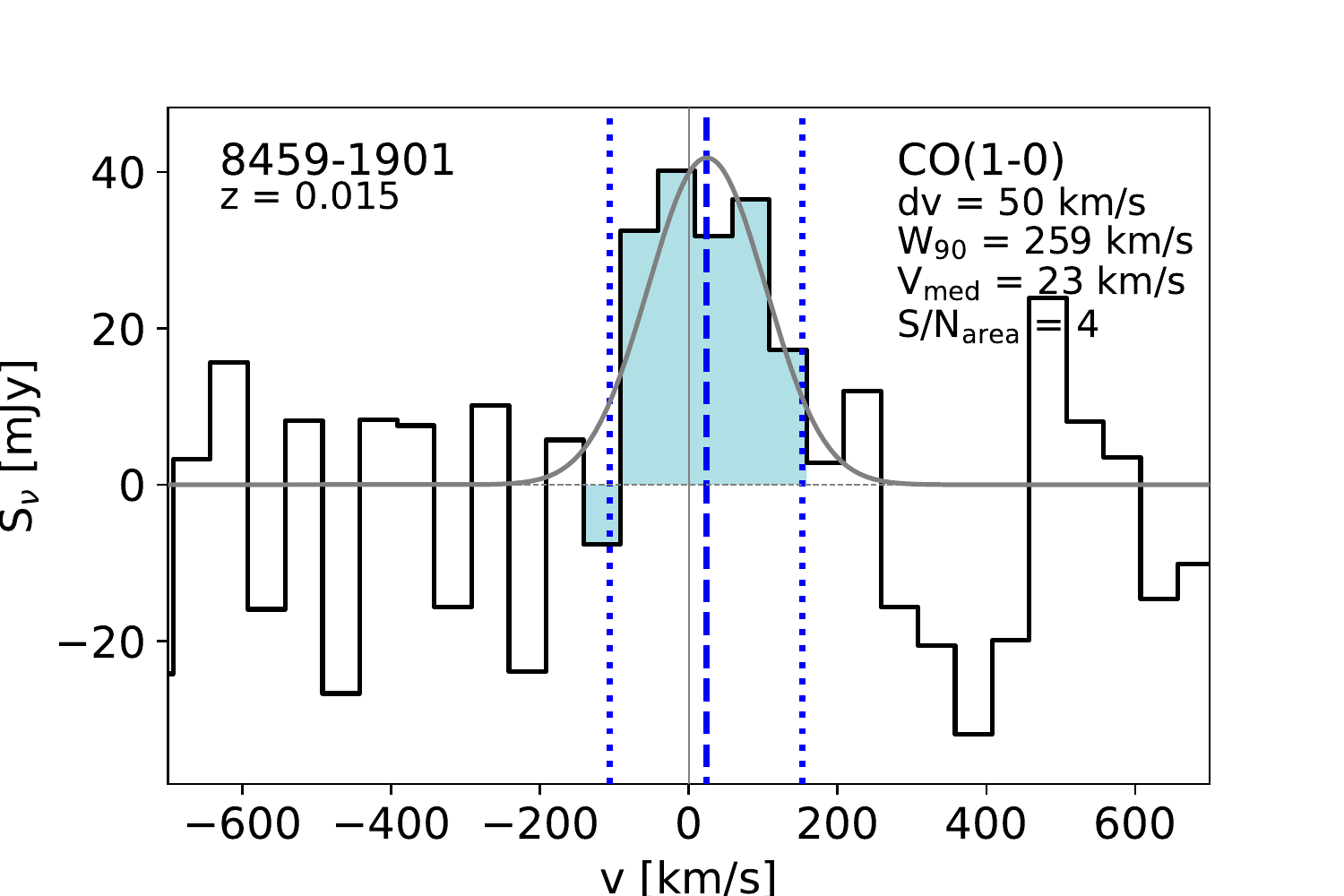}
\hspace{0.4cm}  \centering  \includegraphics[width = 0.17\textwidth, trim = 0cm 0cm 0cm 0cm, clip = true]{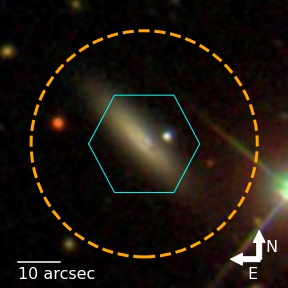} \includegraphics[width = 0.29\textwidth, trim = 0cm 0cm 0cm 0cm, clip = true]{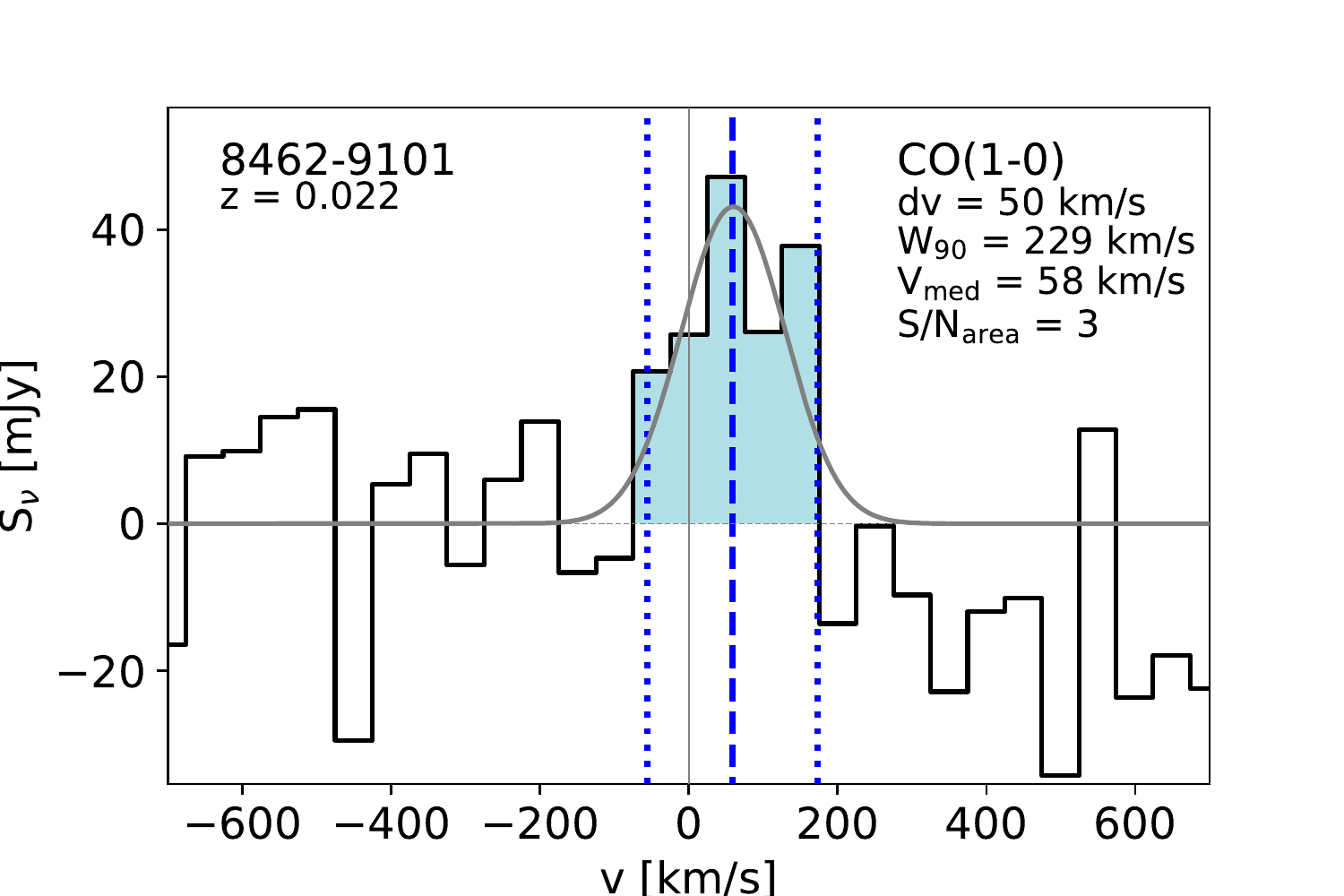}
\end{figure*}

\begin{figure*}    
 \centering  \includegraphics[width = 0.17\textwidth, trim = 0cm 0cm 0cm 0cm, clip = true]{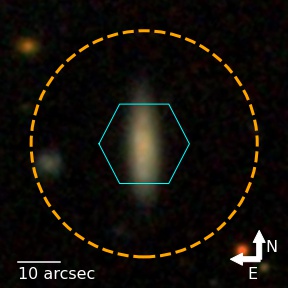} \includegraphics[width = 0.29\textwidth, trim = 0cm 0cm 0cm 0cm, clip = true]{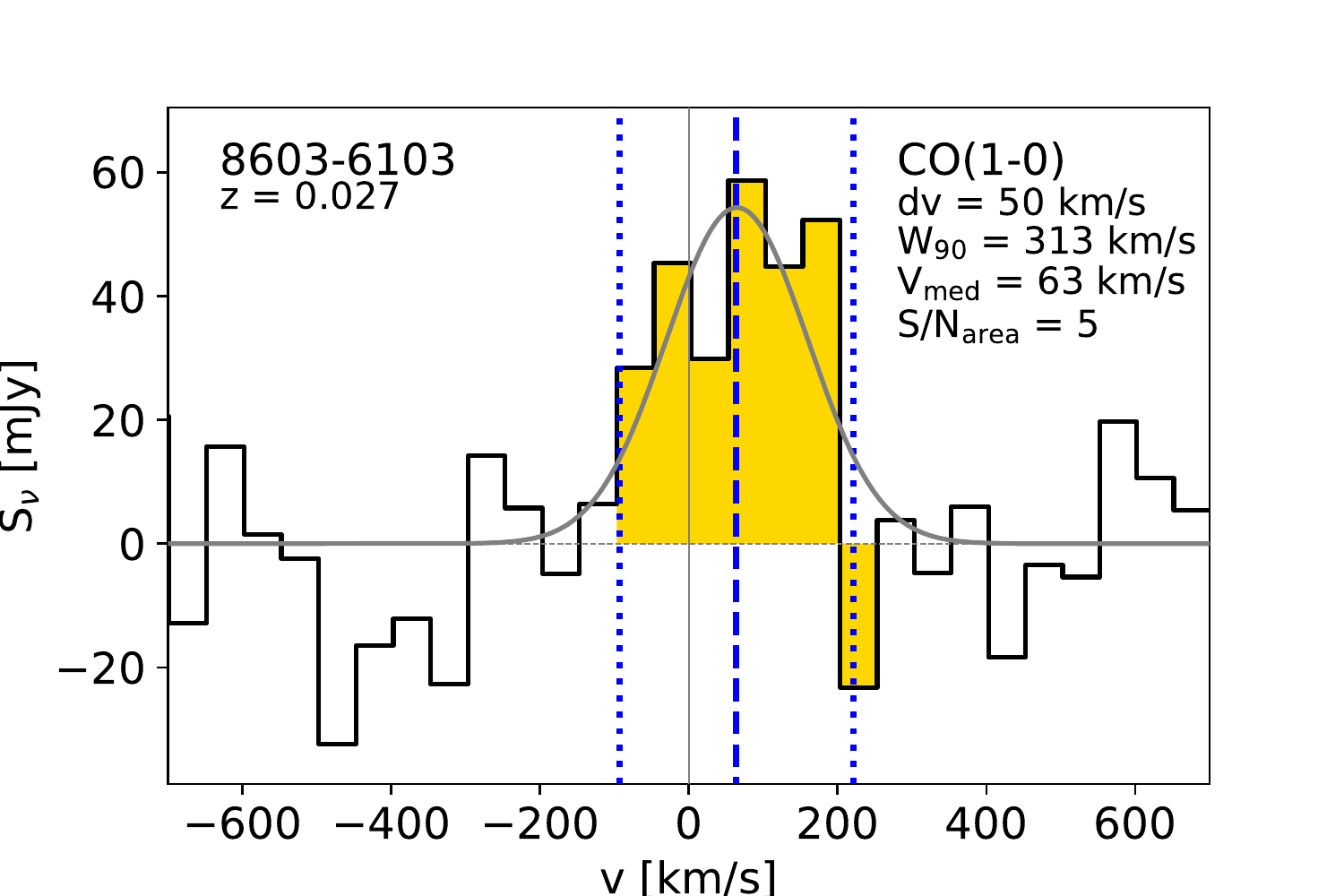}
\hspace{0.4cm}  \centering  \includegraphics[width = 0.17\textwidth, trim = 0cm 0cm 0cm 0cm, clip = true]{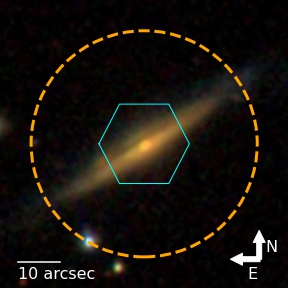} \includegraphics[width = 0.29\textwidth, trim = 0cm 0cm 0cm 0cm, clip = true]{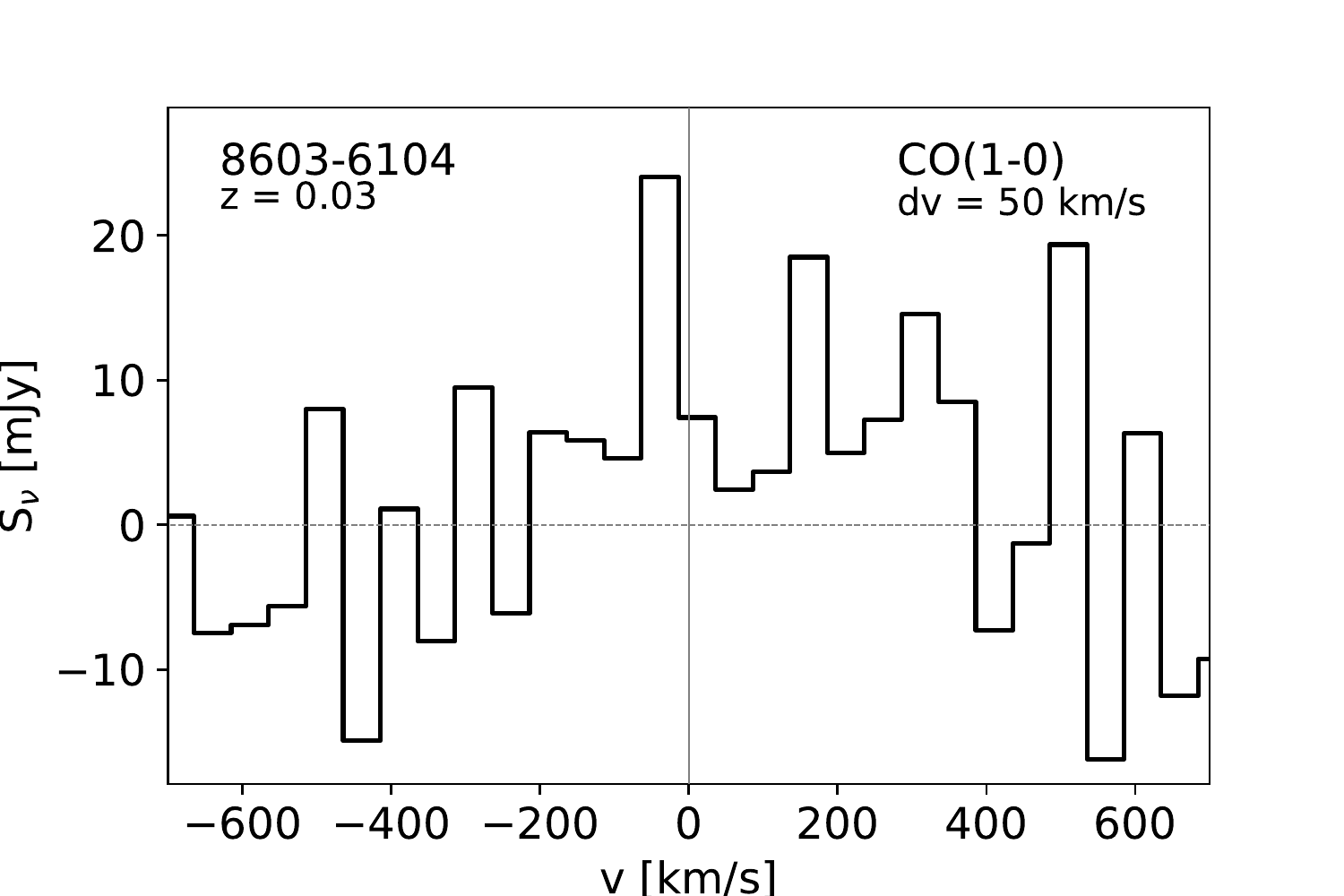}
\end{figure*}

\begin{figure*}    
 \centering  \includegraphics[width = 0.17\textwidth, trim = 0cm 0cm 0cm 0cm, clip = true]{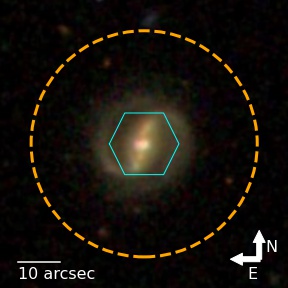} \includegraphics[width = 0.29\textwidth, trim = 0cm 0cm 0cm 0cm, clip = true]{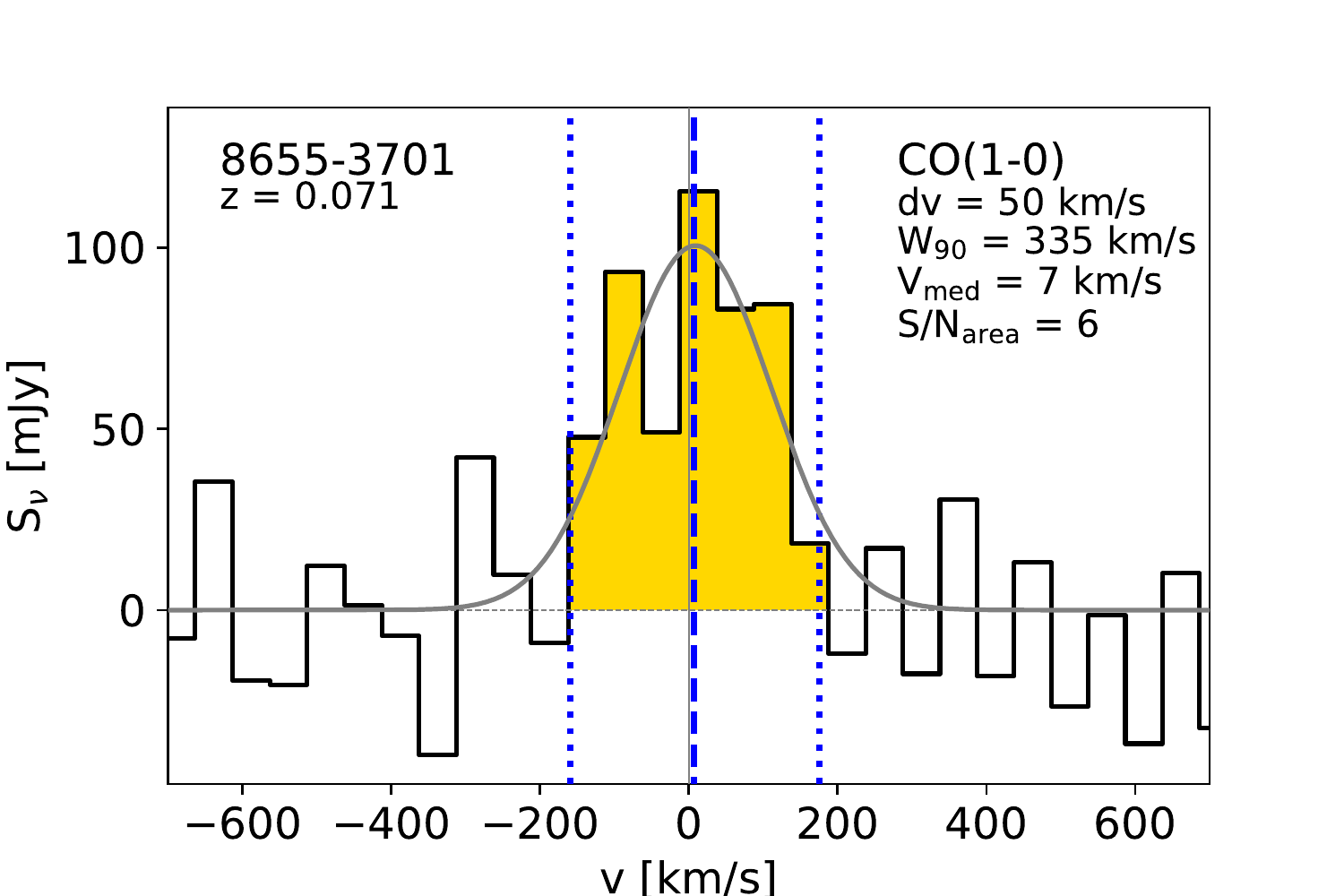}
\hspace{0.4cm}  \centering  \includegraphics[width = 0.17\textwidth, trim = 0cm 0cm 0cm 0cm, clip = true]{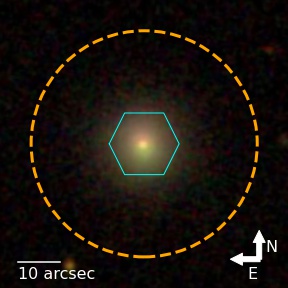} \includegraphics[width = 0.29\textwidth, trim = 0cm 0cm 0cm 0cm, clip = true]{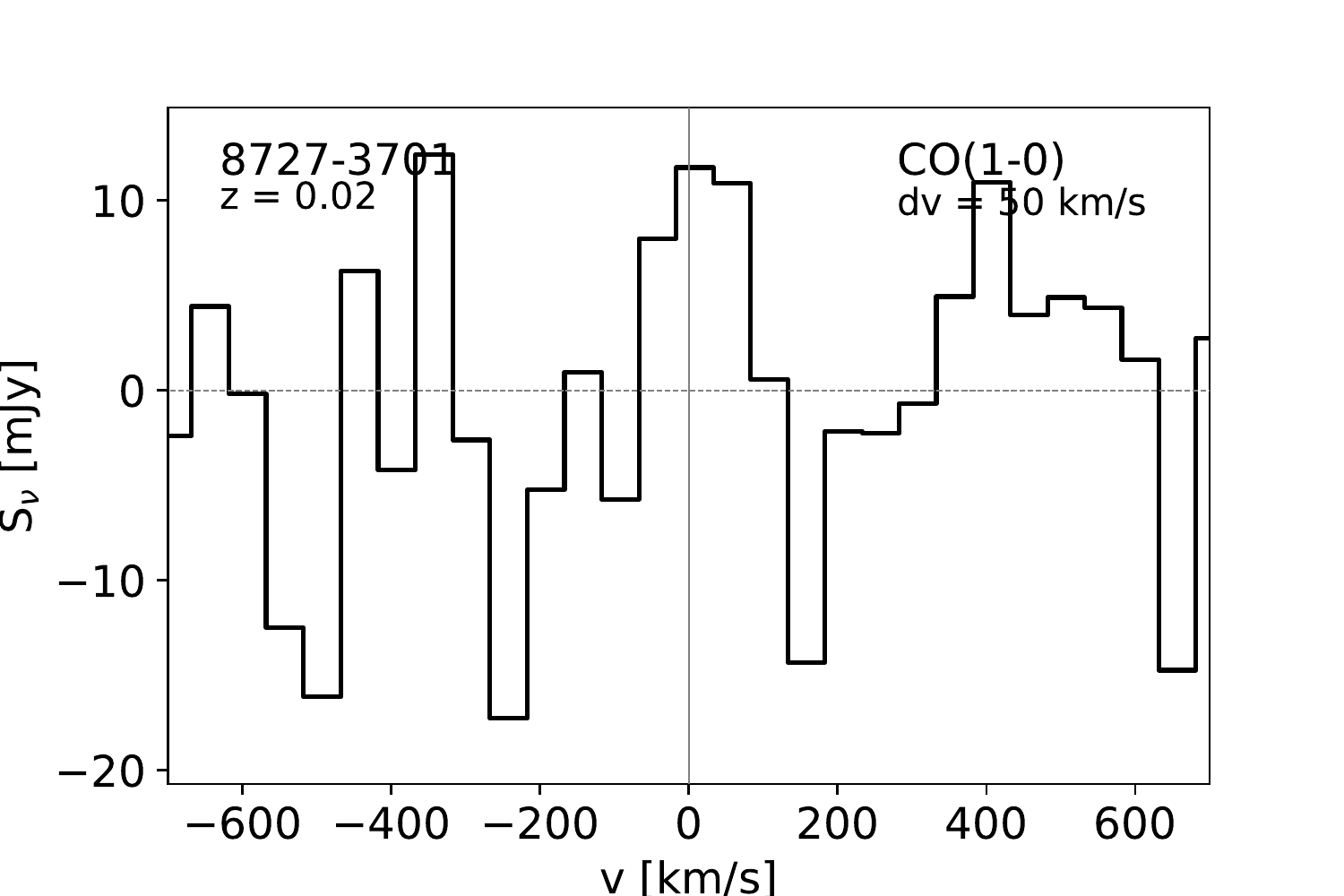}
\end{figure*}

\begin{figure*}   
   \ContinuedFloat  
 \centering  \includegraphics[width = 0.17\textwidth, trim = 0cm 0cm 0cm 0cm, clip = true]{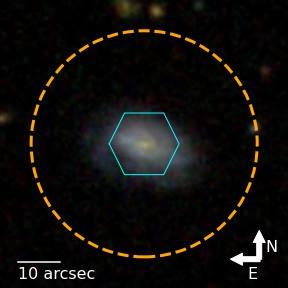} \includegraphics[width = 0.29\textwidth, trim = 0cm 0cm 0cm 0cm, clip = true]{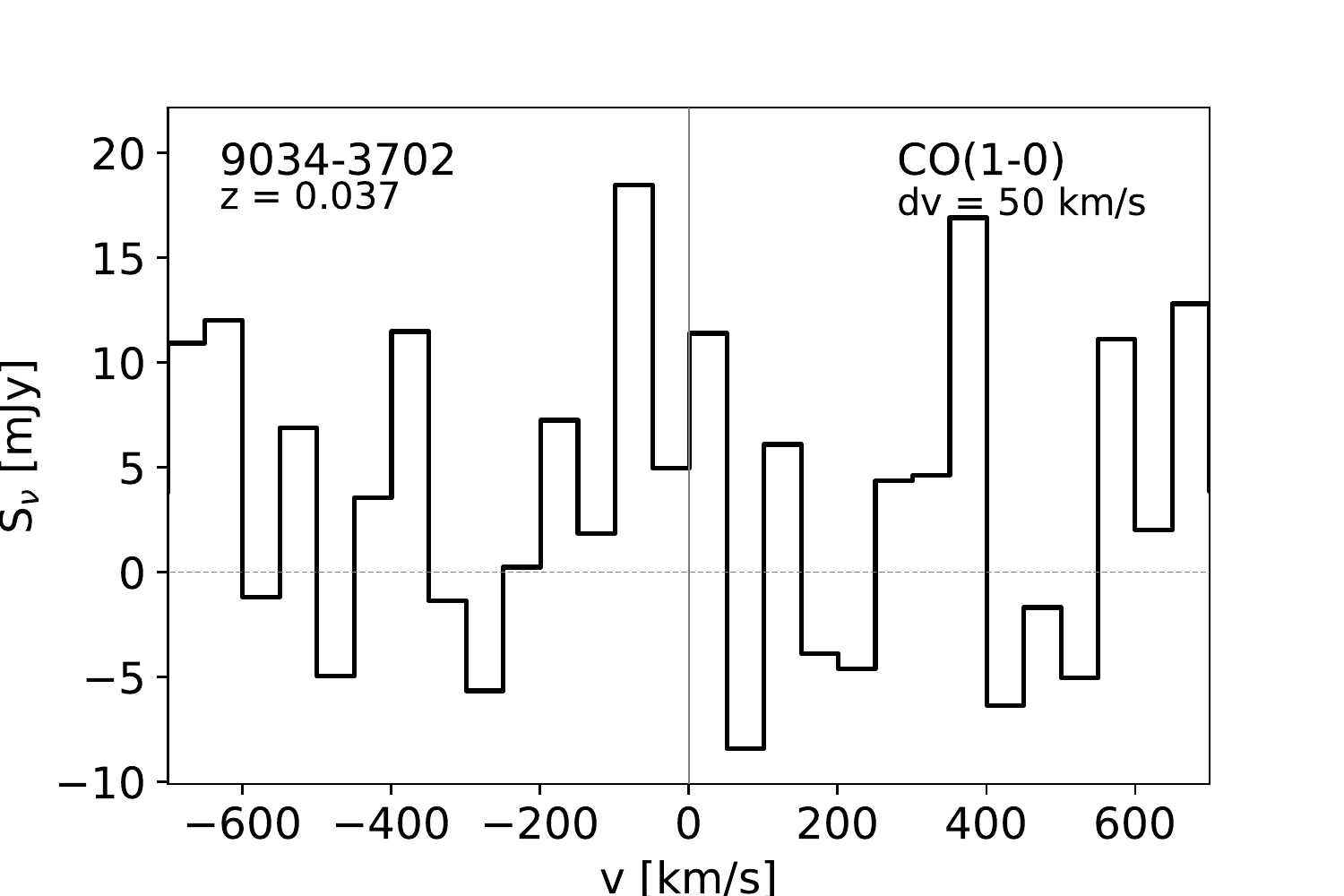}
\hspace{0.4cm}  \centering  \includegraphics[width = 0.17\textwidth, trim = 0cm 0cm 0cm 0cm, clip = true]{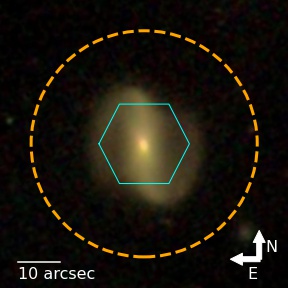} \includegraphics[width = 0.29\textwidth, trim = 0cm 0cm 0cm 0cm, clip = true]{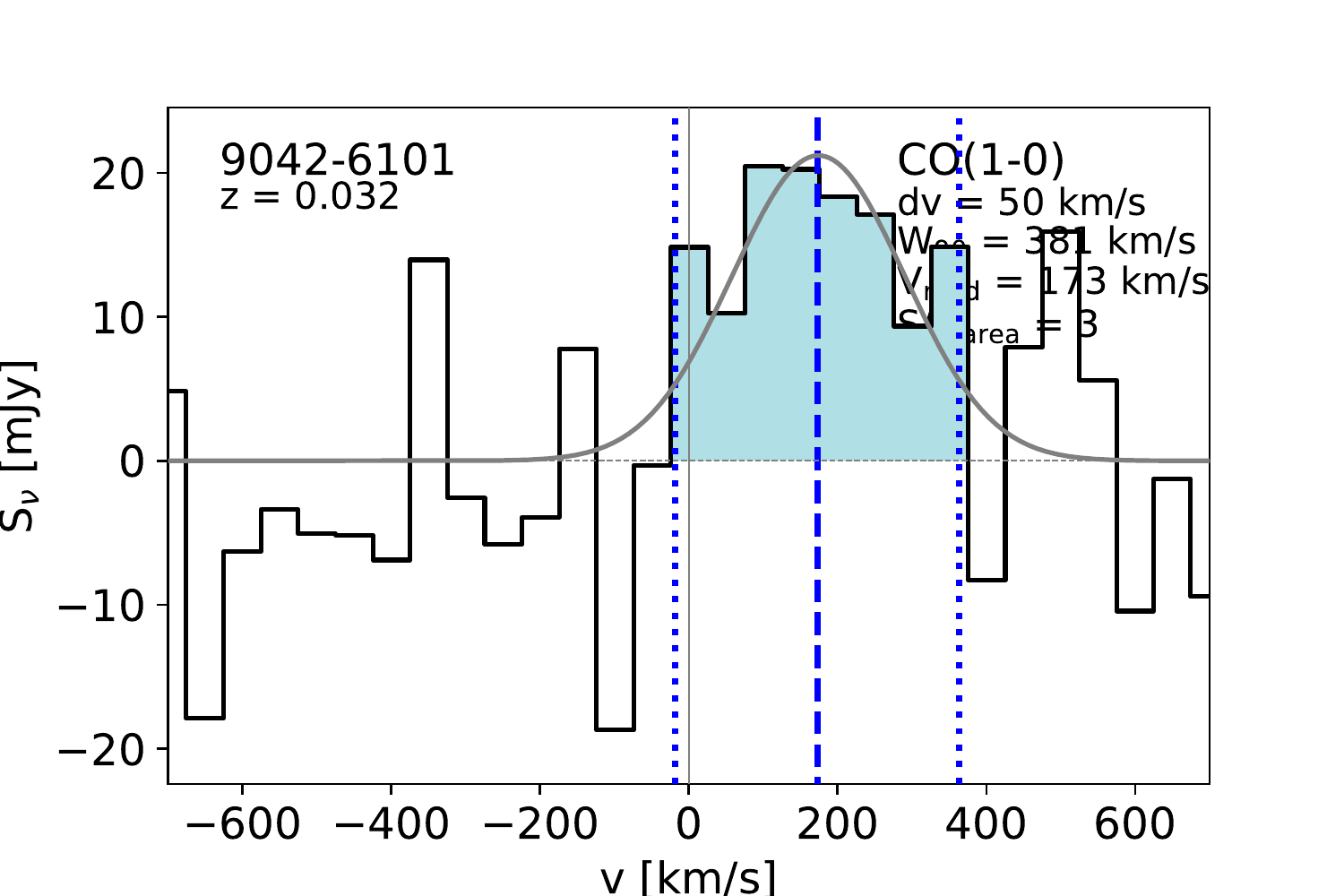}
 \caption{continued.}
\end{figure*}

\begin{figure*}    
 \centering  \includegraphics[width = 0.17\textwidth, trim = 0cm 0cm 0cm 0cm, clip = true]{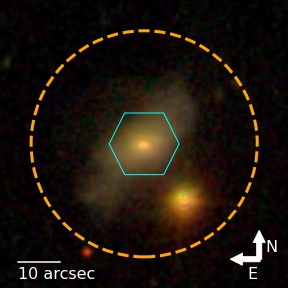} \includegraphics[width = 0.29\textwidth, trim = 0cm 0cm 0cm 0cm, clip = true]{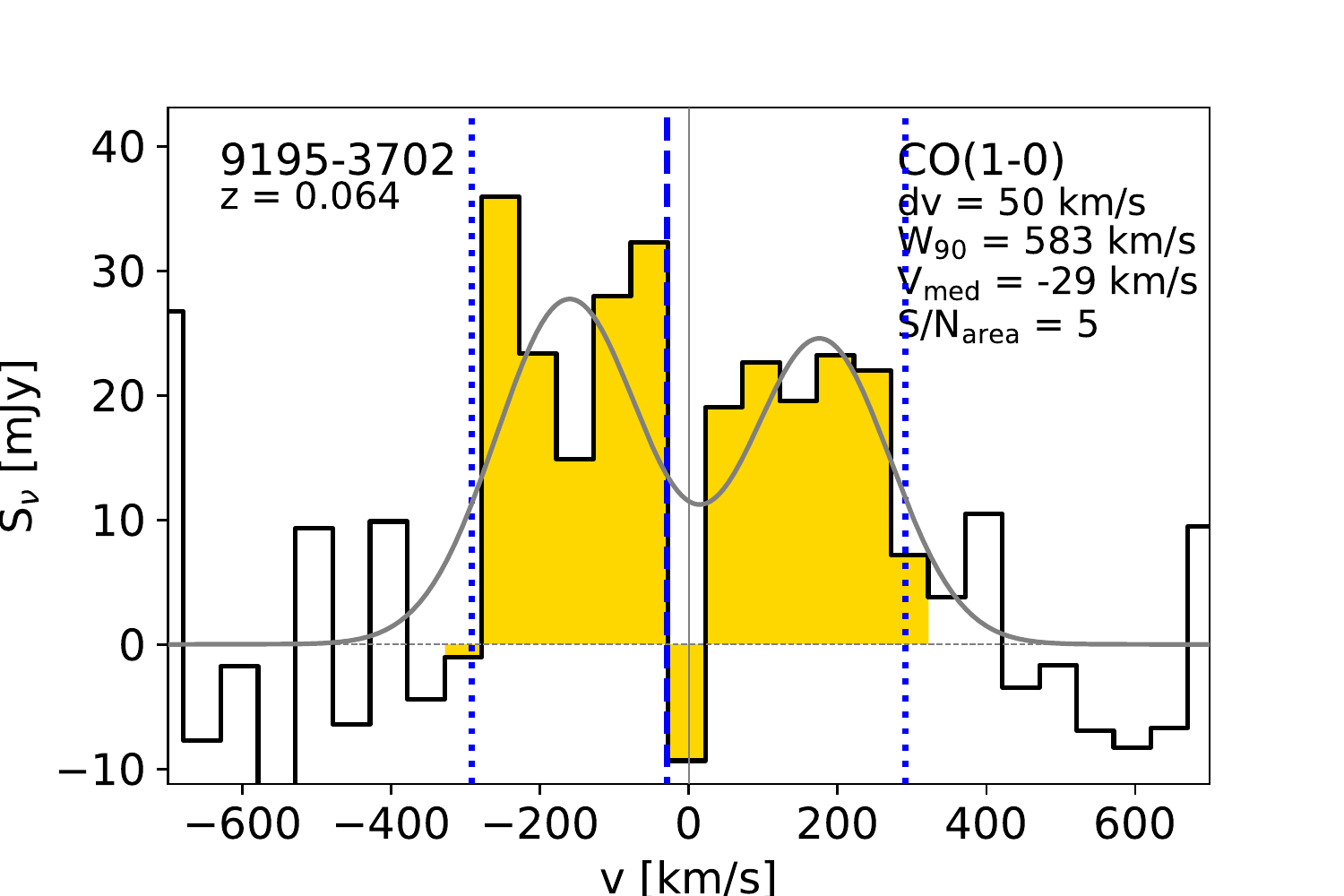}
\hspace{0.4cm}  \centering  \includegraphics[width = 0.17\textwidth, trim = 0cm 0cm 0cm 0cm, clip = true]{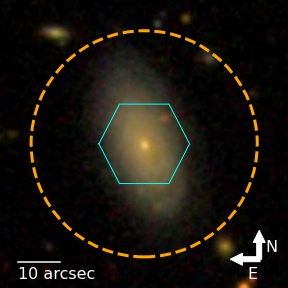} \includegraphics[width = 0.29\textwidth, trim = 0cm 0cm 0cm 0cm, clip = true]{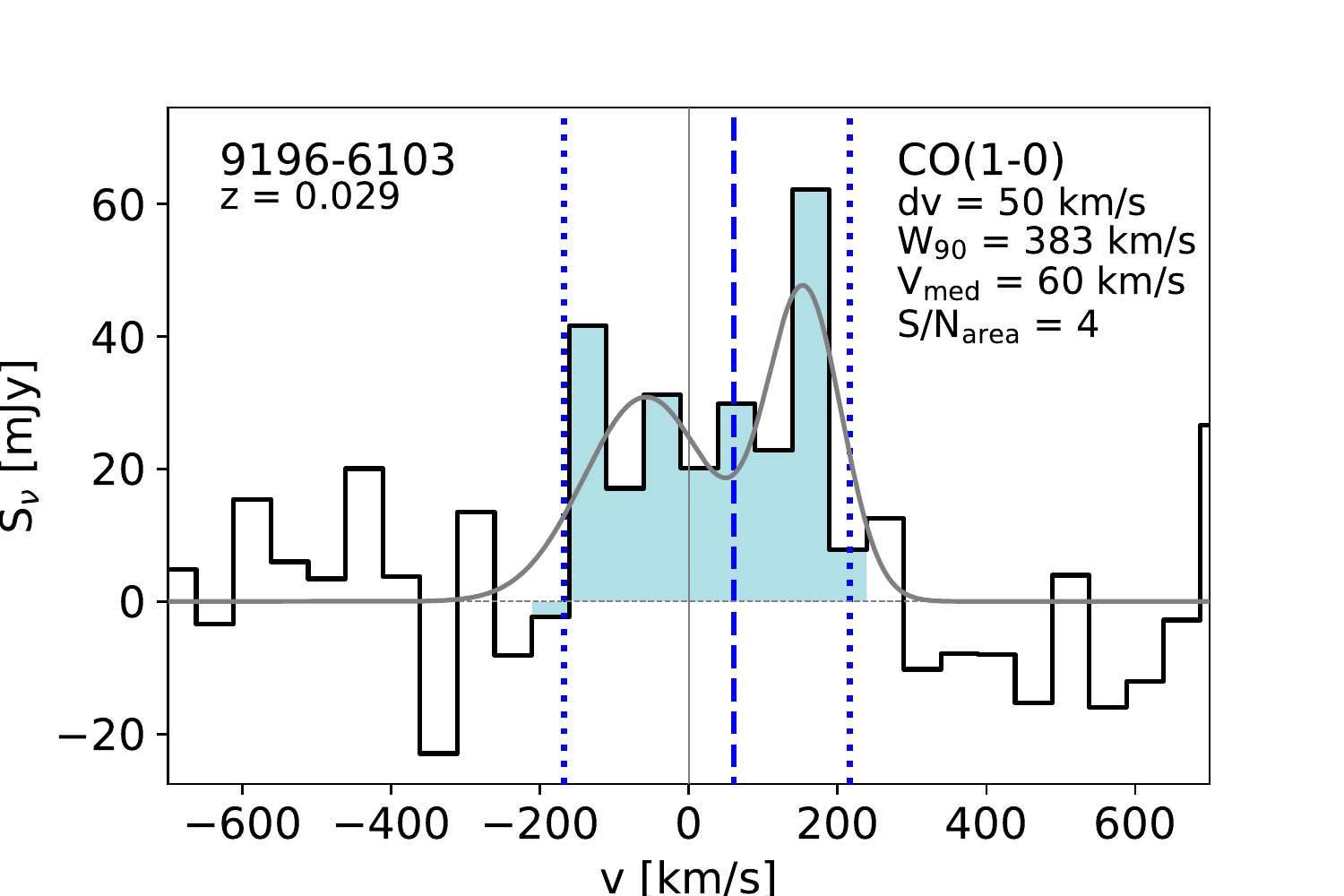}
\end{figure*}

\begin{figure*}    
 \centering  \includegraphics[width = 0.17\textwidth, trim = 0cm 0cm 0cm 0cm, clip = true]{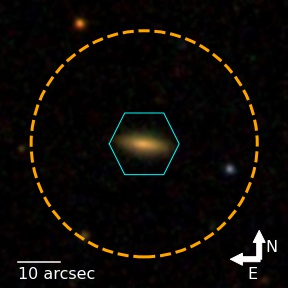} \includegraphics[width = 0.29\textwidth, trim = 0cm 0cm 0cm 0cm, clip = true]{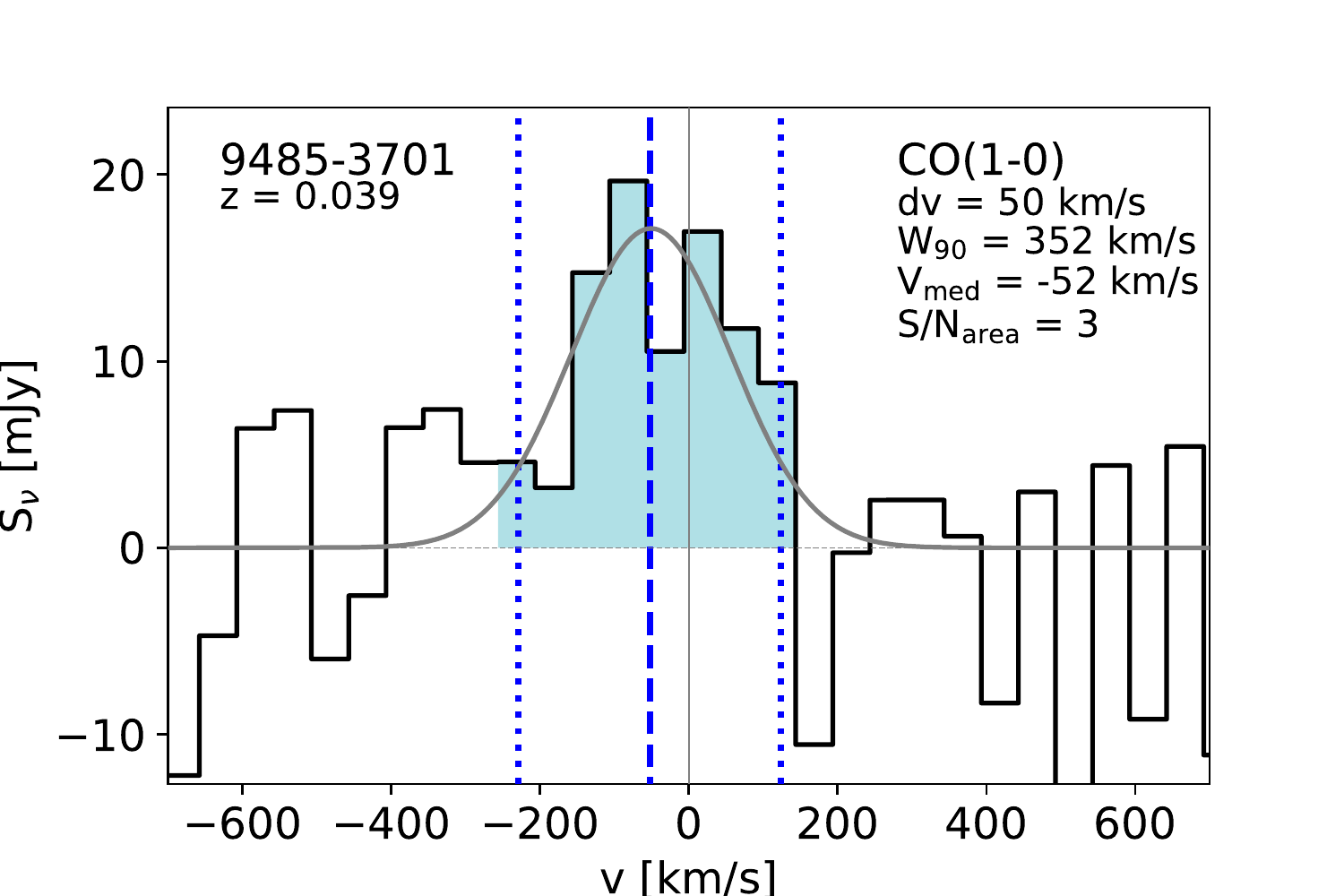}
\hspace{0.4cm}  \centering  \includegraphics[width = 0.17\textwidth, trim = 0cm 0cm 0cm 0cm, clip = true]{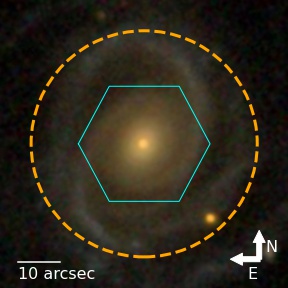} \includegraphics[width = 0.29\textwidth, trim = 0cm 0cm 0cm 0cm, clip = true]{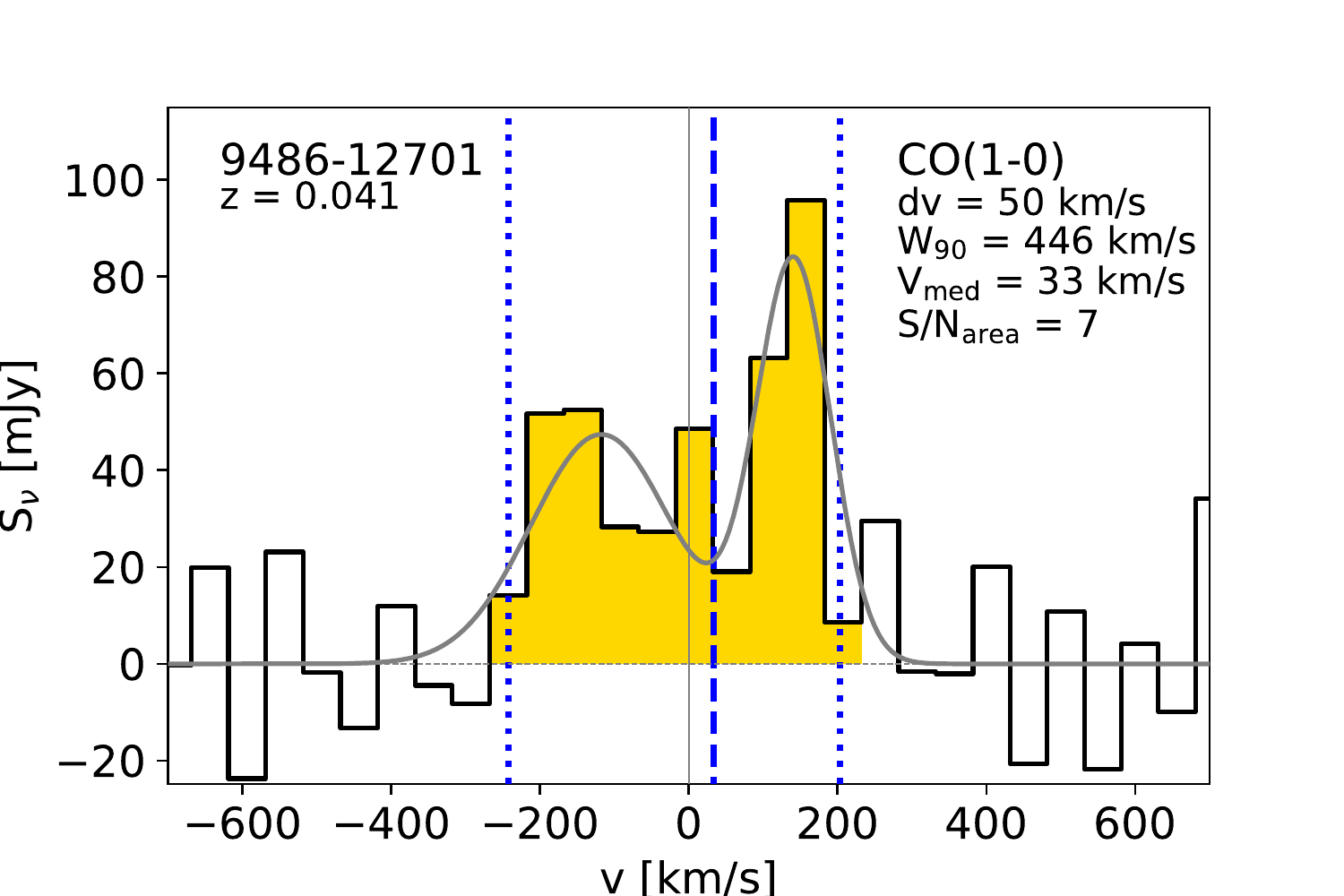}
\end{figure*}

\begin{figure*}    
 \centering  \includegraphics[width = 0.17\textwidth, trim = 0cm 0cm 0cm 0cm, clip = true]{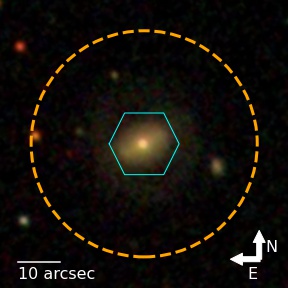} \includegraphics[width = 0.29\textwidth, trim = 0cm 0cm 0cm 0cm, clip = true]{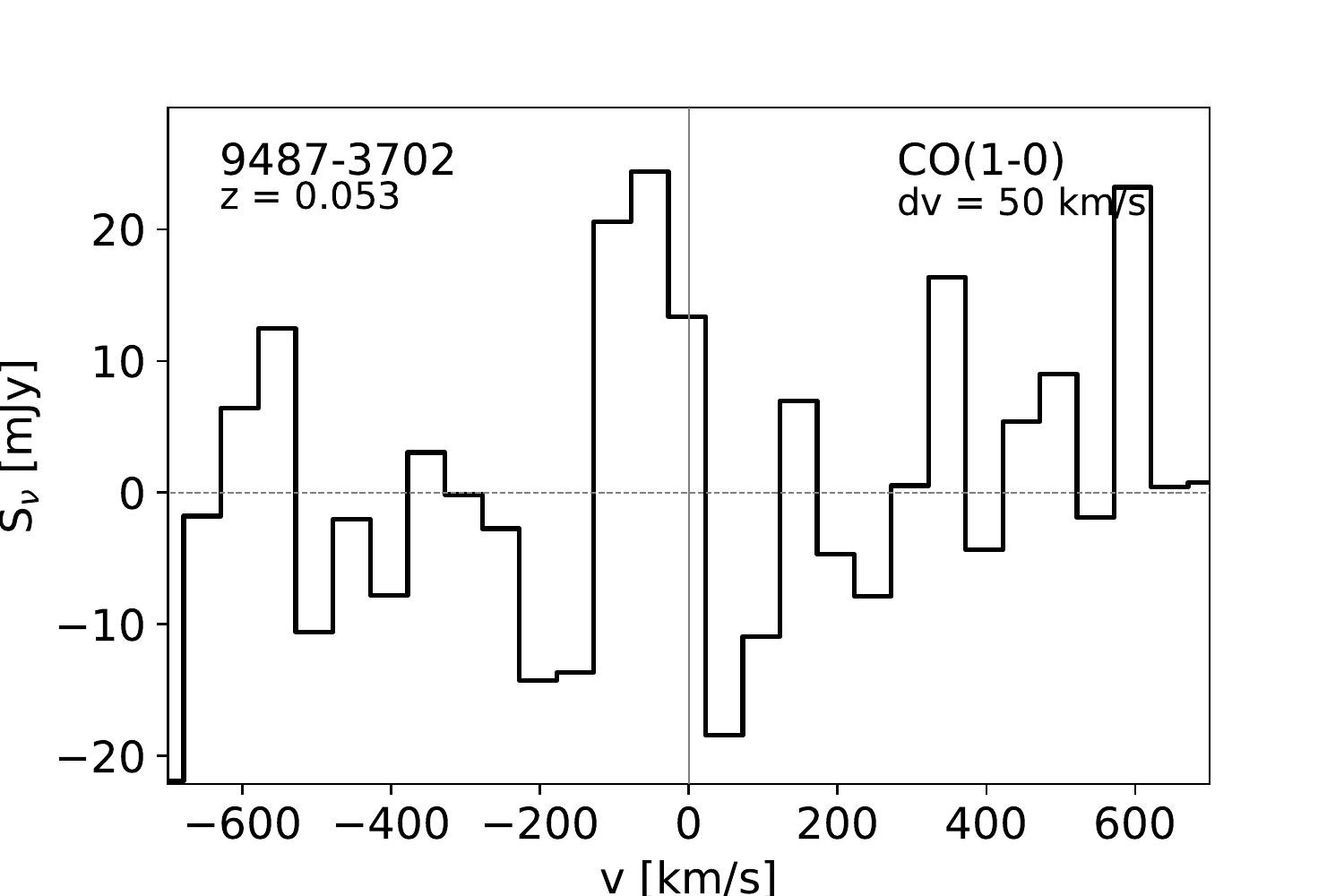}
\hspace{0.4cm}  \centering  \includegraphics[width = 0.17\textwidth, trim = 0cm 0cm 0cm 0cm, clip = true]{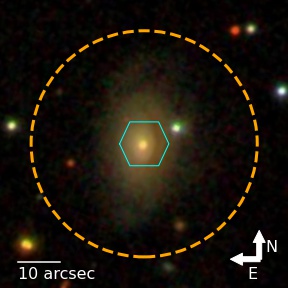} \includegraphics[width = 0.29\textwidth, trim = 0cm 0cm 0cm 0cm, clip = true]{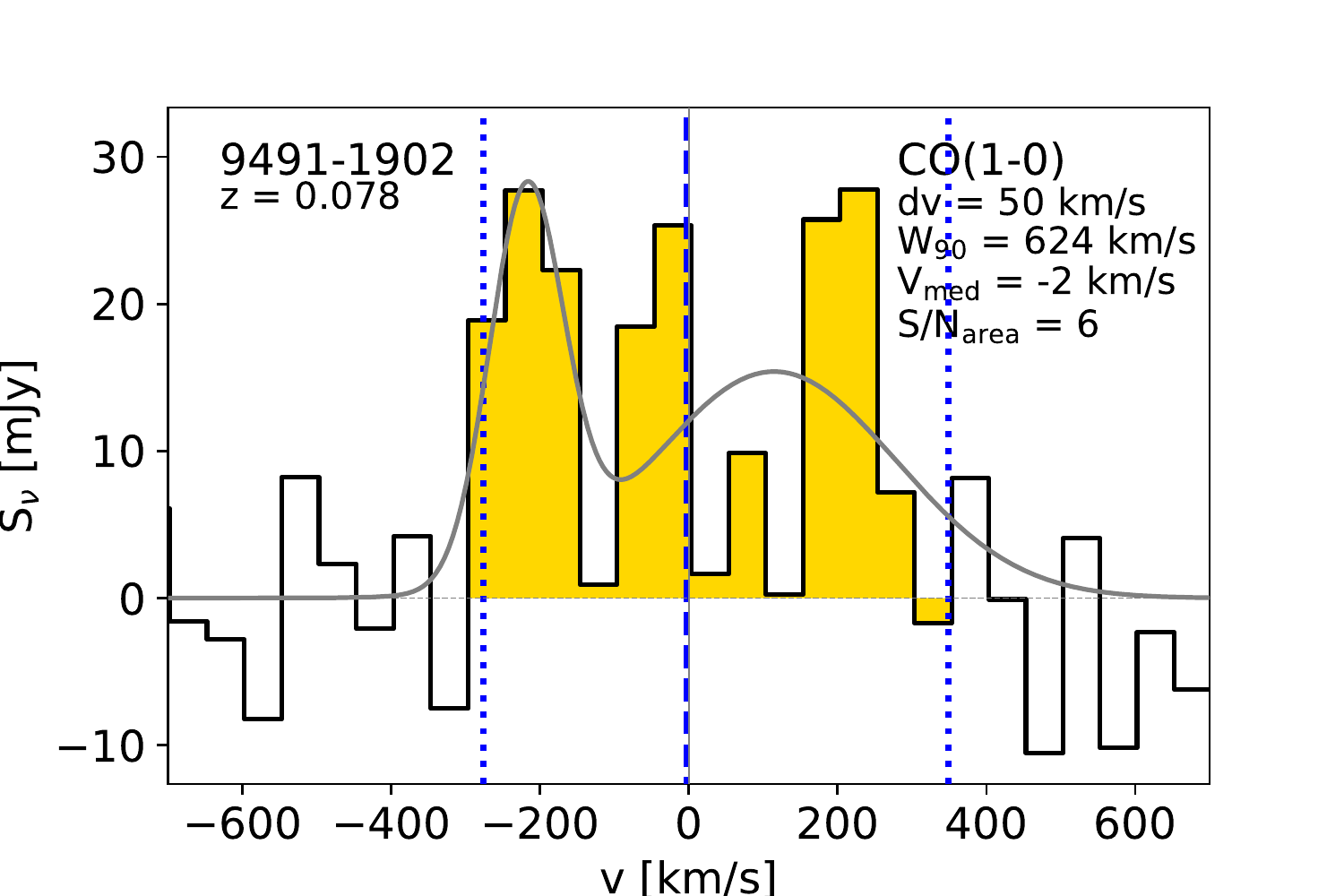}
\end{figure*}

\begin{figure*}    
 \centering  \includegraphics[width = 0.17\textwidth, trim = 0cm 0cm 0cm 0cm, clip = true]{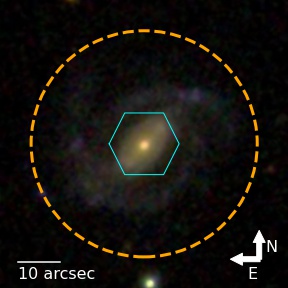} \includegraphics[width = 0.29\textwidth, trim = 0cm 0cm 0cm 0cm, clip = true]{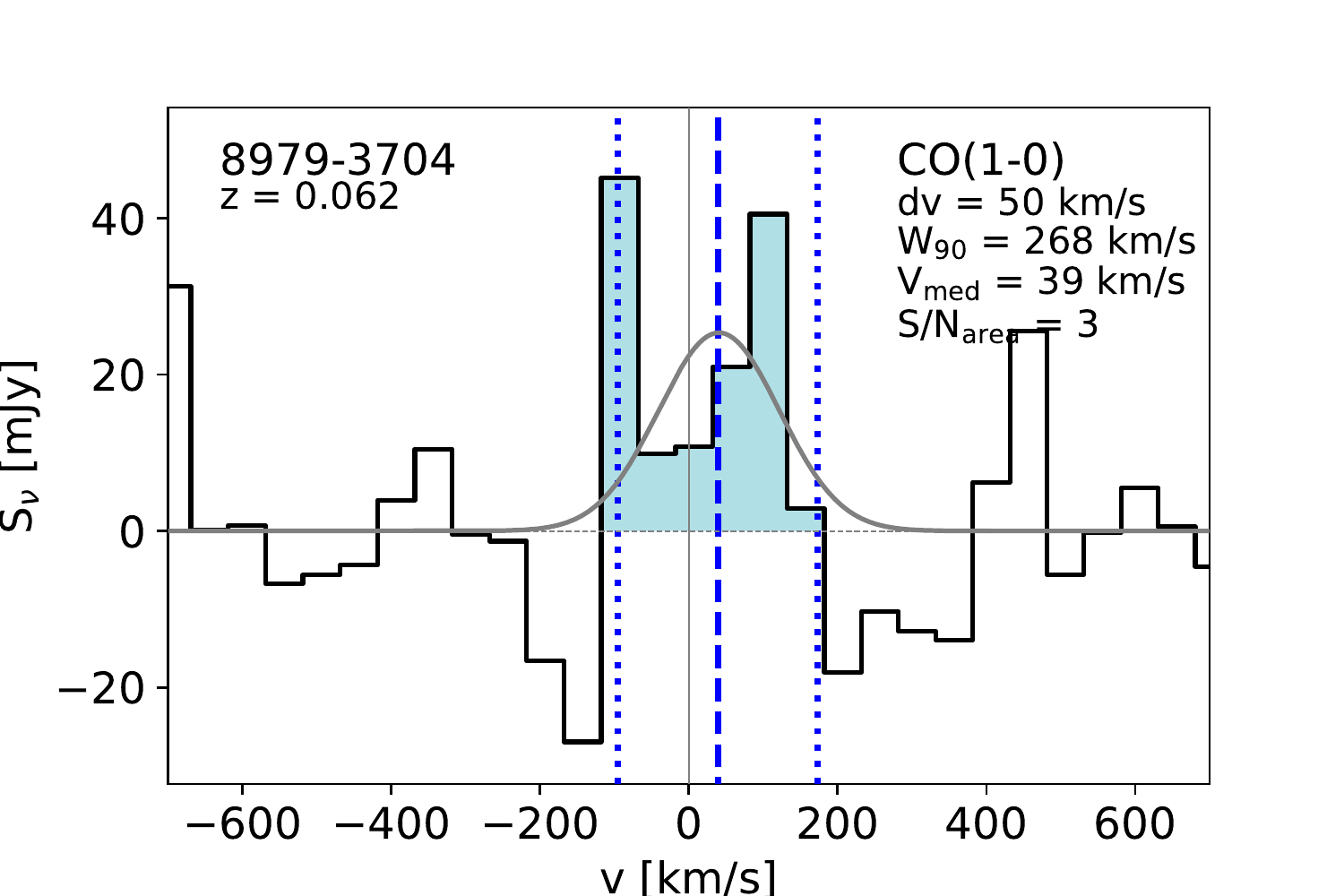}
\hspace{0.4cm}  \centering  \includegraphics[width = 0.17\textwidth, trim = 0cm 0cm 0cm 0cm, clip = true]{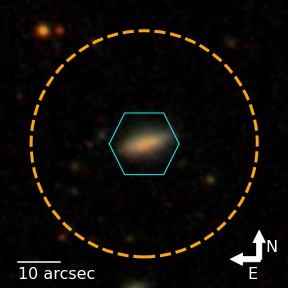} \includegraphics[width = 0.29\textwidth, trim = 0cm 0cm 0cm 0cm, clip = true]{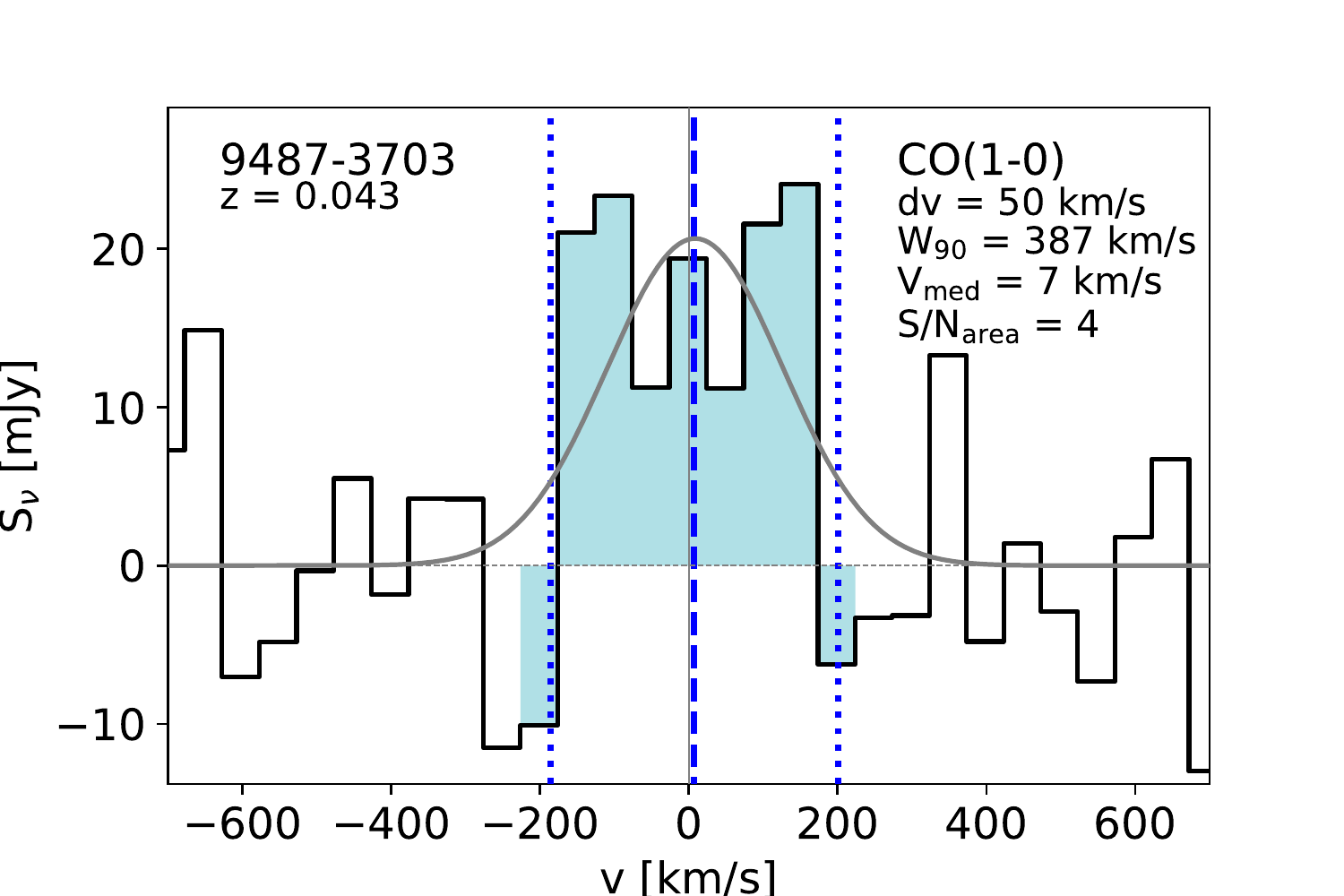}
\end{figure*}

\begin{figure*}    
   \ContinuedFloat  
 \centering  \includegraphics[width = 0.17\textwidth, trim = 0cm 0cm 0cm 0cm, clip = true]{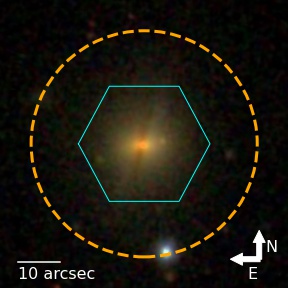} \includegraphics[width = 0.29\textwidth, trim = 0cm 0cm 0cm 0cm, clip = true]{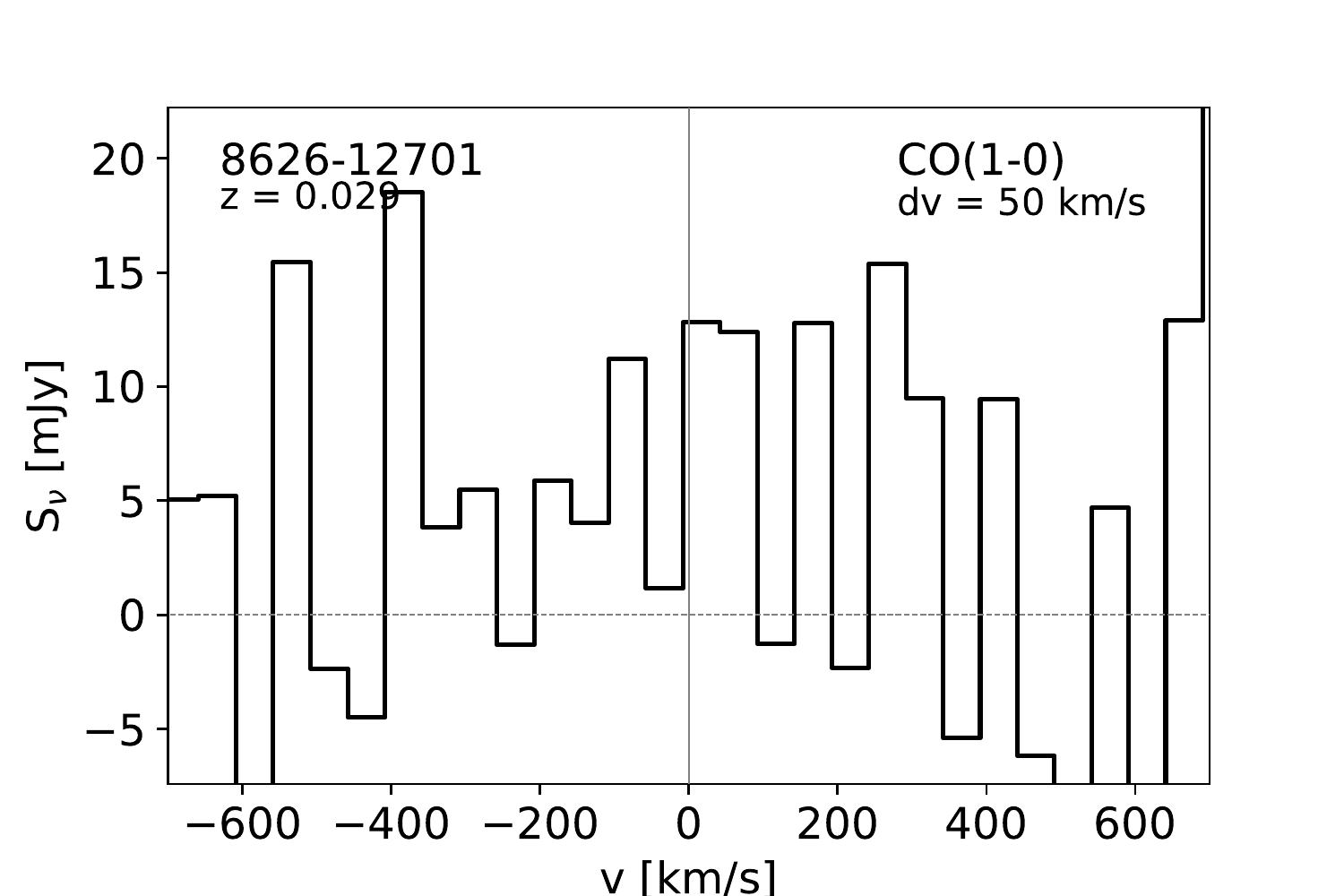}
  \caption{continued.}
\end{figure*}

\clearpage
\label{lastpage}
\end{document}